\documentclass[11pt,a4paper]{article}
\usepackage{hyperref}
\usepackage{enumitem}
\usepackage{tikz-cd}
\usepackage[utf8]{inputenc}
\usepackage{graphicx}  \usepackage{mathptmx}
\usepackage{amsmath, amssymb,lgreek}
\numberwithin{equation}{section}
\usepackage{gensymb}

\newcommand{\id}[1]{\ensuremath{\mathrm{id}}}

\newcommand{\Q}{\mathbb{Q}}

\newcommand{\quar}{\mbox{\footnotesize $\frac{1}{4}$}}
\newcommand{\half}{\mbox{\footnotesize $\frac{1}{2}$}}

\newcommand{\qm}{quantum mechanics}
\newcommand{\er}{\eqref}
 
\newcommand{\beq}{\begin{equation}}
\newcommand{\eeq}{\end{equation}} 
\newcommand{\bea}{\begin{eqnarray}}
\newcommand{\eea}{\end{eqnarray}} \newcommand{\nn}{\nonumber}

\newcommand{\ovl}{\overline}

\newcommand{\wed}{\wedge}
 \newcommand{\til}{\tilde}
\newcommand{\raw}{\rightarrow}

 \newcommand{\Raw}{\Rightarrow}

\newcommand{\lraw}{\leftrightarrow}
\newcommand{\LRaw}{\Leftrightarrow}

\newcommand{\ot}{\otimes} 
\newcommand{\la}{\langle} \newcommand{\ra}{\rangle}
\newcommand{\ran}{\mathrm{ran}}
 
\newcommand{\x}{\times}

\newcommand{\ssb}{{\sc ssb}}

\newcommand{\ca}{C*-algebra} 
 \newcommand{\rep}{representation}

\newcommand{\Hs}{Hilbert space} 
 
 \newcommand{\cci}{C^{\infty}_c}

   \newcommand{\vna}{von
Neumann algebra}

\newcommand{\CA}{\mathcal{A}} 
\newcommand{\al}{\alpha} 
\newcommand{\gm}{\gamma} \newcommand{\Gm}{\Gamma}
\newcommand{\dl}{\delta} \newcommand{\Dl}{\Delta}
\newcommand{\ep}{\epsilon} \newcommand{\varep}{\varepsilon}

\newcommand{\lm}{\lambda} \newcommand{\Lm}{\Lambda}
\newcommand{\rh}{\rho} \newcommand{\sg}{\sigma}
\newcommand{\Sg}{\Sigma} \newcommand{\ta}{\tau} \newcommand{\ph}{\phi}
\newcommand{\Ph}{\Phi} \newcommand{\phv}{\varphi}
\newcommand{\ch}{\ch} \newcommand{\ps}{\psi} 
\newcommand{\om}{\omega} \newcommand{\Om}{\Omega}
\newcommand{\ups}{\upsilon}
\newcommand{\daw}{\!\downarrow}

\newcommand{\inv}{^{-1}}

\newcommand{\Tr}{\mbox{\rm Tr}\,}

\newcommand{\GS}{\mathfrak{S}}

\renewcommand{\t}{\mathfrak{t}}

\newcommand{\CB}{{\mathcal B}}
 \newcommand{\CF}{{\mathcal F}}

 \newcommand{\CM}{{\mathcal M}}
\newcommand{\CN}{{\mathcal N}} 
\newcommand{\CO}{{\mathcal O}} \newcommand{\CP}{{\mathcal P}}

\newcommand{\C}{{\mathbb C}} 
\newcommand{\N}{{\mathbb N}} \newcommand{\R}{{\mathbb R}}
\newcommand{\T}{{\mathbb T}} \newcommand{\Z}{{\mathbb Z}}
  \makeatletter
 
\makeatletter
\def\moverlay{\mathpalette\mov@rlay}
\def\mov@rlay#1#2{\leavevmode\vtop{%
   \baselineskip\z@skip \lineskiplimit-\maxdimen
   \ialign{\hfil$\m@th#1##$\hfil\cr#2\crcr}}}
\newcommand{\charfusion}[3][\mathord]{
    #1{\ifx#1\mathop\vphantom{#2}\fi
        \mathpalette\mov@rlay{#2\cr#3}
      }
    \ifx#1\mathop\expandafter\displaylimits\fi}
\makeatother

\newtheorem{definition}{Definition}[section]
\newtheorem{lemma}[definition]{Lemma}
\newtheorem{theorem}[definition]{Theorem}
\newtheorem{proposition}[definition]{Proposition}

\newtheorem{corollary}[definition]{Corollary}
\newtheorem{exercise}[definition]{Exercise}
\newcommand{\bex}{\begin{exercise}}
\newcommand{\eex}{\end{exercise}}
\newcommand{\X}{\mathcal{X}}\newcommand{\F}{\mathcal{F}}
\newcommand{\Y}{\mathcal{Y}}

\newcommand{\LDP}{\textsc{ldp}}
\newcommand{\DLR}{\textsc{dlr}}
\newcommand{\ML}{Martin-L\"{o}f}
\renewcommand{\Pr}{\mathrm{Prob}}
\newcommand{\tr}{\mathrm{tr}}

\newcommand{\tto}{$\mathrm{II}_1$}
\newcommand{\tti}{$\mathrm{II}_{\infty}$}

\newcommand{\QED}{\null\nobreak\hfill\ensuremath{\square}}%
\topmargin = - 1 cm \textheight = 23 cm \textwidth = 15 cm
\usepackage[symbol]{footmisc}
\renewcommand{\thefootnote}{\fnsymbol{footnote}}
\begin{document}\pagenumbering{arabic} \setlength{\unitlength}{1cm}\cleardoublepage
\date\nodate
\begin{center}
\begin{Huge}
\textbf{The Okinawa Lectures on Entropy}
\end{Huge}
\bigskip\bigskip\bigskip

\begin{Large}
 Klaas Landsman\vspace{10pt}
 \end{Large} 
 
 \bigskip\bigskip
 \begin{large}
 	Theoretical Sciences Visiting Program, Okinawa Institute of  Science  and Technology Graduate University, Onna, 904-0495, Japan \\ \vspace{2mm}
  Radboud Center for Natural Philosophy  \\ \vspace{1mm}
Institute for Mathematics, Astrophysics, and  Particle Physics\\ \vspace{1mm}
Radboud University, Nijmegen, The Netherlands\\ \vspace{1mm}
\texttt{landsman@math.ru.nl} \hspace{25pt}  \texttt{rcnp.science.ru.nl}
\end{large}
\vspace{30pt}

 \fbox{\emph{Dedicated to the memory of Nico Hugenholtz (1924--2026)}}
\end{center}

\bigskip\bigskip

 \begin{abstract} 
\noindent
After a historical introduction, the most important classical and quantum entropies are introduced as constructions in classical and quantum probability theory.  Classical entropies give the rate of exponential decay of large deviations from the mean value of some fluctuating quantity. This is studied in what is called large deviation theory, whose foundations we explain (including  central theorems of Sanov, Cram\'{e}r, G\"{a}rtner--Ellis, and Varadhan), and  illustrate in some applications to both Boltzmannian and Gibbsian statistical physics. 
Quantum entropies superficially connect to classical entropies at the formula level, but more deeply do so via the crucial role of entropy in statistical hypothesis testing. The classical (relative) Kullback--Leibler entropy, its quantum counterpart introduced by Umegaki, as well as their ``deformations'' proposed by R\'{e}nyi  all fit naturally in this context.
Though they historically came first, the Shannon and von Neumann entropies are best seen as special cases of relative entropies (where the prior is uniform).
Quantum entropy faces the new problem of defining and computing the relative entropy of a pair of states on a subsystem, here  formalized as a \vna.  This is straightforward only for so-called type I \vna s; the general case requires modular (aka Tomita--Takesaki) theory, which provides the framework for the relative quantum entropies introduced by Araki and Uhlmann (these encompass both the Kullback--Leibler and Umegaki entropies as special cases). To (re)define and compute these entropies in terms of density operators and traces, further constructions are needed, namely
   Haagerup's noncommutative $L^p$ spaces. Von Neumann algebras also provide the setting for the  Connes--St\o rmer--Narnhofer--Thirring entropy, which is a quantum version of the Kolmogorov--Sinai entropy in dynamical systems and ergodic theory, to which we also provide an introduction. This course was originally inspired by, and should be relevant to,
black hole thermodynamics, although we discuss neither this application nor the second law. The course tries to be both mathematically rigorous and interesting to theoretical physicists. Prerequisites are undergraduate probability theory, functional analysis, and quantum theory.

\end{abstract}\newpage
\begin{footnotesize}
\tableofcontents
\end{footnotesize}
\thispagestyle{empty}
\renewcommand{\thefootnote}{\arabic{footnote}}
\newpage \setcounter{footnote}{0}
\section*{Preface}\addcontentsline{toc}{section}{Preface}
These notes are an extended version of ten two-hour graduate lectures on entropy  given in the Spring of 2026 at \textsc{oist}, the  \emph{Okinawa Institute of Science and Technology}.\footnote{I am immensely grateful to Jonas Fischer, Philipp Hoehn, Yu-Jhen Lin, Nic Shannon, Reiko Toriumi, and Harry Wilson for their hospitality and help in arranging my 6-month stay at \textsc{oist} including these lectures; to the participants for their lively interest, questions, and friendship, notably Gon\c{c}alo Araujo Regado, William Horowitz, Torbj\"{o}rn Lundh, Ben Knepper, and Bilyana Tomova;  and to my beloved Edith de Jong for her role in making this period unforgettable.}
The aim of these notes is to bring a wider audience ranging from theoretical and mathematical physicists to philosophers of science up to date on some of the various mainstream \emph{definitions} of entropy and their properties. Taking a classical as well as a quantum point of view on entropy in a single text highlights both what is common and what is different; this also applies to the relevant mathematical background  in classical and quantum probability theory  we review from scratch, 
 notably large deviations and von Neumann algebras, respectively. 
 Nothing in this course is original, except perhaps for the arrangement of the material, the general approach, and some proofs.\footnote{Thanks to Max Bloemers, Silvester Borsboom, Nino Dekkers, and Jacques Janse van Rensburg for corrections!} My basic sources were:
\begin{small}
\begin{description}
\item[] Austin, T. (2017). \emph{Math 254A: Entropy and Ergodic Theory}. \url{https://www.math.ucla.edu/~tim/entropycourse.html}. 
\item[]   Cover, T.M.,  Thomas, J.A. (2006). \emph{Elements of Information Theory. Second Edition} (Wiley). 
\item[] Dembo, A., Zeitouni, A. (1998). \emph{Large Deviations: Techniques and Applications. Second Edition} (Springer). 
\item[]  Dorlas, T.C. (2021).
\emph{Statistical Mechanics: Fundamentals and Model Solutions, Second Edition} (CRC).
\item[] Ellis, R.S. (1995). An overview of the theory of large deviations and applications to statistical mechanics. \emph{Scandinavian Actuarial Journal} 1, 97--142.
\item[]  Hayashi, M. (2017). \emph{Quantum Information Theory: Mathematical Foundation. Second Edition} (Springer). 
\item[] Hiai, F. (2021a). \emph{Lectures on Selected Topics in von Neumann Algebras} (EMS Press).
\item[] Hiai, F. (2021b). \emph{Quantum f-Divergences in von Neumann Algebras:
Reversibility of Quantum Operations} (Springer).
\item[]     Jak\v{s}i\'{c}, V. (2018). Lectures on entropy. Part I. \url{https://arxiv.org/pdf/1806.07249}.
\item[]   Jak\v{s}i\'{c}, V.,  Ogata, Y., Pillet, C. A.,  Seiringer, R. (2012). Quantum hypothesis testing and non-equilibrium statistical mechanics. \emph{Reviews in Mathematical Physics} 24:1230002.
\item[]  Khatri, S., Wilde, M.M. (2024). \emph{Principles of Quantum Communication Theory: A Modern Approach}.
\url{https://arxiv.org/abs/2011.04672}. 
\item[] Landsman, K.,  van Weert, Ch.G. (1987).  Real-and imaginary-time field theory at finite temperature and density. \emph{Physics Reports} 145, 141--249. 
\item[] Landsman, K. (2017). \emph{Foundations of Quantum Theory: From Classical Concepts to Operator Algebras} (Springer).
\url{https://link.springer.com/content/pdf/10.1007/978-3-319-51777-3.pdf}.  
\item[] Neshveyev, S., St\o rmer, E. (2006). \emph{Dynamical Entropy in Operator Algebras} (Springer).  
\item[] Ohya, M., Petz, D. (1994). \emph{Quantum Entropy and its Use} (Springer). 
\item[] Rassoul-Agha, F., Sepp\"{a}l\"{a}inen, T. (2015). \emph{A Course on Large Deviations with an Introduction to Gibbs Measures} (AMS).
\item[]  Shields, P.C. (1996). \emph{The Ergodic Theory of Discrete Sample Paths} (AMS).
\item[] Sorce, J. (2023). An intuitive construction of modular flow. \emph{Journal of High Energy Physics} 2023:79.
\item[]   Takesaki, M. (2002/2003). \emph{Theory of Operator Algebras. Volumes I, II, III} (Springer).
\item[]   Uffink, J. (2001). Bluff your way in the second law of thermodynamics. \emph{Studies in History and Philosophy of Modern Physics} 32, 305--394.
\end{description}
\end{small}
More detailed references are provided throughout these notes, typically in footnotes.

  Our most glaring omission (apart from some historical comments in the Historical Introduction) is the absence of the second law of thermodynamics. This needs some justification. Although a famous quote by Eddington even puts it ahead of all other laws of physics,\footnote{`The law that entropy always increases--the second law of thermodynamics--holds, I think, the supreme position among the laws of nature. If someone points out to you that your pet theory of the universe is in disagreement with Maxwell's equations--then so much the worse for  Maxwell's equations. If it is found to be contradicted by observation--well, these experimentalists bungle things some times. But if your theory is found to be against the second law of thermodynamics I can give you no hope; there is nothing for it but to collapse in deepest humiliation.' (Arthur S. Eddington, \emph{The Nature of the Physical World}, 1935).} and the novelist C.P. Snow compared it to works of Shakespeare,\footnote{`A good many times I have been present at gatherings of people who, by the standards of the traditional culture, are thought highly educated and who have with considerable gusto been expressing their incredulity at the illiteracy of scientists. Once or twice I have been provoked and have asked the company how many of them could describe the Second Law of Thermodynamics. The response was cold: it was also negative. Yet I was asking something which is the scientific equivalent of: Have you read a work of Shakespeare's?’ 
(C.P. Snow, \emph{The Two Cultures}, 1959).}
this second law is in fact not very well understood at all, including even its very formulation.\footnote{The \emph{locus classicus} for this critical view is Uffink (2001). See also Roberts (2022). A very precise analysis of the second law is given by Lavis \& Frigg (2025), but their analysis is restricted to phenomenological thermodynamics (as opposed to its statistical underpinning) and even so it remains based on a specific choice of axioms,  going back to Carath\'{e}odory (1909), and more recently taken up by Lieb \& Yngvason (1999, 2000).  See also the Introduction below.}  For example, in the view taken here classical entropy originates in fluctuations around (and hence off) equilibrium, whereas thermodynamical  entropy is an equilibrium concept.\footnote{As argued by Brown \& Uffink (2001), return to equilibrium is
\emph{prior} to the second law, which should then \emph{follow} since such a return is typically an entropy-increasing or dissipative process. Although the pull of fluctuations is in the oppositie direction, the two are related by
fluctuation-dissipation theorems. See Darrigol (2023) for a historical perspective and for example Maes (1999) and
 Cuneo, Jak\v{s}i\'{c}, Pillet, \&  Shirikyan (2026) for technical developments. }
Even the most spectacular incarnation of the second law, namely the area law for black holes and the ensuing identification of the entropy of a black hole with the area of its event horizon (times a constant), remains puzzling.\footnote{The original references are Hawking (1971,1972) and Bekenstein (1972, 1973), although it is now clear that the area law was first found by Penrose, who explained it to Hawking in 1970 (Seife, 2021, pp. 278-279). Bekenstein pushed for a thermodynamic interpretation of this  black hole entropy, with ensuing ``laws of black hole thermodynamics'', right from the start, whereas Hawking and his allies resisted this interpretation and spoke of the ``laws of black hole mechanics'' (Bardeen, Carter, and Hawking, 1973). But after his discovery of the radiation now named after him, Hawking made a U-turn towards the former view. 
See  van Dongen \& Landsman (2026) for a detailed history. Although the fact that black holes may be treated as
thermodynamic systems confirms the  universal applicability and validity of thermodynamics--see Einstein's quote below \er{Cl1law}--this is surely one of the most fascinating and unexpected developments in modern physics. But then again, this may be less surprising if we see thermodynamics as a theory of the split between useful/observable and useless/unobservable contributions to energy (Roberts, 2025), like work versus heat in 19th century thermodynamics, and degrees of freedom separated by an event horizon in black hole physics.
} 
This
context, typically enriched by ``holography'', has recently led to an explosion of interest  in von Neumann algebras and their entropies among (former) string theorists, along the mathematical lines covered in these notes.\footnote{The papers initiating this development were Witten (2022) and Leutheusser \& Liu (2023), written in the opposite order; see also the review by Liu (2025).
Recent work  includes Faulkner \& Speranza (2024), 
 De Vuyst, Eccles,  Hoehn, \& Kirklin (2025), Fewster et al.\ (2025), Kudler-Flam,  Leutheusser, \& Satishchandran  (2025), 
 Kudler-Flam,  Leutheusser,  Rahman,  Satishchandran, \&  Speranza (2025), 
 and
 Klinger, Kudler-Flam, \& Satishchandran (2026). 
 \label{Kudler} }

Finally, another very important aspect of entropy not included here is its role in partial differential equations, which originated with  Boltzmann (1872) 
in connection with the second law and the more general phenomenon of irreversibility (which the increase of entropy tries to capture).\footnote{
In this approach irreversible macroscopic evolution equations emerge from the most probable microscopic configurations.
For Boltzmann's work see Darrigol (2018) and Uffink (2024). On irreversibility in the 19th century see Brush (1974) and van Strien (2013). Villani (2013) is a nice  introduction  from a \textsc{pde} perspective.  See also Zuchowski (2024) and Dekkers \& Landsman (2026). \emph{Gradient flow} relates entropic  \textsc{pde} theory to large deviations  (Peletier, 2026). The latest mathematical news in this area is the proof of the Boltzmann equation (and hence of the increase of entropy)  for arbitrary times by Deng, Hani, \& Ma (2024), rewarded with a well-deserved Fields Medal for Yu Deng in July 2026.   }
\smallskip

 \section{Historical Introduction}
 \begin{quote}
\begin{small}
You should call it ``entropy'' for two reasons: First, the function is already used in thermodynamics under that name; second, and more importantly, most people don't know what entropy really is, and if you use the word ``entropy'' in an argument you will win every time.\footnote{Myron Tribus got this anecdote first-hand from Shannon in 1961 (Levine \& Tribus, 1978, pp.\ 2--3).}\\ \mbox{} \hfill
 (John von Neumann to Claude Shannon, 1940)
\end{small}
\end{quote} Let us start with a historical introduction to the idea of entropy, focusing on the contributions by Clausius (and the role of thermodynamics), Boltzmann,
Gibbs, Einstein, von Neumann, Shannon, R\'{e}nyi, and Kolmogorov, 
 who set the stage for all later developments. 
To begin, the very concept op \emph{entropy} was introduced by Clausius (1865) in one of the founding papers of thermodynamics.\footnote{See Brush (1974, 1976, 2003), Truesdell (1980),  von Plato (1994), Darrigol (2003), M\"{u}ller (2007), Uffink (2001, 2007), Emch \& Liu (2013), Weinberger (2013), Gaudenzi (2019),  Saslow (2020),  and Norton (2022) for history and  analysis of thermodynamics and 19th century statistical physics. Apart from entropy,  thermodynamics also gave the concept of \emph{energy} a central role in physics; see e.g.\ Smith (1998) for (especially) the British side, Wegener (2009) 
for  (especially) the German side, and Elkana (1974), Harman (1982), and Coppersmith (2010) in general. \label{HTrefs}} This is done in his eq.\ (59), which remains a cornerstone of the theory, and reads \emph{verbatim}:\footnote{In view of its historic importance we just  copy his formula, but we now know better than to write $dQ$, which is \emph{not} an exact 1-form; indeed this fact (within the mathematics of his time) was Clausius's whole point in introducing $S$.}
\beq
dS=\frac{dQ}{T}, \label{C59}
\eeq
 where $Q$ is heat and $T$ absolute temperature.  This is preceded by the comment that $dQ/T$  is a `complete differential of a quantity that only depends on the state of a body at a particular instant' and hence (like energy) is a `function of state', as opposed to depending on some path towards that state (which is the case for both heat and work, whose \emph{sum} is 
 the path-independent energy).
  
 This formula had a long pedigree, starting with the work of Carnot (1824) on the efficiency of heat engines (such as a steam engine), which (with hindsight) is often seen as the beginning of thermodynamics. Like most of his contemporaries, Carnot believed in the so-called \emph{caloric theory of heat} in which heat is a substance and as such is a conserved quantity. Hence he compared heat engines  to water mills with heat in the role of water, flowing from high to low temperature regions much as water flows from high to low altitudes due to gravity. Although in the next few decades the caloric theory of heat fell from grace and conservation of heat  was replaced by conservation of energy (which became mainstream around 1850), as we still have it, Carnot's memoir contained a result and also a kind of reasoning that did stand the test of time, namely his description of heat engines as running in cycles and his conclusion that \emph{reversible} cycles have maximal efficiency (see below) and are universal in the sense that this maximal efficiency only depends on the temperatures at which heat comes in and goes out. Moreover, the structure of his arguments, namely (albeit informal) proofs by contradiction, in his case based on the impossibility of a \emph{perpetuum mobile}, was also influential and would be repeated by all leading  thermodynamicists after him.
 
 Thus thermodynamics--a word coined by Thomson  (= Kelvin)  as late as 1849--became the science of the conversion of heat into work and \emph{vice versa}, and more generally physics became ``the science of energy'', of which heat and work were recognized as special cases. Using conservation of energy instead of conservation of heat, and crucially making use of
 Thomson's absolute temperature scale $T$ (which he introduced in 1851), in 1854 Thomson and Clausius both rewrote Carnot's fundamental insight on heat engines running in reversible cycles as the formula
 \beq
\frac{Q_1}{T_1}+\frac{Q_2}{T_2}=0, \label{CKC}
\eeq
 where $Q_1$ is the amount of heat that in each cycle comes in at temperature $T_1$ (for example from burning coal) and $Q_2$ is \emph{minus} the amount of heat that leaves the engine  at temperature $T_1$. In between, the initial heat 
 generates some work $W$ (such as driving a wheel or a pump), so that 
 \beq
 W=Q_1+Q_2, \label{firstlaw}
 \eeq
  by conservation of energy.\footnote{In a steam or car engine one has $W>0$, but in a heat pump work is added ($W<0$) and heat is produced.}
 Eq.\ \er{CKC} holds for a \emph{reversible} cycle, and \emph{mutatis mutandis} 
  Carnot's conclusion that reversibility is necessary and sufficient for maximizing the  efficiency could still be derived through proofs by contradiction, provided the efficiency is defined by   
  $\eta=W/Q_1$. Eqs.\ \er{CKC} and \er{firstlaw} turn this into its maximum value for reversible cycles, namely the textbook formula
 \begin{equation}
\eta =1-\frac{T_2}{T_1}.
\end{equation}
In their proofs,  Thomson and Clausius used different but (allegedly) equivalent principles replacing Carnot's impossibility of a \emph{perpetuum mobile}. Paraphrasing their own words:\footnote{Quoted from Uffink (2002), p.\ 328, who also explains why the textbook proofs of their equivalence are dubious. Having said this, Steane (2016), chapter 8, is a good account of these principles and their equivalence. }
\begin{quote}
\begin{small}
It is impossible to perform a cyclic process with no other
result than that heat is absorbed from a reservoir, and work is performed. (Thomson)

It is impossible to perform a cyclic process which has no
other result than that heat is absorbed from a reservoir with a low temperature
and emitted into a reservoir with a higher temperature. (Clausius)
\end{small}
\end{quote}
These were early versions of the second law of thermodynamics. 
Both Thomson and Clausius had the idea of generalizing \er{CKC}, which applies to what we now call a Carnot cycle, to 
\begin{equation}
\Sigma_i\frac{Q_i}{T_i}=0,
\end{equation}
for a chain of cycles of heat engines or even more general processes involving heat and work, upon which Clausius in 1854 moved to  ``infinitesimal'' processes and wrote, for reversible cycles, 
\begin{equation}
\oint \frac{dQ}{T}=0.\label{Cle}
\end{equation}
Clausius  also derived \er{Cle} from his own version of the second law above, which he took to  imply
\begin{equation}
\oint \frac{dQ}{T}\leq 0,\label{Cleb}
\end{equation}
for arbitrary cycles. If the cycle is reversible one obtains both $\leq 0$ and $\geq 0$ by reversing it, yielding \er{Cle}.
The latter  implies \er{C59} from a theorem stating that given \er{Cle} such a function $S$ must exist.\footnote{Different approaches to thermodynamics differ in many ways, one of which is the question whether $\delta Q=TdS$ is always true (where we have now written the heat 1-form as $\delta Q$ rather than Clausius's $dQ$, which in modern differential geometry is wrong), and if so on which axioms this should be based; or 
only holds along reversible paths, as Clausius had it, with $\dl Q\leq TdS$ along possibly irreversible thermodynamically admissible paths (where we say that some relation between 1-forms holds along a path if it holds upon applying the 1-forms in question to  tangent vectors along such a path).
The latter approach is explained and defended for example by Steane (2016) and Dorlas (2021). This allows a formulation of the second law to the effect that $dS\geq 0$ along adiabatic paths (where $\dl Q=0$).  Even so, the fundamental relation $dU=TdS-pdV$ (the first law of thermodynamics as introduced by Clausius) holds along any path. Steane's  argument is that it is initially derived along reversible paths from the energy conservation law $dU=\delta Q + \delta W$ (where $W$ is the 1-form representing work),
 but is then seen to be a relation between functions of state that always holds.
 On the other hand, Carath\'{e}odorian  axiomatic approaches to thermodynamics (Carath\'{e}odory, 1909; Lieb \& Yngvason, 1999, 2000; Lavis \& Frigg, 2025) are based on adiabatic inaccessibility axioms leading to an entropy function $S$ such that $\dl Q=TdS$ is always true. In that case $S$ is trivially constant along adiabatic paths, and the second law must be sought elsewhere.  A nice axiom implying this 
is the one introduced by Jauch (1972), stating that a cyclic process without heat exchange cannot perform any work. See
Roberts (2026) for a corrected proof that this implies $\delta Q=TdS$.}
 
 A few pages after his equation \er{C59} Clausius introduces his neologism ``entropy'', as follows:
\begin{quote}
\begin{small}
Sucht man f\"{u}r $S$ einen bezeichnenden Namen, so k\"{o}nnte man, \"{a}hnlich wie von der Gr\"{o}sse $U$ gesagt ist, sie sei der W\"{a}rme - und Werkinhalt des K\"{o}rpers. Da ich es aber f\"{u}r besser halte, die Namen derartiger f\"{u}r die Wissenschaft wichtiger  Gr\"{o}ssen aus den alten Sprachen zu entnehmen, damit sie unver\"{a}dert in allen neuen Sprachen angewandt werden k\"{o}nnen, so schlage ich vor, die  Gr\"{o}sse $S$ nach dem Griechischen Worte\begin{greek} <h trop`h\end{greek}, die Verwandlung, die\\  E n t r o p i e des K\"{o}rpers zu nennen. Das Wort E n t r o p i e habe ich absichtlich dem Worte E n e r g i e m\"{o}glichst \"{a}hnlich gebildet, denn die beiden  Gr\"{o}ssen, welche durch diese W\"{o}rter benannt werden sollen, sind ihren physikalischen Bedeutungen nach einander so nahe verwandt, dass eine gewisse Gleichartigkeit in der Benennung mit zweckm\"{a}ssig zu seinn scheint. (Clausius, 1867, p.\ 34).\footnote{`If one is looking for a descriptive name for $S$, one could, in a similar way to what has been said about the quantity $U$, say that it is the heat and work content of the body. However, since I think it is better to take the names of such important scientific terms from the old languages, so that they can be used without modification in all new languages, I suggest that the term $S$ be used after the Greek words \begin{greek} <h trop`h\end{greek}, the transformation, the e n t r o p y of the body. I have deliberately made the word e n t r o p y as similar as possible to the word e n e r g y , because the two quantities which are to be named by these words are so closely related in their physical meanings that a certain similarity in the naming seems appropriate.' Translation by the author.  The page number is from the 1867 reprint.}
\end{small}
\end{quote}
The end of this paper is very famous, expressing the (main) two laws of thermodynamics:\footnote{
1) `T h e\: e n e r g y\: o f\: t h e\: w o r l d\: i s\: c o n s t a n t.\\  \mbox{} \hspace{15pt} 2)
T h e\: e n t r o p y\: o f\: t h e\: w o r l d\: t e n d s\: t o w a r d s\: a\: m a x i m u m.'}

\begin{quote}
\begin{small}
\begin{description}
\item[ 1)]  D i e\:  E n e r g i e \: d e r \: W e l t \:  i s t\: c o n s t a n t.
\item[2)] D i e\:  E n t r o p i e \: d e r \: W e l t \: s t r e b t\: e i n e m\: M a x i m u m\: zu.\end{description}
\end{small}
\end{quote}
Similar claims had been made by Thomson already around 1850 (without using the word `entropy'),
who added a  religious connotation to the effect that the first law reflected the eternity of God whereas the second  applied to the non-divine realm, for example by quoting from Psalm 102:
\begin{quote}
\begin{small}
Of old hast thou laid the foundation of the earth; and the heavens are the work of thy hands; They shall perish, but thou shalt endure, yea, all of them shall wax old like a garment; as a vesture shalt thou change them, and they shall be changed. But thou art the same, and thy years shall have no end.
\end{small}
\end{quote}
Although in the grandiose generality stated by Clausius these laws, originally inspired by steam engines as they were,  are not really backed by the precise formalism of thermodynamics at all, the alleged consequence of the ``heat death'' of the universe made a huge impression at the time.\footnote{For a critique see Uffink (2001). The general context, religious and otherwise, is explored in Kragh (2008). }
In any case, by adding to the introduction of entropy the ensuing reformulation of the first law as
\begin{equation}
dU=TdS-pdV,\label{Cl1law}
\end{equation}
Clausius (1865) had magnificently brought thermodynamics into its more or less its final form:\footnote{Clausius (1865) does not literally state  \er{Cl1law},
but it trivially follows from his very first, unnumbered equation $dQ=dU+AdW$, where $A$ is the `caloric equivalent to work' (i.e.\ a unit conversion constant  omitted in later literature), later numbered as (57) in the equivalent version $dU=dQ-dw$, then either 
his unnumbered eq.\ $dW=pdv$ above (16) or (16) itself, stating that $dw=Apdv$, and of course his (59), which is our \er{C59}. }
\begin{quote}\begin{small}
Eine Theorie ist desto eindrucksvoller, je gr\"{o}sser die Einfachheit ihrer Pr\"{a}missen ist, je verschiedenartigere Dinge sie verknüpft, und je weiter ihr Anwendungsbereich ist. Deshalb der tiefe Eindruck, den die klassische Thermodynamik auf mich machte. 

Es ist die einzige physikalische Theorie allgemeinen Inhaltes, von der ich \"{u}berzeugt bin, dass sie im Rahmen der Anwendbarkeit ihrer Grundbegriffe niemals umgestossen werden wird.\footnote{A theory is the more impressive, the greater the simplicity of its premises is, the more different kind of things it relates, and the more extended its area of applicability. Therefore, classical thermodynamics made a deep impression on me. Thermodynamics is the only physical theory of universal content concerning which I am convinced that, within the framework of the applicability of its basic concepts, it will never be overthrown.  (Einstein, 1949, p.\ 33)}
 (Einstein, 1949, p.\ 32) \end{small}
\end{quote}

Subsequently, the upcoming kinetic theory of gas (which among other things replaced the caloric theory of heat) suggested 
a microscopic underpinning of thermodynamics. Following Maxwell's first attempts at probabilistic reasoning in this context,\footnote{Apparently, Maxwell was not interested in entropy and hence made no attempt at some probabilistic explanation of it (Darrigol, (2003, 2018). Though a pioneer of kinetic gas theory, too, even Clausius was not involved in this. 
} including his well-known formula for the velocity distribution in equilibrium, this was taken up by Boltzmann in a series of papers of which we discuss the two main (and deservedly most famous) ones.
\begin{itemize}
\item 
 Boltzmann (1872) is one of the founding papers of kinetic gas theory; in particular, it contains the \emph{Boltzmann equation}.\footnote{Of the references in footnote \ref{HTrefs},  Darrigol (2018) and Uffink (2024) specifically concern Boltzmann.} For us, Already the opening Summary  contains the formula
\begin{equation}
E = \int_0^{\infty} f(x,t)\left[ \log\left(\frac{f(x,t)}{x}\right)-1\right]dx, \label{BolE}
\end{equation}
where $f(x,t)$ is the `number of molecules  having energy $x$ at time $t$'.
This is the first occurrence of the combination ``$p_i \log p_i$'' (usually with a minus sign omitted by Boltzmann) that characterizes practically all versions of entropy that deserve the name.\footnote{The questions how Boltzmann arrived at \er{BolE} remains a matter of speculation. Darrigol (2018), pp.\ 491--493, gives three possibilities. First, Boltzmann (1871)
contains a  discussion of the canonical ensemble, and his formula (18) for the entropy of a polyatomic gas may be rewritten in the now standard form $\int\rh\log\rh$; perhaps Boltzmann saw this between his 1871 and 1872 papers. 
Second, he may have systematically searched for a functional of the distribution function $f(x,t)$ that returns the entropy of a gas in the special case where $f$ is the Maxwell distribution. Finally, Boltzmann may have guessed the existence of a functional of $f$ whose value decreases when $f$ solves the Boltzmann equation. In any case, finding eq.\ \er{BolE} as a solution is `perhaps 
the most astonishing jump in all of Boltzmann's constructions.' (Darrigol, 2018, p.\ 491).
 We are indebted to Olivier Darrigol for correspondence on this topic. 
}
 Boltzmann introduced it as a tool for proving irreversibility: what is now known as his $H$-theorem (after a later renaming of a simplified version of $E$ to $H$) states that if $f$ solves the Boltzmann equation, then $E(f)$ `can never increase but must always decrease or remain constant.' \item
 Of equal importance to the (hi)story of entropy is Boltzmann (1877), in which he initiated the combinatorial and probabilistic approach to entropy that became one of the sources of the large deviations approach advocated in these notes. In particular, it contains a Clausius-like ``maximum entropy principle'', according to which `the most likely state', identified with an equilibrium state, is found by minimizing a (negative) entropy  
\beq
f\mapsto M'(f) =\int_0^{\infty} f(x)\log f(x) dx,
\eeq
subjects to constraints that determine the total particle number and total energy, viz.
\begin{align}
n=\int_0^{\infty} f(x)dx && L=\int_0^{\infty} xf(x)dx.
\end{align}
Here, as in Boltzmann (1872), $f$ is
still regarded as a probability distribution function of singe-particle energy, but the expression has clearly been simplified compared to \er{BolE} and later in the paper Boltzmann defines the entropy  with the ``usual'' sign, this time regarding $f(x,y,z,u,v,w)$ as a function of position $\vec{x}=(x,y,z)$ and velocity $\vec{v}=(u,v,w)$.
\end{itemize}
Although Boltzmann's  second (1877) paper looks quite different from the first (1872), and  for Boltzmann himself was a fresh start,\footnote{See Darrigol (2018) and Uffink (2007, 2024).}
 from a modern mathematical point of view these papers are  closely related in that both are based on a coarse-graining technique using what we now call an \emph{empirical measure},
 which as we will see plays a central role in large deviation theory:\footnote{  \label{fn15} 
 The following passage is extracted from Dekkers \& Landsman (2025).
  The notation in both \er{deff} and \er{LN0} is as follows.   For any  measure space $X$ (equipped with a $\sg$-algebra $\Sg$ we suppress), let $\Pr(X)$ be the
  space of probability measures on $X$. Any $x\in X$ defines a  \emph{point measure} $\dl_x\in\Pr(X)$ via $\dl_x(U)=1$ if $x\in U$ and $\dl_x(U)=0$ if $x\notin U$ (for any $U\in\Sg$). 
  In \er{deff} we have $X=\R^{2d}$, and in \er{LN0} we have $X=A^N$. }
  \begin{itemize}
\item If we denote  the space of probability measures on $\R^{2d}$ by $\Pr(\R^{2d})$,
Boltzmann (1872) associated a one-particle distribution function $f_N\in \Pr(\R^{2d})$ to a
 microscopic $N$-particle configurations
$(\mathbf{z}_1,  \ldots  \mathbf{z}_{N})\in\R^{2dN}$, where  $\mathbf{z}=(\mathbf{r}, \mathbf{v})$.\footnote{Boltzmann (1872) initially introduced
$f_N$ for homogeneous gases in which the spatial argument $\mathbf{r}$ is absent and the velocity $\mathbf{v}$ is replaced by the kinetic energy $x$, so that `the number of molecules in unit volume whose kinetic energy at time $t$ lies between $x$ and $x+dx$ I will call $f(x,t)dx$' (p.\ 268 of the English translation). Later in the paper he introduces what we call $f_N(\mathbf{z})$ for inhomogeneous gases, with a similar interpretation. \label{f72}} We (re)write his construction as 
\begin{equation}
 f_N (\mathbf{z}_1, \ldots  ,\mathbf{z}_{N}):=
 \frac{1}{N}\Sigma_{n=1}^{N} \dl_{\mathbf{z}_n}. \label{deff}
\end{equation}
If we equip $\R^{2dN}$ with a probability measure $\mathbb{P}_N$, then, again from a modern point of view,  $f_N$ is a  random variable \emph{on} $(\R^{2dN}, \mathbb{P}_N)$ taking values \emph{in} $\Pr(\R^{2d})$. 
 In physics one prefers probability \emph{distributions} to \emph{measures}:
one would like to write some $\mu\in\Pr(\R^{2d})$ as \beq
d\mu(\mathbf{z})= \rh(\mathbf{z})d\mathbf{z},
\eeq
 where $\rh$ is a probability \emph{distribution} on $\R^{2d}$. This formally works for $\mu=\dl_{\mathbf{z}_n}$ if we take 
$\rh(\mathbf{z})=\dl(\mathbf{z}-\mathbf{z}_n)$, the Dirac delta-function, so that  we may also write \er{deff} as
\begin{equation}
f_N (\mathbf{z};\mathbf{z}_1, \ldots  ,\mathbf{z}_{N})= \frac{1}{N}\Sigma_{n=1}^{N} \dl(\mathbf{z}-\mathbf{z}_n).\label{deff3}
\end{equation}
\item  
Boltzmann (1877) uses the same idea: in modern notation, let $A^N$ be the set of $N$-particle configurations $\sg:\{1, \ldots, N\}\raw A$, where $A$ is a finite set (whose elements Boltzmann took to be discrete energy levels),\footnote{Putting his energy unit $\varep=1$ for simplicity, Boltzmann (1877) has $A=\{0, 1, \ldots, p\}$. Writing $N$ for his $d$, his `complexions' $(k_1, \ldots, k_N)$ therefore correspond to our microstates $\sg\in A^N$.  His `distribution of states' (\emph{Zustandsverteilung}) $(N_0, \ldots, N_p)$  corresponds to our $Np_N$, so that $N_k=p_N(k)=\Sigma_{n=1}^{N} \dl_{\sg(n)k}$ is the number of molecules with energy $k$. \label{B77}
} 
 and $N$ is once again the number of particles. One now has a discrete version $p_N$ of the  empirical measure $f_N$ in \er{deff}, which  coarse-grains $\sg\in A^N$ into
 \begin{align}
 p_N(\sg):= \frac{1}{N}\Sigma_{n=1}^{N} \dl_{\sg(n)}. \label{LN0}
 \end{align}
Thus $p_N(\sg)$ is a probability distribution on $A$, and once again, if we equip $A^N$ with some probability measure $\mathbb{P}_N$, then $p_N$  is a  random variable \emph{on} $(A^N, \mathbb{P}_N)$
taking values \emph{in} $\Pr(A)$. Since $A$ is finite, the move from  measures to  distributions is trivial:\footnote{If $X$ is finite (and $\Sg$ is just the power set of $X$), then any probability \emph{measure} $\mathbb{P}$ on $X$ is equivalently given by a probability \emph{distribution} (or \emph{mass}) $P$, where $\mathbb{P}(U)=\Sigma_{x\in U} P(x)$ and conversely $P(x)=\mathbb{P}(\{x\})$.
  In that case, the probability distribution corresponding to the point measure $\dl_x$ is the Kronecker delta $\dl_x(y)=\dl_{xy}$. }
the analogue of \er{deff3}  is now simply
 given by $p_N(a;\sg)= \frac{1}{N}\Sigma_{n=1}^{N} \dl_{\sg(n)a}$, where $\dl_{ba}$ is the usual Kronecker delta. 
\end{itemize}
Ironically, perhaps the most famous outcome of Boltzmann's work on entropy is a formula
 \beq
 S=k\log\, W \label{BPE} 
 \eeq 
 that he never wrote down, although it  appears on his tombstone in the \emph{Wiener Zentralfriedhof}!\footnote{ The formula was  put on Boltzmann's grave in 1938, when it was relocated (and his widow joined him). Eq.\ \er{BPE} does come from the asymptotics Boltzmann derived in his 1877 paper (cf.\ footnote \ref{BEold}), but it was first stated in this simple form by Planck; see eq.\ (3) in Planck (1901) and \S134 in Planck (1906), \S134. Planck here also introduced ``Boltzmann's constant'' $k$. However, Boltzmann (1898), p.\ 172, does give a special case of \er{BPE}. Anticipating Einstein's approach to entropy reviewed below, Planck adds that \er{BPE} should be seen as the \emph{definition} of the probability $W$, which he claims to be inaccessible otherwise. See  Hoyer (1980) for the transition from Boltzmann to Planck.
} 

Gibbs (1902) is usually seen as the beginning of modern statistical mechanics, or at least as a \emph{new} beginning, after Boltzmann; their differences still remain to be  sorted out and understood.\footnote{ Sklar (1993) is a good starting point. 
Frigg \& Werndl  (2024ab) and Wallace (2024) are more recent views.  }
 Gibbs introduces the microcanonical, canonical, and grand canonical ensembles and defines the entropy (in a certain ensemble defined by a probability measure $\rh$ on an $N$-particle phase space) as minus the average value (under $\rh$) of what he calls the `index of probability' $\log\rh$ of $\rh$, without, incidentally, ever writing down an expression like $\rh\log\rh$.\footnote{`the average index of probability \emph{with its sign reversed} corresponds to entropy.' (Gibbs, 1902, p.\ 50). The notation $\rh$ seems of later use; Gibbs (1902) always writes down explicit expressions for his probability measures.} Though Gibbs (1902) had its roots in Boltzmann (1871), his streamlined version of classical statistical mechanics has set the stage for at least a century. In modern notation, and for simplicity (in tune with most of the rest of these notes) using a finite space space $\Om$, and stripping all unnecessary detail,
Gibbsian statistical mechanics  culminates in the following variational principles for the free energy and the entropy:\footnote{
These corresponds to Theorems III and II in Chapter XI of Gibbs (1902),  respectively.}
\begin{align}
F(U)&=\inf_{P}\{\la U\ra_{P}-S(P)\}; \label{FU}\\
S(P)&=\inf_U\{\la U\ra_{P}-F(U)\}. \label{Sp}
\end{align}
The first infimum is taken over all  probability distributions $P\in\Pr(\Om)$ on $\Om$, the second over all functions $U:\Omega\raw\R$, and the \emph{free energy} $F(U)$, \emph{partition function} $Z(U)$,   and \emph{entropy} $S(P)$ are 
 \begin{align}
  F(U)&:=-\log\, Z(U); &&  Z(U):=\Sigma_{\om\in \Om}e^{-U(\om)};  
&& S(P):=-\Sigma_{\om\in\Om} P(\om)\log P(\om).\label{RE}
 \end{align}
 As we shall see, the variational principles \er{FU} - \er{Sp} state that free energy and entropy are Fenchel transforms of each other (cf.\ Appendix \ref{AppC}),\footnote{For this and other reasons (e.g.\ his study of phase diagrams) Gibbs is often credited with recognizing the crucial role of convexity in statistical mechanics. See especially the  famous Introduction by A.S. Wightman to Israel (1979).} which fact will fall into place in the theory of large deviations; indeed, next to Boltzmann and Einstein one should also include Gibbs among the sources of this theory from physics. 
In our simple situation where $\Om$ is assumed finite, the infimum in \er{FU}  is  attained uniquely (so it is a minimum) by the \emph{canonical (Gibbs) distribution}
\begin{equation}
P_U(\om)=\frac{1}{Z(U)}e^{-U(\om)}.
\end{equation}
In physics one writes $U=\beta H$, where $\beta\in\R$ is a constant, related to the temperature by $\beta=1/T$, and for a given Hamiltonian $H$ one writes $F(\beta)=F(\beta H)/\beta$, and similarly $P_{\beta}:=P_{\beta H}$, i.e.,
\begin{align}
P_{\beta}(\om):=\frac{1}{Z(\beta)}e^{-\beta H(\om)}; &&  Z(\beta):=\Sigma_{\om\in \Om}e^{-\beta H (\om)}=e^{-\beta F(\beta)}. \label{gibbs}
\end{align}
Computing the minimum in \er{FU} using \er{RE} by straightforward calculus gives
 \begin{equation}
F(\beta)=\la H\ra_{P_{\beta}}-T S(P_{\beta}), \label{sc}
\end{equation}
which is a version of the equation ``$F=E-TS$'' familiar from equilibrium thermodynamics.\footnote{In thermodynamics  entropy is a function of energy $U\in\R$ rather than temperature $T=1/\beta$, such that $\partial S/\partial U=\beta$, see \er{Cl1law}; but solving this for $U=U(T)$ and inverting it to obtain $T=T(U)$ (if possible) makes $S(P_{\beta})$ a function of $U$.}

The key difference between the statistical mechanics of Boltzmann and of Gibbs lies in the role of coarse-graining, which is crucial in the former but is \emph{prima facie} absent in the latter. Large deviation theory applies to both, but it is much more complicated for Gibbs, cf.\ Theorem \ref{CramerGibbs}.

The next major contributor to the concept of entropy was  Einstein, whose  work  before 1905 mainly concerned thermodynamics and statistical mechanics, on which topics he continued to work throughout his career.\footnote{See  \emph{The Collected Papers of Albert Einstein}, especially Volumes 1, 2, and 3, available online at
\url{https://einsteinpapers.press.princeton.edu}.
See also Stachel et al., (1990), Klein et al.\ (1994), and  Uffink (2006).} 
As Darrigol (2003), p.\ 10 put it, `Einstein focused on the fluctuations around equilibrium that were negligible for Boltzmann and non-existent for Planck.' In particular, Einstein (1910) turned the Boltzmann--Planck formula \er{BPE}
on its head by reading it as 
\beq
W=e^{S/k},
\eeq
and interpreting $W$ as the probability of energy fluctuations around equilibrium. 
Einstein's argument is not probabilistic but empirical: like Planck (1901), he is dissatisfied with the formula \er{BPE} for the entropy because in practice one may not know the microscopic theory from which to compute $W$ in sufficient detail, whereas if one knows the entropy phenomenologically,  it gives information about the microscopic theory. As we shall see, the theory of large deviations combines and rederives  Einstein's insight that the probability of fluctuations is exponential in the entropy with Boltzmann's asymptotics for computing such probabilities, and hence in so far as inspiration from physics is concerned large deviation theory also has its roots in Einstein's work.\footnote{Much later, the work of Lanford (1973) on statistical mechanics also played an important role in this respect.}

 The next statue in our entropic Hall of Fame is that of von Neumann, who,  in the third paper of his trilogy that once and for all established the mathematical structure of quantum mechanics (von Neumann, 1927abc),\footnote{See Petz (2001)  for von Neumann's work on entropy, and R\'{e}dei \& St\"{o}ltzner (2001),  Bacciagaluppi (2022), Landsman (2022),   and Duncan \&Janssen (2023)
 for his contributions to quantum mechanics in general. Briefly,  at the age of 23 von Neumann defined the abstract concept of a Hilbert space, which previously had only appeared in examples, and developed the spectral theory of bounded as well as
unbounded normal operators on a \Hs. 
Thus he  saw that Schr\"{o}dinger's wave functions were unit vectors in a
\Hs\ of $L^2$ type, and that  Heisenberg's observables were linear operators
on a dif{f}erent \Hs, of $\ell^2$ type. A unitary transformation between these spaces then
provided the the mathematical equivalence between wave mechanics and matrix mechanics. 
von Neumann (1932) is still in print. 
 }  gave us the first quantum-mechanical formula for
 entropy.  Namely,
 \begin{equation}
\mathsf{S}=-Nk\, \mathrm{Spur}(U \ln U), \label{vN1}
\end{equation}
where $U$ is his notation for a density operator (which he had introduced in the second paper), Spur is the trace, $k$ is Boltzmann's constant, and $N$ is the number of particles in an ideal gas whose entropy he specifies; for the corresponding single-particle entropy we would now write 
\beq
S(\rh)=-\Tr(\rh\log\rh). \label{vNnow}
 \eeq
 His argument for \er{vN1}  has been described as `a remarkable feat of mathematical legerdemain',\footnote{See Duncan (2025),  p.\ 27.  Von Neumann's later (1932) argument, also outdated, is reviewed by Petz (2001).} which even von Neumann himself replaced by a different argument
 in his  book from 1932.
 
What has been of lasting value is von Neumann's insight that the eigenvalues  $(w_{\mu})$ of $U$ (repeating degenerate ones) are probabilities,\footnote{That is, $w_{\mu} \geq 0$ and  $\Sigma_{\mu}w_{\mu}=1$. For this and \er{vN2} see von Neumann (1927c),  page 287.}
 in terms of which
  \begin{equation}
\mathrm{Spur}(U \ln U)= \Sigma_{\mu}w_{\mu}\ln w_{\mu}.\label{vN2}
\end{equation}
Here is his argument. Using an ensemble interpretation of probability, he assumes that a mixture 
\beq
U=\Sigma_{\mu} w_{\mu}P_{\ps_{\mu}}\label{vN3}
\eeq
of orthogonal states $(\psi_{\mu})$ with weights $(w_{\mu})$ corresponds to a \emph{classical} ideal gas of $N$ particles, 
with volume $V$ at some  temperature $T$ (which drops out of the calculation), consisting of $w_1N$ quantum systems in state $\psi_1$, $w_2N$ quantum systems in state $\psi_2$, etc.,\footnote{In  eq.\ \er{vN3} $P_{\ps_{\mu}}$ is the projection onto $\ps_{\mu}$, i.e., $P_{\ps_{\mu}}=|\ps_{\mu}\ra\la\ps_{\mu}|$ in Dirac notation.}
each of which is contained in a massive box so that the gas behaves classically. The entropy of this gas is supposed to be the sum of the (so far unknown) quantum entropy $\mathsf{S}\equiv \mathsf{S}(U)$ of $U$ and the classical entropy $S(N,V,T)$ of this ideal gas. To compute 
$\mathsf{S}(U)$, von Neumann takes a similar gas in which all $N$ quantum states equal some pure state $\phv$, which he assumes can be reversibly and hence isentropically transformed into any of the $\ps_{\mu}$. He then  encloses some
such $\phv$ into a volume $w_1 V$, the next into $w_2 V$, etc., transforms $\phv$ in $w_1V$ into $\psi_1$ (also in $w_1V$), etc., and finally expands each $w_{\mu}V$ to $V$ so as to obtain the desired gas of $w_{\mu}N$ quantum systems in state $\psi_{\mu}$ for each $\mu$, all contained in $V$. Its entropy is $\mathsf{S}(U)+S(N,T,V)$, and it can only differ from the  entropy $\mathsf{S}(P_{\phv})+S(N,T,V)$ of the original gas by the entropy $\Sigma_{\mu}\Dl S_{\mu}$ of the isothermal expansion of the volumes  $w_{\mu}V$ to $V$, for each $\mu$. The latter is computed for an ideal gas and gives 
\beq
\Dl S_{\mu}=kw_{\mu}\ln(V/w_{\mu}V)=-Nkw_{\mu}\ln w_{\mu},
\eeq
 so that
\begin{equation}
\mathsf{S}(U)+S(N,T,V)=\mathsf{S}(P_{\phv})+S(N,T,V)-Nk\Sigma_{\mu}w_{\mu}\ln w_{\mu}.\label{vN4}
\end{equation}
Finally, von Neumann assumes that pure quantum states have zero entropy, i.e., $\mathsf{S}(P_{\phv})=0$, so that \er{vN4} and \er{vN2} yield \er{vN1}. Of course, this assumption is then returned by \er{vN1} or \er{vNnow}.

Most interesting generalizations of \er{vNnow}  have been formulated in the language of \vna s,\footnote{This is also true for the quantum relative entropy \er{QKL} introduced by Umegaki (1962), since, although the formula does not show this, his proposal came from his work on conditional expectations in \vna s.} which von Neumann introduced in a series of papers published between 1930 and 1949,  partly in collaboration with his assistant Murray.\footnote{See Kadison (1982), Petz \&  R\'{e}dei (1995), and Landsman (2017),  Appendix C, for references and  information.
}  His motivation for doing so included \qm,
 functional analysis, measure theory, ergodic theory, and \rep\ theory, all of which fields in turn benefited from 
 their interaction with operator algebras. But even beyond providing the mathematical language for quantum entropies of infinite systems, as we will review in some detail in these notes,\footnote{See  R\'{e}dei (1996) and Valente (2008) for background to and somewhat different formulations of our thesis.}
 we suspect that von Neumann also rejoiced in the fact that whereas the usual density operators $\rh$ on an infinite-dimensional \Hs\ $H$ do not include a maximum entropy or uniform prior state $\rh_m$ (which in finite dimension is $\rh_m=1_H/\dim(H)$, where $1_H$ is the unit operator on $H$ and $\dim(H)$ is the dimension of $H$), some infinite-dimensional \vna s do, namely his favourite $\mathrm{II}_1$ factors, simply via
 \beq
 \rh_m=1_H.
 \eeq

As illustrated by our opening quote, von Neumann's influence on the history of entropy even included a crucially important role of entropy in modern society, namely in information theory.\footnote{ Guizzo (2003) and Gleick (2011) are histories of information theory.}
In the founding paper of this field,\footnote{The connection between entropy and information was made earlier by Szilard (1929) in connection with Maxwell's demon. See  Earman \& Norton (1998, 1999) and Maroney (2009) for a critical philosophical analysis.}
  Shannon (1948) asked how much ```choice'' is involved in the selection of an event' that is drawn from a finite  space of events with probabilities $p_1, \ldots, p_n$, or `how uncertain we are of the outcome'. His answer is what is now called the \emph{Shannon entropy} (which he quite rightly attributes to Boltzmann):\footnote{Shannon (1948) gives no reference but mentions `Boltzmann's famous $H$ theorem.'   } 
 \begin{equation}
S_{2}(p):=-\Sigma_{a\in A} p(a)\log_{2} p(a).\label{SE}
\end{equation}
In other words, $S_2(p)$ is the expectation value under $p$ of the $p$-dependent ``information'' function
  \begin{equation}
 I_2(a)=\log_2(1/p(a)). \label{info}
\end{equation}
Some first intuition for these formulae comes from taking a flat distribution $p=f$, defined by
\begin{equation}
f(a):=1/|A|, \label{flat}
\end{equation}
in which case one obtains 
\beq
S_2(f)=\log_2 |A|, \label{logA}
\eeq
which is a poor man's version of Planck's formula \er{BPE}.  In the further special case where 
\beq
|A|=2^k,
\eeq
 so that $S_2(f)=k$, the Shannon entropy \er{SE} gives the minimal number of bits necessary to encode the elements $a\in A$ (or, equivalently, the minimal number of yes-no questions one needs to ask to figure out which element the opponent has in mind in a systematic procedure not relying on guesswork and luck). The general case is covered by Shannon's  coding theorems, which we will briefly discuss in \S\ref{Ecoding}.  The quantity \er{info}, which as we have seen was already used by Gibbs (1902), may also be interpreted 
 as a quantification of the   \emph{surprise} brought by an outcome $a\in A$ upon sampling a probability space $(A,p)$. The idea is that the surprise should be great if the outcome is unlikely, suggesting a ``surprise function'' $s(a)\sim 1/p(a)$. But one likes this function to be additive on pairs of independent events, which rather suggests \er{info}. 
Thus $S_2(p)$ is also the ``average surprise'' inherent in $p$.   In particular, if $p=p_b$ for some $b\in A$ (where $p_b(a)=p_{ab}$), then $S_2(p)=0$, and indeed there is no surprise at all since the outcome is certain to be $a$. On the other hand, the flat or uniform probability distribution \er{flat}  maximizes the average surprise:
\bex\label{boundsSc}
Show that $0\leq S(p)\leq \log(|A|)$ for any $p\in\Pr(A)$, and also that:
\begin{enumerate}
\item  $S(p)=0$ iff $p=p_b$ for some $b\in A$.
\item $S(p)= \log(|A|)$ iff $p=f$.
\end{enumerate}\eex
Apart from some similar heuristic motivation for the use of the logarithm, Shannon also gave an axiomatic characterization of \er{SE}, of which we give the following very slightly different and streamlined version,
where $\Pr(A)$ denotes the set of all probability distributions on $A$:\footnote{Shannon tacitly assumed that
$S_2(p)$  is defined by the same formula for any $A$. Khinchin (1957) made this explicit.}
\begin{enumerate}
\item  $S_2(p)\equiv S_2(p_1, \ldots, p_n)$ is continuous in the probabilities $p_i\equiv p(a_i)$, where $n=|A|$.
\item If $p(a)=1/n$ for all $a\in A$, then $g(n):= S_2(1/n, \ldots, 1/n)$ satisfies $g(n+1)\geq g(n)$ for all $n$.
\item If $p'$ is obtained from $p$ by omitting all events with zero probability, then $S_2(p')=S_2(p)$.
\item  If $r\in\Pr(C)$, where $C$ is finite like $A$, and we coarse-grain   $C=\bigsqcup_{a\in A} U_a$ into disjoint events $U_a\subset C$, with associated distribution $p\in\Pr(A)$ given by $p(a)=\Sigma_{c\in U_a} r(c)$, then 
\begin{equation}
S_2(r)=S_2(p)+\Sigma_{a\in A} p(a) S_2(r|a), \label{Khin0}
\end{equation}
where the conditional probability  $r|a\in\Pr(C)$ is given by $r|a(c)=r(c|U_a)$.\footnote{The usual  conditional probability $r(c|U_a)$ equals zero if $c\notin U_a$ and equals $r(c)/r(U_a)$ if $c\in U_a$.}
\item $S_2(\half,\half)=1$. 
\end{enumerate}
\bex
Show that  \er{SE} satisfies  conditions 1--5, and that conversely 1--5 imply \er{SE}.
\eex

As far as we know this marked the beginning of the axiomatic study of entropy.\footnote{See Acz\'{e}l \& Dar\'{o}czy (1975), Klir (2006), Csisz\'{a}r (2008), and Leinster (2022).  One has \emph{split additivity} and \emph{subadditivity} characterizations of entropy. Let $\CP$ be the set of all probability distributions on finite sets, and assume that $H:\CP\raw[0,\infty)$ is not identically zero and is such that for any $A$ and any bijection 
 $f:A\raw B$ and any
$p\in\Pr(A)$ we have $H(p\circ f)=H(p)$. Axioms 1, 3, 4, and 5 (with $S_2\leadsto H$), of which 4 is  split additivity, then
yield $H=S_2$ (Jak\v{s}i\'{c}, 2018, Theorem 3.4). 
 A set of axioms based of the second sort  that also yields $H=S_2$ is  given by 1, 3, and 5, but 4 replaced by the subadditivity property
$H(r)\leq H(p)+H(q)$ with equality iff $r=p\x q$
(Jak\v{s}i\'{c}, 2018, Theorem 3.5). } Yet:\footnote{`This theorem' refers to his Theorem 2, which  characterizes
 his formula  \er{SE} for entropy by his axioms.} 
 \begin{quote}
\begin{small}
This theorem and the assumptions required for its proof, are in no way necessary for the present theory. It is given chiefly to lend a certain plausibility to some of our later definitions. The real justification of these definitions, however, will reside in their implications. (Shannon, 1948, p.\ 393)
\end{small}
\end{quote}
We agree! That said, the  important entropy introduced by R\'{e}nyi (1961) was originally motivated by axiomatics.  R\'{e}nyi's one-parameter family of entropies $R_{t}$, where  $t\in (0,\infty)\backslash \{1\}$, is given by 
\begin{equation}
R_{t}(p):=\frac{1}{1-t}\log_2 \Sigma_{a\in A} p(a)^{t}, \label{RE1}
\end{equation}
where $p\in \Pr(A)$ is a probability distribution on a (finite) set $A$. The limit $t\raw 1$ recovers $S_2(p)$, so we may define $R_{t}$ for all $t>0$ by \er{RE1} for $t\neq 1$ and $R_1=S_2$.  To arrive at this expression, R\'{e}nyi weakened Shannon's
axiom 4 above by taking  $C=A\x B$ and $U_a=\{a\}\x B$. The probability distribution $p\in\Pr(A)$ in \er{Khin0} 
and the conditional probability $q|a\in\Pr(B)$  are then given by
\begin{align}
 p(a)=\Sigma_{b\in B} r(a,b); &&
 q|a(b)=r(a,b)/p(a),
\end{align}
  provided that $p(a)>0$; if $p(a)=0$ then the term $p(a) S_2(q|a)$ in the sum over $a$ is omitted. If $A$ and $B$ are  independent, i.e., $r=p\x q$ in the sense that $r(a,b)=p(a)q(b)$, then $q|a=q$, and \er{Khin0} implies that
  \emph{$S_2$ is additive on independent probabilities}, that is,
  \begin{equation}
S_2(r)= S_2(p)+S_2(q).\label{subadditive}
\end{equation}
  This property is strictly weaker than 
 \er{Khin0}: R\'{e}nyi's point was that mere additivity on independent probabilities  allows \er{RE1}, and, as R\'{e}nyi noted, even some other, less interesting possibilities.
 
 As a bridge to Kolmogorov--Sinai entropy, the last to be discussed in this  introduction, and as a means of visualizing the contexts of Boltzmann and Shannon,  consider  the following diagram:\smallskip
\begin{center}
 \includegraphics[width=0.95\textwidth]{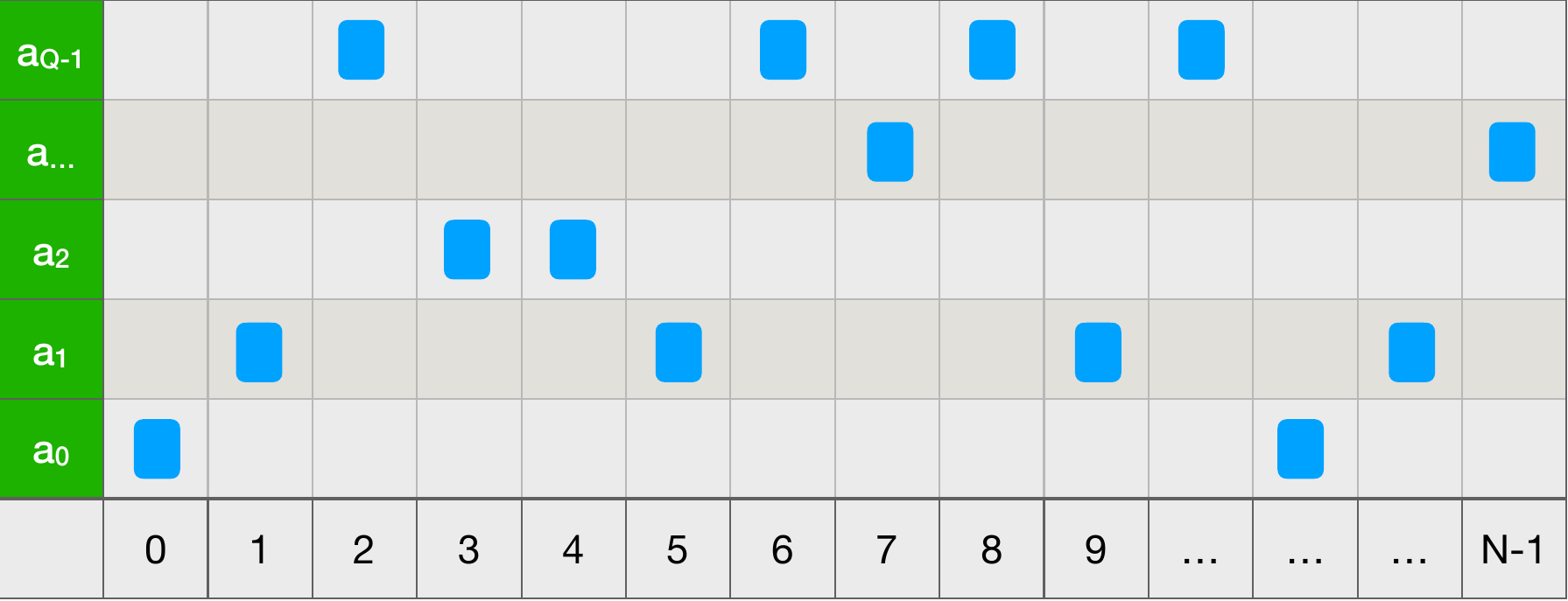}
\end{center}\medskip

Here $N\in\N_*=\{1,2,3,\ldots\}$ is some \emph{large} natural number (we will often consider the limit $N\raw\infty$), whereas $Q=|A|\in\{2,3,\ldots\}$ is the cardinality of a given finite set 
$A=\{a_0, \ldots, a_{|A|-1}\}$. The case of small $|A|$ is already interesting: even $A=\{0,1\}$ and hence $|A|=2$,
 shows most of the richness of the theory. The following set-theoretic notation may require some getting used to:
 \begin{definition}
 \begin{enumerate}
\item The natural number $N\in\N$ is identified with the set $N=\{0, 1, \ldots, N-1\}$.
\item The Cartesian product $A^N$ is interpreted as the set of all functions $\sg:N\raw A$.
\item Such a function is also called a \emph{string} over $A$, having length $\ell(\sg)\equiv |\sg|=N$. 
\item 
 We write either $\sg(n)$ or $\sg_n$ for its value at $n\in N$, and sometimes write $\sg$ as $\sg_0\sg_1\cdots \sg_{N-1}$. 
 \item Likewise, $A^\N$ is the set of all functions $s:\N\raw A$, and such an $s$ is called a \emph{sequence} over $A$.
 \item  For $s\in A^{\N}$, we write $s_{|N}$ for $s_0s_1\cdots s_{N-1}\in A^N$ (to be  distinguished from $s_n\equiv s(n)\in A$).
\end{enumerate}
\end{definition}
In particular, if $A=2=\{0,1\}$, then $\sg$ is a binary string, most simply seen as a possible outcome of $N$ coin tosses.
 Thus a \emph{string} $\sg$ is \emph{finite}, whereas a \emph{sequence} $s$ is \emph{infinite}. The set of all binary sequences is therefore denoted by $2^{\N}$; it is one of the most important infinite sets in this course. 
 
We have already seen two interpretations of the above diagram:
\begin{itemize}
\item In \emph{statistical mechanics} \`{a} la
Boltzmann (1877), $N$ is the number of (distinguishable) particles in a gas, and  $a\in A$ labels some property a single particle may have. For example, for non-interaction particles one may  think of $a\in A$ as a label of their energy $E_a\in\R$. Spin chains also fall under this formalism, where $a\in A$ labels some internal degree of freedom at site $n$. In statistical mechanics $\sg\in A^N$ is called a  \emph{microstate} of the gas (or spin chain, etc.). 
    \item In \emph{information theory} as created by Shannon (1948), $N$ is the number of letters drawn from an alphabet $A$ by sampling a given probability distribution $p
\in\mathrm{Prob}(A)$, the space of all probability distributions on $A$. 
So each ``microstate'' $\sg\in A^N$ is a word with $N$ letters.
\end{itemize}
More generally, our diagram shows a sample path of
some discrete-time stochastic process with a finite state space.\footnote{This interpretation again goes back to Kolmogorov (1933). See Appendix \ref{MarC} for details as well as the general case.} That is, one has some probability space  $(\Om,\Sg, P)$, a finite set $A$, and a sequence 
$X_n:\Om\raw A$ of measurable maps, where $n\in N$ (with eventually $N\raw\infty$, so that $n\in \N$).
A \emph{sample path} of such a process is just an $A$-valued string  $(X_n(\om))_{n\in N}$.  Moreover, 
the \emph{Kolmogorov representation theorem} states that without loss of generality one may even assume that
$\Om= A^N$ and  $X_n(\sg)=\sg_n$, so that $P\in\Pr(A^N)$. Thus the previous two cases fall under this general scope. 

\noindent Kolmogorov (1958) gave a another new twist to our diagram by interpreting it as follows.\footnote{See  Sinai (1989)  Charpentier, Lesne, \& Nikolski (2007) for Kolmogorov's contributions to dynamical systems, entropy, and ergodic theory. Textbooks include  Collet \& Eckmann (2006),  Castiglione \emph{et al.} (2008), and Viana \& Oliveira (2016). Kolmogorov's  motivation was to find a new invariant for dynamical systems. The underlying formalization of probability theory also goes back to Kolmogorov (1933), as does algorithmic randomness (Appendix \ref{EKR})!} 
\begin{itemize}
  \item A \emph{dynamical system} (in the probabilistic sense we use) is a triple $(X,P,T)$, 
  where $(X, P)$ or more precisely $(X,\Sg, P)$ is a probability space (where $\Sg$ is some $\sg$-algebra on $X$), and 
  \beq
  T:X\raw X \label{TXX00}\eeq
   is a measurable (but not necessarily invertible) map,   required to preserve $P$ in the sense that 
  \beq P(T\inv B)=P(B), \label{PTB00}
  \eeq
  for any measurable $B\subseteq X$ ($B\in\Sg$). Here  $T\inv(B):=\{x\in X\mid T(x)\in B\}$; 
  if $T$ is invertible, this set coincides with the image $\{T\inv(x)\mid x\in B\}$ of $B$ under the inverse map $T\inv: X\raw X$. 
\item A \emph{partition} of $(X,\Sg)$ is  a collection of measurable subsets $\pi=(U_a)_{a\in A}$ of $X$ such that
\begin{align}
P\left(\bigcup_a U_a\right)=1; && P(U_a\cap U_b)=0 \:\:\: (a\neq b).
\end{align}
 \item  A microstate  $\sg\in A^N$ now encodes a coarse-grained trajectory \emph{of a single particle} that 
  starts at some point $x\in U_{\sg(0)}$ at $t=0$, then hops, in discrete time $t\in \N$,  to $Tx\in U_{\sg(1)}$ at $t=1$, \ldots, on to  $T^nx\in U_{\sg(n)}$ at $t=n$, etc., ending at $T^{N-1}x\in U_{\sg(N-1)}$ at
time $t=N-1$. Defining
 \begin{equation}
 U_{\sg}:= \{x\in X\mid x\in U_{\sg(0)}, Tx\in U_{\sg(1)}, \ldots T^{N-1}x\in U_{\sg(N-1)}\},   \label{Xsg01} 
\end{equation}
we may then identify the probability of the coarse-grained path labeled by $\sg$ with $P(U_{\sg})$.
 \end{itemize}
 Kolmogorov's idea was now to define an entropy of a dynamical system $(X,P,T)$ by starting from
       \begin{align}  
 H^{(N)}_P(\pi)&:=-\Sigma_{\sg\in A^N} P(U_{\sg})\log P(U_{\sg});\label{preKS1}\\
 h_P(\pi) &:= \lim_{N\raw\infty} \frac{1}{N} H^{(N)}_P(\pi),
 \end{align}
where the limit in question turns out to exist.  But $h_P(\pi)$ clearly 
depends on the  partition $\pi$. If one wants to see the effect of some particular partition this may be desirable, but otherwise 
the \emph{Kolmogorov--Sinai} (or \emph{metric}) \emph{entropy} $h(X,P,T)$ of $(X,P,T)$ is defined as 
 the supremum of $h_P(\pi)$ over all finite measurable partitions $\pi$ of $X$. This  is  computable by the \emph{Kolmogorov--Sinai theorem} (Theorem \ref{KSGen}), and it turns out to be  useful: two  examples are the role of  $h(X,P,T)$  as an \emph{invariant of dynamical systems} (which was Kolmogorov's original motivation), and \emph{Pesin's theorem} (or formula) relating it to the sum of the positive Lyapunov exponents of the flow.\footnote{See, for example, Katok \& Hasselblatt (1995), Supplement, or Collet \& Eckmann (2006), Theorem 6.17.}

All entropies discussed in this course (and most entropies altogether) derive from those introduced by Clausius, Boltzmann, Gibbs, Einstein, von Neumann, Shannon, R\'{e}nyi, or Kolmogorov.  That said, a crucial innovation, central to our course,  has been a move from single-variable entropies to relative ones (which have two slots). Since, apart from giving appropriate references, the history of this move remains to be written and hence is beyond the scope of this course, 
we have finished our historical introduction and move to an outline of the course and its structure.
\section{Overview of the course}\label{section2}
Boltzmann's (alleged) formula ``$S=k \log W$'' is misleading in that it hides the crucial fact that
\emph{all (well-defined) classical entropies are a function of some quantity:} for example, the Clausius entropy is a function of \emph{energy};\footnote{And typically also of volume,  particle number, and perhaps other quantities, yet energy comes first.} both Boltzmann's entropies from 1872 and 1877 are functions of \emph{probability measures} (or distributions), as is the Gibbs entropy from 1902, where the sets or spaces on which these probabilities are defined are different in each case. \emph{Et cetera}. As we see it, the key to understanding (and computing) classical entropies  is that the argument of some specific entropy points to some fluctuating quantity 
whose large deviations from its equilibrium value (in a  sense to be explained shortly) decay exponentially at a rate controlled by the associated entropy. 

This point of view, which as we have seen goes back to Einstein, has been made  rigorous through the theory of large deviations.\footnote{This theory has roots in both physics and mathematics. 
As a chapter in probability theory, inspired by earlier results by Cram\'{e}r and Sanov (to be discussed shortly) as well as by Brownian motion and other aspects of probability theory not discussed in these notes,   this theory was developed mainly by Donsker and Varadhan in a series of papers in the mid 1970s, referred to and summarized in the lecture notes by Varadhan (1984, 2016). This crossed independent work on mathematically rigorous statistical mechanics by Lanford (1973),  indebted to Ruelle (1969) and others, which of course stands in the long tradition of statistical physics and entropy partly reviewed in our historical introduction. These two apparently independent sources
 led to a synthesis in the 1980s, reflected for example in the leading textbooks on large deviations by Ellis (1985),  Dembo \& Zeitouni (1998), den Hollander (2000), and Rassoul-Agha \& Sepp\"{a}l\"{a}inen (2015).
See also  Dorlas (2021) for an excellent statistical mechanics course based on large deviations.} 
The foundation of this theory (at least in our setup), which also connects well to the work of Boltzmann, is \emph{Sanov's theorem}. Inspired by \er{deff} and \er{LN0}, in the context of our diagram above, and hence for finite $A$,  consider  the following objects:
 \begin{align}
 L_N: A^N\raw \Pr(A); && 
L_N(\sg)=\frac{1}{N}\Sigma_{n=0}^{N-1} \dl_{\sg_n};\label{ST11}\\
T_N: \Pr(A)\raw \mathcal{P}(A^N); && T_N(p):=\{\sg\in A^N\mid L_N(\sg)=p\},
\end{align}
where, for any $a\in A$, the point measure  $\dl_a\in\Pr(A)$ gives $\dl_a(B)=1$ if $a\in B$ and $\dl_a(B)=0$ if $a\notin B$, where $B\subseteq A$. Here $L_N$ is called the \emph{empirical measure}  and 
$T_N$ is called the \emph{type class}. The former reads the configuration of squares in our diagram and computes the corresponding relative frequency (seen as a probability) that some square enters $a\in A$.
The latter \emph{starts} from a  probability distribution on $A$ and assembles all microstates whose empirical probability  equals the given one. First, a computation going back to Boltzmann (1877) recovers the Shannon entropy \er{SE} via
\begin{equation}
\lim_{N\raw\infty}\frac{1}{N} \log |T_N(p_N)|=S(p):=-\Sigma_{a\in A} p(a)\log p(a), \label{arche0}
\end{equation}
for any sequence $(p_N)$  in $\Pr(A)$ such that $p_N\in L_N(A^N)$ and 
 $p_N\raw p$ in $\Pr(A)$ weakly (here we use base-$e$ logarithms and $\log\equiv \ln$).
Furthermore, the strong law of large numbers gives
\begin{equation}
\lim_{N\raw\infty}L_N(s)=q, \label{24n}
\end{equation}
weakly, for any prior $q\in\Pr(A)$ and $q^\N$-almost every sequence $\in A^{\N}$, where we regard $L_N$ as a function from $A^{\N}$ to $\Pr(A)$ just as in \er{ST11}, so that $L_N(s)$ only depends on the initial segment $s_{|N}$.

Sanov's theorem studies the ``large'' fluctuations of $L_N$ around this limit for finite $N$ (but asymptotically in $N$). 
One of the lessons of large deviation theory is that entropies with a single argument like $S(p)$ are usually special cases of so-called \emph{relative entropies}, which have two arguments of the same kind. And it is these relative entropies that appear in deeper asymptotic results.

 In the case at hand, the  relative entropy is the \emph{Kullback--Leibler distance} (or \emph{divergence}):\footnote{The original reference is Kullback \& Leibler (1951). This entropy was rediscovered by Jauch \& Baron (1972).}
\begin{equation}
S_q(p)\equiv S(p, q):=\Sigma_{a\in A} p(a)\log\left(\frac{p(a)}{q(a)}\right), \label{KLD}
\end{equation}
which is defined as stated  provided $q(a)=0$ implies $p(a)=0$ (i.e.,  $p\ll q$, or: $p$ is absolutely continuous w.r.t.\ $q$); if not, we put $S(p, q)=\infty$. For the uniform prior $q=f$, see \er{flat}, we have
\beq
S(p,f)=-S(p)+\log |A|. \label{KLflat}
\eeq 
Sanov's theorem then states that if $q(a)>0$ for all $a\in A$ and $U\subseteq\Pr(A)$ is open,\footnote{Here we once again use
the weak topology on $\Pr(A)$, defined in the next section.}
 then
\begin{equation}
\lim_{N\raw\infty} \frac{1}{N} \log q^N(L_N\in U)=-
 \inf_{p\in U} S(p, q),\label{135}
\end{equation}
where $q^N$ is the product (Bernoulli) probability measure on $A^N$ induced by $q$, that is,
\begin{equation}
q^N(\sg):=q(\sg_0)\cdots q(\sg_{N-1})=\prod_{n=0}^{N-1}q(\sg_n). \label{defqN}
\end{equation}
 If $q\in U$, then the right-hand side of \er{135} vanishes, since obviously $S(q, q)=0$, 
  and this forces $q^N(L_N\in U)\raw 1$, consistent with \er{24n}. But if $q\notin U^-$ (the closure of $U$),  then 
   $S(p, q)>0$  for all $p\in U^-$ (we will see later that $S(p, q)\geq 0$ with equality iff $p=q$), and hence
  $\inf_{p\in U} S(p, q)>0$. Thus $q^N(L_N\in U)$ exponentially decays in $N$, i.e.,
for some $d>0$ and sufficiently large $N$ we have
\begin{equation}
q^{N}(L_N\in B)\leq e^{-dN}.\label{4.17b}
\end{equation}
Moreover, similar to the Gibbs variational princples \er{FU} - \er{Sp} we have a \emph{Fenchel dual pair}
 \begin{align}
 \Pi_q(\phv)&=\sup_p\{ \la \phv\ra_p-S_q(p)\}; \label{0FDDd} \\
S_q(p) &=\sup_{\phv}\left\{\la \phv\ra_p-\Pi_q(\phv)\right\}, \label{0FDDi} 
\end{align}
where the first supremum is taken over all  probability distributions $p\in\Pr(A)$ on $A$, the second  over all functions $\phv:A\raw\R$, and the pressure $\Pi_q$ is defined in terms of a partition function $Z_q$ by
\begin{align}
\Pi_q(\phv):= \log Z_q(\phv); &&
Z_q(\phv):=\la e^{\phv} \ra_q=\Sigma_{a\in A} q(a) e^{\phv(a)}. \label{0FandP} 
\end{align}

The reason \er{135} is called a \emph{large deviation} result is clearer from the 
 oldest  example of large deviation theory,  namely Cram\'{e}r's theorem.\footnote{The original source is Cram\'{e}r (1938), but the version of Theorem \ref{cramer} goes back to Chernoff (1952).  }
 Keeping $A$ finite for simplicity, equipped with a prior $q\in\Pr(A)$ as before, take some function 
$E:A\raw\R$, which is conceptually a real random variable on $(A,q)$. Just think of a coin toss, where $A=\{0,1\}$ and $E(a)=a$. Then the average  
\begin{equation}
S_N(\sg):=\frac{1}{N}\Sigma_{n=0}^{N-1} E(\sg_n)\label{0defXNov}
\end{equation}
 is  a real random variable on the  probability space $(A^N,q^N)$. If similarly to \er{24n} we (re)define \er{0defXNov}  on $A^{\N}$ equipped with the Bernoulli measure $q^\N$, then by the strong law of large numbers  $S_N(s)$ converges to   $\la E\ra_q\equiv \mu$ for $q^\N$-almost every $s\in A^{\N}$.
  One then has three interesting regimes:
\begin{enumerate}
\item The small fluctuations around the limit $\mu$ are described  by the \emph{weak law of large numbers}:
\begin{equation}
\forall_{\varep>0}\, \lim_{N\raw\infty} q^N\left(-\varep\leq S_N-\mu\leq \varep\right)=1. \label{wlln5}
\end{equation}
\item The $O(N^{-1/2})$ fluctuations of $S_N-\mu$ are described by the \emph{central limit theorem}:
\begin{equation}
\lim_{N\raw\infty} q^N\left(\frac{a}{\sqrt{N}}\leq S_N-\mu\leq \frac{b}{\sqrt{N}}\right)=\frac{1}{\sqrt{2\pi\sg^2}}\int_a^b dx\, e^{-x^2/2\sg^2}, \label{clt}
\end{equation}
where $\sg^2=\la E^2\ra_q-\la E\ra^2_q$ is the variance of $E$ (which is finite for finite $A$).
\item The $O(1)$ or ``large'' fluctuations, then, are described by \emph{Cram\'{e}r's  theorem}:
\begin{equation}
\lim_{N\raw\infty} \frac{1}{N} \log q^N(a\leq S_N-\mu\leq b)= -\inf_{x\in [\mu+a,\mu+b]} I_q(x),\label{78a}
\end{equation}
where the rate of this exponential decay $I_q$ is defined by a ``maximum entropy principle":
\begin{align}
I_q(x):=\inf\{S_q(p)\mid p\in\Pr(A), \la E\ra_p=x\}.
\label{IMEPI}
\end{align}
\end{enumerate}
And also in this case we alternatively have a  dual pair description of the rate function $I_q$, namely
\begin{align}
\Pi_q(t)&=\sup_{x\in\R}\{xt-I_q(x)\};\\
I_q(x)&=\sup_{t\in\R}\{xt-\Pi_q(t)\},  \label{I0qC1}
\end{align}
where once again the rate function is given by  pressure- and partition-function like objects, viz.
\begin{align}
 \Pi_q(t):=\log Z_q(t); && Z_q(t):= \la e^{tE}\ra_q= \Sigma_{a\in A} q(a) e^{tE(a)}. \label{0tildeP}
\end{align}

The course, then,  starts with asymptotics of both single-argument (Boltzmann--Shannon)  and double-argument (relative) entropies, culminating--after a brief and easy detour on Shannon's simplest coding theorems which also rely on these asymptotics--in the theorems of Sanov and Cram\'{e}r. On the basis of these results we then discuss the abstract and general theory of large deviations,
covering for example key theorems by Varadhan, Bryc, G\"{a}rtner--Ellis, and Kifer. 
We typically \emph{prove} results for finite sets $A$, and merely \emph{state} their generalization to Polish spaces, which form a wide class of topological and hence measure spaces on which probability theory works well. 

It would be nice if there were a quantum theory of large deviations, but so far this has not been found. In the literature one finds various ``quantum Sanov'' theorems, but (with some very few and recent exceptions that remain to be widely accepted) these are in fact quantum versions of \emph{Stein's lemma}, which is a result on the error rate of asymmetric hypothesis testing. This lemma may be derived from Sanov's theorem, but it is not  equivalent to it. Nonetheless, Stein's lemma and some closely related asymptotic results named after Chernoff and Hoeffding provide a beautiful bridge between classical and quantum entropies in the sense that these results have adequate quantum versions that play the same role in quantum hypothesis testing. Even apart from the link with quantum theory, hypothesis testing provides a very nice application of large deviation theory. 

The idea is simple. Given a finite set $A$ from which we repeatedly draw elements, we  try to determine the probability distribution $p\in\Pr(A)$ supposed to control the sampling. We only consider the case where just two hypotheses are considered, say $p=p_0$ and $p=p_1$. 
 Let $\mathsf{H}_i$ be the hypothesis that $p_i$ is true ($i=0,1$).  If we only draw once and get $a\in A$, 
 it seems reasonable to:\footnote{One may accept either hypothesis if $p_0(a)=p_1(a)$; we choose to accept  $\mathsf{H}_1$ in that case. Also, $B^c=A\backslash B$. }
 \begin{itemize}
\item accept $\mathsf{H}_0$ (i.e., $p=p_0$)  if $p_0(a)>p_1(a)$;
\item accept $\mathsf{H}_1$ (i.e., $p=p_1$) if $p_0(a)\leq p_1(a)$;
\end{itemize}
This is a special case of a \emph{test} $T\subset A$ that guides our choice based on the outcome of the draw: if $a\in T$ is drawn we accept $\mathsf{H}_0$, whereas $a\notin T$ points to $\mathsf{H}_1$. The above procedure corresponds to  
\beq
T^*=\{a\in A \mid p_0(a)>p_1(a)\}.\label{NPT}
\eeq
This may be repeated $N$ times, which is the same as a single ``draw'' $\sg$ from $A^N$. We  assume that the different draws from $A$ are independent and identically distributed (i.i.d.), so that
 $\mathsf{H}_i$ states that $p^N_i$ is true ($i=0,1$), and a test $T_N\subset A^N$ means that drawing $\sg\in T_N$ makes us choose $\mathsf{H}_0$, etc.
Alas, any (finite) test is liable to false inferences, whose probabilities we wish to compute. In particular:
\begin{itemize}
\item $p^N_0(T_N^c)$ is the  \emph{false positive}  probability that $\mathsf{H}_1$ is accepted on the basis of the observations $\sg$, namely in case that $\sg\in T_N^c$,  although in fact $\mathsf{H}_0$ is true (which explains the $p^N_0$).
\item $p^N_1(T_N)$ is  the  \emph{false negative} probability that $\mathsf{H}_0$ is accepted from  $\sg$, although $\mathsf{H}_1$ is true. 
\end{itemize}
All kinds of combinations of these error probabilities 
have been considered. A false negative conclusion (in which for example an actual disease is not treated) is usually more damaging than a false positive one (in which case unnecessary treatment is given), so one should make $p^N_1(T_N)$ as small as possible without trivially making it zero  by taking $T_N=\emptyset$. A more sophisticated procedure, then, is to accept an error  $0<\varep<1$ in $p^N_0(T_N^c)$ and then minimize $p^N_1(T_N)$, by defining
\begin{equation}
\beta_N^{\varep}(p_0,p_1):=\inf_{T_N\subset A^N}\{p^N_1(T_N)\mid p^N_0(T_N^c)<\varep\}, \label{2.22I}
\end{equation}
Stein's lemma then yields a spectacular operational appearance of the relative entropy \er{KLD}, viz.
\begin{equation}
\lim_{N\raw\infty}\frac{1}{N}\log \beta^{\varep}_N(p_0,p_1)=-S(p_0, p_1). \label{CSLI}
\end{equation}
This can be proved using large deviation theory, for example from Sanov's theorem, and in practice is taken to be one of the main guides for interpreting quantum entropies and justifying specific expressions thereof, which  are initially guessed on the basis of their classical counterparts.\footnote{As we have seen, von Neumann made a similar, now forgotten effort at an operational justification of \er{vNnow}.}

In quantum mechanics on finite-dimensional \Hs s $H$, comparable to  finite sets $A$ in the classical case, one simply replaces the probability space $(A,p)$ by a pair $(H,\rh)$, where $\rh\in D(H)$ is a density matrix, i.e., $\rh\geq 0$ and $\Tr(\rh)=1$. 
The (binary) hypothesis testing situation is similar to the classical case, in the sense that one wishes to find out through tests if some unknown $\rh\in D(H)$ is equal to either $\rh_0$ or $\rh_1$, where now an old-school test is a \emph{projection} $T:H\raw H$
(i.e., $T$ is linear and $T^2=T^*=T$), whose measurement, by orthodoxy, comes down to drawing not some \emph{arbitrary} unit vector $\psi$ in $H$ (up to a phase, so really:  a one-dimensional projection $|\psi\ra\la\psi|$), but a unit \emph{eigenvector} of $T$. For a nontrivial projection (that is, $T\neq 0$ and $T\neq 1_H$), hence with spectrum $\sg(T)=\{0,1\}$,  this just leaves two possibilities, $T\psi=\psi$ (eigenvalue $\lm=1$) and $T\psi=0$ (eigenvalue $\lm=0$). As a small notational nuisance the first confirms $\mathsf{H}_0$ and the latter $\mathsf{H}_0$, but otherwise the classical story can be copied, \emph{mutatis mutandis}. In particular, an $N$-fold measurement corresponds to measuring a projection $T_N$ on $H^{\ot N}\equiv H^N$, 
and so by the Born rule:
\begin{itemize}
\item $\Tr_{H^N}(\rh^N_0(1_{H^N}-T_N))$  is the false positive probability that $\mathsf{H}_1$ is accepted although  $\mathsf{H}_0$ is true.
\item $\Tr_{H^N}(\rh^N_1T_N)$ is  the false negative probability that $\mathsf{H}_0$ is accepted although $\mathsf{H}_1$ is true. 
\end{itemize}
Hence analogously to \er{2.22I}, but now taking  the infimum  over all projections on $H^N$,
we put \begin{equation}
\beta_N^{\varep}(\rh_0,\rh_1):=\inf_{T_N}\{\Tr_{H^N}(\rh^N_1T_N)
\mid \Tr_{H^N}(\rh^N_0(1_{H^N}-T_N))<\varep\}. \label{2.22IQ}
\end{equation}
In fact, since modern quantum measurement theory is based on so-called \textsc{povm}s (\emph{Positive Operator Valued Measures}),\footnote{Standard textbooks about quantum mechanics based on \textsc{povm}s are De Muynck (2002) and Busch et al. (2016).} one could take the infimum over all operators $0\leq T_N\leq 1_{H^N}$ (sometimes called \emph{effects}), but it turns out that the ensuing formulae are the same, basically because the infimum is in fact attained by a projection. 
Either way, an early succes of quantum hypothesis testing was the \emph{quantum Stein lemma},\footnote{See
   Hiai \& Petz (1991 and Ogawa \& Nagaoka (2002). For surveys see Hayashi (2017) and Khatri \& Wilde (2024).} which gave a perfect  quantum analogue of \er{CSLI}, namely
   \begin{equation}
\lim_{N\raw\infty}\frac{1}{N}\log \beta^{\varep}_N(\rh_0,\rh_1)=-S(\rh_0, \rh_1),\label{2.25I}
\end{equation}
where the \emph{relative von Neumann entropy} (due to Umegaki) is defined for $\rh,\sg\in D(H)$ by 
\begin{equation}
S(\rh,\sg):=\Tr(\rh(\log\rh -\log\sg)), \label{Umegaki}
\end{equation}
at least if $\ker(\sg)\subseteq\ker(\rh)$; if not,\footnote{We also define $\rh\log\sg=0$ on $\ker(\sg)$ as well as $\rh\log\rh=0$ on $\ker(\rh)$. Thus $\ker(\sg)\subseteq\ker(\rh)$ is the quantum analogue of the condition $q\ll p$ in the definition of the classical relative entropy $S(p, q)$.
The relative von Neumann entropy was introduced by
Umegaki (1962) as part of his study on conditional expectations on \vna s. 
} we put $S(\rh\|\sg):=\infty$. Compare the classical version \er{KLD}.

An important application of quantum hypothesis testing is to \emph{entanglement detection}, in which  
\beq
H=H_A\ot H_B\equiv H_{AB}, \label{HABI}
\eeq
 and one tries to find out if some unknown density matrix $\rh_{AB}\in D(H_{AB})$ is either \emph{separated}, i.e., 
 \beq
 \rh_{AB}=\Sigma_i\rh_A^{(i)}\ot \rh_B^{(i)},\label{Scc0}\eeq
  for certain $\rh_A^{(i)}\in D(H_A)$ and $\rh_B^{(i)}\in D(H_B)$, or \emph{entangled}, which means that it is not separated. If 
  \beq
  \rh_{AB}=|\psi_{AB}\ra\la\psi_{AB}| \label{EPS}
  \eeq
   is pure, the separated case  corresponds to $\psi_{AB}=\psi_A\ot\ps_B$ for certain $\ps_A\in H_A$ and $\psi_B\in H_B$. 
   The most famous examples of entangled pure states are the \emph{Bell states}, where $H_A=H_B=\C^2$, viz.\footnote{Here we use the physicists' notation where $(|0\ra, |1\ra)$ is the standard basis of $\C^2$ and $|00\ra\equiv |0\ra\ot|0\ra$, etc.}
   \begin{align} 
 \psi_{\pm} := \frac{1}{\sqrt{2}}(|01\ra\pm |10\ra; && \varphi_{\pm}:= \frac{1}{\sqrt{2}}(|00\ra\pm |11\ra. \label{Bellstates0}
\end{align}
     Given some entangled density matrix $\rh_1\in D(H_{AB})$, we still take $\mathsf{H}_1$ to be the hypothesis $\rh_{AB}=\rh_1$, but replace $\mathsf{H}_0$  by  the hypothesis that $\rh_{AB}$ is an arbitrary separated state,\footnote{Of course, one could also state that $\rh_{AB}$ is an arbitrary entangled state.} (so that we try to guess if $\rh_{AB}$ is either separated or equal to a  \emph{given} entangled  density matrix).\footnote{Eq.\ \er{HI2025} is due to  Hayashi \&  Ito (2025).
  See also Lami (2025ab) and follow-up papers by  Hayashi and Lami.   }
 Let us replace \er{2.22IQ} by 
  \begin{equation}
\beta_N^{\varep}(\rh_1):=\inf_{0\leq T_N\leq 1_{H^N}}\left\{\Tr_{H^N}(\rh^N_1T_N)
\mid \sup_{\rh_0\in S_{AB}} \{
\Tr_{H^N}(\rh^N_0(1_{H^N}-T_N))\}<\varep\right\},\label{2.22QHI}
\end{equation}
where $S_{AB}$ is the set of all separated density matrices on $H_{AB}$. Then eq.\ \er{2.25I} nontrivially implies
\begin{equation}
\lim_{N\raw\infty}\frac{1}{N}\log \beta^{\varep}_N(\rh_1)=-\inf_{\rh\in S_{AB}} S(\rh, \rh_1).\label{HI2025}
\end{equation}
Formally involving a rate function $I_{\rh_1}(\rh)=S(\rh, \rh_1)$, eq.\ \er{HI2025}
 looks like a large deviation result
\begin{equation}
\lim_{N\raw\infty}\frac{1}{N}\log P_1^N(\rh_{AB}\in S_{AB})=-\inf_{\rh\in S_{AB}} I_{\rh_1}(\rh),
\end{equation}
where the fluctuating quantity  is the density matrix $\rh_{AB}$, 
and $P_1^N(\rh_{AB}\in S_{AB})$ is the  probability 
 that the (wrong) conclusion $\rh_{AB}\in S_{AB}$ (i.e., $\rh_{AB}$ is separated)  is the  \emph{outcome} of an optimal sequence of $N$ repeated measurements of $T_N$ in the entangled state $\rh_1^N$ (so that $\rh_{AB}$ is  \emph{in fact} not separated),
 under the constraint on the sequence $(T_N)$ that if $\rh_{AB}$ were in fact separated, then the probability of wrongly concluding that it equals $\rh_1$ and hence is entangled is bounded by some $0<\varep<1$. 

Given an entangled pure state \er{EPS}, most physicists would compute its \emph{entanglement entropy} 
 \begin{align}
E(\ps_{AB}):= S(\rh_A); && \rh_A:=\Tr_{H_B}(\rh_{AB}), \label{EABI}
\end{align}
where $\Tr_{H_B}\rh_{AB}$ is the partial trace of $\rh_{AB}$ over $H_B$, which results in a density matrix on $H_A$ (this strange construction will fall into place the moment we do algebraic quantum theory--it is just the quantum analogue of taking marginals in classical probability theory). It is a nontrivial fact, though easily proved from the Schmidt decomposition of $\rh_{AB}$, that for any pure state \er{EPS},
 \begin{align}
S(\rh_A)=S(\rh_B); && \rh_B:=\Tr_{H_A}(\rh_{AB}), \label{SRAI}
\end{align}
so that we were justified on omitting a further index $A$ on $E(\ps_{AB})$, which is unambiguously defined.
   
 From the point of view of entropy, a key difference between classical and quantum probability is that although $S(\rh_{AB})=0$ if $\rh_{AB}$ is pure, as in \er{EPS}, unless $\ps_{AB}=\ps_A\ot\ps_B$, in other words: if $\ps_{AB}$ is entangled, we have
 $E(\ps_{AB})>0$. For example, for the Bell states \er{Bellstates0} we have $\rh_A=\rh_B=\half\cdot 1_2$ and hence $E(\ps_{\pm})=E(\phv_{\pm})=\log 2$.
 This cannot happen in classical probability theory, where pure states (point measures) always reduce to pure states and all of these have zero entropy. 

Although we will briefly explain such results and will provide the necessary background, most readers are probably familiar with them and instead our main goal is to answer two questions:
\begin{enumerate}
\item How do the quantum entropies \er{vNnow} and \er{Umegaki} and related constructions (as well as their applications to e.g.\  hypothesis testing) generalize to possibly infinite quantum systems?\footnote{In some sense even a particle moving in $\R^3$ is an infinite system, since its associated Hilbert space is infinite-dimensional. But here one can write down \er{vNnow} and \er{Umegaki} \emph{verbatim}, adding some elementary functional analysis.}
\item What replaces the unifying role of the theory of large deviations for classical entropies?
\end{enumerate}
A toolkit for both questions is the theory of \vna s, where the second question more specifically relies on what is called the 
\emph{modular theory}  of such algebras  (also known as \emph{Tomita--Takesaki theory}). This, in turn,  culminated in Connes's classification of hyperfinite factors  (which we will not need) and in Haagerup's theory of noncommutative $L^p$-spaces (which does play an important role in quantum theories of entropy).
Thus a significant part of this course consists of an introduction to \vna s including these more advanced developments, focusing as much as possible on aspects relevant to entropy. Since even basic textbooks on \vna s tend to get lost in technicalities, our  goal lies in conceptual clarification (at no loss of rigour) rather than in providing complete proofs and doing the most general cases; we will be happy if the reader ``sees what is going on'' and feels encouraged to find  the details elsewhere.

Moreover, our  approach is not very much in the spirit of the way \vna s were originally introduced in mathematical physics by von Neumann himself and especially Haag,\footnote{See his textbook (Haag, 1992), and also his recollections (Haag, 2010). The author had the extremely good fortune of being a postdoc in Hamburg during 1993--1994, the last year in which Haag was around almost daily before retiring.} who was initially motivated by quantum field theory and later also by quantum statistical mechanics. It is rather based on the idea that operator algebras (i.e., \vna s as well as the more general C*-algebras, which we also need at certain points and therefore will introduce, albeit in less detail)
form \emph{the right language for combining classical and quantum theory into a single mathematical framework}.\footnote{See Landsman (1998, 2017). Many mathematical people in quantum information theory also takes this view.}
This perspective makes it quite easy to motivate the use of operator algebras initially from the study of finite systems, with a seamless generalization to infinite ones, once the right definitions are in place. In particular, probability measures on some space seen as states in classical physics and density operators on some \Hs\ seen as states in quantum theory become unified via the algebraic concept of a state. Partial traces  also fall into place.

  A \emph{\vna},\footnote{Here being
\emph{unital} means that $M$ contains the unit operator on $H$, and being a \emph{$\mbox{}^*$-subalgebra} means that
$M$ is a vector space (within $B(H)$) that is closed under operator addition, multiplication, and taking adjoints.}
 then, is a unital $\mbox{}^*$-subalgebra $M\subseteq B(H)$ of the algebra $B(H)$ of all bounded operators on some \Hs\ $H$, having a special property that turns out to be so natural and powerful that it can be defined in many seemingly very different ways, e.g.:\footnote{If not used as a definition the first bullet is von Neumann's \emph{bicommutant theorem}, his first result in the subject.}
\begin{itemize}
\item $M''=M$, where $M':=\{S\in B(H)\mid ST=TS \mbox{ for all } T\in M\}$, and $M''=(M')'$;
\item $M=N'$, where $N\subset B(H)$ is some subset closed under involution (adjoint);\footnote{Here one may think of $N=U(G)$ where $G$ is some group and $U:G\raw B(H)$ is a unitary \rep\ on $H$. This encourages the point of view under which \vna s consist of symmetries, i.e., $G$-invariant operators.}
\item $M$ is closed under all kinds of natural operator limits, such as strong and weak ones;\footnote{
We have $T_{\lm}\raw T$ \emph{weakly} if $\la\phv, T_{\lm}\ps\ra\raw \la\phv, T\ps\ra$ for all $\phv,\psi\in H$, and  \emph{strongly} if $T_{\lm}\ps\raw T\ps$ for all $\psi\in H$,}
\item $M$ is a Banach space that is the dual of some Banach space (called its \emph{predual} $M_*$).\footnote{Thus it is assumed that $M$ is already a C*-algebra, defined as a norm-closed $\mbox{}^*$-subalgebras of $B(H)$, with unit.}
\end{itemize}
This gives \vna s all kinds of pleasant features, such as an abundance of projections. If $H$ is finite-dimensional,\footnote{In finite dimension we usually write $L(H)$ for $B(H)$, the space of all bounded linear operators on $H$, since in that case \emph{B}oundedness is automatic and \emph{L}inearity is the property to be stressed.}
 then one and hence all of these properties automatically hold if $M$ is a unital $\mbox{}^*$-algebra within $B(H)\equiv L(H)$. Given our motivation we will place some perhaps unusual emphasis on the interplay between commutative and noncommutative \vna s, and already here it is good to remark that commutative \vna s always take the form $L^{\infty}(X,\mu)$, where $(X,\mu)$ is some measure space, and under some natural assumptions (such as maximality  of $M$ and separability of $H$) such an $M$ can always be taken to act on $H=L^2(X,\mu)$ in the natural way, that is, by multiplication operators.\footnote{And it can even be concluded that $(X,\mu)$ is a standard Borel space, which is easy to work with.} On the other hand, the ``most noncommutative'' \vna s are so-called \emph{factors},  defined by the condition
\begin{equation}
M\cap M'=\C\cdot 1_H. \label{factor}
\end{equation}In other words, a factor has a trivial center, just consisting of multiples of the unit (where a commutative \vna\ 
$M$ coincides with its center). 
The simplest factor is $M=B(H)$ for any \Hs\ $H$. The next simplest case, central to quantum information theory,  arises if $H=H_{AB}$ is a tensor product \er{HABI}, where now $H_A$ and $H_B$ may be infinite-dimensional, and 
\beq
M=B(H_A)\ot 1_{H_B}, \label{naivefactor}
\eeq
 that is, the set of all operators $T_A\ot 1_{H_B}$ with $T_A\in H_A$. Then a simple computation shows that
 \beq
M'= 1_{H_A}\ot B(H_B), \label{naivefactor2}
\eeq
 and hence \er{factor} follows. 
 Surprisingly,  there are innumerably many factors that are not of this ``factorized'' form; in other words, eq.\ \er{naivefactor} implies \er{factor} but not \emph{vice versa}; this is the main source of the inexhaustible richness of \vna s. Again in other words, in so far as they are described by \vna s, 
 subsystems of a quantum system need not come from tensor products; and conversely,  combining systems is richer than forming tensor products. This has dramatic  consequences for entanglement, for which we first need another key definition.
 
 A \emph{state} on $M$ is a linear functional $\om:M\raw \C$ that is \emph{positive} in that $\om(S^*S)\geq 0$ for all $S\in M$, and \emph{normalized} in the sense that $\om(1_H)=1$, where the unit opertor $1_H$ on $H$ is also the algebraic unit $1_M$ of $M$. Special states are given by density operators $\rh\in D(H)$ (which means $\rh\geq 0$ and $\Tr(\rh)=1$, as in finite dimension, with a more complicated notion of the trace, though) via
\begin{equation}
\om(T)=\Tr(\rh T). \label{239I}
\end{equation}
Such states are called \emph{normal}, and these states span the predual $M_*$; indeed, if $M=B(H)$ then $M_*$ consists of all trace-class operators on $H$ (so $M_*=B_1(H)$ in that case).
\emph{Any} state is continuous in the norm-topology on $M$ (inherited from the operator norm on $B(H)$), with 
$\|\om\|=\om(1_H)$,
 but \emph{normal} states satisfy an additional continuity condition: they are also \emph{$\sg$-weakly continuous}, where the \emph{$\sg$-weak topology} on $M$ is simply the weak topology on $M$ in its capacity of being 
the Banach dual to $M_*$. That is, we have  $T_{\lm}\raw T$ $\sg$-weakly iff $\om(T_{\lm})\raw \om(T)$ for all $\om\in M_*$.\footnote{Equivalently, 
$\Tr(\rh T_{\lm})\raw \Tr(\rh T)$ for all $\rh\in D(H)$, but we will see shortly that this is somewhat misleading.}

Although the $\sg$-weak topology may seem very technical, it is crucial for physics. The \emph{normal} states are  the \emph{physical} ones; in quantum theory they come from density operators (as opposed to non-normalizable ``states'' \`{a} la Dirac) as we have seen, and in classical physics normal states are probability measures (as opposed to merely finitely additive pseudomeasures).\footnote{One should not confuse non-normal states with \emph{weights}, which may be infinite, like the trace on $B(H)$.}
 Moreover, the requirement of normality has  consequences for \emph{pure} states, a concept  defined by von Neumann via convexity,\footnote{Remarkably, both operator algebras and large deviations heavily rely on convexity theory, which we will not teach in this course, but simply rely on by summarizing the main aspects in an appendix and at special spots in the main text.}  which naturally comes down to unit vectors in elementary quantum theory on finite-dimensional \Hs s and to points in phase space in the classical case. Namely, \emph{normal pure states typically do not exist on the \vna s used in quantum field theory and the quantum statistical mechanics of infinite systems.}
  One reason is that a unit vector $\psi\in H$, which defines a pure state
 $\om(T)=\la \psi, T\ps\ra$ on $B(H)$, by the very same formula usually defines a \emph{mixed} state on $M\subseteq B(H)$.
 In the tensor product  situation \er{naivefactor} studied above a pure state $\ps_{AB}$ restricts to a pure state on $M$ iff it is separated (i.e., $\psi_{AB}=\ps_A\ot\psi_B$), but  many interesting \vna s, even factors, are not factors in a tensor product, and it is precisely in  such cases that no unit vector in $H$ restricts to a pure state on $M$ (which would have been normal if it did).
 
 The following (closely related) point is extremely important: for a general \vna\ $M\subseteq B(H)$, computing the von Neumann entropy \er{vNnow} of a normal state $\om$ on $M$ from some density operator $\rh\in D(H)$ that represents it via \er{239I}, or likewise computing the relative entropy \er{Umegaki} of a pair of normal states from density operators on $H$,
 faces a serious problem: \smallskip
 
 \noindent\emph{unless $M=B(H)$, such an operator $\rh$ ``representing''  $\om$ is not unique, and different choices may lead to vastly different entropies.}
 For example, take \er{EPS} with \er{EABI}, so that for $T_A\in L(H_A)$,
\begin{equation}
\om_A(T_A):=\Tr_{H_A}(\rh_A T_A)=
\Tr_{H_{AB}}(\rh_{AB} (T_A\ot 1_{H_B}))=\Tr_{H_{AB}}\left(\left(\rh_{A}\ot \frac{1_{H_B}}{\dim(H_B)}\right) T_A\ot 1_{H_B}\right).
\label{241I}\end{equation}
This shows two profoundly different ways of associating a density matrix on $H_{AB}$ to $\om_A$, of which the first, $\rh_{AB}$, gives zero von Neumann entropy to $\om_A$, whereas the second, $\rh_{A}\ot (1_{H_B}/\dim(H_B))$,  gives the correct value $S(\rh_A)$;
indeed, $\rh_A$ is the \emph{intrinsic} density operator associated to $\om_A$.
An important goal of this course is therefore to find some operator $\rh$ ``canonically associated'' to $\om$, and define the entropy of $\om$ via $\rh$ by a formula like \er{vNnow}. Likewise for two normal states, whose relative entropy we would like to define via some more sophisticated version of \er{Umegaki}. This goal will be achieved via Haagerup's theory of noncommutative $L^p$-spaces over \vna s.

The more general entropies that are needed to deal with all these problems arise from the main technical development in the theory of \vna s since its inception by von Neumann himself in the 1930, namely \emph{modular theory}. It is well known, but worth repeating,\footnote{The author heard this story from Winnink in 1989, with some more details in Haag (2010), quoted below.}
 that this theory was initiated by Tomita in order to study the commutant of tensor products of \vna s, but the striking originality of his construction was not matched by the absolute standards of rigour in the field, and his two papers from 1967 have remained unpublished.  This was remedied by his fellow Japanese operator algebraist Takesaki, who restructured the theory and rigorously proved all of Tomita's claims in 1970. This edifice, which in its entirely took place in pure mathematics, crossed work in mathematical physics on quantum statistical mechanics by Haag, Hugenholtz, \& Winnink (1967), who from a completely different point of view were led to essentially the same mathematical structure and hence also gave the first physical interpretation of the brand new Tomita--Takesaki theory.\footnote{Subsequently, many other applications of modular theory to physics were discovered, for which Borchers (2000) is a good starting point. In fact, almost all rigorous work in the interface of quantum field theory and statistical mechanics, including Hawking radiation, and even most rigorous work in quantum field theory itself, is now based on this theory.}
 Here is Haag's recollection of the birth of this theory:
 \begin{quote}
\begin{small}
 In Spring 1967, there was a conference in Baton Rouge, Louisiana, important for
a variety of reasons. It was a meeting between mathematicians who represented the
heritage of von Neumann, with physicists who used C*-algebras and von Neumann-algebras in quantum field theory and quantum statistical mechanics. In Baton Rouge
the first surprise was a thick manuscript by the Japanese mathematician Tomita.
When Nico [Hugenholtz] had looked over it he said: ``If this is true then our paper (Haag, Hugenholtz, \& Winnink, 1967)
is just a special case of a much more general situation''. I asked Dick Kadison what
he thought about it. He was not impressed: ``I have seen such papers before. You
start reading them and find the first mistake on page 15. It can be easily mended
but then the next one comes on page 35; it also can be mended with some eﬀort.
Finally on page 57 you come across one which cannot be mended. No, I would not
advise you to get into that''. Luckily Masamichi Takesaki was not dissuaded but
delved into Tomita's paper, simplifying the arguments, closing gaps until finally there
resulted the Tomita--Takesaki theory (Takesaki, 1970), one of the major advances in
operator algebras, opening the door for many subsequent developments. For me this
was a manifestation of the miracle called prestabilized harmony between mathematical
structures which are pure products of the mind, with important areas of physics.
Certainly neither Tomita nor Takesaki were thinking of quantum statistical mechanics.
But their first central theorem stating that any faithful state of a von Neumann
algebra defines an automorphism group called ``modular automorphisms'' and an anti-unitary conjugation, mapping the algebra onto its commutant, was a generalization of
our discussion of thermodynamic equilibrium states in [our  1967 paper]. In fact our paper
could now be summarized by saying that the equilibrium state is a faithful state
whose modular automorphisms are the time translations. The modular conjugation
is simply related to time reversal. (Haag, 2010, p.\ 289)
\end{small}
\end{quote}
Let us try to give a summary of the Tomita--Takesaki  theory from these two perspectives.
\begin{itemize}
\item \emph{Physics perspective}.\footnote{See  Haag, Hugenholtz, \& Winnink (1967), and also Landsman \& van Weert (1987), and Sorce (2023). From now on, we replace the quantum information theory notation $S, T$ for operators by $A, B$, as usual for operator algebras.} 
First, simply consider a \emph{finite-dimensional} \Hs\ $H$. Let 
\beq
h:H\raw H
\eeq
 be some self-adjoint operator, seen as a Hamiltonian, let $\beta=1/kT$ be  the inverse temperature, and $\om$  the state on $M=B(H)$  defined via \er{239I} by the canonical density matrix
\begin{align}
\rh=\frac{e^{-\beta h}}{Z}; && Z=\Tr(e^{-\beta h}). \label{242I}
\end{align}
The associated \textsc{gns}-construction $(H_{\om},\pi_{\om}, \Om_{\om})$ is equivalent to letting the:\footnote{Briefly: any (normal) state $\om$ on a C*-algebra (or a \vna) $\CA$ gives rise to a \rep\ $\pi_{\om}:\CA\raw B(H_{\om})$ on a \Hs\ $H_{\om}$ containing a cyclic unit vector $\Om_{\om}$ (which means that
$\pi_{\om}(\CA)\Om_{\om}$ is dense in $H_{\om}$),
such that $\om(A)=\la \Om_{\om}, \pi_{\om}(A)\Om_{\om}\ra$.  In case that
 $\om$ is faithful (in that $\om(A^*A)=0$ implies $A=0$) and $\CA$ is finite-dimensional and unital (which applies here), we simply have $H_{\om}=\CA$ with inner product $\la A,B\ra:=\om(A^*B)$, and $\pi_{\om}(A)B=AB$, with
  $\Om_{\om}=1_\CA$. Hence in our example we have $H_{\om}=B(H)$ with inner product $\la A,B\ra=\Tr (\rh A^*B)$.
  and $\Om_{\om}=1_\CA H$. However, Haag, Hugenholtz, \& Winnink took the  unitarily equivalent realization described above.}
\begin{itemize}
\item   \Hs\ be $H_{\om}=B(H)$  with Hilbert--Schmidt inner product $\la A,B\ra=\Tr (A^*B)$;
\item  \rep\ be $\pi_{\om}(A)B=AB$, i.e. $\pi_{\om}(A)=L(A)$ with $L(A)B=AB$;
\item  cyclic unit vector be $\Om_{\om}=\rh^{1/2}$ (which here is also separating for $\pi_{\om}$),\footnote{This means that $\pi_{\om}(A)\Om_{\om}=0$ iff $A=0$, which is true because $\om$ is faithful.}
 so that
 \begin{align}
\om(A)=\la \Om_{\om}, \pi_{\om}(A)\Om_{\om}\ra && (A\in B(H)). \label{243I}
\end{align}
\end{itemize}
Then the commutant $\pi_{\om}(M)'$ is anti-isomorphic to $\pi_{\om}(M)=L(M)$ itself via $L(A)\lraw R(A)$ with $R(A)B=BA$; equivalently, if we define an antilinear (and in fact antiunitary) operator 
\begin{align}
J:H_{\om}\raw H_{\om}; && JA:=A^*, 
\end{align}
then
$R(A)=JL(A)J$ and $A\mapsto JAJ$ is an anti-isomorphism from $\pi_{\om}(M)$ to $\pi_{\om}(M)'$.

Moreover, if we define Heisenberg-picture dynamics via automorphisms $\al_t:M\raw M$ through 
\begin{equation}
\al_t(A)=e^{ith} A e^{-ith},\label{245I}
\end{equation}
so that $\al_t$ is linear and satisfies $\al(A^*)=\al_t(A)^*$ and $\al_t(AB)=\al_t(A)\al_t(B)$, then
\begin{equation}
\om(\al_t(A))=\om(A), \label{246I}
\end{equation}
for all $A\in M$, and this implies that one may define a new ``Hamiltonian'' $h_{\om}$ on $H_{\om}$ such that
\begin{align}
\pi_{\om}(\al_t(A))=e^{ith_{\om}} \pi_{\om}(A) e^{-ith_{\om}};  && e^{ith_{\om}}\Om_{\om}=\Om_{\om}; &&
&& h_{\om}\Om_{\om}=0.
\end{align}
Namely, using the fact that $\Om_{\om}$ is cyclic for $\pi_{\om}$ as well as  Stone's theorem to extract $h_{\om}$, put
\beq
e^{ith_{\om}} \pi_{\om}(A) \Om_{\om}:=\pi_{\om}(\al_t(A))\Om_{\om}, \label{248I}
\eeq
for which \er{246I} is obviously necessary. 
It is easy to show that, at least in this simple  setting,
\begin{equation}
h_{\om}=L(h)-R(h). \label{249I}
\end{equation}
Using basic functional analysis  this  works for infinite-dimensional $H$ and $M=B(H)$, too, as long as $\rh$ exists (which is the case if $h$, possibly unbounded now,  had discrete spectrum with finite multiplicities); now take $H_{\om}=B_2(H)$, the \Hs\ of Hilbert--Schmidt operators, which is a two-sided ideal in $B(H)$, so that $L(A)$ and $R(A)$ remain well defined. Furthermore, since $\rh\in B_1(H)$ is trace class, its square root $\rh^{1/2}\in B_2(H)$ duly lies in $H_{\om}$.

Except for \er{248I}, this can even be generalized to arbitrary \vna s if the right replacement of \er{242I}, which makes no direct sense in infinite volume, is found. Since the idea goes back to Kubo, Martin, and Schwinger,  this replacement is called the \emph{\textsc{KMS} condition}.\footnote{See Kubo (1957) and Martin \& Schwinger (1959). Haag, Hugenholtz \& Winnink (1967) were also inspired by earlier work of Araki \& Woods (1963) on the construction of non-type I factors using quantum statistical mechanics.}
The starting point remains some time-evolution, which in  quantum mechanics is given by some Hamiltonian $h$ with associated  unitary evolution \er{245I}, but in the thermodynamics limit is given directly by some 1-parameter automorphism group $\al_t$ of $M$, typically obtained as the infinite-volume limit of a unitary evolution driven by local Hamiltonians (but not itself 
generated by an underlying Hamiltonian, which in infinite volume does not exist). The \textsc{KMS} condition, then, states that for any $A,B\in M$ the two-point function
\begin{equation}
f(z):=\om(A\al_z(B))
\end{equation}
initially defined for real values $t$ of $z$,
has an analytic extension to the strip $0<\mathrm{Im}(z)<\beta$ with continuous boundary values 
at $z=t\in\R$ and $z=t+i\beta$ (again with $t\in \R$) given by
\begin{align}
f(t)=\om(A\al_t(B)); &&
f(t+i\beta)=\om(\al_t(B)A). \label{KMS1}
\end{align}
For finite-dimensional $H$ (and with a bit more  analysis also for infinite-dimensional $H$) the \textsc{KMS} condition can be \emph{checked} from \er{239I} and \er{242I} by a simple calculation, but the  leap of faith is to \emph{define} equilibrium states by this condition. It can  be shown from this condition that $\om$ is faithful and time-invariant, and that the associated  \textsc{gns}-construction has exactly the same features as in the finite-dimensional case.\footnote{This approach typically starts from a C*-algebra $A$ of (quasi-) local observables and so one has to complete $\pi_{\om}(A)$ into a \vna, for example by taking the bicommutant $\pi_{\om}(A)''$, which is then anti-isomorphic to $\pi_{\om}(A)'$.} A crucial fact here is that whereas in finite systems the solution to the \textsc{KMS} condition (for given dynamics) is unique, namely \er{242I}, in infinite systems it is not, which is one way to understand phase transitions.\footnote{Here Bratteli \& Robinson (1981) is a standard reference. It should be mentioned, though, that understanding phase transitions and  spontaneous symmetry breaking 
only in infinite systems is seriously misleading, as in ``Nature'' it is finite systems that break symmetry and undergo phase transition; this possibility even seems blocked by the kind of uniqueness theorems for equilibrium states mentioned in the main text (which under additional assumptions also apply to ground states)! See Tasaki (2020) and  van de Ven,  Groenenboom,  Reuvers, \&  Landsman (2020) for a way out.}
\item \emph{Math perspective}.
Let $M\subseteq B(H)$ be a \vna\ with  a \emph{cyclic} and \emph{separating} unit vector $\Om$, that is, $M\Om:=\{A\Om, A\in M\}$ is dense in $H$, and $A\Om=0$ implies $\Om=0$ for all $A\in M$, respectively.\footnote{
 The case explained after \er{249I},  with better notation so that $M=B(H')$ acts on $H=B_2(H')$ by left-multiplication, and $\Om=\rh^{1/2}$ for any \emph{invertible} $\rh\in D(H')$, i.e., 
 $\rh\in B_1(H)$, $\rh>0$, and $\Tr(\rh)=1$, is an example. If $\dim(H)=\infty$ the inverse $\rh\inv$, if it exists, is typically unbounded, but this does not spoil anything. See Exercise \ref{inverseunbounded}. } Tomita's main idea was to define an \emph{antilinear} operator $S$ on $H$ by
 \begin{align}
SA\Om:=A^*\Om; && (A\in M). \label{defSTT}
\end{align}
Since $\Om$ is separating for $M$ this operator is well defined,
and since $\Om$ is cyclic for $M$ it is defined on a \emph{dense} domain $M\Om\subset H$. Note that $S$ is antilinear because of the adjoint $A^*$. Furthermore, $S$ may be unbounded, but  we ignore this complication for now.\footnote{It means that for certain inclusions $M\subset B(H)$ the operator  $S$ cannot be extended from its initial 
 domain $M\Om$ 
to all of $H$ by continuity, but on the positive side this initial domain 
\emph{can} always  be slightly extended so as to make $S$ closed; this is enough for it to have  a unique polar decomposition. 
 If $M$ is a factor, then $S$ is in fact bounded except for type III.} The key point is that $S$ (or more precisely its closure) has a unique \emph{polar decomposition}
 \begin{align}
S=J\Dl^{1/2}; && \Dl=S^*S, \label{polarS}
\end{align}
where $J$ is anti-unitary and $\Dl$ is self-adjoint (and possibly unbounded).\footnote{See footnote \ref{polarfn}. In general the first factor, here $J$, is just a partial isometry, but here it is (anti) unitary.}
In our example,
 \begin{align}
 J\xi=\xi^*; &&  \Dl \xi=\rh \xi\rh\inv && (\xi\in H).
 \end{align}
Here, and in general, neither $J$ nor $\Dl$ (even if it is bounded) lies in $M$. Nonetheless, the first key theorem of the Tomita--Takesaki theory is that the \emph{modular automorphisms}
\begin{align}
\sg_t(A):=\Dl^{it}A\Dl^{-it} && (t\in\R), \label{mgTT}
\end{align} 
map $A\in M$ into some $\sg_t(A)\in M$,  and hence \er{mgTT}  \emph{defines a time-evolution on $M$}.
It also follows from the definition of these objects that, writing $\Dl=e^{h_{\om}}$ and hence $\Dl^{it}=e^{ith_{\om}}$,
\begin{align}
J\Om=\Om; && \Dl^{it}\Om=\Om; && h_{\om}\Om=0,
\end{align}
and $h_{\om}$ is well defined as a possibly unbounded operator on $H$. If $M$ is a factor, it turns out that a split like \er{249I} only exists if $M$ is type I or II, in which case the modular group consists of \emph{inner} automorphisms, given by a unitary group $U_t\in M$ via $\sg_t(T)=U_t T U_t^*$.
For type III the modular group is always outer (i.e., not inner), but (for all types) modular groups coming from different choices of $\Om$ differ by inner automprhisms. This insight, to which we will return, was one the keys to Connes's classification of hyperfinite type III factors. 
Finally, the Tomita--Takesaki yields an anti-isomorphism of $M$ and its commutant $M'$ via
\begin{equation}
M'=JMJ.
\end{equation}
\end{itemize}

Comparing the earlier physics perspectives with the math one:  in the former one obtains a \emph{state from the dynamics} via \er{239I} and \er{242I}, 
whereas in the latter  one obtains \emph{the dynamics from a state} via \er{defSTT}, \er{polarS}, and \er{mgTT}. As we have seen, in the physics construction state and dynamics are related by the KMS condition \er{KMS1}. This is also the case in the math perspective: defining
\begin{equation}
\om(A):=\la\Om, A\Om\ra, \label{omTTT}
\end{equation}
which makes sense for all $A\in B(H)$, and with  $A\in M$ gives a  faithful normal state $\om$ on $M$, it turns out that
the pair $(\om,\sg_t)$ defined by \er{omTTT} and \er{mgTT} satisfies the KMS condition at $\beta =-1$.  Conversely, if some dynamics on $M\subset B(H)$ satisfies  the KMS condition at $\beta =-1$ for the state \er{omTTT} defined by some cyclic and separating vector unit $\Om\in H$, then this dynamics  must coincide with the modular group \er{mgTT} defined by $\Om$. Thus the physics-mathematics circle closes:\footnote{The strange value $\beta=-1$ in the the KMS condition comes from the fact that in our example $\Dl^{it}=e^{ith}$ with $\Om=\rh^{1/2}$ gives $\rh=e^h/\Tr(e^h)$, which is the physicists' \er{242I} with $\beta =-1$. If one starts from \er{242I}, then 
$\Dl=e^{-h/\beta}$.
}
\begin{quote}
\begin{small}
The relation between this
``KMS'' condition and the modular operator of Tomita's Hilbert algebras, discovered
by M.\ Takesaki and M.\ Winnink, remains one of the deepest points of contact between
physics and pure mathematics. (Connes, 1994, p.\ 14)
\end{small}
\end{quote}

Let us now briefly explain how entropies arise in this context. The key idea of relativizing the operator \er{defSTT} by invoking a second vector  goes back to Connes and was  applied to entropy independently (and somewhat differently) by Araki and Uhlmann.\footnote{The original references are Araki (1976, 1977) and Uhlmann (1977). See also Ohya \& Petz (1994).} With $M\subseteq B(H)$ as before, we now assume we have two unit vectors $\Om_1, \Om_2\in H$ that are both cyclic and separating for $M$ (this will be weakened later on), with associated normal faithful states $\om_1, \om_2$ defined as in \er{omTTT}. Now define a \emph{relative modular operators} $S_{12}$ and $S_{21}$ on $H$ by
 \begin{align}
S_{12}A\Om_2:=A^*\Om_1; && S_{21}A\Om_1:=A^*\Om_2 && (A\in M), \label{defSTTrel}
\end{align}
with associated polar decompositions as in \er{polarS}, so that in particular we have
\begin{align}
\Dl_{12}=S_{12}^*S_{12}; && \Dl_{21}=S_{21}^*S_{21}.
\end{align}
In our running example, in which we now set $\Om_i=\rh_i^{1/2}$ for $i=1,2$, these are given by 
\begin{align}
\Dl_{12}A=\rh_1 A\rh_2\inv;  && \Dl_{21}A=\rh_2A\rh_1\inv && (A\in H=B_2(H')). \label{262I}
\end{align}
 Araki's relative entropy may then be defined in two equivalent ways:
 \begin{equation}
S(\om_1,\om_2):= \om_2(\Dl_{12}\log \Dl_{12})=-\om_1(\log\Dl_{21}). \label{Araki11}
\end{equation}
It follows from \er{262I} that in our example this definition returns the Umegaki entropy \er{Umegaki}, and we will also find that the commutative case gives the classical relative entropy \er{KLD}, even for much more general measure spaces than finite ones. But the point is that we now have a well-defined relative entropy for (even not necessarily faithful) normal states on any \vna. 

 So far, so good, but how are we actually going to compute these entropies? To this end it would be nice to have some version of \er{Umegaki}, which requires \emph{intrinsic}  density operators and traces for arbitrary \vna s.  This sounds impossible, since only type I \vna s have the former and 
 type III ones even lack the latter, but nonetheless our goal is accomplished  by 
Haagerup's  theory of noncommutative spaces $L^p(M)$ over an arbitrary \vna\ $M$, which relies on the Tomita--Takesaki theory.\footnote{The original reference is Haagerup (1979). The detailed exposition by Terp (1981) was the basis of later reviews  by Goldstein \& Labuschagne (2020) and Hiai (2021a). See also the useful summary in Haagerup, Junge, \& Xu (2010). \label{Haageruprefs}}
 The ordinary (commutative) theory associates Banach spaces $L^p(X,\mu)$ to measure spaces $(X,\mu)$ for $p\in[1,\infty]$, 
whose elements are (appropriate equivalence classes of) measurable functions $f:X\raw \C$ for which
$\int_X d\mu(x) |f(x)|^p<\infty$ with norm 
\begin{align}
\|f \|_p:= \left(\int_X d\mu(x) |f(x)|^p\right)^{1/p} && (1<p<\infty),
\end{align} 
whereas $L^{\infty}(X,\mu)$ consists of all measurable functions that are bounded (outside some set of measure zero), equipped with the  supremum-norm (again up to sets of measure zero). Within this:
\begin{itemize}
\item $L^2(X,\mu)$ is a \Hs;
\item $L^{\infty}(X,\mu)$ is a commutative \vna;
\item $L^1(X,\mu)$ is (isomorphic to) the predual of $L^{\infty}(X,\mu)$.
\item $L^{\infty}(X,\mu)$ canonically acts on $L^2(X,\mu)$ (namely by multiplication operators).
\end{itemize}
That is, $L^1(X,\mu)^*\cong L^{\infty}(X,\mu)$ as Banach spaces, where the star denotes the Banach dual.\footnote{If $B$ is a Banach space $B^*$ consists of all linear functionals $\phv:B\raw \C$ for which $\sup\{|\phv(f)|, f\in B, \|f\|\leq 1\}<\infty$, with norm $\|\phv\|$ given by this supremum.
If $B=L^1(X,\mu)$, then $B^*\cong L^{\infty}(X,\mu)$ via $\phv(f)=\int_X d\mu\, \phv f$. If $p\inv + q\inv=1$ with $p\geq 1$,
the isomorphism $L^p(X,\mu)^*\cong L^q(X,\mu)$ is defined similarly (this doesn't work for $p=\infty$).
} This gives a  realization of the predual, similar to isomorphism $B(H)_*\cong B_1(H)$ via \er{239I}.

Haagerup's theory constructs Banach space $L^p(M)$ for any $1\leq p\leq \infty$, with similar virtues:
\begin{itemize}
\item $L^2(M)$ is a \Hs\ with Hilbert--Schmidt type inner product $\la \xi,\eta\ra=\tr(\xi^*\eta)$;
\item $L^{\infty}(M)$ is isomorphic to $M$ itself (and hence is a \vna\ in its own right);
\item $L^1(M)$ is isomorphic to the predual of $L^{\infty}(M)$;
\item $L^{\infty}(M)$ --and therefore $M$-- canonically acts on $L^2(M)$ (namely by left multiplication).
\end{itemize}
Here the pairing between $L^1(M)$ and $L^{\infty}(M)$, hence effectively between $M_*$ and $M$, is done via 
\begin{equation}
\om(A)=\mathrm{tr}(\rh \pi(A)), \label{265I}
\end{equation}
where $\pi(A)\in L^{\infty}(M)$ is the image of $A\in M$ under the isomorphism in the second bullet,  $\rh\in L^1(M)$ is  \emph{canonically} associated to $\om\in M_*$ under the isomorphism in the third bullet, cf.\ \er{239I}, and the (Haagerup) trace $\tr$  is a highly nontrivial object to be defined when we get there.  To see how $\om$ defines $\rh$, note that
$L^2(M)$ brings $M$ acting on it via $\pi$ in \emph{standard form}: this means, in particular, that any normal state $\om$ on $M$ can be canonically  ``purified'' by some unit vector $\Om\in L^2(M)$ via 
\beq
\om(A)=\la\Om, \pi(A)\Om\ra,
\eeq
cf.\ \er{omTTT}, 
 which $\Om$ is \emph{unique} subject to conditions \emph{intrinsic to} $M$. Using the first bullet, this  gives $\rh=\Om^2$, or $\Om=\rh^{1/2}$, so that it looks like type I, cf.\ \er{242I} - \er{243I}.
 Consequently, the Haagerup construction  gives the usual formulas for entropy. For example, we  can now express \er{Araki11} as
\begin{equation}
S(\om_1,\om_2)=\mathrm{tr}(\rh_1(\log\rh_1-\log\rh_2)), \label{ArakiHiai}
\end{equation}
at least when $\ker(\rh_2)\subseteq\ker(\rh_1)$, putting  $S(\om_1,\om_2)=\infty$ if this is not the case.
This should make $S$ computable once we know $\rh_1$ and $\rh_2$ and the trace $\tr$, and solves the problem discussed after \er{239I}:  if $M\subset B(H)$ is some given subsystem of an ambient quantum theory, then density operators $\rh\in D(H)$ implementing some  normal state $\om\in M_*$ via $\om(T)=\Tr(\rh T)$ are not unique, in a lethal way in so far as defining (relative)  entropies is concerned. But now we have $\rh\in L^1(M)$ and \er{265I}. 
 
The  construction of the $L^p(M)$ spaces is quite technical,\footnote{Unfortunately, unless $p=\infty$,  all $L^p(M)$ spaces consist of \emph{unbounded} operators on the \Hs\ $H\ot L^2(\R)$ on which $N$ canonically acts. For $p=1$ this implies that the ``density operators'' $\rh$ in \er{265I} are also unbounded. Nonetheless, eq.\ \er{ArakiHiai} stands, and should make physically relevant entropies computable without ugly limiting constructions, which are supposed to be taken care of already in the construction of the trace in \er{265I}.}
 especially because of the prominent role of unbounded operators. For now, let us just say that it is based on  the crossed product
\begin{equation}
N=M \rtimes_{\sg^{ \om}}\R
\end{equation}
of $M$ by the action of $\R$ on $M$ via the modular automorphism group $\sg^{ \om}$ associated to an arbitrary faithful normal state $ \om$ on $M$.\footnote{This construction therefore depends on a chosen realization of $M\subset B(H)$ with cyclic and separating vector, but this dependence is weak (all choices being unitarily equivalent) and easy to keep track of.} Such crossed products play an important role in the classification theory of factors,
 and are also increasingly popular in theoretical physics, where the action of $\R$ is supposed to describe some observer or associated ``quantum reference frame'' (see footnote \ref{Kudler}).
 \smallskip
 
The  material summarized so far, expanded in \S\S\ref{FAP} - \ref{LDGT} and \S\S\ref{FromCQ} - \ref{MWM},
was more or less what was covered in the  Okinawa lectures. To give a more complete picture of entropy, in these notes we have added some further material on entropy in statistical physics (\S\S\ref{ASP} -- \ref{LDGM}) as well as on entropy in classical and quantum dynamical systems (see \S\ref{EDS} and \S\ref{CNTE}, respectively). Furthermore, one finds appendices on convexity, stochastic processes \& Markov chains, ergodic theory, and Kolmogorov randomness. Convexity theory is often used in both large deviation theory and \vna s, in support of which Appendix \ref{AppC} is primarily  a list of results without proofs and as such falls far short of a proper course; it is best
 consulted while reading the main text.
The other
extra material may be omitted at a first reading. Nonetheless, we recommend studying it:
\begin{itemize}
\item Classical statistical physics and its link to thermodynamics provides an important area of application of large deviation theory; indeed, this theory arose partly from statistical physics. As we shall see, this applies to both Boltzmannian and Gibbsian approaches, whose commonalities and differences neatly fall into place from the perspective of large deviations.\footnote{Regrettably, the philosophy of physics literature on statistical physics has so far ignored large deviation theory. Consequently, the differences between the Boltzmann and Gibbs approaches are often vastly exaggerated.  }
\item The Kolmogorov--Sinai entropy of a classical dynamical system, already briefly discussed in the Historical Introduction, is one of the central entropies of at least mathematical physics. Introducing also its quantum version in the form of the Connes--St\o rmer--Narnhofer--Thirring entropy expands our discussion of the role of \vna s in theories of entropy.
\item Markov chains (which we only discuss in the very simplest yet already rich setting of  discrete time and  finite state spaces) are the simples stochastic processes and as such provide the general interpretation of our diagram following \er{subadditive} in the Historical Introduction. Some of the results in large deviation theory, such as Sanov's theorem, extend from the i.i.d.\ setting to Markov chains, much as these generalize laws of large numbers from that setting.
\item Ergodic theory provides a setting for various results about entropy, such as the  Shannon--McMillan--Breiman theorem we cover earlier on, and also sheds light on the irreducibility assumption on Markov chains that often needs to be made to get clean results. Convexity theory also interacts beautifully with ergodic theory (as it does with most of our material). 
\item Kolmogorov randomness is an intriguing refinement of (classical) probability theory that tries to capture 
 randomness and chaos, and as such it is natural to expect that entropy should play a role in it. Indeed it does, and the goal of our discussion is just to explain that role.
\end{itemize}
That said, the classical theory covered in \S\S\ref{FAP} - \ref{ASP} may be cut off at \S\ref{LDGT} by omitting the Kolmogorov--Sinai entropy as well as the statistical physics part. Furthermore, the quantum theory of entropy discussed in \S\S\ref{FromCQ} -\ref{CNTE} logically stands alone, although some motivation may then  be lacking; moreover, as already mentioned, we see it as  enlightening to study classical and quantum theories of almost anything in conjunction, so as to highlight both what differs and what agrees.
 
 Finally, we have included numerous exercises. These are intended to help readers digest the  pretty abstract material by (mostly) practicing with the notation and the arguments in the main text, and are of course highly recommended. Depending on the student, their difficulty ranges from ``almost trivial'' to ``doable with a serious amount of work'';  perhaps the only exception is the most important exercise in the course and indeed one of the reasons it was offered, viz.\ no.\ \ref{superex}. 
 \begin{center}
 \fbox{\emph{Enjoy!}}\end{center}
\section{Asymptotic properties of entropy}\label{FAP}
After this historical and conceptual introduction we  turn to our main topic. Let $A$ be a \emph{finite} set, with number of elements $|A|\geq 2$ (keeping in mind that the binary case $A=\{0,1\}$ is already quite interesting), and let $N>1$ be a natural number. Any interpretation of $A$ and $N$ explained after our diagram in the Historical Introduction may be kept in mind (plus the simple example of a coin toss for $A=\{0,1\}$, where $N$ is the number of throws). Also, recall the notation introduced after the diagram. Further to this, a \emph{probability distribution} on $A$ is a map $p:A\raw\R^+$ that satisfies 
\beq
\Sigma_{a\in A} p(a)=1,
\eeq
so that $p(a)\in [0,1]$. We denote the space of all  probability distributions $p$ on $A$ by $\Pr(A)$, if necessary  equipped with the weak topology (see \S\ref{Sanovchapter}); for finite $A$ this topology simply means that $p_n\raw p$ in $\Pr(A)$ iff $p_n(a)\raw p(a)$ for all $a\in A$. Note that $\Pr(A)$ is a convex set (in $\R^{|A|}$).
\begin{itemize}
\item 
We  already saw Boltzmann's coarse-graining strategy via the \emph{empirical measure}
\begin{align}
 L_N: A^N\raw \Pr(A); && L_N(\sg):=\frac{1}{N}\Sigma_{n=0}^{N-1} \dl_{\sg(n)},
 \label{LNb}
\end{align}
where for any $a\in A$, such as $a=\sg(n)$,  the probability distribution $\dl_a\in\Pr(A)$ is defined by $\dl_{a}(b)=\dl_{ab}$. 
Since $L_N(\sg)$ is a convex sum of elements of $\Pr(A)$, also $L_N(\sg)\in\Pr(A)$.
Although we defined
  $L_N$  as a function $L_N: A^N\raw \Pr(A)$, we often see it as a function
  \begin{align}
 L_N: A^{\N}\raw \Pr(A); && L_N(s):=\frac{1}{N}\Sigma_{n=0}^{N-1} \dl_{s(n)}, 
 \label{LNbN}
\end{align}
on the set $A^{\N}$ of infinite $A$-valued sequences. Clearly,
the value  $L_N(s)$  depends on $s_{|N}$ only. 

  Define
 $\Pr_N(A)$ as the subset of $\Pr(A)$ consisting of all  $p\in\Pr(A)$ for which 
 \beq
 p(a)\equiv p_a=n_a/N, \label{pana}
 \eeq for some $n_a\in\{0, 1, \ldots, N\}$, necessarily with $\Sigma_{a\in A} n_a=N$. It is then clear that 
 \begin{equation}
L_N(\sg)\in\Pr_N(A)\subset\Pr(A).
\end{equation}
 \item
In the opposite direction, we also already saw the \emph{type class} $T_N(p)\subset A^N$ of $p\in\Pr(A)$:
\begin{equation}
T_N(p):=\{\sg\in A^N\mid L_N(\sg)=p\}.\label{TNp}
\end{equation}
Thus $T_N$ is a function $T_N:\Pr(A)\raw\CP(A^N)$, with $T_N(p)=\emptyset$ if $p\notin \Pr_N(A)$. Trivially,
\begin{align}
 \sg\in T_N(p) && \LRaw && L_N(p)=\sg.
\end{align}
\end{itemize}
Recall the Boltzmann--Shannon entropy \er{SE}, now in base $e$, i.e., using $\log\equiv\ln\equiv\log_2$, so that
\begin{equation}
S(p):=-\Sigma_{a\in A} p(a)\log p(a). \label{Aentropy}
\end{equation}
The first appearance of this  formula in this context follows from a simple computation, cf.\ \er{defqN}:
\begin{align}
 \sg\in T_N(p) &&  \Raw && 
p^N(\sg)=e^{-NS(p)}. \label{PNpsg} 
\end{align}
Boltzmann (1879) was the first to compute the cardinality of $T_N(p)$:  if \er{pana} holds, then
\begin{equation}
|T_N(p)|=\frac{N!}{\prod_{a\in A} (Np_a)!}=\frac{N!}{\prod_{a\in A}n_a!}.\label{combi}
\end{equation}
One (too sophisticated) way to  see this is to pick some $\sg\in T_N(p)$, which is possible since $p\in\Pr_N(A)$.  Then any  $\sg'\in T_N(p)$ can be obtained from $\sg$ by a permutation $\pi$ of the $N$ particles, i.e.,  $\sg'=\sg\circ\pi$. Define an equivalence relation $\sim$ on the permutation group $\mathfrak{S}_N$ on $N$ symbols by
\begin{align}
\pi\sim\pi'&&  \mathrm{iff} && \sg\circ\pi=\sg\circ\pi'.
\end{align}
 Then 
$T_N(p)|=|\mathfrak{S}_N/\sim|=|\mathfrak{S}_N/S_{\sg}|=|\mathfrak{S}_N|/|S_{\sg}|$,
where $S_{\sg}=\{\pi\in \mathfrak{S}_N\mid \sg\circ\pi=\sg\}$ is the stabilizer of $\sg$.  Since
$|\mathfrak{S}_N|=N!$ and $|S_{\sg}|=\prod_{a\in A}n_a!$, this gives \er{combi}. 
Its asymptotics   gives our second appearance of the entropy \er{Aentropy}, as follows:\footnote{\label{BEold}   Boltzmann (1877)  interpreted the ``$W$'' in \er{BPE} as 
$W(p)=|T_N(p)|$. Hence
$\lim_{N\raw\infty}\frac{1}{N}  \log W(p)=S(p)$.}
for any $p\in\Pr(A)$ and any sequence $(p_N)$ such that  $p_N\in\Pr_N(A)$ and $p_N\raw p$ (weakly) we have
\begin{equation}
\lim_{N\raw\infty}\frac{1}{N} \log |T_N(p_N)|=S(p). \label{arche}
\end{equation}
\bex\label{Ex3.1}  Prove  \er{PNpsg}, \er{combi} (in your own way), and
\er{arche}. \emph{Hint:}  For \er{arche}, use Stirling's formula $\log (n!)=n\log n -n+O(\log n)$ for both $n=N$ and $n=Np_a$  in the middle  term of \er{combi}.
\eex 
Eqs.\  \er{PNpsg} and \er{arche} suggest the following picture of the probability space $(A^N,p^N)$ for large $N$:
\smallskip

\noindent \emph{Since each $\sg\in T_N(p)$ has the same probability $p^N(\sg)=e^{-NS(p)}$ and $|T_N(p)|\approx e^{NS(p)}$, it must be that $p^N(T_N(p))\approx 1$, although the size of $T_N(p)$ within $A^N$ is exponentially small (unless $p=f$).
}
\smallskip

\noindent Here $f$ is the \emph{flat} probability distribution on $A$ (also called \emph{fair} or \emph{uniform}) , i.e., 
\begin{align}
f(a)=1/|A| && (a\in A).
\end{align}
This is Shannon's \emph{Asymptotic Equipartition Property} (AEP), of which a more precise form will be given shortly. 
The last point may neatly be shown via the \emph{relative entropy} \er{KLD}, which we recall:
\begin{align}
 S_q(p)\equiv S(p, q)&:=\Sigma_{a\in A} p(a)\log\left(\frac{p(a)}{q(a)}\right)\:\:\: \mbox{ if } p\ll q \label{KL1};\\
S(p, q)&:=\infty\:\:\: \mbox{ otherwise}, \label{KL2}
\end{align}
where, at least in the case that $A$ is finite considered here, the `absolute continuity'
 $p\ll q$ simply means that $q(a)=0$ implies $p(a)=0$, in which case the corresponding term in 
\er{KL1} is zero. 
\bex \label{ex2}
For $A=\{0,1\}$, identify $\Pr(A)$ with  $[0,1]$ via $p\mapsto p(1)$. Taking $q=1$, show that 
\begin{align}
S_1(p)=\infty \hspace{25pt} (0\leq p<1); && S_1(1)=0.\end{align}
\eex
On the other hand,  $p\ll q$ holds for any strictly positive $q$, such as  $q=f$, in which case, cf.\ \er{KLflat},
\begin{equation}
S(p, f)=-S(p)+\log|A|. \label{Dfflat} 
\end{equation}
So if $|T_N(p)|\approx e^{NS(p)}$ as suggested by \er{arche}, then 
\begin{equation}
p^N(T_N(p))=\frac{|T_N(p)|}{|A|^N}\approx e^{N(S(p)-\log|A|)}=e^{-NS(p, f)}. \label{3.17ja}
\end{equation}
It follows from the next proposition that $S(p, f)\geq 0$ with equality $S(p, f)= 0$ iff $p=f$, so that 
$S(p, f)>0$ whenever $p\neq f$. In that case $T_N(p)$ is indeed exponentially small within $A^N$ as $N\raw\infty$.
\begin{proposition}[Gibbs inequality]\label{gibbsin}
 We have $S(p, q)\geq 0$, with equality $S(p, q)= 0$ iff $p=q$. 
  \end{proposition}
\bex \label{Sconvexex}
\begin{enumerate}
\item  Prove this  from Jensen's inequality \er{Jensen}
with $f(x)=-\log x$,  $F(a)=q(a)/p(a)$, and $P=p$, assuming $p\ll q$. Then treat the case \er{KL2}. 
\item  Prove that $S(p,q)$ is both \emph{convex} in $p$ and \emph{jointly convex} in $(p,q)$ in the sense that
\begin{align}
S(tp_1+(1-t)p_2, tq_1 + (1-t) q_2)\leq tS(p_1,q_1)+(1-t)S(p_2,q_2) && (t\in [0,1]).\label{Spqconvex}
\end{align}
\end{enumerate}
\eex
If $p=f$, then $S(p)=\log|A|$,  and hence $p^N(\sg):=|A|^{-N}=e^{-NS(p)}$ not just for all $\sg\in T_N(p)$ but for all $\sg\in A^N$.
The property  $|T_N(p)|\approx e^{NS(p)}=|A|^N=|A^N|$ remains  nontrivial, but since  $T_N(p)$ is no longer exponentially small within $A^N$ (but asymptotically equal it) the AEP loses its thrust.

We now give a more precise version of the AEP, which overcomes earlier handwaving and also removes the awkward 
sequence $(p_N)$ in \er{arche}.
Instead of $T_N(p)$, consider the larger set
 \begin{equation}
T_{N,\dl}(p):=\{\sg\in A^N\mid  e^{-N(S(p)+\dl)}\leq  p^N(\sg)\leq  e^{-N(S(p)-\dl)}\}, \label{TEN}
\end{equation}
where $\dl>0$, so that $T_N(p)\subset T_{N,\dl}(p)$ because of \er{PNpsg}.  Here are the key facts about $T_{N,\dl}(p)$.
\begin{theorem}[Asymptotic Equipartition Property] For any $\dl>0$ we have:
\begin{align}
\lim_{N\raw\infty} p^N(T_{N,\dl}(p))&=1; \label{pone2}\\
\lim_{\dl\downarrow 0} \lim_{N\raw\infty}\frac{1}{N} \log |T_{N,\dl}(p)|&=S(p); \label{archedelta}\\
(1-\dl)e^{N(S(p)-\dl)} \leq |T_{N,\dl}(p)|&\leq e^{N(S(p)+\dl)}, \label{est33}
\end{align}
 where the upper bound is true for any $N$, whereas the lower bound holds for sufficiently large $N$. 
\end{theorem}
The interpretation of this result is the same as in the easier case $T_N(p)$  above, assuming $p\neq f$:

\smallskip

\noindent \emph{For large $N$ the set $T_{N,\dl}(p)$ has about $e^{N S(p)}$ elements, which is just an exponentially small 
fraction $e^{-NS(p, f)}$ of the total number $|A|^N$ of elements of $A^N$. And yet $T_{N,\dl}(p)$
 carries almost all of the probability (under $p^N$), on top of which each  $\sg\in T_{N,\dl}(p)$ has approximately 
the same probability }
\beq
p^N(\sg)\approx e^{-NS(p)}.\eeq
\emph{Hence for  large $N$ the set $A^N$ splits into two parts:
$T_{N,\dl}(p)$ and its complement, whose elements may have wildly varying individual probabilities but which as a whole can be ignored under $p^N$.}
\smallskip

\noindent\emph{Proof}.
Consider the Information (or ``surprise'') function and its average
\begin{align}
I_p: A\raw\R; && I_p(a)=\log(1/p(a)); && \la I_p\ra_p\equiv \Sigma_{a\in A} p(a)I_p(a)=S(p), \label{321}
\end{align}
The associated random variables on the probability space $(A^N,p^N)$ defined by
\begin{align}
X_n: A^N\raw \R; && X_n(\sg)=I_p(\sg_n)
\end{align}
 are  i.i.d. Hence, using \er{321} for the mean = limiting value,  the weak law of large numbers gives 
 \begin{equation}
\lim_{N\raw\infty}  p^N\left( \left|\frac{1}{N}\Sigma_{n=0}^{N-1}X_n- S(p)\right|<\delta\right)=1. \label{wlln1}
\end{equation}
\bex
Show that $ \left|\frac{1}{N}\Sigma_{n=0}^{N-1}X_n- S(p)\right|<\delta$ iff $\sg\in T_{N,\dl}(p)$, making \er{pone2} the same as \er{wlln1}.
\eex
Hence \er{pone2}  follows from the  weak law or large numbers (see supplement below).
The proof of the upper bound in \er{est33}, which implies \er{archedelta},  is almost trivial:  by  definition \er{TEN} we have 
\begin{align}
1=p^N(A^N)\geq p^N(T_{N,\dl}(p))=\Sigma_{\sg\in T_{N,\dl}(p)}p^N(\sg)\geq |T_{N,\dl}(p) |e^{-N(S(p)+\dl)}.
\end{align}
The proof of the lower bound in \er{est33} relies on \er{pone2}, which for $\dl>0$ gives $M=M(\dl)$ such that
\begin{equation}
p^N(T_{N,\dl}(p))> 1-\dl, \label{CWLL}
\end{equation}
for all $N>M$. A similar computation to the one just given then yields the lower bound in \er{est33}:
\begin{align*}
1-\dl<p^N(T_{N,\dl}(p))=\Sigma_{\sg\in T_{N,\dl}(p)}p^N(\sg)\leq |T_{N,\dl}(p) |e^{-N(S(p)-\dl)}. \tag*{$\Box$}
\end{align*}

There is also a \emph{strong} version of \er{pone2},
 called the  \emph{Shannon--McMillan--Breiman theorem}:
\begin{align}
 \lim_{N\raw\infty} \frac{1}{N}\log  p^{N}(s_{|N})=-S(p), \label{SMB}
\end{align}
for $p^\N$-almost all $s\in A^\N$. This removes the $\delta$-buffer $T_{N,\dl}(p)$ around $T_N(p)$ and states that the remarkable formula \er{PNpsg}, originally proved for $\sg\in T_N(p)$, is asymptotically generic. 
Eq.\ \er{SMB} has the same proof as \er{wlln1}, except for the fact that we now use the strong law of large numbers.
See also Theorem \ref{SMBs} for a generalization of \er{SMB} from the perspective of dynamical systems. 

The relative entropy \er{KL1}  also appears naturally in asymptotics,  in what will be our first step in large deviation theory. The key to this will be a generalization of \er{PNpsg}: for any $p,q\in\Pr(A)$,
\begin{align}
\sg\in T_N(p) && \Raw &&
q^N(\sg)=e^{-N(S(p)+S(p, q))}.
 \label{2.13} \end{align}
This implies a result that is valid for all $p,q\in\Pr(A)$ but trivial (i.e.\ $0=0$) for $p\notin \Pr_N(A)$:
\begin{equation}
 q^N(L_N=p)=q^N(T_N(p))=\Sigma_{\sg\in T_N(p)} q^N(\sg)=|T_N(p)|e^{-N(S(p)+S(p, q))}. \label{328}
\end{equation}
Eq.\  \er{arche} suggests $|T_N(p)|\approx e^{NS(p)}$, as in our first version of the AEP. Since this term cancels the term $e^{-NS(p)}$ in \er{328}, one is led to an estimate that will be taken up in \S\ref{Sanovchapter}: given \er{pana}, we have
\begin{equation}
 q^N(L_N=p)=N!\prod_{a\in A}\frac{q(a)^{n_a}}{n_a!} \approx e^{-NS(p, q)}. \label{329}
\end{equation}
\bex
Prove \er{2.13}.   Prove both parts of \er{329},  taking the second approximation to mean
\begin{align}
\lim_{N\raw\infty}\frac{1}{N} \log (q^N(L_N=p))=-S(p,q). \label{archeq}
\end{align}
\eex
\subsection*{Supplement: Laws of large numbers for i.i.d.\ variables}\addcontentsline{toc}{subsection}{Supplement: Laws of large numbers for i.i.d.\ variables}
 Here we give a brief summary of \er{wlln1} and related results; see Appendix \ref{MarC} for more details on the measure-theoretic side.
In what follows the underlying probability space is $(A^\N, \mathcal{F}, q^\N)$, where $\mathcal{F}$ is the smallest $\sg$-algebra on $A^{\N}$ that makes all  evaluation maps $X_n(s)=s_n$ measurable, for $s\in A^{\N}$ and $n\in\N$. Using Carath\'{e}odory's extension theorem,\footnote{See e.g.\ Dudley (1989) or Klenke (2020), and our Appendix \ref{MarC}.
 This construction goes back  to Kolmogorov.} it can  be shown that (probability) measures on $\mathcal{F}$, such as
 $q^\N$, are determined by their values on  the so-called \emph{cylinder sets}
 \begin{align}
 [\sg]_N:=\{s\in A^{\N}\mid s_{|N}=\sg\}=\sg A^\N && (\sg\in A^N), \label{cylinder}
 \end{align}
 that is, $[\sg]_N$ consists of all $s\in A^\N$ that start with $\sg$.
 On such sets one defines 
 \beq
 q^\N([\sg]_N):=q^N(\sg).
 \eeq 
 As in \er{0defXNov}, take $E:A\raw\R$, and, for each $N\in\N$, define random variables $S_N$ on $(A^\N, \mathcal{F}, q^\N)$ by
 \begin{align}
S_N: A^\N\raw\R; &&
S_N(s):=\frac{1}{N}\Sigma_{n=0}^{N-1} E(s_n).\label{SsumE}
\end{align}
Then the strong  law of large numbers  states that  for $q^\N$-almost every $s\in A^\N$ one has
  \beq\lim_{N\raw\infty} S_N(s)=\la E\ra_q:=\Sigma_{a\in A} q(a)E(a). \label{SLLN}
  \eeq 
 The simplest example is  $A=2=\{0,1\}$, $q=f$,  and $E(a)=a$, so that 
for $f^\N$-almost every $s\in 2^\N$,
  \beq\lim_{N\raw\infty} S_N(s)=\half.
  \eeq 
The empirical measures \er{LNb} and \er{LNbN} take values in $\Pr(A)$ instead of $\R$, but if we  define $p\in\Pr(A)$ 
  by its real values $p(b)$, $b\in A$,  taking $E_b(a)=\dl_{ab}$, with average $\la E_b\ra_q=q(b)$,
the strong laws for each $E_b$  neatly combine into the statement that  for $q^\N$-almost every $s\in A^\N$,
  \beq\lim_{N\raw\infty} L_N(s)=q. \label{SLLNSanov}
  \eeq 
  Although weaker than the strong law,  the \emph{weak law of large numbers} is often more useful:
\begin{theorem}\label{WLLN}
If random variables $(X_n)$ taking values in $A\subset\R$ are i.i.d.\ by $p\in\Pr(A)$, then
\begin{equation}
\lim_{N\raw\infty}  p^N\left( \left|\frac{1}{N}\Sigma_{n=0}^{N-1}X_n-\la X_0\ra_p\right|<\delta\right)=1, \label{wlln1bis}
\end{equation}
for all $\delta>0$, provided $\la |X_0|\ra_p<\infty$. If in addition $\la X^2_0\ra_p<\infty$, then one has the explicit bound
\begin{equation}
p^N\left( \left|\frac{1}{N}\Sigma_{n=0}^{N-1}X_n-\la X_0\ra_p\right|\geq \delta\right)\leq\frac{\mathrm{Var}_p(X_0)}{N\delta^2}.\label{wlln2}
\end{equation}
\end{theorem}
Here $\mathrm{Var}_p(X):=\la X^2\ra_p-\la X\ra^2_p$  is the variance of a random variable $X$ with law $p$, as usual, and
 we could have written $p^\N$ instead of $p^N$ (and the i.i.d.\ assumption could also be relaxed). 
Eq.\  \er{wlln1bis} can be proved either from the strong law,\footnote{See e.g.\ Klenke (2014), Remark 5.13, based on Fatou's lemma.} or from the more powerful pointwise (Birkhoff) ergodic theorem, see Theorem \ref{ET} later on.\footnote{If \er{wlln1} fails there is $\dl>0$ such that
for each $M$ there is $N>M$ for which $P(|f_N-f^*|)\geq \delta)\geq\dl$. This blocks  $f_N\raw f^*$ in $L^1$ (or in any $L^p$, $1\leq p<\infty$). Given Theorem \ref{ET}.2, this is a proof by contradiction of \er{wlln1}.}
If it applies (i.e.\ for finite variance), eq.\  \er{wlln2} also implies \er{wlln1bis}. The latter follows in turn from \emph{Chebyshev's inequality} 
\beq
p(|X-\la X\ra_p|\geq \delta)\leq \mathrm{Var}_p(X)/\delta^2,
\eeq
 plus the (easily verifiable)  fact that the assumption that the random variables  $X_n$ be i.i.d.\ gives
\beq
\mathrm{Var}_p\left(N\inv \Sigma_{n=0}^{N-1}X_n\right)=N\inv\mathrm{Var}(X_0).
\eeq
\section{Entropy and coding theory}\label{Ecoding}
 As Shannon  realized, the AEP has interesting consequences for coding (or data compression). 

 A \emph{binary code} of $A^N$ (for some finite alphabet $A$ and $N\geq 1$) is an injective map 
 \beq
 C^{(N)}:A^N\raw 2^*,
 \eeq 
 where $2^*:=\cup_{N\in\N_*} 2^N$ is the set of all finite binary strings ($\N_*=1,2,\ldots$).
 Given some probability distribution $p^{(N)}$ on $A^N$, the \emph{average length} (i.e.\  number of bits) of a codeword is defined by
  \begin{equation}
 \la\ell(C^{(N)})\ra_{p^{(N)}}:=\Sigma_{\sg\in A^N} p^{(N)}(\sg)\ell(C^{(N)}(\sg)). \label{defCN}
\end{equation}
We restrict ourselves to \emph{memoryless sources}; this means that $A^N$ is distributed by $p^N$ for some $p\in\Pr(A)$. The simplest possibility is then to choose a code $C:A\raw 2^*$ and extend this  to $A^N$ by
   \begin{align}
 C^{(N)}\equiv C^N:A^N\raw 2^*; && C^N(\sg):=C(\sg_0)\cdots C(\sg_{N-1}). \label{concat}
\end{align}
In that case the average length \er{defCN} is given by 
\beq
\la \ell(C^N)\ra_{p^N}=N\Sigma_{a\in A} p(a) \ell(C(a))=N \la\ell(C)\ra_p.\label{concatmean}
\eeq
Another possibility is to encode $A^N$ using some list.  This
requires approximately 
\beq
\log_2(|A|^N)=N\log_2(|A|)
\eeq
 bits (which is exact if $|A|$ is a power of two), which gives
\begin{align}
\lim_{N\raw\infty}\frac{1}{N}  \la\ell(C^{(N)}_{naive})\ra_{p^N}&=\log_2(|A|)=S_2(f), \label{naC}
\end{align}
cf.\ \er{logA}. But if $p\neq f$  we can do better if we use the AEP. Encode the elements of $A^N$ by:\footnote{We follow Cover \& Thomas (2006), \S3.2.}
\begin{enumerate}
\item Encoding elements of $T_{N,\dl}(p)\subset A^N$ naively via some list. By \er{est33} the subset $T_{N,\dl}(p)$ has approximately $e^{NS(p)}=2^{NS_2(p)}$ elements and hence 
this requires  $NS_2(p)+2$ bits: one extra 
bit since $S_2(p)$ may not be an integer and we incorporate $\dl>0$, and another extra bit to add a prefix 0 indicating that we have a string in
$T_{N,\dl}(p)$.
\item Also encode the complement $A^N\backslash T_{N,\dl}(p)$ by a list,  adding a prefix 1. Taking $A^N$ as an upper bound of the size of $A^N\backslash T_{N,\dl}(p)$, this takes $NS_2(f)+2=N\log_2(|A|)+2$ bits, cf.\ \er{logA}.
\end{enumerate}
Call this coding  $C^{(N)}_{\dl}$, for some $\dl>0$. We may estimate the average codeword length in $C^{(N)}_{\dl}$ by
\begin{align}
\la\ell(C^{(N)}_{\dl})\ra_{p^N}&=\Sigma_{\sg\in A^N} p^N(\sg)\ell(C^{(N)}_{\dl}(\sg))\nn \\ &=\Sigma_{\sg\in T_{N,\dl}(p)} p^N(\sg)\ell(C^{(N)}_{\dl}(\sg))+
\Sigma_{\sg\notin T_{N,\dl}(p)} p^N(\sg)\ell(C^{(N)}_{\dl}(\sg))\nn \\ 
&\leq \Sigma_{\sg\in T_{N,\dl}(p)} p^N(\sg)\cdot NS_2(p)
+
\Sigma_{\sg\notin T_{N,\dl}(p)} p^N(\sg)\cdot NS_2(f) +O(1)\nn \\ 
&\leq  p^N( T_{N,\dl}(p))\cdot NS_2(p)+  p^N(A^N\backslash T_{N,\dl}(p))\cdot NS_2(f) +O(1)\nn \\ 
&\leq  N(S_2(p) + \dl S_2(f)) +O(1), \label{340}
\end{align}
where for the last inequality we used the weak LLN \er{CWLL}, or rather its equivalent version
\begin{equation}
p^N(A^N\backslash T_{N,\dl}(p))\leq\dl,
\end{equation}
which is valid for sufficiently large $N$. In
 the limit one has a clean result
\begin{equation}
\lim_{\dl\raw 0}\lim_{N\raw\infty}\frac{1}{N}  \la\ell(C^{(N)}_{\dl})\ra_{p^N}\leq S_2(p).\label{cleanr}
\end{equation}
On the other hand, the very definition of the coding scheme $C^{(N)}_{\dl}$ 
gives the  inequalities
  \begin{align}
\la\ell(C^{(N)}_{\dl})\ra_{p^N}&=\Sigma_{\sg\in T_{N,\dl}(p)} p^N(\sg)\ell(C^{(N)}_{\dl}(\sg))+
\Sigma_{\sg\notin T_{N,\dl}(p)} p^N(\sg)\ell(C^{(N)}_{\dl}(\sg))\nn \\ &\geq\Sigma_{\sg\in T_{N,\dl}(p)} p^N(\sg)\ell(C^{(N)}_{\dl}(\sg))
\geq \Sigma_{\sg\in T_{N,\dl}(p)} p^N(\sg) \cdot NS_2(p)\nn \\ &=p^N(T_{N,\dl}(p))\cdot  NS_2(p)\nn \\ &\geq (1-\dl)
\cdot  NS_2(p), \label{344}
\end{align}
again using the weak law of large numbers \er{CWLL}. Hence  \er{cleanr} may be supplemented with 
\begin{equation}
\lim_{\dl\raw 0}\lim_{N\raw\infty}\frac{1}{N}  \la\ell(C^{(N)}_{\dl})\ra_{p^N}\geq S_2(p),\label{3422}
\end{equation}
so that for the  coding just described the average length per letter is asymptotically given by
\begin{equation}
\lim_{\dl\raw 0}\lim_{N\raw\infty}\frac{1}{N}  \la\ell(C^{(N)}_{\dl})\ra_{p^N}= S_2(p).\label{cleanr2}
\end{equation}
Compared with \er{naC},  since $S_2(p)\leq S_2(f)$ with equality iff 
 $p=f$, we see that (unless $p=f$) the improved coding $C^{(N)}_{\dl}$ considerably compresses the average length of long messages. 
 
It is instructive to see what happens if one tries to construct a similar  coding based on some alternative subset $B_{\dl}\subset A^N$ for which 
\beq
\lim_{N\raw\infty} p^N(B_{\dl})=1, \label{instructive}
\eeq
 cf.\ \er{pone2}, as is the case for $B_{\dl}=T_{N,\dl}(p)$. If \er{instructive} holds, then
 for large $N$ one can again achieve 
\beq
p^N(B_{\dl}) >1-\dl,
\eeq
 cf.\ \er{CWLL}, which is the key to \er{344}.
For any probability space $(X,P)$ and $A\subset X$ and $B\subset X$ such that $P(A)>1-\varep_1$ and $P(B)=1-\varep_2$ one has $P(A\cap B)> 1-\varep_1-\varep_2$;
  to see this, note that 
  \begin{align}
  P(A\cap B)=P(A)-P(A\backslash B); && P(A\backslash B)\leq P(X\backslash B)=1-P(B).
  \end{align} Therefore, \beq
  P(T_{N,\dl}(p)\cap B_{\dl})> 1-2\dl,
  \eeq
   so that for small $\dl>0$ the set $B_{\dl}$ is similar to $T_{N,\dl}(p)$ and we are back to the previous situation.\footnote{This argument is taken from Cover \& Thomas, Theorem 3.3.1.}
 
Moreover, we  cannot do any better because of Shannon's \emph{noiseless coding theorem}. We only discuss this theorem for codings of the type \er{concat}, so that we only need to talk about  
 codes \beq
 C:A\raw 2^*.\eeq
  Furthermore, we only consider 
\emph{prefix codes} (aka \emph{instantaneous codes}),  defined as follows.
\begin{definition}\label{SNCTD}\begin{enumerate}
\item 
We say that $\sg'\in 2^*$ is a \emph{prefix} of $\sg\in 2^*$ (or of $s\in2^\N$),  written $\sg'\prec\sg$ (or $\sg'\prec s$)
 if $\sg=\sg'\ta$ for some $\ta\in 2^*$ (or $s=\sg' t$ for some
 $t\in 2^\N$).
\item A subset $S\subset 2^*$ is a \emph{prefix set} if no $\sg'\in S$ is a prefix of any $\sg\in S$.
\item  A map $C:A\raw 2^*$ is a \emph{prefix code} if it is injective and $\{C(a), a\in A\}$ is a prefix set in $2^*$.
\end{enumerate}
  \end{definition}
   In other words, in a prefix code $C$, for any $a,b\in A$  there is no $\ta\in 2^*$ such that $C(b)=C(a)\ta$.
 It is easy to see that prefix codes are examples of the larger class of uniquely decodable codes.
 \begin{lemma}[\emph{Kraft inequality}]\label{KraftIn}
 \begin{enumerate}
\item 
  Any  prefix subset $S\subset 2^*$ 
 satisfies
 \begin{equation}
\Sigma_{\sg\in S}2^{-\ell(\sg)}\leq 1.\label{Kraft1}
\end{equation}
\item Consequently (since $C(A)\subset 2^*$ is a prefix set), for any prefix code $C:A\raw 2^*$ we have
\begin{equation}
\Sigma_{a\in A}2^{-\ell(C(a))}\leq 1.\label{Kraft}
\end{equation}
\item Conversely,  for any (at most countable) set $A$ and subset $\{\ell_a\}_{a\in A}$ of $\N$ that satisfies 
\beq
\Sigma_{a\in A} 2^{-\ell_a}\leq 1,
\eeq  there exists a prefix code $C:A\raw 2^*$ for which the codeword lengths satisfy
\beq
\ell(C(a))=\ell_a.
\eeq
\end{enumerate}
 \end{lemma}
\emph{Proof.}
Recall the cylinder set $[\sg]_{\ell(\sg)}=\{s\in 2^\N\mid s_{|\ell(\sg)}=\sg\}$. Since $S$  is a prefix set, we have
\beq
[\sg]_{\ell(\sg)}\cap [\sg']_{\ell(\sg')}=\emptyset,
\eeq
 whenever $\sg\neq\sg'$ (check this!). In terms of the flat prior $f$ on $2=\{0,1\}$, i.e.\ $f(0)=f(1)=\half$, and the ensuing Bernoulli measure $f^\N$ on $2^\N$, we have $f^\N([\sg]_{\ell(\sg)})=2^{-\ell(\sg)}$, and hence
 \begin{equation}
\Sigma_{\sg\in S}2^{-\ell(\sg)}=\Sigma_{\sg\in S}f^\N([\sg]_{\ell(\sg)})=f^\N\left(\Sigma_{\sg\in S}[\sg]_{\ell(\sg)}\right )\leq f^\N(2^\N)=1. 
\end{equation}
\bex Prove the converse, that is, prove part 3 of Lemma \ref{KraftIn}.\footnote{See e.g.\ Cover \& Thomas, Theorem 5.2.1, or Austin (2017), Lecture 3, Theorem 4.1.} 
\eex
  \begin{theorem}\label{NCT}
  \begin{enumerate}
\item Any prefix code satisfies $\la \ell(C)\ra_p\geq 
S_2(p)$, cf.\ \er{concatmean}.
\item There exist  prefix codes $C$ that satisfy
\beq
S_2(p)\leq \la \ell(C)\ra_p\leq S_2(p)+1.\label{boundsS}
\eeq
\end{enumerate} 
  \end{theorem}
 In the rare cases where  $p(a)=2^{-k(a)}$ for some integer $k(a)\in\N$, codes exist for which 
 \beq
 \ell(C(a))=I_2(a),
 \eeq
  for each $a\in A$; this is the only way to achieve $\la \ell(C)\ra_p=S_2(p)$, cf.\ Cover \& Thomas, \S 5.4. 
  Thus the information $I_2(a)$ in $a$ is roughly the length of the codeword $C(a)$ in an optimal coding $C$. 
  
  This should of course also apply if we take $(A^N,p^N, C^N)$ instead of $(A,p,C)$. Indeed, we have
 \begin{equation}
S_2(p^{N})=-\Sigma_{\sg\in A^N} p^{N}(\sg)\log_2 p^{N}(\sg)=NS_2(p),\label{NH2}
\end{equation}
and combining this with \er{concatmean} we see that Theorem \ref{NCT} consistently describes both the first and the second case. Of course, taking $(A^N, P^{(N)}, C^{(N)})$ for $(A,p, C)$ one is not restricted to 
the choices $P^{(N)}=p^N$ and $C^{(N)}=C^N$:  hence Theorem \ref{NCT} also covers  general sources and ``block codes''.
\smallskip

\noindent\emph{Proof.} Part 1 is an exercise. 
Part 2 follows by construction: for the \emph{Shannon code} $C_S$, order 
\begin{align}
A=(a_1, \ldots, a_{|A|}); && p(a_k)\geq p(a_{k+1}),
\end{align}and
 take $C_S(a_k)$ to be the first $\left \lceil{I_2(a_k)}\right \rceil$ bits from the binary expansion of $\Sigma_{l=1}^{k-1}p(a_l)$. Then trivially
\beq
\ell(C_S(a))=\left \lceil{I_2(a)}\right \rceil,
\eeq
 i.e.\ the smallest integer larger than or equal to $I_2(a)=-\log_2(p_a)$, cf.\ \er{info}. This gives the bound 
 \beq
 \la\ell(C_S)\ra_p\leq S_2(p)+1.
 \eeq\vspace{-5mm}
\bex 
Prove Theorem \ref{NCT}.1. Hint: Jensen's inequality for the convex function $x\mapsto -\log_2(x)$.
\eex
 The Shannon code is mainly of theoretical interest; in practice other codes $C$ are used with 
 \beq
 \la\ell(C)\ra_p\leq  \la\ell(C_S)\ra_p.\eeq
 The famous \emph{Huffman code} achieves the lowest possible value of   $\la\ell(C)\ra_p$ among all prefix codes.
 \bex
 Look up and describe this code. 
\eex

We now move to noisy coding, using the parlance of quantum information theory.   Alice  (the sender) holds a finite set or alphabet $A$ and wants to send encoded messages to Bob (the receiver). We still work with a memoryless source, so that as before $A$ carries a probability distribution $p\in\Pr(A)$, expanded to $p^N\in\Pr(A^N)$. Then the following distinction can be made:
\begin{itemize}
\item The noiseless case  just discussed was based on a map $C_N:A^N\raw 2^*$ (where $N$ is fixed) providing binary codewords $C_N(\sg)$   of \emph{variable} length for each message $\sg\in A^N$ of \emph{fixed} length. This map was assumed to be a prefix code and hence was 
uniquely decodable, so that Bob's decoding did not play a role. This is what made the procedure \emph{noiseless}. Shannon's noiseless coding theorem (Theorem \ref{NCT})  stated that the optimal average length of Alice's codewords per symbol (i.e.\ $\x 1/N$) is approximately $S_2(p)$.
\item  The noisy case  works with codewords of \emph{fixed} length $\ell_N$ (again for given $N$), based on an encoding map $E_N:A^N\raw 2^{\ell_N}$. Bob's decoding now enters explicitly as a map $D_N:2^{\ell_N}\raw A^N$, so that we obtain a combined map $D_N\circ E_N:A^N\raw A^N$. Define the \emph{fault probability}
\begin{equation}
\mathcal{F}_N(\ell_N, E_N, D_N):=p^N(D_N\circ E_N\neq \mathrm{id})=p^N(\{\sg\in A^N\mid D_N\circ E_N(\sg)\neq\sg\}), \label{7.1}
\end{equation}
which is  the probability that Bob's decoding of Alice's codeword for $\sg$ is wrong. 
The goal is then to find, for some give error tolerance  $\dl\in (0,1)$, the smallest possible integer $\ell_N$ so that maps $E_N$ and $D_N$ exist for which the fault probability remains acceptable, in  that
\beq
\mathcal{F}_N(\ell_N, E_N, D_N)\leq\dl.\label{7.2}
\eeq
\end{itemize}
The answer to this optimization problem is \emph{Shannon's noisy coding theorem}:
\begin{theorem}\label{NCT2}
Given  $(A,p)$ and  $\dl\in (0,1)$, for all except finitely many $N\in\N$, we have:
\begin{itemize}
\item If $\ell_N> S_2(p^N)$, then 
 \er{7.2} can be achieved for some coding scheme 
 \beq
 A^N\stackrel{E_N}{\raw}2^{\ell_N} \stackrel{D_N}{\raw}A^N.
 \eeq
\item  If $\ell_N< S_2(p^N)$, then 
 \er{7.2} fails for all coding schemes $(E_N, D_N)$ of this kind.\footnote{Equivalently:  for all $\dl\in (0,1)$,
 there exists  $N_0\in\N$ such that  for all $N> N_0$ and all $\ell_N> NS_2(p)$ there are  $(E_N, D_N)$ such that  \er{7.2} holds;  but if for any $N_0$ there is $N> N_0$ such that
    $\ell_N< S_2(p^N)$, then  \er{7.2} fails for all $(E_N, D_N)$. }
\end{itemize}
\end{theorem}
Defining the  \emph{rate} initially as $R_N=\ell_N/N$, by \er{NH2}  this case distinction corresponds to $R_N> S_2(p)$ and $R_N< S_2(p)$, and since these bounds are independent of $N$ we might as well write $R> S_2(p)$ and $R< S_2(p)$, in terms of which $\ell_N=NR$ recovers the description in terms of lengths. 
Since  $\dl$ is arbitrary, it follows that one may find $(E_N, D_N)$ with 
 $\ell_N> S_2(p^N)$ for all $N$, such that
\begin{equation}
\lim_{N\raw\infty} \mathcal{F}_N(\ell_N, E_N, D_N)=0. \label{7.4}
\end{equation}
Compare with Shannon's noiseless coding theorem, according to which optimal noiseless codes (i.e., $\dl=0$) with \emph{varying} length $\ell_N$ of codewords (for given $N$), have \emph{average} length 
\beq
\la\ell_N\ra\approx S_2(p^N),
\eeq as $N\raw\infty$. In the noisy case, then,  this is the optimal \emph{fixed} length of codewords (for given $N$). 

\noindent But  if the codewords of the schemes 
$A^N\stackrel{E_N}{\raw}2^{\ell_N} \stackrel{D_N}{\raw}A^N$ (almost) all have length $\ell_N< S_2(p^N)$, then 
\begin{equation}
\lim_{N\raw\infty} \mathcal{F}_N(\ell_N, E_N, D_N)=1.\label{7.5}
\end{equation}
We now sketch a proof of Theorem \ref{NCT2}, based on the AEP, in which we switch from $\dl$ to $\varep$, since $\dl$ is already in use, cf.\ \er{7.2}. Thus for any $\varep>0$, we define $T_{N,\varep}(p)\subset A^N$ by
\begin{equation}
T_{N,\varep}(p):=\{\sg\in A^N\mid  2^{-N(S_2(p)+\varep)}\leq  p^N(\sg)\leq  2^{-N(S_2(p)-\varep)}\}. \label{TEN2}
\end{equation}
Then one has, from (2.41) and the proof of (2.40) from (2.48), respectively,
\begin{align}
(1-\varep)2^{N(S_2(p)-\varep)} &\leq |T_{N,\varep}(p)|\leq 2^{N(S_2(p)+\varep)}; \label{est33b}\\
p^N(T_{N,\varep}(p))& \geq 1-\frac{\sg^2}{N\varep^2}, \label{pone22}
\end{align}
 where in \er{est33b} the upper bound is true for any $N$, whereas the lower bound holds for sufficiently large $N$, and $\sg$ in \er{pone22} is a constant independent of $\varep$ and $N$ (it does depend on $p$). 
We start with the ``successful'' case $\ell_N> S_2(p^N)$, i.e., $R>S_2(p)$. Take  
\beq\varep=\half(R-S_2(p))>0,\eeq in which case
\begin{equation}
N(S_2(p)+\varep)=\half (S_2(p^N)+\ell_N)<\ell_N,
\end{equation} so that
 it follows from \er{est33b} that $|T_{N,\varep}(p)|< 2^{\ell_N}$. Therefore, Alice is able to define her encoding map
 $E_N: A^N\raw 2^{\ell_N}$ such that it is injective on $T_{N,\varep}(p)\subset A^N$, so that Bob can trivially construct his decoding map $D_N: 2^{\ell_N}\raw A^N$ in such a way that $D_N\circ E_N(\sg)=\sg$ for all $\sg\in T_{N,\varep}(p)$, namely by sending $E_N(\sg)$ to $\sg$, which is unique by injectivity of $E_N$ on $T_{N,\varep}(p)$. Thus for all $N\geq \sg^2/\varep^2\dl$,
 \begin{align}
 \mathcal{F}_N(\ell_N, E_N, D_N)=p^N(\{\sg\in A^N\mid D_N\circ E_N(\sg)\neq\sg\})
 \leq p^N(\sg\notin T_{N,\varep}(p)\leq \frac{\sg^2}{N\varep^2}\leq\dl,
\end{align}
where we used \er{pone22}. This proves the first case. 
 The proof for  $\ell_N< S_2(p^N)$ is an exercise. \QED
\bex
Prove the case  $\ell_N< S_2(p^N)$ of Theorem \ref{NCT2}, using the \emph{lower} bound in \er{est33b}.
\eex
 \section{Large deviations: Sanov's theorem}\label{Sanovchapter}
 After this intermezzo  we turn to large deviation theory. An appropriate starting point is
 \emph{Sanov's theorem}, which  is a more powerful  version of \er{archeq}, in which $L_N=p_N$ is replaced by $L_N\in B$:\footnote{Here $\lim\inf_{N\raw\infty}r_N=\lim_{N\raw\infty} \inf_{M\geq N}\{r_M\}$ and $\lim\sup_{N\raw\infty}r_N=\lim_{N\raw\infty} \sup_{M\geq N}\{r_M\}$, where $(r_N)$ is some real sequence. 
 These limits always exist (but may be $\pm\infty$). Furthermore, $(L_N\in B)$ is short for $\{\sg\in A^N\mid L_N(\sg)\in B\}$.
}
\begin{theorem}\label{sanov}
Let $q\in\Pr(A)$. For any (weakly measurable) subset $B\subset\Pr(A)$ one has
\begin{align}
- S_q(\mathring{B}) \leq \lim\inf_{N\raw\infty} \frac{1}{N} \log q^N(L_N\in B)
\leq  \lim\sup_{N\raw\infty} \frac{1}{N} \log q^N(L_N\in B)\leq - S_q(B^-).  \label{eqLDP1} 
\end{align}
Equivalently, for any closed subset $F\subset\Pr(A)$ and any open subset $U\subset\Pr(A)$ one has
\begin{align}
\lim\sup_{N\raw\infty} \frac{1}{N} \log q^N(L_N\in F)&\leq - S_q(F);\label{LDPl}\\
\lim\inf_{N\raw\infty} \frac{1}{N} \log q^N(L_N\in U)&\geq - S_q(U).\label{LDPu}
\end{align}
\end{theorem}
 Here, for any function $I: \mathcal{X}\raw (-\infty,\infty]$ and $B\subset \mathcal{X}$,  given above by $\mathcal{X}=\Pr(A)$, we write
\begin{equation}
I(B):=\inf_{x\in B} \{I(x)\},\label{Binf}
\end{equation}
so that $S_q(B)=\inf_{p\in B} S_q(p)$, etc., with $S_q(p)\equiv S(p, q)$, cf.\ \er{KL1}. The values $S_q(\mathring{B})=\infty$ and  $S_q(B^-)=\infty$ are allowed (where the former does not imply the latter, see Exercise \ref{ex7}).
\bex
Prove the equivalence between \er{eqLDP1} and the conjunction \er{LDPl} - \er{LDPu}.
\eex
Yet both versions look quite technical! The situation greatly simplifies if $B$ satisfies the condition
\begin{equation}
S_q(\mathring{B})=S_q(B^-), \label{condition}
\end{equation}
where  $\mathring{B}$ is the interior of $B$ and $B^-$ is its closure (with respect to the weak 
 topology on $\Pr(A)$). 
 For example,  since for finite $A$ the function $p\mapsto S_q(p)$ is finite and  continuous on its domain $\mathcal{D}_{S_q}$, 
 i.e., the subset of $\Pr(A)$ where $p\ll q$ for some fixed prior $q\in\Pr(A)$, eq.\ \er{condition}  is satisfied if 
\beq
B\subseteq (\mathring{B})^-\subseteq \mathcal{D}_{S_q}, \label{opencl}
\eeq
cf.\ Definition \ref{conbasicdef}.2. The second inclusion holds for any $B$ if we assume that $q(a)>0$ for all $a\in A$ (a typical case is $q=f$), since in that case $p\ll q$ for all $p$ and hence $S_q$ is finite on all of $\Pr(A)$, that is, 
$\mathcal{D}_{S_q}=\Pr(A)$. Irrespective of the second inclusion, the first inclusion in \er{opencl} holds, for example, if
  $B$ is either an open  $\varep$-ball $\mathcal{B}_{\varep}(p')$ or its closure $\mathcal{B}_{\varep}(p')^-$, 
 for some $\varep>0$, i.e., 
  \begin{align} 
  \mathcal{B}_{\varep}(p')=\{p\in \Pr(A)\mid d(p,p')<\varep\}; && d(p,p'):=\Sigma_{a\in A}|p(a)-p'(a)|.
  \end{align} 
\begin{corollary}\label{sanovC}
Let $q\in\Pr(A)$. For any (measurable) $B\subseteq\Pr(A)$ such that \er{condition} holds, we have
 \begin{equation}
\lim_{N\raw\infty} \frac{1}{N} \log q^N(L_N\in B)= - S_q(B).\label{611}
\end{equation}
\end{corollary}
This follows at once from Theorem \ref{sanov}, since under \er{condition} the outer ends of \er{eqLDP1} coincide, 
so that  $\lim\inf$ and $\lim\inf$ are equal to each other and hence to the  limit. 
\newpage

Eq.\ \er{condition} 
avoids some pathologies. For example, taking $B=\{p\}$ for some $p\in\Pr(A)$ that is not in any $\Pr_N(A)$ gives $(L_N\in B)=(L_N=p)=\emptyset$ and hence $q^N(L_N\in B)=0$, so that the left-hand side of \er{611} is $-\infty$ for all $N$, whereas the right-hand side equals $-S_q(p)$, which is finite provided $p\ll q$, cf.\ \er{KL1} - \er{KL2}. This is  what complicated  our motivational eq.\ \er{329}.
On the other hand, eq.\ \er{611} may even be valid in some cases where \er{condition} does not hold. For example:
\bex \label{ex7}
Using the notation of Exercise \ref{ex2}, again take $q=1$ and show that for $B=[0,1)$,
\begin{align}
S_1(\mathring{B})=S_1(B)=\infty; && S_1(B^-)=0,
\end{align}
and also argue that  $q^N(L_N\in B)=0$ for all $N$, so that \er{611} is valid in this case as $-\infty=-\infty$.
\eex
Before proving Theorem \ref{sanov}, let us explore some consequences of its Corollary \ref{sanovC}. First, 
the simplest illustration is as always on $A=\{0,1\}$, where we identify  
\begin{align}
\Pr(\{0,1\})\cong[0,1]; && p\lraw p(1)\equiv x, \label{431}
\end{align} cf.\ Exercise \ref{ex2}.
On this identification we have a simple expression for the empirical measure, viz.
\begin{align} L_N: 2^N\raw [0,1]; && L_N(\sg)=\frac{1}{N}\Sigma_{n=0}^{N-1} \sg_n.
\end{align}
We now take $q=f$. Using \er{Dfflat} and \er{Aentropy}, and  denoting $S_f(p)$ by $I(x)$, we obtain
\begin{equation}
I(x)=(1-x)\log (1-x)+x\log x +\log 2. \label{Ix1}
\end{equation}
\begin{center}
 \includegraphics[width=0.4\textwidth]{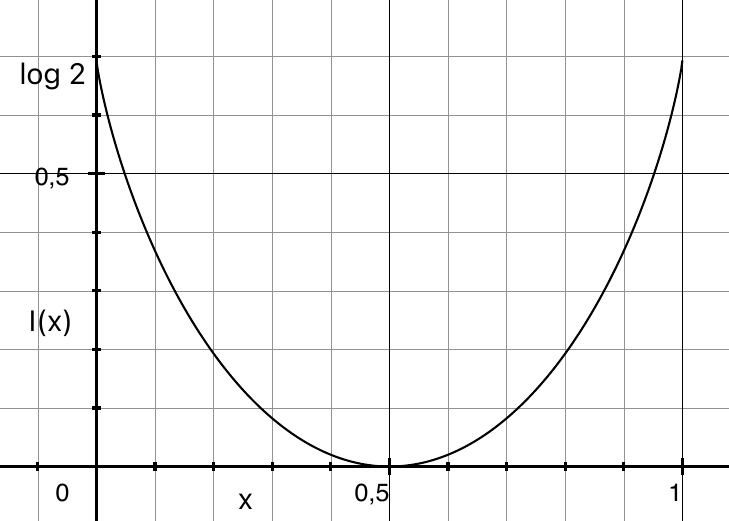}
\end{center}
We see from this graph that if $x>\half$, then $I([x,1])=I(x)$, so that \er{611} gives, for example, 
 \begin{align}
\lim_{N\raw\infty} \frac{1}{N} \log f^N(L_N\geq x)= - I(x)>0 && (x>\half),\label{IxSanov}
\end{align}
so that $f^N(L_N\geq x)\raw 0$ exponentially. 
But if $\half\in B$, then $I(B)=0$ and hence $f^N(B)\raw 1$.

Second,  we show how Corollary \ref{sanovC} relates to the strong law of large numbers.
  \begin{lemma}\label{L4.3}
  Let $(r_N)$ be a sequence in $\R$ for which $\lim_{N\raw\infty} \frac{1}{N} \log r_N=-c$ with $c>0$. Then
there exist $d>0$ in $\R$ and $M>0$ in $\N$ such that for all $N>M$ we have $ r_N\leq e^{-dN}$.
  \end{lemma}
\bex
Prove this  (most easily by contradiction).
\eex 
To appreciate the following argument please recall \er{SLLNSanov}, stating that $L_N\raw q$ in a suitable sense. 
Assume \er{611} and take $B$ closed such that $q\notin B$.  Then  $S_q(B)>0$  by the Gibbs inequality in Proposition  \ref{gibbsin}. It follows from Lemma \ref{L4.3} that for some $d>0$ and sufficiently large $N$ we have
\begin{equation}
q^N(L_N\in B)\leq e^{-dN}.\label{4.17}
\end{equation}
Thus for large $N$,  the probability that $L_N$ takes values $p$ that ``deviate largely'' from its limit value $q$ (in the sense that $p$ lies outside some open $\varep$-ball around $q$) is exponentially damped in $N$.
 
 On the other hand, take $q\in B$, assuming \er{611}. 
 Then we  know from  \er{SLLNSanov} that 
 \begin{equation}
\lim_{N\raw\infty} q^N(L_N\in B)=1.\label{4.18}
\end{equation}
In fact, the strong law of large numbers \er{SLLNSanov}, and hence \er{4.18},   follows from Theorem \er{sanov}.
\smallskip

\noindent\emph{Proof of  the strong law of large numbers \er{SLLNSanov}}. 
Take $\varep>0$ and define an event
$C_N\subset A^\N$ by 
\beq
C_N= (L_N\notin \mathcal{B}_{\varep}(q))\equiv \{s\in A^\N\mid L_N(s)\notin \mathcal{B}_{\varep}(q)\}.\eeq
 Since
$q\notin \mathcal{B}_{\varep}(q)^c$ by construction and $ \mathcal{B}_{\varep}(q)^c$ is closed,
we have
 $S_q(\mathcal{B}_{\varep}(q)^c)>0$ by Proposition  \ref{gibbsin},  so that Lemma \ref{L4.3}  guarantees that $\Sigma_{N=1}^{\infty} q^N(C_N)<\infty$. 
By the Borel--Cantelli lemma,\footnote{\noindent Recall that the  Borel--Cantelli lemma states that if $(X,\Sg,P)$ is a probability space and $(B_N)$ is a sequence in $\Sg$ for which $\Sigma_N P(B_N)<\infty$, then  almost surely $x\in X$ lies in only finitely many $B_N$.} $q^\N$-a.e.\ $s\in A^{\N}$ lies in only finitely many $C_N$. Hence for each $\varep>0$ and $q^\N$-a.e.\ $s\in A^{\N}$ we have $\lim_{N\raw\infty} L_N(s) \in \mathcal{B}_{\varep}(q)$, so that, letting $\varep\raw 0$, we have
$\lim_{N\raw\infty} L_N(s)=q$ almost surely. \QED
\smallskip

 \noindent \emph{Proof of Theorem \ref{sanov}}. We prove the version  \er{LDPl} -  \er{LDPu}.
 The key to the proof is \er{328}, i.e., 
 \begin{equation}
 q^N(L_N=p)=|T_N(p)|e^{-N(S(p)+S_q(p))}. \label{328bis}
\end{equation}
 \begin{itemize}
\item \emph{Proof of upper bound} \er{LDPl}. This is usually the easier one, also here. It relies on the estimate
\begin{equation}
 |T_N(p)|\leq e^{NS(p)},\label{413}
\end{equation}
which follows from \er{PNpsg} and the following computation:
\beq
1\geq p^N(T_N(p))=\Sigma_{\sg\in T_N(p)} p^N(\sg)=\Sigma_{\sg\in T_N(p)}e^{-NS(p)}=|T_N(p)| e^{-NS(p)},
\label{inter}\eeq
 which gives $|T_N(p)| e^{-NS(p)}\leq 1$ and hence \er{413}. Combining
 \er{328bis} and \er{413} 
 yields
 \begin{equation}
 q^N(L_N=p)\leq e^{-NS_q(p)},
\end{equation}
from which we continue to estimate
 \begin{align}
 q^{N}(L_N\in F) &= \Sigma_{p\in F\cap\, \Pr_N(A)} q^N(L_N=p) \leq  \Sigma_{p\in F\cap\, \Pr_N(A)} e^{-NS_q(p)}
 \leq  \Sigma_{p\in F\cap\, \Pr_N(A)} e^{-NS_q(F)}
 \nn \\
 &\leq |F\cap\, \Pr_N(A)|e^{-NS_q(F)}\leq |\Pr_N(A)|e^{-NS_q(F)} \nn \\ &\leq 
 (N+1)^{|A|} e^{-NS_q(F)}. \label{416}
 \end{align}
 For the  second inequality, since $S_q(p)\geq S_q(F):=\inf_{p\in F}\{ S_q(p)\}$ we have 
\beq
e^{-N S_q(F)}\leq e^{-N S_q(F^-)}.
\eeq
For the last inequality, we use the fact that 
\beq
|\Pr_N(A)|\leq (N+1)^{|A|},\label{factthat}
\eeq
 which follows because in the probabilities $p(a)=n_a/N$ comprising
$\Pr_N(A)$ each number $n_a\in\{0, 1, \ldots, N\}$ can only take $N+1$ possible values. Eq.\ \er{416} at once gives  \er{LDPl}.
\item \emph{Proof of lower bound} \er{LDPu}. The role of \er{413} is now played by the estimate
\begin{align}
 |T_N(p)|\geq 
(N+1)^{-|A|} e^{NS(p_N)}; && (p_N\in\Pr_N(A)),  \label{est1}
\end{align}
to be proved later on. 
Using \er{est1} in \er{328bis} once again removes $S(p_N)$, so that we obtain
\begin{align}
 q^N(L_N=p_N)\geq (N+1)^{-|A|} e^{-NS_q(p_N)} &&  (p_N\in\Pr_N(A)).  \label{estq}
\end{align}
Combining the estimates
\er{413} and \er{est1} gives the sandwich 
\begin{equation}
 (N+1)^{-|A|} e^{NS(p)}\leq|T_N(p)|\leq e^{NS(p)}, \label{est1est2}
\end{equation}
where the lower bound is  valid for $p\in\Pr_N(A)$ only. 

Assume that $S_q(U)<\infty$, for otherwise \er{LDPu} is trivially satisfied. Since
$U\subset\Pr(A)$ is open,  by  definition of an infimum, for any
 $\varep>0$
   there is some $p\in U$ such that 
\begin{align}
S_q(p)<S_q(U)+\varep/2 && \Raw && e^{-NS_q(p)}>e^{-N(S_q(U)+\varep/2)}.
\end{align}  Furthermore, 
for any $p\in\Pr(A)$ we can find a sequence $(p_N)$ in $\Pr_N(A)$ that (weakly) converges to $p$, 
and since $U$ is open, for large enough $M$ we have $p_N\in U$ for all $N>M$. By continuity of $S_q$ (on its domain) we can also find $M'$ such that $|S_q(p_N)-S_q(p)|<\varep/2$ for all $N>M'$, so that for all $N>\max(M,M')$ we have 
$e^{-NS_q(p_N)}>e^{-N(S_q(U)+\varep)}$. Hence
\begin{equation}
 q^N(L_N\in U)\geq q^N(L_N=p_N) \geq (N+1)^{-|A|} e^{-NS_q(p_N)}
\geq (N+1)^{-|A|} e^{-N(S_q(U)+\varep)}.
\end{equation}
Taking the limit in \er{LDPu} and letting $\varep\raw 0$ at the end then gives  \er{LDPu}. 

We still need to prove  \er{est1}. This  uses the estimate, for any $p,q\in\Pr_N(A)$, 
\begin{equation}
p^N(T_N(q))\leq p^N(T_N(p)). \label{use}
\end{equation}
This may be unsurprising, but it is actually hard to prove (see below).
Also, trivially,
\beq
A^N=\bigcup_{p\in\Pr_N(A)} T_N(p). \label{trAN}
\eeq
Using \er{use}, \er{factthat},  \er{trAN}, and \er{inter}, 
we obtain  \er{est1} by the following argument:
\begin{align}
1&=p^N(A^N)=p^N\left(\bigcup_{q\in\Pr_N(A)} T_N(q)\right)=\Sigma_{q\in\Pr_N(A)}p^N(T_N(q))
\leq \Sigma_{q\in\Pr_N(A)}p^N(T_N(p)) \nn \\
&= |\Pr_N(A)| p^N(T_N(p))\leq (N+1)^{|A|}p^N(T_N(p))=(N+1)^{|A|}|T_N(p)| e^{-NS(p)}.
\end{align}
 Finally, to prove  \er{use} we compute:
\begin{align}
 p^N(T_N(q))&=\Sigma_{\sg\in T_N(q)} p^N(\sg)=\Sigma_{\sg\in T_N(q)} \prod_{a\in A}p(a)^{Nq(a)}=
 |T_N(q)| \prod_{a\in A}p(a)^{Nq(a)}  \label{2.12.1}  \\ &= \frac{N!}{(Nq_{a_1})!\cdots (Nq_{a_{|A|}})!}\prod_{a\in A}p(a)^{Nq(a)},
\end{align}
 where we used \er{2.13} with $p$ and $q$ swapped, as well as \er{combi} with $p\leadsto q$. Therefore,
 \begin{equation}
\frac{ p^N(T_N(p))}{ p^N(T_N(q))}=\prod_{a\in A} \left( \frac{(Nq_a)!}{(Np_a)!}\cdot p_a^{N(p_a-q_a)}\right).
\end{equation}
The inequality $(m!/n!)\geq n^{m-n}$ then yields \er{use}.\QED

\end{itemize}
\bex Prove  this inequality and supply the ensuing final step towards \er{use}. 
\eex

Sanov's theorem also holds for Polish spaces (cf.\ the supplement below), provided we appropriately generalize our concepts  used so far for finite $A$. In particular, we replace probability distributions by probability measures (which we still call $p$ or $q$), and
define $S_q(p)\equiv S(p,q)$  by
   \begin{align}
S_q(p)&:=\int_A dq\, \frac{dp}{dq}\log\left(\frac{dp}{dq}\right)
 \:\:\: \mbox{ if }  p\in\Pr(A)\: \mathrm{and}\:\: p\ll q; \label{DefSq1}\\
S_q(p)&:=\infty\:\:\: \mbox{ otherwise} .\label{DefSq2}
\end{align}
Here $dp/dq$ is the Radon--Nikodym derivative,\footnote{\label{RNder} For any two measures $p,q$ on the same $\sg$-algebra $\Sg$ we say that $p$ is \emph{absolutely continuous} with respect to  $q$, written $p\ll q$, iff $q(\Dl)=0$ implies $p(\Dl)=0$ for $\Dl\in\Sg$; for finite $A$ this just means that $q(a)=0$ implies $p(a)=0$.  In that case the \emph{Radon--Nikodym derivative} $\rh =dp/dq$ exists and satisfies $\int_X dp\, f=\int_X dq\, \rh f$ for any $\Sg$-measurable function $f:X\raw \R\cup\{\infty\}$; for finite $A$ we have $dp/dq(a)=p(a)/q(a)$ if $q(a)>0$ and $dp/dq(a)=0$ if $q(a)=0$.}
which exists because of the clause $p\ll q$. Proposition \ref{gibbsin} and Exercise \ref{Sconvexex} remain valid in this generality:
the Gibbs inequality $S_q(p)\geq 0$, with equality iff $p=q$, still  follows from Jensen's inequality \er{Jensen} for the strictly convex function $x\log x$, and convexity of $p\mapsto S_q(p)$ follows from Theorem \ref{FTDq}, especially \er{FDD1},  below.\footnote{Joint convexity \er{Spqconvex} also holds, with a similar proof as for finite $A$ (this is not needed for Sanov's theorem).} Theorem \ref{sanov} then holds \emph{verbatim}, but unfortunately the proof is beyond the scope of these notes.\footnote{See e.g.\ Rassoul-Agha \& Sepp\"{a}l\"{a}inen (2015), \S5.2,  Dembo \& Zeitouni, \S6.2,
 or the particularly neat proofs by  Csisz\'{a}r (2006) and  Baldasso,  Oliveira,  Pereira, \& Reis (2023), which are based on the finite case via partitions.}

For a statistical physics perspective on the rate function $S_q$ in Sanov's theorem, first take a look at the \emph{Fenchel transform} \er{FeT} in Appendix \ref{AppC}.  In the optimal case where $f:X\raw(-\infty, \infty]$ is convex and lsc,  in which case $f^*: Y\raw (-\infty,\infty]$ is automatically convex and lsc, too, we have
\begin{align}
 f(x)= \sup_{\phv\in Y}\{\phv(x)-f^*(\phi)\}; &&
 f^*(\phv)=\sup_{x\in X}\{\phv(x)-f(x)\}. \label{FeTd}
 \end{align}
 Our aim is to prove these relations, for Polish spaces $A$, by taking:
 \begin{itemize}
\item $X=\mathcal{M}(A)$,  the linear space of \emph{finite signed Borel measures} $p$ on $A$, of which the space $\Pr(A)$
of probability measures on $A$ is a closed convex subspace.\footnote{For finite $A$ we have $X=\R^{|A|}$,  the linear space of all maps $p:A\raw\R$, which as we have seen contains $\Pr(A)$.}
\item $Y=C_b(A)$, the linear space of all bounded real-valued continuous functions $E:A\raw\R$.\footnote{For finite $A$ these are all  functions $E:A\raw\R$, and $Y$ is the algebraic or even Banach dual space of $X$.}
 The dual variable $\phv$ in \er{FeTd} is therefore $E\in C_b(A)$, and the pairing $\phv(x)$ is here given by
  \begin{equation}
 E(p)\equiv \la p,E\ra=\int_A dp\, E=\Sigma_{a\in A} p(a)E(a),\label{435}
\end{equation}
where the last expression of course applies to the finite case only. 
\item $f=S_q$ for some fixed prior $q\in\Pr(A)$, where the variable $x$ in  \er{FeTd}  is now $p\in\mathcal{M}(A)$,
and we extend $S_q$, initially defined on $\Pr(A)\subset \mathcal{M}(A)$ only, to all of $\mathcal{M}(A)$ by putting
\begin{align}
S_q(p)=\infty && (p\notin\Pr(A)). \label{438}
\end{align}\end{itemize}
\bex \label{Sqpex}
 For finite $A$, show that the function $p\mapsto S_q(p)$ as defined  on  $\mathcal{M}(A)$ by \er{KL1} - \er{KL2} and \er{438} is convex and continuous where $S_q$ is finite, and convex and lsc everywhere on  $\mathcal{M}(A)$.
\eex
This is also true for Polish spaces $A$, a fact that may either be used as a lemma for  the following key theorem, or, as in our approach, is a corollary of it. Either way, for any $q\in\Pr(A)$ we first define  the \emph{pressure} $\Pi_q: C_b(A)\raw \R$  in terms of a \emph{partition function} $Z_q:C_b(A)\raw \R$ by
 \begin{align} 
 \Pi_q(E):= \log Z_q(E); &&
Z_q(E):=\la e^{E} \ra_q=\int_A dq(a)\, e^{E(a)}. \label{FandP} 
\end{align}
\begin{lemma}\label{lsccp}
The function $f\mapsto \Pi_q(f)$ on $C_b(A)$ is convex and lsc (in norm). 
\end{lemma}
\emph{Proof.} Lower semicontinuity of $\Pi_q$   will be proved from Fatou's lemma.\footnote{This states that for any continuous (or even lsc) function $F:\R\raw [0,\infty]$ and weakly convergent sequence $P_n\raw P$ in $\Pr(\R)$ we have 
$\lim\inf_{n\raw\infty} \int_{\R} dP_n\, F\geq \int_{\R} dP\, F$.}
For $f\in C_b(A)$ the probability $q\in\Pr(A)$ induces a unique probability $P\in\Pr(\R)$ such that
\beq
\Pi_q(f)=\log\int_A dq(a)\, e^{f(a)}=\log\int_{\R} dP(x) e^x,
\eeq
viz.\ $P(B)=q(f\inv(B))$, where $B\subset\R$ is (Borel) measurable. Likewise, $f_n$ determines $P_n$.
If $f_n\raw f$ in $C_b(A)$, then $P_n\raw P_n$ weakly in $\Pr(\R)$, whence
 Fatou's lemma with $F(x)=\exp(x)$ gives
\begin{equation}
\lim\inf_n\, \Pi_q(f_n)= \lim\inf_n\,  \log\int_{\R} dP_n(x) e^x\geq  \log\int_{\R} dP(x) e^x=\Pi_q(f),
\end{equation}
using also continuity and monotonicity of log. Then 
\er{lscliminf} makes $\Pi_q$ lower semicontinuous.
 \bex \label{exHol} Prove  convexity of $\Pi_q(f)$  from H\"{o}lder's inequality. \QED  \eex
 \begin{theorem}\label{FTDq}
The Fenchel transform $S_q^*: C_b(A)\raw (-\infty, \infty)$ of $S_q: \mathcal{M}(A)\raw(-\infty, \infty]$ is given by
\beq
S_q^*=\Pi_q.
\eeq
Thus $S_q$ and $\Pi_q$, as defined by \er{DefSq1} - \er{DefSq2} and \er{FandP}, respectively, form a Fenchel dual pair
 \begin{align}
S_q(p)&=\sup_{E\in C_b(A)}\{\la E\ra_p-\Pi_q(E)\} && (S_q=\Pi_q^*);\label{FDD1}  \\
\Pi_q(E)&=\sup_{p\in \mathcal{M}(A)}\{ \la E\ra_p-S_q(p)\} && (\Pi_q=S_q^*). \label{FDD2} 
\end{align}
\end{theorem}
  \emph{Proof}.\footnote{We follow Dembo \& Zeitouni, Lemma 6.2.13 and Rassoul-Agha \& Sepp\"{a}l\"{a}inen (2015), Theorems 5.4 and 5.6.} 
  We prove \er{FDD2}. Eq.\  \er{FDD1} then follows from Theorem \ref{FDT}, which applies because of Lemma \ref{lsccp}.
As promised, eq.\ \er{FDD1} makes  $S_q$ is convex and lsc, cf.\ Definition \ref{defF} etc.

Because of \er{438}, eq.\  \er{FDD2}, the one to be proved, is obviously equivalent to
\begin{equation}
\Pi_q(E)=\sup_{p\in \Pr(A)}\{ \la E\ra_p-S_q(p)\},\label{FDD2b} 
\end{equation}
in which we in turn may further restrict the supremum to those $p\in\Pr(A)$ for which $p\ll q$. Then by the Radon--Nikodym theorem there is a Radon--Nikodym derivative $\nu:=dp/dq$ in $L^1(A,q)$, with $\nu\geq 0$ and  $\int dq\ \nu=\int dp=1$. Hence eq.\  \er{FDD2b}, now the one to be proved, becomes
\begin{equation}
\log \int_A dq\, e^f=\sup_{\nu\in L^1(A,q), \nu\geq 0,\int dq\, \nu=1}\left\{\int_A dq\, \nu(f-\log \nu)\right\}.\label{theone}
\end{equation}
We prove  $\leq$ and $\geq$ to give =. Taking $\nu=e^f/\int_A dq\, e^f$ in the curly brackets on the right-hand side gives the left-hand side, which proves $\leq$.
Conversely, Jensen's inequality \er{Jensen} for $-\log$ gives
\begin{equation}
\log \int_A dq\, e^f\geq \log \int_A dq\, 1_{\nu>0}\, \nu\cdot\frac{e^f}{\nu}=\log \int_{\{a\in A\mid \nu(a)>0\}} dp\, \frac{e^f}{\nu}\geq  \int_{\{a\in A\mid \nu(a)>0\}} dp\, (f-\log \nu).
\end{equation}
Since $dp=dq\, \nu$, this gives \er{theone} with $\geq$ instead of $=$, which gives
 \er{FDD2b}.  \QED

\begin{proposition}\label{GDS}
For finite $A$, the supremum in \er{FDD2b} is  attained uniquely by $p=q_E$, defined by
\begin{align}
q_E(a):=\frac{q(a)e^{E(a)}}{Z_q(E)}; && Z_q(E)= \Sigma_{a\in A} q(a) e^{E(a)}. \label{GibbsSanov}
\end{align}
\end{proposition}
This is called the \emph{Gibbs distribution}.
In statistical mechanics one puts $E=-U=-\beta h$ with
 $\beta=1/kT$ and  $h$ the Hamiltonian, and $q$ is the flat prior $q=f$. With the latter choice it is conventional to omit the prefactor $f(a)=1/|A|$ from $Z_f(\beta h)\equiv Z(\beta)$, so that the Gibbs distribution becomes
 \begin{align}
p_E(a):=\frac{e^{E(a)}}{Z(\beta}; && Z(\beta)=\Sigma_{a\in A} e^{-\beta h(a)}.
 \end{align} 
 \emph{Proof of Proposition \ref{GDS}.} We compute the supremum using calculus. As a warm-up, to practice with the  constraint $p\in\Pr(A)$,  let us first maximize the entropy $S(p)$ as given by \er{Aentropy} but defined on $\mathcal{M}(A)=\R^{|A|}$ rather than on $\Pr(A)$,
 subject to $p\in\Pr(A)$. Ignoring the ensuing constraint $p(a)\geq 0$ for the moment,  we impose $\Sigma_a p_a=1$ using a Lagrange multiplier, so that we extremize 
 \beq
 g(p)=S(p)+\lm\left(\Sigma_a p_a-1\right).
 \eeq
Putting $\partial g/\partial p_a=0$ gives $p_a=e^{\lm-1}$, upon which $\Sigma_a p_a=1$ gives $e^{\lm-1}=1/|A|$, i.e, 
$p=f$; the positivity conditions $p_a\geq0$ hold automatically. Since 
\beq
\frac{\partial^2 S}{\partial p_a\partial p_b}(f)=-|A|\dl_{ab},
\eeq
 this extremum is a maximum. To compute \er{FDD2b}, our real problem, one likewise extremizes 
 \beq
 h(p)=\la p,E\ra-S_q(p)+\lm\left(\Sigma_a p_a-1\right)=\Sigma_{a\mid q_a>0} p_a(\log q_a-\log p_a +E_a+\lm)-\lm,
 \eeq
 where we may restrict the sum to those $a$ for which $q_a>0$ since if $q_a=0$ and $p\ll q$ then  $p_a=0$, whilst 
 $q_a=0$ and $p_a>0$ gives $S_q(p)=\infty$, and such $p$ cannot contribute to the supremum in \er{FDD2b}.
Putting $\partial h/\partial p_a=0$ gives $p_a=q_ae^{E_a}e^{\lm-1}$, which, using $\Sigma_a p_a=1$, gives
\er{GibbsSanov}. \QED
\smallskip

Form this, it is instructive to also (re)prove \er{FDD1} - \er{FDD2} for finite $A$. The latter is easy: substituting \er{GibbsSanov} in $h(p)$ yields $\Pi_q(E)$, and since the second derivative of $h(p)$ at $p=q_E$ equals 
\beq
\frac{\partial^2 h}{\partial p_a\partial p_b}(q_E)=
-\frac{\dl_{ab} Z_q(E)}{q_ae^{E_a}}<0,
\eeq
 for those $a$ with $q_a>0$, and equals zero otherwise, the extremum is indeed a maximum. 

For \er{FDD1}, take
\beq
 k(E):=\Sigma_a E_ap_a-\log \Sigma_b q_be^{E_b}.\label{willik} 
\eeq
First assume that $p\in\Pr(A)$ and that $p\ll q$, i.e.,  $q_a=0$ implies $p_a=0$. Solve 
 \begin{align}
 \frac{\partial k(E)}{\partial E_a}=p_a-\frac{q_ae^{E_a}}{\Sigma_b q_be^{E_b}}=0, \label{gnul}
 \end{align}
 which gives $p_a=q_ae ^{E_a}/\Sigma_b q_b e^{E_b}$, cf.\ 
 \er{GibbsSanov}, but now meant to solve for the $E_a$ in terms of the $p_a$.
   If $q_a>0$, this equation gives 
 \beq
 E_a=\log(p_a/q_a)+\log\left(\Sigma_b q_be^{E_b}\right),
 \eeq
whilst $q_a=0$ simply returns our assumption $p_a=0$. Either way,  a fortunate cancellation  gives \begin{equation}
k(\ovl{E})=\Sigma_{a\in A} p(a)\log\left(\frac{p(a)}{q(a)}\right),
\end{equation}
at the value $\ovl{E}$ of $E$ where \er{gnul} holds. If $q(a)=0$, then $p(a)=0$ and the corresponding term in the sum vanishes. This is also the case if $p(a)=0$ whilst $q(a)>0$. The Hessian of $k$ at $\ovl{E}$ is
\begin{equation}
 \frac{\partial^2 k(\ovl{E})}{\partial E_a\partial E_b}=-p_a\dl_{ab}+p_ap_b. \label{thetwo}
\end{equation}
This matrix is negative-semidefinite,\footnote{Its negative
$M_{ab}=p_a\dl_{ab}-p_ap_b$ is \emph{diagonally dominant}, i.e.\ for each $a$ we have $|M_{aa}|\geq\Sigma_{b\neq a}|M_{ab}|$; indeed we have equality for each $a$, since $|M_{aa}|=M_{aa}=p_a-p_a^2$ and 
$\Sigma_{b\neq a}|M_{ab}|=p_a\Sigma_{b\neq a}p_b=p_a\Sigma_{b}p_b-p_a^2=p_a-p_a^2$.
Since it is also symmetric with non-negative diagonal entries (as $p_a\geq p_a^2$ for $0\leq p_a\leq 1$), $M$ is positive semi-definite.
}
 and hence $\ovl{E}$ is a maximum of $k$, so that \er{willik} follows.
 
  We finally remove the restrictions $p\in\Pr(A)$ and  $p\ll q$ under which we proceeded so far:
 \begin{itemize}
\item If $p\in\Pr(A)$ but $p\ll q$ fails, then
 there is some $a$ for which $q(a)=0$ and $p(a)>0$. In that case \er{gnul} has no solution and $k(E)\raw\infty$ as $E^a\raw\infty$, making $\sup_{E} k(E)$ infinite.
\item  If $p\notin\Pr(A)$, there are two cases:
\begin{enumerate}
\item If at least one $p_a<0$, then obviously \er{gnul} has no solution for that value of $a$ (since $q_a\geq 0$), and once again $k(E)\raw\infty$, but this time for $E^a\raw-\infty$. 
\item If all $p_a\geq 0$ but $\Sigma_a p_a\neq 1$, then 
  \er{gnul} has no solution either, since if it had one, then 
  \beq
  \Sigma_a p_a=\Sigma_a \frac{q_ae^{E_a}}{\Sigma_b q_be^{E_b}}=1.
  \eeq
\end{enumerate}
\end{itemize}

All in all, we have verified \er{FDD2}, including all exceptional values of $p$ where $S_q(p)=\infty$. \QED
\smallskip

\subsection*{Supplement: Polish spaces and standard Borel spaces} \addcontentsline{toc}{subsection}{Supplement: Polish spaces and standard Borel spaces}
 A topological space $A$ is called \emph{Polish} if its  topology is metrizable and as such $A$ is  homeomorphic to a complete and separable metric space (for compact spaces metrizability and separability are equivalent).\footnote{  Dembo \& Zeitouni (1998), Appendix D, Takesaki (2002), Appendix, and  Rassoul-Agha \& Sepp\"{a}l\"{a}inen (2015), Appendix B.4, are useful summaries of probability theory on Polish spaces.}  We equip the space $\Pr(A)$ of Borel probabiliity measures on some Polish space $A$ with the \emph{weak topology}, which
 is most easily defined by stating what convergence in this topology means;
since $A$ is metrizable,
 it is enough to define what convergence of sequences means:\footnote{The official way would be to either use nets or define the weak topology directly, namely by saying that it is generated by the $\varep$-balls $U_{f,x,\varep}:=\{p\in\Pr(A)\mid |\int_A dp\, f-x|<\dl\}$, where $f\in C_b(A)$, $x\in\R$ and $\varep>0$. .}
 \begin{definition}\label{weak2c}
 For a sequence $(p_n)$ in $\Pr(A)$ we say that $p_n\raw p$ weakly in $\Pr(A)$  iff \begin{center}
 $\int_A dp_n\, f\raw \int_A dp\, f$ for each $f\in C_b(A)$.\end{center}
 \end{definition}
 If $A$ is compact and metrizable, then so is $\Pr(A)$, and if $A$ is Polish, then so is $\Pr(A)$.\footnote{A metric  on $\Pr(A)$ returning the weak topology is the \emph{L\'{e}vy--Prohorov metric} $d_P$, defined in terms of the metric $d$ on $A$, as follows. 
For any $F\subset A$,  and any $\varep>0$, define
$F^{\varep}:=\{a\in A\mid \exists_{b\in F}: d(a,b)<\varep\}=\bigcup_{a\in F}\mathcal{B}_{\varep}(a)$,
where $\mathcal{B}_{\varep}(a)=\{b\in A\mid d(a,b)<\varep\}$.
 Then 
$d_P(p,q):=\inf\{\varep>0\mid \forall_{F\subset A, F\mathrm{closed}}: p(F)\leq q(F^{\varep})+\varep\}$.}
 The \emph{Portmanteau theorem} gives various  criteria for $p_n\raw p$ weakly, of which a particularly useful one is:  $p_n\raw p$ weakly iff $p_n(B)\raw p(B)$ for any measurable $B\subset A$ for which 
$p(B^-\backslash \mathring{B})=0$. 

If $A$ is finite this simplifies.\footnote{This simplification inspires the so-called \emph{$\tau$-topology} on 
$\Pr(A,\Sg)$ for general probability spaces $(A,\Sg)$, defined as the coarsest topology for which each map $p\mapsto p(\Dl)$ from $\Pr(A)$ to $\R$ is continuous, $\Dl\in\Sg$. This is finer than the weak topology and in general is not  metrizable.
See Dembo \& Zeitouni (1998), p.\ 263, and Csisz\'{a}r (2006), \S 1.}
 Let $A^*=\{f:A\raw\R\}\cong \R^{|A|}$ with the usual topology, and let $\Pr(A)\subset A^*$ inherit this topology, in which it is closed and bounded, and therefore compact. The relative Euclidean topology is equivalent with the weak topology, and convergence in $\Pr(A)$ is simply pointwise, that is, $p_n\raw p$ weakly $p_n(a)\raw p(a)$ for each $a\in A$. 

 In defining $\Pr(A)$ for a Polish space by (tacitly) taking the $\sg$-algebra $\Sg$ on which our probability measures $p:\Sg\raw[0,1]$  to be the smallest one containing all open sets, we have moved from Polish spaces to so-called \emph{standard Borel spaces}, which may be defined precisely as those measure spaces that arise from Polish spaces in the said way. Such measure spaces may be classified up to isomorphism.\footnote{\label{Kuratowski1} An \emph{isomorphism} between two measure spaces $(A_1,\Sg_1)$ and $(A_2,\Sg_2)$ is a bijection $f:A_1\raw A_2$ such that $f\inv(\Dl_2)\in\Sg_1$ for each $\Dl_2\in\Sg_2$ and $f(\Dl_1)\in\Sg_2$ for each $\Dl_1\in\Sg_1$ (where $f(\Dl_1)$ is conceptually $g\inv(\Dl_1)$ with $g=f\inv$).
 } It is a deep result that this classification is incredibly simple:\footnote{The original source is Kuratowski (1933).
 An accessible recent treatment is Preston (2008).}
 \begin{itemize}
\item Two standard Borel spaces  $(A_1,\Sg_1)$ and $(A_2,\Sg_2)$ are isomorphic iff $A_1\cong A_2$ as sets.\footnote{That is, if $A_1$ and $A_2$ have the same cardinality, which is the case iff there is bijection $f:A_1\raw A_2$.}
\item Up to isomorphism, this only leaves $A=N$ for some $N\in\N$ (with $\Sg=\CP(N)$); $A=\N$ (again with $\Sg=\CP(N)$); and $A=[0,1]$ (with $\Sg$ the Borel $\sg$-algebra generated by the open sets).
\end{itemize}
Even the  probability measures on standard Borel spaces can be classified, up to subsets of measure zero.\footnote{Extending  footnote \ref{Kuratowski1},  $(A_1,\Sg_1, \mu_1)$ and $(A_2,\Sg_2,\mu_2)$ are isomorphic if 
there is a bijection $f: A_1\raw A_2$ as in  footnote \ref{Kuratowski1} that also satisfies $\mu_2(\Dl_2)=0$ $\Raw$ $\mu_1(f\inv\Dl_2)=0$ and $\mu_1(\Dl_1)=0$ $\Raw$ $\mu_2(f\Dl_1)=0$ ($\Dl_1\in\Sg_1, \Dl_2\in\Sg_2$).} 
To state the result   we  put $\N$ and its subsets 
$N=\{0,1, \ldots, N-1\}$ and $\N$ inside $[0,1]$ as  $2^{-\N}\subset[0,1]$.
Up to isomorphism, all possibilities may then be realized on $A=[0,1]$, and they are:
\begin{enumerate}
\item $\mu=\mu_N$ with $\mu_N(2^{-n})=1/N$,  supported on $2^{-N}:=\{2^{-n}\mid n=0, 1, \ldots, n-1\}\subset[0,1]$;
\item $\mu=\mu_{\N}$ with  $\mu_\N(2^{-n})=2^{-n-1}$, supported on $2^{-\N}:=\{2^{-n}\mid n\in\N\}\subset[0,1]$;
\item $\mu=\mu_L$, i.e.,  Lebesgue measure;
  \item  $\mu=t\mu_L+(1-t)\mu_N$, for some $N\in\N$ and $0<t<1$;
  \item $\mu=t\mu_L+(1-t)\mu_\N$, for some $0<t<1$.
  \end{enumerate}
  Yet classifying up to isomorphism hides the riches of examples. Case 3, the most interesting one, may also be realized in a seemingly very different way  by taking $A=2^\N$ with the $\sg$-algebra $\mathcal{F}$ generated by the cylinder sets \er{cylinder} and the Bernoulli measure $f^\N$.
The binary expansion \er{xomega},
 provides a bijection $\beta: 2^\N\stackrel{\cong}{\raw}[0,1]$ via $s\mapsto x\equiv\beta(s)$, with $\mu_L\circ\beta\inv=f^\N$; see Exercise \ref{diffex}.\footnote{To make $\beta$ a bijection one has to deal with the dyadic rationals, defined as 
 the fractions of the kind $m\cdot 2^{-n}$ for $m,n\in\N_*$ subject to $0<m<2^n$. These have  two different binary expansions, the first ending with an infinite sequence of zeros (following their natural  \emph{finite} binary expansion) and the second ending with an infinite sequence of ones. For example,  $\half=\Sigma_{n=1}^{\infty}2^{-(n+1)}$ has binary expansions 
 $10000\cdots$ and $01111\cdots$.
 One of these has to be removed from $2^\N$.\label{dyadicfn} }
\section{Classical hypothesis testing}\label{Cht}
In this section we clarify the meaning 
of Sanov's theorem through an important application, keeping $A$ finite.\footnote{In that case the $\sg$-algebra $\Sg$ is tacitly assumed to be the power set $\CP(A)$ of $A$, making all its subsets measurable.}
Suppose we do not know the probability distribution $p$ in the probability space $(A,p)$. An $N$-fold drawing $\sg\in A^N$ from $A$ gives an estimate $L_N(\sg)$ of $p$, which by any law of large numbers \er{SLLNSanov} gives ever better approximations to $p$ as $N\raw\infty$. But what happens for finite $N$? A systematic way of approaching this is the theory of \emph{hypothesis testing}. To start, suppose we have two candidates, $p_0$ and $p_1\neq p_0$, for $p$. For example, $p_0$ could be $f$ (``the coin is fair''). Let  $\mathsf{H}_i$ be the hypothesis that $p=p_i$  ($i=0,1$).
We want to make a well-motivated choice for  $\mathsf{H}_0$ or $\mathsf{H}_1$ on the basis of our finding $\sg\in A^N$.  It looks reasonable to choose $\mathsf{H}_0$ if $L_N(\sg)$ is ``closer'' to $p_0$ than to $p_1$ (and \emph{vice versa}), according to some distance function $D$ on $\Pr(A)$, in other words:
\begin{align}
D(L_N(\sg),p_0)< D(L_N(\sg),p_1) && \Raw&& \mathsf{H}_0; && D(L_N(\sg),p_1)\leq D(L_N(\sg),p_0) && \Raw&& \mathsf{H}_1, 
\label{51}
\end{align} 
where the  $<$ and $\leq$ signs could have been swapped: if 
$D(L_N(\sg),p_0)=D(L_N(\sg),p_1)$, 
the choice between $\mathsf{H}_0$ and  $\mathsf{H}_1$ given $\sg$ is arbitrary.  
But what to choose for $D$? We propose to take
\begin{equation}
D(p,q)=S(p, q).\label{52}
\end{equation}
One reason is that  this is a course on entropy,\footnote{It should be mentioned that $S$ is not a metric since it even fails to be symmetric (it also fails the triangle inequality), yet the Gibbs inequality  (Proposition \ref{gibbsin}) is a promising start for regarding $S(p,q)$ as a ``distance'' between $p$ and $q$.} but there is a better one: eq.\ \er{51} is now simply
\begin{align}
p_0^N(\sg)>p_1^N(\sg) && \Raw&& p=p_0; && p_0^N(\sg)\leq p_1^N(\sg)  && \Raw&& p=p_1.\label{55}
\end{align}
As usual in statistics, define the \emph{log-likelihood function} for the hypotheses $p=p_0$ and $p=p_1$ by
 \begin{align}
 \mathsf{LL}_N:A^N\raw\R; &&
\mathsf{LL}_N(\sg):=\frac{1}{N}\log\left(\frac{p_0^N(\sg)}{p_1^N(\sg)}\right), \label{defLLN}
\end{align}
where for simplicity we assume $p_i(a)>0$ for all $a\in A$.\footnote{If not,  the formulae below may contain $\pm\infty$ but they are still correct if \er{KL2} is duly taken into account.} We then easily find
\begin{equation}
\mathsf{LL}_N(\sg)=S(L_N(\sg), p_1)-S(L_N(\sg), p_0)\equiv S_1(L_N(\sg))-S_0(L_N(\sg)), \label{LLND}
\end{equation}
where we use the notation $S_i(p)=S(p,p_i)$ for $i=01,1$. Hence \er{51} is also the same as
\begin{align}
\mathsf{LL}_N(\sg)>0 && \Raw&& \mathsf{H}_0; && \mathsf{LL}_N(\sg)\leq 0 && \Raw&& \mathsf{H}_1.
\label{54}\end{align}
While \er{55} seems self-explanatory, some  implicit assumption has been made here, namely treating $\mathsf{H}_0$ and $\mathsf{H}_1$ on equal footing (called \emph{symmetric} hypothesis testing). To quantify this assumption, we  introduce the concept of a \emph{test}: for given $N$ this is simply a subset $T_N\subset A^N$ 
such that:
\begin{align}
\sg\in T_N && \Raw&& \mathsf{H}_0; && \sg\notin T_N && \Raw&& \mathsf{H}_1.
\label{556}\end{align}
For example, the famous \emph{Neyman--Pearson test} $T_N=T_N^*(u_N)$ is defined for any $u_N\in\R$ by
  \beq
T_N^*(u_N):= (\mathsf{LL}_N>u_N)\equiv \left\{\sg\in A^N \mid \mathsf{LL}_N(\sg)>u_N\right\}=(p_0^N>e^{Nu_N}p_1^N). \label{LLN}
\eeq
From this, we see at once that \er{54} and \er{55} correspond to $T_N=T^*_N(0)$, which by \er{LLND} equals
\begin{equation}
T_N^*(0)=\{\sg\in A^N\mid S_1(L_N(\sg))> S_0(L_N(\sg))\}.\label{TNstar}
\end{equation}
We will now show that this test treats $\mathsf{H}_0$ and $\mathsf{H}_1$ symmetrically in giving the statistical \emph{errors}
that are possible (and unavoidable) equal weight. This analysis also 
prepares us for the asymmetric case (which will correspond to $u_N\neq 0$). First, the two possible errors and their probabilities are:
\begin{itemize}
\item The  \emph{false-positive} probability that $\mathsf{H}_1$ is accepted on the basis of an outcome $\sg$ that happens to lie in $T_N^c$, \emph{although in fact $\mathsf{H}_0$ is true} (which would have been confirmed if $\sg\in T_N$), is
\begin{equation}
\al_N(T_N)=p_0^N(T_N^c)=1-p_0^N(T_N).\label{defalfaN}
\end{equation}
\item The  \emph{false-negative} probability that $\mathsf{H}_0$ is accepted on the basis of an outcome $\sg$ that happens to lie in $T_N$, \emph{although in fact $\mathsf{H}_1$ is true} (which would have been confirmed if $\sg\in T^c_N$), is
\beq
\beta_N(T_N)=p^N_1(T). \label{defbetaN}
\eeq
\end{itemize}
This terminology comes from  applications where $\mathsf{H}_0$ states the \emph{absence}  of some threat, 
whereas $\mathsf{H}_1$ states its \emph{presence}.\footnote{ The terms \emph{type I} and \emph{type II} error are also used for the first and the second situation,  respectively. In medical applications, where one may even think of $N=1$,  $\al_N(T_N)$ is  the probability that the tested patient appears to have some disease but does not in fact have it,  whereas $\beta_N(T_N)$ is the probability that the test suggests that the patient does \emph{not} have the disease whilst actually having it.
This error is typically more dangerous, since the necessary treatment is then withheld from the patient, whereas unnecessary treatment, though often not quite harmless, usually causes  less damage.} One would of course like both errors to be zero, but if $p_0\neq p_1$ (as we assume), this is impossible: 
the only test $T_N$ that makes $\al_N(T_N)=0$ is $T_N=A^N$, in which case $\beta_N(T_N)=1$ (and \emph{vice versa}). 
But one can make $T_N$ optimal, depending on how one weighs the two errors. In the symmetric case
 $\mathsf{H}_0$ and $\mathsf{H}_1$ are treated symmetrically, and hence one minimizes
\beq
\gm_N(T_N):=\al_N(T_N)+\beta(T_N).\eeq
\begin{lemma}\label{NPL}
For any test $T_N\subset A^N$ one has $\gm_N(T^N)\geq \gm_N(T_N^*)$, where $T_N^*\equiv T_N^*(0)$, cf.\ \er{LLN}, that is,
\begin{equation}
\gm_N:=\inf_{T_N\subset A^N}\{\gm_N(T_N)\}=\gm(T_N^*).\label{defgmN}
\end{equation}
Or: if $\al_N(T_N)\leq \al_N(T_N^*)$, then  $\beta_N(T_N)\geq \beta_N(T_N^*)$, and  if $\beta_N(T_N)\leq \beta_N(T_N^*)$, then $\al_N(T_N)\geq \al_N(T_N^*)$.
\end{lemma}
This is the simplest case of the so-called \emph{Neyman--Pearson lemma}.\footnote{The test $T^*$ is not a unique minimizer of
$\gm$; for example, one could equally well take $\til{T}_*=\{a\in A \mid p_0(a)\geq p_1(a)\}$.} The proof is easy.\footnote{In fact, it is so easy that this can't be the whole story. Indeed, the full Neyman--Pearson lemma concerns a broader class of tests. To briefly explain this, note that subsets $T_N\subset A^N$ are equivalent to their characteristic functions $1_{T_N}:A^N\raw \{0,1\}$, and conversely any $f:A^N\raw \{0,1\}$ is given by $f=1_{T_N}$ for some $T_N\subset A^N$.
The more general test are functions $T_N:A^N\raw [0,1]$ so that if $\sg\in A^N$ is drawn, then $\mathsf{H}_0$ is merely accepted with probability $T_N(\sg)$. See for example Poor (1994),  \S II.D. In quantum theory this corresponds to using \textsc{povm}s instead of projections, cf.\ \er{POVM}.\label{Poor}} 
\smallskip

\emph{Proof}.
We put $N=1$ and invite the reader to throw in $N$ as appropriate.
\begin{align}
\gm(T)&=1-p_0(T)+p_1(T)=1+\Sigma_{a\in T}(p_1(a)-p_0(a))\nn \\ &=1+\Sigma_{a\in T\cap T^*}(p_1(a)-p_0(a))+\Sigma_{a\in T\cap (T^*)^c}(p_1(a)-p_0(a))\nn \\ &\geq 1+\Sigma_{a\in T\cap T^*}(p_1(a)-p_0(a))\geq 1+\Sigma_{a\in T^*}(p_1(a)-p_0(a))=\gm(T^*),
\end{align}
since by definition of $T^*$ we have $p_1(a)-p_0(a)\geq 0$ for all $a\in T\cap (T^*)^c$.
\QED
\bex Adapt this proof to arbitrary values of $N$.\eex
This lemma provides the  justification of our criterion \er{54} - \er{55}, for it shows that $T_N^*$ minimizes the total error if the two hypotheses are treated symmetrically. 
The next point is that since
 \beq
 \gm_{N+1}(T_N\x A)=\gm_N(T_N),
 \eeq
  for any
if $T_N\subset A^N$, we have  $\gm_{N+1}\leq\gm_N$, and hence
it is  interesting to study  the asymptotic behaviour of $\gm_N$ as $N\raw\infty$. The key result in this direction is \emph{Chernoff's theorem}, which  involves a new (relative) entropy-like  function we aptly call the \emph{Chernoff entropy}. This  entropy is defined by
\begin{align}
C(p_0,p_1):=-\inf_{t\in(0,1)} S_{t}(p_0,p_1); &&  S_{t}(p_0,p_1):= \log\left(\Sigma_{a\in A} p_0(a)^{1-t}p_1(a)^{t}\right). \label{CIF}
\end{align}
Here we  used $S_t(p_0,p_1)$ for $t\in (0,1)$, but in principle it is defined for any $t\in\R$. Equivalently,
\begin{equation}
C(p_0,p_1)=\sup_{t\in(0,1)}\{1-t)R_t(p_0,p_1)\},
\end{equation}
where the 
relative \emph{R\'{e}nyi entropy} is defined for $t\in\R$ by
\begin{align}R_t(p,q)&:=\frac{1}{t-1}\log\left(\Sigma_{a\in A} p(a)^{t}q(a)^{1-t}\right) \hspace{30pt} (t\neq 0,1);\label{RRE}\\
R_0(p,q)&=0;\hspace{130pt}
R_1(p,q):=S(p,q),
\label{LRE}
\end{align}
so that $R_0(p,q)=\lim_{t\downarrow 0}R_t(p,q)$ and $R_1(p,q)=\lim_{t\uparrow 1}R_t(p,q)$.
For $t\in (0,1)$ it behaves nicely:\footnote{See Jak\v{s}i\'{c} (2018), Prop.\ 4.16, for proofs and further good properties.}
\begin{enumerate}
\item $R_t(p,q)\geq 0$, with $R_t(p,q)=0$ iff $p=q$, and $R_t(p,q)=\infty$ iff $p$ and $q$ have disjoint supports.
\item The map $(p,q)\in\Pr(A)\x\Pr(A)\raw R_t(p,q)$ is continuous and jointly convex, cf.\ \er{Spqconvex}.
\end{enumerate}
The Chernoff entropy $C(p_0,p_1)$ inherits all  these properties, in particular,
\beq
C(p_0,p_1)\geq 0, \label{Cgnul}
\eeq
with equality iff $p_0=p_1$, 
 and hence it may also be seen as a distance  between $p_0$ and $p_1$ in $\Pr(A)$: unlike the relative entropy $S(p,q)$, the function
$C(p,q)$  is even symmetric by \er{CIF}. 
\bex Prove \er{Cgnul}. \emph{Hint:} First, rewrite 
\begin{equation}
\Sigma_{a\in A} p_0(a)^{1-t}p_1(a)^{t}=\Sigma_{a\in A} p_0(a)\left(\frac{p_1(a)}{p_0(a)}\right)^t=
\la (p_1/p_0)^t\ra_{p_0}. \label{1.24} \end{equation}
Show that $x\mapsto x^t$ is concave for $t\in [0,1]$ (and convex for all other $t\in\R$), and use Jensen's inequality 
to make the left-hand side of \er{1.24} $\leq 1$ for $t\in [0,1]$ (and $\geq 1$ otherwise). Conclude that $S_t(p_0, p_1)\leq 0$  for $t\in [0,1]$ (and $\geq 0$ otherwise). Prove the iff from the strict version of Jensen.
\eex
\begin{theorem}[Chernoff]\label{Chernoff}
With $\gm_N\equiv \gm_N(p_0,p_1)$ as defined in \er{defgmN}, we have
\begin{equation}
\lim_{N\raw\infty}\frac{1}{N}\log \gm_N(p_0,p_1)=-C(p_0,p_1). \label{CMP} 
\end{equation}\end{theorem}
Even for finite $N$ one has the \emph{Chernoff bound}, 
\begin{equation}
 \gm_N\leq e^{-NC(p_0,p_1)},\label{Chbound}
\end{equation}
which follows from the proof below and the observation that $\gm_N\leq e^{NS_{t}(p_0,p_1)}$
for $t\in(0,1)$ trivially implies \er{Chbound}.
We will prove Theorem \ref{Chernoff} together with the corresponding results for \emph{asymmetric hypothesis testing}, in which we fix an error bound on $\al_N$ and make the best of $\beta_N$ given this bound.

\noindent
As already mentioned,  one may make the false-negative error $\beta_N$ as small as one likes by decreasing $T_N$ (up to $\beta_N(T_N)=0$ for $T_N=\emptyset$), but this increases the false-positive  error $\al_N$ (up to $\al_N=1$), which may be too much even if missing  $\mathsf{H}_1$ is more harmful than missing  $\mathsf{H}_0$. Instead, we define
\begin{align}
\beta_N^{\varep}(p_0,p_1)&:=\inf_{T_N\subset A^N}\{\beta_N(T_N)\mid \al_N(T_N)\leq \varep\} \hspace{75pt} (0<\varep<1);  \label{9.21}\\
\beta^*(p_0,p_1)&:= \inf_{(T_N)}\left\{ \lim_{N\raw\infty} \frac{1}{N}\log \beta_N(T_N)\mid T_N\subset A_N,
\lim_{N\raw\infty}\al_N(T_N)<1
\right\};\label{9.18bis}\\
\beta_*(p_0,p_1)&:= \inf_{(T_N)}\left\{ \lim_{N\raw\infty} \frac{1}{N}\log \beta_N(T_N)\mid T_N\subset A_N,
\lim_{N\raw\infty}\al_N(T_N)=0
\right\}, \label{9.18}
\end{align}
where in \er{9.18bis} and \er{9.18}  the infimum over tests  is taken not for each $N$ separately, as in \er{9.21}, after which the limit $N\raw\infty$ is taken,  but over all \emph{families} $(T_N)$ of tests \emph{for which the limits inside the curly brackets exist}.  The relative entropy \er{KL1} - \er{KL2} now makes a spectacular comeback:
\begin{theorem}[Stein]\label{Stein}
With $\beta^{\varep}_N$ as in \er{9.21}, we have, independently of $\varep\in (0,1)$:
\begin{align}
\lim_{N\raw\infty}\frac{1}{N}\log \beta^{\varep}_N(p_0,p_1)&=-S(p_0, p_1); \label{Stein1}\\
\beta^*(p_0,p_1)=\beta_*(p_0,p_1)
&=-S(p_0, p_1). \label{Stein2}
\end{align}
\end{theorem}

The third ``regime'' makes the bound $\varep$ on $\al_N(T_N)$ in \er{9.21} $N$-dependent by defining 
\begin{align}
\beta_N(r, p_0,p_1):=\inf_{T_N\subset A^N}\{\beta_N(T_N)\mid \al_N(T_N)\leq e^{-Nr}\}; && (r>0). \label{9.21b}
\end{align}
\begin{theorem}[Hoeffding]\label{Hoeffding}
With $\beta_N(r, p_0,p_1)$ as in \er{9.21b}, we have 
\beq
\lim_{N\raw\infty}\frac{1}{N}\log \beta_N(r, p_0,p_1)=-H(r,p_0,p_1),\label{HB}
\eeq
where the \emph{Hoeffding entropy} $H(r,p_0,p_1)$ is defined by
 \beq
H(r,p_0,p_1):=\sup_{t\in(0,1)} \left\{ \left( \frac{1-t}{t}\right)\cdot (R_t(p_1,p_0)-r)\right\}.
 \label{defH01}
\eeq
Moreover, the following  dichotomy arises: 
\begin{itemize}
\item 
For $0<r<S(p_1,p_0)$ we have $H(r,p_0,p_1)>0$, so that
\begin{equation}
\lim_{N\raw\infty} \beta_N(r,p_0,p_1)=0, \label{firstcasec}
\end{equation}
\item For $r\geq S(p_1,p_0)$ we have $H(r,p_0,p_1)=0$, and
\begin{equation}
\lim_{N\raw\infty} \beta_N(r,p_0,p_1)=1. \label{secondcasec}
\end{equation}
\end{itemize}
In \er{secondcasec} the
approach to 1 is measured by the ``dual'' Hoeffding entropy 
\beq
 H^*(r,p_0,p_1):= \sup_{t>1}\left\{\left(\frac{t-1}{t}\right)\cdot (r-R_t(p_1,p_0))
\right\},
\eeq
according to
\begin{equation}
\lim_{N\raw\infty}\frac{1}{N}\log(1- \beta_N(r,p_0,p_1))=-H^*(r,p_0,p_1),
\end{equation}
where $H^*(r,p_0,p_1)>0$ if $r>S(p_1,p_0)$, whilst $H^*(r,p_0,p_1)=0$ for $0<r\leq S(p_1,p_0)$.
\end{theorem}
So the message of Hoeffding is that we cannot restrict the false-positive error $\al_N(T_N)$ too much (i.e., make $r$ too large) without damaging--in fact ruining--the false-negative  error $\beta_N(T_N)$.\footnote{The threshold $r=S(p_1, p_0)$ is similar to the one $R= S(p)$ in source coding, see Theorems \ref{NCT} and \ref{NCT2}.
} Here we just stated a simple form of Hoeffding's theorem, which resembles the other two theorems as much as possible. Analogously to \er{9.18bis} etc.\ one may alternatively define
\begin{align}
h(r,p_0,p_1)&:= \inf_{(T_N)}\left\{ \lim_{N\raw\infty} \frac{1}{N}\log \beta_N(T_N)\mid T_N\subset A_N,
\lim_{N\raw\infty}\frac{1}{N}\log \al_N(T_N)\leq -r
\right\}\label{Hoef1},
\end{align}
and analogous expressions with $\lim$ replaced by $\lim\sup$ and $\lim\inf$, so that the theorem states that
\begin{equation}
h(r,p_0,p_1)=-H(r,p_0,p_1),\label{HBb}
\end{equation}
and similarly for the $\lim\sup$ and $\lim\inf$ variations (which  coincide), with the bonus statement that the infimum in \er{Hoef1} is actually achieved, so that for $r>0$ there is  a sequence $(T_N)$ for which 
\begin{align}
\lim_{N\raw\infty}\frac{1}{N}\log \beta_N(T_N)=-H(r,p_0,p_1); && \lim_{N\raw\infty}\frac{1}{N}\log \al_N(T_N)= -r.
\end{align}
\bex Prove the claims about $H(r,p_0,p_1)$ after the two bullets in Theorem \ref{Hoeffding}, so that
\begin{equation}
H(r,p_0,p_1)\geq 0.
\end{equation}
\eex

The rate functions on the right-hand sides of \er{CMP}, \er{Stein1}, and \er{HB} play a double role: they primarily give the rate of exponential decay of the error (or difficulty) in distinguishing $p_0$ and $p_1$, but precisely for that reason they also provide a distance measure between $p_0$ and $p_1$: for the closer these are, the harder it must be to distinguish them, and the error will only decrease slowly.

All three cases have a similar proof, in which Sanov's theorem rewrites the sought limit as:\footnote{Eq.\ \er{64111} holds provided $p_0\ll p_1$, since otherwise $S(p_0,p_1)=\infty$. See Exercise \ref{newStein}.}
\begin{align}
\lim_{N\raw\infty}\frac{1}{N}\log \gm_N(p_0,p_1)&=\inf_{p\in\Pr(A)} \{S(p,p_1) \mid S(p,p_1)> S(p,p_0)\};\\
\lim_{N\raw\infty}\frac{1}{N}\log \beta^{\varep}_N(p_0,p_1)&=\inf_{p\in\Pr(A)} \{S(p,p_1) \mid S(p,p_1)> S(p,p_0)+S(p_0,p_1)\};
 \label{64111}\\
\lim_{N\raw\infty}\frac{1}{N}\log \beta_N(r, p_0,p_1)&=\inf_{p\in\Pr(A)} \{S(p,p_1) \mid  S(p,p_0)\leq r\}. \label{keyHoeffding}
\end{align}
This needs to be shown of course, after which the infima on the right must actually be computed.
 \subsection*{Proof of Chernoff's theorem}
 From Lemma \ref{NPL} and eqs.\  \er{TNstar},  \er{defalfaN}, and \er{defbetaN}, recalling that $T_N^*\equiv T_N^*(0)$, 
 we have 
 \begin{align}
\gm_N&=\al_N(T_N^*)+\beta_N(T_N^*);\label{536}\\
\al_N(T_N^*)
  &=p_0^N(\{\sg\in A^N\mid S(L_N(\sg), p_1)\leq  S(L_N(\sg), p_0)\}) 
 = p_0^N(L_N\in C_*^c); \\
  \beta_N(T_N^*)&=
p_1^N(\{\sg\in A^N\mid S(L_N(\sg), p_1)> S(L_N(\sg), p_0)\})=p_1^N(L_N\in C_*), \label{538}
\end{align}
where we abbreviate
 \begin{align}
C_*&:=\{p\in \Pr(A)\mid S(p, p_1)> S(p, p_0)\};\\
C_*^c&=\{p\in \Pr(A)\mid S(p, p_1)\leq  S(p, p_0)\}.  \label{TNstar2}
\end{align}
The asymptotics of $\gm_N$ then follow from  Sanov's theorem. For our open and closed sets $C_*$ and $C_*^c$, this theorem applies in its simplest form \er{611}. This gives
 \begin{align}
\lim_{N\raw\infty} \frac{1}{N} \log p_0^N(L_N\in C_*)& =-\inf\{S(p, p_0)\mid p\in C_*^c\};\label{9.15}\\
\lim_{N\raw\infty} \frac{1}{N} \log p_1^N(L_N\in C_*)& =-\inf\{S(p, p_1)\mid p\in C_*\}.\label{9.16}
\end{align}
 We now  show that 
 \begin{align}
\inf \{S(p, p_0)\mid p\in C_*^c\}&=\inf \{S(p, p_0)\mid p\in \partial C_*\};\label{Jac0}\\
\inf\{S(p, p_1)\mid p\in C_*\}&= \inf\{S(p, p_1)\mid p\in \partial C_*\},\label{Jac0b}
\end{align}
where $\partial C_*=\partial C_*^c$ is the common (topological) boundary of the regions $C_*$ and $C_*^c$, i.e.,
\begin{equation}
\partial C_*=\{p\in \Pr(A)\mid S(p, p_1)= S(p, p_0)\}.
\end{equation}
If we abbreviate $S(p, p_i)$ as $S_i(p)$, for $i=0,1$,
eqs.\ \er{Jac0} - \er{Jac0b} are the same as 
\begin{align}
\inf_{p\in\Pr(A)} \{S_0(p) \mid S_0(p)\geq S_1(p)\}&=\inf_{p\in\Pr(A)} \{S_0(p) \mid S_0(p)= S_1(p)\}; \label{Jac1}\\
\inf_{p\in\Pr(A)} \{S_1(p) \mid S_1(p)>S_0(p)\}&=\inf_{p\in\Pr(A)} \{S_1(p) \mid S_1(p)= S_0(p)\}. \label{Jac1b}
\end{align}
Let us first note that the infimum is in fact attained, since $C_*^c$ is a closed set in the compact space $\Pr(A)$. 
We prove \er{Jac1} by contradiction;\footnote{This proof was kindly supplied by Jacques Janse van Rensburg.}
the proof of \er{Jac1b} is  the same (and omitted). 
 Suppose, in violation of \er{Jac1}, that some $q\in\Pr(A)$ satisfies the following assumptions:
\begin{align}
S_0(q)>S_1(q); && S_0(q)\leq S_0(p) \mbox{ for all } p \mbox{ for which }\: S_0(p)=S_1(p).\label{assraa}
\end{align}
Now consider the line segment 
\begin{align}
q_{\lm}:=\lm p_0+(1-\lm)q; && \lm\in [0,1], \label{qlam}
\end{align}
which  starts at $q_0=q$ and ends at $q_1=p_0$.  At $\lm=0$ we have $S_0(q_0)>S_1(q_0)$ by assumption (since $q_0=q$), whereas at $\lm=1$ we have $S_0(q_1)<S_1(q_1)$, as $S_0(q_1)=0$ whilst $S_1(q_1)=S_1(p_0)>0$.
Hence by continuity the line \er{qlam} must intersect the $S_0=S_1$ hyperplane at some value $\lm=\mu\in (0,1)$, where
$S_0(\mu)=S_1(\lm_0)$. 
However, a simple computation shows that the non-negative function $\lm\mapsto S_0(q_{\lm})$ has strictly negative derivative for all $\lm\in [0,1)$.
\bex
Perform this computation.
\eex
 Hence $\lm\mapsto S_0(q_{\lm})$ decreases from $S_0(q_0)=S_0(q)>0$ 
to $S_0(q_1)=S_0(p_0)=0$ as $\lm$ grows. Thus 
\beq
S_0(q_{\mu})< S_0(q_0)=S_0(q).
\eeq
 But taking $p=q_{\mu}$ in \er{assraa} 
gives $S_0(q)\leq S_0(q_{\mu})$. Hence \er{Jac1} and its equivalent \er{Jac0} follow. 
\smallskip

 Minimizing $S(p, p_0)$ and  $S(p, p_1)$ over $\partial C_*$ can be done 
 with Lagrange multipliers for both constraints $S_0(p)=S_1(p)$ and $\Sigma_a p_a=1$. To state the result, define
 $p_{\lm}\in\Pr(A)$,  $\lm\in [0,1]$,  by
\begin{align}
p_{\lm}(a):=\frac{ p_0(a)^{1-\lm}p_1(a)^{\lm}}{Z_{\lm}(p_0,p_1)}; && Z_{\lm}(p_0,p_1)=\Sigma_{a\in A}p_0(a)^{1-\lm}p_1(a)^{\lm}.
 \label{plm}
\end{align} 
 Our notation is consistent, since  $\lm=0$ in \er{plm} reproduces the original $p_0$, and likewise for $p_1$. Indeed, $\lm\mapsto p_{\lm}$ is a curve from $p_0$ to $p_1$.
This curve intersects $\partial C_*$ at some $\ovl{\lm}\in (0,1)$ where
\beq
S(p_{\ovl{\lm}}, p_1)= S(p_{\ovl{\lm}}, p_0).\label{DD}
\eeq
Straightforward calculus shows that $p=p_{\ovl{\lm}}$ minimizes both $p\mapsto S_0(p)$ and $p\mapsto S_1(p)$ on $\Pr(A)$ subject to the constraint $S_0(p)=S_1(p)$. In addition, $\lm=\ovl{\lm}$ minimizes the function 
\beq
\lm\mapsto \Sigma_{a\in A} p_0(a)^{1-\lm}p_1(a)^{\lm}, \label{660e}
\eeq
and hence also minimizes its logarithm (see exercise). The existence of $\ovl{\lm}$ follows by considering
\begin{equation}
f(\lm):=\Sigma_{a\in A} p_0(a)^{1-\lm}p_1(a)^{\lm}\log\left(\frac{p_0(a)}{p_1(a)}\right),
\end{equation}
for $\lm\in[0,1]$. Then \er{DD} is the same as $f(\ovl{\lm})=0$. Now \er{KL1} and the Gibbs inequality give
 \begin{align} f(0)=S(p_0,p_1)> 0; && f(1)=-S(p_1,p_0)<0,
 \end{align}
  so that continuity and the intermediate value theorem guarantee that $\ovl{\lm}$ exists. Moreover, since
  \begin{equation}
f'(\lm)=-\Sigma_{a\in A} p_0(a)^{1-\lm}p_1(a)^{\lm}\left(\log\left(\frac{p_0(a)}{p_1(a)}\right)\right)^2<0,
\end{equation}
$f$ is strictly decreasing from $\lm=0$ to $\lm=1$, so that the minimizer $\ovl{\lm}$ is also unique.
 Thus 
 \begin{equation}
S(p_{\ovl{\lm}}, p_1)= S(p_{\ovl{\lm}}, p_0)=\min_{0\leq\lm\leq 1}S_{\lm}(p_0,p_1)=\inf_{0<\lm< 1}S_{\lm}(p_0,p_1)=-C(p_0,p_1).
 \label{9.20}
\end{equation}
Theorem \ref{Chernoff} now follows from eqs.\ \er{536} to \er{538}, \er{9.15},  \er{9.16}, and \er{9.20}. \QED
\bex Prove the three claims stated just before  \er{660e}.\eex
\subsection*{Proof of Stein's theorem} 
The proof of both  Stein and Hoeffding  relies on  the following generalization  of Lemma \ref{NPL}:
\begin{lemma}\label{NPP}
Let $u_N\in\R$.
For any test $T_N\subset A^N$, where $N$ is fixed,  we have, cf.\ \er{LLN}, 
\begin{equation}
\inf_{T_N\subset A^N}\{\al_N(T_N)+e^{Nu_N}\beta_N(T_N)\}=\al_N(T_N^*(u_N))+e^{Nu_N}\beta_N(T_N^*(u_N)). \label{1.3}
\end{equation}
Consequently:
\begin{itemize}
\item  If $\al_N(T_N)\leq \al_N(T_N^*(u_N))$, then  $\beta_N(T_N)\geq \beta_N(T_N^*(u_N))$;
\item If $\beta_N(T_N)\leq \beta_N(T_N^*(u_N))$, then $\al_N(T_N)\geq \al_N(T_N^*(u_N))$.
\end{itemize}
\end{lemma}
\bex
Prove  Lemma \ref{NPP} (for example, from Lemma \ref{NPL}). 
\eex
To proceed we now assume $p_0\ll p_1$ until Exercise \ref{newStein}, so that in particular $S(p_0,p_1)<\infty$. 
\begin{lemma}\label{lem612}
For any open interval $(a,b)$ with $0<a<b<1$ and sufficiently large $N$  there is  $u_N\in\R$ such that $\al_N(T_N^*(u_N))\in (a,b)$ (i.e., 
 there is  $M\in\N$ such that this holds for all  $N>M$).
 \end{lemma}
See below for a complete proof; the idea is that from \er{LLN} we have  
\beq
\al_N(T_N^*(u_N))=p_0^N(\mathsf{LL}_N\leq u_N)=p_0^N(\{\sg\in A^N\mid 
p_0^N(\sg)\leq e^{Nu_N} p_1^N(\sg)\}). \label{666}
\eeq
 For $u_N\raw -\infty$ this  forces $p_0^N(\sg)\raw 0$, whereas for $u_N\raw \infty$ any $\sg$  satisfies the inequality, so that
 \begin{align}
\lim_{u_N\raw-\infty}\al_N(T_N^*(u_N))=0; &&
\lim_{u_N\raw\infty}\al_N(T_N^*(u_N))=1;
\end{align}
in fact, it easily follows from \er{666} that there is an $N$-independent compact interval $[u_-,u_+]$ such that $\al_N(T_N^*(u_N))=0$ for $u_N<u_+$ and 
 $\al_N(T_N^*(u_N))=1$ for $u_N>u_-$, so that one only needs to take values $u_N\in [u_-,u_+]$ into account.
The point is that the function $f(x)=\al_N(T_N^*(x))\in [0,1]$ is nondecreasing by construction, so that the claim would follow from the intermediate value theorem \emph{if $f$ were continuous}. It isn't: but the saving grace is that the values of $p_0^N(B_N)$, where $B_N$ runs over all subsets of $A^N$, become ``increasingly dense'' in $[0,1]$ as $N\raw \infty$. 
Granting this,  for the $\varep\in (0,1)$ in the statement of Theorem \ref{Stein}, take  $0<\dl<\varep<1$ and then in the above lemma take
\beq
(a,b)=(\varep-\delta, \varep),
\eeq 
so that for sufficiently large $N$ there is some  $u_N\in\R$ such that, in a form useful for a later step, 
 \begin{align}
0<\varep-\dl< \al_N(T_N^*(u_N))< \varep<1 && \LRaw &&  0< \varep-\dl<p_0^N(\mathsf{LL}_N\leq u_N)< \varep<1.\label{9.311}
\end{align}
Hence given some test $T_N$ with $0<\al_N(T_N)<\varep-\dl$, there exists a  Neyman--Pearson test $T_N^*(u_N)$ for which  $\al_N(T_N)< \al_N(T_N^*(u_N))<\varep$, so that  the first claim after \er{1.3} in Lemma \ref{NPP}  gives 
\beq
\beta_N(T_N^*(u_N))\leq \beta(T_N).\label{9.53}
\eeq
 The infimum in \er{9.21} may therefore be computed from  Neyman--Pearson tests $T_N^*(u_N)$ alone via
 \begin{equation}
\beta^{\varep-\dl}_N(p_0,p_1)=\inf_{u_N}\{p_1^N(\mathsf{LL}_N> u_N)\},\label{563}
\end{equation}
 where the infimum is over all $u_N\in\R$ such that \er{9.311} is satisfied.\footnote{Note that \er{563} holds for sufficiently large $N$ depending on $\dl$.} Next, by definition of the infimum in \er{563}, for any $N$ there is some $u_N\in [u_-,u_+]$ such that  \er{9.311} holds, and
 \begin{align}
 p_1^N(\mathsf{LL}_N> u_N)-1/N < \beta^{\varep-\dl}_N(p_0,p_1); &&  p_1^N(\mathsf{LL}_N> u_N) \geq \beta^{\varep-\dl}_N(p_0,p_1).\label{672a}
 \end{align}
This gives a sequence $(u_N)_N$ in $[u_-,u_+]$ for which \er{9.311} and \er{672a} hold. Since $[u_-,u_+]$ is compact this sequence has an accumulation point, which, as we now show,  condition \er{9.311} forces to be unique and hence to be the unique limit of $(u_N)$, see \er{PBC} below. To this end,
 rewrite $\mathsf{LL}_N$ as 
\begin{align}
\mathsf{LL}_N=\frac{1}{N}\Sigma_{n=0}^{N-1}E_n; &&  E_n(\sg)=E(\sg_n);   && E(a)=\log\left(\frac{p_0(a)}{p_1(a)}\right).
\label{1.19}\end{align}
By \er{1.19} and the strong law of large numbers,
$\mathsf{LL}_N$ converges $p_i^{\N}$-almost surely  to its average:
\begin{align} \la E\ra_{p_0}=S(p_0, p_1); && \la E\ra_{p_1}=-S(p_1, p_0),
\end{align}
of which only the first part will be used (we  give the second part for clarity). Hence (\emph{idem dito}),
\begin{align}
\mathsf{LL}_N\raw S(p_0, p_1) \:\:\: p_0^{\N}\mbox{-a.s.}; && \mathsf{LL}_N\raw -S(p_1, p_0) \:\:\: p_1^{\N}\mbox{-a.s.}
\label{1.20}
\end{align}
Strong convergence  implies convergence in probability, that is, for any $\varep>0$ we have
\begin{align}
\lim_{N\raw\infty} p_0^N (|\mathsf{LL}_N-  S(p_0, p_1)|>\varep)=0; \label{9.24}\\
\lim_{N\raw\infty} p_1^N (|\mathsf{LL}_N + S(p_1, p_0)|>\varep)=0. \label{9.24bis}
\end{align}
\bex 
Prove (for example,  by contradiction) that \er{9.311} and \er{9.24} enforce 
 \begin{equation}
\lim_{N\raw\infty} u_N=S(p_0, p_1). \label{PBC}
\end{equation}
\eex
 Eqs.\ \er{672a} and \er{PBC} and both inner and outer regularity of the Bernoulli measure $p^\N$ then  give
\begin{align}
\lim\inf_{N\raw\infty}\frac{1}{N}\log \beta^{\varep-\dl}_N(p_0,p_1)&\geq \lim\inf_{N\raw\infty}\frac{1}{N}\log p_1^N(\mathsf{LL}_N> S(p_0,p_1));\label{571}\\
\lim\sup_{N\raw\infty}\frac{1}{N}\log \beta^{\varep-\dl}_N(p_0,p_1)&\leq\lim\sup_{N\raw\infty}\frac{1}{N}\log p_1^N(\mathsf{LL}_N> S(p_0,p_1)).\label{572}
\end{align}
We may now either let $\dl\raw 0$, as we shall do, or  keep $\dl$, since the final result will be independent of $\varep$ and hence of $\varep-\dl$; our choices were such that both $\varep\in(0,1)$ and $\varep-\dl\in(0,1)$. In any case, 
we are now in a similar situation as in the proof of Chernoff's theorem. Using \er{LLND}, we see that
\begin{align}
p_1^N(\mathsf{LL}_N> S(p_0,p_1))&=p_1^N(L_N\in S_*);\\
S_*&:=\{p\in\Pr(A)\mid S(p,p_1)>S(p,p_0) +S(p_0,p_1).
\end{align}
As in \er{9.16}, Sanov's theorem gives equality of the right-hand sides of \er{571} - \er{572}, so that
\begin{equation}
\lim_{N\raw\infty} \frac{1}{N} \log p_1^N(L_N\in S_*) =-\inf_{p\in\Pr(A)}\{S(p, p_1)\mid S(p,p_1)>S(p,p_0) +S(p_0,p_1)\}.
\end{equation}
Its solution is now immediate from the Gibbs inequality $S(p,p_0)\geq 0$ with equality iff $p=p_0$. Hence the infimum is attained at $p=p_0$, with value $S(p_0,p_1)$, 
and \er{Stein1} follows for $p_0\ll p_1$.
\bex Prove  \er{Stein1}  in case that $p_0\ll p_1$ fails, and hence $S(p_0,p_1)=\infty$.\label{newStein}
\eex
\emph{Proof of Lemma \ref{lem612}.} 
To make the idea of the proof stated  after the lemma precise, we note that
$f$ is locally constant except for jumps $\Dl_N f(x)$ at  $x$ where $p_0^N(\sg)=e^{Nx} p_1^N(\sg)$ for some $\sg\in A^N$, with \begin{equation}
\Dl_N f(x)=p_0^N(\{\sg\in A^N\mid 
p_0^N(\sg)= e^{N x} p_1^N(\sg)\}). \label{669}
\end{equation}
We first study the simple case where $A=\{0,2\}$, $p_0=f$, and $p_1\neq p_0$ does not matter; for concreteness' sake, take $p_1(0)=1/4$ and $p_1(1)=3/4$. The argument of $p_0^N$ in \er{669} has $
\left(
\begin{array}{c}
N \\ k  
\end{array}
\right)$ elements, for some $k\in \{0, 1, \ldots, N\}$, which equals the number of terms $\sg_n=0$ in $\sg=\sg_0\cdots \sg_{N-1}$, which number in turn is determined by the condition $p_1^N(\sg)=e^{-Nx} p_0^N(\sg)=e^{-Nx} 2^{-N}$.
Taking $N$ even for simplicity, it is well known that the largest value of this binomial coefficient is at $k=N/2$. This is a special case of \er{combi}, where $A=\{0,1\}$ and $p=f$.
To estimate this largest possible value 
\beq
\Dl_N f(x)=2^{-N}\left(
\begin{array}{c}
N \\ N/2
\end{array}
\right),
\eeq
 the leading part of Stirling's approximation, see Exercise \ref{Ex3.1} is not enough, as it produces \beq
\left(
\begin{array}{c}
N \\ N/2
\end{array}
\right)\approx 2^N,
\eeq
 but the next term,
\beq
\log (n!)=n\log n -n+\half \log(2\pi n)+O(1/n), \label{6722}
\eeq
 gives $\Dl_N f(x)\approx N^{-1/2}$. Hence even the largest jump size goes to zero as $N\raw\infty$, uniformly in $x$.

The general case is best treated in a different way.\footnote{This part of the proof is due to Jacques Janse van Rensburg.}
We rewrite $\Dl_N f(x)$ as
\begin{equation}
\Dl_N f(x)=p_0^N(\mathsf{LL}_N=x),
\end{equation}
and, given \er{1.19}, use the central limit theorem, where we write  $\mu\equiv S(p_0,p_1)$ as usual for the average, cf.\ \er{1.20}, to   infer that for any $\dl'>0$ and $b>0$ and sufficiently large $N>M(\dl')$ one has
\begin{equation}
\Dl_N f(x)\leq \half\dl'+\frac{1}{\sqrt{2\pi}}\int_{x-\mu-b}^{x-\mu+b} dy\,  e^{-y^2/2}.
\end{equation}
Since $b>0$ is arbitrary one may choose it such that the integral is $\leq\half\dl'$, again uniformly in $x$.\QED
\smallskip

The other two cases \er{Stein2} require a different proof, since following our  approach to proving \er{Stein1} would not imply \er{PBC}. It is quite instructive to follow a different strategy, which avoids Sanov's theorem and also works for  \er{Stein1}.
This strategy is based on the following two lemmas:
 \begin{lemma}\label{CHiaiPetz}
   For  any $\dl>0$ there is a sequence of tests $(T_N)$, $T_N\subset A^N$, such that
 \begin{align}
   \lim_{N\raw\infty} p_0^N(T_N^c)&=0\label{CHP2}; \\
 \lim\sup_{N\raw\infty} \frac{1}{N} \log p_1^N(T_N) &\leq -S(p_0, p_1)+\dl. \label{CHP1}
 \end{align}
 \end{lemma}
 \begin{lemma}\label{CON}
If some sequence $(T_N)$ of tests $T_N\subset A^N$ satisfies $\lim_{N\raw\infty} \al^N(T_N)<1$, then 
\beq
 \lim\inf_{N\raw\infty} \frac{1}{N} \log p_1^N(T_N) \geq -S(p_0, p_1).\label{580}
 \eeq
\end{lemma}
We first reprove \er{Stein1} from these lemmas. First, Lemma \ref{CHiaiPetz}  yields the upper bound 
\begin{equation}
\lim\sup_{N\raw\infty}\frac{1}{N}\log \beta^{\varep}_N\leq 
-S(p_0, p_1),\label{9.30}
\end{equation} since, for fixed $0<\varep<0$,  because of \er{CHP2} 
one can find $N(\varep)$ such that $\al_N(T_N)<\varep$ for all $N>N(\varep)$, so that, given that $\beta_N^{\varep}\leq 
p_1^N(T_N)$, one has
\beq
\lim\sup_{N\raw\infty} \frac{1}{N} \beta_N^{\varep}\leq -S(p_0, p_1)+\dl.
\eeq
Being true for all $\dl>0$, this yields  \er{9.30}.
 Lemma \ref{CON} implies the corresponding lower bound 
 \begin{equation}
\lim\inf_{N\raw\infty} \frac{1}{N} \log \beta_N^{\varep}\geq -S(p_0, p_1).\label{9.77}
\end{equation}
 Towards \er{Stein2},  note that the various conditions on $\al_N$ in \er{9.21} - \er{9.18} are  ordered in strength by
\begin{align}
\al_N\raw 0&&\Raw && \al_N<\varep &&\Raw && \al_N<1,
\end{align}
at least for sufficiently large $N$ in so far as the middle one is concerned (which is enough for proving our limits).  Denoting the left-hand side of \er{Stein1} by $\beta_{\varep}$, this gives 
$\beta^*\leq \beta_{\varep}\leq \beta_*$.
 Lemma \ref{CHiaiPetz} gives $\beta_*\leq -S(p_0,p_1)$, whereas Lemma \ref{CON} gives 
$\beta^*\geq -S(p_0,p_1)$, so that we obtain
\begin{equation}
-S(p_0,p_1)\leq \beta^*\leq \beta_{\varep}\leq\beta_*\leq -S(p_0,p_1).
\end{equation}
Hence \er{Stein2} follows, provided we are able to prove Lemmas \ref{CHiaiPetz} and \ref{CON}.\smallskip

\noindent \emph{Proof of Lemma \ref{CHiaiPetz}.} Tests $T_N$ accomplishing \er{CHP2} - \er{CHP1}  are the
\emph{relative entropy typical sets}
\begin{align}
T_{N,\dl}(p_0,p_1)&:=\left\{\sg\in A^N\mid p_1^N(\sg)e^{N(S(p_0, p_1)-\dl)}\leq p_0^N(\sg)\leq  p_1^N(\sg)e^{N(S(p_0, p_1)+\dl)}\right\}\nn \\
&=
\{\sg\in A^N\mid |\mathsf{LL}_N(\sg)-S(p_0, p_1)|<\dl\},
\label{DAEP}
\end{align}
cf.\ \er{defLLN},
where $\dl>0$, as in Lemma \ref{CHiaiPetz}. These tests are inspired by our ``AEP sets'' \er{TEN}, i.e., 
\begin{align}
T_{N,\dl}(p)&:=\left\{\sg\in A^N\mid  e^{-N(S(p)+\dl)}\leq  p^N(\sg)\leq  e^{-N(S(p)-\dl)}\right\}\nn \\
&= \{\sg\in A^N\mid |\mathsf{L}_N(\sg)+ S(p)|\leq \dl\},  \label{AEPset}
\end{align}
where $\mathsf{L}_N(\sg):=(\log p_N(\sg))/N$. 
One has bounds similar to \er{est33b} for the AEP, viz.\footnote{See Cover \& Thomas (2006), Theorem 11.8.2, for a proof.}
\begin{equation}
(1-\dl)e^{-N(S(p_0, p_1)+\dl)}\leq  p_1^N (T_{N,\dl}(p_0,p_1))\leq e^{-N(S(p_0, p_1)-\dl)},\label{9.26}
\end{equation}
where the upper bound is true for any $N$ whereas the lower bound holds for large enough $N$.
The upper bound gives \er{CHP1}, whereas \er{CHP2} follows from \er{9.24}. \QED\smallskip

\emph{Proof of Lemma \ref{CON}.} 
 Let $\lim_{N\raw\infty} \al^N(T_N)=\varep$, so that $\varep<1$ by assumption.
We start out as in the (first) proof above, 
which for any given $T_N$ such that $0<\al_N(T_N)<\varep$ provides a  test $T^*_N(u_N)\equiv T^*_N$ (do not confuse this with $T^*_N(0)$) satisfying \er{9.311} and \er{9.53}. We then have
\begin{align}
\beta_N(T_N)&\geq \beta_N(T_N^*)=p_1^N(T_N^*)\geq p_1^N(T_N^*\cap T_{N,\dl}(p_0,p_1))=\Sigma_{\sg\in T_N^*\cap T_{N,\dl}(p_0,p_1)} p_1^N(\sg)\nn \\
&\geq e^{-N(S(p_0, p_1)+\dl)}\Sigma_{\sg\in T_N^*\cap T_{N,\dl}(p_0,p_1)} p_0^N(\sg)\nn \\ &= e^{-N(S(p_0, p_1)+\dl)} p_0^N(T_N^*\cap T_{N,\dl}(p_0,p_1)), \label{9.74} 
\end{align}
where now $0<\dl<\varep<1$ is arbitrary and we used the upper bound in the first line of \er{DAEP}. Now 
$p_0^N(T_N^*)>(1-\varep)$ because $\al_N(T^*_N)<\varep$, and $p_0(T_{N,\dl}(p_0,p_1))>1-\delta$ by \er{9.24}, so that 
\begin{equation}
p_0^N(T_N^*\cap T_{N,\dl}(p_0,p_1))> 1-\varep-\delta.\label{9.75} 
\end{equation}
Combining \er{9.74} and \er{9.75} and letting $\delta\raw 0$ at our given value $\varep>0$ gives 
\er{580}.\QED
\subsection*{Partial proof of Hoeffding's theorem}
\noindent Theorem \ref{Hoeffding} is  difficult to prove.\footnote{See Jak\v{s}i\'{c} (2018), \S 4.8, for a complete proof.} The key step is \er{keyHoeffding},
whose  lower bound
\begin{equation}
\lim_{N\raw\infty}\frac{1}{N}\log \beta_N(r) \geq -\inf_{p\in\Pr(A)}\{S(p, p_1)\mid S(p, p_0)\leq r\}\label{HL},
\end{equation}
we prove, so that the reader at least gets an idea of what is going on. The remaining equality
\begin{equation}
\inf_{p\in\Pr(A)}\{S(p, p_1)\mid S(p, p_0)\leq r\}=-\sup_{t\in(0,1)} \left\{ \left( \frac{1-t}{t}\right)\cdot (R_t(p_1,p_0)-r)\right\}, \label{HLb}
\end{equation}
on the other hand, is a straightforward exercise:\footnote{The answer is in Jak\v{s}i\'{c} (2018), Proposition 4.20.
His $P, Q, R$ are our $p_1, p_0, p$ and his $R_{\al}$ is our $p_{1-\lm}$.}
\bex
Prove \er{HLb}.
\eex
The proof of  \er{HL} will  be done for 
$r\in (0, S(p_1, p_0))$. This implies the general case $r>0$: 
\begin{enumerate}
\item The left-hand side of \er{HL}
 is $\leq 0$ and increases with $r$ by definition, since the constraint 
 \beq
 p_0^N(T_N)\geq 1-e^{-Nr}\eeq
   requires larger tests $T_N$ as $r$ and hence $1-e^{-Nr}$ grows, which also increases $p_1^N(T_N)$.
 \item  The right-hand side of \er{HLb}, call it  $R(r)$, vanishes for $r\geq S(p_1, p_0)$, with 
 \beq
 \lim_{r\uparrow S(p_1, p_0)}R(r)=0.\eeq
\end{enumerate}
 Though still  quite remote from a proof of the dichotomy that forms the second part of the theorem, we hope that this argument at least shows the role of the critical value $r=S(p_1, p_0)$.
 
To prove \er{HL} for $r\in (0, S(p_1, p_0))$, then, take some
\begin{equation}
u\in (-S(p_1, p_0), S(p_0, p_1)), \label{9.62}
\end{equation} 
 and consider  the Neyman--Pearson tests
\begin{equation}
T^*_N(u)=\{\sg\in A^N\mid \mathsf{LL}_N>u\}.
\end{equation}
Using \er{LLND}, we have
\begin{align}
\al_N(T_N^*(u))&=p_0^N(S(L_N(\cdot), p_1)-S(L_N(\cdot), p_0)\leq u);\label{alfaHoef}\\
\beta_N(T_N^*(u))&=p_1^N(S(L_N(\cdot), p_1)-S(L_N(\cdot), p_0)> u). \label{betaHoef}
\end{align}
Sanov's theorem for \er{alfaHoef}  turns the bound $\al_N(T_N^*(u))< e^{-Nr}$
 into
\begin{equation}
\inf_{p\in \Pr(A)}\{S(p, p_0)\mid S(p, p_0)\geq S(p, p_1) -u\}\geq r,
\end{equation}
which implies that if $S(p, p_0)\geq S(p, p_1) -u$, then $S(p, p_0)\geq r$. This gives the inclusion
\begin{equation}
\{p\in \Pr(A)\mid S(p, p_0)\geq S(p, p_1) -u\}\subset \{p\in \Pr(A)\mid S(p, p_0)\geq r\},
\end{equation}
whose contrapositive is the opposite inclusion
\begin{equation}
\{p\in \Pr(A)\mid S(p, p_0)< S(p, p_1) -u\}\supset \{p\in \Pr(A)\mid S(p, p_0)< r\}.
\label{Hoefcontra}
\end{equation}
Sanov's theorem for \er{betaHoef} gives
\begin{equation}
\lim_{N\raw\infty}\frac{1}{N} \log \beta_N(T^*(u))=-\inf_{p\in \Pr(A)}\{S(p, p_1)\mid S(p, p_1)> S(p, p_0)+u\}.
\end{equation}
Now \er{Hoefcontra} gives
\begin{equation}
\inf_{p\in \Pr(A)}\{S(p, p_1)\mid S(p, p_1)> S(p, p_0)+u\}\leq \inf_{p\in \Pr(A)}\{S(p, p_1)\mid  S(p, p_0)< r\},
\end{equation}
since by \er{Hoefcontra} the infimum on the left-hand side is taken over a possibly larger set than on the right-hand side.
Replacing $<r$ by $\leq r$ by continuity, this yields the lower bound \er{HL}. \QED\medskip

So far we have  treated the ``atomic'' case where $\mathsf{H}_i$ states that $p=p_i$, for $i=0,1$. We close this section with a brief discussion on  \emph{compound hypothesis testing}, in which  we allow a hypothesis $\mathsf{H}_i$ to be $p\in B_i$ for some subset $B_i\in\Pr(A)$. For example, the null hypothesis $\mathsf{H}_i$ could still be atomic, e.g., $p=f$ (``the coin is fair''), but it could be tested against a compound hypothese 
to the effect that say $d(p,f)\geq \varep$ for some $\varep>0$ and some metric $d$ on $\Pr(A)$. 

 Fortunately, as long as $A$ is finite all three theorems (Chernoff, Stein, Hoeffding) for the atomic case imply corresponding results for the compound case, as follows.\footnote{See Mosonyi,   Szil\'{a}gyi, \& Weiner (2021), Theorem III.7, and references therein to earlier results like that.}  Define 
 \begin{align}
 \al_N(T_N,B_0)&:=\sup_{p_0\in B_0} p_0^N(T_N^c); \hspace{50pt}  \beta_N(T_N,B_1):=\sup_{p_1\in B_1} p_1^N(T_N);\\
 \gm_N(B_0,B_1)&:=\inf_{T_N}\{ \al_N(T_N,B_0) + \beta_N(T_N,B_1)\};\\
 \beta_N^{\varep}(B_0,B_1)&:=\inf_{T_N}\{\beta_N(T_N, B_1)\mid \al_N(T_N,B_N)\leq \varep\};\label{Steincompound}\\
 \beta_N(r, B_0,B_1)&:=\inf_{T_N}\{\beta_N(T_N,B_1)\mid \al_N(T_N,B_0)\leq e^{-Nr}\},
 \end{align}
so that one studies the worst-case errors (here $0<\varep<1$ and $r>0$ as before). We then have:
 \begin{align}
 \lim_{N\raw\infty}\frac{1}{N}\log  \gm_N(B_0,B_1)&= -C(B_0,B_1); \label{5109C}\ \\
 \lim_{N\raw\infty}\frac{1}{N}\log  \beta_N^{\varep}(B_0,B_1)&=-S(B_0,B_1);\label{5109}\\
 \lim_{N\raw\infty}\frac{1}{N}\log  \beta_N(r, B_0,B_1)&=-H(r,B_0,B_1),\label{5109H}
  \end{align}
  where for each of the relative entropies $D=C$, $D=S$, and $D=H(r)$ we put
  \begin{equation}
D(B_0,B_1):=\inf_{p_0\in B_0, p_1\in B_1}\{ D(p_0,p_1)\},
\end{equation}
which again reflects the worst-case scenario, because the smaller the rate $c$ in $e^{-cN}$,  the worse the error. 
  For example, taking $B_1=\{p_1\}$ and using \er{611} transforms \er{5109} into
  \begin{equation}
 \beta_N^{\varep}(B,q)=-S_q(B)= \lim_{N\raw\infty}\frac{1}{N}\log  q^N(L_N\in B), \label{5112}
\end{equation}
which has led people to call  quantum versions of \er{5112} ``quantum Sanov's theorems'', on the grounds that the l.h.s.\ of \er{5112} \emph{can} be generalized to quantum theory whereas the r.h.s.\ can\emph{not}. See \S\ref{QHTsec}.
Another interesting result, still  for the case that $\mathsf{H}_1$ is atomic, is as follows:\footnote{See Bjelakovi\'{c} et al.\ (2005) who also prove this classical  result directly (as well as its quantum version).}
\begin{theorem}\label{CMP2005C}
For any  $B\subset \Pr(A)$ there are tests $(T_N)$ such that for all  $p\in B$ and $q\in\Pr(A)$:
\begin{align}
\lim_{N\raw\infty}p^N(T_N^c)=0; && 
\lim_{N\raw\infty}\frac{1}{N}\log p_1^N (T_N)=-S_q(B).
\end{align}
\end{theorem}
Taking $B=\{p_0\}$ sharpens Lemma \ref{CHiaiPetz} so as to achieve $\dl=0$, and hence
Stein's theorem follows from Theorem \ref{CMP2005C}  and Lemma \ref{CON}. 
Conversely, Theorem \ref{CMP2005C} follows from Stein's theorem.\footnote{If $B_0$ satisfies \er{condition} and $p_1$ is faithful, then the second part  follows from Sanov's theorem by taking $T_N=(L_N\in B_0)$. If not, as in the case $B_0=\{p_0\}$, one should take a decreasing sequence of closed sets $B_N\subset \Pr(A)$ such that $B_N\raw B_0^-$ in a suitable sense; for example, if $p_1$ is faithful it suffices to take a sequence $(\varep_N)$ such that $\varep_N>0$ and $\varep_n\raw 0$, and, in some metric $d$ on $\Pr(A)$,
$B_N=\{p\in\Pr(A)\mid d(p,p_0)\leq \varep_N\:\mbox{ for at least one } p_0\in B_0\}$.
Then take $T_N=(L_N\in B_N)$. For more general $p_1$ some further refinements are necessary. See Bjelakovi\'{c} et al.\ (2005), Lemma 1.}

Of course, one may generalize Theorems \ref{Chernoff}, \ref{Stein}, and \ref{Hoeffding} in various other directions.\footnote{See for example  Lami (2025b), Cuneo et al.\ (2026), and the entire body of literature  on quantum hypothesis testing,
which contains classical hypothesis testing as a special case, cf.\ \S\ref{QHTsec} and references given there.
}
 \section{Large deviations:  Cram\'{e}r's theorem}\label{LDCT}
 After Sanov's theorem \ref{sanov}, the next major result in large deviation theory (and historically even the first) is
Cram\'{e}r's theorem (originally from 1938). In physics,  Sanov is about the fluctuation of one-particle distribution functions, whereas Cram\'{e}r is about (for example) energy fluctuations. It is basically a  corollary of Sanov's theorem, though some crucial extra information is added that replaces the infima in \er{eqLDP1} - \er{Binf}, which are typical for large deviations theorems, by pointwise values of the rate function. 
 We first discuss this theorem for finite $A$ and will later generalize.\footnote{ We roughly follow Dembo \& Zeitouni, \S 2.1.2 with some arguments taken from  Jak\v{s}i\'{c} et al (2012), \S 2.2.}
  
 We  pick an injective ``energy'' function $E:A\raw\R$ and a prior $q\in\Pr(A)$, and return to  \er{SsumE}. The strong law of large numbers \er{SLLN}  states that $S_N(s)\raw \la E\ra_q$
for   $q^\N$-almost every $s\in A^\N$.
  Cram\'{e}r's theorem  describes the probability of large deviations from $\la E\ra_q$ via the function
    \begin{align}
I_q(x)&:=\inf\{S_q(p)\mid p\in\Pr(A), \la E\ra_p=x\}; \label{IMEP}\\
I_q(x)&:=\infty\:\: \mathrm{if\:\: no}\:\:  p\in\Pr(A)\:\: \mathrm{with}\:\: \la E\ra_p=x\:\: \mathrm{exists}.\label{IMEP2}
\end{align}
Note that \er{IMEP2} in fact follows from \er{IMEP}, since, as a bizarre fact of logic,  $\inf\,\emptyset=\infty$.\footnote{Let $S\subset [-\infty,\infty]$. Then $z=\inf S$ iff: (1) $\forall_{s\in S}( z\leq s)$; (2) $\forall_w((\forall_{s\in S}(w\leq s)) \raw w\leq z)$. Now $\forall_{s\in S} F(s)$ is an abbreviation of $\forall_s (s\in S\raw F(s))$. If $S=\emptyset$, then $s\in S$ is false and hence any implication of it is true, so (1) is always true. Likewise, in (2) the antecedent $\forall_{s\in S}(w\leq s)$ is always true, and so to make (2) true the conclusion $w\leq z$ must be true for all $w$. This forces $z=\infty$, i.e.\  $\inf\,\emptyset=\infty$. Similarly, $\sup\,\emptyset=-\infty$. }
\bex
Show that in the special case  $A=\{0,1\}$,  $E(a)=a$,  $q=f$, 
our function equals
\begin{align}
 I_f(x)&= (1-x)\log (1-x)+x\log x +\log 2 &&(x\in [0,1]); \label{IFCTa}\\
I_f(x)&=\infty && (x\notin [0,1]).\label{IFCT}
\end{align}\eex
\begin{center}
 \includegraphics[width=0.35\textwidth]{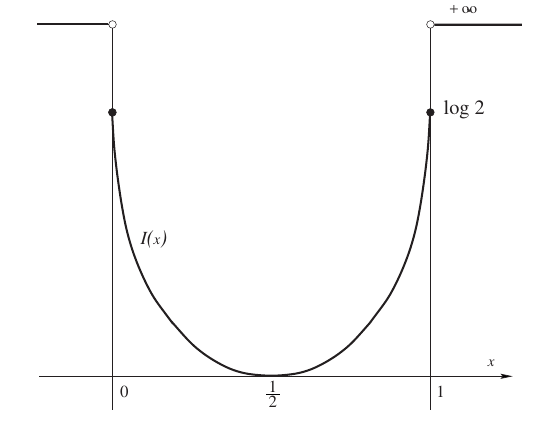}
\end{center}
This picture looks similar to the one below \er{Ix1}, and indeed \er{IFCTa} and \er{Ix1} are identical.\footnote{The picture is copied from Olivieri \& Vares, (2004), p.\ 4; the $I$ in the picture is our $I_f$. }  The difference between $I(x)$ in \er{Ix1} and $I_f(x)$ in \er{IFCTa} - \er{IFCT} is that the latter is defined on all of $\R$, by making it infinite where $I(x)$ was not defined yet, i.e., for $x\notin[0,1]$. But even this difference is straightened out if we extend $I(x)$ from $\Pr(2)\cong [0,1]$ to $\mathcal{M}(2)\cong \R$ as explained after \er{FeTd}. Note that $I_f$ is convex and lsc; to \emph{see} this, just observe that its epigraph is convex and closed.\footnote{A function $f:X\raw (-\infty, \infty]$ is convex iff its epigraph $\mathrm{epi}(f):= \{(x,t)\in \mathcal{D}_f\x\R\mid f(x)\leq t\}$
is convex and it is lsc iff its epigraph is closed. Here $\mathcal{D}_f=\{x\in X\mid f(x)<\infty\}$ is the domain of $f$.
See Appendix \ref{AppC}.} The reason for this apparent coincidence is that, as we will see, for $A=\{0,1\}$ Cram\'{e}r's theorem does not only follow from Sanov's but is in fact equivalent to it. Indeed, in that case we have
\beq
S_N(s)= (L_N(s))(1),
\eeq
and so if we identify $p\in\Pr(2)$ with $p(1)\in [0,1]$, we simply have $S_N=L_N$.
  \begin{theorem}[Cram\'{e}r] \label{Cr0} Let $E:A\raw\R$ be injective and let $q\in\Pr(A)$. For
   $x\geq \la E\ra_q$ we have
   \begin{align}
\lim_{N\raw\infty} \frac{1}{N} \log q^N(S_N\geq x)=-I_q(x) && (x\geq \la E\ra_q). \label{79}
\end{align}
  \end{theorem}  
  This is the most basic form of the theorem. A useful equivalent version (change some signs) is
  \begin{align}
\lim_{N\raw\infty} \frac{1}{N} \log q^\N(S_N\leq x)=-I_q(x) && (x\leq \la E\ra_q).  \label{79b}
\end{align}
Our favourite special case $A=\{0,1\}$ and $q=f$ may  be proved directly, as follows:\footnote{See McKean (2014), \S 2.4.1.}
\bex
Prove that
 \begin{align}
\lim_{N\raw\infty} \frac{1}{N} \log f^N(S_N\geq x)=-I_f(x) && (x\geq \half), \label{easycramer}
\end{align}
 for example along the following lines: 
\begin{enumerate}
\item Show that \er{easycramer} holds for $x=\half$, $x>1$ (both trivial) and $x=1$ (easy). 
\item So $x\in (\half,1)$ remains. Show that in this open interval,\footnote{Here $\lceil y\rceil:=\min\{k\in\Z\mid k\geq y\}$, for $y\in\R$, i.e.\ the smallest integer above $y$.}
\begin{equation}
 f^N(S_N\geq x)=2^{-N} \Sigma_{k=\lceil Nx\rceil}^N 
\left(
\begin{array}{c}
 N  \\
k
\end{array}
\right).
 \end{equation}
\item Show that for $x\in (\half,1)$,
\begin{equation}
\left(
\begin{array}{c}
 N  \\
\lceil Nx\rceil
\end{array}
\right)
\leq  \Sigma_{k=\lceil Nx\rceil}^N 
\left(
\begin{array}{c}
 N  \\
k
\end{array}
\right)
 \leq (N+1)\cdot 
\left(
\begin{array}{c}
 N  \\
\lceil Nx\rceil
\end{array}
\right),\end{equation}
and hence that
\begin{equation}
2^{-N} \left(
\begin{array}{c}
 N  \\
\lceil Nx\rceil
\end{array}
\right)
\leq   f^N(S_N\geq x)\leq 2^{-N} (N+1)\cdot  \left(
\begin{array}{c}
 N  \\
\lceil Nx\rceil
\end{array}
\right). \label{515}
\end{equation}
\item
Using an appropriate version of Stirling's approximation for $n!$, prove \er{easycramer} from \er{515}.
\end{enumerate}
\eex
Compared to Sanov, eq.\ \er{79} does not contain an infimum of the rate function. Here is why:
\begin{proposition}\label{prop61}
\begin{enumerate}
\item The rate function $x\mapsto I_q(x)$ is convex and lsc.
\item $I_q$ has a unique minimum at $x_0=\la E\ra_q$, where  $I_q(x_0)=0$.
\item $I_q(x)>0$ for all $x\neq x_0$.
\item Wherever $I_q$ is finite it is
  increasing if $x> x_0$ and  decreasing for $x< x_0$.
\end{enumerate}
\end{proposition} 
\emph{Proof}. 
Convexity of $p\mapsto S_q(p)$ also makes $x\mapsto I_q(x)$ convex:  for $t\in (0,1)$ and $x,y\in\R$  we have
\begin{align}
I_q(tx+(1-t)y)&\leq \inf\{S_q(tp'+(1-t)p'')\mid p',p''\in\Pr(A), \la E\ra_{tp'+(1-t)p''}=tx+(1-t)y \}\nn \\
&\leq \inf\{S_q(tp'+(1-t)p'')\mid p',p''\in\Pr(A), \la E\ra_{p'}=x, \la E\ra_{p''}=y \}\nn \\
&=\inf\{tS_q(p')+(1-t)S_q(p'')\mid p',p''\in\Pr(A), \la E\ra_{p'}=x, \la E\ra_{p''}=y \}\nn \\
&= \inf\{tS_q(p')\mid p'\in\Pr(A), \la E\ra_{p'}=x\}\nn \\ &+\inf\{
(1-t)S_q(p'')\mid p''\in\Pr(A), \la E\ra_{p''}=y \}\nn \\
&=tI_q(x)+1-t I_q(y).
\end{align}
\bex Prove, from lower semicontinuity of $p\mapsto S_q(p)$, that $x\mapsto I_q$ is also lsc.
\eex
Clearly, $I_q(x)\geq 0$ because $S_q(p)\geq 0$ by Proposition \ref{gibbsin},  and 
taking $x=\la E\ra_q\equiv x_0$ in \er{IMEP} gives $I_q(x_0)=0$.  
Proposition \ref{gibbscon} below implies that 
$\la E\ra_p=\la E\ra_q$ iff $p=q$, so that this minimum at $x_0$ is unique. Convexity of 
  $I_q$ forbids other local minima, which implies the remaining claims. \QED
  \smallskip
  
\noindent  Before proving (and extending) Theorem \ref{Cr0}, let us continue the study of $I_q$.  We  assume  $q(a)>0$ for each $a\in A$, but if \er{IMEP2} is  taken into account the conclusions do not depend on this.\footnote{Proposition \ref{gibbscon} and its  proof are taken from Austin (2017), Lecture 11, and Borwein \& Zhu (2005), \S4.7.} Let
\begin{align}
E_-:=\min_a E(a)=E(a_-); && E_+:=\max_a E(a)=E(a_+), \label{defEplusmin}
\end{align} 
where $a_{\pm}$ are unique since $E$ was assumed to be injective. 
 For $\beta\in\R$, define $q_{\beta}\in\Pr(A)$ by
\begin{align}
q_{\beta}(a):=\frac{q(a)e^{-\beta E(a)}}{Z_{\beta}}; && Z_{\beta}:= \Sigma_{b\in A} q(b)e^{-\beta E(b)}.\label{BoF}
\end{align}
We  extend these (Boltzmann) probabilities to $\beta\pm\infty$ by defining the point measures $q_{\pm\infty}\in\Pr(A)$:
\beq
q_{\pm\infty}(a_{\mp})=1.
\eeq
\bex
Obtain (instead of define) these by taking the (weak) limits $\beta\raw\pm\infty$ in \er{BoF}.
\eex
\begin{proposition} \label{gibbscon}
For each $x\in [E_-,E_+]$ there exists a unique $\beta(x)\in [-\infty,\infty]$ (where $\beta(x)\in (-\infty,\infty)$ for 
$x\in (E_-,E_+)$ and $\beta=\pm\infty$ for $x=E_{\mp}$) 
 such that the infimum in \er{IMEP}
is a unique minimum at $p=q_{\beta(x)}$, that is, $\la E\ra_{q_{\beta(x)}}=x$ and $\la E\ra_p\neq x$ for all $p\neq q_{\beta(x)}$.
 In terms of this, $I_q(x)$ is given by
\begin{align}
I_q(x)&= S_q(q_{\beta(x)})=-\beta x-\log Z_{\beta} && (E_-<x<E_+); \label{FTS}\\
I_q(E_{\pm})&=S_q(q_{\mp\infty})=-\log q(a_{\mp\infty}) && (x=E_{\pm}); \label{FTS2}\\
I_q(x)&= \infty && (x\notin [E_-,E_+]). \label{FTS3}
\end{align}
\end{proposition}
\emph{Proof}.\footnote{
 Proposition \ref{gibbscon} is a special case of the following result. 
Let $A$ be a Polish space. If $C\subset\Pr(A)$ is closed and convex and such that
$S_q(C)=S_q(\mathring{C})<\infty$,
 there is a unique $p\in C$ that minimizes $S_q(p)$ within $C$. For the conditional probability measures $P_N\in\Pr(\Pr(A))$ defined by $P_N(B):=q^\N(L_N\in B\mid L_N\in C)$ (i.e.\ $P_N=q^\N(\cdot\mid C)$) we have  
$\lim_{N\raw\infty} P_N=\dl_p$
exponentially fast, in that for any weak nbhd $U\in\CO_p(\Pr(\Pr(A)))$ there is $b>0$ such that
$p_N(U^c)\leq e^{-bN}$.
 See e.g.\  Dembo \& Zeitouni, Theorem 3.3.3, or Rassoul-Agha \& Sepp\"{a}l\"{a}inen (2015), \S 5.3.}
The case $x\notin (E_-,E_+)$ is an exercise. Assume $E_-<x<E_+$. Since $\la E\ra_{q_{\beta}}\raw E_{\pm}$ as $\beta\raw \mp\infty$ by Laplace or directly, the desired value of $\beta$ for which 
\beq
\la E\ra_{q_{\beta}}=x
\eeq
 exists 
  by continuity of $\beta\mapsto \la E\ra_{q_{\beta}}$ and the intermediate value theorem.
      Eq.\ \er{FTS} then follows from a simple computation based on \er{BoF} and \er{KL1}.  
      To prove uniqueness of the minimizer $q_{\beta}$, first note from \er{KL2} that a possible other minimizer  $p\in\Pr(A)$ must satisfy  $p\ll q$ and hence
  $p\ll q_{\beta}$, so that the Radon--Nikodym derivative $f(a)=p(a)/q_{\beta}(a)$  (which for finite $A$ is just a quotient) exists. Then, by \er{KL1}, we obtain (exercise)
  \begin{equation}
  S_q(p)=S_{q_{\beta}}(p)+S_q(q_{\beta}), \label{weobtain}
\end{equation}
where we used the constraint $\la E\ra_p=x$ under which $p$ is supposed to minimize $S_q(p)$, and \er{FTS}.
By Proposition \ref{gibbsin} we conclude that $S_q(p)\geq S_q(q_{\beta})$ with equality iff $p=q_{\beta}$.\QED
\bex
Prove \er{weobtain} for $E_-<x<E_+$. Also, prove the proposition for $x\notin (E_-,E_+)$.
\eex
The existence and uniqueness part of the Proposition \ref{gibbscon} is a special case of the following:\footnote{We omit the proof; see  Dembo \& Zeitouni, Theorem 3.3.3, or Rassoul-Agha \& Sepp\"{a}l\"{a}inen, \S 5.3.}
\begin{proposition}\label{gibbscon2}
Let $A$ be a Polish space. If $C\subset\Pr(A)$ is closed and convex and such that
\beq
D_q(C)=D_q(\mathring{C})<\infty,
\eeq
 there is a unique $p\in C$ that minimizes $D_q(p)$ within $C$. For the conditional probability measures $P_N\in\Pr(\Pr(A))$ defined by $P_N(B):=q^{\N}(L_N\in B\mid L_N\in C)$, i.e.\ $P_N=q^{\N}(\cdot\mid C)$,  we have  
\begin{equation}
\lim_{N\raw\infty} P_N=\dl_p
\end{equation}
exponentially fast, in that for any weak nbhd $U\in\CO_p(\Pr(\Pr(A)))$ there is $b>0$ such that:\footnote{Eqs.\ \er{FTS} and \er{eqLDP2} merely imply this for $N\raw\infty$.}
\begin{equation}
p_N(U^c)\leq e^{-bN}.
\end{equation}
\end{proposition}

For $0<\beta<\infty$ we follow the physicists in writing  
\begin{align}
\beta =1/T; && S_q(q_{\beta})=-S; && x=U; && \log Z_{\beta}=-\beta F.
\end{align} Eq.\ \er{FTS} then gives the free energy  $F=U-TS$.
If in view of Proposition \ref{gibbscon} we write the second member of the Fenchel duality \er{FDD1} - \er{FDD2} as
\begin{equation}
-\log Z_q(-\beta E)=\inf_{p\in\Pr(A)}\{\beta \la E\ra_p+S_q(p)\}, \label{nieuwZ}
\end{equation}
where $Z_q$ is defined by \er{FandP}. Noticing that $ Z_q(-\beta E)$ equals $Z_{\beta}$  in \er{BoF}, 
we conclude that not only the constrained infimum in \er{IMEP}  but also the unconstrained one \er{nieuwZ} is attained at $p=q_{\beta}$.\smallskip

\emph{Proof of Theorem \ref{Cr0}.} We  prove from Sanov's theorem that for (measurable)  $C\subset \R$, we have
\begin{align}
- I_q(\mathring{C})\leq \lim\inf_{N\raw\infty} \frac{1}{N} \log q^N(S_N\in C)
\leq  \lim\sup_{N\raw\infty} \frac{1}{N} \log q^N(S_N\in C)\leq - I_q(C^-). \label{eqLDP2} 
\end{align}
Similarly to the statement of Sanov's theorem, one may replace \er{eqLDP2} by the equivalent pair
 \begin{align}
\lim\sup_{N\raw\infty} \frac{1}{N} \log q^N(S_N\in F)&\leq - I_q(F) \:\:\: (F\subset\R\: \mathrm{closed});\label{LDPC1}\\
\lim\inf_{N\raw\infty} \frac{1}{N} \log q^N(S_N\in U)&\geq - I_q(U)  \:\:\: (U\subset\R\: \mathrm{open}).\label{LDPC2}
\end{align}
And again as in Sanov, if $C\subset \R$ is such that $I_q(\mathring{C})=I_q(C^-)$, cf.\ \er{Binf}, then  \er{eqLDP2} trivially implies
\begin{equation}
 \lim_{N\raw\infty} \frac{1}{N} \log q^N(S_N\in C)=-I_q(C). \label{SNLDPsimple}
\end{equation}
This last equation applies to $C=[x,\infty)$ for  $x\geq \la E\ra_q$, as follows from a case distinction:
\begin{enumerate}
\item If $x<E_+$, then $x\in (E_-,E_+)$, where  $I_q$  is continuous, so that  $I_q(\mathring{C})=I_q(C^-)$.
\item If $x>E_+$, then $I_q(y)=\infty$ for all $y\in C$ by 
Proposition \ref{gibbscon}, and hence $I_q(\mathring{C})=I_q(C^-)=\infty$. Eq.\ \er{SNLDPsimple} then holds in the form $-\infty=-\infty$. 
\item If $x=E_+$, then $I_q(\mathring{C})=\infty$, but $I_q(C^-)=I_q(E_+)$. However, eq.\ \er{SNLDPsimple} can be checked directly: since $q^N(S_N\in C')=0$ whenever $C'\cap [E_-, E_+]=\emptyset$, we have
$(S_N\in C)=(S_N=E_+)$, which forces the event $\sg^+$ defined  by $\sg^+_n=a_+$ for all $n$, so that
$q^N(S_N\in C)=q^N(\sg^+)=q(a_+)^N$, and hence $ \frac{1}{N} \log q^N(S_N\in C)=\log (q(a_+))$. Eq.\ \er{SNLDPsimple} then follows from \er{FTS2}. 
\end{enumerate}
Finally, if $C=[x,\infty)$, then $I_q(C)=I_q(x)$ by Proposition \ref{prop61}, so that eq.\ \er{SNLDPsimple} implies \er{79}.

To prove \er{eqLDP2} from Sanov we re-express  $S_N$ in
 terms of the empirical measure \er{LNb}:
\begin{align}
\la E\ra_{L_N(\sg)}&=\Sigma_{a\in A} L_N(\sg)(a)E(a)=\frac{1}{N}\Sigma_{a\in A}\Sigma_{n=0}^{N-1}\dl_{\sg_n a} E(a)=\frac{1}{N}\Sigma_{n=0}^{N-1} E(\sg_n)=S_N(\sg),
\end{align}
i.e.\  $S_N(\sg)$ is  the expectation  of $E$ in $L_N(\sg)\in\Pr(A)$. Hence for every subset $C\subset\R$ we have
\begin{align}S_N\in C&& \mathrm{iff} && L_N\in B:=\{p\in\Pr(A)\mid \la E\ra_p\in C\}.\label{BC}
\end{align}
By basic set theory and the definition of an infimum we obtain
\begin{align}
\{p\in\Pr(A)\mid \la E\ra_p=C\}&= \{p\in\Pr(A)\mid \la E\ra_p=x,\, x\in C\};\\ \Raw
\inf\{S_q(p)\mid p\in\Pr(A), \la E\ra_p\in C\}&=\inf_{x\in C} \{S_q(p)\mid p\in\Pr(A), \la E\ra_p=x\}. \label{new839}
\end{align}
Eq.\  \er{611} gives the asymptotics of $q^N(L_N\in B)$. Combining this with \er{BC} and \er{new839} gives
\begin{align}
S_q(B)&=\inf_{p\in B}\{S_q(p)\}=\inf\{S_q(p)\mid p\in\Pr(A), \la E\ra_p\in C\}\nn \\
&= \inf_{x\in C} \{S_q(p)\mid p\in\Pr(A), \la E\ra_p=x\}= \inf_{x\in C}I_q(x)=I_q(C). 
\end{align}
Hence \er{SNLDPsimple} follows from \er{611}, which is enough for \er{79}, so we are done.\footnote{To also show that
\er{eqLDP2} follows from  \er{eqLDP1}, note that since $p\mapsto \la E\ra_p$ is  (weakly) continuous, if $B\subset\Pr(A)$ corresponds to $C\subset\R$ via \er{BC}, then $C$ is open if $B$ is open and hence \er{LDPC2} follows from \er{LDPu}. The derivation of \er{LDPC1}  from \er{LDPl} needs some more care and follows from the detailed properties of $I_q$ as stated. }\QED\smallskip

In  the same way as for Sanov's theorem, one also obtain statements like:
 \begin{corollary}\label{cramercor}
 \begin{enumerate}
\item If $\la E\ra_q\notin C^-$, then there is $d>0$ such that for sufficiently large $N$ one has
\beq
q^\N(S_N\in C)\leq e^{-dN}.\eeq
\item On the other hand, if  \er {SNLDPsimple} holds and $\la E\ra_q\in C$, then
\begin{equation}
 \lim_{N\raw\infty}  q^\N(S_N\in C)=1, 
\end{equation}
and hence the strong law of large numbers \er{SLLN} holds.
\end{enumerate}
  \end{corollary}
  
  Again similarly to Sanov, an important aspect of Cram\'{e}r's theorem lies in the following fact:
 \begin{proposition} \label{cramer0}
 The rate function $I_q$ is alternatively given as a Fenchel transform:
\begin{align}
I_q(x)&=\sup_{t\in\R}\{xt-\til{\Pi}_q(t)\} && (I_q=\til{\Pi}_q^*); \label{IqC1}\\
\til{\Pi}_q(t)&=\sup_{x\in\R}\{xt-I_q(x)\} && (\til{\Pi}_q=I_q^*), \label{ptft0}
\end{align}
where the ``pressure'' $\til{\Pi}_q(t)$ corresponding to the  ``energy'' function $E:A\raw\R$ is equal to
\begin{align}
 \til{\Pi}_q(t):=\log \til{Z}_q(t); && \til{Z}_q(t):= \la e^{tE}\ra_q= \Sigma_{a\in A} q(a) e^{tE(a)}. \label{tildeP}
\end{align} Moreover, the supremum over $t$ in \er{IqC1} is uniquely attained by $t=-\beta(x)$, cf.\ Proposition \ref{gibbscon}.
\end{proposition}
It follows that $\til{\Pi}_q$, like $I_q$,  is convex and lsc; for finite $A$ the pressure is even continuous, which is trivial from \er{tildeP}; but all of this also generalizes to Polish spaces in the obvious way.
\bex
Verify \er{IqC1} in the  example at the beginning of this section. First show that
\begin{align}
I_f(x)&=\log 2+
\sup_{t\in\R}\left\{xt-\log (1+e^t)\right\}. \label{Ihalf}
\end{align}
Then consider the function $f_x(t):= xt-\log (1+e^t)$. Prove the following cases:
\begin{itemize}
\item For $x\in (0,1)$ the function $t\mapsto f_x(t)$ has a unique maximum at $t=t(x)$, at which value:
\beq
f_x(t(x))= (1-x)\log (1-x)+x\log x.
\eeq
\item For $x=0$ our function assumes no \emph{maximum} but has a finite \emph{supremum}, given by 
\beq
\sup_{t\in\R}\{-\log\left(1+e^t\right)\}=-\inf_{t\in\R}\{\log\left(1+e^t\right)\}=
-\lim_{t\raw-\infty}\log \left(1+e^t\right)=0.
\eeq
\item Similarly for  $x=1$: no maximum but  a finite supremum, this time given by 
\beq
\sup_{t\in\R} \{t-\log\left(1+e^t\right)\}=\lim_{t\raw\infty}(t- \log\left(1+e^t\right))=0.
\eeq
\item For $x\notin[0,1]$, the function $t\mapsto f_x(t)$ has neither extrema nor asymptotes, and
\begin{equation}
\sup_{t\in\R}\{f_x(t)\}=\infty.
\end{equation}
\end{itemize}
\eex
\emph{Proof of Proposition \ref{cramer0}.}
The equivalence between \er{IMEP} and \er{IqC1} is a central feature of large deviation theory and statistical mechanics. We first sketch an informal ``proof'', based on the Fenchel transform \er{FDD1} - \er{FDD2} and the Lagrange multiplier method for computing a constrained extremum like \er{IMEP}. As in \er{willik}, for finite $A$ eq.\  \er{FDD1} simply reads
\begin{equation}
S_q(p)=\sup_{E\in\R^{|A|}}\left\{ \Sigma_a E_ap_a-\log \Sigma_b q_be^{E_b}\right\}.
\end{equation}
Furthermore, because of \er{DefSq2} we may take the infimum in \er{IMEP} over all $p:A\raw\R$ instead of all $p\in\Pr(A)$, i.e.\ we may forget the conditions $p(a)\geq 0$ and $\Sigma_a p(a)=1$. Calling the Lagrange multiplier $t$,  the constrained extremum in \er{IMEP} is therefore found (if it exists) by minimizing 
\begin{equation}
f(p,t):= S_q(p)-t\left(\Sigma_a E_a p_a-x\right)=\sup_{E\in\R^{|A|}}\left\{ \Sigma_a (E_a-t E_a)\cdot p_a-\log \Sigma_b q_be^{E_b}\right\}+tx.
\end{equation}
Minimizing $f$ with respect to $p$ by moving $\partial /\partial p_a$ inside the $\sup_{E}$, which effectively 
replaces $\inf_p\sup_{E}$ by $\sup_{E}\inf_p$ and is illegal without further arguments, 
 and then putting $\partial f/\partial p_a=0$
 gives $E_a=t E_a$, which clears the first term after the sup. Substituted into the second term, it also gives  
 \beq
 \Sigma_b q_b e^{E_b}=\la e^{tE}\ra_q.
 \eeq
  Minimizing with respect to $t$ returns the constraint $\Sigma_a E_ap_a=x$, but if we write this second minimization as $\inf_t$, we see at once that $I_q$ as defined in \er{IMEP} equals $I_q$ as stated in \er{IqC1}. 
  
 A  proof of the equality of \er{IMEP} and \er{IqC1} without the above handwaving, and valid for Polish spaces $A$,\footnote{For Polish spaces the statement of Theorem \ref{Cr0} is the same, and the sum in \er{tildeP} is replaced by an integral.}
  is based on \emph{Fenchel--Rockafellar} duality, cf.\ Theorem \ref{FRDT} and Corollary \ref{forMEP}:
 \begin{proposition}\label{FDentropy} 
For a Banach space  $X$, let $f:X\raw(-\infty,\infty]$ be  lsc and convex, and $T:X\raw \R^d$  linear and continuous. If $x\in\R^d$ lies in the interior (or the core) of $T(\mathcal{D}_f)$, then
\begin{equation}
\inf_{p\in X}\{f(p)\mid Tp=x\}=\sup_{t\in\R^d}\{\la t,x\ra -f^*(T^*t)\}, \label{FDeq}
\end{equation}
and the supremum is even a maximum (i.e.\ it is attained). Here $f^*$ is the Fenchel transform of $f$.
\end{proposition} 
This applies to \er{IMEP} since as noted before we may replace $p\in\Pr(A)$ by $p:A\raw\R$. Hence 
in Proposition \ref{FDentropy} we may take $X=\R^{|A|}$, $f=S_q$, $d=1$, and $T: \R^{|A|}\raw\R$ defined by 
\beq
Tp=\la E\ra_p=\Sigma_a E_ap_a.
\eeq Then $f^*:Y=\R^{|A|}\raw(-\infty,\infty]$ is given by $S_q^*=\Pi_q$, see \er{FDD2}, and $T^*:\R\raw \R^{|A|}$ is given by 
\beq
T^*t=tE,
\eeq
 because $\la T^*t,p\ra:=\la t, Tp\ra$ gives $\Sigma_a p_a (T^*t)_a=t\Sigma_a E_ap_a$, and hence
$(T^*t)_a=tE_a$. Note  that 
\beq
\Pi_q(tE)=\til{\Pi}_q(t),
\eeq
with abuse of notation,
 where the left-hand side is defined in \er{FandP} whilst the right-hand side comes from \er{tildeP}. Thus the equality in \er{FDeq} is almost the equality between \er{IMEP} and \er{IqC1}; except that \er{FDeq} is only stated for $x\in\mathrm{int}(T(\mathcal{D}_f))$. Assuming once again that $q_a>0$ for all $a\in A$ for simplicity, we have $\mathcal{D}_f=\Pr(A)$ and hence 
 \begin{align}
 T(\mathcal{D}_f)=
 [E_-, E_+]; && 
\mathrm{int}(T(\mathcal{D}_f))=(E_-, E_+).
\end{align}
  Eq.\ \er{FDeq} therefore establishes  the equality between \er{IMEP} and \er{IqC1} for all $x\in (E_-, E_+)$.
Extending this analysis to $x\notin (E_-, E_+)$
is  similar to the  case $A=2$, $E(a)=a$, $q=f$:
\begin{itemize}
\item  For the boundary points $x=E_{\pm}$ we first recall that  $E_{\pm}=E(a_{\pm})$. 
If $x=E_-$, the constraint $\la E\ra_p=E_-$ can  only be satisfied by $p(a_-)=1$, so that $S_q(p)=-\log q(a_-)$. But this is also the supremum of \er{alsosup} 
 for $x=E_-$, which is reached as $t\raw -\infty$. The constraint $\la E\ra_p=E_+$ can only be satisfied by $p(a_+)=1$, so that $S_q(p)=-\log q(a_+)$; which is  the supremum of \er{alsosup} for $x=E_+$, but this time as $t\raw\infty$.
 \item For $x> E_+$ or $x< E_-$ we have, using \er{IMEP},
$I_q(x)=\inf\emptyset=\infty$. But using \er{IqC1} we also find $I_q(x)=\infty$ by an argument similar to the one between \er{thetwo} and \er{theone}. See  exercise.\QED
\end{itemize}
 \bex Show that for $x> E_+$ or $x< E_-$ the function 
\beq
t\mapsto tx-\til{\Pi}_q(t)=tx-\log\Sigma_a q_a e^{tE_a} \label{alsosup}
\eeq
 has no maximum \emph{and} that its supremum is infinite.
 \eex
To close this section we state a  general (though still one-dimensional) form of Cram\'{e}r's theorem:
\begin{theorem}\label{cramer}
Let $(X_n)$ be i.i.d.\ $\R$-valued random variables distributed by $Q\in\Pr(\R)$ such that
\begin{align}
\til{Z}_Q(t):=\la e^{tX_1}\ra_Q=\int_\R dQ(x)\, e^{tx}<\infty \hspace{10pt} (t\in\R); && \mu:=\la X_1\ra_Q<\infty.
 \label{Fq}
\end{align}
Then the sums $S_N:=\frac{1}{N}\Sigma_{n=0}^{N-1} X_n$ satisfy 
   \begin{align}
\lim_{N\raw\infty} \frac{1}{N} \log Q^N(S_N\geq x)=-I_Q(x) && (x\geq \mu), \label{79bis}
\end{align}
 where the rate function $I_Q$ is defined by
 \begin{align}
I_Q(x)&=\sup_{t\in\R}\{xt-\til{\Pi}_Q(t)\}; &&  \til{\Pi}_Q(t):=\log \til{Z}_Q(t).\label{655}
\end{align}
\end{theorem}
Compare with \er{79} and \er{tildeP}. Our previous version corresponds to the special case where each $X_n$ is given as $X_n:A^\N\raw \R$ with $X_n(s)=E(s_n)$ for some finite set $A$ and (injective) function $E:A\raw\R$, and $Q=q\circ E\inv$ for some $q\in\Pr(A)$.
By a theorem of Kolmogorov,\footnote{See e.g.\ Dudley (1989), Theorem 12.1.2 or  Klenke, 2020, Theorem 14.39.}  replacing $A$ by $\R$  (which is Polish) in this construction gives the general case (up to isomorphism in an appropriate sense), where $s:\N\raw\R$ is now  a real sequence.  The alternative form \er{IMEP} - \er{IMEP2} of the rate function is also valid in general. 
Theorem \ref{cramer} then follows from Sanov's theorem for Polish spaces plus Fenchel--Rockafeller duality as discussed above.\footnote{For various other proofs see Cerf \& Petit (2010), Rassoul-Agha \& Sepp\"{a}l\"{a}inen (2015), \S2.4, Den Hollander (2000), \S I.3, Olivieri \& Vares (2004), or Klenke (2020), \S23.1, Theorem 23.3.} Yet it is instructive to outline a proof for a special case  ``opposite'' to the case of finite $A$, namely where the support of $Q$ is all of $\R$.\footnote{In other words, the smallest interval $(a,b)\subset\R$ where $Q(a,b)=1$ is $\R$.} The simplest example is the standard Gaussian $dQ(x)=dx(2\pi)^{-1/2}e^{-x^2/2}$, for which \er{655} gives
\begin{equation}
I_Q(x)=x^2/2,
\end{equation}
so  that for $x>0$  the fluctuations $Q^N( S_N\geq x)$ exponentially decay like $\exp(-Nx^2 /2)$, i.e.,
\begin{equation}
\lim_{N\raw\infty} \frac{1}{N} \log Q^N(S_N\geq x)=-x^2/2.\label{LFG}
\end{equation}
\bex Prove Theorem \ref{cramer} directly for this case.\footnote{Use the fact that here $S_N$ is also normal, with mean $\mu=0$ but  variance $\sg=1/N$.}
\eex
On the other hand, the central limit theorem with $\mu=0$ and $\sg=1$, cf.\ \er{clt}, gives,
\begin{equation}
\lim_{N\raw\infty} Q^N\left( S_N\geq \frac{x}{\sqrt{N}}\right)=\frac{1}{\sqrt{2\pi}}\int_x^{\infty} dy\, e^{-y^2/2}, \label{clt2}
\end{equation}
which reinforces the idea that \er{LFG} with $x>0$ concerns \emph{large} fluctuations, that is, away from the mean $x=0$ by $O(1)$ in $N$, as opposed to \emph{small} fluctuations that are $O(1/\sqrt{N})$ away from the mean. The latter have a Gaussian distribution, whereas the former are typically  exponentially damped.
\smallskip

\emph{Proof of Theorem \ref{cramer}.\footnote{We follow Bucklew (1990), Chapter II.}}
First, without any assumptions on $Q$ we have the same  properties of $I_Q$ that were  discussed in Proposition \ref{prop61}, but now they have to be proved from \er{655}:  \begin{enumerate}
\item \emph{Positivity} of $I_Q$ follows from the fact that  $\til{\Pi}_Q(0)=\log 1=0$, so that $t\mapsto xt-\til{\Pi}_Q(t)$ assumes the value zero in any case, so that its supremum over $t$ must be $I_Q(x)\geq 0$.
\item \emph{Convexity} and \emph{lower semicontinuity}  of   $I_Q$ follow because these functions are
 \emph{defined} as Fenchel transforms, see Definition \ref{DefF} and following text (briefly: they are convex and lsc because they are suprema of affine--hence convex--and continuous--hence lsc--functions). 

\item Jensen for the convex function $x\mapsto e^{x}$ gives 
\beq
\la e^{tX_1}\ra_Q\geq e^{t\la X_1\ra_Q}=e^{t\mu},
\eeq
 and hence $\til{\Pi}_Q(t)\geq t\mu$ for all $t$, so that $\mu t-\til{\Pi}_Q(t)\leq 0$ and hence 
$I_Q(\mu)=\sup_t\{\mu t-\til{\Pi}_Q(t)\}\leq 0$. Hence $I_Q(\mu)=0$ by point 1. This local minimum is a global minimum since $I$ is convex, see Proposition \ref{PropA13}. It is also unique: the minimum value $I_Q(\mu)=0$ is assumed for 
$t=0$, and if (first) $x>\mu$, then since $\til{\Pi}_Q'(0)=\mu$ and $t\mapsto \til{\Pi}_Q(t)$ is continuous (even $C^1$),
for sufficiently small $t$ we have $\til{\Pi}_Q'(t)< x$. Hence if we define 
 \beq
 g_x(t):=xt-\til{\Pi}_Q(t),
 \eeq  we have  $g_x'(t)=x-\til{\Pi}_Q'(t)>0$ and so $g_x(t)>g_x(0)=0$. Consequently, $I_Q(x)=\sup_t g_x(t)>0$. Similarly, if $x<\mu$, then  $\til{\Pi}_Q'(t)> x$ so that $g_x'(t)<0$ for small $t$, so that again $g_x(t)>0$ and hence $I_Q(x)>0$. Thus
 \begin{align}
 I_Q(\mu)=0; && I_Q(x)>0\:\:\:\:\: (x\neq\mu).
 \end{align} 
 \item Convexity of $I_Q$ and uniqueness of its minimum at $x=\mu$ also yield property 4 of Proposition \ref{prop61}, that is, on its domain $I_Q$  is
  increasing if $x> \mu$ and  decreasing for $x< \mu$.
  \item The following property will also be used in the remainder of the proof:  defining
  \begin{equation}
 I^+_Q(x):= \sup_{t\geq 0}\{xt-\til{\Pi}_Q(t)\},
\end{equation}
  for $x\geq \mu$ we have
  \begin{align}
  I_Q(x)= I^+_Q(x) && (x\geq \mu). \label{659}
\end{align}
that is, one may take the supremum in \er{655} over $t\geq 0$ rather than $t\in\R$. Recall from point 1 that  $\til{Z}_Q(t)\geq e^{t\mu}$, so that
 for $t\leq 0$ and $x\geq\mu$ one has  $\til{Z}_Q(t)\geq e^{tx}$ and hence $xt-\til{\Pi}_Q(t)\leq 0$. For $t=0$ one has $x\cdot 0-\til{Z}_Q(0)=0$. Hence for $x\geq\mu$ the supremum $\sup_{t\in\R}$ in \er{655} must be reached for $t\geq 0$, so that $I_Q(x)=I^+_Q(x)$ for $x\geq\mu$.  A similar argument shows that  for $x\leq\mu$ the supremum $\sup_{t\in\R}$ in \er{655} must be reached for $t\leq 0$, and if we also define
   \begin{equation}
I^-_Q(x):= \sup_{t\leq 0}\{xt-\til{\Pi}_Q(t)\},
\end{equation}
 then the proof will show that for any $x\in\R$ one has
 \begin{align}
 \lim_{N\raw\infty} \frac{1}{N} \log Q^{N}(S_N\geq x)&=-I^+_Q(x);\\
  \lim_{N\raw\infty} \frac{1}{N} \log Q^{N}(S_N\leq x)&=-I^-_Q(x).
 \label{78}\end{align}
\item  Though not necessary for the proof, the useful Fenchel inversion formula 
 \begin{equation}
\til{\Pi}_Q(t)=\sup_{t\in\R}\{xt-I_Q(x)\}
\end{equation}
follows from Theorem \ref{FDT} provided we can show that $\til{\Pi}_Q$ is convex and lsc, which we can:
convexity of $\log \til{Z}_Q$ follows from H\"{o}lder's inequality, as in Exercise \ref{exHol}, whilst
continuity of $\til{Z}_Q$ and hence of $\til{\Pi}_Q=\log \til{Z}_Q$ follows from Lebesgue monotone convergence.
 \end{enumerate}
 We  first prove the upper bound (without any assumption on $Q$). 
   \begin{align}
\lim\sup_{N\raw\infty} \frac{1}{N} \log Q^N(S_N\geq x)\leq -I^+_Q(x) && (x\geq \mu). \label{79tris}
\end{align}
By the exponential Chebyshev inequality \er{CMcor}, for $x\in\R$ and $t\geq 0$ we have, as the $X_n$ are i.i.d., 
\begin{equation}
Q^N(S_N\geq x)\leq e^{-tx}\la e^{tS_N}\ra_{Q^N}=e^{-tx} \left\langle e^{tX_1/N}\right\rangle_Q^N=
e^{-tx}(\til{Z}_Q(t/N))^N,
\end{equation}
so that, replacing $t\geq 0$ by $t/N$ and using $x\geq\mu$ and then \er{659} in the last step, 
\begin{equation}
\frac{1}{N} \log Q^{N}(S_N\geq x)\leq -(tx-\til{\Pi}_Q(t))\leq I_Q^+(x)=I_Q(x). 
\end{equation}
The proof of the corresponding lower bound
   \begin{align}
\lim\inf_{N\raw\infty} \frac{1}{N} \log Q^N(S_N\geq x)\geq -I_Q(x) && (x\geq \mu), \label{79q}
\end{align}
is much more difficult. It is here that we use 
the simplification brought about by the assumption that the support of $Q$ is $\R$: in that case (see exercise) the supremum in \er{655} is  attained, i.e.,
\begin{equation}
I_Q(y)=t_yy-\til{\Pi}_Q(t_y),\label{ty}
\end{equation}
for some $t_y$, and if $y\geq \mu$, then $t_y\geq 0$ by \er{659}. 
\bex
Prove that if $\mathrm{supp}(Q)=\R$, then \er{ty} is the case.
 \eex
This makes $I_Q$ finite on all of $\R$, and, by convexity, also continuous on $\R$ (see Proposition \ref{BV-2-1-12}). Now take $y>x$ and $\dl>0$ such that $x<y-\dl$. Trivially, 
  $Q^{N}(S_N\geq x)\geq Q^N(S_N\in (y-\dl,y+\dl))$. Hence \er{79q} follows if for any $y>\mu$ and $0<\dl<y-\mu$
  we can show that
   \begin{align}
\lim\inf_{N\raw\infty} \frac{1}{N} \log Q^N(S_N\in (y-\dl,y+\dl))\geq \til{\Pi}_Q(t_y)-t_yy.\label{79qrs}
\end{align}
This is done via new probability measures $Q_t$ on $\R$ defined by the Radon--Nikodym derivative
   \begin{align}
\frac{dQ_t(x)}{dQ(x)}=\frac{e^{tx}}{\til{Z}_Q(t)}. \label{670}
\end{align}
Note that $Q_0=Q$. 
Since $t_y$ in \er{ty} solves $y=\frac{d}{dt}\log\til{Z}_Q(t)_{|t=t_y}$, taking $t=t_y$ in \er{670}  we obtain
\begin{equation}
\int_{\R} dQ_{t_y}(x)\, x=\frac{1}{\til{Z}_Q(t_y)}\int_{\R} dQ(x)\, e^{tx}x=\frac{d}{dt} \log\til{Z}_Q(t)_{|t=t_y}=
y, \label{meanC}
\end{equation}
cf.\  \er{Fq}.
Using the random variable $Y(x)=x$ on $\R$ distributed by $Q_{t_y}$ and passing to the corresponding i.i.d.\ family $(Y_n)$, 
eq.\ \er{meanC} and the weak law of large numbers give, for any $\varep>0$, 
\begin{equation}
\lim_{N\raw\infty} \int_{R^N} dQ_{t_y}(x_0)\cdots dQ_{t_y}(x_{N-1}) 1_{|\frac{1}{N}\Sigma_{n=0}^{N-1}x_n-y|<\varep}=1,
\label{672}
\end{equation}
For $0<\varep<\dl$ we use that for any $t\geq0$, $|\frac{1}{N}\Sigma_{n=0}^{N-1}x_n-y|<\varep$ implies  $tN(y+\varep)\geq t\Sigma_{n=0}^{N-1}x_n$
and hence 
$e^{tN(y+\varep)}\geq e^{t\Sigma_{n=0}^{N-1}x_n}$, and hence $1\geq e^{-tN(y+\varep)}e^{t\Sigma_{n=0}^{N-1}x_n}$. We
 subsequently \er{670} to estimate 
\begin{align}
Q^N(S_N\in (y-\dl,y+\dl))&= \int_{\R^N} dQ(x_0)\cdots dQ(x_{N-1}) 1_{|\frac{1}{N}\Sigma_{n=0}^{N-1}x_n-y|<\dl}\nn\\
&\geq  \int_{\R^N} dQ(x_0)\cdots dQ(x_{N-1}) 1_{|\frac{1}{N}\Sigma_{n=0}^{N-1}x_n-y|<\varep}\nn\\
&\geq e^{-t_yN(y+\varep)}\int_{\R^N} dQ(x_0)\cdots dQ(x_{N-1}) e^{t_y\frac{1}{N}\Sigma_{n=0}^{N-1}x_n}1_{|\frac{1}{N}\Sigma_{n=0}^{N-1}x_n-y|<\varep}\nn \\
&= e^{-t_yN(y+\varep)}\til{Z}_Q(t_y)^N
\int_{\R^N} dQ_{t_y}(x_0)\cdots dQ_{t_y}(x_{N-1})1_{|\frac{1}{N}\Sigma_{n=0}^{N-1}x_n-y|<\varep}.\label{673}
\end{align}
We then invoke  \er{672} to obtain the following estimate:
\beq
\lim\inf_{N\raw\infty} \frac{1}{N} \log Q^N(S_N\in (y-\dl,y+\dl))\geq \log\til{Z}_Q(t_y)-t_y(y+\varep).
\eeq
Letting $\varep\downarrow 0$  gives \er{79qrs}, and hence \er{79q} and hence, with \er{79tris}, Cram\'{e}r's theorem \er{79bis}. \QED
\section{Large deviations:  General theory}\label{LDGT}
The theorems of
 Sanov and Cram\'{e}r were the first \emph{large deviation principles} (\LDP). These involve:
 \begin{itemize}
\item  \emph{A sequence $(X_N)$ of random variables} $X_N\in\X$,  where $\X$ is some
some regular topological space, and $X_N$ is distributed by $P_N\in\Pr(\X)$.
This is short for $X_N:\Om\raw\X$ (or more generally:  $X_N:\Om_N\raw\X$), 
 where $\Om$ (or $\Om_N)$ carries a probability measure $\mathbb{P}$ (or  $\mathbb{P}_N$) so that for (measurable) $B\subset\X$ we then have $P_N(B)=\mathbb{P}(X_N\in B)$ (or  $P_N(B)=\mathbb{P}_N(X_N\in B)$). 
\begin{itemize}
\item For Sanov, $X_N$ is the empirical measure $L_N:A^\N\raw\Pr(A)$, see \er{LNbN}, and $\mathbb{P}=q^\N$
is the Bernoulli measure   on $\Om=A^\N$, given some prior $q\in\Pr(A)=\X$.
\item  For  Cram\'{e}r, $X_N$ is $S_N:A^\N\raw\R$, see \er{SsumE}, with distribution $P_N$, this time on $\X=\R$, similarly determined by the Bernoulli measure $q^\N$  on $A^\N$.
\end{itemize}
\item  \emph{A lower semicontinuous (but not necessarily convex) rate function} $I:\X\raw[0,\infty]$.\footnote{Lower semicontinuity can be imposed without loss of generality: if  $I:\X\raw[0,\infty]$ satisfies \er{GLDPlu}, then the maximal lsc function $\til{I}$ majorized by $I$, obtained or defined by
$\til{I}(x)=\lim\inf_{y\raw x} I(y)$, 
 also satisfies \er{GLDPlu}, see Rassoul-Agha \& Sepp\"{a}l\"{a}inen (2015), \S 2.2.
Convexity does not always hold, but in our examples it usually does.}  Lower semicontinuity implies that for any $s\in [0,\infty)$ the epigraph 
\beq
K_s:=\{x\in\X\mid I(x)\leq s\}
\eeq
 is closed; if it is also compact for all $s$, then $I$ is called \emph{tight} or \emph{good}.\footnote{$\X$ is usually Polish, in which case compact sets are closed, and hence a tight rate function is automatically lsc.}
\begin{itemize}
\item For Sanov this is the relative entropy \er{KL1} - \er{KL2}, or more generally \er{DefSq1} - \er{DefSq2}.
It is tight, which 
is easy for finite $A$, and in general  requires  functional analysis.\footnote{See supplement to \S\ref{Sanovchapter} for the topology on $\Pr(A)$. For finite $A$, the space $\Pr(A)\subset A^*$ is compact. For given prior $q\in\Pr(A)$ the domain $\mathcal{D}_{S_q}=\{p\in\Pr(A)\mid p\ll q\}$ is closed in $\Pr(A)$ and hence is also compact: for if $p_n\ll q$, then $q(a)=0$ enforces $p_n(a)=0$; if $p_n\raw p$ weakly this implies $p(a)=0$. Furthermore, by Proposition \ref{gibbsin}.2, the function
 $S_q$ is continuous on  its domain $\mathcal{D}_{S_q}$, so that the set $\{p\in\Pr(A)\mid S_q(p)\leq s\}$ is closed and hence compact. 
  This is also true in general (i.e.\ if $A$ is Polish), but with a very technical proof; see Dembo \& Zetouni, Lemma 6.2.12. The key result from functional analysis used in the proof is the Eberlein--Smulian theorem. }
\item For Cram\'{e}r this is the function defined by \er{IMEP} - \er{IMEP2}, which is  tight for finite $A$.\footnote{This follows from Theorem \ref{contraction} below. For general $A$ it is tight provided $\til{\Pi}_q(t)<\infty$ in a nbhd of $t=0$, see Rassoul-Agha \& Sepp\"{a}l\"{a}inen, p.\ 29 and  Dembo \& Zetouni, Lemma 2.2.5. }
\end{itemize}
\item \emph{Estimates on the asymptotic properties of $P_N$}, which may be given either in the form:\footnote{See Theorem \ref{sanov} for Sanov  and eqs.\ \er{LDPC1} - \er{LDPC2} for Cram\'{e}r.}
\begin{align}
\lim\sup_{N\raw\infty} \frac{1}{N} \log P_N(F)\leq -I(F) && (F\subset\X\: \mathrm{closed});\label{GLDPl}
\\
\lim\inf_{N\raw\infty} \frac{1}{N} \log P_N(U)\geq - I(U) && (U\subset\X\: \mathrm{open}),
\label{GLDPu}
\end{align} where for  $B\subset\X$ we define $I(B):=\inf_{x\in B}I(x)$;
or in the equivalent form, for any  $B\subset\X$:
\begin{align}
- I(\mathring{B}) \leq \lim\inf_{N\raw\infty} \frac{1}{N} \log P_N(B)
\leq  \lim\sup_{N\raw\infty} \frac{1}{N} \log P_N (B)\leq -I(B^-),  \label{GLDPlu}
\end{align}
where $\mathring{B}$ and $B^-$ are the interior and closure of $B$, respectively. Here $P_N(X_N\in B)\equiv P_N(B)$.\end{itemize}
\begin{definition}\label{defLDP}
If these bullets apply, we say that $(\X, X_N, P_N, I)$, or $(\X, P_N, I)$, satisfies an \LDP.
\end{definition}
\bex
Show that \ \er{GLDPl} - \er{GLDPu} and \er{GLDPlu} are equivalent.
\eex 
Further properties of the rate function $I$ like convexity of tightness will be mentioned explicitly.
 If
\begin{equation}
I(\mathring{B})=I(B^-),\label{6.4} 
\end{equation}
which is a condition on both $B$ and $I$, 
    eq.\ \er{GLDPlu} obviously implies the direct  estimate
\begin{equation}
\lim_{N\raw\infty} \frac{1}{N} \log P_N(B)= - I(B).\label{611G}
\end{equation}\vspace{-5mm}
\begin{proposition}
The rate function in an \LDP\ is unique, and is given by
\begin{equation}
I(x)=\sup_{U\in \mathcal{O}_x(\X)}\{-\lim\inf_{N\raw\infty} \frac{1}{N}\log P_N(U)\}, \label{86}
\end{equation}
where $\mathcal{O}_x(\X)$ is the set of open neighbourhoods of $x$. 
\end{proposition}
\emph{Proof.} 
Since $x\in U$, eq.\ \er{GLDPu} immediately gives 
\beq	
I(x)\geq\inf_{y\in U} I(y) =I(U)\geq -\lim\inf_{N\raw\infty} \frac{1}{N}\log P_N(U).
\eeq
Taking the supremum over all $U\in \mathcal{O}_x(\X)$ gives $\geq$ after $I(x)$ in \er{86}. 
For $\leq$, 
take $t<I(x)$ and
use Lemma \ref{lsclemma} to find  $V_0\in \mathcal{O}_x(\X)$ such that 
$t<I(y)$ for all $y\in V_0$. Since $\X$ is regular one can separate $x$ and the closed set $F=\{x\in\X\mid I(x)\leq t\}$ by open sets, say $V_1\ni x$ and $V_2\supset F$, respectively, i.e.\ $V_1\cap V_2=\emptyset$. Let $W=V_0\cap V_1$.
Now $V_2^c=\X\backslash V_2$ is closed and $W\subset V_2^c$, so also $W^-\subset V_2^c$. By construction
  $I(x)>t$ on $V_2^c$, and so   $I(x)>t$ on $W^-$. Hence \er{GLDPl} gives
  \begin{align}
\sup_{U\in \mathcal{O}_x(\X)}\left\{-\lim\inf_{N\raw\infty} \frac{1}{N}\log P_N(U)\right\}&\geq -\lim\inf_{N\raw\infty} \frac{1}{N}\log P_N(W)\nn \\ &\geq -\lim\inf_{N\raw\infty} \frac{1}{N}\log P_N(W^-)
\geq -\lim\sup_{N\raw\infty} \frac{1}{N}\log P_N(W^-)\nn \\
&\geq I(W^-) =\inf_{x\in W^-}I(x)\geq t.
\end{align}
Thus $y\geq t$ is true for all $t<s:=I(x)$. Hence $y\geq s$, which gives \er{86} with $\leq$ instead of $=$. 
\QED
\begin{proposition}
If $(\X, X_N, P_N, I)$ satisfies an \LDP\ with tight rate function $I$:
\begin{enumerate}
\item The zero set $I\inv(\{0\})\subset\X$ is compact and nonempty.
\item If $B^-\cap I\inv(\{0\})=\emptyset$ for some measurable set $B\subset\X$, then
\begin{equation}
\lim_{N\raw\infty} P_N(B)=0. \label{NB0}
\end{equation}
\end{enumerate}
\end{proposition}
\bex
Prove part 1.
\eex
\emph{Proof of part 2.} If $B^-\cap I\inv(\{0\})=\emptyset$, then $I(B^-)>0$: if $\inf_{x\in B^-}I(x)=0$, then this infimum would equal $\inf_{x\in B^-\cap K_s}I(x)$ for any $s>0$, and since $B^-\cap K_s$ is compact and $I$ is lsc, the infimum is attained at some $x\in B^-$, contradicting the  assumption on $B$. Then \er{GLDPlu} enforces \er{NB0}. \QED\smallskip

To go beyond the two cases we have discussed, which mathematically are based on i.i.d.\ variables and physically describe non-interacting particles, we now discuss some general techniques and results.
One of these, which implicitly was already used in deriving Cram\'{e}r's theorem from Sanov's, is the powerful \emph{contraction principle}, which creates a new \LDP\ from a given one.\footnote{We combine Dembo \& Zeitouni, \S4.2.1, and Rassoul-Agha \& Sepp\"{a}l\"{a}inen (2015), \S 3.1, who  give a more general result:  if $I$ is not tight 
$(\Y, f\circ X_N, P_N\circ f\inv, \til{J})$ satisfies an \LDP, where $\til{J}$ is  the maximal lsc function majorized by $J$. }
\begin{theorem}\label{contraction}
Let $f:\X\raw\Y$ be a continuous map between Hausdorff spaces, and suppose $(\X, X_N, P_N, I)$ satisfies an \LDP\ with tight rate function $I$. Define $Y_N=f\circ X_N$ as an $\Y$-valued random variable, with probability $Q_N=P_N\circ f\inv$, and define a  function $J:\Y\raw[0,\infty]$ by
\begin{align}
J(y):=\inf\{I(x)\mid x\in\X, f(x)=y\}\:\:\: (y\in\mathrm{range}(f)); && J(y)=\infty\:\:\: (y\notin\mathrm{range}(f)).
 \label{Jcon}
\end{align}
Then the push-forward data $(\Y, Y_N, Q_N, J)$ satisfies an \LDP\ with a tight rate function $J$.
\end{theorem}
\emph{Proof.} For the main idea, first assume \er{611G} and ignore topology and possible infinities.  As in the proof of Cram\'{e}r from Sanov, we take $C\subset \Y$ and find $B\subset\X$ such that
$Y_N\in C$ iff $X_N\in B$: indeed,
\begin{align}
B=\{x\in\X\mid f(x)\in C\}\equiv f\inv(C); && \Raw && Q_N(C)=P_N(B), && J(C)= I(B). \label{6.11}
\end{align}
The first two are trivial. For the last equality, compute
\begin{align}
I(B)&=\inf_{x\in B}\{I(x)\}=\inf\{I(x)\mid x\in \X,  f(x)\in C\}\nn \\ &=\inf\{I(x)\mid x\in \X,y\in C,f(x)=y\}\nn \\
&=\inf_{y\in C}\inf\{I(x)\mid x\in\X, f(x)=y\}\nn \\ &=\inf_{y\in C}\{J(y)\}=J(C).
\end{align}
Provided \er{6.4} implies $J(\mathring{C})=J(C^-)$, eqs.\ \er{611G} and \er{6.11}  then immediately give
\begin{equation}
\lim_{N\raw\infty} \frac{1}{N} \log Q_N(C)= 
\lim_{N\raw\infty} \frac{1}{N} \log P_N(B)= - I(B)=-J(C).
\end{equation}
More generally,
eqs.\ 
 \er{GLDPl} - \er{GLDPu} transfer from $(X_N)$ to $(Y_N)$ almost by definition, since continuity of $f$ guarantees that for $C\subset \Y$ open/closed also $B=f\inv(C)\subset\X$ is open/closed. For example,
 \begin{align}
\lim\sup_{N\raw\infty} \frac{1}{N} \log (P_N\circ f\inv (F'))\leq I(f\inv(F'))& =\inf_{x\in X\mid x\in f\inv(F')}I(x)=\inf_{x\in X\mid f(x)\in F'}I(x)\nn \\ &=\inf_{x\in X\mid f(x)=y, y\in F'}I(x)=\inf_{y\in F'}J(y)=J(F'),
\end{align}
where $F'\subset\Y$ is closed, and similarly for $U'\subset\Y$ open. Now, for $s\in\R$, we claim that
\beq
\{y\in\Y\mid J(y)\leq s\}=
f(\{x\in\X\mid I(x)\leq s\}) . \label{AisB}
\eeq
Assuming this for the moment, the argument of $f$ is a compact set because $I$ is tight. Since
the continuous image of a compact set is compact, eq.\ \er{AisB} makes the set on the left-hand side compact, so that $J$ is tight and hence also lsc. 
To prove \er{AisB} we write it as $A=B$. Then $y\in B$ iff there is $x\in\X$ such that $f(x)=y$ and $I(x)\leq s$, whereas $y\in A$ iff $\inf_{x\in\X\mid f(x)=y} I(x)\leq s$. Hence $B\subseteq A$.  For the converse inclusion, first note that  only those $y$ contribute to $A$ for which $J(y)<\infty$, so that $f\inv(\{y\})$ is non-empty, and closed by continuity of $f$ and the fact that singletons in Hausdorff spaces are closed.  Suppose $y\in A$ and $y\notin B$. Then $I(x)>s$ for all $x\in f\inv(\{y\})$. But consider 
 \beq
 K_n:=\{x\in f\inv(\{y\})\mid I(x)\leq s+1/n\}, \label{810}
 \eeq
  which is compact because $I$ is tight and $f\inv(\{y\})$ is closed. The assumptions $y\in A$ and $y\notin B$ imply that infinitely many $K_n$ are nonempty. Since the $K_n$ are nested, the intersection of the nonempty $K_n$ is not empty, so there is an $x\in f\inv(\{y\})$ with $I(x)\leq s$, contradicting $I(x)>s$. Hence $y\in A$ implies $y\in B$, i.e.\ $A\subseteq B$, and with the earlier $B\subseteq A$ we conclude that $A=B$, which is \er{AisB}. \QED
\smallskip

Let us rehearse how this applies to the theorems of Sanov and Cram\'{e}r: in the former (Theorem \ref{sanov}) we have $\X=\Pr(A)$ and $X_N=L_N$, and in the latter (Theorem \ref{Cr0}) we have $\Y=\R$ and $X_N=S_N$. For given $E:A\raw\R$, the function $f:\Pr(A)\raw\R$ is given by $f(p)=\la E\ra_p$.
\smallskip

Another important way to generate a new \LDP\ from an old one is \emph{Varadhan's theorem}:
\begin{theorem}\label{varadhan}
Let $(\X, X_N, P_N, I)$ satisfy an \LDP\ with $\X$ Polish and $I$ tight. Let $F:\X\raw\R$ be continuous
 and bounded from above.\footnote{Various (contrived) weaker conditions suffice, such as 
 $\lim\sup_{N\raw\infty} \frac{1}{N} \log \int_{\X}dP_N(x)e^{\gm NF(x)}<\infty$ for some $\gm>1$, or
 $\lim_{M\raw\infty} \lim\sup_{N\raw\infty} \frac{1}{N} \log \int_{F(x)\geq M}dP_N(x)e^{NF(x)}-\infty$
 (Rassoul-Agha \& Sepp\"{a}l\"{a}inen, \S 3.2; Dembo \& Zeitouni, \S4.3).}
 Then  the  perturbed probability measures  on $\X$ given by
\begin{align} 
dQ_N(x)=\frac{1}{Z_N} dP_N(x)\, e^{NF(x)}; && Z_N=\int_{\X}dP_N(x)e^{NF(x)}\label{815}
\end{align}
satisfy an \LDP\ on $\X$ with rate function $I_F:\X\raw[0,\infty]$ given by
\begin{equation}
I_F(x):=I(x)-F(x)-\inf_{y\in\X}\{I(y)-F(y)\}.
 \label{812}
\end{equation}
\end{theorem}
Note that the third term guarantees that $\inf_x\{I_F(x)\}=0$. If  the distribution $P_N$ of $X_N:\Omega_N\raw\X$ originates in some $\mathbb{P}_N\in\Pr(\Om_N)$ via $P_N=\mathbb{P}\circ X_N\inv$, as in all our examples, 
 then by definition
\beq
\la e^{N F}\ra_{P_N}=
\int_{\X}dP_N(x)\, e^{NF(x)}=\la e^{N F(X_N)}\ra_{\mathbb{P}_N}=\int_{\Om_N} d\mathbb{P}_N(\om)\, e^{N F(X_N(\om))}.
 \label{830a}
\eeq Part of the proof is the following lemma, which is a major result by itself (also due to Varadhan): 
 \begin{lemma} \label{Varlemma} In the setting of Theorem \ref{varadhan}
  we have ``Laplacian'' asymptotics
 \begin{equation}
\lim_{N\raw\infty} \frac{1}{N} \log \int_{\X}dP_N(x)e^{NF(x)}=\sup_{x\in\X}\{F(x)-I(x)\}. \label{811}
\end{equation}
\end{lemma}
Note that  the steepest descent method for fixed probability measures is a special case: if $P_N$ is independent of $N$ we have $I(x)=0$ and hence \er{811} reproduces the usual Laplacian asymptotics
\begin{equation}
\lim_{N\raw\infty} \frac{1}{N} \log \int_{\X}dP(x)e^{NF(x)}=\sup_{x\in\X}\{F(x)\}. \label{811b}
\end{equation}
Heuristically, eq.\ \er{811b} follows from the idea that as $N\raw\infty$ the integral is dominated by arbitrarily small regions $\Dl_i$ around the  values $x_i$ of $x$ for which $F(x_i)$ is largest and equal to $\sup F$. This
gives 
\beq
\int_{\X}dP(x)e^{NF(x)}\approx \Sigma_i P(\Dl_i)e^{N\sup F},
\eeq
Since $0<\Sigma_i P(\Dl_i)<1$ is just a constant, this gives \er{811b}.
 Similarly, to argue heuristically for \er{811} we use the \LDP\ satisfied by $P_N$, that is, $dP_N(x)\approx e^{-NI(x)}$ at least in small regions where $I(x)$ is close to its infimum in that region. This replaces $F(x)$ by $F(x)-I(x)$ in the above argument. Similarly, in \er{815} the large-$N$ asymptotics of $Q_N$ comes from those of $P_N$ corrected by $NF(x)$ so that $e^{-NI(x)}$ for $P_N$ is replaced by $e^{-N(I(x)-F(x))}$ times $1/Z_N$ for $Q_N$. Then, roughly speaking, 
 \beq
 \lim_{N\raw\infty} \frac{1}{N} \log Q_N(B)\approx -(I(B)-F(B))- \lim_{N\raw\infty} \frac{1}{N} \log Z_N.
 \eeq
 The first term on the right-hand side is $-I_F(B)$. The second follows from \er{811} and the obvious
 \begin{equation}
\sup_{x\in\X}\{F(x)-I(x)\}=-\inf_{y\in\X}\{I(y)-F(y)\}.
\end{equation}
\emph{Proof of Lemma \ref{Varlemma}.}
Assume $F\leq 0$ (if $F\leq K$, replace $F$ by $F-K$). For $i=1, \ldots, n^2$, define
\begin{equation}
G_i:=\left\{x\in\X\mid -\frac{i}{n}\leq F(x)\leq -\frac{i-1}{n}\right\}.
\end{equation}
Each $G_i$ is closed, $F$ varies at most by $1/n$ within each $G_i$, and $F\leq -n$ on $G=\X\backslash\cup_i G_i$. Thus
\begin{align}
\int_{\X}dP_N(x)e^{NF(x)}&\leq \Sigma_i P_N(G_i)e^{NF^*(G_i)}+P_N(G)e^{-nN}\leq n\max_i \{P_N(G_i)e^{NF^*(G_i)}\}+e^{-nN}\nn \\ &\leq
(n+1)\max\{ e^{-nN}, \max_i \{P_N(G_i)e^{NF^*(G_i)}\}\},
\end{align}
where $F^*(G_i):=\sup_{x\in G_i}F(x)$, analogous to   $I(B)\equiv I_*(B)=\inf_{x\in B}I(x)$.
Eq.\ \er{GLDPl} then gives
\begin{align}
\lim\sup_{N\raw\infty} \frac{1}{N} \log \int_{\X}dP_N(x)e^{NF(x)}
\leq \max\{-n,\sup_{x\in\X } \{F(x) -I(x) +1/n\}\}. \label{820}
\end{align}\vspace{-2.5mm}
\bex Prove this.\eex
Letting $N\raw\infty$ gives 
 \begin{equation}
\lim\sup_{N\raw\infty} \frac{1}{N} \log \int_{\X}dP_N(x)e^{NF(x)}\leq\sup_{x\in\X}\{F(x)-I(x)\}. \label{811u}
\end{equation}
To prove a lower bound, take $x_0\in\X$ and $\dl>0$ arbitrary, with ensuing open set
\begin{equation}
U_{\dl}(x_0):=\{x\in\X\mid F(x)>F(x_0)-\dl\}, 
\end{equation}
which is nonempty since $x_0$ lies in it. This time using \er{GLDPu}, we estimate
 \begin{align}
\lim\inf_{N\raw\infty} \frac{1}{N} \log \int_{\X}dP_N(x)e^{NF(x)}\geq  F(x_0)-I(x_0)-\dl. \label{825}
\end{align}
Since this true for any $x_0\in\X$ and $\dl>0$, we obtain
 \begin{equation}
\lim\inf_{N\raw\infty} \frac{1}{N} \log \int_{\X}dP_N(x)e^{NF(x)}\geq\sup_{x\in\X}\{F(x)-I(x)\}, \label{811l}
\end{equation}
which together with \er{811u} implies \er{811}. \QED\smallskip

 \noindent \emph{Proof of Theorem \ref{varadhan}.} A similar proof yields \er{812}: the  term $I(x)-F(x)$ comes from $P_N$ and $F(x)$, whilst the  supremum comes from $Z_N$, exactly as in \er{811}.\QED
\bex Prove \er{825}. 
\eex
In a  stronger version of Varadhan's theorem,\footnote{See e.g.\ Rassoul-Agha \& Sepp\"{a}l\"{a}inen (2015), \S 3.2.} the assumption that $F$ is bounded is replaced by the mere existence of some constant $\al>0$ such that
\begin{equation}
\sup_N\left( \int_{\X}dP_N(x)e^{\al NF(x)}\right)^{1/N}<\infty.
\end{equation}
This version has a corollary to the effect that under additional assumptions 
the rate function is a Fenchel transform. Assume that $\mathcal{Y}$ is a (real) topological vector space in separating duality with $\X$, and in \er{811} take $F(x)=\la y,x\ra$ for $y\in\mathcal{Y}$, where $\la\cdot,\cdot\ra$ is the pairing between $\X$ and $\mathcal{Y}$.
\begin{corollary}\label{RA425}
Let $(\X, X_N, P_N, I)$ satisfy an \LDP\ with $\X$ Polish and $I$ convex,\footnote{Recall that our rate functions are lsc by definition.}  and assume
\begin{equation}
\sup_N\left( \int_{\X}dP_N(x)e^{N\la y,x\ra}\right)^{1/N}<\infty,  \label{826N}
\end{equation}
for all $y\in\mathcal{Y}$. Then the \emph{pressure} 
\beq
\Pi: \mathcal{Y}\raw (-\infty,\infty],\eeq defined by
\begin{equation}
\Pi(y):=\lim_{N\raw\infty}\log\left(\left( \int_{\X}dP_N(x)e^{N\la y,x\ra}\right)^{1/N}\right)=
\lim_{N\raw\infty}  \frac{1}{N} \log \int_{\X}dP_N(x)e^{N\la y,x\ra},
\end{equation}
exists and is convex and lsc, 
and one has a Fenchel duality $I=\Pi^*$ and $\Pi=I^*$. This remains true if the \LDP\  holds on a closed convex subset $\mathcal{C}\subset \X$ and $I$ is extended to $\X$ by $I(x)=\infty$ on $\X\backslash\mathcal{C}$.
\end{corollary}
We will not prove this, but explain its use in our two pet cases (the second as an exercise). 

 First, for Sanov's theorem, as before take  $\X=\mathcal{M}(A)$, the space of signed measures on $A$ with subspace $\mathcal{C}=\Pr(A)$, and  $\mathcal{Y}=C_b(A)$. We take  $A$ finite, where this means, as before,
 \begin{align}
 \X=A^*=\{p:A\raw\R\}; && \Y=\{E:A\raw\R\}; && \la E, p\ra=\Sigma_{a\in A} p(a)E(a).
 \end{align}
 Moreover, $P_N$ is initially the probability measure on $\Pr(A) \subset \mathcal{M}(A)$ defined on  $B\subset \mathcal{M}(A)$ by
 \beq
P_N(B):=q^N\circ L_N\inv(B\cap\Pr(A)),
\eeq
so that $P_N(B)=0$ whenever $B\subset \mathcal{M}(A)\backslash\Pr(A)$, assuming that $B$ is weakly measurable. Then:
\begin{align}
 \int_{\X}dP_N(p)e^{N\la E,p\ra}&=\int_{\Pr(A)} dP_N(p)e^{N\Sigma_a p(a) E(a)}=\Sigma_{\sg\in A^N} q^N(\sg)e^{N\Sigma_a (L_N(\sg))(a)E(a)}\nn \\ &= \Sigma_{\sg\in A^N} q^N(\sg)e^{\Sigma_a\Sigma_{n=0}^{N-1}\dl_{\sg(n)a}E(a)}= \Sigma_{\sg\in A^N} q^N(\sg)e^{\Sigma_{n=0}^{N-1}E(\sg(n))}\nn \\ &=\Sigma_{\sg_0, \ldots, \sg_{N-1}\in A}
 \prod_{n=0}^{N-1} q(\sg(n)) e^{E(\sg(n))}=\left(\Sigma_{a\in A} q(a) e^{E(a)}\right)^N, \label{VarSan}
\end{align}
so that $\Pi(E)=\Pi_q(E)$, cf.\ \er{FandP}. Thus we obtain the rate function $I=S_q$ as a Fenchel transform systematically from Corollary \er{RA425}, rather than as a somewhat random fact. 
\bex
Compute the pressure $\Pi(y)$ for Cram\'{e}r's theorem, taking $\X=\mathcal{Y}=\R$, and rederive the rate function
$I_q$ from the above Corollary, cf.\ \er{IqC1}.
\eex

\emph{Bryc's theorem} is a converse to Varadhan's theorem. A sequence of probability measures $(P_N)$ on $\X$ is called  \emph{exponentially tight} if for each $0<s<\infty$ there a compact set $C_s\subset\X$ such that 
\begin{equation}
 \lim\sup_{N\raw\infty} \frac{1}{N} \log P_N(\X\backslash C_s)<-s.
\end{equation}
This suggests that for large $N$ the probability $P_N$ is concentrated on a compact set. It can be shown that if $\X$ is Polish and we have an \LDP\ with tight rate function,  then this function is automatically tight.\footnote{See Rassoul-Agha \& Sepp\"{a}l\"{a}inen (2015), Theorem 2.21}  Bryc's theorem works in the opposite direction:
\begin{theorem}\label{bryc}
Let $(P_N)$ be an exponentially tight sequence of probability measures on a Polish space $\X$. Suppose that for each $F\in C_b(\X)$ the following limit exist:
\begin{equation}
P(F):=\lim_{N\raw\infty} \frac{1}{N} \log \int_{\X}dP_N(x)e^{NF(x)}<\infty.
\end{equation}
Then $(\X,P_N,I)$ satisfies an \LDP\ with exponentially tight (and hence tight) rate function 
\begin{equation}
I(x):=\sup_{F\in C_b(\X)}\{F(x)-P(F)\}.\label{815t} 
\end{equation}
\end{theorem}
We will not use this result and  omit the proof.\footnote{See Rassoul-Agha \& Sepp\"{a}l\"{a}inen, \S3.3, or Dembo \& Zeitouni, \S4.4.}
If $F$ is a linear function (in which case it  cannot be bounded),  eq.\ \er{815t} looks like a Fenchel transform. This situation is covered by the important  \emph{G\"{a}rtner--Ellis theorem},  which similarly
gives existence of an \LDP\ \emph{assuming the existence of the partition function}.\footnote{
The first part below,  the usual version of the G\"{a}rtner--Ellis theorem, is stated here under
 relatively strong assumptions, following Ellis (1995), Theorem 5.1.  See Rassoul-Agha \& Sepp\"{a}l\"{a}inen (2015), \S 12.2, and  Dembo \& Zeitouni, \S 2.3 and \S 4.5.3, and den Hollander (2000), \S V.2, for versions under weaker assumptions.
The Cram\'{e}r-like second part
follows  Jak\v{s}i\'{c},   Ogata, Pillet, and  Seiringer (2012), \S2.2. Touchette (2009), Appendix C, gives interesting heuristics.
 } In statistical physics it allows interacting and correlated particles.

Let   $(P_N)$ be a sequence of probability measures on $\X=\R$, with ``partition function''
\begin{equation}
 Z_N(t) := \int_{\R}dP_N(x)e^{Ntx}; \label{PFGE}
\end{equation}
this function is lsc and convex (by H\"{o}lder's inequality) and takes values in $(-\infty,\infty]$. 
These measures usually  originate in a sequence of random variables $X_N:\Omega_N\raw\R$ on some sequence  $(\Om_N, \mathbb{P}_N)$ of probability spaces via $P_N=\mathbb{P}_N\circ X_N\inv$, in which case
\begin{align}
 Z_N(t)=\la e^{N tX_N}\ra_{\mathbb{P}_N}&=\int_{\Om_N}d\mathbb{P}_N(\om)\,  e^{NtX_N(\om)};   \label{830}\\
P_N(A)\equiv  P_N(x\in A)&=\mathbb{P}(X_N\in A)\equiv P_N(\{\om\in\Om\mid X_N(\om)\in A\}), \hspace{50pt} (A\subset\R).
  \label{830b}
\end{align}
\begin{theorem}[G\"{a}rtner--Ellis]\label{GE}
 In the  context of \er{PFGE} or \er{830} - \er{830b},
 suppose that
 \begin{align}
\Pi(t):=\lim_{N\raw\infty} \frac{1}{N} \log Z_N(t) \label{816}
\end{align} exists and is $C^1$ for all $t\in\R$. Then the $P_N$ or $X_N$ satisfy an \LDP\ with tight and convex rate function
\begin{align}
I(x):=\sup_{t\in\R}\{tx-\Pi(t)\}.
 \label{IGE1}\end{align}
 Moreover, noting that $\Pi'(0)=\lim_{N\raw\infty}\la X_N\ra_{\mathbb{P}_N}$,
 one has the special cases
 \begin{align}
\lim_{N\raw\infty}\frac{1}{N}\log \mathbb{P}(X_N\geq x)& =-I(x) && (x\geq \Pi'(0));\label{649}\\
\lim_{N\raw\infty}\frac{1}{N}\log \mathbb{P}(X_N\leq x)& =-I(x) && (x\leq \Pi'(0)).\label{650}
\end{align}
\end{theorem}
 Cram\'{e}r's theorem  \ref{Cr0} is  a special case of the G\"{a}rtner--Ellis theorem, where
 \begin{align}
 \Om_N&=A^N; \hspace{55pt} \mathbb{P}_N=q^N; \hspace{55pt}  X_N=S_N; \\
 Z_N(t)&=\Sigma_{\sg\in A^N} q^N(\sg)e^{NtS_N(\sg)}=
\Sigma_{\sg_0, \ldots, \sg_{N-1}\in A}
 \prod_{n=0}^{N-1} q(\sg(n)) e^{tE\sg(n)}=\left( \Sigma_{a\in A} q(a)e^{tE(a)}\right)^N \nn \\ &=
\la e^{tE}\ra_q^N,\label{9.34}
 \end{align}
  cf.\ \er{VarSan}. In that case, the ``pressure'' \er{816} trivially exists,  cf.\ \er{tildeP}:
 \begin{align}
\Pi(t):=\lim_{N\raw\infty} \frac{1}{N} \log Z_N(t)= \log \la e^{tE}\ra_q=\Pi_q(t).\label{816C}
\end{align} 
\bex Also rederive Sanov's theorem for finite $A$ from Theorem \ref{GE}.
\eex
\emph{Sketch of proof of Theorem \ref{GE}.}
Eq.\ \er{649} is not only equivalent \er{650}, but also to the  \LDP\ expressed in the main claim. 
On the one hand,  eq.\ \er{649}  follows from the  \LDP\  by the reasoning in point 3 after \er{Piprime}: for $x\geq \Pi'(0)$ the rate function $I(x)$ is increasing and hence the right-hand side of \er{649}, which by the LDP is $\inf_{y\geq x} I(x)$, equals $I(x)$.  Conversely, some lengthy topological reasoning reduces the general case to this special case.\footnote{See for example Bucklew (1990), Chapter II.B.} 
Thus we only need to prove \er{649}. 

Assuming $x\geq \Pi'(0))$, the upper bound
\beq \lim\sup_{N\raw\infty}\frac{1}{N}\log \mathbb{P}(X_N\geq x) \leq -I(x) \label{649ub}
\eeq
follows  as for Cram\'{e}r's theorem: for any $t\geq 0$ the exponential Chebyshev inequality \er{CMcor} gives
\begin{equation}
\frac{1}{N}\log \mathbb{P}(X_N\geq x)\leq -\frac{t}{N}+\frac{1}{N} \log Z_N(t/N),
\end{equation}
in which we replace $t/N$ by $t$ and the take the limit $N\raw\infty$, which by assumption exists and via \er{816} gives \er{649ub}. The lower bound
\beq \lim\inf_{N\raw\infty}\frac{1}{N}\log \mathbb{P}(X_N\geq x) \geq -I(x) \label{649lb}
\eeq
is also proved similarly to Cram\'{e}r's theorem: as in \er{670},  introduce  probability measures $P_{t,N}$ via
 \begin{align}
 \frac{dP_{t,N}}{dP_N(x)}(x):= \frac{e^{Ntx}}{Z_N(t)}. \end{align}
Differentiability of  $\Pi(t)$ is now used to obtain existence of the mean value of $X_N$ under $P_{t,N}$, i.e., 
\beq
\frac{d\Pi(t)}{dt}=\lim_{N\raw\infty}\int dP_{t,N}(x) x=\lim_{N\raw\infty}\la X_N\ra_{P_{t,N}}.\label{Pt1}
\eeq
Differentiability is also used to prove the counterpart of \er{ty} in the proof of Theorem \ref{cramer}. However,
since we are not in the i.i.d.\ case (that is the whole point of the G\"{a}rtner--Ellis theorem!) we do not have an immediate weak law of large numbers of the kind that states that for $\varep>0$,
\begin{equation}
\lim_{N\raw\infty}P_{t,N}(|X_N-\la X_N\ra_{P_{t,N}}|<\varep)= 1. \label{835}
\end{equation}
We now cheat a little bit and assume that $\Pi(t)$ is even \emph{twice} differentiable.\footnote{Some reasoning from convex analsis (= ``calculus without differentiability'') avoids this extra assumption. Bryc (1993) shows that if the assumption \er{816} holds for all \emph{complex} $t\in\C$ with $|t|<\varep$, for some $\varep>0$, 
and $\la X_N\ra=0$ for all $N$, 
then the random variables $\sqrt{N}X_N$ converge in law to a normally $N(0,\sg)$ distributed one, with $\sg^2=\Pi''(0)$.} In that case we have
\beq
\frac{d^2\Pi(t)}{dt^2}=\lim_{N\raw\infty}N(\la X_N^2\ra_{P_{t,N}}-\la X_N\ra_{P_{t,N}}^2)=\lim_{N\raw\infty}N\cdot \mathrm{Var}_{P_{t,N}}(X_N), \label{Ptt}
\eeq
so that the usual proof of the weak law of large numbers (rehearsed after Lemma \ref{CMlemma}) also applies here and yields \er{835}. Nonetheless, we have to overcome a problem with  the computation \er{673}: this essential step in the proof of Cram\'{e}r's theorem \ref{cramer} was based on the i.i.d.\ assumption. As a stopgap,  here  is  a ``formal'' (i.e., physicists') computation, which using  Laplacian asymptotics \`{a} la Theorem \ref{varadhan} can be made rigorous. This is based on a suggestive use of the asymptotics
 \begin{equation}
dP^t_N(x)\stackrel{N\raw\infty}{\approx}  dP_N(x)\cdot e^{N(tx-\Pi(t))},
\end{equation}
which in turn is based on the existence of \er{816}. We may then (``formally'' = informally) write
\begin{align}
P_N(X_N\in (y-\varep, y+\varep))&=\int_{y-\varep}^{y+\varep} dP_N(x)\nn \\ &= \int_{y-\varep}^{y+\varep} dP_N^t(x)\, \frac{dP_N^(x)}{dP_N^t(x)} \nn \\ &\approx\int_{y-\varep}^{y+\varep} dP_N^t(x)\, e^{-N(tx-\Pi(t))} \nn \\ &\approx  e^{-N(ty-\Pi(t))} P_N^t(X_N\in (y-\varep, y+\varep)).\label{836}
\end{align}
To proceed, here is a small intermezzo. If  $d\Pi(t)/dt=x$ has a solution $t=t_0$, then, since \er{Ptt} shows that $d^2\Pi(t)/dt^2\geq 0$, the function $t\mapsto tx-\Pi(t)$ has at least a local maximum at $t=t_0$; but since the function in question is concave,\footnote{The proof that $t\mapsto \Pi(t)$ is convex is practically the same as in  Theorem \ref{cramer}; the limit $N\raw\infty$ preserves convexity. }
 this maximum is even global, so that 
 \begin{equation}
t_0x-\Pi(t_0)=\sup_{t\in\R}\{tx-\Pi(t)\}=I(x). 
\end{equation}
  So if we now take $y=d\Pi(t)/dt$ in \er{836}, then using \er{Pt1} and \er{835} we obtain
 \begin{equation}
 \lim_{N\raw\infty} \frac{1}{N} \log P_N(X_N\in (y-\varep, y+\varep))=-I(x).
\end{equation}
The same arguments as in the proof of Theorem \ref{cramer} then yield \er{649lb}. \QED
\smallskip

The simplest application of the G\"{a}rtner--Ellis theorem beyond i.i.d.\ variables is provided by irreducible Markov chains,\footnote{Our presentation closely follows Dembo \& Zeitouni, \S3.1.} cf.\ Appendix \ref{MarC} for background. We realize a real-valued stochastic process $(X_n)$ with finite state space $A$ as in Theorem \ref{KolRT}, so that $\Om=A^\N$ and $X_n(s)=E(s(n))$ for some
(injective) ``energy'' function  $E:A\raw \R$ (one could also 
embed $A\subset\R$ straight away). Assume the $(X_n)$ form an irreducible stationary Markov chain
with corresponding irreducible stochastic matrix $(P_{ab})$  giving the transition probabilities, with Perron--Frobenius eigenvalue $\rh$ (see Theorem \ref{limlemma}). Then
\begin{equation}
P(t)_{ab}:= P_{ab}e^{t E(b)} \label{Pabt}
\end{equation}
is also an irreducible stochastic matrix, which  has a unique Perron--Frobenius eigenvalue $\rh(t)$. 
\begin{theorem}\label{CTMC}
The ``average energy'' $S_N:=\frac{1}{N}\Sigma_{n=1}^{N-1} X_n$ satisfies an \LDP\ with tight rate function
\begin{align}
I(x):=\sup_{t\in\R} \{xt- \log\rh(t)\}; && (I=(\log\rh)^*).\label{84000}
\end{align}
\end{theorem}
\emph{Proof.} By Theorem \ref{GE} we are ready if we can show that
\begin{equation}
\lim_{N\raw\infty} \frac{1}{N} \log \la e^{Nt S_N}\ra_P =\log\rh(t), 
\label{GEMC}
\end{equation}
where the expectation value is with respect to the probability measure $P$ on $A^{\N}$ defining the given Markov chain. Using \er{SMC} we may therefore compute 
\begin{align}
\la e^{Nt S_N}\ra_P=
\Sigma_{a_0}p(a_0) e^{tE(a_0)}P(t)^{N-1}_{a_0a_{N-1}}.
\end{align}\vspace{-5mm}
\bex Prove this.\eex
Eq.\ \er{GEMC} therefore follows from Theorem \ref{limlemma}, notably from \er{PFTlog}. Finally,
since again by Theorem \ref{limlemma} the special eigenvalue $\rh(t)$ is a non-degenerate root of the characteristic polynomial of $P(t)$, it is
 clear from \er{Pabt} that $\rh(t)$ and hence $\log \rh(t)$ exists for all $t\in\R$ and is $C^1$ in $t$.\QED 
 \bex
 Check that Theorem \ref{CTMC} implies Theorem \ref{cramer0}.
 \eex
 
 Theorem \ref{CTMC} is a generalization of Cram\'{e}r's theorem from i.i.d.\ variables to certain Markov chains. Sanov's theorem can equally well be generalized in the same direction, with $A$ finite:
 \begin{theorem}\label{SanovMC}
 Let $(A^\N, P)$ and the usual random variables $(X_n:A^\N\raw A)$ given by $x_n(s)=s_n$ define an an irreducible stationary Markov chain with transition probabilities $P_{ab}$. Then the  empirical measures $(L_N)$ satisfy an \LDP\ with tight rate function $I_P:\Pr(A)\raw[0,\infty]$ given by
 \begin{align}
I_P(p)&=\sup_{u\gg 0}\left\{\Sigma_{j=1}^d p_j\log\left(\frac{u_j}{(uP)_{j}}\right)\right\}\label{KLDAM}.
\end{align}
 \end{theorem}
 Here $u\gg 0$ means that $u_i>0$ for each $i$, ad we write $p_j=p(a_j)$ as well as $(uP)_j=\Sigma_i u_i P_{ij}$.\smallskip
 
 \emph{Proof.}
  For finite $A$ this can  be proved from  Theorem \ref{CTMC} (and hence ultimately from the G\"{a}rtner--Ellis theorem),  which we first need to generalize from $\R$ to $\R^d$; this is easily done with practically the same proof.  Instead of $E:A\raw\R$ we now work with $E:A\raw\R^d$,
 we have $x\in\R^d$ and $t\in\R^d$,  in \er{Pabt} we replace $tE_b$ by $\la t, E(b)\ra$,
where $\la\cdot,\cdot,\ra$ is the standard inner product in $\R^d$ (rather than some expectation value), and similarly, in \er{84000} we replace  $xt$ by $\la x,t\ra$. 
  We take $d=|A|$ and embed $\Pr(A)$ in $\R^d$  by arbitrarily ordering the elements $a\in A$ as $(a_1, \ldots, a_d)$ and hence identifying $p\in\Pr(A)$ with $(p(a_1), \ldots, p(a_d))\in\R^d$. 
  Then define 
  \begin{align}
  E(a)=\dl_a\in\Pr(A)\subset\R^d; && E(a_1)= (1, 0, \ldots, 0), \ldots, E(a_d)=(0, \ldots, 0,1),\label{Ehah}
  \end{align}
  that is, $E(a_i)$ is just the unit vector $e_i$ in $\R^d$; the labeling $a\in A$ has been changed to $i\in\{1,\ldots, d\}$.
  Hence
  $X_n:A^\N\raw\R^d$ is given by  $X_n(s)=\dl_{s(n)}$, so that $S_N=L_N$ with $L_N(s)\in\R^d$. 
 
 By  Theorem \ref{CTMC} and \er{Ehah}, which gives $\la t, E(a_j)\ra= t_j$, we  obtain an \LDP\ for large deviations from this average with tight rate function
   \begin{equation}
I_P(p)=\sup_{t\in\R^d}\{\la p,t\ra-\log\rh(t)\}, \label{SanMC}
\end{equation}
where $\rh(t)$ is the Perron--Frobenius eigenvalue of the matrix (where $i,j=1, \ldots, d=|A|$)
\beq
P(t)_{ij}=P_{ij}e^{t_j}.  \label{Pet} 
\eeq 
As an exercise, let us recover the rate function $I(p)=S_q(p)=S(p,q)$ in Theorem \ref{sanov}. In the i.i.d.\  case with prior $q\in\Pr(A)$, here given as $q\in\R^d$ via its components $q_i=q(a_i)$, we have 
\begin{align}
P_{ij}=q_j; && P(t)_{ij}=q_je^{t_j}.
\end{align} The second matrix has an eigenvector $u(t)$ with components $u_i(t)=q_ie^{t_i}$ (no sum) and eigenvalue
\beq
\lm(t)=\Sigma_i q_i e^{t_i}.
\eeq
Since $u\gg 0$ (in the sense that $u_i>0$ for each $i$), we can use \er{PFTlog} to compute $\rh(t)$, yielding 
\beq
\rh(t)=\lm(t).
\eeq
Our earlier result \er{FDD1}, proved around \er{willik}, then gives 
\beq
I_P=S_q.\eeq

To avoid confusion we denote $I_P$ as defined in \er{SanMC} by $I_P'$ and  show that $I_P=I_P'$.
For arbitrary but fixed $p$, we take an arbitrary  $u\gg 0$ and show that the point  $t\in\R^d$  defined by
 \begin{equation}
t_j=\log\left(\frac{u_j}{(uP)_j}\right) \label{tj}
\end{equation}
leads to $\rh(t)=1$ and hence $\log\rh(t)=0$ (to see that $t_j$ is well defined, a simple corollary of the definition of an irreducible stochastic matrix is that $u\gg 0$ implies $uP\gg 0$).
 This gives 
 \beq
 \la p,t\ra-\log\rh(t)=\la p,t\ra=\Sigma_{j=1}^d p_j\log\left(\frac{u_j}{(uP)_{j}}\right).
 \eeq
  Hence  \er{SanMC} gives $I_P'(p)\geq \Sigma_{j=1}^d p_j\log\left(\frac{u_j}{(uP)_{j}}\right)$, and since $u$ was arbitrary \er{KLDAM} implies 
\beq
I_P'(p)\geq I_P(p).
\eeq
Finally, $uP(t)=u$, as is easily checked from \er{tj}, and so using $v=u$ in \er{PFTlog} gives $\rh(t)=1$.

Conversely, fix $t\in\R^d$ and take $u\gg 0$ to be the left eigenvector of $P(t)$ (in Theorem \ref{limlemma}
$u$ is unfortunately called $p$). Using \er{KLDAM}, \er{Pet},  $uP(t)=\rh(t)u$, and $\Sigma_jp_j=1$, we  find
\begin{align}
I_P(p)&\geq \Sigma_{j=1}^d p_j\log\left(\frac{u_j}{(uP)_{j}}\right)=-\Sigma_{j=1}^d p_j\log\left(\frac{(uP)_{j}}{u_j}\right)
=\Sigma_l p_j t_j -\Sigma_{j=1}^d p_j\log\left(\frac{(uP(t))_{j}}{u_j}\right)\nn \\ &=
\la p,t\ra-\log\rh(t).
\end{align}
Since $t$ was arbitrary, we may take the supremum over all $t\in\R^d$ to obtain 
$I_P(p)\geq I'_P(p)$.\QED

We close this section with a powerful result due to Kifer (1990) that generalizes Sanov's theorem to  correlated (non i.i.d.) and interacting partcles, much as the G\"{a}rtner--Ellis did so for Cram\'{e}r's theorem. 
Here $(\Om,\F,Q)$ is a probability space, $X$ is a compact metric space, and 
\beq
Y_N: \Om\raw\Pr(X)
\eeq
 is a sequence of random variables. Here one may think of the following special cases:
\begin{itemize}
\item  $\Om=A^\N$, $X=A$, and $Y_N=L_N$ (the empirical measure), see \er{LNb};
\item $X=\Om$ with continuous (or   Borel) map $T:X\raw X$ and the \emph{occupational measure}
\beq
Y_N(x) :=\frac{1}{N}\Sigma_{n=0}^{N-1} \dl_{T^nx}.\label{occm}
\eeq
\end{itemize} 
 For any $E\in C(X)$ one then defines  the \emph{partition function} and \emph{pressure} by 
\begin{align}
Z_N(E)&:= \left\la e^{N \la E\ra_{Y_N}}\right\ra_Q=\int_{\Om}dQ(\om)\, e^{N \la E\ra_{Y_N}};\\
\Pi_Q(E)&:= \lim_{N\raw\infty}\frac{1}{N} \log Z_N(E), \label{9217}
\end{align}
where the existence of the limit must be assumed.
\begin{theorem}[Kifer] \label{KiT}
If the pressure \er{9217} exists and is finite for each $E\in C(X)$, then:
\begin{enumerate}
\item The map $E\mapsto \Pi(E)$ is convex and continuous from $C(X)$ with sup-norm to $\R$, and gives rise to a Fenchel dual pair (in which the conjugate  $I_Q:\Pr(X)\raw[0,\infty]$ is convex and lsc):
\begin{align}
I_Q(P)&=\sup_{E\in C(X)}\{ \la E\ra_P-\Pi_Q(E)\}; \label{Kifer1}\\
\Pi_Q(E)&=\sup_{P\in\Pr(X)}\{\la E\ra_P-I_Q(P)\}. \label{Kifer2}
\end{align}  
\item The \LDP\ upper bound \er{GLDPl} holds with rate function \er{Kifer1}:  for closed $F\subset \Pr(X)$,
\beq
\lim\sup_{N\raw\infty}\frac{1}{N} \log Q(Y_N\in F)\leq -I(F). \label{Kifer3}
\eeq
\item If for $E$ in a suitable dense subspace of $C(X)$ 
the supremum in \er{Kifer2} is attained by a \emph{unique} $P\in\Pr(X)$, then
the corresponding lower \LDP\ bound \er{GLDPu} also holds.\footnote{This dense subspace is the linear span of a countable set of functions $(f_n)$ in $C(X)$,  each with $\| f_n\|_{\infty}=1$. The ensuing  probability measure $P$ may be called an \emph{equilibrium state} for $f$.} 
\end{enumerate}
\end{theorem}
\emph{Partial proof.} We just prove the fairly easy upper bound.\footnote{We follow Kifer (1990). His proof of the lower bound is prohibitively difficult and unsuitable for this course.} The case $I(F)\leq 0$ is trivial,  since the left-hand side of \er{Kifer3} is $\leq 0$. Suppose $0<I(F)<\infty$. Pick
 $\varep>0$. Then
 \begin{equation}
F\subset \{P\in\Pr(X)\mid I_Q(P)> I_Q(F)-\varep\}=\bigcup_{E\in C(X)} \Gamma_{\varep}(E).\label{Kiferlastig}
\end{equation}\vspace{-5mm}
\bex Prove this.\eex
Hence the open sets $\Gamma_{\varep}(E)$ cover $F$, and since $F$ (being a closed subset of the weakly compact set $\Pr(X)$) is compact, it has a finite subcover $(\Gamma_{\varep}(E_a))_{a\in A}$. Using \er{Kiferlastig} and \er{CMineq} gives
\begin{align}
Q(Y_N\in F)&\leq \Sigma_{a\in A} Q(Z_n\in \Gamma_{\varep}(E_a))=\Sigma_{a\in A} Q(\{Y_N\in \Pr(X)\mid
\la E_a\ra_{Y_N}> \Pi_Q(E_a)+I_Q(F)-\varep\})\nn \\
&\leq \Sigma_{a\in A} e^{-N(\Pi_Q(E_a)+I_Q(F)-\varep)} \la e^{N \la E_a\ra_{Y_N}}\ra_Q=
e^{-N (I_Q(F)-\varep)} \Sigma_{a\in A} e^{-N\Pi_Q(E_a)}Z_N(E_a).
\end{align}
Taking $(1/N)$ times the logarithm of this inequality and taking $\lim\sup_{N\raw\infty}$ makes the very last two terms  cancel in view of \er{Kifer2}, leaving $-(I_Q(F)-\varep)$. Letting $\varep\raw 0$ then gives \er{Kifer3}. 

The case $I(F)=\infty$ is similar; instead of $\Gm_{\varep}(E)$ we now take 
\begin{equation}
\Gamma^N(E):=\{P\in\Pr(X)\mid \la E\ra_P-\Pi_Q(E)> N\},\label{9230}
\end{equation}
and likewise show that $F\subset \bigcup_{E\in C(X)} \Gamma^N(E)$. The same argument  gives
$-N$ instead of the previous $-(I_Q(F)-\varep)$, and so letting $N\raw\infty$ gives 
$\lim\sup_{N\raw\infty}\frac{1}{N} \log Q(Y_N\in F)\leq -\infty$, hence $=-\infty$.
 \QED\smallskip
 
Kifer also gives a   nice application of this result  to dynamical systems on compact Riemannian manifolds $(X,g)$,
 where $Q$ is the corresponding measure $dQ(x)=c\cdot d^nx \sqrt{g(x)}$, normalized to $Q(X)=1$, and $T$ (now at least $C^2$) is \emph{expanding} in the sense that there exists $\gm>0$ such that 
 \begin{equation}
\| D_xT^N\xi\|\geq \gm e^{\gm N}\|\xi\|,
\end{equation}
for all $\xi\in T_xM$, $x\in M$, and $N>0$, where $D_xT: T_xM\raw T_xM$ is the derivative of $T:M\raw M$. If $J(x)$ is the Jacobian of $D_xT$ and $H(X,P,T)$ is the Kolmogorov--Sinai entropy of $(X,T,P)$, see \S\ref{EDS} below, then the assumptions of Theorem \ref{KiT} are met, and the 
rate function \er{Kifer1} is given by
\begin{equation}
I_Q(P)=\la \log J\ra_P-H(X,P,T).
\end{equation}
 \section{Large deviations and statistical physics: Easy examples}\label{ASP}
 Statistical physics is one of the areas in which large deviation theory originated, and indeed it still provides interesting applications of this theory. 
 The simplest model of classical statistical physics is arguably a baby model of a \emph{paramagnet}, consisting of $N\raw\infty$ frozen non-interacting spin-$\half$ particles,\footnote{See Dorlas (2021), Chapters 19--21.} whose  microstates are $\sg\in 2^N$
 (the dimension of space does not matter). An external magnetic field splits the ground state energy (taken to be zero for simplicity) and hence gives rise to two energy levels $\pm\ep$ per particle. In the setting of 
 Cram\'{e}r's theorem,  see \S\ref{LDCT},  we take 
 \begin{align}
 A=\{0,1\}; && q=f; && E(0)=-\ep; && E(1)=\ep, \label{beginPM}
 \end{align}
 i.e.\ $q(0)=q(1)=\half$, so that in Proposition \ref{cramer0} and especially \er{tildeP} we have 
 \begin{align}
\tilde{Z}_f(t)=\half\left(e^{t\ep}+ e^{-t\ep}\right)=\cosh(t\ep); && \til{\Pi}_f(t)= \log \cosh\t\ep).
\end{align}
\bex 
Show from \er{IqC1}, quite similarly to the computation leading to \er{IFCT}, that 
\begin{align}
I_f(x)&= \sup_{t\in\R}\{tx-\log\cosh (t\ep)\} \nn \\
&=\half(1+x/\ep)\log (1+x/\ep)+\half (1-x/\ep)\log (1-x/\ep) & (x\in [-\ep,\ep]);\nn \\
&= \infty &  (x\notin [-\ep,\ep]).  \label{sprint}
\end{align}\eex
Changing $x$ to $u$ since it is an energy, the  corresponding (Clausius) entropy is now defined by
\begin{equation}
S_C(u):= -I_f(u) +\log 2,\label{9.4y}
\end{equation}
where a constant $\log|A|=\log 2$ has been added to make $S_C$ positive: 
 this corresponds to redefining the partition function for the flat prior by $\tilde{Z}_f(t)=\Sigma_{a\in A} e^{t E(a)}$, as is usual in physics. Thus we obtain 
\begin{equation}
S_C(u)= -\frac{1+(u/\ep)}{2} \log \left(\frac{1+(u/\ep)}{2}\right)-\frac{1-(u/\ep)}{2} \log \left(\frac{1-(u/\ep)}{2}\right). \label{Dore2} \end{equation}
This formula  is valid
\emph{verbatim} for $u\in (-\ep,\ep)$, and by taking limits also for $u=\pm \ep$; if $u\notin [-\ep,\ep]$, then $S_C(u)=-\infty$. For $\ep=1$ our entropy function looks like this:\footnote{Figure copied from Dorlas (2021), Figure 21.1, page 139.}
\begin{center}
 \includegraphics[width=0.6\textwidth]{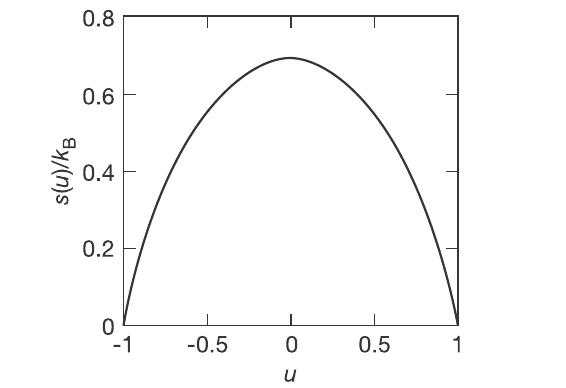}
\end{center}
 Bypassing large deviation theory--or perhaps justifying the identification of \er{9.4y} with entropy--eq.\ \er{Dore2} may also be derived directly from a  Boltzmannian conception of entropy, here defined as
\begin{align}
S_C(u)&:=\lim_{N\raw\infty}\frac{1}{N} \log |\{E_N=u_N)\}|; \label{105SM} \\
 \{E_N=u_N)\}&:=\{\sg\in 2^N\mid S_N(\sg)=u_N\}; && S_N(\sg)=\frac{1}{N} \Sigma_{n=0}^{N-1} E(\sg_n),
\end{align}
where  $(u_N)$ is a sequence converging to $u\in (-1,1)$ such that $u_N=S_N(\sg)$ for some $\sg\in A^N$, i.e., 
\beq
u_N=\ep\cdot\frac{2M-N}{N}\label{usubN},
\eeq
 for some $M=0, \ldots, N$, so that $M$ spins are in state $\sg=1$ with energy $+1$ and $N-M$ spins are in  $\sg=0$ with energy $-1$.  Hence
\begin{equation}
|\{E_N=u_N)\}|=
\left(
\begin{array}{c}
 N     \\
 M  
\end{array}
\right). \label{ENuN}
\end{equation}
\bex
Compute \er{105SM} from \er{ENuN}, and Stirling's approximation, and thus recover \er{Dore2}.
\eex
Eq.\ \er{Dore2} implies all thermodynamic properties of this baby paramagnet. In the absence of heat we  have $du=TdS$ and hence the (inverse) temperature $T$ is given by $T\inv=dS/du$, yielding
\begin{align}
\frac{1}{T}=\frac{1}{2\ep}\log\left(\frac{\ep-u}{\ep+u} \right)&&\LRaw && u=-\ep\tanh \left(\frac{\ep}{T} \right). \label{Tu}
\end{align}
The key macroscopic quantity is the \emph{magnetization} $m$ as a function of $T$. For finite $N$
one defines 
\begin{equation}
m_N(u_N)=-\frac{\mu}{\ep}u_N, \label{muu} 
\end{equation}
see \er{usubN}, where $\mu$ is the magnetic moment. upon which $m(T)$ is defined as $\lim_{N\raw\infty} m_N(u)$ with $u$ solved for $T$. Combining \er{Tu} and \er{muu} then gives the famous formula:\footnote{Note that negative energies and temperatures are allowed here. See Dorlas (2021), Remark 21.1.}
\begin{equation}
m(T)=\mu\tanh \left(\frac{\ep}{T} \right). \label{wnfm} 
\end{equation}

The above treatment corresponds to the \emph{microcanonical ensemble}, in which the energy is fixed. One can also use the \emph{canonical ensemble}, in which only the average energy is fixed.  Returning to general $A$ and $q\in\Pr(A)$, the  product probability  $q^N$ on $A^N$ is then replaced by a  distribution
\begin{align}
\mathbb{P}_{N,\beta}(\sg)=\frac{1}{Z_{N}(\beta)}e^{-\beta H_{N}(\sg)}; && Z_N(\beta):=\Sigma_{\sg\in A^N} e^{-\beta H_{N}(\sg)},  \label{10punt1}
\end{align}
 where $\beta>0$ and $H_N:A^N\raw \R$ is a function defined ``uniformly'' for each $N$. What this means depends on the kind of interaction, as explained below. In any case, the ``pressure'', defined as 
 \begin{align} 
p(\beta)=\lim_{N\raw\infty}\frac{1}{N} \log Z_{N}(\beta), \label{pinN}
\end{align}
should exist, and with it the ``free energy'' $f(\beta):=-p(\beta)/\beta$, so that, in our earlier sense of $\approx$,
\beq
Z_N(\beta)\approx e^{Np(\beta)}=e^{-N\beta f(\beta)}.\eeq
 In the simplest case, which encompasses both no interaction and so-called mean-field interactions, we take an injective function $E:A\raw\R$ as before, with associated average $S_N:A^N\raw \R$ defined by \er{SsumE}, and a further function $h: [E_-,E_+]\raw\R$, where $E_{\pm}$ were defined in \er{defEplusmin}. This $h$ is typically  the restriction of some polynomial  $h:\R\raw\R$ (abusing notation), and we finally put 
\begin{align}
H_N:A^N\raw \R: && H_N=Nh(S_N). \label{localh} 
\end{align}
The simplest nontrivial example is then given by
\begin{align}
h(x)=x; && h(S_N)=S_N; && H_N(\sg)=\Sigma_{n=0}^{N-1} E(\sg_n). \label{HNCr}
\end{align}
In this case the partition function and the pressure are easily computed (do it!) as
 \begin{align}
Z_N(\beta)=\left(\Sigma_{a\in A}e^{-\beta E(a)}\right)^N; &&  p(\beta)=\log\left(\Sigma_{a\in A}e^{-\beta E(a)}\right).
\label{ZNbeta1}
\end{align}
In this case the right-hand side of \er{pinN} is independent of $N$ and no limit is needed.

We know from Cram\'{e}r's theorem \ref{cramer0} how to compute the probability of the fluctuations of the energy $h(S_N)$ \emph{with respect to} $q^N$, in which case we also have laws of large numbers to the effect that $h(S_N) \raw \la h(S_N)\ra_q$ is various ways, at least for linear functions $h$. 
But in statistical physics we often need  the probability of the fluctuations of  $h(S_N)$ \emph{with respect to} $\mathbb{P}_{N,\beta}$, and as we shall see, already for quadratic functions $h$ there may not even be a straightforward large of large numbers.

 In the linear case \er{HNCr} there are two ways to proceed, of which the first one is a bit easier.
 \begin{enumerate}
\item We derive a \LDP\ from the G\"{a}rtner--Ellis theorem
 \ref{GE} with $d=1$ and  for $X_N=S_N$.
In \er{816} we substitute \er{830} with $\Om_N=A^N$ and $\mathbb{P}=\mathbb{P}_{N,\beta}$. In terms of \er{ZNbeta1} this gives 
\begin{align}
\Pi(t)&=\lim_{N\raw\infty} \frac{1}{N} \log 
\la e^{Nt X_N}\ra_{P_N}=\lim_{N\raw\infty} \frac{1}{N} \log 
\la e^{Nt S_N}\ra_{\mathbb{P}_{N,\beta}}\nn \\ &=\lim_{N\raw\infty} \frac{1}{N} \log  \left(\frac{1}{Z_N(\beta)}\Sigma_{\sg\in A^N}e^{-N(\beta-t)S_N(\sg)}   \right)\nn \\
&=p(\beta-t)-p(\beta).
\end{align}
Theorem \ref{GE} therefore applies: the random variables $X_N=S_N$  satisfy a \LDP\ with respect to $\mathbb{P}_{N,\beta}$ with convex (and tight)  rate function $I_E:\R\raw[0,\infty]$, given as a Fenchel transform 
\end{enumerate}
\begin{align}
I_E(u)&=\sup_{t\in\R}\{tu-\Pi(t)\}=  \sup_{t\in\R}\{tu-p(\beta-t)\}+p(\beta)\nn \\ &=\beta u+p(\beta)-\inf_{t\in\R}\{tu+p(t)\}.
\label{IEGE}
\end{align}
 \begin{enumerate}[resume]
\item Alternatively, we may combine  Cram\'{e}r's Theorem \ref{cramer0} and part 2 of Varadhan's Theorem \ref{varadhan}. In  Varadhan's theorem we take $P_N$ to be the probability measure on $\X=\R$ induced by
$S_N:A^N\raw\R$ from the flat prior $f^N$ on $A^N$, where $f(a)=1/|A|$ for each $a\in A$ as usual and hence $f^N(\sg)=|A|^{-N}$ for $\sg\in A^N$ (so that  Cram\'{e}r's Theorem applies), and also put
\beq
F(x)=-\beta x. \label{959}
\eeq 
Note that $S_N$ takes values in $[E_-,E_+]$, so that  $F:[E_-,E_+]\raw\R$, which is bounded. Hence Theorem \ref{varadhan} applies. The aim of the following exercise is to complete this approach by deriving the \LDP\ mentioned in approach 1 for the $S_N$ with respect to  $\mathbb{P}_{N,\beta}$.
\end{enumerate}
\bex 
\begin{enumerate}
\item Similarly  to \er{830}, show that
  the probability measure $Q_N$ on $\R$ defined in \er{815} with \er{959} equals the distribution of $S_N:A^N\raw\R$ induced by the probability measure $\mathbb{P}_{N,\beta}$ on $A^N$. 
  \item Show that \er{815} and  \er{10punt1} are related by
\begin{equation}
Z_N=|A|^{-N} Z_N(\beta).
\end{equation}
\item Show  that $\Pi_f(t)$ as defined in  \er{tildeP} with $q=f$ is related to $p(\beta)$ in \er{pinN} by
\beq \Pi_f(t)=p(-t)-\log|A|. \label{1010}
\eeq
\item According to Theorem \ref{cramer0} and especially \er{1010},  the  $(S_N)$ satisfy a \LDP\ with respect to $\mathbb{P}_N=f^N$ with rate function  $I_f(u)=\Pi_f^*(u)$.
Combine this with Theorem \ref{varadhan}.2, especially with \er{812} where $I=I_f$, to prove the  \LDP\ stated in  approach 1 above, with $I_F=I_E$ in \er{812} given by \er{IEGE}. 
 Hint: use  Fenchel duality \er{A19} to compute the  infimum in \er{812}.
\end{enumerate}
\eex
The equilibrium properties of the system are found by minimizing $I_E(u)$, which as we saw in \er{IEGE}, or indeed from the general setup, is a function of $\beta$. If $I_E$ is convex, as is the case here, it has a unique minimum at some energy $u=u(\beta)=-p'(\beta)$. The simplest physically relevant example is the model for paramagnetism explained above from \er{beginPM} onwards.
\bex
 Show that $p(t)=\log\cosh (t\ep)$ in this model, and consequently, that 
 \beq
 I_E(u)=I_f(u)+\beta u+p(\beta),
 \eeq where $I_f(u)$ is given in \er{sprint}.
Computing $u(\beta)=-p'(\beta)$, conclude that
 \begin{equation}
 u(\beta)=-\ep\tanh (\beta\ep).
\end{equation}
\eex 
Using \er{muu}, this recovers \er{wnfm}. Hence the two approaches discussed so far
are consistent; this is a special case of the \emph{equivalence of the micro-canonical ensemble and the canonical ensemble}.

\emph{Mean-field models} are defined by taking some polynomial $h:\R\raw\R$ in \er{localh}. Such models may also be defined on $\Z^d$, but they are insensitive to the dimension $d$ and may therefore as well be defined on $\Z$ or  $\N$. We only treat the case $A=\{0,1\}$ with $E(0)=-1$ and $E(1)=1$. It is then easier to work with $A=\{-1,1\}$ and $E(a)=a$. In that case, our $S_N$ in \er{SsumE} just comes down to 
\beq
S_N(\sg)= \frac{1}{N}\Sigma_{n=0}^{N-1}\sg_n.
\eeq 
The \emph{Curie--Weiss model}, the mean field model \emph{par excellence},  is defined by the function 
\begin{align}
 h(x)=-(\half Jx^2+Bx), \label{hCW} 
\end{align}
where $J>0$ and $B\in\R$ are constants. In other words, we have
\begin{align}
H_N(\sg) =-N(\half J S_N(\sg)^2+BS_N(\sg)) =-\frac{J}{2N}\Sigma_{m,n=0}^{N-1} \sg_m\sg_n-B\Sigma_{n=0}^{N-1} \sg_n. \label{921}
\end{align}
The Curie--Weiss model is an approximation to the Ising model, in which for any given site $x\in\Z^d$  the pair interaction of $\sg_x$ \emph{with its nearest neighbours} is approximated  by the
an interaction with the ``block spin'' $S_N$ that averages the value of \emph{all other spins}. We now need a key concept.
\begin{definition}\label{states}
\begin{enumerate}
\item A \emph{state} on a configuration space $A^N$ is  probability distribution  on $A^N$. 
\item A \emph{Gibbs state} on $A^N$ at inverse temperature $\beta$ with respect to a Hamiltonian $H_N:A^N\raw\R$ is a  probability distribution \er{10punt1}.
\item  A \emph{ground state} on $A^N$ with respect to a Hamiltonian $H_N$  is a minimizer $\sg\in A^N$ of $H_N$.
\end{enumerate}\end{definition}
The apparent discrepancy between point 3 and the other two is overcome by reinterpreting a ground state $\sg\in A^N$ as a point measure $\dl_{\sg}\in\Pr(A)$; we will often tacitly identify the two.\footnote{See \S\ref{FromCQ} for  details. 
The concept of a \emph{state} was arguably first introduced in the context of thermodynamics by Sadi Carnot; see Coopersmith (2015), Chapter 12.  It  generally denotes a specification of the ``relevant'' facts of the matter about some physical system at some given time (and hence the concept of a state is predicated on the concept of ``now'', mistaken as this may well be).
These facts ideally give a complete description of the system, such as the momentary positions and velocities (or momenta) of all particles in a gas, or the exact configuration $\sg\in A^N$. We earlier used the term \emph{microstates} for this,
which term now makes sense in the way explained just after Definition \ref{states} for ground states. 
Boltzmann and Gibbs began to see states as \emph{probability distributions} on the set of microstates of the system).} 
 Gibbs states are unique by definition (one needs the limit $N\raw\infty$ to get some physically relevant non-uniqueness), but ground states, henceforth denoted $\sg^{(0)}$,  may or may not be unique. This is already clear from the Hamiltonian  \er{921}.
For $B=0$,  either $\sg^{(0)}_n=1$ for all $n$, or  $\sg^{(0)}_n=-1$ for all $n$ are ground states; we denote these by $\sg_{\pm}^{(0)}$, respectively. 
 This is an example of \emph{spontaneous symmetry breaking} (\ssb), in that the $\Z_2$-symmetry of the the Hamiltonian $H_N$ given by $\sg\mapsto -\sg$ is not respected by the ground state; what does happen is that this symmetry maps  $\sg_{+}^{(0)}$ to $\sg_{-}^{(0)}$ and \emph{vice versa}. For $B\neq 0$ the spins align with the sign of $B$, i.e.\ the ground state is unique and given by $\sg_{+}^{(0)}$ if $B>0$ and by $\sg_{-}^{(0)}$ if $B<0$. In those cases the Hamiltonian itself  breaks the $\Z_2$ symmetry, which maps $S_N$ to $-S_N$  (this is called \emph{explicit} as opposed to \emph{spontaneous} symmetry breaking). 

Identifying
  $\sg_{\pm}$ with the point measures $\dl_{\sg_{\pm}}\equiv\dl_{\pm}$, 
  one easily  derives the (weak) limits
 \begin{align}
\lim_{\beta\raw\infty}\mathbb{P}_{N,\beta}&=\dl_0:= \half(\dl_++\dl_-) && (B=0);\label{h0limit}\\
\lim_{\beta\raw\infty}\mathbb{P}_{N,\beta}&=\dl_+ && (B>0);\\
\lim_{\beta\raw\infty}\mathbb{P}_{N,\beta}&=\dl_-&& (B<0); \\
\lim_{\beta\raw0}\mathbb{P}_{N,\beta}&=f^N && (\mbox{all cases}).
\label{hilimit}
\end{align}
Here $f$ is the flat prior on $\{-1,1\}$.
Neither the zero temperature  limit state \er{h0limit} nor its infinite $T$ counterpart  \er{hilimit}  breaks the $\Z_2$ symmetry; as a limit of the symmetric states $\mathbb{P}_{N,\beta}$ the former could not do so, and it is remarkable that even for $B\neq 0$ the latter even \emph{restores} the symmetry! 

What happens at $0<\beta<\infty$ and at $N\raw\infty$?
To answer this, we note  that $S_N$ is an \emph{order parameter} for the $\Z_2$ symmetry, in  that its expectation value vanishes in symmetric states and is non-zero  in asymmetric states. For example,
\begin{align}
\la S_N\ra_{\dl_{\pm}}=\pm 1; && \la S_N\ra_{\dl_0}=0; && \la S_N\ra_{P_{\Lm,\beta}}=0.
\end{align}
So we study the random variables $(S_N)$ as $N\raw\infty$, not even knowing if a law of large numbers applies! Fortunately, this is a clean application of Theorem \ref{varadhan}, which gives:
\begin{proposition}\label{rateS}
The random variables $(S_N)$ satisfy an \LDP\ with respect to the law induced by
$\mathbb{P}_{N,\beta}$ with Hamiltonian \er{921},
 with lsc (and tight) \emph{but not necessarily convex}  rate function
\begin{align}
I_{CW}(x)&=I_0(x)+\beta h(x)-C(\beta) \:\:\: (x\in [-1,1]); && I_{CW}(x)=\infty \:\:\: (x\notin [-1,1])\label{IFx} \\
I_0(x)&=\half((1-x)\log(1-x)+(1+x)\log(1+x)); &&
C(\beta)=\inf_{y\in\R}\{I_0(y)+\beta h(y)\}.\label{930}
\end{align}
Consequently, there are three cases, where $Q_N$ is the probability measure of $S_N$ on its range $[-1,1]\subset \R$ induced by the probability distribution $\mathbb{P}_{N,\beta}$ on $A^N$ (cf.\ Theorem \ref{varadhan}):\footnote{The limits \er{932} and \er{934} are taken in the weak topology of $\Pr(\R)$ with respect to the Borel structure. We state without proof that for $\beta\leq J\inv$ we also have $S_N\raw 0$  strongly with respect to the Gibbs measure on $A^\N$ induced by the finite-size probability measures $\mathbb{P}_{N,\beta}$. On the other hand, for $\beta>J\inv$ there is neither a weak nor a strong LLN!}
\begin{itemize}
\item  $B=0$ and $\beta\leq J\inv$ (i.e.\ $T\geq J$):  $I_{CW}$ has a unique minimum $I_{CW}(x_0)=0 $ at $x_0=0$,  and  
\beq
\lim_{N\raw\infty} Q_N= \dl_0.\label{932}
\eeq
this is equivalent to $S_N\raw 0$ as in the weak law of large numbers, i.e., for all $\ep>0$,
\begin{equation}
\lim_{N\raw\infty} Q_N(S_N\in (-\ep,\ep))=1.\label{933}
\end{equation}
\item $B=0$ and $\beta >J\inv$ (i.e.\ $T< J$):  the model has a phase transitions in the sense that the rate function $I_{CW}$ has two degenerate minima $I_{CW}(x_{\pm})=0$ at $x_{\pm}=m_{\pm}(\beta)\neq 0$, and hence 
\beq
\lim_{N\raw\infty} Q_N=\half( \dl_{m_+(\beta)} +\dl_{m_+(\beta)}). \label{934}
\eeq
\item $B\neq 0$ and arbitrary $\beta$ and $J$: same as the first case, except that now the
unique minimum $I_{CW}(x_0)=0 $ lies at some at $x_0=m(\beta, B)\neq 0$, so that 
\beq
\lim_{N\raw\infty} Q_N= \dl_{m(\beta, B)}.\eeq
\end{itemize}
\end{proposition} 
\emph{Proof.} We see from \er{10punt1} and \er{921} that $S_N$ is distributed according to $Q_N$ as defined in \er{815}, with $P_N$ the flat prior on $\{-1,1\}^N$ and $F(x)=-\half Jx^2-Bx$. Therefore, taking into account the switch from $A=\{0,1\}$ to $A=\{-1,1\}$, from \er{IFCTa} - \er{IFCT} and \er{812} we obtain \er{IFx} - \er{930}.
Tightness of the rate function $I_{CW}$ is clear from the fact that $I_0$ is lsc, $F$ is even continuous, and the domain $[-1,1]$ of $I_0$ and hence of $I_{CW}$ is compact (so that its closed subsets are compact). 
It is easiest to find the following  results by plotting $I_0$ and $F$, and using the fact that for $x\in(-1,1)$, 
\begin{align}
I_{CW}'(x)= \half\log\left(\frac{1+x}{1-x}\right)-\beta (Jx+B); &&
I_{CW}''(x)=\frac{1}{1-x^2}-\beta J.
\end{align}
Therefore,  $I_{CW}'(x)=0$ iff one of the most famous equations in mean field theory holds, namely
\begin{equation}
x=\tanh(\beta (Jx+B)).\label{famous}
\end{equation}
\begin{itemize}
\item For $0<\beta J\leq 1$ and any $B\in\R$, eq.\ \er{famous} has a unique solution  (which is $x=0$ for $B=0$ and whose sign otherwise equals the sign of $B$), which minimizes $I_{CW}$.  See figure below.\footnote{Copied from Rassoul-Agha \& Sepp\"{a}l\"{a}inen  (2015), page 47, Figure 3.1. Their $h$ is our $B$. }
\item For $\beta J>1$ and $B=0$ one finds two solutions $x_{\pm}$ of \er{famous}, related by $x_-=-x_+$, at each of which $I_{CW}''(x_{\pm})>0$, and hence $I_{CW}$ has two degenerate minima (and no maxima). 
\item
For $\beta J>1$ and $B\neq 0$ there may be up to three solutions  of \er{famous}; the number depends on the relative size of $J$ and $B$; e.g.\ for $B\ll J$ there are three zeros. But only the one with the same sign as $B$ (which always exists) corresponds to an absolute minimum of $I_{CW}$. 
\end{itemize}
\bex Prove these properties. \eex
\begin{center}
 \includegraphics[width=0.95\textwidth]{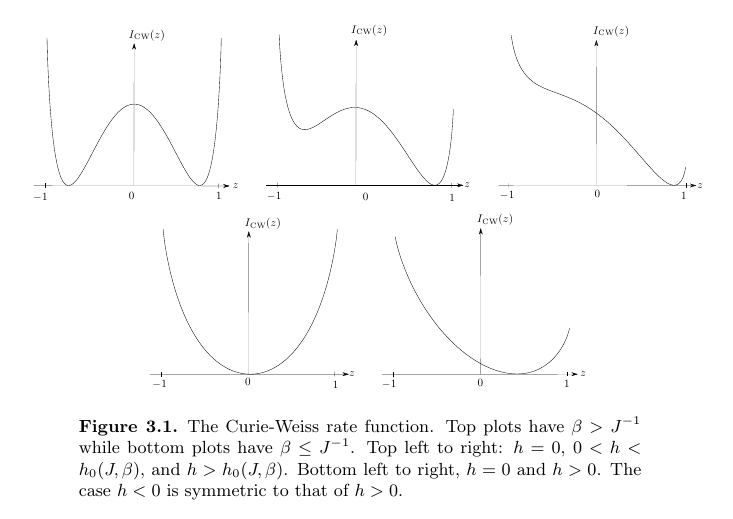}
\end{center}
The remaining claims then follow from the \LDP. In the first case (i.e.\ $B\neq 0$ etc.), eq.\ \er{GLDPl} gives
\begin{align}
\lim\sup_{N\raw\infty} \frac{1}{N} \log Q_N(S_N\notin (m(\beta, B)-\ep,m(\beta, B)+\ep))&\leq -I_0((-\infty, m(\beta, B)-\ep]\cup[  m(\beta, B)+\ep, \infty])\nn \\ & <0,
\end{align}
since $I_{CW}(m(\beta, B)=0$ is the unique minimum of $I_{CW}\geq 0$ and $I_{CW}$ is continuous on $[-1,1]$ (since both $I_0$ and $F$  are continuous). Hence $Q_N(S_N\notin (m(\beta, B)-\ep,m(\beta, B)+\ep))\raw 0$, which implies \er{933}.
The equivalence between \er{933} and \er{932} follows from the Portmanteau theorem.\footnote{See the supplement to \S\ref{Sanovchapter}.
This theorem applies, as $\partial((m(\beta, B)-\ep,m(\beta, B)+\ep))=\{m(\beta, B)-\ep, m(\beta, B)+\ep\}$, and since $\ep>0$ we have $\dl_{m(\beta, B)}(\{m(\beta, B)-\ep, m(\beta, B)+\ep\})=0$.}

In the second case (i.e.\ $B=0$ and $\beta J>1$), similar reasoning from  \er{GLDPl} gives 
\begin{equation}
\lim_{N\raw\infty} Q_N(S_N\in ((m_+(\beta)-\ep,m_+(\beta)+\ep)\cup (m_-(\beta)-\ep,m_-(\beta)+\ep)
))=1.\label{937}
\end{equation}
The Portmanteau theorem then gives weak convergence of $Q_N$ to some convex combination of the point measures $\dl_{m_{\pm}(\beta)}$; upon which the $Z_2$ symmetry of $Q_N$ forces \er{934}. \QED\smallskip

\noindent
 \emph{Note that  the rate function $I_{CW}$ is not convex if $J\beta>1$, i.e.,  when there is a phase transition.}\smallskip

Proposition \ref{rateS} describes fluctuations in the average magnetization $S_N$, which is enough to determine the phase structure of the model, as stated. It is also interesting (and  easy) to study the fluctuations of the energy $h(S_N)$ in the Curie--Weiss model, see \er{localh}. Both as a step towards this and as a goal in itself, we note that in our model the pressure defined in \er{pinN}  and hence the free energy exist and are computable via Theorem \ref{varadhan}. Writing $s_0=-I_0$, cf.\ \er{930}, we obtain
 \begin{equation}
p(\beta):=\lim_{N\raw\infty} \frac{1}{N} \log Z_N(\beta)=
-\inf_{x\in[-1,1]}\{I_0(x)+\beta h(x)\}=
\sup_{x\in[-1,1]}\{s_0(x)-\beta h(x)\}, \label{pCW}
\end{equation} where, according to  \er{ZNbeta1}, $Z_N$ is given by 
\begin{equation}
Z_N(\beta)=\Sigma_{\sg\in A^N}e^{-\beta Nh(S_N)}.\label{ZNbeta2}
\end{equation}
 Theorem \ref{contraction} now applies, because in our model the energy is a function of the spins whose rate function we  know. Since
 $p(\beta)$ is minus the constant $C$ in \er{930}, see also \er{812}, we obtain:
 \begin{proposition}\label{rateE}
The random variables $h(S_N)_N$ satisfy an \LDP\ as $N\raw\infty$, with tight rate function
\begin{equation}
I_E(u)=\inf_{x\in[-1,1]}\{I_0(x)+\beta h(x)\mid h(x)=u\}+p(\beta).\label{IECW2}
\end{equation}
\end{proposition} 
Hence for the corresponding entropy $s_E(u):=-I_E(u)$ we obtain
\begin{equation}
s_E(u)=\sup_{x\in[-1,1]}\{s_0(x)\mid h(x)=u\}-(\beta u+p(\beta)). \label{se1}
\end{equation}
As in the previous section, we may alternatively compute $I_E$ via Theorem \ref{GE}; this once agin gives \er{IEGE}, where this time $p(\beta)$ is given by \er{pCW}. 
\bex \label{Ex68}
Carry this out: show that the rate function computed from \er{812} equals the one obtained from 
\er{IGE1}; in other words, that \er{IEGE} with \er{pCW} coincides with \er{IECW2} with \er{930}.
\eex
\section{Gibbs measures}\label{GM}
The previous section saw  our first  encounter with the \emph{canonical ensemble} in classical statistical mechanics, which gives a systematic way to describe equilibrium states of classical spin systems (and also of continuous systems, which we will not discuss here) and the associated fluctuations of important physical quantities such as the energy. 
 The mathematical formalism for this is provided by the theory  of \emph{Gibbs measures}, which extends the construction \er{10punt1} to infinite $N$ (and more generally to infinite volume).\footnote{This theory 
 was developed in the 1960s by Dobrushin, Lanford, Ruelle, and others. See for example Simon (1993), Keller (1998), Ruelle (2004),  Georgii (2011), and Friedli \& Velenik (2018. For very useful summaries see also Georgii (1993) and Van Enter, Fern\'{a}ndez, \& Sokal (1993).  Georgii (2011) also describes the link with large deviations, as do Rassoul-Agha \& Sepp\"{a}l\"{a}inen (2015).  Most of these authors use the formalism of \emph{specifications}, which we avoid, at some cost to the generality of our treatment and at greater cost, unfortunately, to the proofs we  give. } The following two aspects of this theory may be separated:
\begin{enumerate}
\item The definition of equilibrium states in infinite volume, based on the assumption that in finite volume such states are described by the canonical ensemble.
\item The study of large deviations of (notably) the energy in such states (which replace Bernoulli measures in this respect), and the identification of the role of entropy and free energy. 
\end{enumerate}
We start with the first point. We often generalize $\N$ to $\Z^d$ (where $d=1,2,3,\ldots$), and in what follows $\Lm\subset\Z^d$ \emph{is always finite}. For our introductory treatment it is even enough to take one of:
\begin{align}
\N\supset \Lm&=\Lm_N:=\{0, \ldots, N-1\},\label{LmNmf} \\
\Z^d\supset\Lm&=\Lm_N:=\{x=(x_1, \ldots, x_d)\in\Z^d\mid |x_i|\leq N, i=1, \ldots, d\}, \label{rectZ} 
\end{align}
so that the number of points $|\Lm_N|$  in $\Lm_N$ is given by  $|\Lm_N|=N$ in the first case, and  in the second by 
\beq
|\Lm_N|=(2N+1)^d.
\eeq
For $\Z^d$, all large deviation theory so far then remains valid if we replace $N$ by $|\Lm_N|$ whenever some quantity is scaled by $N$ or $1/N$. 
With $A$ a finite set, as before, we often write \begin{align}
\Om:=A^{\Z^d}; && \Om_{\Lm}:=A^{\Lm}, \label{920}
\end{align} 
and similarly $\Om=A^\N$. Hence $\Om_{\Lm}$
 is finite and is equipped with the $\sg$-algebra $\mathcal{P}(\Om_{\Lm})$. Measure theory on $\Om$ is based on the \emph{cylindrical  $\sg$-algebra $\mathcal{F}$}. We give the definition for $\Z^d$, but the adaptation to $\N$ should be obvious.\footnote{See also Definition \ref{defF} and surrounding material.}
  For each $x\in\Z^d$, the evaluation map  $\om_x:\Om\raw A$ is given by
  \beq
  \om_x(s)= s(x).
  \eeq 
 \begin{definition}\label{defsigmaF} 
 Let $A$ be a finite set.\footnote{The definition is almost the same for Polish $A$, except that
 $\CP(A)$ below is replaced by the Borel $\sg$-algebra on $A$.}
 \begin{enumerate}
\item  $\mathcal{F}$ is
the  smallest $\sg$-algebra  $\Om$ for which the evaluation maps $\om_x$ are measurable (with respect to  the maximal $\sg$-algebra $\mathcal{P}(A)$ on $A$) for all $x\in\Z^d$.
\item For \emph{finite} $\Lm\subset\Z^d$, $\mathcal{F}_{\Lm}$ is
the  smallest $\sg$-algebra for which $\om_x$ is measurable  for all $x\in\Lm$.
\item For  \emph{finite} $\Lm\subset\Z^d$,  $\mathcal{T}_{\Lm}$ is the smallest $\sg$-algebra making all maps $\om_x$ with $x\notin\Lm$ measurable.
\item  The
  \emph{tail $\sg$-algebra} $\mathcal{T}$ is defined by
$\mathcal{T}:=\bigcap_{\Lm}\mathcal{T}_{\Lm}$ (intersection over all finite $\Lm$).
\end{enumerate}
\end{definition}
Equivalently, the ``local''  $\sg$-algebra $\mathcal{F}_{\Lm}$
 consist of all cylinder sets 
 \begin{align}
[\Dl]_{\Lm}:=\{\om\in\Om\mid \om_{|\Lm}\in\Dl\} &&  (\Dl\subset A^\Lm), \label{102}
 \end{align}
 where $\om_{|\Lm}: \Lm\raw A$ is the restriction of $\om:\Z^d\raw A$ to $\Lm$. Then $\mathcal{F}$ equals the
 \emph{quasi-local}   $\sg$-algebra 
generated by $\bigcup_{\Lm}\mathcal{F}_{\Lm}$, where $\Lm$ runs over all finite subsets of  $\Z^d$. 
For finite $A$    we have a bijection
  \begin{align}
 \mathcal{P}(A^{\Lm})\cong \mathcal{F}_{\Lm} && \Dl\lraw [\om_{|\Lm}\in\Dl].
  \end{align}
Hence $\mathcal{F}_{\Lm}$ is a finite $\sg$-algebra,  generated by ``atoms'' 
 (i.e.\ its smallest mutually disjoint elements)
  \begin{align}
[\sg]:=\{\om\in\Om\mid \om_{|\Lm}=\sg\}; &&  (\sg\in A^\Lm), \label{102a}
 \end{align}
in the sense that  $\mathcal{F}_{\Lm}$ consists of all unions and intersections of these atoms. A function $f:\Om\raw\R$ is $\mathcal{F}_{\Lm}$-measurable iff it is constant on each atom $[\sg]$, which means that $f(\om)$ is independent of the values of $\om:\Z^d\raw A$ outside $\Lm$. In that
case, for any (probability) measure on $\mathcal{F}_{\Lm}$ we have\footnote{This follows from integration theory. An integral
$\int_{\Om} dP(\om)\, f(\om)$ for some $\Sg$-measurable function $f:\Om\raw[0,\infty]$ is defined as the supremum 
over all finite sums  $\Sigma_i c_i P(B_i)$, where $0\leq \sum c_i 1_{B_i}\leq f$  and $B_i\in\Sg$, which may be taken to be disjoint. If $\Sg$ is finite,
then it contains disjoint atoms $A_i$ whose union is $\Om$. In that case $f=\Sigma_i c_i1_{A_1}$ exactly and hence $\int dP\,  f=\Sigma_i c_i P(A_i)$, where $c_i$ equals the constant value of $f$ on $A_i$. Note that any finite $\sg$-algebra is a power set.
 }
\beq
\int dP\,  f=\Sigma_{\sg\in A^{\Lm}} P([\sg]) f(\sg), \label{intsumP}
\eeq
 where we slightly incorrectly write $f(\sg)$ for the constant value $f(\om)$ at any $\om\in[\sg]$. 

A given physical theory is then specified by a family of (\emph{fixed})  local \emph{Hamiltonians} 
\beq
H_{\Lm}:A^{\Lm}\raw\R.
\eeq
It is desirable to derive these from some $\Lm$-independent object 
that specifies the dynamics. We have already seen this for mean-field theories. Another uniform way to write down a large class of Hamiltonians $H_{\Lm}$ for different regions $\Lm$ describes local interactions via so-called \emph{potentials}:\footnote{We (unusually) allow the empty set $M=\emptyset$, so that constants may be added to the Hamiltonian via $\Phi_{\emptyset}$.}
\begin{definition}\label{D91}
A \emph{potential} $\Phi=(\Phi_M)_M$ is a set of  $\mathcal{F}_M$-measurable functions 
\beq
\Phi_M:\Om\raw\R,\label{PhiR}
\eeq
 indexed by the set of all \emph{finite} subsets $M\subset\Z^d$, such that:
 \begin{enumerate}
 \item $\Phi_M$ only depends on $s_{|M}$ (so that we may also write $\Phi_M:A^M\raw\R$). 
\item  $\Phi$ is translation-invariant in the sense that 
\beq
\Phi_{M+x}=\Phi_M\circ \theta_x\eeq
 for each $x\in\Z^d$, where 
\begin{align}
\theta_x:\Z^d\raw\Z^d; &&
\theta_x(y)=x+y. \label{925}
\end{align}
\item $\Phi$ satisfies the summability condition
\begin{equation}
\|\Phi\|:=\Sigma_{0\in M\subset\Z^d} \sup_{\sg\in\Om}|\Phi_{M}(\sg)|<\infty.\label{sumPhi}
\end{equation}
\end{enumerate}\end{definition}
In many realistic models (like the Ising model) the sum in \er{sumPhi} only has a finite number of terms, so that the condition is automatically satisfied. Mean-field models have no potential.

In the simple case of  \emph{free boundary conditions}, the local Hamiltonians are then given by
\begin{align}
H_{\Lm}: A^{\Lm}\raw\R; && H_{\Lm}\equiv H^{\mathrm{free}}_{\Lm}=\Sigma_{M\subset\Lm}\Phi_M.\label{Hfree}
\end{align} Other boundary conditions will be studied later.
In view of \er{PhiR} it would  be more precise to write $H_{\Lm}:\Om\raw\R$ in \er{Hfree}, too, and then observe that because of condition (a) in Definition \ref{D91} indeed $H_{\Lm}$ only depends on $s_{|\Lm}$.
{The} archetypical example is the \emph{Ising model}, in which the spins take values in $A=\{-1,1\}$ and the potential is given by
\begin{align}
\Phi_{\{x\}}(\sg)&=-B\sg_x; \nn \\ \Phi_{\{x,y\}}(\sg)&=-J\sg_x\sg_y \:\:(\mbox{if } \|x-y\|_1=1); \nn \\
 \Phi_M&=0\:\:\mbox{(otherwise)}, \label{PhiIsing} 
\end{align}
where $B\in\R$ (representing a possible external magnetic field) and $J>0$  (in the ferromagnetic case) are constants, and $\|x-y\|_1=1$ means that $x$ and $y$ are ``nearest'' neighbours, denoted n.n., since the $\|\cdot\|_1$ norm on $\R^d\supset\Z^d$ is used (so ``diagonal'' neighbours are excluded). Thus
\begin{equation}
H_{\Lm}=-J\Sigma_{x,y\in\Lm,\, x,y\: \mathrm{n.n.}} \sg_x\sg_y-B \Sigma_{x\in\Lm}\sg_x. 
\end{equation}

In statistical mechanics these (\emph{fixed}) Hamiltonians are combined with a (\emph{variable})  \emph{temperature} 
\beq
k_BT=\beta\inv,
\eeq where $k_B$ is
Boltzmann's constant   taken to be unity in what follows, such that the Bernoulli  measures $q^{\Lm}$ on $A^{\Lm}$ (and in the limit on $A^\N$ etc.) are replaced by the  \emph{local Gibbs measures}
\begin{align}
P_{\Lm,q, \beta}(\{\sg\})&:=\frac{1}{Z_{\Lm}(\beta)}q^{\Lm}(\sg)e^{-\beta H_{\Lm}(\sg)}\:\:\: (\sg\in A^{\Lm});   \label{GML1aa}  \\ Z_{\Lm,q}(\beta)&:=\Sigma_{\sg\in A^{\Lm}} q^{\Lm}(\sg)e^{-\beta H_{\Lm}(\sg)}. \label{GML1a} 
 \end{align}
 The resemblance between \er{GML1aa} - \er{GML1a}  and Proposition \ref{gibbscon}, especially eq.\ \er{BoF}, is striking.\footnote{Dorlas (2021), Chapter 24, and others even justify the local Gibbs measures on this basis. Physically speaking Proposition \ref{gibbscon} describes non-interacting particles; a very large system of interacting particles may be divided into $N$  large subsystems whose components interact with each other whilst boundary interactions between the various subsystems may be neglected. Looking at each subsystem as a particle, eq.\   \er{BoF} suggests \er{GML1}.}
  We  also introduce the global \emph{pressure} $\Pi_{\Lm,q}(\beta)$ and \emph{free energy} $F_{\Lm,q}(\beta)$ by
\begin{align}
\Pi_{\Lm,q}(\beta) := \log Z_{\Lm,q}(\beta);&& F_{\Lm,q}(\beta):=-\beta\inv \Pi_{\Lm,q}(\beta),
\end{align}
so that
\begin{align}
Z_{\Lm,q}(\beta)=e^{-\beta F_{\Lm,q}(\beta)}.
\end{align}
In case of a flat prior $q=f$ on $A$ (i.e.\ $f(a)=1/|A|$ for each $a\in A$),  also  each microstate $\sg\in\Om_{\Lm}$ is  \emph{a priori} equiprobable, with probability
\beq
f^{\Lm}(\sg)=1/|\Om_{\Lm}|=|A|^{-|\Lm|}.  \label{936}
\eeq 
The Gibbs measure \er{GML1a} modifies this equiprobability via the Boltzmann factor  $e^{-\beta H_{\Lm}(\sg)}$, but at $\beta=0$, i.e.\ $T=\infty$, we recover \er{936}. Thus (high) temperature introduces (high)  randomness.\footnote{Dekkers \& Landsman (2025)  analyze this idea in detail in connection with chaos.}  It is customary to omit the factor $f^{\Lm}$ in the above expressions, so that, writing $P_{\Lm,q,\beta}=P_{\Lm,\beta}$ etc.,
 \begin{align}
P_{\Lm,\beta}(\{\sg\})&=\frac{1}{Z_{\Lm}(\beta)}e^{-\beta H_{\Lm}(\sg)}; \\  Z_{\Lm}(\beta)&=\Sigma_{\sg\in A^{\Lm}} e^{-\beta H_{\Lm}(\sg)}= e^{\Pi_{\Lm}(\beta)}=e^{-\beta F_{\Lm}(\beta)} \label{GML1} 
 \end{align}
 We recall that the relative entropy and the Shannon entropy, now indexed by $\Lm$, are defined by
 \begin{align}
 S_{\Lm}(P,Q)&=\Sigma_{\sg\in \Om_{\Lm}} P(\sg)\log\left(\frac{P(\sg)}{Q(\sg)}\right);\\
 S_{\Lm}(P)&=-\Sigma_{\sg\in \Om_{\Lm}} P(\sg)\log P(\sg), \label{SLmP}
 \end{align}
where $\Om_{\Lm}$ is finite. 
For the flat prior $Q=f^{\Lm}$ we infer from \er{936} that these  are related by
 \begin{equation}
S_{\Lm}(P,f^{\Lm})=-S_{\Lm}(P)+|\Lm|\log|A|. \label{940}
\end{equation}
The following result is a special case of \er{FDD2}, in which we replace $A$ by $A^{\Lm}$, $q$ by $q^{\Lm}$, and $f$ by $-\beta H_{\Lm}$, 
but it is instructive to state  and prove it directly:
 \begin{proposition} \label{P92} For any finite $\Lm\subset\Z^d$,   $H_{\Lm}:\Om_{\Lm}\raw\R$,   $\beta\in\R$, and  $q\in\Pr(A)$, we have
 \begin{equation}
\Pi_{\Lm,q}(\beta)=-\inf_{P\in\Pr(\Om_{\Lm})}\{\beta\la H_{\Lm}\ra_{P}+S(P,q^{\Lm})\},\label{941}
\end{equation}
where the infimum is a minimum that is uniquely attained at $P=P_{\Lm,q,\beta}$. The flat prior $q=f$ gives
 \begin{equation}
\Pi_{\Lm}(\beta)=
\sup_{P\in\Pr(\Om_{\Lm})}\{S_{\Lm}(P)-\beta\la H_{\Lm}\ra_{P} \},
\label{943}
\end{equation}
where the supremum is uniquely attained at the Gibbs measure $P=P_{\Lm,\beta}$, cf.\ \er{GML1}.\end{proposition}
 For $\beta>0$, eqs.\ \er{941} and \er{943} evidently give the  thermodynamical formulae ``$F=E-TS$'':
\begin{align}
F_{\Lm,q}(\beta)&=\inf_{P\in\Pr(\Om_{\Lm})}\{\la H_{\Lm}\ra_{P}+\beta\inv S(P, q^{\Lm})\};\\
F_{\Lm}(\beta)&= \inf_{P\in\Pr(\Om_{\Lm})}\{\la H_{\Lm}\ra_{P}-S_{\Lm}(P)\}.
\end{align}
Thus at nonzero temperature \emph{Gibbs measures minimize the free energy}, much as ground states (see below) minimize the energy at zero temperature.
\bex Prove Proposition \ref{P92}.\eex

As a special case Proposition \ref{gibbscon}, in which we simply replace $A$ by $A^{\Lm}$ and $E$ by $H_{\Lm}$, we see:
\begin{proposition}\label{P93}
For any 
\beq
U\in (\min_{\sg} H_{\Lm}(\sg), \max_{\sg} H_{\Lm}(\sg)),
\eeq there is a unique $\beta\in\R$ such that
\begin{align}
\la  H_{\Lm}\ra_{P_{\Lm,\beta}}&=U\label{947} ;\\
S_{\Lm}(P_{\Lm,\beta})&=\sup\{S_{\Lm}(P)\mid P\in\Pr(\Om_{\Lm}), \la  H_{\Lm}\ra_{P}=U\}.\label{948} 
\end{align}
\end{proposition}
\emph{Proof.}
Also here a direct proof is instructive (though  unnecessary). The function $\beta\mapsto
\Pi_{\Lm}(\beta)$ is smooth  (in the finite system at hand it is even analytic), with first and second derivatives
\begin{align}
\frac{d}{d\beta} \Pi_{\Lm}(\beta)=-
\la  H_{\Lm}\ra_{P_{\Lm,\beta}}; &&\frac{d^2}{d\beta} \Pi_{\Lm}(\beta)=\mathrm{Var}_{P_{\Lm,\beta}}(H_{\Lm})\geq 0,
\end{align}
 with equality near the end iff $H_{\Lm}$ is constant (in which case the proposition is trivially true). 
Thus the continuous function $\beta\mapsto \la H_{\Lm}\ra_{P_{\Lm,\beta}}$ is strictly decreasing. Hence 
\begin{align}
\lim_{\beta\raw-\infty}  \la H_{\Lm}\ra_{P_{\Lm,\beta}}=\max_{\sg} H_{\Lm}(\sg)); && \lim_{\beta\raw\infty}  \la H_{\Lm}\ra_{P_{\Lm,\beta}}=\min_{\sg} H_{\Lm}(\sg)),
\end{align}
from which the first claim follows. To prove \er{948}, we now fix $\beta$ at the value yielding \er{947} for given $U\in\R$, and 
use the variational principle \er{943} to estimate, for any $P$ with $\la  H_{\Lm}\ra_{P}=U$,
\begin{equation}
S_{\Lm}(P)-\beta U=S_{\Lm}(P)-\beta \la  H_{\Lm}\ra_{P}\leq S_{\Lm}(P_{\Lm,\beta})-\beta \la  H_{\Lm}\ra_{P_{\Lm,\beta}}= S_{\Lm}(P_{\Lm,\beta})-\beta U,
\end{equation}
 where in the last step we used \er{947}. Hence $S_{\Lm}(P)\leq S_{\Lm}(P_{\Lm,\beta})$, whence  \er{948}.\QED
\smallskip

The point is  to study the limit  $\Lm\raw\N$ or $\Lm\nearrow\Z^d$  in a suitable sense, which we here  take to be $\Lm=\Lm_N$ and $N\raw\infty$, hoping to see for example phase transitions  and other interesting  phenomena through lack of uniqueness of limiting measures. If these limits exist, a key role is played by
 \begin{align} 
p(\beta):=\lim_{N\raw\infty}\frac{1}{|\Lm_N|} \log Z_{\Lm_N}(\beta); && f(\beta):=
\lim_{N\raw\infty}\frac{1}{|\Lm_N|} F_{\Lm_N}=-\frac{p(\beta)}{\beta}, \label{pandf}
\end{align}
called the \emph{pressure} and the \emph{free energy}, respectively. It is hard to miss the formal analogy with (the G\"{a}rtner--Ellis) Theorem \ref{GE}, which indeed will be relevant to what follows.

We now define global Gibbs measures as extensions of the local expressions \er{GML1a} or \er{GML1} to $\Om=A^{\Z^d}$, where for simplicity we still assume that $A$ is finite (but the formalism is much more general).
 The main feature will be that such extensions are not necessarily unique. 
We first  include \emph{boundary conditions} in \er{Hfree} and  \er{GML1}. As before, $\Lm$ \emph{is  a finite sublattice of $\Z^d$}. 

Fix $\eta\in\Om$. For any $\Lm$ and $\sg\in A^{\Lm}$,  define $\om=\sg\eta\in\Om$ by $\om(x)=\sg(x)$ for $x\in\Lm$ and $\om(x)=\eta(x)$ for $x\notin\Lm$ (some people therefore write
$\sg\eta_{|\Lm^c}$ instead of $\sg\eta$; our notation is easier provided we only write $\sg\eta$ iff it is clear that
$\sg\in A^{\Lm}$ and we realize that \emph{within the expression} $\sg\eta$ we interpret $\eta$ as $\eta_{|\Lm^c}$).
Given a potential $\Phi$ (see Definition \ref{D91}) and some $\Lm$, extend \er{Hfree} to 
\begin{equation}
H^{\eta}_{\Lm}(\sg)=\Sigma_{M\cap\Lm\neq\emptyset}\Phi_M(\sg\eta), \label{Hgen}
\end{equation}
i.e.\ the sum includes all finite $M\subset\Z^d$ that overlap with the given $\Lm$.  For example, for the Ising model we may have $x\in\Lm$, $\|x-y\|_1=1$, but $y\notin\Lm$, so that \er{PhiIsing} includes terms $-J\sg_x\eta_y$, which are absent from \er{Hfree}. Here one may think of $\eta(x)=\pm 1$ for all $x$. Then introduce probabilities
 \begin{align}
P^{\eta}_{\Lm,\beta}(\sg)&:=\frac{1}{Z^{\eta}_{\Lm}(\beta)}e^{-\beta H^{\eta}_{\Lm}(\sg)}\:\:\: (\sg\in A^{\Lm}); \label{991a} \\
Z^{\eta}_{\Lm}(\beta)&=\Sigma_{\sg\in A^{\Lm}} e^{-\beta H^{\eta}_{\Lm}(\sg)}
=e^{-\beta F^{\eta}_{\Lm}(\beta)}; \label{991} \\ 
P^{\eta}_{\Lm,\beta}(B)&:= \Sigma_{\sg\in A^{\Lm}\mid \sg\eta\in B} P^{\eta}_{\Lm,\beta}(\sg) =
\frac{1}{Z^{\eta}_{\Lm}(\beta)}\Sigma_{\sg\in A^{\Lm}}1_B(\sg\eta)
e^{-\beta H^{\eta}_{\Lm}(\sg)}
\hspace{50pt} (B\in\mathcal{F}).
 \label{GMLeta} 
 \end{align}
 The expression $P^{\eta}_{\Lm,\beta}(\sg)$ only depends on $\sg\in A^{\Lm}$ and $\eta_{|\Lm^c}$
(with $\Lm^c:=\Z^d\backslash \Lm$), since $H_{\Lm}^{\eta}(\sg)$ and hence also
$Z_{\Lm}^{\eta}(\beta)$ only depend on $\eta_{|\Lm^c}$.
 For later use,  note that $P^{\om}_{\Lm,\beta}(\{\sg\eta\})$ equals $P^{\eta}_{\Lm,\beta}(\sg)$ iff $\eta=\om$ outside $\Lm$, and vanishes otherwise. Using \er{Hfree}, we  symbolically use ``$\eta=\mathrm{free}$'' to mean
   \begin{align}
P^{\mathrm{free}}_{\Lm,\beta}(\sg)&:=\frac{1}{Z^{\mathrm{free}}_{\Lm}(\beta)}e^{-\beta H^{\mathrm{free}}_{\Lm}(\sg)};   && Z^{\mathrm{free}}_{\Lm}(\beta)=\Sigma_{\sg\in A^{\Lm}} e^{-\beta H^{\mathrm{free}}_{\Lm}(\sg)}
.\label{991free}
  \end{align}
 In the following definition $\Phi$ and $\beta$ (which  is often absorbed into $\Phi$ or $H_{\Lm}$) are fixed (although the dependence on $\Phi$ is suppressed in the notation).\footnote{
 The tail $\sg$-algebra $\mathcal{T}_{\Lm}$ that will appear was defined in Definition \ref{defsigmaF}.
 If $\Sg_0\subset\Sg$ is a sub-$\sg$-algebra of some $\sg$-algebra $\Sg$ on $X$, and $f:X\raw \R$ is $\Sg$-measurable, then the \emph{conditional expectation} $E_P(f\mid \Sg_0)$ is the unique \emph{$\Sg_0$-measurable} function on $X$ that satisfies
$\int_B dP\, E_P(f\mid \Sg_0)=\int_B dP\, f$
for each $B\in\Sg_0$ (here uniqueness is up to other possibilities that coincide with some given choice $P$-a.e.).
Conceptually,  $E_P(f\mid \Sg_0)$ is a coarse-grained version of $f$.   Eq.\ \er{DLR1} or \er{DLR2} is called the \emph{\DLR-equation} (after Dobrushin, Lanford, and Ruelle, who proposed it).  The mere \emph{existence} of some probability measure 
$P_{\Lm^c}$ on $\Om_{\Lm^c}$ for which \er{DLR3} holds, for each finite $\Lm\subset\Z^d$ and $\sg\in \Om_{\Lm}$, is equivalent to the conditions in Definition \ref{D96}, see Ruelle (2004), \S 1.5.} 
 \begin{definition}\label{D96}
 A probability measure $P_{\beta}\in\Pr(\Om)$ is a \emph{Gibbs measure} at inverse temperature $\beta\in\R$ iff for all finite $\Lm\subset\Z^d$  and $B\in \mathcal{F}$ (i.e.\ the cylindrical $\sg$-algebra on $\Om=A^{\Z^d}$) we have
 \begin{equation}
P_{\beta}(B\mid \mathcal{T}_{\Lm})(\om)=P^{\om}_{\Lm,\beta}(B),\label{DLR1}
\end{equation}
for $P_{\beta}$-almost each $\om\in\Om$, or, equivalently (given the definition of a conditional expectation):
\begin{equation}
P_{\beta}(B)=\int_{\Om}dP_{\beta}(\om)\, P^{\om}_{\Lm,\beta}(B).\label{DLR2}
\end{equation}
 \end{definition}
This looks obscure. One simplification arises if we take $B=[\sg]_{\Lm}$ for some $\sg\in A^{\Lm}$. In that case the condition $\sg\om\in B$ holds for arbitrary $\om\in\Om$, and since for given  $\sg\in A^{\Lm}$ the expression
  $P^{\om}_{\Lm,\beta}([\sg])=P^{\om}_{\Lm,\beta}(\sg)$ only depends on $\om_{\Lm^c}$ (as already noted), eq.\ \er{DLR2} reads, with $P_{\Lm^c}=\pi_{\Lm^c}\inv P_{\beta})$,
  \begin{equation}
 P_{\beta}([\sg])=\pi_{\Lm}\inv P_{\beta}(\sg)=\int_{\Om_{\Lm^c}} dP_{\Lm^c} (\om) P^{\om}_{\Lm,\beta}(\sg).
 \label{DLR3} 
\end{equation}

Another special case that may clarify the meaning of \er{DLR2} is $B=\{\sg\eta\}$ for fixed $\sg\in \Om_{\Lm}$ and $\eta\in\Om$, but unfortunately, like for any point,  in that case $P_{\beta}(B)=0$. To remedy this, 
 we replace $\Z^d$ by a finite sublattice ${\Lm'}\subset\Z^d$ that contains $\Lm$ (i.e., $\Lm\subset {\Lm'}$),  and replace condition \er{DLR2}  by
 \begin{equation}
P_{\Lm'}(B)=\Sigma_{\om\in \Om_{\Lm'}} P_{\Lm'}(\om)P^{\om}_{\Lm,\beta}(B).\label{DLR2N}
\end{equation}
If we now put $B=\{\sg\eta\}$, where  $\eta\in \Om_{\Lm'}$, we see from \er{GMLeta} that only terms in which $\om_{|{\Lm'} \backslash\Lm}=\eta$ contribute to the sum in \er{DLR2N}, which therefore comes down to
\begin{equation}
P^{(\Lm')}_{\beta}(\sg\eta)=\Sigma_{\om\in \Om_{\Lm'}\mid \om_{|{\Lm'} \backslash\Lm}=\eta} P^{(\Lm')}_{\beta}(\om)P^{\eta}_{\Lm,\beta}(\sg).
\end{equation}
Using the elementary conditional expectation $P(B\mid C)=P(B\cap C)/P(C)$, this is the same as:\footnote{The left-hand side is short for $P^{(\Lm')}_{\beta}(B\mid C)$ with $B=\{\om\in\Om_{\Lm'}\mid \om_{|\Lm}=\sg\}$ and
$C=\{\om\in\Om_{\Lm'}\mid  \om_{|{\Lm'} \backslash\Lm}=\eta\}$.}
\begin{equation}
P^{(\Lm')}_{\beta}(\om_{|\Lm}=\sg\mid \om_{|{\Lm'} \backslash\Lm}=\eta)=P^{\eta}_{\Lm,\beta}(\sg).\label{Ruelle}
\end{equation}
 In words: \emph{the conditional probability that $\om$ coincides with $\sg$ on $\Lm$ given its value $\eta$ outside $\Lm$ equals the local Gibbs probability of $\sg$  for the Hamiltonian with boundary condition $\eta$. }
 \bex \label{Ruellex} 
 In the above situation:
 \begin{enumerate}
\item Using \er{Hgen}, where $H_{\Lm'}$  has free boundary conditions (so $\eta =$ ``free''), show that the difference $H_{\Lm'}(\sg\eta)-H_{\Lm}(\sg\eta)$ is independent of $\sg\in A^{\Lm}$ for any $\eta$.\footnote{Note that the  independence condition in point 1 is reasonable (except in mean-field models), since $H_{\Lm'}(\sg\eta)-H_{\Lm}(\sg\eta)$ contains the interactions between spins in $\Lm'\backslash\Lm$, which are given by $\eta$. }
\item Use this independence to show that   eq.\ \er{Ruelle} is solved by
 \begin{align}
P^{(\Lm')}_{\beta}(\om)=\frac{1}{Z_{\Lm'}}e^{-\beta H_{\Lm'}(\om)}\:\:\: (\om\in A^{\Lm'}); 
  &&
Z_{\Lm'}=\Sigma_{\om'\in A^{\Lm'}} e^{-\beta H_{\Lm'}(\om)}.
\end{align}
\item Now assume that $\Lm'\subset\Z^d$, and show that \er{Hgen} satisfies the independence condition in point 1 for any boundary condition $\eta'\in\Om$ (not to be confused with  $\eta$ in $\sg\eta$).
\item Use this fact to show that   eq.\ \er{Ruelle} is solved by
$P^{(\Lm')}_{\beta}= P_{\Lm',\beta}^{\eta'}$.
\end{enumerate}
\eex
Although \er{Ruelle} is only valid for \emph{finite} $\Lm'$ (as otherwise $P^{(\Lm')}_{\beta}(\om_{|{\Lm'} \backslash\Lm}=\eta)=0$),  the  point is that:
\begin{center} \emph{A global Gibbs measure is defined by specifying conditional probabilities for all finite sublattices $\Lm$ of $\Z^d$ and  boundary conditions outside these sublattices in the form of local Gibbs measures}. \end{center}
Another way to (re)write \er{DLR1} - \er{DLR2} is to realize that each $\Lm\subset\Z^d$ induces a factorization
\begin{align}
\Om\stackrel{\cong}{\longrightarrow}\Om_{\Lm}\x \Om_{\Lm^c}; && \om\mapsto(\om_{|\Lm},\om_{|\Lm^c}),
\end{align}
where $\Lm^c=\Z^d\backslash\Lm$. Any two measures $\mu_1, \mu_2$ on measure spaces $X_1, X_2$
induce a product measure $\mu_1\x\mu_2$ on the Cartesian product $X_1\x X_2$ by extending $\mu_1\x\mu_2(B_1\x B_2):=\mu_1(B_1)\mu_2(B_2)$, where $B_i\subset X_i$ (assumed measurable).\footnote{See e.g.\ Dudley (1989), \S 4.4.} 
Any positive measurable  function $f: X_1\x X_2\raw\R^+$  induces a further measure $f\cdot  (\mu_1\x\mu_2)$ on $X_1\x X_2$. Take:
\begin{align}
 \Lm\subset\Z^d\:\: \mbox{finite}; && X_1=\Om_{\Lm}; && X_2=\Om_{\Lm^c}; && \mu_1=\dl_{\Lm},
 \end{align}
  i.e.,  the counting measure $\dl_{\Lm}(\sg)=1$ for each $\sg\in \Om_{\Lm}$),  $\mu_2=\pi_{\Lm^c}\inv P$ for some $P\in\Pr(\Om)$, where $\pi_{\Lm^c}:\Om\raw \Om_{\Lm^c}$ is projection onto the second coordinate, 
and finally 
\begin{align}
f=P_{\Lm,\beta}; && P_{\Lm,\beta}(\om)=P^{\om_{|\Lm^c}}_{\Lm,\beta}(\om_{|\Lm}),
\end{align}
cf.\ \er{991a} and subsequent comment. The \DLR-equation \er{DLR1} or \er{DLR2} then comes down to
\begin{equation}
P_{\beta}=P_{\Lm,\beta}\cdot(\dl_{\Lm}\x \pi_{\Lm^c}\inv P_{\beta}).
\end{equation}
\emph{Warning}:\footnote{We owe this comment to Friedli \& Velenik (2018), pp.\ 269--270.} The local Gibbs measures $P_{\Lm,\beta}^{\eta}$ are not the marginals $\pi_{\Lm}\inv P_{\beta}$ of some
global Gibbs measure $P_{\beta}$, not even for free boundary conditions. Neither are they marginals  of some other local Gibbs measure $P_{\Lm',\beta}^{\eta'}$ for some finite $\Lm'\supset\Lm$. Thus although \emph{in principle} a global Gibbs measure $P_{\beta}$ could be defined by its marginals, as  in  Kolmogorov's extension Lemma \ref{KolET}, \emph{in practice} such a definition is useless since these marginals can only be computed if we already know $P_{\beta}$. 

Having said this, local Gibbs measures do converge to global ones! Here is the situation:\footnote{See Ruelle (2004), Theorem 1.9; Georgii (2011), Theorems (14.5) and (14.15); 
Rassoul-Agha \& Sepp\"{a}l\"{a}inen  (2015), Theorem 7.24; 
 We use the weak topology on $\Pr(\Om)$, see Definition \ref{weak2c}.
 The Portmanteau theorem  is very useful here, since for $\Om=A^{\Z^d}$ we have $P_n\raw P$ weakly
 iff  $P_n(B)\raw P(B)$ for any cylinder sets $B\subset \Om$. Indeed, in the product topology on $\Om$ cylinder sets are clopen and hence satisfy $B^-\backslash \mathring{B}=\emptyset$, so that
$P(B^-\backslash \mathring{B})=0$ for any $P$. \label{PMT2}
 }
\begin{theorem}\label{5theorem}
Fix a potential $\Phi$ and an inverse temperature $\beta\in\R$. 
\begin{enumerate}
\item For any boundary condition $\eta\in\Om$, or for a free boundary condition in which $H_{\Lm}^{\eta}$ is given by \er{Hfree}, the sequence $(P^{\eta}_{\Lm_N,\beta})_N$ of local Gibbs measures on $\Om_{\Lm_N}$ has a  subsequence that converges to some global Gibbs measure $P_{\beta}$ in the following sense: for any finite $\Lm$ and from sufficiently large $N$ such that $\Lm\subset\Lm_N$ onwards we have, cf.\ \er{defpiGF},
\begin{equation}
 \pi_{\Lm}\inv P_{\beta}=\lim_{N\raw\infty} \pi_{\Lm_N\Lm}\inv P^{\eta}_{\Lm_N,\beta}.
\end{equation}
In particular, there exists a limit Gibbs state for free boundary conditions, denoted by $P_{\beta}^{\mathrm{free}}$.
\item The set $\mathcal{G}(\beta\Phi)$ of all (global) Gibbs states is the closed convex hull of these limit states.\footnote{Like $\Pr^T(X)$ for compact metrizable $X$, the set $\mathcal{G}(\beta\Phi)$  is  a \emph{simplex}, see Definition \ref{Choquet}. }
\item The set $\mathcal{G}(\beta\Phi)$ is a nonempty compact convex subspace of $\Pr(\Om)$. 
\item The extreme boundary $\partial_e \mathcal{G}(\beta\Phi)$ consists of all $P\in \mathcal{G}(\beta\Phi)$ that are trivial on $\mathcal{T}$. 
\item The extreme boundary $\partial_e (\mathcal{G}(\beta\Phi)\cap \Pr^{\Z^d}(\Om))$ consists of all 
$P\in \mathcal{G}(\beta\Phi)$ that are trivial on $\mathcal{I}$ (and hence are ergodic
 Gibbs measures); in other words, we have
 \begin{equation}
\partial_e (\mathcal{G}(\beta\Phi)\cap \Pr^{\Z^d}(\Om))=\mathcal{G}(\beta\Phi)\cap \partial_e\Pr^{\Z^d}(\Om).
\end{equation}
\end{enumerate}
\end{theorem}
Here $\mathcal{T}$ is the tail $\sg$-algebra, cf.\ Definition \ref{defsigmaF}. 
The invariant $\sg$-algebra $\mathcal{I}$ and the last claim are explained at the end of
 Appendix \ref{ErgApp} on ergodic theory (which is a key to probability theory). 
In particular,  the $\Z^d$-action \er{925} on $\Z^d$ by translation induces a $\Z^d$-action $\theta^*$ on $\Om=A^{\Z^d}$ by 
\begin{align}
\theta^*_x: \Om\raw\Om; &&
\theta^*_x(\om)_y=\om_{x+y} && \left(x\in\Z^d,\, \om\in\Om\right).\label{Zdaction}
\end{align}
We  say that $P\in\Pr(\Om)$ is \emph{translation-invariant} if $P(\theta^*_xB)=P(B)$ for all $B\in\mathcal{F}$ and $x\in\Z^d$, and generically write $\Pr^{\Z^d}(\Om)$ for the set of translation-invariant probability measures on $\Om$.
 
  Even for finite $A$,  the case $\Om=A^{\Z^d}$ is rich enough  to give rise to a strange structure of the compact convex set $K=\Pr^{\Z^d}(\Om)$ of invariant probability measures on $\Om$: it is always a \emph{Poulsen complex}, in the sense that $\partial_e K$ is (weakly) dense in $K$ (which for finite $\Om$ would be unheard of).\footnote{See Georgii (2011), Theorem (14.12), and Simon (2011), Example 9.7, reviewed below. See also Definition \ref{Choquet}.}
 
 Here is an example of this phenomenon, for $d=1$ and $A=\{0,1\}$, so that, seen as a dynamical system
 $(\Om,P,T)$, the map $T$ is the shift map $T=S$,  \er{ULS}, now defined for all $n\in\Z$. For some given $S$-invariant $P\in\Pr^S(\Om)$, we construct ergodic probability measures $P_N\in \partial_e \Pr^S(\Om)$ such that $P_N\raw P$ (weakly). This construction is based on writing $\Z$ as a disjoint union, for given $N>0$,
 \begin{align}
\Z=\bigsqcup_{k\in\Z} I_k; && I_k:= [k(2N+1)-N, k(2N+1)+N],
\end{align}
so that each blok $I_k$ in the partition has length $2N+1$. This in turn induces a decomposition
\begin{equation}
\{0,1\}^{\Z}=\times_{k\in\Z}\{0,1\}^{I_k}.
\end{equation}
Given $P\in\Pr^S(\Om)$, for fixed $N>0$ we then define $P_{\Lm_N}\in\Pr(\Om_{\Lm_N})$, with $\Lm_N=[-N,N]$ and $\Om_{\Lm}=A^{\Lm}$, as usual, i.e.\ by 
\beq
P_{\Lm_N}(\sg)=P([\sg])=P(\{\om\in\Om\mid \om_{|\Lm_N}=\sg_{|\Lm_N}\}),
\eeq for any $\sg\in\Om_{\Lm_N}$. Noting that $\Lm_N=I_0$ , we then transfer $P_{\Lm_N}$ to any  $\Om_{I_k}$ by translation:  define
\beq
P^{(k,N)}_{I_k}(\sg')=P_{I_0}(S^{k(2N+1)}\sg'),
\eeq
where  $\sg'\in\Om_{I_k}$ and hence $S^{k(2N+1)}\sg'\in \Om_{I_0}$. Finally, define our measures  $P_N$ on $\Om$ via
\begin{align}
P_N:=\frac{1}{2N+1}\Sigma_{n=-N}^N S^{-n}Q_N; && Q_N=\times_{k\in\Z}P^{(k,N)}_{I_k}.\label{9115}
\end{align}
The reason for this move is that the product measure $Q_N$ is not $S$-invariant, but merely periodic, in that $S^{2N+1}Q_N=Q_N$ by construction (translation by an  block $I_0$ of length $2N+1$). 
\bex Prove that
$P_N$ is $S$-invariant.
\eex
Since ``$\lim_{N\raw\infty}[-N,N]=\Z$'', it is natural that $P_N\raw P$.
A  proof uses the fact that $\cup_M\mathcal{F}_{\Lm_M}$ is dense in $C(\Om)$, and so to prove
$P_N\raw N$ weakly it suffices that $|P_N(f)-P(f)|\raw 0$ for all $M$ and all $f\in   \mathcal{F}_{\Lm_M}$ (whose elements are continuous!). Indeed, for $N>M$ one shows that
\begin{equation}
|P_N(f)-P(f)|\leq 2\cdot \frac{2M+1}{2N+1}\|f\|_{\infty}.
\end{equation}
 The proof that  $P_N$ is ergodic (cf.\ Definition \ref{defE1}) relies on a nice lemma (of we we only need half):
 \begin{lemma}\label{simonlemma}
 Let $(X,P,T)$ a dynamical system with $X$ compact and $T$ invertible,  and consider\begin{equation}
f_{\Lm_N}(x):=\frac{1}{2N+1}\Sigma_{n=-N}^N f(T^nx), \label{ETsum2}
\end{equation}
for $f\in L^2(X,P)$.
Then $P$ is ergodic iff for all $f$ in a dense subspace of $L^2(X,P)$ we have
\beq
\lim_{N\raw\infty}\la |f_{\Lm_N}|^2\ra_P=|\la f\ra_P|^2. \label{9118}
\eeq
  \end{lemma}
  \bex Prove this lemma, and then  check that \er{9118} with $P\leadsto P_N$ holds.
  \eex 
 Using \er{9115} then gives $\lim_{L\raw\infty} \la f_{\Lm_L}^2\ra_{P_N}=\la f\ra_{P_N}^2$, which is  \er{9118}. Since $M$ was arbitrary and $\cup_M\mathcal{F}_{\Lm_M}$ is dense in $L^2(\Om,P')$ for any probability measure $P'$ defined on the cylindrical $\sg$-algebra generated by the $\mathcal{F}_{\Lm_M}$,  $P_N$ is ergodic by Lemma \ref{simonlemma}. Since $P_N\raw P$ and $P\in\Pr^S(\{0,1\}^{\Z})$ was arbitrary, we have shown that the compact convex set
$\Pr^S(\{0,1\}^{\Z})$ is a Poulsen simplex!
 
 As in the local case, the unbiased Bernoulli measure $P=f^{\Z^d}$ is  a Gibbs measure at $\beta=0$ (i.e.\ $T=\infty$). To see this, take $\Lm'\subset\Z^d$ finite, and a generic cylinder set defined by  $\eta'\in A^{\Lm'}$:
\beq
B=[\eta']:=\{\om\in\Om\mid \om_{|\Lm'}=\eta'\}\in\mathcal{F}_{\Lm'}.
\eeq
Then on the \emph{Ansatz} $P_0=f^{\Z^d}$ 
the left-hand side of \er{DLR2} equals 
\beq
f^{\Z^d}([\eta'])=\prod_{x\in\Lm'}q(\eta'_x)=\prod_{x\in\Lm'}|A|\inv=|A|^{-|\Lm'|}.
\eeq
 Now we make a (logically unnecessary but pedagogically useful) case distinction:
\begin{enumerate}
\item If $\Lm\cap\Lm'=\emptyset$, then 
\beq
P^{\om}_{\Lm,0}([\eta'])=1,
\eeq
 since 
\beq
Z_{\Lm}^{\om}(0)=|A^{\Lm}|=|A|^{|\Lm|} \label{ZZZ}
\eeq
 in \er{991}, whilst the second sum over $\sg$ in \er{GMLeta} also includes all $\sg\in A^{\Lm}$. 
The right-hand side of \er{DLR2} therefore equals
\beq
\int_{\om_{\Lm'}=\eta'} df^{\Z^d}(\om)=f^{\Z^d}([\eta'])=|A|^{-|\Lm'|}.
\eeq
\item If $M:=\Lm\cap\Lm'\neq\emptyset$, then \er{ZZZ} of course remains true, but this time 
the second sum over $\sg$ in \er{GMLeta} is constrained to those $\sg\in A^{\Lm}$ that satisfy 
$\sg\om\in [\eta']$, which implies
\begin{align}
\sg_x=\eta'_x \:\:\: \:\:\: (x\in M); && \om_x=\eta'_x \:\:\: \:\:\: (x\in \Lm'\backslash M).\label{9104}
\end{align}
 By the first equation  the second sum  in \er{GMLeta} has  only $|A^{\Lm\backslash M}|=|A|^{|\Lm|-|M|}$ terms, so that
\beq
P^{\om}_{\Lm,0}([\eta'])= \frac{1}{Z_{\Lm}^{\om}(0)}\Sigma_{\sg\in A^{\Lm}\mid \sg\om\in [\eta']}e^0=
\frac{|A|^{|\Lm|-|M|}}{|A|^{|\Lm|}}=|A|^{-|M|}. \label{9105} 
\eeq 
The second equation in \er{9104}, which the reader should verify from the explanation preceding \er{Hgen},  implies that the integral over $\om$ on the right-hand side of \er{DLR2} is constrained to $\om\in    \Lm'\backslash M$, and hence equals $|A|^{-(|\Lm'| -|M|)}$ times the result of \er{9105}, so that 
 \begin{equation}
\int_{\Om}dP_{\beta}(\om)\, P^{\om}_{\Lm,\beta}([\eta'])=|A|^{-(|\Lm'| -|M|)}\cdot |A|^{-|M|}=|A|^{-|\Lm'|},
\end{equation}
and once again we see that $P_0=f^{\Z^d}$ is a Gibbs measure. Note that case 1 follows from  2.
\end{enumerate}

Stationary Markov chains also fall under this scope,  at least under a condition that is even stronger than irreducibility, namely \emph{strict positivity} ($P_{ab}>0$) of  transition probabilities \er{defTP} (and finite $A$). In view of Theorem \ref{limlemma} this gives a unique stationary probability $P\in\Pr(A^\N)$ such that the underlying stochastic process has joint probabilities \er{SMC}. Realizing the process on 
\beq
\Om=A^\N,
\eeq
cf.\ Theorem  \ref{KolRT}),
which we now interpret as a spin chain, 
the pair potential defined by
\begin{align}
\Phi_{n,n+1}(\sg):=-\log P_{\sg(n),\sg(n+1)} ; && \Phi_M=0\:\:\: (M\neq\{n,n+1\}),\label{9108}
\end{align}
turns out to have a unique global Gibbs measure at $\beta=1$, which equals $P$. 
 This only works if the Markov chain in question is indexed by $n\in\Z$ instead of $n\in\N$, which extension can trivially (and uniquely) be achieved by exploiting stationarity, cf.\ \er{defTP}: we add random variables $X_n$ also for $n<0$ and define the ensuing ``new'' joint probabilities bu putting, assuming  $m>0$ and $k>-m$,
\begin{equation}
P(X_{-m}=a_{-m}, \ldots, X_k=a_k, \ldots):=P(X_{0}=a_{-m}, \ldots, X_{k+m}=a_k, \ldots).
\end{equation}
Short of proving this,\footnote{See Keller (1998), \S5.4,  Rassoul-Agha \& Sepp\"{a}l\"{a}inen  (2015), Example 7.21, or  Br\'{e}maud (2020), Theorem 10.1.9.} we just verify that $P$ as given by \er{SMC} satisfies the \DLR-equation \er{DLR2}, and even this we only do in a special case (which clearly shows what is going on): we take
\begin{equation}
B=(X_0=a_0, \ldots, X_N=a_N)\equiv\{\om\in\Om\mid X_0(\om)=a_0, \ldots, X_N(\om)=a_N\},\label{B9110}
\end{equation}
and take $\Lm=\Lm_M=\{-M, \ldots, M\}$ for some $M>N$, cf.\ \er{LmNmf}. The constraint $\sg\om\in B$ that appears in \er{DLR2} with \er{GMLeta} then amounts to $\sg_n=a_n$ for $n=0, \ldots, N$, and no constraint on $\om$ at all. In computing \er{GMLeta} we therefore sum only over $\sg_{-M}$ to $\sg_{-1}$ and over $\sg_{N+1}$ to $\sg_M$. Using \er{GMLeta}, \er{9108}, and \er{Hgen} we find
\begin{align}
\Sigma_{\sg\in A^{\Lm_M}\mid \sg\om\in B}e^{-H^{\om}_{\Lm_M}(\sg)}&=P^{M+1}_{\om_{-M-1}a_0} P_{a_0a_1}\cdots P_{a_{N-1}a_N}P_{a_N \om_{M+1}}^{M-N+1};\label{9111}\\
 Z^{\om}_{\Lm_M}(1)&=\Sigma_{\sg_{-M}, \ldots, \sg_M}P_{\om_{-M-1}\sg_{-M}}\cdots P_{\sg_M\om_{M+1}}\nn \\ &=P^{2M+2}_{\om_{-M-1}\om_{M+1}}.
\end{align}To obtain \er{9111} it may be useful to note the intermediate step where the left-hand side equals
$$
 \Sigma_{\sg_{-M}, \ldots, \sg_{-1}; \sg_{N+1}, \ldots, \sg_M}P_{\om_{-M-1}\sg_{-M}}P_{\sg_{-M}\sg_{-M+1}}\cdots
 P_{\sg_{-1}a_0} P_{a_0a_1}\cdots P_{a_{N-1}a_N} P_{a_N\sg_{N+1}}\cdots P_{\sg_M\om_{M+1}}.$$
Hence for $B$ in \er{B9110},  $\Lm=\Lm_M$ ($M>N$), $\beta=1$, and $\Phi$ from \er{9108} the r.h.s.\ of \er{DLR2} equals
\begin{equation}
\int_{\Om}dP_1(\om)\, P^{\om}_{\Lm_M,1}(B)=P_{a_0a_1}\cdots P_{a_{N-1}a_N}\int_{\Om}dP_1(\om)\,\frac{P^{M+1}_{\om_{-M-1}a_0}\,P_{a_N \om_{M+1}}^{M-N+1}}{P^{2M+2}_{\om_{-M-1}\om_{M+1}}}.\label{9113}
\end{equation}
We are checking that \er{DLR2} holds for $P_1=P$, the probability measure for the given Markov chain. Thus we put $P_1=P$ in \er{9113} and use the fact that stationarity of the Markov chain implies that the right-hand side of \er{9113} is independent of $M$ (as long as $M>N$). We may therefore let $M\raw\infty$, and use \er{limP2} from Theorem \ref{limlemma} (which applies because we  assumed that $P_{ab}>0$). After some further analysis (of the second expression below) which we omit,\footnote{See Keller, Proof of Theorem 5.4.1.} this gives 
\begin{align}
P^{M+1}_{\om_{-M-1}a_0}\raw p(a_0); && \frac{P_{a_N \om_{M+1}}^{M-N+1}}{P^{2M+2}_{\om_{-M-1}\om_{M+1}}}\raw \frac{p(\om_{M+1})}{p(\om_{M+1})}= 1.
\end{align} 
Thus \er{9113} equals $p(a_0)P_{a_0a_1}\cdots P_{a_{N-1}a_N}$, which by \er{SMC} equals $P(B)$. We conclude that \beq
P_1=P\eeq satisfies \er{DLR2} and hence is a Gibbs measure. 
  \section{Large deviations and Fenchel duality for Gibbs measures} \label{LDGM}
We now extend Proposition \ref{P92} and the closely related duality \er{FDD1} - \er{FDD2} to infinite systems. This is technically difficult and we just outline the main steps and ideas.\footnote{References for this chapter are  Keller (1998), Chapter 5; Ruelle (2004), Introduction and Chapters 3 and 4; Georgii (2011), Chapter 15 (with bibliographical notes); Rassoul-Agha \& Sepp\"{a}l\"{a}inen  (2015), Chapters 7 and 8.} We then give a large deviations perspective on these results, generalizing Sanov's theorem and Cram\'{e}r's theorem to Gibbs measures. Though complex, this theory has achieved a remarkable degree of perfection.
    
The relative entropy  is a good place to start. In the context of ergodic theory (and Theorem \ref{5theorem} shows that Gibbs measures do fit this context), the original definition \er{DefSq1} - \er{DefSq2} is problematic  in view of Lemma \ref{VOLemma}: for if $p$ in this definition is ergodic and $q$ is $T$-invariant (or, in the Gibbs context, $\Z^d$-invariant as we often assume), then $S_q(p)$ is either zero (if $q=p$) or infinite (if $q\perp p$). Fortunately, this can be remedied by a limit construction, which also clarifies the relationship between the relative entropy  and the Kolmogorov--Sinai entropy to be discussed in \S\ref{EDS}.

 For a general measure space $(\Om,\Sg)$ we fix a prior $Q\in\Pr(\Om)$; if $\Om=A^{\Z^d}$ (with $\Sg=\mathcal{F}$) this will initially be the  prior $Q=f^{\Z^d}$  induced by the flat prior  $q=f$ on $A$, and later will be a Gibbs measure. Since we need to vary--and therefore explicitly include--the $\sg$-algebra $\Sg$ on which $P$ and $Q$ are defined, we restore our original notation $S(P, Q)$,  and add $\Sg$ as a suffix. Thus:
\begin{align}
S_{\Sg}(P, Q)&:=\int_{\Om} dP\, \log\left(\frac{dP}{dQ}\right)_{\Sg}
 \:\:\: \mbox{ if } P\ll Q \mbox{ on }\, \Sg \label{KLD4};\\
S_{\Sg}(P, Q)&:=\infty\:\:\: \mbox{ otherwise}, \label{KLD4b}
\end{align}
where $P$ and $Q$ \emph{are defined on $\Sg$}, and also the Radon--Nikodym derivative $dP/dQ$
\emph{is defined with respect to} $\Sg$ and hence \emph{must be measurable for $\Sg$}. 
For example,  if $\Om=A^{\Z^d}$ (for finite $A$) and $\Sg=\F_{\Lm}$, in which case we write 
$S_{\Lm}$ for $S_{\F_{\Lm}}$, 
 for general $Q\in\Pr(A^{\Z^d})$ and $P\ll Q$ we  have
\begin{align}
\left(\frac{dP}{dQ}\right)_{\F_{\Lm}}(\om)&=\frac{P([\om_{|\Lm}])}{Q([\om_{|\Lm}])}; \label{intsum0}\\
S_{\Lm}(P, Q)&=\Sigma_{\sg\in A^{\Lm}} P([\sg])  \log\left(\frac{P([\sg])}{Q([\sg])}\right), \label{intsum}
\end{align}
cf.\ \er{102a} for  $[\sg]$, which also implies the notation $[\om_{|\Lm}]=\{\om'\in A^{\Z^d}\mid \om'_{|\Lm}=\om_{|\Lm}\}$ used here.
\bex Prove \er{intsum0}  from \er{intsumP}
and the definition of the Radon--Nikodym derivative.\footnote{See footnote \ref{RNder}.}
\eex

And similarly for any measure space $(\Om,\Sg)$ whose underlying $\sg$-algebra $\Sg$ is generated by a finite (measurable) partition of $\Om$. Even in general, we may write:\footnote{See Georgii (2011), \S 15.1.}
\begin{equation}
S_{\Sg}(P, Q)=\sup_{\Pi}\Sigma_{U\in\Pi} P(U)  \log\left(\frac{P(U)}{Q(U)}\right),
\end{equation}
where the supremum is taken over all finite $\Sg$-measurable partitions $\Pi$ of $\Om$. 
Compare \er{defKSE}!

Of course, for fixed $\Sg$ this refined relative entropy  has the properties stated in Proposition \ref{gibbsin}.1. and is lsc, but it now acquires an additional monotonicity property if we vary $\Sg$:
\begin{proposition}\label{p910}
If $\Sg_1\subset\Sg_2$ are $\sg$-algebras on $\Om$, then 
\beq
S_{\Sg_1}(P, Q)\leq S_{\Sg_2}(P, Q).\eeq
\end{proposition}
Since $S_{\Sg}$ will be a negative entropy, see eg.\ \er{9141} below, or refer to its earlier (and also later) identification as a rate function, and $\Sg_1\subset\Sg_2$ means that the information in $\Sg_2$ is finer (greater) than the information in $\Sg_1$ (think of the extreme case $\Sg_1=\{\Om,\emptyset\}$), 
this proposition states that the entropy of less information is greater than entropy of more information, as expected.\footnote{In quantum information theory this would be called an \emph{information processing inequality}.} \smallskip

\noindent\emph{Proof}. The proof relies on switching between $dP/dQ$  defined with respect to $\Sg_2$ and w.r.t.\ $\Sg_1$. Denoting the former by $\nu_2$ and the latter by $\rho_1$, they are related by
\begin{equation}
\rho_1=E_Q(\rho_2\mid\Sg_1), \label{9138}
\end{equation}
as can be checked from the definitions of the Radon--Nikodym derivative and the conditional expectation,  respectively.  Eq.\ \er{9138} leads to the identity
 \begin{align}
 S_{\Sg_1}(P, Q)=\la \phv(\rh_1)\ra_Q
 =\la \phv\circ E_Q(\rh_2\mid \Sg_1)\ra_Q, && 
 \end{align} 
 in terms of the (strictly) convex function $\phv(x)=x\log x$.  Jensen for conditional expectations gives
 \begin{equation}
\la \phv\circ E_Q(\rh_2\mid \Sg_1)\ra_Q\leq \la E_Q(\phv\circ \rh_2\mid \Sg_1)\ra_Q=\la \phv(\rh_2)\ra_Q= S_{\Sg_2}(P, Q).
\end{equation}
\bex
Show that  \er{intsum0} satisfies \er{9138} with $\Sg_1=\F_{\Lm_1}$ and  $\Sg_2=\F_{\Lm_2}$ for $\Lm_1\subset\Lm_2$.
\QED \eex

 The following lemma paves our way to the thermodynamic limit.\footnote{ We follow Georgii (2011), Proposition (15.10). Lemma
 \ref{dqp1} is similar  to Kolmogorov's inequality \er{subadE}. Similarly, Lemma \ref{dqp2} resembles 
 Lemma \ref{subadlemma}, although the latter has the inequality in the opposite direction and has an infimum where  Lemma \ref{dqp2} has a supremum. But  Lemma \ref{subadlemma} applies to the Kolmogorov--Sinai entropy, which is the negative of the relative entropy  to which  Lemma \ref{dqp2} applies, cf.\ \er{9141}. } 
\begin{lemma}\label{dqp1} 
For $\Om=A^{\Z^d}$, $P\in \Pr(\Om)$, any prior $Q=q^{\Z^d}$, and any finite $\Lm_1,\Lm_2\subset\Z^d$ one has
\begin{equation}
S_{\Lm_1}(P, Q)+S_{\Lm_2}(P, Q)\leq 
S_{\Lm_1\cup\Lm_2}(P, Q)+S_{\Lm_1\cap\Lm_2}(P, Q). \label{9156}
\end{equation}
\end{lemma}
\bex Prove this lemma.\eex
This lemma (with  $\Lm_1\cap\Lm_2=\emptyset$) makes the following adaptation of Fekete's lemma relevant:
\begin{lemma}\label{dqp2} 
Suppose we have numbers $a_{\Lm}\in[0,\infty]$ indexed by rectangles $\Lm\subset\Z^d$, i.e.\ finite sublattices of the form 
\begin{align}
\Lm=\prod_{i=1}^d [m_i,n_i] && (m_i,n_i\in \Z), \end{align}
 with intervals taken in $\Z$, that: 
\begin{enumerate}
\item satisfy $a_{\Lm_1}+a_{\Lm_2}\leq a_{\Lm_1+\Lm_2}$ whenever  $\Lm_1\cap\Lm_2=\emptyset$;
\item are  translation-invariant in the sense that $a_{\Lm+x}=a_{\Lm}$ for all $x\in\Z^d$.
\end{enumerate}
Then 
\beq
\lim_{N\raw\infty} \frac{a_{\Lm_N}}{|\Lm_N|} =\sup_{\Lm} \frac{a_{\Lm}}{|\Lm|},
\eeq
where the supremum is taken over all rectangles.\footnote{See \er{rectZ} for the definition of the special rectangles $\Lm_N$.}
In particular the limit 
 exists in $[0,\infty]$.\end{lemma}
\emph{Proof.}\footnote{See Rassoul-Agha  \& Sepp\"{a}l\"{a}inen  (2015), page 86, or  Georgii (2011), Lemma (15.11). } Take $C<\sup_{\Lm}a_{\Lm}/|\Lm|$, so there exists some $\Lm$ such that $a_{\Lm}/|\Lm|>C$.
For given $N$, fill $\Lm_N$ with as many translates of $\Lm$ as possible, say $k_N$ copies, so that  
\beq\lim_{N\raw\infty}  \frac{|\Lm_N|}{k_N|\Lm|}=1.
\eeq Using this property, iterating properties 1 and 2, and the property $k_N|\Lm|\leq |\Lm_N|$  then gives
\begin{equation}
\sup_{\Lm}\frac{a_{\Lm}}{|\Lm|}\geq \lim\inf_{N\raw\infty} \frac{a_{\Lm_N}}{|\Lm_N|}=\lim\inf_{N\raw\infty} \frac{a_{\Lm_N}}{k_N|\Lm_N|}\geq \frac{a_{\Lm}}{|\Lm|}>C.
\end{equation}
Letting $C\raw \sup_{\Lm}a_{\Lm}/|\Lm|$ gives 
$\lim\inf_{N\raw\infty} \frac{a_{\Lm_N}}{|\Lm_N|}=\sup_{\Lm}\frac{a_{\Lm}}{|\Lm|}$, and trivially we also have 
\begin{align}
\lim\sup_{N\raw\infty} \frac{a_{\Lm_N}}{|\Lm_N|}&\leq \sup_{\Lm}\frac{a_{\Lm}}{|\Lm|}; \\
\lim\sup_{N\raw\infty} \frac{a_{\Lm_N}}{|\Lm_N|}&\geq \lim\inf_{N\raw\infty} \frac{a_{\Lm_N}}{|\Lm_N|}= \sup_{\Lm}\frac{a_{\Lm}}{|\Lm|}.
\end{align}
Hence 
\beq
\lim\inf_{N\raw\infty} \frac{a_{\Lm_N}}{|\Lm_N|}=\lim\sup_{N\raw\infty} \frac{a_{\Lm_N}}{|\Lm_N|}=\sup_{\Lm}\frac{a_{\Lm}}{|\Lm|},
\eeq  so that $\lim_{N\raw\infty} \frac{a_{\Lm_N}}{|\Lm_N|}$ exists and equals their common value $\sup_{\Lm}\frac{a_{\Lm}}{|\Lm|}$. \QED
\begin{proposition}\label{defdPQ}
For $Q=q^{\Z^d}$ and any translation-invariant $P\in\Pr^{\Z^d}(A^{\Z^d})$ the expression
\begin{equation}
s(P,Q)\equiv s_Q(P):=\lim_{N\raw\infty}\frac{S_{\Lm_N}(P, Q)}{|\Lm_N|}  \label{dQP1} 
\end{equation}
exists and equals the supremum of $S_{\Lm}(P, Q)/|\Lm|$ over all rectangles $\Lm$. Furthermore:
\begin{enumerate}
\item The function $P\mapsto s_Q(P)$ is lsc and affine.
\item For any $c\geq 0$ the level sets $\{ P\in\Pr^{\Z^d}(A^{\Z^d})\mid s_Q(P)\leq c$ are compact.
\end{enumerate}
\end{proposition}
We  call $s_Q(P)$  the \emph{mean relative entropy} (over the lattice $\Z^d$). So far we defined it only for translation-invariant probability measures $P\in\Pr^{\Z^d}(A^{\Z^d})$, but if necessary $s_Q$ may be extended to all $P\in\Pr(A^{\Z^d})$ or even all $P\in C(A^{\Z^d})^*$ by complementing \er{dQP1} with
\begin{align} s_Q(P)=\infty; && (P\notin\Pr^{\Z^d}(A^{\Z^d})). \label{dQP2} 
\end{align}
The existence of the limit in \er{dQP1} follows at once from Lemmas \ref{dqp1} and \ref{dqp2}. The limit function $s_Q(\cdot)$ being lsc follows from the fact each approximant is lsc and the limit is a supremum. 
This property remains valid if we add \er{dQP2}. It is surprising, however, that 
$s_Q(\cdot)$ is not just \emph{convex}, like its approximants, but even \emph{affine}.
\bex
 Prove concavity  (i.e.\ the opposite inequality to convexity) of $P\mapsto s_Q(P)$.\eex
We omit the technical proof that its level sets are compact.\footnote{See Georgii (2011), Proposition (15.14), or Rassoul-Agha  \& Sepp\"{a}l\"{a}inen  (2015), Proposition 6.8.} \QED\smallskip

The function $s_Q$ appears as the rate function in a version of Sanov's theorem.\footnote{We follow Rassoul-Agha  \& Sepp\"{a}l\"{a}inen  (2015), Theorem 6.13. The result is valid for general Polish spaces $A$.}  This originally applied to the case $\Om=A^{\N}$, cf.\ Theorem \ref{sanov}, and described large fluctuations of the  empirical measure \er{LNb}, which lies in $\Pr(A)$. This is easily generalized to  $\Om=A^{\Z^d}$, in which case we put
\begin{equation}
L_N(\om)=\frac{1}{|\Lm_N|} \Sigma_{x\in\Lm_N}\dl_{\om(x)},\label{9167}
\end{equation}
and for a given prior $Q=q^{\Z^d}$ we then have $L_N(\om) \raw q$ for $Q$-a.e.\ $\om\in\Om$, weakly in $\Pr(A)$. The function $S_q$ is then the rate function for large fluctuations of $L_N$ around its limit value $q$.

A \LDP\ for probability measures on $A^{\Z^d}$ instead of $A$ starts  from the \emph{empirical field}
\begin{equation}
R_N(\om):=\frac{1}{|\Lm_N|} \Sigma_{x\in\Lm_N}\dl_{\om\circ \theta_x},\label{RNom}
\end{equation}
where $(\om\circ\theta_x)_y:=\om_{x+y}$, cf.\ \er{925}, which takes values in $\Pr(\Om)$. Since the 
$\Z^d$-action on $\Om=A^{\Z^d}$ is ergodic, if $A$ is compact (so that $\Om$ is compact in its canonical product topology), the generalized pointwise ergodic theorem  \er{ETMZd}  gives, weakly for $ q^{\Z^d}$-a.e.\ $\om\in\Om$,
\beq
\lim_{N\raw\infty} R_N(\om)= q^{\Z^d}.\label{9168}
\eeq 
Unfortunately, $R_N(\om)$ depends on values of $\om$ outside $\Lm_N$ and hence is not $\F_{\Lm_N}$-measurable. Moreover, it is not translation invariant. Both issues are remedied at one stroke, as follows.

First, restrict $\om$ to $\Lm_N$ and then extend $\om_{|\Lm_N}:\Lm_N\raw A$ from $\Lm_N$ to all of $\Z^d$ by making it periodic: that is, we fill up $\Z^d$ with copies of $\Lm_N$ by writing $z=(z_1, \ldots, z_d)\in\Z^d$ uniquely as 
$z=x+y$ with $x\in\Lm_N$ and $y_i=k_i\cdot (2N+1)$ for $i=1, \ldots, d$ and $k_i\in\Z$, and  defining
\beq
\om_N(z):=\om(x).
\eeq This only depends on $\om_{|\Lm_N}$ and hence gives the duly $\F_{\Lm_N}$-measurable \emph{periodized empirical field}
\begin{equation}
\til{R}_N(\om):=\frac{1}{|\Lm_N|} \Sigma_{x\in\Lm_N}\dl_{\om_N\circ \theta_x}.\label{deftildeR}
\end{equation}
By the Portmanteau theorem, $P_N\raw P$ weakly in $\Pr(\Om)$ iff $P_N(B)\raw P(B)$ in $\R$ for any cylinder set $B\in\F_{\Lm}$ (see footnote \ref{PMT2}). Since $|R_N(\om)(B)- \til{R}_N(\om)(B)|\raw 0$,\footnote{If $B\in\F_{\Lm}$ then, since $\om(x+y)=\om_N(x+y)$ whenever $x+y\in\Lm_N$, for fixed $x\in\Lm_N$ we have $\dl_{\om\circ\theta_x}(B)=\dl_{\om_N\circ\theta_x}(B)$ provided $x+y\in\Lm_N$  for all $y\in\Lm$. The points $x\in \Lm_N$ for which this fails (i.e.\ for which there is some $y\in\Lm$ for which $x+y\notin\Lm_N$) mus lie within a distance $\mathrm{diam}(\Lm)$ of the boundary of $\Lm_N$ and hence the number of such points
  divided by $|\Lm_N|$ goes to zero as $N\raw\infty$. The ensuing convergence $|R_N(\om)(B)\raw \til{R}_N(\om)(B)|\raw 0$ is even uniform in $\om$.} eq.\ \er{9168} gives
  \beq
\lim_{N\raw\infty} \til{R}_N(\om)= q^{\Z^d}.\label{9169}
\eeq 
We are now in a position to state Sanov's theorem for lattices (where $A$ is Polish):\footnote{The ergodic theorem leading to \er{9168} and \er{9169} require $A$ to be compact, but Theorem \ref{sanovZd} does not.
}
\begin{theorem}\label{sanovZd}
The periodized empirical field  \er{deftildeR} on $\Om=A^{\Z^d}$, with a prior $Q=q^{\Z^d}\in\Pr(\Om)$ for some $q\in\Pr(A)$,  satisfies a \LDP\ with tight rate function $I=s_Q$, cf.\ \er{dQP1}.
\end{theorem}
We omit the proof, which is very lengthy even by the standards of large deviation theory.\footnote{See Georgii (2011), Theorem (15.45) or  Rassoul-Agha  \& Sepp\"{a}l\"{a}inen  (2015), Theorem 6.13. We follow the former in stating the theorem just for the  periodized empirical field and not for the empirical field \er{RNom}. The latter states it for both $R_N$ and $\til{R}_N$ using the same rate function $s_Q$ defined by \er{dQP1} - \er{dQP2}, but this is suspicious for $R_N$, since the samples $R_N(\om)$ are not translation invariant. Although the limit measure $q^{\Z^d}$ is translation invariant, fluctuations around this limit for finite $N$ need not be, and yet $s_Q$ gives these zero probability because of \er{dQP2}. This is only appropriate for $\til{R}_N$, since each sample $\til{R}_N(\om)$ is translation invariant and hence $\til{R}_N$ cannot fluctuate into non-translation invariant  measures.}

Our next job is the construction of an infinite-volume limit of the pressure for some potential $\Phi$, see Definition \ref{D91} (we have included translation invariance in the definition). 
\begin{proposition}\label{Ruelleprop}
For a  general prior $Q=q^{\Z^d}$ and $\beta\in\R$, the \emph{pressure} defined by
\begin{align}
\pi_q(\beta\Phi)&:= \lim_{N\raw\infty}  \frac{1}{|\Lm_N|}\log Z_{\Lm_N,q}^{\eta}(\beta); \\
Z_{\Lm,q}^{\eta}(\beta)&:=
 \left\la e^{-\beta H^{\eta}_{\Lm}} \right\ra_Q=\Sigma_{\sg\in A^{\Lm}} q^{\Lm}(\sg)e^{-\beta H^{\eta}_{\Lm}(\sg)}, \label{9166}
\end{align}
where $H^{\eta}_{\Lm}$ in \er{Hgen}, exists, is finite, and is independent of the boundary condition $\eta\in \Om$.
\end{proposition}
Here free boundary conditions, with Hamiltonian \er{Hfree}, are included. 
See also  \er{tildeP} and \er{991a}, and \er{GML1a}. For a flat prior $q=f$ one  omits 
$q^{\Lm}(\sg)=|A|^{-|\Lm|}$ from  \er{9166}, so that
\begin{align}
\pi(\beta\Phi):= \lim_{N\raw\infty}  \frac{1}{|\Lm_N|}\log \Sigma_{\sg\in A^{\Lm_N}} e^{-\beta H^{\eta}_{\Lm_N}(\sg)}= \pi_f(\beta\Phi)+\log|A|.
\label{9166f}
\end{align}
\emph{Proof.} 
We  prove this proposition for pair potentials with finite range and free boundary conditions; the general case involves some more limiting arguments but is based on the same idea.\footnote{We follow Dorlas (2021), based on  lectures by N.M. Hugenholtz (in which  the factor $R^d$ in \er{9176} is missing). The general case is in Ruelle (2004), Theorem 3.4; Georgii (2011), Theorem (15.30); or Rassoul-Agha  \& Sepp\"{a}l\"{a}inen  (2015), Proposition 6.14.  The special case of the Ising model is also instructive, see Friedli \& Velenik (2018), \S 3.2.1.
} Thus
\begin{equation}
H_{\Lm}(\sg)=\Sigma_{x\in\Lm}\Ph_{\{x\}}(\sg_x)+\Sigma_{y\in\Lm, \|x-y\|\leq R}\Phi_{\{x,y\}}(\sg_x,\sg_y),
\end{equation}
where $R<\infty$, $\|\cdot\|$ is any norm on $\R^d$ (restricted to $\Z^d$), and $
\Phi_{\{x,y\}}$ depends on $x-y$ only (since $\Phi$ is  translation-invariant). In particular, the sum in \er{sumPhi} is finite and the Ising model  \er{PhiIsing} clearly falls within this class. For simplicity we  put $\beta=1$ and take the flat prior, so that we
use
 \begin{align}
 H_{\Lm}&=\Sigma_{M\subseteq\Lm}\Phi_M; \\ Z_{\Lm}&=\Sigma_{\sg\in A^{\Lm}}e^{-H_{\Lm}(\sg)};  \\ \pi(\Phi)&= \lim_{N\raw\infty}  \frac{1}{|\Lm_N|}\log Z_{\Lm_N}.
\end{align}

Take $N\in\N_*$ and the ensuing cube $\Lm_N$, see \er{rectZ}, so that 
\beq
|\Lm_N|=(2N+1)^d.
\eeq
Then for any $n\in\N_*$ pick $n$ disjoint cubes 
\begin{align}
C_k(N)=\prod_{i=1}^d [m_i,n_i] && (k=1, \ldots, n),
\end{align}
  with equal       sides $n_i-m_i=2N+1$ for each $i=1, \ldots, d$. Their union 
 $\cup_{k=1}^n C_k(N)$ has volume 
 \beq
 |\cup_{k=1}^n C_k(N)|=n(2N+1)^d.
 \eeq
  Then
 \begin{equation}
\lim_{N\raw\infty}  \left(\frac{1}{|\Lm_N|}\log  Z_{\Lm_N}-\frac{1}{|\cup_{k=1}^n C_k(N)|}\log Z_{\cup_{k=1}^n C_k(N)}\right)=0, \label{9170}
\end{equation}
\emph{uniformly in $n$}. For  $N_1, N_2\in\N_*$ the region $\Lm_{N_1N_2}$ consists of $n_1=N_2^d$ copies of $\Lm_{N_1}$, suitably translated to become disjoint (and similarly  it consists of $n_2=N_1^d$ copies of $\Lm_{N_2}$). Since  convergence in \er{9170} is uniform in $n$, for any $\varep >0$ can find $N_0$ such that for all $N_1\geq N_0$ and any (fixed) $N_2$,
\begin{equation}
\left|   \frac{1}{|\Lm_{N_1}|}\log  Z_{\Lm_{N_1}}- \frac{1}{|\Lm_{N_1N_2}|}\log  Z_{\Lm_{N_1N_2}}
\right|< \varep/2,
\end{equation}
and likewise with $N_1$ and $N_2$ swapped. By a standard $\varep/2$ argument, for $N_1,N_2\geq N_0$,
\begin{equation}
\left|   \frac{1}{|\Lm_{N_1}|}\log  Z_{\Lm_{N_1}}- \frac{1}{|\Lm_{N_2}|}\log  Z_{\Lm_{N_2}}\right|<\varep.
\end{equation}\vspace{-5mm}
\bex Prove this. \eex
Hence  $(|\Lm_N|\inv \log Z_{\Lm_N})_N$ is a Cauchy sequence in $\R$, which must converge, namely to $\pi(\Phi)$.

 It remains to prove \er{9170}.  The idea is that the boundary interactions between the cubes in the union $\cup_{k=1}^n C_k(N)$ divided by the volume vanish in the limit, whilst the  interactions within the cubes cancel the first term in  \er{9170}. To make this precise, we decompose the energy as
 \begin{align}
H_{\cup_{k=1}^n C_k(N)}(\sg)&=\Sigma_{k=1}^n H_{ C_k(N)}(\sg)+\Sigma_{k,l=1, l\neq k}^nI_{kl}(\sg);\label{9173} \\
I_{kl}(\sg)&=\Sigma_{x\in C_k(N),\, y\in C_l(N), \, \|x-y\|\leq R} \Phi_{\{x,y\}}(\sg_x,\sg_y).\label{9174}
\end{align}
Recalling  that $\Phi_{\{x,y\}}$ depends on $\|x-y\|$ only, the finite expression
\begin{equation}
M(x):=\Sigma_{y\in\Z^d, \|x-y\|\leq R}\max_{\sg_x\sg_y\in A}\{| \Phi_{\{x,y\}}(\sg_x,\sg_y)|\}
\end{equation}
in fact does not depend on $x$, so $M(x)=M$. For fixed $k$, now count the number of pairs $(x,y)$ for which 
$x\in C_k(N)$ and $\|x-y\|\leq R$, so that certainly all $y\in C_l(N)$, $l\neq k$ with  $\|x-y\|\leq R$ are included, as required by \er{9173} - \er{9174}. 
 For simplicity assume $R\in\N$. For any cube $C_L$ whose sides have length $L\gg R$ there are at most $L^d-(L-2R)^d$ points $x\in C_L$ within a distance $R$ from the boundary of $C_L$; the others cannot interact with points outside $C_L$. Each of the former points can interact with at most $R^d$ points $y\notin C_L$. 
With $L=2N+1$ in our case, this 
 implies the  bound.
\begin{equation}
\left| \Sigma_{l=1, l\neq k}^nI_{kl}(\sg)\right| \leq R^d(L^d -(L-2R)^d)M.\label{9176}
\end{equation}
Using the inequality $L^d -(L-2R)^d\leq 2dRL^{d-1}$, this gives
\beq \left| \Sigma_{k,l=1, l\neq k}^nI_{kl}(\sg)\right| \leq 2ndR^{d+1}(2N+1)^{d-1}M.\label{9177}
\eeq
Furthermore, translation invariance gives 
\begin{equation}
\Sigma_{\sg\in A^{\cup_{k=1}^n C_k(N)}} e^{-\Sigma_{k=1}^n H_{ C_k(N)}(\sg)}=(Z_{\Lm_N})^n.
\end{equation}\vspace{-5mm}
\bex Prove this. \eex
Combining all of this, and recalling that 
\beq
|\cup_{k=1}^n C_k(N)|=n(2N+1)^d=n|\Lm_N|, \label{ncancels} 
\eeq
we obtain
\begin{equation}
\left|\frac{1}{|\Lm_N|}\log  Z_{\Lm_N}-\frac{1}{|\cup_{k=1}^n C_k(N)|}\log Z_{\cup_{k=1}^n C_k(N)}\right|\leq
\frac{2dR^{d+1}M}{2N+1}, \label{9180}
\end{equation}
since the $n$ in \er{9177} cancels the $n$ in \er{ncancels} whilst the $(2N+1)^{d-1}$ in \er{9177} combines with the $(2N+1)^d$ in \er{ncancels} to leave the  factor $(2N+1)^{-1}$ in \er{9180}. This proves \er{9170}. \QED\smallskip

Having infinite-volume versions of the entropy and pressure, respectively, we also seek an infinite-volume version of the Fenchel duality \er{FDD1} - \er{FDD2}. The main issue is to figure out the infinite-volume analogue of the term $\la f\ra_P$ in \er{FDD2}, and its relationship to $\Phi$. To this end, define
\begin{align}
f_{\Phi}: \Om\raw\R; && f_{\Phi}:=\Sigma_{0\in M\subset\Z^d} \frac{\Phi_M}{|M|},
\end{align}
where $M$ is finite. For example, for the Ising model \er{PhiIsing} we have
\begin{equation}
f_{\Phi}(\sg)=-J\Sigma_{x\in \Z^d,\|x\|_1=1} \sg_x\sg_0-B\sg_0.
\end{equation}
For any $P\in\Pr^{\Z^d}(A^{\Z^d})$ and boundary condition $\eta\in\Om$ (including $\eta=\mathrm{free}$) we then have:\footnote{See Georgii (2011), Theorem (15.23). This is even true for sequences $(\eta_N)$ of boundary conditions.}
\begin{equation}
\la f_{\Phi}\ra_P=\lim_{N\raw\infty} \frac{1}{|\Lm_N|} \la H^{\eta}_{\Lm_N}\ra_P, \label{Thus} 
\end{equation}
 Thus $f_{\Phi}$ is the average energy per site, as already suggested by the definition. Confusingly, this object leads some authors to define the local Hamiltonians with free boundary conditions by
 \begin{equation}
H'_{\Lm}:=\Sigma_{x\in\Lm} f_{\Phi}\circ\theta_x,
\end{equation}
cf.\ \er{925},
which may not  coincide with \er{Hfree}, although for the Ising model it does. However,\footnote{See Rassoul-Agha  \& Sepp\"{a}l\"{a}inen  (2015), Lemma 8.2, for a more detailed statement and proof of \er{kleineo}.}
\beq
\lim_{N\raw\infty}\frac{1}{|\Lm|} \|H^{\eta}_{\Lm_N}-H'_{\Lm_N}\|_{\infty}=0, \label{kleineo}
\eeq 
and hence in studying the energy as $N\raw\infty$ one may use any expression; this is
the key to \er{Thus}.

Similarly, using $f_{\Phi}$ instead of $\Phi$ gives rise to an alternative versions of the pressure, namely
\begin{align}
\pi'_q(f)&:= \lim_{N\raw\infty}  \frac{1}{|\Lm_N|}\log Z'_{\Lm_N,q}(f); \\
Z'_{\Lm,q}(f)&:=
 \left\la e^{\Sigma_{x\in\Lm}f\circ\theta_x} \right\ra_Q=\Sigma_{\sg\in A^{\Lm}} q^{\Lm}(\sg)e^{\Sigma_{x\in\Lm}f\circ\theta_x(\sg)}.\label{9166g}
\end{align}
where $f\in C_b(\Om)$ (i.e.\ $f\in C(\Om)$ if $\Om$ is compact). Fortunately, it can be shown that
\begin{equation}
\pi_q(\Phi)=\pi'_q(-f_{\Phi}).\label{9196}
\end{equation}
 We return to Fenchel duality.\footnote{Here we assume that $A$ is finite, in which case $f:\Om\raw\R$ is continuous iff it is \emph{quasi-local}, i.e., a sup-norm-limit of \emph{local functions}, which in turn are defined as the functions $f$ that are $\F_{\Lm}$-measurable for some finite $\Lm\subset\Z^d$, which is the case if $f=\pi_{\Lm}^*f_{\Lm}$ for some
 $f_{\Lm}:\Om_{\Lm}\raw\R$ (so that $f(\om)$ only depends on $\om_{|\Lm}$); local functions are  continuous.} 
 In terms of the pressure $\pi_q'$, a version close to \er{FDD1} - \er{FDD2} is  
\begin{align}
s_Q(P)&=\sup_{f\in C(\Om)}\{\la f\ra_P-\pi'_q(f)\} ;\label{FDD10}  \\
\pi'_q(f)&=\sup_{P\in\Pr^{\Z^d}(\Om)}\{ \la f\ra_P-s_Q(P)\}; \label{FDD20} 
\end{align}
 recall that $s_Q(P)$ has only been defined for translation-invariant probabilities $P\in \Pr^{\Z^d}(\Om)$, cf.\ Proposition \ref{defdPQ}, where evidently also $Q=q^{\Z^d}\in \Pr^{\Z^d}(\Om)$ for any prior $q\in\Pr(A)$.
Hence
\begin{equation}
\pi_q(\beta\Phi)=-\inf_{P\in\Pr^{\Z^d}(\Om)}\{ \beta\la f_{\Phi}\ra_P+s_Q(P)\}; \label{FDD30} 
\end{equation}
or,  in terms of the entropy  and free energy
\begin{align}
s(P):=-s_{f^{\Z^d}}(P); &&
f(T)=-T\pi_f(\Phi/T), \end{align}
where $T>0$, in the equivalent formulation
 \begin{equation}
f(T)=\inf_{P\in\Pr^{\Z^d}(\Om)}\{\la f_{\Phi}\ra_P-T s(P)\}.\label{FDD31} 
\end{equation}
\emph{Proof of  \er{FDD10} - \er{FDD20}.} These imply all other versions. The key is Corollary \ref{RA425}, in which 
we replace $\N$ by $\Z^d$ and hence $N$ by $|\Lm_N|$ as appropriate, and take $\X=C(\Om)^*$ with closed convex subspace $\Pr^{\Z^d}(\Om)$, and $\mathcal{Y}=C(\Om)$.
The \LDP\ comes from Theorem \ref{sanovZd} and we do not need to assume \er{826} since we already proved that the pressure exists. Moreover, we note that 
\begin{equation}
\Sigma_{x\in\Lm_N}f\circ\theta_x=|\Lm_N|  \la f\ra_{R_N}, \label{justS}
\end{equation}
so that the partition function \er{9166g} may be rewritten accordingly:
\begin{equation}
Z'_{\Lm_N,q}(f)= \left\la e^{\Sigma_{x\in\Lm_N}f\circ\theta_x} \right\ra_{q^{\Z^d}}= \left\la e^{|\Lm_N|  \la f\ra_{R_N}} \right\ra_{q^{\Lm_N}}=\int_{\Pr(\Om)}dP_N(x) e^{|\Lm_N|\la y,x\ra},\label{Zprimeq}
\end{equation}
where $P_N$ is the probability measure on $\Pr(\Om)$ induced by $R_N:\Om\raw \Pr(\Om)$, and furthermore,
\begin{align}
q^{\Z^d}\in\Pr(\Om); && y=f_{\Phi}\in C(\Om); &&  x=R_N; && \la y,x\ra =\la y\ra_x.
\end{align}
To complete the justification for invoking Corollary \ref{RA425}, we resolve the tension between the  periodized empirical field $\til{R}_N$ in
 Theorem \ref{sanovZd} and the empirical field $R_N$ appearing in the definition of the pressure \er{9166} with \er{justS}.  The argument between \er{deftildeR} and \er{9169} implies that
\begin{align}
\pi'_q(f)&= \lim_{N\raw\infty}  \frac{1}{|\Lm_N|}\log \left\la e^{\Sigma_{x\in\Lm}f\circ\theta_x} \right\ra_Q=
  \lim_{N\raw\infty}  \frac{1}{|\Lm_N|}\log \left\la e^{|\Lm_N|  \la f\ra_{R_N}} \right\ra_Q\nn \\ &=  \lim_{N\raw\infty}  \frac{1}{|\Lm_N|}\log \left\la e^{|\Lm_N|  \la f\ra_{\til{R}_N}} \right\ra_Q,
\end{align}
so that in \er{Zprimeq} we may replace $R_N$ by $\tilde{R}_N$, and in its wake replace $P_N$ by the probability measure $\til{P}_N$  on $\Pr(\Om)$ induced by $\til{R}_N:\Om\raw \Pr(\Om)$. Thus \er{FDD10} - \er{FDD20} is a special case of the dual pair $I=\Pi^*$ and $\Pi=I^*$ in Corollary \ref{RA425}, with $I=s_Q$ and $\Pi=\pi'_q$. \QED\smallskip
  
  Let us finally explain  the  relationship between large deviations and variational principle for Gibbs measures. To be on the safe side we assume that $A$ is compact and metrizable,\footnote{Most results are valid for more general Polish spaces, or even more generally, see especially Georgii (1993, 2011).} and as usual we write $\Om=A^{\Z^d}$, as well as $\Pr^{\Z^d}(\Om)$ for the (compact convex) space of translation-invariant probability measure on $\Om$ (in the weak topology). Also, $\mathcal{G}(\beta\Phi)$ is the (convex compact) set of all (global) Gibbs states for given potential $\Phi$ at inverse temperature $\beta$,  based on the flat prior $Q=f^{\Z^d}$ (the results are easily generalized to more general priors $Q=q^{\Z^d}$, though). The  structure of this set was outlined in Theorem \ref{5theorem}. In particular, the unique Gibbs measure $P_{\beta}^{\mathrm{free}}$ from part 1 of  Theorem \ref{5theorem} is translation invariant and  plays the role of a reference Gibbs measure.\footnote{It may or may not be the only element of $\mathcal{G}(\beta\Phi)$.}

The following equivalence of quite different things  is one of the main results of the theory.\footnote{See Ruelle (2004), Theorem 4.2; Rassoul-Agha  \& Sepp\"{a}l\"{a}inen  (2015), page 127; Georgii (2011), Theorems (15.30) and (15.39).}
\begin{theorem} \label{DLRVP}
Let $P\in \Pr^{\Z^d}(\Om)$.
The following statements are equivalent: 
\begin{enumerate}
\item $P\in \mathcal{G}(\beta\Phi)$.
\item $P$ attains the infimum in \er{FDD30}, so that,  in terms of  entropy, 
\begin{align}
\pi_f(\beta\Phi)=-\beta \la f_{\Phi}\ra_P +s(P); && s(P):=-s_{f^{\Z^d}}(P). \label{Pattains}
\end{align}
\item $s(P, P_{\beta}^{\mathrm{free}})=0$.
\item $s(P, P_{\beta})=0$ for all translation-invariant Gibbs measures $P_{\beta}\in \mathcal{G}^{\Z^d}(\beta\Phi)$.
\end{enumerate}
\end{theorem}
This theorem is an infinite-volume version of Proposition \ref{P92}, with the difference that in finite volume the (local) Gibbs measure for free boundary conditions is unique; each different boundary condition comes with its own version of Proposition \ref{P92}. In infinite volume (global) Gibbs measures need not be unique, but according to the above theorem they all lead to the same pressure (as we had already seen), and, perhaps more unexpectedly, they all have zero mean relative entropy ``distance'' to the reference Gibbs measure $P_{\beta}^{\mathrm{free}}$. More generally, it can be shown that
\begin{equation}
s(P, P_{\beta})=-s(P) + \beta \la f_{\Phi}\ra_P +\pi_f(\beta\Phi), \label{9205}
\end{equation}
for any $P\in \Pr^{\Z^d}(\Om)$  and translation-invariant Gibbs measure $P_{\beta}\in \mathcal{G}(\beta\Phi)$, for some given potential $\Phi$. Since the right-hand side of \er{9205} is independent of the choice of $P_{\beta}\in \mathcal{G}(\beta\Phi)$,  the equivalence between statements 2 and 4 in Theorem \ref{DLRVP} already follows. We also see that 
\begin{equation}
s(P_{\beta}', P_{\beta})=0
\end{equation}
for any two translation-invariant Gibbes measures (as always, defined for the same potential $\Phi$).

Eq.\ \er{9205} follows from the  finite-volume case plus limiting arguments.\footnote{See  Rassoul-Agha  \& Sepp\"{a}l\"{a}inen  (2015), Theorem 8.3.} For simplicity we take $P_{\beta}=P_{\beta}^{\mathrm{free}}$. Using \er{991free}, \er{KLD4},  \er{intsum}, and \er{Hfree}, for finite $\Lm\subset\Z^d$ we have
\begin{align}
S_{\Lm}(P, P_{\beta}^{\mathrm{free}})=
-S_{\Lm}(P)+\beta \la H^{\mathrm{free}}_{\Lm}\ra_{P_{\Lm}} + \log Z^{\mathrm{free}}_{\Lm}(\beta), \label{9200}
\end{align}
where for $\sg\in A^{\Lm}$ we put 
\begin{align}
P_{\Lm}(\sg):=P([\sg]); && S_{\Lm}(P):=-\Sigma_{\sg\in A^{\Lm}} P_{\Lm}(\sg) \log P_{\Lm}(\sg).
\end{align} Using \er{Thus}, \er{9196},  \er{9166}, and of course Proposition \ref{defdPQ}, we obtain \er{9205}.
\bex Prove \er{9200}.\eex

We now turn to large deviations, first of the periodized empirical field $\til{R}_N$, and then of the energy.  Theorem \ref{sanovZd} describes the fluctuations of  $\til{R}_N$
 against a prior (and hence limit value) $Q=q^{\Z^d}$. Continuing the story \er{9205}, using a  Gibbs measure as prior (etc.) similarly gives:\footnote{See Georgii (2011), Theorem (15.45), for a proof.}
\begin{theorem}\label{sanovGibbs}
For any  prior $P_{\beta}\in \mathcal{G}^{\Z^d}(\beta\Phi)$,
the periodized empirical field  \er{deftildeR}  satisfies a \LDP\ with tight rate function $I(P)=s(P, P_{\beta})$; this function is  the same for all $P_{\beta}\in \mathcal{G}^{\Z^d}(\beta\Phi)$.
\end{theorem}
The last claim is evident from \er{9205}, according to which we could equivalently have defined
\begin{equation}
I(P)=-s(P) + \beta \la f_{\Phi}\ra_P +\pi_f(\beta\Phi).
\end{equation}
This shows that whereas for a flat prior $Q=f^{\Z^d}$ the rate function was essentially 
minus the Shannon entropy $s(P)$, changing the prior to a Gibbs measure $Q=P_{\beta}$ 
 induces some extra terms that are of course determined by the potential $\Phi$, like $P_{\beta}$ itself. 
 The zeroes of $I(P)$ form the compact convex space $\mathcal{G}^{\Z^d}(\beta\Phi)$ of  translation-invariant Gibbs measures, which may be degenerate. 

Using the contraction principle (Theorem \ref{contraction}) we also obtain a \LDP\ for the energy.\footnote{Theorem  \ref{CramerGibbs} is a special case of Corollary (15.48) in Georgii (2011).}
Compared to Cram\'{e}r's Theorem \ref{cramer0}, the random variables $S_N$ are now replaced by 
\beq
h_N:=\frac{H_{\Lm_N}^{\eta}}{|\Lm_N|}, \label{hNH}
\eeq 
see \er{Hgen}, although because of  \er{Thus} the functions $h_N=f_{\Phi}$ give the same result.
\begin{theorem}\label{CramerGibbs}
For any  prior $P_{\beta}\in \mathcal{G}^{\Z^d}(\beta\Phi)$,
the energies $(h_N)$ satisfy a \LDP\ with  rate function 
\begin{align}
I_E:\R\raw[0,\infty]; && I_E(u)=\inf\{I(P), \, P\in \Pr^{\Z^d}(\Om)\mid \la f_{\Phi}\ra_P=u\},
\end{align}
with  $I_E(u)=\infty$ if no $P$ exists with $\la f_{\Phi}\ra_P=u$. Furthermore, $I_E$ is convex and tight, and we have
\begin{equation}
I_E(u)= \sup_{t\in\R}\{ tx-\pi_f((\beta-t)\Phi)\}+\pi_f(\beta\Phi)= -\inf_{t\in\R} \{\pi_f(t\Phi)+tu\} +\beta u+\pi_f(\beta\Phi). \label{9213}
\end{equation}
Finally, the unique zero of $I_E$ lies at $u=\la f_{\Phi}\ra_{P_{\beta}}$.
\end{theorem}
This is similar to  Proposition \ref{rateE} and Exercise \ref{Ex68}, and  similar comments apply: we may rewrite 
\begin{align}
I_E(u)=s_{\beta}(u)-s_{\mathrm{eq}}(u); && s_{\beta}(u)=s_{\mathrm{eq}}(u(\beta))+\beta(u-u(\beta)),
\end{align}
where $s_{\mathrm{eq}}(u)$ is the first term on the right-hand side of \er{9213}, identified with the equilibrium entropy at a temperature $\beta(u)$ uniquely determined by $u$ via 
\beq
\la f_{\Phi}\ra_{P_{\beta(u)}}=u, \label{9215} 
\eeq
whilst  
\beq
s_{\mathrm{eq}}(u(\beta))=\pi_f(\beta\Phi)+\beta \la f_{\Phi}\ra_{P_{\beta}},
\eeq cf.\ \er{Pattains}, in which we write $\la f_{\Phi}\ra_{P_{\beta}}=u(\beta)$; in contrast with \er{9215}, this time the energy $u(\beta)$ is determined by the inverse temperature $\beta$, rather than the other way round. Thus 
\beq
I_E(u(\beta))=0,
\eeq
 confirming the last claim in Theorem \ref{CramerGibbs}. All in all, we may interpret  
$ -I_E=s_{\mathrm{eq}}-s_{\beta}$ as a nonequilibrium entropy, normalized so that it vanishes at the equilibrium (energy) value of its argument, given the background Gibbs measure at given $T\inv=\beta$.
\section{Entropy of classical dynamical systems}\label{EDS}
The theory  just explained  used a new kind of entropy \er{dQP1} defined by a limiting construction.
The underlying idea was introduced by 
 Kolmogorov (1958) in the context of dynamical systems; from our point of view he found a third way of looking at our diagram in the Introduction.\footnote{See  Sinai (1989) and more generally  Charpentier, Lesne, \& Nikolski (2007) for Kolmogorov's closely related contributions to dynamical systems, entropy, and ergodic theory. Relevant textbooks include for example Collet \& Eckmann (2006),  Castiglione \emph{et al.} (2008), and Viana \& Oliveira (2016). } 
 \begin{definition}\label{defDS}
 \begin{enumerate}
\item    A \emph{dynamical system} is a triple $(X,P,T)$,
  where $(X, P)\equiv (X,P,\Sg)$ is a probability space (we suppress the $\sg$-algebra $\Sg$ in our notation whenever possible), and 
  \beq
  T:X\raw X \label{TXX}\eeq
   is a measurable (but not necessarily invertible) map,\footnote{  If $T$ is invertible $T\inv(B)$  coincides with the image $\{T\inv(x)\mid x\in B\}$ of $B$ under the inverse map $T\inv: X\raw X$.}  required to preserve $P$ in that 
  \begin{align}
  P(T\inv B)=P(B); && T\inv(B):=\{x\in X\mid T(x)\in B\},\label{PTB}
  \end{align}
  for any measurable $B\subset X$ ($B\in\Sg$).
  \item   
  A  \emph{partition} of $X$  is  a collection of measurable subsets $(U_a)_{a\in A}$ of $X$ such that
\begin{align}
P\left(\bigcup_a U_a\right)=1; && P(U_a\cap U_b)=0 \:\:\: (a\neq b).  \label{Xdu}
\end{align}
\end{enumerate}
  \end{definition}
Thus $X=\bigcup_{a\in A} U_{a}$ for disjoint $U_a$, so to speak, up to measure zero.  For any natural number $N>0$, an element
 $\sg\in A^N$ corresponds to a \emph{coarse-grained path of a single particle} via the condition
  \begin{align}
  T^nx\in U_{\sg(n)} && (n=0,1,\ldots, N-1, \sg_n\in A).
  \end{align} 
  Hence our particle starts at $x\in U_{\sg(0)}$ at $t=0$, moves to $Tx\in U_{\sg(1)}\subset X$ at $t=1$, etc.,
 and  at time $t=N-1$ finds itself at $T^{N-1}x\in  U_{\sg(N-1)}$. In our deterministic setting the entire coarse-grained path is fixed given its starting point $x\in X$ at $n=0$, so that a  partition \er{Xdu} $P$-a.e.\ defines a map
\begin{align} \xi: X\raw A^{\N}; &&  \xi(x)_n=a\in A \:\:\mbox{ iff } \:\:T^nx\in U_a \:\:\: (n\in\N). \label{defxi}
\end{align}
Hence  a fine-grained path $(x, Tx, T^2x, \ldots)\in X^{\N}$ is coarse-grained to $\xi(x)\in A^{\N}$, which in turn may be truncated  to give a map $\xi_N: X\raw A^N$ defined by $\xi_N(x)= \xi(x)_{|N}\in A^N$, where $\xi(x)_{|N}\in A^N$ is the restriction of $\xi\in A^\N$ to $N\subset \N$.
A key role will be played by   the (unilateral) \emph{shift}
\begin{align}
S: A^{\N}\raw A^{\N}; && (Ss)_n:=s_{n+1} \:\:\: (n=0,1,\ldots), \label{ULS}
\end{align}
because this map satisfies the intertwiner relation
\beq
S\circ \xi=\xi\circ T. \label{Scrucial}
\eeq
Thus the given triple  $(X,P,T)$ is coarse-grained by a new triple $(A^{\N}, P', S)$, 
where  
\beq
P'(B)=P(\xi\inv B).
\eeq
\bex
\begin{enumerate}
\item Prove the relation \er{Scrucial}.
\item Take $X=A^{\N}$ with time-evolution $T=S$ and partition $(U_a)_{a\in A}$ defined by 
\begin{equation}
U_a=\{s\in A^\N\mid U_0=a\}. 
\label{Uapart}
\end{equation}
Show that $\xi=\mathrm{id}$, that is, $\xi(s)=s$ for all $s\in A^\N$.
\end{enumerate}
\eex
If $\xi$ were injective, then, spectacularly, nothing would be lost in coarse-graining! This property is quite rare, but it is (almost by construction) the case for the \emph{doubling map} 
\begin{align}
T_D: [0,1)\raw [0,1); &&
T_D(x)=2x \:\:\: (0\leq x<\half); && T_D(x)=2x-1\:\:\:  (\half\leq x<1), \label{28}
\end{align}
so that $T_D(x)$ takes the fractional part of $2x$. Perhaps surprisingly (because of the factor 2), \emph{because $T_D$ is not invertible} this map preserves Lebesgue measure $\mu_L$, acting as the probability measure on $X=[0,1)$ in charge.  To show this, one only needs to check \er{PTB} for intervals.\footnote{Turning  these special cases into a  proof requires,  for example, the following   theorem: \emph{Let $R\subset\Sg$ be an \emph{algebra}, that is, a subset $\Sg\subset\mathcal{P}(X)$ that contains $X$ and is closed under complementation $A\backslash B$ and unions $A\cup B$ (and hence under intersections $A\cap B$). Let $R$ also contains an increasing sequence $(B_n)$ such that $X=\bigcup_n B_n$.  
If $P(T\inv B)=P(B)$ for all $B\in R$, then $P(T\inv B)=P(B)$ for all $B\in \Sg$. } See for example   Dajani \& Kalle, (2021), Theorem 1.2.1.\label{DK1}
}
\begin{exercise}
For  $0\leq a<b<1$, show that $\mu_L(T_D\inv[a,b])=\mu_L([a,b])=a-b$. 
\end{exercise}
In terms of the binary expansion
\begin{equation}
x=\Sigma_{n=0}^{\infty} s_n2^{-(n+1)},\label{xomega}
\end{equation}
the doubling map $T_D$ corresponds to the unilateral shift \er{ULS} on $2^{\N}$. There are two cases:\footnote{This is true 
provided dyadic rationals in $[0,1)$ are mapped to finite binary sequences (i.e., those ending with infinitely many zeros). Binary expansions are not unique, see footnote \ref{dyadicfn}.  If $x$ is a dyadic rational we should take the $s\in 2^\N$ with infinitely many zeros at the end. This is clear from \er{28}, since every dyadic rational eventually ends up at $x=0$ and stays there.
Similarly, the map $\xi$ in \er{210} below is not quite a bijection, but it is a bijection up to the set of dyadic rationals in $[0,1)$, which has Lebesque measure zero. Alternatively, one could omit elements of $2^\N$ that end with an infinite string of 1's (these are precisely the binary sequences that superfluously represent dyadic rationals).}
\begin{enumerate}
\item 
if $0\leq x<\half$, which corresponds to $s_0=0$, then 
\begin{align}
T_D(x)&=2x=2\cdot \Sigma_{n=0}^{\infty} s_n2^{-(n+1)}=\Sigma_{n=0}^{\infty} s_n2^{-n}=
\Sigma_{n=1}^{\infty} s_n2^{-n}=\Sigma_{n=0}^{\infty} s_{n+1}2^{-(n+1)}\nn \\ &=\Sigma_{n=0}^{\infty}(Ss)_n 2^{-(n+1)}. 
\end{align}
\item If  $\half\leq x<1$, which corresponds to $s_0=1$, then $s_0$ also disappears.\end{enumerate}
\begin{exercise}
Show this.
\end{exercise}
In other words, if we write $T_D':2^\N\raw 2^\N$ for the transfer of the 
 doubling map $T_D:[0,1)\raw [0,1)$ to $2^\N$ via  \er{xomega}, then $T_D'=S$.
 We use this fact to show that if we  use the partition $\{U_0,U_1\}$ of $[0,1)$ given by $U_0=[0,\half)$ and $U_1=[\half,1)$, then the infinite coarse-grained path $\xi(x)\in 2^\N$ equals
  \begin{align}
\xi:[0,1)\raw 2^\N; && 
\xi(x)=s, \label{210}
\end{align}
 where $s$ is defined by \er{xomega}.
 To see this, first look at the partition $\{U_0', U_1'\}$ of $2^\N$ where, cf.\ \er{Uapart} 
\begin{equation}
U_i'=\{s\in\ 2^\N\mid s_0=i\}; \:\:\: ( i=0,1). 
\end{equation} Clearly, $x\in U_i$ iff $s\in U_i'$ for $i=1,2$.
Then $\xi':2^\N\raw 2^\N$, defined as in \er{defxi} by the property 
\beq
\xi'(s)_n=i\in\{0,1\} \:\:\: \mbox{iff} \:\:\: S^ns\in U_i, 
\eeq
is the identity,
  since for $n=1$ the condition $Ss\in U_i$
means that $(Ss)_0=s_1=i$, and likewise $S^ns\in U_i$ means $s_n=i$. Transporting this back to $[0,1)$ gives \er{210}.

A slightly more subtle example where $\xi$ is a bijection is given by the \emph{tent map}
\begin{align}
T_t:[0,1)\raw [0,1); &&  T_t(x)=1-|2x-1|, 
\end{align}
that is, $T_t(x)=2x$ for $0\leq x<\half$ and $T_t(x)=2-2x$ if $\half\leq x<1$. This map also preserves $\mu_L$.
\begin{exercise} Show this.\end{exercise}
The situation is similar to the doubling map, but with a twist: seen as a map
\beq
T_t':2^\N\raw 2^\N
\eeq
  via the binary expansion \er{xomega} of $x$, for $s_0=0$, corresponding to $x\in[0,\half)$, we have \beq
  T_t'=T_D'=S.
  \eeq
   But if $s_0=1$ and hence $x\in [\half,1)$, we have 
   \beq
   T_t'(s)_n=
s_{n+1}':=1- s_{n+1}.
\eeq\vspace{-5mm}
\begin{exercise}
Show this.
\end{exercise}
This  gives a formula for the map 
\beq
\xi_t:[0,1)\raw 2^\N,
\eeq
 relative to the same partition $\{U_0,U_1\}$ as for the doubling map, defined recursively:
 \begin{align}
\xi_t(x)_0=s_0; && \xi_t(x)_n=s_n \:\:\:\: (\xi_t(x)_{n-1}=0); &&  \xi_t(x)_n=s_n' \:\:\:\: (\xi_t(x)_{n-1}=1). 
\end{align}
  This is once again a bijection (up to the dyadic rationals, which may be taken out without loss).
     \begin{center}
 \includegraphics[width=0.5\textwidth]{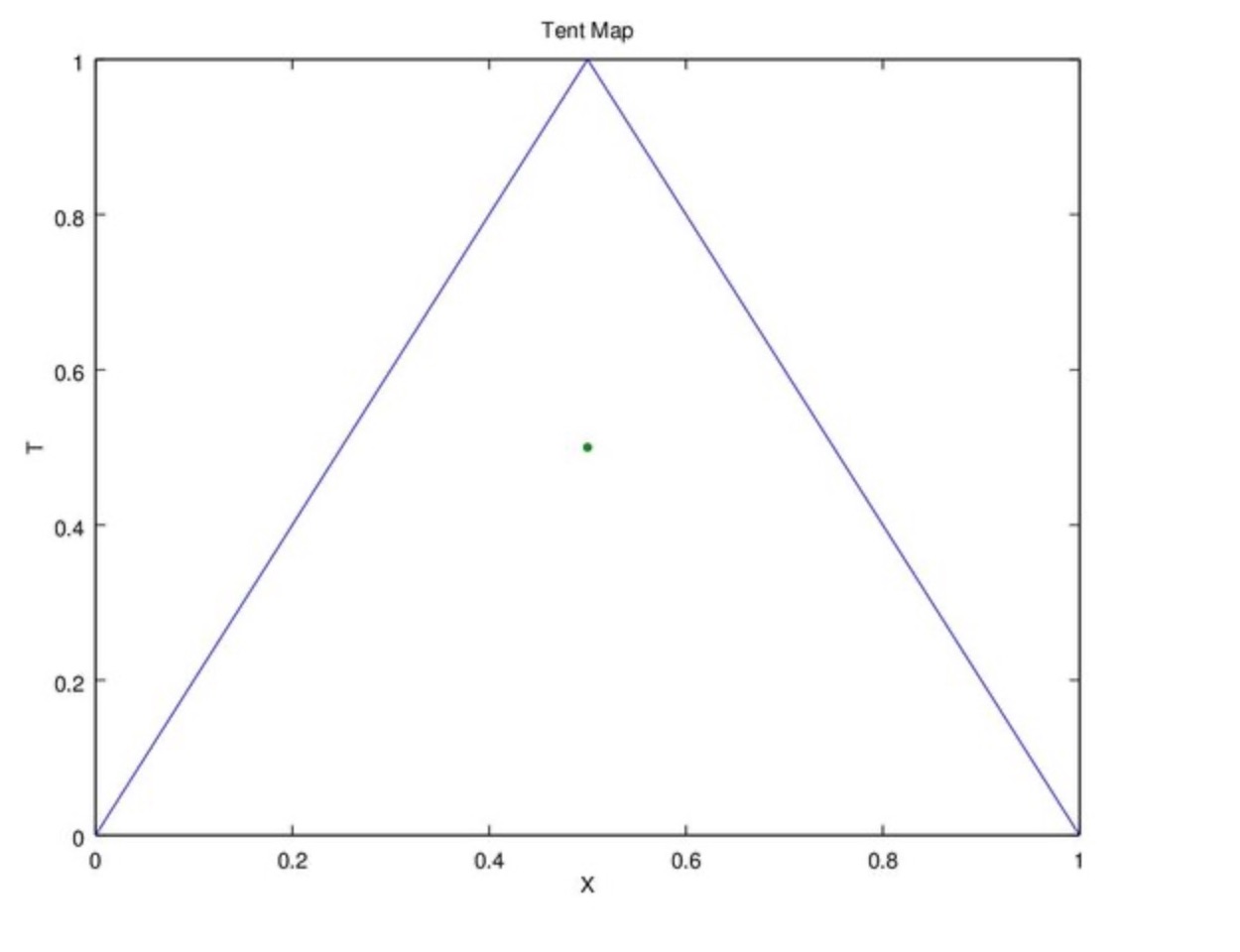}
\end{center}
 
 There are \emph{invertible} $2d$ versions of both maps, both called the \emph{Baker's map}. The $2d$ counterpart of the doubling map is the so-called \emph{unfolded Baker's map}.\footnote{The $2d$ extension of the tent map is given by $T(x,y)=(2x, y/2)$ for $0\leq x<\half$ and  $T(x,y)=(2-2x, 1-y/2)$ for $\half\leq x<1$. Here also $\xi:X\raw 2^{\Z}$ is an isomorphism, differing from the one in the main text for the unfolded version by a twist, as for the tent map. This is called the \emph{folded Baker's map}.} 
  Here \er{28} is extended by
 \begin{align}
 X=[0,1)\x [0,1); \nn \\
 T_B(x,y)=(2x, y/2) \:\:\: (0\leq x<\half); && T_B(x,y)=(2x-1, y/2+1/2) \:\:\: (\half\leq x<1). \label{Bakmap}
 \end{align} 
 This map consists of three stages, visible if we divide the square into two equal columns:\footnote{We follow Shields (1996), \S I..2.b, almost \emph{verbatim}, and also steal two of his pictures, viz.\ Figure I.2.4 and I.2.5.}
 \begin{enumerate}
\item Each column is compressed vertically  by a factor $\half$;
\item Each compressed column is subsequently stretched horizontally by a factor 2;
\item The right rectangle is finally placed on top of the left one (is this really what bakers do?).\vspace{-2.5mm}
\end{enumerate}
  \begin{center}
 \includegraphics[width=0.7\textwidth]{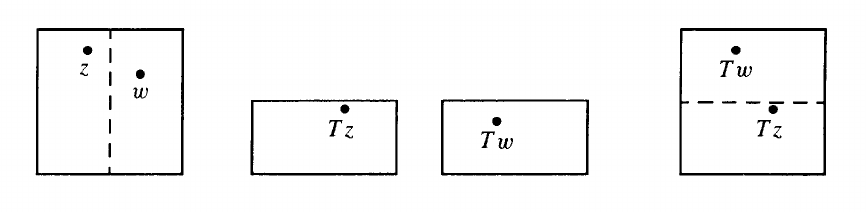}\vspace{-5mm}
\end{center}
By either calculation or visualization, it follows that $T_B$ preserves $2d$ Lebesgue measure.  Short of defining chaos, this is arguably the most chaotic map known: if all points in the  columns
\begin{align}
U_0=\{(0\leq x<\half, 0\leq y<1)\}; && U_1=\{(\half\leq x<1, 0\leq y<1)\} \label{u0u1}
\end{align}
 are marked red and blue, respectively, then the two colours look completely mixed after a dozen or so steps.\footnote{See for example the movie on \url{https://en.wikipedia.org/wiki/Baker's_map}. } The simplest coarse-graining already gives an isomorphism with the shift map, albeit it the two-sided or bilateral version: instead of $2^\N$ we now use $2^\Z$, carrying a shift map
\begin{align}
S: 2^{\Z}\raw 2^{\Z}; && (Ss)_n:=s_{n+1} \:\:\: (n\in\Z). \label{BLS}
\end{align}
We now use the binary expansion \er{xomega} of $x$, and a relabeled binary expansion for $y$:
\begin{equation}
 y=\Sigma_{n=0}^{\infty} s_{-(n+1)}2^{-(n+1)}.
\end{equation}
Thus $(x,y)$ corresponds to a single sequence $s\in 2^\Z$, and it is easy to show that under this correspondence the map \er{Bakmap} is exactly the bilateral shift \er{BLS}.
\begin{exercise}
Show this.
\end{exercise}

 Consequently, the (unfolded) Baker's map is essentially the same as the bilateral Bernoulli shift \er{BLS}.
 Using $\Z$ instead of $\N$, we  replace the subset $N\equiv \{0, 1, \ldots, N-1\}\subset\N$ by 
 \begin{equation}
\Lm_N:=\{-N, \ldots, 0, \ldots, N\},
\end{equation}
and in its wake replace $A^N$ by $A^{\Lm_N}=\{\sg:\Lm_N\raw A\}$,  here with $A=\{0,1\}$, and replace \er{defxi} by
\begin{align} \xi: X\raw A^{\Z}; &&  \xi(x)_n=a\in A \:\:\mbox{ iff } \:\:T^nx\in U_a \:\:\: (n\in\Z), \label{defxiZ}
\end{align}
 which may be truncated to $\Lm_N$ so as to obtain the finite-time coarse-grained path
 \beq
 \xi_{\Lm_N}(x)=\xi(x)_{|\Lm_N}\in A^{\Lm_N}.
 \eeq
Applying this to the Baker's map \er{Bakmap} via  the partition \er{u0u1} leads to the conclusion that the map $\xi$ in \er{defxiZ}, with $A=\{0,1\}$ and $X=[0,1)\x[0,1)$, is a bijection, so that coarse-graining is lossless.  To see this, one again starts with the bilateral shift \er{BLS}, with corresponding partition 
\begin{align}
U_0'=\{s\in 2^\Z\mid s_0=0\}; && U_1'=\{s\in 2^\Z\mid s_0=1\},
\label{U0U1}
\end{align}
and repeats the arguments for the doubling map almost \emph{verbatim}. 

\noindent
 Returning to the general story:  Kolmogorov's idea was to refine the partition \er{Xdu}, written as
  \begin{align}
 \pi&=\{U_a, a\in A\}; && X=\bigcup_a U_a; && P(U_a\cap U_b)=0\:\: (a\neq b); &&
P(\cup_a U_a)=1, \label{parpi}  
 \end{align}
   to a finer partition $ \pi^N$, defined for all $N\geq 1$, by \begin{align}  
   \pi^N&=\{U_{\sg}, \sg\in A^N\}; \label{defpiN}\\
 U_{\sg}&:=\bigcap_{n=0}^{N-1} T^{-n}  U_{\sg_n}= U_{\sg_0}\cap T\inv U_{\sg_1}\cap \cdots \cap T^{-(N-1)}U_{\sg_{N-1}}\nn \\ &
   = \{x\in X\mid x\in U_{\sg_0}, Tx\in U_{\sg_1}, \ldots T^{N-1}x\in U_{\sg_{N-1}}\},
   \label{Xsg0}
 \end{align}
i.e.\ the set of all $x$ whose coarse-grained path $\xi_N(x)$ up to time $N-1$ corresponds to $\sg\in A^N$;
here $T^0=\mathrm{id}_X$ so that $\pi^1=\pi$. Note that $\sg\in A^N$ is just a label identifying such a coarse-grained path, which itself is given by the sequence $(U_{\sg_0}, U_{\sg_1}, \ldots  U_{\sg_{N-1}})$ of elements of the partition $\pi$ in which the particle resides at time $n\in\{0, 1, \ldots, N-1\}$. 
The set $U_{\sg}$  is empty for those $\sg\in A^N$ that (for given map $T$)  are not in the image of the map $\xi_N: X\raw A^{N}$. Such labels $\sg$ should be omitted in $A^N$, so that in general $\pi^N$ corresponds to some subset of $A^N$. Here are (\emph{the}) two extreme cases:
\begin{enumerate}
\item If $T=\mathrm{id}_X$, i.e., $T(x)=x$, then $U_{\sg}\neq\emptyset$  iff $\sg_n=a$ for all $n\in\{0, 1, \ldots, N-1\}$, for some $a\in A$.
\item If $T=S$ for $X=A^\N$, with partition $U_a=\{s\in A^\N\mid s_0=a\}$,
then all $\sg$ occur, since
\begin{equation}
U_{\sg}=[\sg]_N=\{s\in A^\N\mid s_{|N}=\sg\}.
\end{equation}
\end{enumerate}
 We also display a few partitions for the Baker's map (also forward in time), starting with \er{U0U1}:
 \begin{center}
 \includegraphics[width=0.7\textwidth]{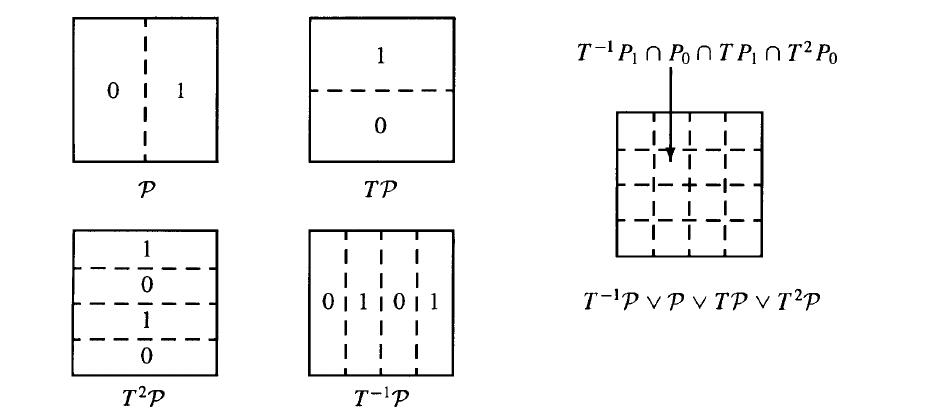}
\end{center}

 Within the image of $\xi_N$,
$\xi_N(x)\in A^N$ bijectively corresponds to the subset $U_{\xi_N(x)}\in \pi^N$. 
If for any partition $\gm$ we write $\gm(x)$ for the element of $\gm$ containing $x$, 
then our notation implies that
 \begin{equation}
\pi^N(x)= U_{\xi_N(x)}. \label{defpiNx}
\end{equation}
The given probability measure $P\in\Pr(X)$ induces a probability distribution $P_N\in\Pr(A^N)$ via
\begin{equation}
P_N(\sg)=P(U_{\sg}),
\end{equation}
interpreted as the probability of some truncated coarse-grained path labeled by $\sg$. This suggests (base-$e$) notions of  information and entropy, as follows. First, for any finite measurable partition $\gm$ of $(X,P)$, suppressing our fixed $X$ in the notation,  we define information and entropy by, cf.\ \er{RE},
\begin{align}
I_P(\gm)&:=\Sigma_{C\in\gm} 1_C \log\, (1/P(C))=-\Sigma_{C\in\gm}1_C \log\, P(C);\label{IPalfa} \\
H_P(\gm)&:=\la I_P(\gm)\ra_P=-\Sigma_{C\in\gm} P(C) \log P(C). \label{ppKS2}
\end{align}
Note that  $I_P(\gm)$ is a (measurable) function on $X$, so that $\gm$ is not its argument but a label, and
\begin{align}
I_P(\gm)(x)&=  -\log\,P(\gm(x));\label{236}\\
  I_P(\pi^N)(x) &= -\log\,P (\pi^N(x))=-\log P( U_{\xi_N(x)})=-\log P_N(\xi_N(x)).   \label{Imany} 
\end{align}
\bex If $S(P_N)$ is the entropy \er{RE} of the probability space $(A^N, \CP(A^N), P_N)$, show that
\beq
  H_P(\pi^N) = S(P_N).\label{HPpiN}
\eeq\eex
 For any two finite partitions $\gm$ and $\beta$ of $X$ we similarly introduce the \emph{conditional information} and \emph{conditional entropy} of $\gm$ given $\beta$ by
\begin{align}
I_P(\gm|\beta)&:= -\Sigma_{C\in\gm, B\in\beta} 1_{C\cap B} \log P(C|B);\label{IPc}
\\
H_P(\gm|\beta)&:=\la I_P(\gm|\beta)\ra_P=
-\Sigma_{C\in\gm, B\in\beta} P(C\cap B) \log P(C|B), \label{cE}
\end{align}
where $P(C|B)=P(C\cap B)/P(B)$ as usual, provided $P(B)>0$. Clearly, for all $x\in X$,
\begin{align}
I_P(\gm)(x)\geq 0; && I_P(\gm|\beta)(x) \geq 0; && H_P(\gm)\geq 0; && H_P(\gm|\beta)\geq 0.\label{groter}
\end{align}
The fact that the set  of all  partitions of some given set (or measure space) $X$ form a lattice
 plays a major role in the study of entropy. If
 $\gm$ and $\beta$ are partitions of $X$, we define $\beta\leq\gm$ if for all $C\in\gm$ there is $B\in\beta$  such that $C\subset B$. In that case $\gm$ is called a \emph{refinement} of $\beta$ (for example, the partition $\gm$ of a unit square into four equal squares of size $\half$ is a refinement of its partition $\beta$ into two equal columns).
 This partial order has suprema $\vee$ and infima $\wed$ (making it a lattice); we only need 
 \begin{equation}
\gm\vee\beta=\{C\cap B\mid C\in\gm, B\in\beta\}.
\end{equation}
For example, in our dynamical system $(X,P,T)$ we have $T^{-n}\pi=\{B\subset X\mid T^n(B)\in\pi\}$ and 
\beq
\pi^N=\vee_{n=0}^{N-1} T^{-n}\pi=\bigcap_{n=0}^{N-1} T^{-n}\pi=\{U_{\sg}, \sg\in A^N\}, \label{pinot}
\eeq
 cf.\ \er{Xsg0}, where $N\geq 1$; we omit those $\sg\in A^N$ for which $U_{\sg}=\emptyset$.
  If $M\geq N$, then $\pi^M\leq \pi^N$. 
  \begin{lemma}\label{KSlemma0}
\begin{enumerate}
\item
For any two partitions $\gm,\beta$ of $X$ we have 
\begin{align}H_P(\beta|\gm)&\leq H_P(\beta); \label{opineq}\\
H_P(\gm\vee\beta)&=H_P(\gm)+H_P(\beta|\gm), \label{opineq2}\\
\Raw\:\: H_P(\gm\vee\beta)&\leq H_P(\gm)+H_P(\beta). \label{subadE}
\end{align}
\item
If $\beta\leq\gm$, then,  for any third partition $\al$ in the third inequality, 
\begin{align}
H_P(\beta)\leq H_P(\gm); && H_P(\beta|\gm)=0;&&
H_P(\al|\gm)\leq H_P(\al|\beta). \label{gamma}
\end{align}
\item
For the information functions we have
\begin{equation}
I_P(\gm\vee\beta)=I_P(\gm)+ I_P(\beta|\gm). \label{ex941}
\end{equation}
\end{enumerate}
\end{lemma}
\begin{exercise}
Prove parts 1 and 2. Hint: for \er{opineq} use Jensen's inequality \er{Jensen}.  \QED
\end{exercise}

For example, in the unit square example (with Lebesgue measure) where $\beta$ consists of two columns and $\gm$ of four equal ``quarter-squares'',
 we have $H_P(\gm)=2$ whilst $H_P(\beta)=1$. Intuitively, if  partition $\gm$ refines partition $\beta$ and hence has more blocks, then revealing the location of a particle by specifying the block of $\gm$ it is in provides more information than  revealing the corresponding block of $\beta$; and hence $\gm$ should have more entropy than $\beta$. Similarly, in dynamical systems, if $M>N$ then  knowing a coarse-grained path from time $t=0$ till time $t=M-1$ gives more information than knowing it only until time $t=N-1$, so that $H_P(\pi^M)\geq H_P(\pi^N)$ as expected. 

We now show that   $H_P(\pi^N)$ as defined in \er{HPpiN} has a  limit 
\beq
h_P(\pi) := \lim_{N\raw\infty} \frac{1}{N} H_P(\pi^N).\label{preKS}
\eeq The key to its existence is the subadditivity property
\begin{equation}
 H_P(\pi^{M+N})\leq H_P(\pi^N)+ H_P(\pi^{M}),\label{KSsa}
\end{equation}
which follows from Lemma \ref{KSlemma0}. To see this,  for $0\leq k\leq N-1$  extend the notation \er{pinot} to 
\begin{align}
\pi_k^N:=\vee_{n=k}^{N-1} (T^{-n}\pi)=\bigcap_{n=k}^{N-1} (T^{-n}\pi), \label{pi1}
\end{align}
so that  e.g.\ $\pi_0^N=\pi^N$ and $\pi_1^{N+1}=T\inv \pi^N$.
Eqs.\ \er{ppKS2} and \er{PTB},  and subsequently  \er{subadE}, give
\begin{align}
H_P( \pi^{N+M}_M)&=H_P(\pi^N); \hspace{35pt} \Rightarrow  \\
 H_P(\pi^{M+N})&=H_P(\pi^N\vee \pi_N^{M+N}) \leq H_P(\pi^N)+ H_P(\pi_N^{M+N})=
H_P(\pi^N)+ H_P(\pi^{M}).
\end{align}\vspace{-5mm}
\begin{lemma}[Fekete]\label{subadlemma} 
If $(a_N)$ is a subadditive sequence in $\R^+$ in the sense that $a_{M+N}\leq a_M+a_N$, then $\lim_{N\raw\infty} (a_N/N)$ exists (its value may be $\infty$) and equals $\inf_N (a_N/N)$. 
\end{lemma}
\begin{exercise} Prove this lemma (you may look at the internet or elsewhere!).
\end{exercise} 
Eq.\ \er{KSsa} and lemma \ref{subadlemma} give the existence of  \er{preKS}. Alternatively, we can prove that
\begin{align}
h_P(\pi) =\lim_{N\raw\infty} H_P(\pi|\pi^N_1).\label{HPalt}
\end{align}
First, the limit exists, because for $M>N$ the partition $\pi^M_1$ is obviously finer than  $\pi^N_1$ (i.e.\
$\pi^M_1\leq \pi^N_1$), so that  $H_P(\pi|\pi^M_1)\leq H_P(\pi|\pi^N_1)$ by \er{gamma}. Moreover,
by definition \er{IPc} we have 
\beq
I_P(\gm|\beta)\geq0, \eeq
and hence  $H_P(\gm|\beta)\geq0$ by \er{cE}, so that
the sequence $(H_P(\pi|\pi^N_1))_N$ is non-increasing and bounded below, so that it converges. This also implies  that its limit equals its C\'{e}saro limit, that is, 
\begin{equation}
\lim_{N\raw\infty} H_P(\pi|\pi^N_1)=\lim_{N\raw\infty}\frac{1}{N} \Sigma_{j=1}^{N} H_P(\pi|\pi^j_1).
\end{equation}
Eq.\ \er{HPalt} then follows immediately from \er{preKS} and another nontrivial formula, viz.
\begin{equation}
H_P(\pi)+\Sigma_{j=1}^{N-1} H_P(\pi| \pi_1^{j+1}) =H_P(\pi^N) .\label{250}
\end{equation}\vspace{-5mm}
\begin{exercise}
Prove this  by induction.
\end{exercise}
\begin{definition}
The \emph{Kolmogorov--Sinai} (or \emph{metric}) \emph{entropy} of the dynamical system $(X,P,T)$ is
\begin{equation}
h(X,P,T):= \sup_{\pi}\, h_P(\pi),\label{defKSE}
\end{equation}
where the supremum (which may be infinite) is taken over all finite measurable partitions of $X$.
\end{definition}
 Here the right-hand side tacitly depends on $T$ via \er{preKS} and \er{Xsg0}. The reason this lofty quantity is often accessible lies in the crucial \emph{Kolmogorov--Sinai theorem}:
\begin{theorem}\label{KSGen}
If $\pi$ is a finite partition of $X$ whose  refinements $\pi^N$ generate the $\sg$-algebra $\Sg$, i.e.,
\begin{equation}
\Sg=\sg\left(\bigcup_{n=0}^{\infty}T^{-n}\pi\right)=\sg(\{\pi^N, N\in\N_*\}),
\label{genF}
\end{equation}
on which the probability measure $P$ was defined in the first place, then
\begin{equation}
h(X,P,T)=h_P(\pi). \label{KSlemma}
\end{equation}
\end{theorem}
We omit the proof;\footnote{See e.g.\ Viana \& Oliveira, Theorem 9.2.1.} the main idea is that the assumption means that as $N\raw\infty$ the refinements $\pi^N$ of  $\pi$ generate a maximally refined \emph{measurable} partition of $X$: the cells of $\pi^N$ could not eventually go beyond the elements of the given $\sg$-algebra $\Sg$. By Lemma \ref{KSlemma0}.2, entropy increases under refinement and hence a maximally refined partition should reach the sup in \er{defKSE}. 

 The first nontrivial case where this entropy was computed from Theorem \ref{KSGen} was the unilateral shift \er{ULS}, which gives the same result as the \emph{Bernoulli shift}, i.e.\ the same map $S$ but now defined on double-sided sequences $s\in A^{\Z}$ instead of $A^{\N}$. We recall that the  Bernoulli measure $p^{\N}$ on $A^\N$  is
 defined on the $\sg$-algebra generated  by the cylinder sets \er{cylinder}
 by  
 \beq
 p^{\N}([\sg]_N):=p^N(\sg), \label{pompN}
 \eeq
uniquely extended  to the $\sg$-algebra $\Sg$ generated by these cylinder sets.\footnote{See Definitions \ref{defsigmaF} or \ref{defF} for more information.}
Similarly, $\sg\in A^{[-N,N]}$ defines $[\sg]_N=A^{-\N}\sg A^{\N}$, the set of all sequences in $A^{\Z}$ with middle segment $\sg$. The probability measure $p^{\Z}$ is then defined analogously on the 
  $\sg$-algebra generated  by these cylinder sets, where $N\in\N_*$ and $\sg\in A^{[-N,N]}$. Using Theorem \ref{KSGen}, Sinai (1959) found the beautiful result
\begin{equation}
h(A^{\N}, \, p^\N,\, S)=h(A^{\Z},\, p^{\Z},\, S)=
S(p):=-\Sigma_{a\in A} p(a)\log p(a). \label{SinaiSp}
\end{equation}
To see this, let $X=A^{\N}$, $P=p^{\N}$, and $T=S$, cf.\ \er{ULS}; the double-sided case is similar. Define \beq
U_a:=[a]_1=\{s\in A^{\N}\mid s_0=a\},\label{opening}
\eeq
which gives  the starting partition $\pi=\{U_a\}_{a\in A}$ of $A^\N$, cf.\ \er{defpiN} - \er{parpi}. It follows that \begin{align}
U_{\sg}=[\sg]_N; &&  \pi^N=\{[\sg]_N, \sg\in A^N\}, \label{UsgsgN} 
\end{align}
 cf.\ \er{Xsg0}. Therefore, $\pi$ is such that \er{genF} and hence \er{KSlemma} holds. Now finish:
\bex Prove \er{UsgsgN}  and on that basis derive the key step in the proof of \er{SinaiSp}, namely
\begin{equation}
H_{p^{\N}}(\pi^N)=
NS(p). \label{KeySinai} 
\end{equation}
Finally, use this result with  \er{preKS} and \er{KSlemma} to derive \er{SinaiSp}.
\eex

To close this section we show that the relative entropy $S_{\Lm}(P,Q)$  for lattice systems in classical statistical mechanics, defined in \er{intsum}, is a generalized Kolmogorov--Sinai  entropy of a suitably generalized 
dynamical system. To define what this means in the present context, we still start from Definition \ref{defDS}, in which we take $X=\Om=A^{\Z^d}$, as in \er{920}, with $\Sg=\mathcal{F}$ as in Definition \ref{defsigmaF}, and $P$ any translation-invariant probability measure on $(\Om,\CF)$, as explained after \er{Zdaction}. The map $T$ in \er{TXX} is first replaced by the ensuing $\N$-action $\N\x X\raw X$ on $X$ (or $\Z$-action $\Z\x X\raw X$  if $T$ is invertible), i.e., $(n,x)\mapsto T^nx$, and then generalized to the $\Z^d$-action on $\Om$ defined in  \er{Zdaction}. 

To start the machinery, for the partition  \er{parpi} of $\Om$ we  take  the  partition with elements 
\begin{align}
\pi=(U_a)_{a\in A}; && 
U_a=\{\om\in A^{\Z^d}\mid\om_0=a\}.\label{canpart}
\end{align}
Since $N$ in \er{Xsg0} is now replaced by $\Lm_N$, or more generally by $\Lm$,  see \er{rectZ}, 
 we now refine $\pi$ to
 \begin{align} \pi^{\Lm}=\{U_{\sg}, \sg\in A^{\Lm}\};  &&
  U_{\sg}:=\bigcap_{x\in\Lm}\theta^*_{-x}
  U_{\sg_x},\label{Xsgz}
 \end{align}
 see again \er{Zdaction}.
In $d=1$, with $\Lm=\{0, \ldots, N-1\}$, this actually reproduces \er{Xsg0}. 
\begin{proposition}
Using  \er{canpart}, the entropy $S_{\Lm}(P)$ in \er{SLmP} and its limit \er{dQP1} are given by
\begin{align}
S_{\Lm}(P)=H_P(\pi^{\Lm}); && s(P):= \lim_{N\raw\infty} \frac{S_{\Lm_N}(P)}{|\Lm_N|}=h_P(\pi).
\label{9141}
\end{align}
\end{proposition}
\emph{Proof.}
The elements of $\pi^{\Lm}$ are precisely the cylinder sets, that is, for $\sg\in A^{\Lm}$ we have
\beq
U_{\sg}=[\sg]_{\Lm}=\{\om\in A^{\Z^d}\mid\om_{|\Lm}=\sg\},\label{9142}
\eeq
so that 
 \beq
 H_P(\pi^{\Lm})=-\Sigma_{\sg\in A^{\Lm}} P([\sg]_{\Lm})\log P([\sg]_{\Lm}).\label{9142b}
 \eeq 
 \bex Prove \er{9142} and \er{9142b}. \eex
 The  Radon--Nikodym derivative in \er{KLD4} by definition must be $\F_{\Lm}$-measurable and must satisfy
 \begin{equation}
P(B)=\int_B dQ(\om)\, \frac{dP}{dQ}(\om), \label{9143}
\end{equation}
for any $B\in \F_{\Lm}$, i.e.\ for any cylinder set $[\sg]_{\Lm}$ as in \er{9142}. By \er{intsum0}, for $Q=f^{\Z^d}$ 
 this equals
 \begin{equation}
\frac{dP}{dQ}(\om)=\frac{P([\om_{|\Lm}])}{Q([\om_{|\Lm}])}=|A|^{|\Lm|}P([\om_{|\Lm}]),
\end{equation}
cf.\ \er{intsum0}.
Since $P\ll f^{\Z^d}$ for any $P$, we may use \er{KLD4}. Since $P$ is defined on $\F_{\Lm}$, we obtain:
\begin{align}
S_{\Lm}(P,f^{\Lm})&=\int_{\Om} dP(\om)\, \log\left(  |A|^{|\Lm|}P([\om_{|\Lm}])\right)=|\Lm|\log|A|+
\int_{\Om} dP(\om)\, \log P([\om_{|\Lm}])\nn \\ &=|\Lm|\log|A|+\Sigma_{\sg\in A^{\Lm}} P([\sg]_{\Lm})\log P([\sg]_{\Lm})\nn \\ &= |\Lm|\log|A| -H_P(\pi^{\Lm}).
\end{align}
With \er{940} this gives the first part of \er{9141}. The second part is obvious from \er{preKS}. \QED
\section{Bridge: From classical to quantum via C*-algebras}\label{FromCQ}
We now move from classical to quantum theories of entropy. At least in mathematical physics, it has been clear that operator algebras form a powerful and flexible mathematical language for quantum theory. As we shall see, quantum entropies confirm this. 
The operator-algebraic formalism for quantum theory is usually motivated by difficulties with the \Hs\ formalism  when it tries to describe infinite systems, like quantum field theory and quantum statistical mechanics in the thermodynamics limit.\footnote{See Haag (1992). Historical milestones include Haag \& Kastler (1964) and  Haag, Hugenholtz, \&  Winnink (1967).} Our approach, especially in the context of entropy, is rather based on the idea that operator algebras provide a unified language for classical and quantum theory.\footnote{This point of view is developed in Landsman (1998, 2017) and is also standard in quantum information theory.}

We start with a finite set $X$, seen classically as a configuration space.\footnote{When  $X$  is a manifold it is better interpreted as a phase space such as $X=T^*Q$, the cotangent bundel of a configuration space $Q$, although even in that case the quantum theory is usually based on $L^2(Q)$ rather than on $L^2(X)$.} One may think of $X=A^N$ for some  finite set $A$ and some $N\in\N$. Theoretical physics then postulates the following.
\begin{center}
\textbf{Classical physics of a finite system}
\end{center}
\begin{itemize}
\item In classical physics, observables are (arbitrary)  \emph{functions} $f:X\raw\R$, whereas states are \emph{probability distributions} on $X$, i.e.\ functions $p:X\raw[0,1]$ satisfying 
$\Sigma_{x\in X} p(x)=1$.
 All predictions eventually come from \emph{expectation values} of observables $f$ in  states $p$, given by
\begin{equation}
\la f\ra_p=\Sigma_{x\in X} p(x) f(x). \label{CEV}
\end{equation}
The \emph{state space} of the system is the set $\Pr(X)$ of all probability distributions on $X$.
The points $x\in X$ correspond to special (``pure'') states $p=\dl_x$ (where $\dl_x(y)=\dl_{xy}$), in which
\begin{equation}
\la f\ra_{\dl_x}=f(x).
\end{equation}
The \emph{spectrum} $\sg(f)$ of $f$ is its range, i.e., $\sg(f)=\{f(x),x\in X\}\subset\R$. For any $\lm\in\sg(f)$,
\begin{equation}
X_{\lm}=\{x\in X\mid f(x)=\lm\}=f\inv(\{\lm\})
\end{equation}
is the \emph{spectral subset} (of $X$) for $\lm\in\sg(f)$,  with characteristic function $1_{X_{\lm}}$. By \er{CEV},
\begin{align}
\la f\ra_p=\Sigma_{\lm\in\sg} \lm p(f=\lm); &&
p(f=\lm):=p(X_{\lm})=
\la 1_{X_{\lm}}\ra_p.\label{82}
\end{align}
If we randomly sample $(X,p)$, this is  the sum of $p(x)$ over all $x\in X$ where $f(x)=\lm$. If $\lm$ is nondegenerate in that there is a unique $x$ such that $f(x)=\lm$, then \er{82} implies that
\begin{equation}
p(f=\lm)=p(x).\label{82p}
\end{equation}
\item Two systems described by sets $A$ and $B$ are combined via the \emph{Cartesian} product 
 $C=A\x B$. States $p\in\Pr(A)$ and $q\in \Pr(B)$ naturally combine to form a state  $r\in\Pr(C)$ via 
 \begin{equation}
r(a,b)=p(a)q(b). \label{ABind}
\end{equation}
But not nearly all states $r\in\Pr(C)$ arise in this way. States $r$ in which \er{ABind} fails are called \emph{correlated} (so this is a property of states, not of systems). Conversely, any state $r\in\Pr(C)$ induces states $p=r_{|A}\in \Pr(A)$ and $q=r_{|B}\in\Pr(B)$ by taking marginals, defined by 
\begin{align}
p(a):=\Sigma_{b\in B} r(a,b); && q(b):=\Sigma_{a\in A} r(a,b). \label{marginals}
\end{align}
Note that pure states $r=\dl_{(a,b)}$ satisfy \er{ABind}, with $p=\dl_a$ and $q=\dl_b$, and hence have pure states (viz.\ $\dl_a$ and $\dl_b$) as marginals. 
If we start from an arbitrary  state $r\in\Pr(C)$, take marginals \er{marginals}, and recombine these via \er{ABind}, we recover the given state $r$ iff it was given by  \er{ABind} in the first place. If the state is correlated, taking marginals loses information.
\end{itemize}
\begin{center}
\textbf{Quantum physics of a finite system}
\end{center}
\begin{itemize}
\item Let $H=\ell^2(X)$ be the \Hs\  of all functions $\psi:X\raw\C$ with inner product
\begin{equation}
\la\ps,\phv\ra=\Sigma_{x\in X} \ovl{\ps(x)}\phi(x)\label{l2ip}.
\end{equation}
Observables are \emph{linear maps}  $T:H\raw H$ such that $T^*=T$, whilst states are \emph{density matrices}, i.e., linear maps $\rh: H\raw H$ that satisfy $\rh\geq 0$ and $\Tr(\rh)=1$. Here the positivity condition $\rh\geq 0$ means that $\la\ps,\rh\ps\ra\geq 0$ for all $\psi\in H$, and the \emph{trace} of an operator $T$ is defined by
\beq\Tr(T)=\Sigma_i \la \ups_i, T\ups_i\ra,
\eeq 
where  $(\ups_i)$ is an arbitrary basis of $H$ (by which we always mean an \emph{orthonormal} basis). A nontrivial argument yields independence of this sum of the choice of this basis, so that $\Tr(T)$ is well defined.
All predictions eventually come from the expectation values
\begin{equation}
\la T\ra_{\rh}=\Tr(\rh T). \label{QEV}
\end{equation}
The \emph{state space} of the system is the set $D(H)$ of all density matrices. Each unit vector $\psi\in H$ defines a special  \emph{pure} state $\rh_{\psi}\in D(H)$ (in which the phase of $\psi$ is famously lost),\footnote{Vectors are simply written as $\ps\in H$, rather than Dirac's $|\psi\ra$, and we  write $\la\ps,\phv\ra$ for the inner product of $\ps$ and $\phv$, instead of Dirac's $\la\ps|\phv\ra$. We do use Dirac's convenient notation 
$|\ps\ra\la\phv|$ for the operator defined by $|\ps\ra\la\phv|\chi=\la\phv,\chi\ra\ps$. In the special case that $\phv=\psi$ and this is a unit vector, $|\psi\ra\la\psi|$ is the one-dimensional projection onto the subspace $\C\cdot\psi$.} namely
 \beq
\rh_{\psi}=|\psi\ra\la\psi|.\label{89}
\eeq
In such states the expectation values \er{QEV} come down to
\begin{equation}
\la T\ra_{\rh_{\psi}}=\la \psi, T\ps\ra.
\end{equation}
The \emph{spectrum} $\sg(T)$ of $T$ is its set of eigenvalues $\lm\in\R$ (given that $T^*=T$), with  eigenspaces 
\begin{align} 
H_{\lm}=\{\psi\in H\mid T\ps=\lm\ps\} && (\lm\in\sg(T)).
\end{align}
These are linear subspaces of $H$ and hence bijectively correspond to projections 
\begin{align}
E_{\lm}:H\raw H; && E_{\lm}\psi=\ps && (\ps\in H_{\lm}); &&  E_{\lm}\ps=0 && (\ps\in H_{\lm}^{\perp}),\label{defElm}
\end{align}
where $ H_{\lm}^{\perp}=\{\phv\in H| \la \chi,\phv\ra=0\, \forall\,\chi\in H_{\lm}\}$ is the orthogonal complement of $H_{\lm}$.
The subspace $H_{\lm}\subset H$ and the projection $E_{\lm}$ are the quantum analogues of the subset $X_{\lm}\subset X$ and the characteristic function $1_{X_{\lm}}$, respectively. If we now use the spectral resolution of $T$, 
\begin{align}
T=\Sigma_{\lm\in\sg(T)}\lm\cdot E_{\lm}; && 1_H=\Sigma_{\lm\in\sg(T)} E_{\lm},\label{SRTfd}
\end{align}
and use \er{QEV}, we see that quite analogously to the classical case \er{82}  we have
\begin{align}
\la T\ra_{\rh}=\Sigma_{\lm\in\sg(T)}\lm P_{\rh}(T=\lm); && P_{\rh}(T=\lm):=\la E_{\lm}\ra_{\rh}=
\Tr(\rh E_{\lm}).\label{BR1}
\end{align}
Since $E_{\lm}=\Sigma_i |\ps^{(i)}_{\lm}\ra\la\ps^{(i)}_{\lm}|$ for an arbitrary basis 
$(\ps_{\lm}^{(i)})$ of $H_{\lm}$, $i=1, \ldots, \dim(H_{\lm})$, we obtain
\begin{equation}
P_{\rh}(T=\lm)=\Sigma_i \la\ps^{(i)}_{\lm},\rh\ps^{(i)}_{\lm}\ra.\label{817}
\end{equation}
If $\lm$ is nondegenerate, with unique unit eigenvector $\ps_{\lm}$ (up to a phase), this gives
\begin{equation}
P_{\rh}(T=\lm)=\la\ps_{\lm},\rh\ps_{\lm}\ra,\label{817bis}
\end{equation}
which for pure states \er{89} recovers the original \emph{Born rule} from 1926, namely
\begin{equation}
P_{\rh_{\psi}}(T=\lm)=|\la\ps_{\lm},\ps\ra|^2.\label{Born}
\end{equation}
The right-hand side is the \emph{transition probability} from $\ps$ to $\ps_{\lm}$ (more precisely, from $\rh_{\psi}$ to
 $\rh_{\psi_{\lm}}$). 
In general we may use the spectral resolution of $\rh$: since $\rh\in D(H)$, the eigenvalues of $\rh$ (repeating degenerate ones) form a probability distribution $p\in \Pr(A)$ on some set $A$ labeling the eigenvectors 
$(\ups_a)$ of $\rh$, so that $\dim(H)=|A|$ and $\rh\ups_a=p(a)\ups_a$. Hence
\begin{equation}
\rh=\Sigma_{a\in A} p_a|\ups_a\ra\la \ups_a|,\label{sdrho}
\end{equation}
and, still assuming that $\lm\in\sg(T)$ is nondegenerate, eq.\ \er{817} comes down to
\begin{equation}
P_{\rh}(T=\lm)=\Sigma_{a\in A} p_a |\la\ups_a,\ps_{\lm}\ra|^2.\label{817rho}
\end{equation}
Thus the classical probability space $(X,p)$ is replaced by $(H,\rh)$, but (at least according to the Copenhagen interpretation of quantum mechanics)  one cannot simply sample this ``quantum probability space'' without further ado; sampling $(H,\rh)$ is predicated on the choice of some observable $T$ and some experimental setup to measure $T$. The actual measurement of $T$ may still be seen as an act of randomly sampling from $H$, but now the outcome is not an arbitrary unit vector $\psi\in H$, or, more precisely, the corresponding pure state \er{89}, but some unit eigenvector $\ps_{\lm}\in H_{\lm}$ of $T$ (up to a phase), where  $\lm\in\sg(T)$. Applying this to  a maximal self-adjoint operator $T$ (whose spectrum is by definition nondegenerate),  the probability $P_{\rh}(\ps_{\lm})$ of obtaining $\ps_{\lm}$ (up to a phase) in a state $\rh$ equals \er{817bis}, i.e., 
\begin{equation}
P_{\rh}(\ps_{\lm})= \Tr(\rh |\psi_{\lm}\ra\la\psi_{\lm})=\la\ps_{\lm},\rh\ps_{\lm}\ra.
\end{equation}
\item In quantum physics systems are combined via the \emph{tensor product} rather than the Cartesian product, but these are closely related: for finite sets $A$ and $B$ we  \emph{define} the tensor product as
\begin{equation}
\ell^2(A)\ot \ell^2(B):=\ell^2(A\x B).\label{4.1}
\end{equation}
Moreover, we have a canonical bilinear map (that is, linear in each of its two arguments)
\begin{align}
\ot: \ell^2(A)\x \ell^2(B)\raw \ell^2(A)\ot \ell^2(B); && \ot(\psi_A,\psi_B)\equiv \ps_A\ot\ps_B:=\psi_A\psi_B,\label{pot} 
\end{align}
which defines \emph{elementary tensors}. All elements of $\ell^2(A)\x \ell^2(B)$ are finite linear combinations of elementary tensors, since any function $\ps_{AB}:A\x B\raw \C$ is a finite sum of products 
\beq
\ps_{AB}(a,b)=\ps_A(a)\ps_B(b).
\eeq
Moreover, the canonical bases $(\dl_a)_{a\in A}$ of $\ell^2(A)$ and  $(\dl_b)_{b\in B}$ of $\ell^2(B)$ combine to the canonical basis $(\dl_a\ot\dl_b)_{a,b\in A\x B}$ of $\ell^2(A)\ot \ell^2(B)$; indeed, any basis  of $\ell^2(A)$ and basis of 
 $\ell^2(B)$,  combine to a basis of $\ell^2(A)\ot \ell^2(B)$ by forming elementary tensors. Since any finite-dimensional \Hs\ $H_A$ of dimension $\dim(H_A)=|A|$ is isomorphic to $\ell^2(A)$ by choosing a basis $(u_a)_{a\in A}$ and letting $u_a\in H_A$ correspond to $\dl_a\in\ell^2(A)$, this is all you need to know about their tensor products. In particular, either  \er{4.1} or the  basis construction gives
\begin{align}
\dim(H_{AB})=\dim(H_A)\cdot\dim(H_B); &&  H_{AB}:= H_A\ot H_B.
\end{align}
We turn to operators on tensor products, starting with $H_A=\ell^2(A)$ and $H_B=\ell^2(B)$, and hence $H_{AB}=\ell^2(A\x B)$. Then any $T_A\in L(H_A)$ defines an operator called $T_A\ot 1_{H_B}\in L(H_{AB})$  by simply ignoring the second argument $b$ of $\ps_{AB}(a,b)$, that is, by linear extension of 
\begin{equation}
T_A\ot 1_{H_B} (\psi_A\ot \ps_B):=(T_A\psi_A)\ot\ps_B. \label{8230}
\end{equation}
Likewise,  $T_B\in L(H_B)$ induces $1_{H_A}\ot T_B$ by ignoring the  argument $a$. Combining these, 
the  ``elementary tensor product'' of operators  $T_A\ot T_B\in L(H_{AB})$ is given by
 linear extension of 
\begin{equation}
T_A\ot T_B (\psi_A\ot \ps_B):=(T_A\psi_A)\ot (T_B\ps_B),
\end{equation}
and this can be done for any (finite-dimensional) $H_A$ and $H_B$. As can be seen from $\ell^2(A\x B)$,
 any operator on $H_{AB}$ is then a finite linear combination of such elementary tensors $T_A\ot T_B$.

The somewhat mysterious quantum counterpart of taking marginals is the \emph{partial trace}.\footnote{Below, an affine map between convex sets is a map that preserves finite convex combinations.} 
Namely, there are unique affine maps (called the \emph{partial trace} over $H_B$ and $H_A$, respectively)
\begin{align}
 \Tr_{H_B}: D(H_{AB})\raw D(H_A); &&  \Tr_{H_A}:D(H_{AB})\raw D(H_B), \label{8.25}
 \end{align}
whose images  we write as $\rh_A=\Tr_{H_B}(\rh_{AB})$ and 
 $\rh_B=\Tr_{H_A}(\rh_{AB})$, for $\rh_{AB}\in D(H_{B})$. These maps have the defining property that for all $T_A\in L(H_A)$ and $T_B\in L(H_B)$ one has
\begin{align}
\Tr(\rh_A T_A)=\Tr(\rh_{AB}( T_A\ot 1_{H_B})); && \Tr(\rh_B T_B)=\Tr(\rh_{AB}(1_{H_A}\ot T_B)). \label{PT1}
\end{align}
Here the traces without a suffix are over different Hilbert spaces, namely $H_A$, $H_{AB}$, $H_B$, and $H_{AB}$, respectively, but that should be clear from the formula, and writing suffices would compromise the notation in \er{8.25}. The resulting states
are called \emph{reduced density matrices}.

As in the classical case we may combine individual states $\rh_A\in D(H_A)$ and $\rh_B\in D(H_A)$ into a state $\rh_{AB}=\rh_A\ot\rh_B\in D(H_{AB})$ of the combined system, from which we can recover
\begin{align}
\rh_A= \Tr_{H_B}(\rh_A\ot\rh_B); && \rh_B= \Tr_{H_A}(\rh_A\ot\rh_B). \label{recover}
\end{align}
Conversely, we could say that the state $\rh_{AB}$ is correlated if $\rh_{AB}\neq \rh_A\ot\rh_B$, but unlike the classical case there is a sharp distinction within the class of correlated states, namely between those that are merely \emph{classically correlated}, in which case 
\begin{align} 
\rh_{AB}=\Sigma_i\rh_A^{(i)}\ot \rh_B^{(i)} &&   (\rh_A^{(i)}\in D(H_A), \, \rh_B^{(i)}\in D(H_B)),
\end{align}
 see also \er{Scc0}, and those that are \emph{quantum-mechanically correlated}, that is, 
   \emph{entangled}, which is the alternative possibility. The most spectacular cases of entangled states are those in which 
   $\rh_{AB}$ is pure, i.e.,  $\rh_{AB}=|\psi_{AB}\ra\la\psi_{AB}|$, cf.\ \er{EPS}, where $\psi_{AB}\neq\psi_A\ot\ps_B$, and hence $\psi_{AB}$ is a sum of such elementary tensors, as in the Bell states \er{Bellstates0}.
   
This brings us to one of the most profound difference between classical and quantum probability: \emph{the reduced density matrix of a pure state need not be pure}; in fact it is impure (= mixed) whenever $\rh_{AB}$ is entangled. We will return to this in some detail (see for example Theorem \ref{bigdifference}), but for now it may suffice to note that all four of the  Bell states \er{Bellstates0}, which are pure, reduce, on each copy of $\C^2$, to the maximally mixed state 
\beq
\rh_A=\rh_B=\half 1_{\C^2}.\label{Bellreduced2}\eeq
\end{itemize}

Even apart from such discrepancies  the mathematical structures for classical and quantum physics look quite different, and this difference persists (or even gets worse) if we go to manifolds $X$ and infinite-dimensional \Hs s $H$.
Fortunately, there is important common structure, and this is what the ``algebraic'' (more precisely, the operator-algebraic) approach to both delivers.

For a finite set $X$, let  $C(X)$ be the vector space
set of all complex-valued functions on $X$ (with pointwise linear operations). For some finite-dimensional \Hs\  $H$, let $L(H)$ be the vector space of all linear maps $T:H\raw H$. First, each of these is an \emph{algebra} in a natural way:\footnote{All our vector spaces (including Hilbert spaces)  and algebras are \emph{complex}, i.e., defined over $\C$. In a famous paper, Wigner (1960) 
counts the  use of complex numbers in quantum mechanics among one of his examples of the `unreasonable effectiveness of mathematics in the natural sciences', pointing out that `nothing in our experience suggests the introduction of
these quantities' in the first place, and that `the use of complex numbers is in this case not a calculational trick of applied
mathematics but comes close to being a necessity in the formulation of
the laws of quantum mechanics.' }
\begin{itemize}
\item In $C(X)$ the product $fg$ is pointwise, that is, $(fg)(x)=f(x)g(x)$, and is \emph{commutative}.\footnote{All our algebras are associative and, as per the previous footnote, are defined over the complex numbers.}
\item In $L(H)$ the product $ST$ is defined by $(ST)(\psi):=S(T\psi)$, which is \emph{noncommutative}.
\end{itemize}
Second, each of our algebras $\CA$ has a natural  \emph{involution}, i.e.,  a real-linear map $*:\CA\raw \CA$ (written $a\mapsto a^*$) that satisfies $(\lm a)^*=\ovl{\lm} a^*$ and $(ab)^*=b^*a^*$ (which is just $a^*b^*$ if $\CA$ is commutative). 
\begin{itemize}
\item In $C(X)$ the function $f^*$ is defined by pointwise complex conjugation, that is, $f^*(x)=\ovl{f(x)}$.
\item In $L(H)$ the operator $T^*$ is the hermitian conjugate (or adjoint) of $T$, uniquely defined by
\begin{align}
\la\phv, T^*\psi\ra:=\la T\phv,\ps\ra && (\psi,\phv\in H).
\end{align}
\end{itemize}
Third (though less relevant for finite $X$, but crucial in general),  both algebras have a natural \emph{norm}:
\begin{itemize}
\item In $C(X)$ this is the \emph{supremum-norm} or \emph{sup-norm}
\beq
\|f\|_{\infty}=\sup_{x\in X} \{|f(x)|\}.\eeq
\item In $L(H)$ it is the \emph{operator norm} 
\begin{align}
\|T\|:=\sup\{\|T\ps\|, \ps\in H, \|\ps\|\leq 1\}; && \|\psi\|=\sqrt{\la\ps,\ps\ra}.
\end{align}
\end{itemize}
These structures turn both $C(X)$  and $L(H)$ into \emph{C*-algebras}, a first step towards their unification!
\begin{definition}
 A \ca\  is an  algebra $\CA$  with an involution as well as a norm in which $\CA$ is complete (i.e., it is a Banach space),  in which  algebra, involution, and norm are related by 
\begin{align}
\|ab\| \leq& \| a\|\,\| b\|; &&
\|a^*a\|  =  \| a\|^2. 
\end{align}
\end{definition}
\bex
Show that $C(X)$ and $B(H)$ are indeed \ca s (for finite $X$ and $\dim(H)<\infty$).
\eex
\begin{definition}\label{stateCA}
A \emph{state} on a \ca\ $\CA$ with unit is a linear map $\om:\CA\raw\C$ that satisfies
\begin{align}
\om(a^*a)\geq 0; && 
\om(1_\CA)=1, \label{om1a1}
\end{align}
that is, \emph{positivity} and \emph{normalization}.
If $\CA$ does not have a unit,\footnote{In \er{om1} the norm is the usual one on the Banach dual $\CA^*$, i.e., 
$\|\om\|=\sup\{ |\om(a)|, a\in \CA, \|a\|\leq 1\|\}$.}
 then \er{om1a1} is replaced by  \beq
 \|\om\|=1. \label{om1}
 \eeq
\end{definition}
In fact the criterion \er{om1} can also be used if a \ca\ $\CA$ has a unit, in which case we have:
\begin{proposition}
Let $\om:\CA\raw\C$ be a positive  linear map. Then $\om$ is bounded, and 
\beq\|\om\|=\om(1_\CA).\eeq
\end{proposition}
This (algebraic)  notion of a state is what really unifies the probabilistic structures we started with:
\begin{proposition}\label{SRX}\begin{enumerate}
\item For finite sets $X$
there is a bijective correspondence between states $\om$ on $C(X)$ and probability distributions  $p$  on $X$, given by  \begin{align}
p(x)=\om(1_{\{x\}}); && \om(f)=\Sigma_{x\in X} p(x)f(x); && (f\in C(X)). \label{omfsum}
\end{align}
\item For finite-dimensional \Hs s $H$ there is a bijective correspondence between states  on $L(H)$ and density matrices: for any state $\om$ there is a unique 
$\rh\in D(H)$ such that
\begin{equation}
\om(T)=\Tr(\rh T). \label{826}
\end{equation}
\end{enumerate}
\end{proposition}
\bex
Prove this proposition. \QED
\eex

From this proposition we also see the significance and common origin of the familiar marginalization of probability distributions in the classical case and the mysterious partial traces in  quantum mechanics: both correspond to the same operation
of restricting  a state to a subalgebra, applied however to different \ca s, viz.\ those of the form $C(X)$ and $L(H)$, respectively.
\begin{itemize}
\item Classically, the \ca\ $C(A\x B)$ of a combined system has both $C(A)$ and $C(B)$ as subalgebras: the embedding  
\begin{align}
C(A)\hookrightarrow C(A\x B); && f_A\mapsto f_{A}\x 1_B,
\end{align} is just  
\beq
f_{A}\x 1_B(a,b):= f_A(a),
\eeq
 whereas  $C(B)\hookrightarrow C(A\x B)$, $f_B\mapsto 1_A\x f_{B}$,  is given by $1_A\x f_{B}(a,b):= f_B(b)$. If $\om_{AB}$ is the state on $C(A\x B)$ defined by some probability distribution $r\in\Pr(A\x B)$ via \ \er{omfsum}, i.e.,
\beq
\om_{AB}(f_{AB})=\Sigma_{(a,b)\in A\x B} r(a,b) f_{AB}(a,b),
\eeq
then the marginal $p=r_{|A}\in \Pr(A)$ is the probability distribution that corresponds, again via \er{omfsum}, to the state $\om_A$ on $C(A)$ obtained by $\om_A(f_A)=\om_{AB}(f_A\x 1_B)$. Similarly, the marginal $q=r_{|B}\in \Pr(B)$ corresponds to the state $\om_B$ on $C(B)$  obtained by $\om_B(f_B)=\om_{AB}(1_A\x f_A)$.
\item In quantum theory one has canonical embeddings 
\begin{align}
L(H_A) \hookrightarrow L(H_{AB}); && T_A\mapsto T_A\ot 1_{H_B};\\
L(H_B) \hookrightarrow L(H_{AB}); && T_B\mapsto 1_{H_A}\ot T_B,
\end{align}
 see \er{8230},  and the same picture arises:  if a state $\om_{AB}$ on $L(H_{AB})$ corresponds to a density matrix $\rh_{AB}\in D(H_{AB})$ via \er{826}, i.e., $\om_{AB}(T_{AB})=\Tr(\rh_{AB} T_{AB})$,
then the reduced density matrix $\rh_A=\Tr_{H_B}(\rh_{AB})\in D(H_A)$  is the density matrix that corresponds to
the restriction $\om_A$ of $\om_{AB}$ to $L(H_A)$ via $\om_A(T_A)=\om_{AB}(T_A\ot 1_{H_B})$, and likewise the reduced density matrix  $\rh_B=\Tr_{H_A}(\rh_{AB})\in D(H_B)$  is the density matrix that corresponds to
the restriction $\om_B$ of $\om_{AB}$ to $L(H_B)$ via $\om_B(T_B)=\om_{AB}(1_{H_A}\ot T_A)$. This removes the mystery!
\end{itemize}

We return to  \emph{pure states}, which  classically were point distributions $p=\dl_x$
and in the quantum case were one-dimensional projections \er{89}, where $\ps\in H$ is a unit vector. These also fall into a single class of states. To explain this, we need some convexity theory. We denote the state space on a general \ca\ $\CA$, i.e., the set of all states on $\CA$, by $S(\CA)$. By Proposition \ref{SRX} we have 
\begin{align}
S(C(X))\cong \Pr(X); && S(L(H))\cong D(H),\label{827}
\end{align}
at least if $X$ is finite and $H$ is finite-dimensional. 
For simplicity we  assume that our \ca\ $\CA$ has a unit, as is the case for $\CA=C(X)$ and $\CA=L(H)$, even if $X$ is not finite but still compact, and  $H$ is infinite-dimensional (see below).
It then easily follows from the definition that the state space $S(\CA)$ is a convex set (within the vector space $\CA^*$ of all continuous linear functionals on $\CA$).\footnote{This is also true if $\CA$ lacks a unit. But in that case $S(\CA)$ is no longer compact in the weak$\mbox{}^*$-topology.} If we equip $S(\CA)\subset \CA^*$ with the (relative) weak$\mbox{}^*$-topology (so that $\om_n\raw\om$ iff $\om_n(a)\raw \om(a)$ for all $a\in \CA$), as we have already done for $\Pr(X)$, then $S(\CA)$ is even a \emph{compact} convex set.\footnote{This follows from the standard Banach--Alaoglu theorem, since 
$S(\CA)$ is a closed subset of the unit ball in $\CA^*$.}
\bex 
\begin{itemize}
\item Show that $S(C(N+1))\cong \Pr(N+1)$, where $N+1=\{0,1,\ldots, N\}$ as usual,  is isomorphic (as a compact convex set) to the 
  \emph{simplex} $\Dl_N$, defined for all $N\geq  1$ by
 \beq
 \Dl_N=\left\{x\in\R^{N+1}\mid x_i\geq 0, \Sigma_i x_i=1\right\}.\label{A48b}
 \eeq
\item Using the fact that any hermitian $2\x 2$ matrix may be parametrized by $(t,x,y,z)\in\R^4$ as
\beq \rh(t,x,y,z)=\half
\left(
\begin{array}{cc}
  t+ z & x-iy     \\
 x+iy & t-z        
\end{array}
\right), \label{nicely1}
\eeq
show that the state space $S(L(\C^2))$  is isomorphic (as a compact convex set) to the closed unit ball $B^3=\{(x,y,z)\in\R^3\mid x^2+y^2+z^2\leq 1\}$. How does this relate to the simplex $\Dl_1$?
 \end{itemize}
\eex
\begin{definition}\label{defPK}
 The \emph{boundary} $\partial_e K$ of a convex set $K$ consists of all $v\in K$ satisfying the condition:
 \begin{center}
if $v= t w+(1-t)x$ for certain $w,x\in K$ and $t\in (0,1)$, then $v=w=x$. \end{center}
Elements $v\in\partial_e K$ of the boundary  are called \emph{extreme points} of $K$. 
 \end{definition}
 This set  $\partial_eK$ is  called the \emph{extreme} boundary, to distinguish it from the \emph{topological} boundary $\partial K$  in case $K$ is also a topological space.\footnote{The topological boundary of $K\subset \CA^*$, or of any subset $K$ of a topological space $T$,  is defined as $\partial K=K^-\backslash\mathring{K}$, i.e., the closure off $K$ minus its interior (this is the set of all points in $T$ for which every nbhd intersects both $K$ and $T\backslash K$).}
  These two boundaries may or may not coincide.
  
 For example, in the previous exercise it was shown that $\Dl_1\cong [0,1]$, embedded in $\R$, whose extreme and topological boundaries do coincide, and are given by 
 \beq
 \partial_e[0,1]=\partial [0,1]=\{0,1\}.
 \eeq
  But already in the next case, $\Dl_2$, which you have shown to be an equilateral triangle in $\R^2$ with sides of unit length,
  the extreme boundary consist of the three \emph{vertices} whereas the topological boundary consists of its three \emph{sides}.\footnote{The sides of $\Dl_2$ do play a role also in convexity theory. 
A \emph{face} of a convex set $K$ is a nonempty convex subset $F\subseteq K$ such  that: \emph{if $z=tx+(1-t) y$ 
for  $z\in F$ with  $t\in(0,1)$ and $x,y\in K$, then $x,y\in F$.} The faces of our equilateral triangle $\Dl_2$, then, are its vertices, its sides, and $\Dl_2$ as a whole (each extreme point of $K$ is always a face on its own).} This difference persists for $N\geq 2$.
On the other hand,
\begin{equation}
\partial_e B^3=\partial B^3=S^2=\{(x,y,z)\in\R^3\mid x^2+y^2+z^2= 1\}.
\end{equation}
In these examples $K$ was closed. If we remove the extreme boundary in each case we still have a convex set, but now its boundary is empty! But this cannot happen if $K$ is closed, in which case $\partial_e$ is not only nonempty, but ``large enough'' to generate of all $K$ by forming convex combinations of points in $\partial_e K$. In finite dimension this is made precise by \emph{Carath\'{e}odory's theorem}:\footnote{In general the situation is described by the \emph{Krein--Milman theorem}, according to which $\partial_e K$ is always nonempty provided $K$ is compact, and  points in $K$ are at least limits of convex sums of
boundary points. \label{Kreinfn}}
\begin{theorem}\label{Caratheodory}
If $K$ is a nonempty compact convex subset of $\R^{n}$, then $\partial_e K\neq\emptyset$, and each point of $K$ is a convex sum of at most $n+1$ points in  $\partial_e K$.
\end{theorem}
One can clearly see this in  $K=\Dl_N$, now best taken as $\Pr(N+1)$, in which case one trivially has 
  \begin{equation}
p=\Sigma_{x\in X} p_x \dl_x.  \label{bdp}
\end{equation}
What does this look like in our finite (dimensional) classical and quantum cases $C(X)$ and $L(H)$?
The following proposition is  crucial in unifying the notion of a pure state between these cases.
 \begin{proposition}\label{P1310}
 \begin{itemize}
\item The boundary $\partial_e \Pr(X)$ consists of the pure states $\dl_x$, with $x\in X$.
\item The boundary $\partial_e D(H)$ consists of the pure states \er{89}, with $\ps\in H$, $\|\psi\|=1$, that is,
\begin{equation}
\partial_e D(H)=P_1(H), \label{Prho}
\end{equation}
 where $P_1(H)\subset L(H)$ is the set of one-dimensional projections on $H$.
\end{itemize}
  \end{proposition}
  \bex
  Prove this for finite sets $X$ and  finite-dimensional \Hs s $H$.
  \eex 
  We stated this proposition without the condition that the set $X$ be finite and the \Hs\ $H$ be finite-dimensional, and indeed the result holds for all  compact Hausdorff spaces $X$ and all \Hs s $H$, as we will explain in more detail shortly. However, though 
 the first part of \er{827} continues to hold in that generality, the second part does not, and hence the corollary 
 \begin{align}
 \partial_e S(C(X))&\cong X, &&  \om_x(f)=f(x); \label{833} \\
 \partial_e S(L(H))&\cong  P_1(H), && \om_{\psi}(T)=\la\psi, T\psi\ra,\label{834}
 \end{align}
holds for all compact spaces $X$ but only for all \emph{finite-dimensional} \Hs s $H$. In conclusion:
 \begin{itemize}
\item In classical physics on a space $X$, pure states, seen as  points $x\in X$,  bijectively correspond to  points in the boundary $\partial_e S(C(X))$ of the convex state space $S(C(X))\cong\Pr(X)$ via \er{833}.
\item In quantum mechanics \emph{on a finite-dimensional \Hs} $H$, pure states, which as we have seen may equivalently be seen as either unit vectors $\psi\in H$ up to a phase or  as one-dimensional projections $|\psi\ra\la\psi|$,  bijectively correspond to points in the  boundary $\partial_e S(L(H))$ of the convex state space $S(L(H))$ 
via \er{834}.
 \end{itemize}
For infinite-dimensional \Hs\ $H$ one still has \er{Prho}, but the states defined by density operators $\rh\in D(H)$ via \er{826} form a strict subset of the set of all states on $L(H)$, called the \emph{normal} states. We will return to the precise relationship between states and  density operators. 

Despite the common identification of pure states (as seen in theoretical physics) with points in the extreme boundary of the state space (as defined by mathematicians), there is an important qualitative difference between the state spaces of classical and quantum physics. If we return to Theorem \ref{Caratheodory}, we see that in the classical case the convex sum \er{bdp} is unique, that is, there is exactly one way of writing a state as a convex sum of pure states. But in quantum theory the case of $H=\C^2$ already shows that such decompositions are far from unique: take some point $x$ in the interior of $B^3$, draw  \emph{any} straight line segment through $x$,  which intersects the boundary $S^2$ of pure states at antipodal points $N$ and $S$, and conclude that $x$  can be written as a convex sum of $N$ and $S$.
 
Without proofs,\footnote{ These are standard functional analysis. See for example Landsman (2017), on which our discussion is based.}  we now generalize these considerations to larger spaces $X$ and $H$. To keep things relatively simple we assume that our \ca s $\CA$ have units; this poses no restriction on $H$, but it does force $X$ to be a \emph{compact} space.\footnote{We also tacitly assume all our topological spaces to be Hausdorff. There is a  topological mismatch between 
probability theory and commutative \ca s, in that Polish spaces provide the natural setting for the former whereas compact spaces appear in the latter. Polish space need not be compact (for example, any separable \Hs\ is Polish), and compact spaces need not be Polish. The healthy overlap is where $X$ is compact, separable, and metrizable. The assumption that $X$ be compact also removes a nuisance is defining the topology on $\Pr(X)$, which for Polish spaces we took to be the weak topology, whereas in \ca s it is the relative weak$\mbox{}^*$-topology. These coincide if $X$ is compact. In the locally compact (i.e., non-unital) case they are different, since weak$\mbox{}^*$-convergence $\mu_n\raw \mu$ is defined pointwise on $C_0$-functions, whereas
weak convergence is defined pointwise on $C_b$-functions.
} Let us first deal with some  technicalities.
\begin{itemize}
\item For a compact space $X$, probability \emph{distributions} $p:X\raw [0,1]$ on $X$, suitable for the finite case, have to be replaced by (regular Borel) probability \emph{measures} $\mu:\Sg\raw[0,1]$, where $\Sg$ is the $\sg$-algebra of Borel sets in $X$.\footnote{This is the smallest collection of subsets of $X$ that contains all open sets and is closed under complementation and countable unions and intersections. A Borel measure $\mu$ is \emph{regular} if for any $A\in \Sg$ one has $\mu^*(A)=\mu_*(A)=\mu(A)$, where $\mu^*(A)= \inf\{\mu(U)\mid U\supseteq A\}$, in which the infimum is over all \emph{open}  $U\subset X$, and $\mu_*(A) =\sup\{\mu(K)\}$, in which the supremum is over all \emph{compact} $K\subset X$.
 We assume some  familiarity with measure theory and integration theory.
} Furthermore, as the notation suggests, $C(X)$ now stands for the (commutative) algebra of all \emph{continuous} complex-valued functions on $X$.
\item For a general \Hs\ $H$ the algebra $L(H)$ of all linear maps on $H$, suitable for the finite-dimensional case, has to be replaced by $B(H)$, the algebra of all \emph{continuous}  linear maps (operators) on $H$, where an operator is continuous iff it is bounded.\footnote{For $T:H\raw H$ this means that $\sup\{\|T\ps\|, \ps\in H, \|\ps\|\leq 1\}<\infty$. 
If $\dim(H)<\infty$ one has $B(H)=L(H)$.} 
\item Moreover, the trace is much more difficult to define.\footnote{One reason is that not every operator has a finite trace; for example, the  formula \er{defTr} in the main text gives
$\Tr(1_H)=\dim(H)$, which is infinite if $H$ is infinite-dimensional. Another is that
 this sum may no longer be independent of the basis; for suitable $T$ it can even be finite in one basis and infinite in another. This cannot happen if $T\geq 0$.} The correct approach is to define
 \beq
 \Tr(T)=\Sigma_i \la \ups_i, T\ups_i\ra, \label{defTr}
 \eeq
 initially for \emph{positive} operators $T\in B(H)$,\footnote{This  may  still be defined by $\la\ps, T\ps\ra\geq 0$ for all $\ps\in H$, which is equivalent to $T=S^*S$ for some $S\in B(H)$.} for which the sum is independent of the basis, though it may be infinite (take  $T=1_H$). The next step is to introduce the \emph{trace class} $B_1(H)$. This consists of all bounded operators $T\in B(H)$ for which $\Tr(|T|)<\infty$, where \beq
 |T|=\sqrt{T^*T},\eeq
is defined by the continuous functional calculus for the square root. For $T\in B_1(H)$, not necessarily positive, the expression \er{defTr} turns out to be independent of the basis and finite.
 For quantum mechanics it is crucial to define expectations values \er{QEV}, which continue to make sense: the reason is that $B_1(H)$ is a two-sided ideal in $B(H)$, i.e., if $S\in B_1(H)$ and $T\in B(H)$, then $ST\in B_1(H)$ and $TS\in B_1(H)$, with the familiar property
 \beq \Tr(ST)=\Tr(TS).
 \eeq
 \item
 In view of this, we may now define the convex space of density operators $D(H)$ by
 \begin{align}
 D(H):=\{\rh\in B_1(H)\mid \rh\geq 0, \Tr(\rh)=1\},
 \end{align}
 where we may also write $\rh\in B(H)$ instead of $\rh\in B_1(H)$, since given the positivity condition $\rh\geq 0$ the trace is well defined, and asking it to be unity pushes $\rh$ into $B_1(H)$ anyway. 
 Anticipating the theory of \vna s, $B_1(H)$ is a Banach space in the norm
 \begin{equation}
\| T\|_1:=\Tr(|T|), \label{839}
\end{equation}
and $B(H)$ with the usual operator norm is the Banach space dual of $B_1(H)$, in the sense that every continuous (= bounded) linear functional $\phv:B_1(H)\raw \C$ takes the form $\phv=\phv_T$ for some $T\in B(H)$, with $\phv_T(S)=\Tr(ST)$. Thus we have an isometric isomorphism
\begin{equation}
B(H)\cong B_1(H)^*. \label{840}
\end{equation}
\item The \emph{$\sg$-weak topology} on $B(H)$ is the weak$\mbox{}^*$-topology on $B(H)$ as the Banach dual of $B_1(H)$; thus for some net $(T_{\lm})$ in $B(H)$ we have $T_{\lm}\raw T$ in this topology   iff $\Tr(ST_{\lm})\raw\Tr(ST)$ for all $S\in B_1(H)$. This turns out to be equivalent to $\Tr(\rh T_{\lm})\raw\Tr(\rh T)$ for all $\rh\in D(H)$.

A $\sg$-weakly continuous functional on $B(H)$ is called a \emph{normal} functional. Such functionals $\om: B(H)\raw\C$ correspond to operators $S\in B_1(T)$ via $\om_S(T)=\Tr(ST)$.
By the inequality
\begin{align}
|\Tr(ST)|\leq \|S \|_1 \|T\| && (S\in B_1(H), T\in B(H)),
\end{align}
a normal functional is automatically norm-continuous, but not \emph{vice versa} if $\dim(H)=\infty$. 

A normal \emph{state} may equivalently be defined by a condition that looks quite different, namely 
\begin{equation}
\om\left(\Sigma_i E_i\right)=\Sigma_i\om(E_i),\label{normalstates}
\end{equation}
for any orthogonal family $(E_i)$ of projections  (i.e., $E_i^*=E_i$ and  $E_iE_j=\dl_{ij}E_i$); if $H$ is separable,\footnote{ A \Hs\ is \emph{separable} if it has a countable basis.  All separable \Hs\ are isomorphic o $H=\ell^2(\N)$. \label{sepHs} }
 then  such a family must be countable and hence in that case normality is similar to the $\sg$-additivity condition in the definition of a (probability) measure.\footnote{Here  $\Sigma_i E_i$ is defined in the \emph{strong operator topology}, i.e., $\Sigma_i E_i\psi$ converges in $H$ for each $\psi\in H$.}
\item For later use we also introduce the \emph{Hilbert--Schmidt class} 
\beq
B_2(H):=\{T\in B(H)\mid \Tr(T^*T)<\infty\}.
\eeq
This is a Hilbert space and hence a Banach space, with inner product and norm defined by
\begin{align}
\la S,T\ra:=\Tr(S^*T); && \|T\|_2:=\sqrt{\Tr(T^*T)}.\label{HSIP}
\end{align}
Like $B_1(H)$, it is a two-sided ideal in $B(H)$, as will be put to good use in modular theory. 
\item  More generally, for any $1\leq p<\infty$ one has Banach spaces $B_p(H)$, called \emph{Schatten classes}, which are two-sided ideals in $B(H)\equiv B_{\infty}(H)$ (so that really $1\leq p\leq\infty$), defined by
\begin{align}
B_p(H):=\{T\in B(H)\mid \Tr(|T|^p)<\infty\}; && \|T\|_p:= (\Tr(|T|^p))^{1/p}.
\end{align}
Further to \er{840}, let $1\leq p<\infty$ and $1<q\leq \infty$ satisfy $p\inv + q\inv=1$. Then 
\begin{equation}
B_p(H)^*\cong B_q(H), \label{BpHBqH}
\end{equation}
via a bijection like the one explained  between \er{839} and \er{840},  based  on the inequality
\begin{align}
|\Tr(ST)|\leq \|S \|_p \|T\|_q && (S\in B_p(H), T\in B_q(H)),
\end{align}
which also gives $ST\in B_1(H)$. The case $p=1$, $q=\infty$ is \er{840}. The case $p=q=2$ recovers the familiar fact that the \Hs\ $B_2(H)$ of Hilbert--Schmidt operators is self-dual.
\item 
These are noncommutative analogues of the Banach spaces $L^p(X,\mu)$ in measure theory, i.e.,
\begin{align}
L^p(X,\mu):=\left\{f: X\raw\C\mid \int_X d\mu(x)\, |f(x)|^p<\infty\right\};  &&\| f\|_p:= \left( \int_X d\mu(x)\, |f(x)|^p \right)^{1/p}, \label{defLp}
\end{align}
where $(X,\mu)$ is some measure space (suppressing the $\sg$-algebra $\Sg$ on which $\mu$ is defined) and all functions $f$ are supposed to be measurable. One doesn't lose much in having the examples of standard Borel spaces in mind that were listed after Definition \ref{weak2c}, or even just $X=[0,1]$ with Lebesgue measure $dx$.
Eq.\ \er{defLp} holds for  $1\leq p<\infty$. If
\begin{equation}
\| f\|^{\mathrm{ess}}_{\infty}= \inf\{t\in[0,\infty]\mid|f|\leq t\:\:\: \mbox{$\mu$-almost everywhere}\},
 \label{inftynorm}
\end{equation}
where $|f|\leq t$ $\mu$-a.e.\ means that $\mu(\{x\in X\mid |f(x)> t\})=0$, we may also define 
$L^{\infty}(X,\mu)$ as the set of all measurable functions $f: X\raw\C$ for which the above infimum is finite, upon which
$\| \cdot \|^{\mathrm{ess}}_{\infty}$ equips $L^{\infty}(X,\mu)$ with a norm in which it is a Banach space, and we have 
\begin{align}
L^p(X,\mu)^*\cong L^q(X,\mu); && (1\leq p<\infty; \:\: 1<q\leq \infty;\:\:  p\inv + q\inv=1), \label{Lpqd}
\end{align}
where the isomorphism is  as follows: any continuous linear functional $\phv: L^p(X,\mu)\raw\C$ takes the form $\phv=\phv_\rh$ for some $\rh\in L^q(X,\mu)$ such that 
\beq
\phv_\rh(f)=\int_X d\mu\, \rh f.\eeq We then have
\begin{equation}
\left| \int_X d\mu\, \rh f \right|\leq \|f\|_p\|\rh\|_q. \label{863}
\end{equation}
The case $p=q=2$, where $L^2(X,\mu)$ is a self-dual \Hs, may again be singled out, 
Unlike the noncommutative case, where $B_p(H)\subset B(H)$, the spaces $L^p(X,\mu)$ need not be subspaces of $L^{\infty}(X,\mu)$ (e.g.\  if $X=[0,1]$ with $d\mu(x)=dx$, then $1/\sqrt{x}$ is in $L^1$ but not in $L^{\infty}$).
\end{itemize}
We will see (in \S\ref{MWM}) that both  $B_p(M)$ and  $L^p(X,\mu)$ spaces are special cases (up to isomorphism) of the Haagerup $L^p(M)$ spaces over a \vna\ $M$, where $M=B(H)$ leads to the former and $M=L^p(X,\mu)$ to the latter. This will be very important for the quantum theory of entropy. For the moment,
the following theorem generalizes Propositions \ref{SRX} and \ref{P1310} as well as eqs.\ \er{833} - \er{834} to infinite-dimensional \Hs s and compact spaces $X$: 
\begin{theorem} 
\label{theorem88}Let $X$ be a compact space and $H$ an arbitrary \Hs, with associated \ca s $C(X)$ and $B(H)$, and accompanying notion of a state according to in Definition \ref{stateCA}.
\begin{enumerate}
\item The states $\om$ on $C(X)$ bijectively correspond to (regular) probability measures $\mu$ on $X$ via 
 \begin{align}
\om(f)=\int_X d\mu\, f && (f\in C(X)). \label{frommutophv1}
\end{align}
The pure states $\om_x$ bijectively correspond to points $x\in X$ via 
 \begin{align}
\om_x(f)=f(x); && \mu_x=\dl_x, \label{frommutophv2}
\end{align}
where $\dl_x$ is the (Dirac) measure defined on measurable subsets $A\subset X$ by
\begin{align}
\dl_x(A)=1 \hspace{20pt} (x\in A);  && \dl_x(A)=0 \hspace{20pt} (x\notin A). 
\end{align}
Denoting the space of regular (Borel) measures on $X$ by $\Pr(X)$, we therefore have:
\begin{align}
S(C(X))\cong\Pr(X); && \partial_eS(C(X))\cong X.
\end{align}
\item The normal states $\om$ on $B(H)$ bijectively correspond to density operators $\rh\in D(H)$ via 
\begin{align}
\om(T)=\Tr(\rh T) && (T\in B(H)). \label{normal} 
\end{align}
The normal pure  states $\om_{\ps}$ bijectively correspond to unit vectors $\psi\in H$ (up to a phase) via
\begin{align}
\om_{\psi}(T)=\la \ps, T\ps\ra; && \rh_{\psi}=|\psi\ra\la\psi|.\label{855}
\end{align}
Denoting the set of  \emph{normal} states on  $B(H)$ by $S_n(B(H))$, we therefore have:
 \begin{align}
S_n(B(H))\cong D(H); && \partial_eS_n(B(H))\cong P_1(H). 
\end{align}
\end{enumerate}
\end{theorem}
This is the same situation as for finite $X$ and finite-dimensional $H$, where all states are normal. At the end of this section we will say a bit more about non-normal states, which exist if $\dim(H)=\infty$.

We have now seen two kinds of examples of \ca s (with unit): commutative ones $C(X)$ where $X$ is a compact space, and noncommutative ones $B(H)$ where $H$ is a \Hs\ (of dimension $>1$ of course). Moreover, any involutive norm-closed subalgebra of $B(H)$ is trivially a \ca, too.\footnote{Here `involutive' simply means that the algebra $\CA$ is also closed under adjoints, $T\mapsto T^*$, so $T\in \CA$ iff $T^*\in \CA$.}  Let us give an interesting class of examples.
Take some self-adjoint operator $T=T^*\in B(H)$ and define $C^*(T)$ as the smallest \ca\ in $B(H)$ containing $T$ and $1_H$; this \ca\ exists and is given by the norm-closure of the space of all finite polynomials in $T$.
Note that $C^*(T)$ is commutative, since polynomials in $T=T^*$ commute with each other, and this property persists for norm-limits. 
In finite-dimensional \Hs s this simply means that 
\begin{equation}
f(T)=\Sigma_{\lm\in\sg(T)}f(\lm)\cdot E_{\lm}, \label{deffT}
\end{equation}
given the spectral resolution \er{SRTfd}. 
So a commutative \ca\ need not be of the form $C(X)$? 
 If we define the spectrum $\sg(T)$ in the right way,\footnote{If $\dim(H)=\infty$, the spectrum $\sg(T)$ can no longer be defined as the set of eigenvalues of $T$; if we do,  for example the position operator $\hat{x}$ on $L^2([0,1], dx)$, which is bounded, and evidently should takes values in $[0,1]$, would have empty spectrum. Instead, $\sg(T)$  is defined as the set of all $\lm\in\C$ for which the operator $T-\lm\cdot 1_H$ is \emph{not} invertible in $B(H)$. Some of these may be true eigenvalues, since if $T\ps=\lm\ps$ for some nonzero $\psi\in H$, then $\lm\in\sg(T)$; these (proper) eigenvalues form the \emph{discrete} spectrum $\sg_d(T)$. In 
 finite dimension one has $\sg(T)=\sg_d(T)$, but in general $\sg(T)$ contains a nonempty continuous part $\sg_c(T)=\sg(T)\backslash \sg_d(T)$, discovered by Hilbert.  For example, $\sg(\hat{x})=\sg_c(\hat{x})=[0,1]$.} which also makes it compact, 
then we have
\begin{align} 
C(\sg(T))\cong C^*(T) ; && f\mapsto f(T), 
\end{align}
 by the continuous functional calculus: if $f$ is a finite polynomial $p(x)=\sum_n c_n x^n$, then $f(T)$ is given by 
$f(T)=\sum_n c_n T^n$, and if $f$ is any continuous function on the compact set $\sg(T)\subset\R$, then we pick polynomials $p_N$ such that 
 $p_N\raw f$ uniformly on $\sg(T)$ and define $f(T):=\lim_N p_N(T)$ (in norm). So, up to isomorphism,\footnote{Here, as usual, an isomorphism between two \ca s is a bijection that preserves all structure. It is remarkable, though, that preserving the norm is a consequence of preserving the algebraic structure and the involution.}
  our commutative \ca\ $C^*(T)$ does take the form $C(X)$.

Are there any other possibilities? The following two theorems, both due to Gelfand and Naimark and present already in their amazing founding paper of the subject from 1943,  say no:
\begin{theorem}\label{GNT}
\begin{enumerate}
\item Every \ca\ with unit is isomorphic to a norm-closed involutive subalgebra of $B(H)$, for some \Hs\ $H$.
\item Every commutative \ca\ with unit is isomorphic to $C(X)$, for some compact space $X$.
\end{enumerate}
\end{theorem}
This is also true without unit, though in that case the second part needs some change. The proofs are as beautiful as the claims. We here just sketch the main steps under simplifying assumptions.
\begin{enumerate}
\item This is proved by the  \emph{\textsc{gns}-construction}, named after Gelfand, Naimark, and Segal (who found it independently). We assume that our \ca\ $\CA$ has a faithful state $\om$, that is,  $\om(a^*a)>0$ for all $a\neq 0$ (all  non-pathological \ca s have such states). The idea is to create a \Hs\ $H_{\om}$ from $\om$ and $\CA$. Start by putting a sesquilinear form on $\CA$ by 
\begin{equation}
\la a,b\ra_{\om}:=\om(a^*b). \label{IP1473}
\end{equation}
Since $\om$ is faithful, this is an inner product. 
Then complete $\CA$ in the norm coming from this inner product. This completion is $H_{\om}$. For each $a\in \CA$, define a map $\pi_{\om}(a):\CA\raw H_{\om}$ by 
\begin{equation}
\pi_{\om}(a)b:= ab, \label{leftm}
\end{equation}
This operator is bounded and hence can be extended from $\CA$ to all of $H_{\om}$ by continuity. The ensuing map $\pi_{\om}:\CA\raw B(H_{\om})$ is a \emph{representation} of $\CA$ on $H_{\om}$, i.e.\  a  map that preserves all structure. 
It is also injective, and since it is isometric its image $\pi_{\om}(\CA)\subset B(H_{\om})$ is norm-closed. This image is  a copy of $\CA$ within $B(H_{\om})$, which proves part 1 of the theorem.\footnote{The  \textsc{gns}-construction also works for non-faithful states $\om$. The inner product \er{IP1473} is then positive semidefinite, and to make it positive definite one must divide out its null space $\mathcal{N}_{\om}=\{a\in A\mid \om(a^*a)=0\}$ and complete the quotient $\CA/\mathcal{N}_{\om}$ (instead of $\CA$ itself), to a vector space $H_{\om}$, which inherits the form \er{IP1473} but now as an inner product. The action $\pi_{\om}$ of $\CA$ on $H_{\om}$ is still given by \er{leftm}, starting on $\CA$, passing to $\CA/\mathcal{N}_{\om}$, and extending to $H_{\om}$ by continuity.
}
\item Let $\CA$ be our commutative \ca\ with unit. The idea is to construct a compact space $\Sg(\CA)$ from $\CA$, called its
\emph{Gelfand spectrum}, and then define a map $\CA\raw C(\Sg(\CA))$ that turns out to be an isomorphism. 
First, $\Sg(\CA)$ consists of all pure states on $\CA$; equivalently,  $\Sg(\CA)$ may be defined as the set of all nonzero multiplicative linear functionals on $\CA$ (which are automatically continuous). This turns out to be a compact space in the (relative) weak$\mbox{}^*$-topology (in which $\om_{\lm}\raw\om$ iff $\om_{\lm}(a)\raw\om(a)$ for each $a\in \CA$). Then define a map
\begin{align}
\CA\raw C(\Sg(\CA)); && a\mapsto \hat{a}; && \hat{a}(\om):=\om(a),
\end{align}
which is easily seen to preserve all structure, and less easily shown to be a bijection.\footnote{This proof  is conceptually crystal-clear but technically hard; see e.g.\ Landsman (2017), \S C.2 and \S C.3, which also gives the extension to a categorical duality between unital commutative \ca s and compact Hausdorff spaces.}
\end{enumerate}
Two features of the \textsc{gns}-construction (also in a more general form if $\om$ is not faithful) stand out:
\begin{enumerate}
\item The \Hs\ $H_{\om}$ has a \emph{cyclic} vector $\Om_{\om}$ for $\pi_{\om}(\CA)$. This means that 
\begin{equation}
\ovl{\pi_{\om}(\CA)\Om_{\om}}=H_{\om},
\end{equation}
where the bar stands for the closure; in words, any vector in $H_{\om}$ can be arbitrarily closely approximated by vectors of the kind $\pi_{\om}(a)\Om_{\om}$, for some $a\in \CA$. This is  trivial:  $\Om_{\om}=1_\CA$.
\item By a simple computation, for  any $a\in \CA$ we have
\begin{equation}
\om(a)=\la\Om_{\om}, \pi_{\om}(a)\Om_{\om}\ra. \label{877GNS}
\end{equation}
\end{enumerate}
One further property only holds if $\om$ is faithful:
\begin{enumerate}[resume]
\item The vector $\Om_{\om}$ is \emph{separating} for $\pi_{\om}(\CA)$, in that $\pi_{\om}(a)\Om_{\om}=0$ implies $a=0$.
\end{enumerate}

Because of their importance and beauty, we also state some further results about the \textsc{gns}-construction.
 First, as in group theory, 
we  call  a \rep\ $\pi$ of a \ca\ ${\CA}$ on a Hilbert space $H$ \emph{irreducible} if the only closed subspaces $K$ of $H$ that are stable under $\pi({\CA})$ (in the sense that if $\ps\in K$, then $\pi(a)\ps\in K$ for all $a\in \CA$)  are either $K=H$ or $K=\{0\}$.
We then have:\footnote{See Landsman (2017), Theorem C.90, or any textbook, for a proof and further details.}
\begin{proposition}
 The \textsc{gns}-\rep\ $\pi_{\om}$ is irreducible iff $\om$ is pure.\end{proposition}
 And in that case, not only $\Om_{\om}$, but every nonzero vector in $H$ is cyclic for $\pi({\CA})$.
 
Another useful result relates general \rep s to {\sc gns}-\rep s. We call two \rep s $\pi_i:\CA\raw B(H_i)$, $i=1,2$, \index{representations!unitarily equivalent}\emph{unitarily equivalent} if there is a unitary 
\beq
u:H_1\raw H_2\eeq such that
\begin{align}
u\pi_1(a)u^*=\pi_2(a) && \LRaw && u\pi_1(a)=\pi_2(a)u &&(a\in \CA).
\end{align}
\begin{proposition}\label{GNSuneq}
Let $\pi:\CA\raw B(H)$ be a \rep\ for which $H$ has a cyclic unit vector $\Om$ for $\pi({\CA})$, with corresponding linear functional $\om$ on $\CA$ defined by
\begin{equation}
\om(a)=\la\Om,\pi(a)\Om\ra.  \label{ompsi}
\end{equation}
Then $\om$ is a state and the corresponding {\sc gns}-\rep\ $\pi_{\om}$ is unitarily equivalent to $\pi$. 
\end{proposition}
\emph{Proof.} Define  $u:H_{\om}\raw H$ first on $\pi_{\om}(A)\Om_{\om}$ (which is a dense subspace of $H$) by
\beq
u\pi_{\om}(a)\Om_{\om}=\pi(a)\psi.
\eeq 
Using \er{ompsi} and \er{877GNS}, we then obtain
\beq
\|\pi_{\om}(a)\Om_{\om}\|^2=\om(a^*a)=\la\ps,\pi(a^*a)\ps\ra=\|\pi(a)\psi\|^2.
\eeq
This shows that $u$ is well defined as well as isometric, so that it extends to $H_{\om}$ by continuity. Its image is then the closure of $\pi(A)\psi$, which is $H$, since $\psi$ is cyclic by assumption. Thus $u$ is surjective and hence unitary. Finally, 
we compute
\beq
u\pi_{\om}(a)\pi_{\om}(b)\Om_{\om}=\pi(a)\pi(b)\psi=\pi(a)u\pi_{\om}(b)\Om_{\om},
\eeq
so that  $u\pi_{\om}(a)=\pi(a)u$  on the dense space $\pi_{\om}(A)\Om_{\om}$, and thence everywhere.\QED
\subsection*{Supplement: Singular states}\addcontentsline{toc}{subsection}{Supplement: Singular states}
 If $\dim(H)=\infty$, non-normal states on $B(H)$, also called \emph{singular} states, exist. 
It can be shown that a state $\om$ on $B(H)$ is singular iff $\om(E)=0$ for all one-dimensional (and hence all finite-dimensional) projections $E$,\footnote{Equivalently, singular states vanish on any compact operator. See Landsman (2017), Proposition 4.20.} a condition indeed  violated by normal states, cf.\ \er{normal}.  Yet singular states are not as aberrant as the terminology may suggest: such states are routinely  used in the physics literature and are typically denoted by $|\lm\ra$, where $\lm$ lies in the continuous spectrum $ \sg_c(T)$ of some self-adjoint operator $T$ (examples of such  ``improper eigenstates'' are $|x\ra$ and $|p\ra$). 
So take some $\lm\in \sg_c(T)$, and start by humbly defining a to-be-singular state $\om_{\lm}$ on $T$ itself by 
\beq
\om_{\lm}(T):=\lm.\label{856bis}
\eeq
Extend this to the \ca\  $C^*(T)$ generated by $T$ and $1_H$ discussed earlier in this section by 
 \beq
 \om_{\lm}(f(T))=f(\lm).
 \eeq
  The difficult step is then to further extend $\om_{\lm}$ from $C^*(T)$ to $B(H)$. This can always be done using the Hahn--Banach extension theorem of functional analysis, but this is non-constructive and gives a non-unique result. It is enough to prove the  \emph{existence} of singular states, though: the Hahn--Banach extension of $\om_{\lm}$ can be chosen to be pure on $B(H)$, and so if it were normal, it would be given by 
  \begin{align}
  \om_{\lm}(S)=\la\psi , S\psi\ra && (S\in B(H)), \label{1395s}
  \end{align}
   for some unit vector $\psi\in H$. But taking $S=T$, eqs.\ \er{856bis}  and  \er{1395s} would make $\psi$ an eigenvector of $T$, contradicting the assumption that $\lm$ lies in the \emph{continuous} spectrum of $T$.
 
Can we avoid the arbitrariness  in the Hahn--Banach extension theorem?
 In finite dimension, $\lm\in\sg(T)$ is an eigenvalue of $T$, but the above procedure works--though it produces a \emph{normal} pure state $\om_{\ps}$ on $L(H)$, cf.\ \er{855}, where $\ps$ is an eigenvector of $T$. What is instructive, though, is that the pure state extension $\om_{\ps}$ of $\om_{\lm}$ given by \er{855} 
 is unique \emph{provided $T$ has nondegenerate spectrum}: for in that case there is (up to a phase)  just one unit eigenvector $\ps_{\lm}$ of $T$ with eigenvalue $\lm$, and  \er{856bis} and  \er{1395s} force $\om_{\lm}$ to be $\om_{\ps_{\lm}}$. Furthermore, in finite 
 dimension  nondegeneracy of $\sg(T)$ is easily seen to be equivalent to \emph{maximality} of $T$, which means that all operators that commute with $T$ are functions of $T$;  in other words, if $\dim(H)<\infty$, then $T$ is maximal iff 
 \begin{align}
\{T\}'=C^*(T), \label{TCT}
\end{align}
where the \emph{commutant} $Z'$ of any subset $Z\subseteq B(H)$, a concept of great importance,  is defined by
\begin{equation}
Z'=\{S\in B(H)\mid ST=TS\,\forall T\in Z\}. \label{commutant}
\end{equation}
It should then be obvious that $\{T\}'=C^*(T)'$, so that \er{TCT} is equivalent to $C^*(T)'=C^*(T)$. 
 More generally, 
a commutative \ca\ $\CA\subset B(H)$ is  \emph{maximal abelian} if $\CA\subseteq \mathcal{B}\subset B(H)$
for some commutative = abelian \ca\ $\mathcal{B}$ implies $\mathcal{B}=\CA$. This is  equivalent to the property 
\begin{equation}
\CA'=\CA, \label{AA}
\end{equation}
which makes sense for any \ca\ $\CA\subset B(H)$ on any \Hs\ $H$, and as such may be used to \emph{define} maximal abelian \ca s within $B(H)$; note that \er{AA} \emph{implies} that $\CA$ is abelian (and, as we shall see in \S\ref{vNsection}, that $\CA$ is a \vna, i.e., $\CA''=\CA$).

However, if $\dim(H)=\infty$ and $T=T^*$, as soon as $T$ has nonempty continuous spectrum one cannot have $C^*(T)'=C^*(T)$, because the \emph{continuous} functional calculus extends to the \emph{Borel} functional calculus, in which $f(T)$ is  defined for bounded  \emph{measurable} functions $f:\sg(T)\raw\C$. For example, one may take $f=1_{\Dl}$ for any Borel subset $\Dl\subseteq\sg(T)$, which yields the spectral projections $E(\Dl)$ lying at the basis of the spectral resolution that replaces \er{SRTfd} in $\dim(H)=\infty$:
\begin{align}
T=\int_{\sg(T)} dE_{\lm}\, \lm; && 1_H=\int_{\sg(T)} dE_{\lm}. \label{SRT}
\end{align}
These more general functions $f(T)$ still commute with $T$ and even with all of $C^*(T)$, but strictly enlarge the latter. One may now define a new \ca\  $W^*(T)$, located as in 
\beq
C^*(T)\subset W^*(T)\subset B(H),
\eeq
 by  letting $W^*(T)$ consist of all operators $f(T)$, 
where $f$ is a bounded Borel function on $\sg(T)$. This is the same as the closure of $C^*(T)$ in the \emph{strong operator topology} on $B(H)$, where $T_{\lm}\raw T$ for some net $(T_{\lm})$ in $B(H)$  iff $T_{\lm}\ps\raw T\psi$ for each $\psi\in H$; indeed, this is precisely the topology in which the operators $f(T)$ are constructed as limits of polynomials $p_N(T)$. Equivalently, we have
\beq
W^*(T)=C^*(T)'',\label{WTCT}
\eeq
where the \emph{bicommutant} of $Z\subseteq B(H)$ is defined by $Z'':=(Z')'$. See \S\ref{vNsection} below. 
Finally, we define an operator $T=T^*\in B(H)$ to be maximal if $W^*T)$ is a maximal abelian subalgebra of $B(H)$, i.e.,
\begin{equation}
W^*(T)'=W^*(T).\label{WTWT}
\end{equation}
In finite dimension, then, this is true iff $\sg(T)$ is nondegenerate. In infinite dimension, we may now hope that if $T$ is maximal with nonempty continuous spectrum $\sg_c(T)$, and we construct a  state $\om_{\lm}$ from $\lm\in\sg_c(T)$ starting from \er{856bis}, then it has a unique singular pure extension to $B(H)$. More generally,\footnote{In fact this is hardly more general, since a theorem of von Neumann shows that if $H$ is separable and $\CA\subset B(H)$ is a maximal abelian \ca, then $\CA=W^*(T)$ for some $T=T^*\in B(H)$. See Landsman (2017), Theorem B.117.}
 if $\CA\subset B(H)$ is a maximal abelian \ca, we might hope that \emph{any pure state $\om$ on $\CA$ has a unique pure extension to $B(H)$.}
 This turns out to be true if $\om$ is normal, but false if it isn't. To appreciate the incredible subtlety of the situation, consider the following cases:\footnote{Here $\ell^2(\N):=\{f:\N\raw\C\mid\Sigma_{x\in\N} |f(x)|^2<\infty\}$ with inner product $\la f,g\ra:=\Sigma_{x\in\N}\ovl{f(x)} g(x)$, whereas  $\ell^{\infty}(\N)$ consists of all bounded functions $f:\N\raw\C$.
Then $f$ acts as a multiplication operator on $H$ via $\hat{f}\psi(x):=f(x)\psi(x)$. See \er{defmult}. }
\begin{enumerate}
\item $H=L^2(0,1)$ with maximal abelian \ca\ $\CA=L^{\infty}(0,1)$ (as multiplication operators).
\item $H=\ell^2(\N)$ with maximal abelian \ca\ $\CA=\ell^{\infty}(\N)$ (as multiplication operators).
\end{enumerate}
Up to unitary equivalence, these are the only possibilities for maximal abelian \ca s in $B(H)$ on infinite-dimensional separable \Hs\ $H$, except for some combinations, namely:
\begin{enumerate}[resume]
\item $H=L^2(0,1)\oplus\ell^2(\N)$ with maximal abelian \ca\ $L^{\infty}(0,1)\oplus \ell^{\infty}(\N)$;
\item $H=L^2(0,1)\oplus\C^n$ with maximal abelian \ca\ 
$L^{\infty}(0,1)\oplus D_n(\C)$,  for some $n\in\N$,
\end{enumerate}
where $L^2(0,1)=L^2([0,1], dx)$ etc., and 
$D_n(\C)$ consists of the diagonal $n\x n$ matrices. 

Then the statement in italic is false in the first case and true in the second, confirming the so-called \emph{Kadison--Singer conjecture}. Already the first, negative case (due to Kadison and Singer themselves) is very hard  to prove. The proof of the positive second case was a minor sensation.\footnote{See Landsman (2017),  \S4.3, for an introduction, and Stevens (2016) for a full account.}

In sum, the (Dirac) notation $|\lm\ra$ for the improper eigenvector of some self-adjoint operator $T\in B(H)$, with $\lm$ in its continuous spectrum, may be interpreted as a singular state $\om_{\lm}$ on $B(H)$ via 
$S\mapsto \om_{\lm}(S)$, or informally $\la\lm|S|\lm\ra$, $S\in B(H)$,  but this state is only unambiguously defined if firstly 
  $T$ is  maximal, and secondly the pure state extension from $C^*(T)$ to $B(H)$ is unique. 
 \section{Quantum entropy}\label{sec:QE}
 The previous section suggests that the quantum analogue of a finite classical probability space $(A,p)$ is a pair $(H,\rh)$, where $H$ is a finite-dimensional \Hs\ and $\rh\in D(H)$ is a density matrix. Moreover, we saw in the Historical Introduction that
in 1927 von Neumann presciently defined the quantum entropy of a state $\rh\in D(H)$ by, using slightly more modern notation, 
\begin{equation}
S(\rh):=-\Tr(\rh \log\rh). \label{vNe}
\end{equation}
Here the  expression $\rh\log\rh$ is defined by the spectral calculus, see \er{deffT}.
In particular, the spectral resolution \er{sdrho} gives
\begin{align}
\rh\log\rh=\Sigma_{a\in A} p_a\log p_a |\ups_a\ra\la \ups_a|,\label{sdrho2}
\end{align}
where we stipulate that $f(x)=x\log x$ vanishes at $x=0$ (either by continuity or by definition).
Taking the trace in \er{vNe} over the basis $(\ups_a)$ we therefore recover the Shannon entropy \er{Aentropy}:\footnote{The abuse of notation here is hopefully not confusing: the argument of $S$ should point at its definition.}
\beq
S(\rh)=-\Sigma_{a\in A} p_a\log p_a= S(p). \label{vNShannon}
\eeq
In this sense the von Neumann entropy \er{vNe} resembles its classical counterpart. Moreover:
\bex\label{boundsSq}
As in  Exercise \ref{boundsSc}:
show that $0\leq S(\rh)\leq \log(\dim(H))$, with boundary values:
\begin{enumerate}
\item  $S(\rh)=0$ iff $\rh$ is a pure state (so $\rh=|\psi\ra\la\psi|$ for some $\psi\in H$ with $\|\psi\|=1$);
\item $S(\rh)= \log(\dim(H))$ iff $\rh=1_H/\dim(H)$, the ``maximally mixed'' state in $D(H)$.
\end{enumerate}\eex
 But there is a \textsc{big} difference if we combine systems. Take any of the Bell states \er{Bellstates0}, for example $\psi_-$, with ensuing pure state $\rh_{AB}=|\psi_-\ra\la\psi_-|\in D(\C^2\ot\C^2)$. The von Neumann entropy \er{vNe} of any pure state is zero, as we just saw.
Hence $S(\rh_{AB})=0$.  But we already noticed below \er{recover} that the reduced density matrices are
$\rh_A=\rh_B=\half 1_{\C^2}$, see \er{Bellreduced2}. Hence 
\beq
S(\rh_A)=S(\rh_B)=\log 2.
\eeq
Why is this strange? First, below \er{marginals}  we  already saw that in classical probability pure states reduce to pure states, and so the entropy of a reduced pure state vanishes. More generally, 
one classically has $S(p_{AB})\geq S(p_A)$, since the average surprise in revealing $a\in A$ (for which the classical Shannon entropy $S(p)$  is  a measure) is less than the average surprise in learning $(a,b)\in A\x B$, as in the latter case there are more possibilities. But we just showed that the corresponding inequality $S(\rh_{AB})\geq S(\rh_A)$ may fail for the von Neumann entropy. 
 Here is the general situation:
\begin{theorem}\label{bigdifference}
Let $H_{AB}=H_A\ot H_B$, with pure density matrix  $\rh_{AB}\in D(H_{AB})$ given by \er{psdm}
Then $S(\rh_{AB})=0$. For the  reduced density matrices $\rh_A$ and $\rh_B$, see \er{PT1},  we have:
 \begin{enumerate}
\item If $\ps_{AB}=\ps_A\ot\ps_B$ for  unit vectors $\ps_A\in H_A$ and $\ps_B\in H_B$, then 
\begin{equation}
S(\rh_A)=S(\rh_B)=0.
\end{equation}
\item If $\ps_{AB}$ does not consist of a single elementary tensor (and is called \emph{entangled}), then
\begin{equation}
E(\ps_{AB}):= S(\rh_A)=S(\rh_B)>0.\label{eq96}
\end{equation}
\end{enumerate}
\end{theorem}
This $E(\ps_{AB})$ is called the \emph{entanglement entropy} of $\psi_{AB}$. 
Eq.\ \er{eq96}, where both the equality  and the strict positivity are remarkable,  immediately follows from the Schmidt decomposition theorem: 
\begin{proposition}\label{Schmidt}
Any unit vector $\ps_{AB}\in H_{AB}$ (where $H_{AB}\equiv H_A\ot H_B$) may be written as
\begin{equation}
\ps_{AB}=\Sigma_{i=1}^d \lm_i e_{a(i)}\ot u_{b(i)},\label{Schmidtdec}
\end{equation}
where $\lm_i\in\C$ with $\Sigma_i|\lm_i|^2=1$, and $(e_a)$ and $(u_b)$ are \emph{specific} bases of $H_A$ and $H_B$, respectively. 
\end{proposition}
Clearly, $d\leq \dim(H_A)$ and $d\leq\dim(H_B)$. To see the point, note that by definition of the tensor product one always has bases $(e'_a)$ and $(u'_b)$ such that 
\beq
\ps_{AB}=\Sigma_{a,b}\lm_{ab}e'_a\ot u'_b
\eeq
 is a \emph{double} sum; the Schmidt decomposition \er{Schmidtdec}  is a \emph{single} sum, which implies that
\begin{align}
\rh_A=\Sigma_i |\lm_i|^2 |e_{a(i)}\ra\la e_{a(i)}|; 
  && \rh_B=\Sigma_i |\lm_i|^2 |u_{b(i)}\ra\la u_{a(i)}|, \label{exAB}
\end{align}
and hence
\begin{equation}
S(\rh_A)=S(\rh_B)=S(|\lm|^2),
\end{equation}
where $|\lm|^2$ is  the probability distribution $i\mapsto p(i)=|\lm_i|^2$ on the set $\{1, \ldots, d\}$.
Any entangled pure state (in which  \er{Schmidtdec} has more than one term)
therefore reduces to mixed states. 
\bex
Prove Proposition \ref{Schmidt} (the answer may be found all over the place). 
\eex
Conversely: 
\begin{proposition}
Any density matrix $\rh_A\in D(H_A)$ is the reduced density matrix of a pure state: there exists a \Hs\ $H_B$ and a  unit vector $\ps_{AB}\in H_{AB}$ with associated pure density matrix 
\beq
\rh_{AB}=|\psi_{AB}\ra\la\ps_{AB}|, \label{psdm}
\eeq such that  $\rh_A=\Tr_{H_B}(\rh_{AB})$. Both $\ps_{AB}$ and $\rh_{AB}$ are called
 \emph{purifications} of $\rh_A$.
\end{proposition}
\emph{Proof.}
Using the \textsc{gns}-construction explained after Theorem \ref{GNT}: take $A=L(H_A)$ and $\om$ the state $\om_A$ corresponding to $\rh_A$ via \er{826}. If $\om_A$ is faithful, which is the case iff $\rh_A$ is invertible, then $H_{\om_A}=L(H_A)$ with inner product $\la S,T\ra =\Tr(\rh_A S^*T)$,
on which $\pi_{\om_A}$ acts by left multiplication, see \er{leftm}, and $\Om_{\om_A}=1_A$. We now use Proposition \ref{GNSuneq} to change this into a unitarily equivalent \rep\ by changing the inner product on  $H_{\om_A}$ into the Hilbert--Schmidt inner product
\begin{equation}
\la S,T\ra:=\Tr(S^*T),
\end{equation}
cf.\ \er{HSIP}, and changing the cyclic (and separating) vector into 
$\Om=\rh_A^{1/2}$.
Then \er{ompsi} holds. The final move is that for any finite-dimensional \Hs\ $H_A$ one has
\begin{align}
L(H_A)\cong H_A\ot H_A; && |\psi\ra\la\phv|\lraw \psi\ot\phv,
\end{align}
which isomorphism maps  $\pi_{\om_A}(L(H_A))$ to $L(H_A)\ot 1_{H_B}$, and maps $\Om$ into 
\beq
\ps_{AB}=\Sigma_{a\in A}\sqrt{p_a} \ups_a\ot\ups_a,
\eeq
 assuming $\rh_A$ is has spectral resolution  \er{sdrho} with $\rh\leadsto\rh_A$, and $\dim(H_A)=|A|$.

If $\rh$ is not invertible we still have \er{sdrho}, but some $p_a$ are zero. Let $A'\subseteq A$ be the set where $p_a>0$, and 
take $H_B=\ell^2(A')$ with basis $(\dl_a)_{a\in A'}$. Then take
\begin{equation}
\psi_{AB}=\Sigma_{a\in A'} \sqrt{p_a}\, |\ups_a\ra\ot\dl_a,\label{4.38}
\end{equation}
and an easy computation of the partial trace of \er{psdm} over $H_B$ recovers $\rh_A$ via \er{sdrho}. \QED
\smallskip

Clearly, the unit vector $\ps_{AB}$,  with associated  density matrix \er{psdm}, is not the only purification of $\rh_A$. For example, one may also take a third \Hs\ $H_C$, replace $H_B$ above by $H_{BC}$, and replace \er{4.38} by $\ps_{AB}\ot\phv_C$, where $\phv_C$ is any unit vector in $H_C$. Here is the general case:\footnote{For a proof see Walter \& Ozols (2025), Lemma 2.18. Recall that an \emph{isometry} $V:H_B\raw H_C$ is a linear map such that $\la V\ps_B,V\phv_B\ra_{H_C}=\la \ps_B,\phv_B\ra_{H_B}$ for all $\psi_B,\phv_B\in H_B$ (this is the Hilbert space analogue of an injective map between sets). It follows that $V^*V: H_B\raw H_B$ is the identity and $VV^*:H_C\raw H_C$ is the projection onto the image of $V$.}
\begin{proposition}\label{MM2.18}
Let unit vectors $\ps_{AB}\in H_{AB}$ and $\phv\in H_{AC}$ both be purifications of $\rh_A\in D(H_A)$. Wlog,\footnote{This math jargon means: `Without loss of generality'.}  assume $\dim(H_B)\leq\dim(H_C)$. Then there is an isometry $V_{B\raw C}:H_B\raw H_C$ such that 
\begin{equation}
(\mathrm{id}_A\ot V_{B\raw C})\psi_{AB}=\phv_{AC}. 
\end{equation}\end{proposition}

 As in the classical case \er{KL1} - \er{KL2}, we may also define a \emph{relative quantum  entropy}  by
\begin{align}
S(\rh,\sg)&:=\Tr(\rh(\log\rh -\log\sg))
\:\:\: \mbox{ if } \ker(\sg)\subseteq\ker(\rh)
 \label{QKL};\\
S(\rh,\sg)&:=\infty\:\:\: \mbox{ otherwise}, \label{QKLb}
\end{align}
where $\rh,\sg\in D(H)$. In the first line we  define $\rh\log\sg=0$ on $\ker(\sg)$, like $\rh\log\rh=0$ on $\ker(\rh)$. Thus $\ker(\sg)\subseteq\ker(\rh)$ is the quantum analogue of the condition $q\ll p$ in the definition of $S(p, q)$.
\bex Prove that for any $\rh_A,\sg_A\in D(H_A)$ and $\rh_B,\sg_B\in D(H_B)$ we have
\begin{align}
S(\rh_A\ot \rh_B)&=S(\rh_A)+S(\rh_B); \label{Stensor}\\
S(\rh_A\ot\rh_B, \sg_A\ot\sg_B)&=S(\rh_A, \sg_A)+S(\rh_B, \sg_B).
\end{align}\eex

In the spirit of hypothesis testing we write $\rh=\rh_0$ and $\sg=\rh_1$, with spectral resolutions
\begin{align}
\rh_0=\Sigma_{a=1}^d p_0(a) |\ups_a^{(0)}\ra\la \ups_a^{(0)}|; && \rh_1=\Sigma_{b=1}^dp_1(b)|\ups_b^{(1)}\ra\la \ups_b^{(1)}|.
\label{joint01}
\end{align}
Then a naive computation based on \er{QKL} shows that
\begin{equation}
S(\rh_0,\rh_1)=\Sigma_a p_0(a)\left( \log p_0(a)-\Sigma_b |\la \ups_a^{(0)}, \ups_b^{(1)}\ra|^2\log p_1(b)\right), \label{QKLex}
\end{equation}
which may be taken at face value when $\rh_0$ and $\rh_1$ are both faithful (i.e., invertible), in which case $p_0(a)>0$ and $p_1(b)>0$ for all $a,b$. If not, but nonetheless $\ker(\rh_1)\subseteq\ker(\rh_0)$ as in \er{QKL}, then 
\begin{align}
p_1(b)=0 && \Raw && p_0(a)\la \ups_a^{(0)}, \ups_b^{(1)}\ra=0  && \Raw && p_0(a)|\la \ups_a^{(0)}, \ups_b^{(1)}\ra|^2=0
\end{align}
so that all terms in the sum over $b$ in \er{QKLex} for which $p_1(b)=0 $ vanish and $S(\rh_0,\rh_1)$ is finite. 
Otherwise,  \er{QKLex} is  infinite and hence conforms with \er{QKLb}. Thus \er{QKLex} is always correct.
We also note that, albeit  by an unexpectedly difficult proof,\footnote{See e.g.\ Khatri and Wilde (2024), Proposition 7.2} without  any assumption we have
\begin{equation}
S(\rh,\sg)=\lim_{\varep\downarrow 0} S(\rh, \sg+\varep 1_H),
\end{equation}
where the right-hand side is defined by extending \er{QKL}, whose assumption on the kernels is now automatically satisfied for $\varep>0$, to any $\rh\in D(H)$ and $\sg\in L(H)_+$ (i.e., $\sg\geq 0$).

Some properties of $S(p, q)$ generalize to $S(\rh,\sg)$; others don't. From the first category,
\begin{equation}
S(\rh, 1_H/\dim(H))=-S(\rh)+\log(\dim(H)). \label{likeD1}
\end{equation}
Indeed, the density matrix $1_H/\dim(H)$ plays the role of the flat probability distribution $f$. Similarly, the Gibbs inequality $S(p,| q)\geq 0$ is now replaced by the \emph{Klein inequality} 
\begin{equation}
S(\rh,\sg)\geq 0.\label{Klein}
\end{equation}
\bex \label{subex} 
\begin{enumerate}
\item  Prove that for self-adjoint $S,T\in L(H)$ with $\ker(T)\subseteq\ker(S)$ we have:\footnote{If $\ker(T)\subseteq\ker(S)$ we define $S\log T=0$ on $\ker(T)$. Similarly,  $S\log S=0$ on $\ker(S)$.}
\begin{equation}
\Tr(S\log S- S\log T)\geq \Tr(S-T), \label{Klein2}
\end{equation}
with equality iff $S=T$ (often called Klein's inequality, too).
 \item  Prove  \er{Klein} from this inequality.  
 \item 
 From \er{Klein2}, prove that $S$ is concave: for  $t\in(0,1)$ and $\rh,\sg\in D(H)$ we have
\beq
S(t\rh +(1-t)\sg)\geq t S(\rh)+(1-t) S(\sg). \label{Sconc}
\eeq  
Then, if necessary without proof if it takes you too long,  use  the property
\beq
\Tr(S(\log(S+T)-\log S))\geq 0,
\eeq for all $S\geq 0$, $T\geq 0$,
to strengthen \er{Sconc} to:
\begin{equation}
t S(\rh)+(1-t) S(\sg)\leq S(t\rh +(1-t)\sg)\leq t S(\rh)+(1-t) S(\sg) +H(t),
\end{equation}
where $H(t):= -t\log t -(1-t)\log(1-t)$. \end{enumerate}
\eex
Again as in the classical case (where we did not mention this), eq.\ \er{Klein} may be sharpened to:\footnote{This is the \emph{quantum Pinsker inequality}. See  Hayashi (2017), Exercise 3.30, answer on p.\ 143, for a proof.}
\begin{equation}
S(\rh,\sg)\geq \half\|\rh-\sg\|_1^2,
\end{equation}
see \er{839} for the trace norm. 
The main properties of the relative quantum entropy are:
\begin{theorem}\label{Ruskai}
The following properties are all true (and in fact are equivalent to each other):
\begin{enumerate}
\item \emph{Joint convexity:} for arbitrary density matrices (or, more generally, positive operators) $\rh_1, \ldots, \rh_n$
and  $\sg_1, \ldots, \sg_n$, and probabilities $p=(p_1,\ldots, p_n)$, i.e.\ $p_i\geq 0$, $\Sigma_i p_i=1$, we have
\beq
S\left(\Sigma_{i=1}^n p_i\rh_i , \Sigma_{j=1}^n p_j\sg_j\right)\leq \Sigma_{i=1}^n p_iS(\rh_i,\sg_i).\label{102.1}
\eeq
\item \emph{Monotonicity under state reduction:} writing $\rh_A=\Tr_{H_B}\rh_{AB}$, etc., we have
\beq
S(\rh_A,\sg_A)\leq S(\rh_{AB}, \sg_{AB}), \label{102.2}
\eeq 
for all $\rh_{AB},\sg_{AB}\in D(H_{AB})$, or, more generally, all positive operators $\rh_{AB},\sg_{AB}\in L(H_{AB})$.
\item  \emph{Monotonicity under quantum channels:}  if $\Phi: L(H_A)\raw L(H_B)$ is a completely positive and trace-preserving map (see the supplement to this section), then
\begin{align}
S(\Phi(\rh), \Phi(\sg))\leq S(\rh,\sg); && (\rh,\sg\in D(H_A)). \label{102.3}\end{align} 
\end{enumerate}
\end{theorem}
\emph{Proof.\footnote{See the classical papers by Ruskai (2002, 2007) for a different approach, and much more detail.}}
We will prove property 3 from the quantum Stein lemma (Theorem \ref{SteinQ}) at the end of \S\ref{QHTsec}.
 Property 2 then follows from the fact that state reduction is a quantum channel:  replacing 
\begin{align}
H_A\leadsto H_{AB}=H_A\ot H_B; && H_B\leadsto H_A,
\end{align} the map $\Phi: L(H_A)\raw L(H_B)$ in property 3 is taken to be 
$\Tr_B: L(H_{AB})\raw L(H_A)$,   which restricts to $\Tr_B: D(H_{AB})\raw D(H_A)$.  We now show that property 1 follows from property 2.\footnote{It is easily checked that the proofs of properties 1 and 2 extend from density operators to positive operators.} Assume that each $S(\rh_i,\sg_i)$ is finite and hence that $\ker(\sg_i)\subseteq\ker(\rh_i)$, for  otherwise the claim is trivial. We take 
 \begin{align}
 H_A=H; && H_B=\C^n; && \rh_{AB}=\Sigma_i p_i \rh_i\ot |e_i\ra\la e_i|; && \sg_{AB}=\Sigma_i p_i \sg_i\ot |e_i\ra\la e_i|,
\end{align}
where $(e_i)$ is some basis of $\C^n$. Then   \er{102.2} equals \er{102.1}, as simple computations show that
 \begin{align*}
 S(\rh_{AB},\sg_{AB})=\Sigma_{i=1}^n p_iS(\rh_i,\sg_i); && \rh_A=\Sigma_{i=1}^n p_i\rh_i ; && \sg_A=\Sigma_{i=1}^n p_i\sg_i.
\tag*{$\Box$}
 \end{align*}
 \begin{corollary}
 The following \emph{strong subadditivity} properties  are equivalent and true:\footnote{ Note that $1_{H_A}\ot \rh_B)$ and $1_{H_A}\ot \rh_{BC}$ are not density matrices (they are positive but do not trace to 1),
 but the formula \er{QKL} can still be used. Since we assume that all Hilbert spaces are finite-dimensional we could remedy this by rescaling $1_{H_A}\leadsto 1_{H_A}/\dim(H_A)$,  which lies in $D(H_A)$, but we use the stated form since the corollary holds for infinite-dimensional \Hs s, too,}
 \begin{align}
S(\rh_{AB}, 1_{H_A}\ot \rh_B)&\leq S(\rh_{ABC},1_{H_A}\ot \rh_{BC}); \label{SSAS1}\\
S(\rh_{ABC})+ S(\rh_B)&\leq S(\rh_{AB})+S(\rh_{BC}), \label{SSAS2}
\end{align}
where  $\rh_{ABC}\in D(H_{ABC})$, and we use the following notation:
\begin{align}
\rh_{AB}=\Tr_{H_C}\rh_{ABC}; && rh_{BC}=\Tr_{H_A}\rh_{ABC}; && \rh_B=\Tr_A\rh_{AB}.
\end{align} \end{corollary}
\emph{Proof.} The equivalence of \er{SSAS1} and \er{SSAS2} follows from \er{vNe} and \er{QKL}, plus facts like
\begin{align}
\Tr_{H_{AB}} (\rh_{AB}\log (1_{H_A}\ot \rh_B))=\Tr_{H_{AB}} (\rh_{AB}(1_{H_A}\ot \log \rh_B))=\Tr_{H_B}(\rh_B \log\rh_B)=-S(\rh_B),
\end{align} which follows from the definition of the partial trace, cf.\ \er{PT1}, and similarly, 
\beq
\Tr_{H_{ABC}} (\rh_{ABC}\log (1_{H_A}\ot \rh_{BC}))=\Tr_{H_{BC}}(\rh_{BC} \log\rh_{BC})=-S(\rh_{BC}).
\eeq
Eq.\ \er{SSAS1} is a special case of \er{102.2}, with $B\leadsto C$ and then $A\leadsto AB$, so that it reads
\beq
S(\rh_{AB},\sg_{AB})\leq S(\rh_{ABC}, \sg_{ABC}), \label{102.2b}
\eeq 
in which we take $\sg_{ABC}=1_{H_A}\ot \rh_{BC})$.
To derive \er{subadd}, take $\rh_B=1_{H_B}$ (or $\rh_B=1_{H_B}/\dim(H_B)$ if $\dim(H_B)<\infty$) in \er{SSAS2}, use \er{Stensor}, 
and relabel $B\leadsto C$. \QED
\bex
Show that \er{SSAS2} implies subadditivity of the von Neumann entropy: we have
\begin{equation}
S(\rh_{AB})\leq S(\rh_A)+S(\rh_B), \label{subadd}
\end{equation}
with equality iff $\rh_{AB}=\rh_A\ot\rh_B$.  Alternatively, prove \er{subadd} either from \er{Klein2} or from \er{Klein}. 
\eex
   \subsection*{Supplement: Classical and quantum channels}\addcontentsline{toc}{subsection}{Supplement:
   Classical and quantum channels}
Channels play a crucial role in information theory. In the classical case, the idea is that Alice sends Bob messages via a possibly noisy channel. If Alice sends the letter $a\in A$, then ideally Bob receives it as such. In practice, Bob has his own alphabet $B$ and for each $a\in A$ he receives $b\in B$ with probability $p(b|a)$ (the ideal case would be $B=A$ and $p(b|a)=\dl_{ab}$). On the identifications 
\beq
p(b|a)\equiv\CN_{ab}\equiv (\CN(a))(b),
\eeq  a noisy channel  therefore corresponds to  a map $\CN:A\raw \Pr(B)$ that satisfies
\begin{align}
\CN_{ab}\geq 0 \:\:\: (a\in A, b\in B); &&\Sigma_{b\in B}\CN_{ab}=1 \:\:\: (a\in A). \label{N1}
\end{align}
In the spirit of our earlier reformulation of probability distributions $p_A\in \Pr(A)$ in terms of states 
\begin{align}
\om_A: C(A)\raw\C; && \om_A(f_A)=\Sigma_{a\in A} p_A(a)f(a),
\end{align}
 we may also regard $\CN$ as a map (using the same symbol) 
 \begin{align}
 \CN:C(A)\raw C(B); && \CN(f)(b)=\Sigma_{a\in A} \CN_{ab}f(a). \label{6.3}
 \end{align} 
 It follows from  \er{6.3} and the first part of \er{N1} that $\CN$ is \emph{positive}, in the sense that if $f\geq 0$ (pointwise), then $\CN(f)\geq 0$ (pointwise). In addition,  \er{6.3}  and the second part of \er{N1} imply that $f$ is
  \emph{trace preserving}: simply defining $\Tr_A(f)=\Sigma_{a\in A} f(a)$ for $f\in C(A)$, we see that
   \begin{equation}
\Tr_B(\CN(f))=\Tr_A(f).
\end{equation}
\begin{definition}
A \emph{classical channel} is a trace-preserving positive linear map  $\mathcal{N}:C(A)\raw C(B)$.
\end{definition}
Our starting point is then recovered by noting that a  classical channel restricts to an affine map from $\Pr(A)\subset  C(A)$ to $\Pr(B)\subset C(B)$, which is fixed by its values on $\partial_e \Pr(A)\cong A$.
  Conversely, one may see \er{6.3} as as a linear extension of such a map. There is also a dual picture: if $\mathcal{N}:C(A)\raw C(B)$  is positive and \emph{trace-preserving}, then its dual 
 $\CN^*:C(B)\raw C(A)$   is positive and \emph{unital} (i.e., $\Phi(1_B)=1_A$), where $\CN^*$ is uniquely defined by the trace pairing
 \begin{align}
 \Tr_A(f\CN^*(g))=\Tr_B(\CN(f)g) && (f\in C(A), g\in C(B)), 
 \end{align}
It is therefore tempting  to define a \emph{quantum} channel as a positive and trace-preserving linear map  
\beq
\Phi:L(H_A)\raw L(H_B).\eeq 
For one thing, this restricts to an affine map from $D(H_A)\subset L(H_A)$ to $D(H_B)\subset L(H_B)$, and one has a natural passage to a positive and unital dual map $\Phi^*:L(H_B)\raw L(H_A)$, uniquely defined by
 \begin{align}
 \Tr_{H_A}(S\Phi^*(T))=\Tr_{H_B}(\Phi(S)T) && (S\in L(H_A), T\in L(H_B)). 
 \end{align}
But these conditions are not enough for building a successful asymptotic theory of quantum entropy and quantum hypothesis testing (or of 
noisy  quantum communication). Noise may also come from coupling Alice and Bob to some environment $E$, so that $H_A$ and $H_B$ are replaced by $H_{AE}$ and $H_{BE}$, respectively (where $H_{AE}=H_A\ot H_E$, etc.). The initial channel $\Phi$ from Alice to Bob must then be extended to $\Phi\ot\mathrm{id}_E:L(H_{AE})\raw L(H_{BE})$, defined by linear extension of 
\beq
\Phi\ot\mathrm{id}_E(T_A\ot S_E)=\Phi(T_A)\ot S_E.
\eeq
Unexpectedly, if $\Phi$ is  positive (and linear), then $\Phi\ot\mathrm{id}_E$ may not be! 
\bex Find an example of this phenomenon. \eex
This cannot happen in the classical case: if $\CN:C(A)\raw C(B)$ is 
positive (and linear), and we couple $A$ and $B$ to $E$, then positivity of $\CN$  is simply inherited by the canonically extended map 
\begin{align}
\CN\x  \mathrm{id}_E:C(A\x E)\raw C(B\x E); && \CN\x  \mathrm{id}_E(f)(b,e)=\Sigma_{a\in A} \CN_{ab}f(a,e). \label{6.5E}
\end{align}
The trace, though, is preserved by such extensions,  both in the classical and the quantum case. 

To remedy this, we call a linear map $\Phi: L(H_A)\raw L(H_B)$ \emph{completely positive} or \emph{\textsc{cp}} if for any finite-dimensional \Hs\ $H_E$ the extension  $\Phi\ot\mathrm{id}_E:L(H_{AE})\raw L(H_{BE})$ is positive.
We also call it  \textsc{tp} if it is \emph{trace-preserving}, and  \textsc{cptp} if it is both  \textsc{cp} and  \textsc{tp}. This is a golden combination.
\begin{definition}\label{defQC}
A quantum channel is a \textsc{cptp} linear map $\Phi: L(H_A)\raw L(H_B)$.
\end{definition} The dual of a \textsc{cptp} map (under the trace pairing) is then \textsc{cp} and unital.
The study of such maps is a huge field in quantum theory as well as in functional analysis. Here are some examples:\footnote{The  adjoint $W^*:H_B\raw H_A$ of
 $W:H_A\raw H_B$ is the unique operator such that $\la \ps_A, W^*\phv_B\ra_{H_A}=\la W\psi_A, \phv_B\ra_{H_B}$  for all $\ps_A\in H_A$ and $\phv_B\in H_B$. Then $W$ is an isometry if $W^*W=1_{H_A}$, and it is unitary if in addition
 $WW^*=1_{H_B}$. In finite dimension (but only then!), if $H_B=H_A$ (or just $\dim(H_B)=\dim(H_A)$),  then an isometry is automatically unitary.}
\begin{itemize}
\item \emph{Unitary evolution:}  $\Phi: L(H_{A})\raw L(H_B)$ is $\Phi(\rh_A)=U\rh_AU^*$, for some unitary $U:H_A\raw H_B$.
\item \emph{Isometric embedding:}  $\Phi: L(H_{A})\raw L(H_B)$ is $\Phi(\rh_A)=V\rh_AV^*$, for an isometry $V:H_A\raw H_B$.
\item \emph{Conditional expectation:} To define a  map $\Phi: L(H_B)\raw L(H_A)$ in the opposite direction, 
let $H_A\subseteq H_B$ and identify $L(H_A)$ as a subspace of $L(H_B)$ by putting $T_A\psi_B=0$ for all $\psi_B\in H_A^{\perp}$ (whilst $T_A$ is `itself'' on $\ps_B\in H_A$). Using the projection $E:H_B\raw H_B$,  put 
\beq
\Phi(\rh_B)=E\rh_BE.\eeq
\item \emph{Reduction:} $\Phi: L(H_{AB})\raw L(H_A)$ is now
$\Phi(\rh_{AB})=\rh_A\equiv \Tr_{H_B}(\rh_{AB})$,  cf.\ \er{8.25} - \er{PT1}. 
\item \emph{Extension:} $\Phi: L(H_A)\raw L(H_{AB})$ is given by $\Phi(\rh_A)=\rh_A\ot\rh_B$, for some $\rh_B\in D(H_B)$.
\item \emph{Kraus maps:} To define $\Phi: L(H_A)\raw L(H_B)$, take
 finitely many linear maps  $X_i: H_A\raw H_B$ satisfying $\Sigma_i X_i^*X_i=1_{H_A}$ (which guarantees that $\Phi$ is trace preserving), and define
\begin{equation}
\Phi(\rh_A)=\Sigma_i X_i\rh_A X_i^*. \label{Kraus}
\end{equation}
\item \emph{Stinespring maps:} To construct  $\Phi: L(H_A)\raw L(H_B)$ in a different way,  take some ``ancillary'' \Hs\  $H_C$ and an isometry $V:H_A\raw H_{BC}$ (that is,  $V^*V=1_{H_A}$), and define $\Phi$
 by
\begin{equation} 
\Phi(\rh_A)=\Tr_{H_C}(V\rh_A V^*).\label{Stinespring}
\end{equation}
\item  \emph{Lindblad maps:} To define  $\Phi: L(H_A)\raw L(H_B)$ we now take \emph{two} ancillary \Hs s $H_C$ and $H_D$, a unit vector $\ps_D\in H_D$,  and a unitary $U:H_{AD}\raw H_{BC}$, to construct
\begin{equation}
\Phi(\rh_A)=\Tr_{H_C}(U(\rh_A\ot |\ps_D\ra\la\ps_D|) U^*).\label{Stinespring2}
\end{equation}
\end{itemize}
We omit the proof that these maps indeed satisfy the conditions in Definition \ref{defQC}, and also that every quantum channel can be put into any of the  forms \er{Kraus} - \er{Stinespring2}.\footnote{See Theorem 5.3 and Lemma 5.5, as well as Exercise 5.5. for \er{Stinespring}, in Walter \& Ozols (2025).} We also state without (the nontrivial) proof that the composition of \textsc{cp} maps is again \textsc{cp}; the same is true more trivially for \textsc{tp} maps, and hence, as expected,  quantum channels may be composed into quantum channels.   
\section{Quantum hypothesis testing}\label{QHTsec}
As in the classical case, one of the cleanest applications (and operational interpretations) of the relative quantum entropy \er{QKL} - \er{QKLb} arises in hypothesis testing.
The classical problem treated in \S\ref{Cht}, i.e.\
 the question if an unknown $p\in\Pr(A)$ is given by either $p=p_0$ or $p=p_1$, is easily adapted to the quantum setting:\footnote{Except for replacing $A$ by $H$ and $p$ by $\rh$ we try to keep the same notation as in the classical case.}  we now have a (finite-dimensional) Hilbert space $H$ and ask if an unknown density matrix $\rh\in D(H)$ is given by either $\rh=\rh_0$ or $\rh=\rh_1$. In the classical case the answer was determined by the choice of a test $T\subset A$ and the outcome $a\in A$ of a draw, or, for $N>1$, by a test $T_N\subset A^N$ and an outcome $\sg\in A^N$.
 This remains true in the quantum case, where, restricting ourselves to $N=1$ for the moment, tests correspond to (closed) linear subspaces of $H$,  or, equivalently, to the corresponding projections $T$ onto this subspace.\footnote{As in the classical case, this notion of a test will be refined later on.}
 
 There is, however, a fundamental difference between  classical and quantum: classically, sampling $A$ can be done independently of the choice of the test $T$, whereas in quantum theory one cannot just ``sample'' from a \Hs\ $H$ and randomly draw some unit vector $\ps\in H$ (or, more precisely, the corresponding  projection $|\psi\ra\la\psi|$);  instead, one first needs to specify  which operator or observable one measures.
 But this is, of course, the projection $T$ acting as a test, and so the outcome of the sampling is necessarily
  one of its eigenvectors  $\psi\in H_{\lm}$,  yielding the corresponding eigenvalue
   $\lm\in\sg(T)$ with probability $P_{\rh}(T=\lm)=\Tr (\rh E_{\lm})$, cf.\  \er{BR1}.
For a projection $T=T^*=T^2$, assuming $0\neq T\neq 1_H$, we have $\sg(T)=\{0,1\}$, with eigenspaces 
\begin{align}
H_0=T(H)^{\perp}=(1_H-T)H=\ran(1_H-T);  H_1=T(H)=\ran(H).
\end{align}
Hence a measurement of $T$ is a random drawing of some unit vector $\psi$ \emph{in either $H_0$ or $H_1$}, 
with:\footnote{ In particular, if $\rh=|\phv\ra\la\phv|$ and  $\phv\in H_0$, 
 then  $P_{\rh}(T=0)=1$ (and of course $P_{\rh}(T=1)=0$), and if $\phv\in H_1$, 
 conversely $P_{\rh}(T=1)=1$ (and $P_{\rh}(T=0)=0$). For better understanding the analogy with the classical case, recall that in Hilbert space the \emph{orthogonal complement} $T^{\perp}:=\{\ps\in H\mid \la\psi,\phv\ra=0\, \forall \phv\in T\}$ of some closed linear subspace $H_T\subset H$ plays the role of the set-theoretic complement $B^c=A\backslash B$ of a subset $B\subset A$ in the classical setting.  If $T$ is the projection onto $T$, then $1_H-T$ is the projection onto $T^{\perp}$. The  difference between classical and quantum is that whereas $a\in A$ lies in either $B$ or in $B^c$, 
  a vector $\psi\in H$ need not lie in either $T$ or $T^{\perp}$. That said, a convincing classical analogue of the projection $T$ is the indicator (or characteristic) function $1_T$ on $A$, which also assumes the values $0,1$ only.}
 \begin{align}
 P_{\rh}(\psi\in H_0)&= P_{\rh}(T=0)=\Tr(\rh (1_H-T)); \label{11.1HT}\\ P_{\rh}(\psi\in H_1)&=P_{\rh}(T=1)=\Tr(\rh T).\label{11.2}
 \end{align}
Note a  nuisance:\footnote{This also occurs classically if we take a test to be the characteristic function $1_T$ of $T$ instead of $T\subset A$ itself; then $a\in T$ confirming  $\mathsf{H}_0$ is the same as $1_T(a)=1$. See also footnote \ref{Poor}.
 One may  consider more general tests  $T\in [0, 1_H]$, seen as a \textsc{povm} \er{POVM}, see supplement, where
$X=\{0,1\}$, $F(0)=T$, and  $F(1)=1_H-T$. The outcomes $x=0$ and $x=1$ now do confirm 
$\mathsf{H}_0$ and $\mathsf{H}_1$, respectively (avoiding the  nuisance) and hence the probabilities of confirming $\mathsf{H}_0$ and $\mathsf{H}_1$ in a state $\rh$ are $\Tr(\rh T)$ and $\Tr(\rh(1_H-T))$, respectively. In particular,  the error probabilities remain \er{11.4}. }  classically, $a\in T\subset A$ meant accepting the null hypothesis 
 $\mathsf{H}_0$, whereas now an outcome $\psi\in H_1=T(H)$, corresponding to $\lm=1$,  gives 
 $T=1$. Hence \emph{ the outcome  $T=1$ confirms  $\mathsf{H}_0$ ($\rh=\rh_0$), whereas $T=0$ confirms  $\mathsf{H}_1$ ($\rh=\rh_1$)}. By  \er{11.1HT} - \er{11.2} we still have
     \begin{align}
   \al(T)=\Tr(\rh_0 (1_H-T)); && \beta(T)=\Tr(\rh_1 T), \label{11.4}
   \end{align}
   for the false-positive and false-negative probabilities, since by \er{11.2} this expression for $\al(T)$ gives the probability that a ``draw'' $\ps$ from $H$ according to the true state $\rh_0$  nonetheless lies in $H_0$ and hence  confirms $\mathsf{H}_1$, whereas $\beta(T)$ is  the probability that  although the true state is $\rh_1$, we draw some $\ps\in H_1$, which confirms $\mathsf{H}_0$ (to keep the correspondence $H_i\leftrightarrow \mathsf{H}_i$ just change $T$ to $1_H-T$).
   
We now return to  the Neyman--Pearson test \er{55} and epinomous Lemma \ref{NPL}.\footnote{Even if we take generalized tests $T\in [0, 1_H]$ into account the optimal test is given by a good old projection. As in the classical case, $T^*$ still denotes a  Neyman--Pearson test rather than the adjoint of $T$ (which would be $T$ itself.
 } We still take $N=1$ for the moment (which will imply the results for general $N$). We write $T^*$ for $T_1^*$. 
Recall that $T^*(0)\subset A$ 
is the region where $p_0>p_1$; more generally, the test $T^*(u)$ defined by \er{LLN} corresponds to
$p_0>e^u p_1$. The discussion above suggest that the quantum analogue of $T^*(u)$ 
is simply given by the projection onto the linear subspace of $H$ where  $\rh_0>e^u \rh_1$. We denote this by
\begin{equation}
T^*(u):=[\rh_0-e^u\rh_1>0],\label{11.7}
\end{equation}
where for self-adjoint $S=S^*\in L(H)$, we define  $[S>0]$ and $[S\geq 0]$ by the functional calculus: if 
\beq
S=\Sigma_{\lm\in\sg(S)} \lm\cdot E_{\lm}
\eeq is the spectral resolution of $S$, then
\begin{align}
[S> 0]=\Sigma_{\lm\in\sg(S), \lm > 0} E_{\lm}\; && 
[S\geq 0]=\Sigma_{\lm\in\sg(S), \lm\geq 0} E_{\lm}. \label{11.9}
\end{align}
Alternatively, we have a  decomposition $S=S_+-S_-$ with $S_{\pm}\geq 0$ and $S_+S_-=0$, given by
\begin{align}
S=S_+-S_-; && S_+=[S\geq 0]=\Sigma_{\lm\in\sg(S), \lm\geq 0} \lm\cdot E_{\lm}; && S_-=-\Sigma_{\lm\in\sg(S), \lm <0} \lm\cdot E_{\lm}, \label{11.10}
\end{align}
and the connection between \er{11.9} and \er{11.10} is obviously given by
\begin{equation}
S_+=S [S\geq 0]. \label{11.11}
\end{equation}
Also note that by the spectral calculus, here leading to $|S|=\Sigma_{\lm\in\sg(S)}|\lm| E_{\lm}$, 
we have
\begin{equation}
|S|=S_++S_-.\label{Sabs}
\end{equation}
If $\rh_0$ and $\rh_1$ commute and hence can be jointly diagonalized,
say by 
\beq
\rh_i=\Sigma_{a\in A} p_i(a) |e_a\ra\la e_a|,
\eeq
 then
\begin{equation}
1_H-[e^u\rh_1-\rh_0\geq 0]=\Sigma_{a\in T^*} |e_a\ra\la e_a|,
\end{equation}
where
 $(e_a)$ is a basis of $H$, $p_i\in\Pr(A)$, $i=0,1$, and $T^*=(p_0>p_1)$. In that case the quantum Neyman--Pearson test \er{11.7} reduces to the classical one (provided we embed $A$ in $H$ via $a\mapsto e_a$).
 
The key lemma for proving a quantum analogue of Lemma \ref{NPL} is then as follows:
\begin{lemma}\label{trlem}
For any $S^*=S$ we have 
\begin{equation}
\sup_{T\in [0, 1_H]}\{\Tr(ST)\}=\Tr(S_+), \label{preQNPL}
\end{equation}
so that the supremum is in fact achieved by the projection $T=[S\geq 0]$, see \er{11.11}.
\end{lemma}
\bex Prove this lemma.
\eex
From this, a  calculation  using \er{11.11} yields the following  ``quantum Neyman--Pearson lemma'':
 \begin{lemma}[Holevo, Helstrom]
For any $u\in \R$ the error $\al(T)+e^u\beta(T)$ is minimized by \er{11.7}:
  \begin{align}
\gm(u)&:=\inf_{T\in [0, 1_H]}\{\al(T)+e^u\beta(T)\}=\al(T^*(u))+e^u\beta(T^*(u))\nn \\
&=
\Tr(\rh_0[e^u\rh_1\geq\rh_0])+\Tr(\rh_1[\rh_0>e^u\rh_1]).
\label{HHL}
\end{align}
Hence if $\al(T)\leq \al(T^*(u))$, then $\beta(T)\geq\beta(T^*(u))$, and if $\beta(T)\leq \beta(T^*(u))$, then $\al(T)\geq\al(T^*(u))$.
\end{lemma}
Here $\al(T)$ and $\beta(T)$ are defined by \er{11.4}. This follows from \er{preQNPL} via something  similar:\footnote{See  Khatri and Wilde (2024), Theorem 5.3, for a detailed proof.}
\begin{lemma}
For any positive operators $A,B\in L(H)_+$ we have
\beq
\inf_{T\in [0, 1_H]}\{\Tr((1_H-T)A)+\Tr(TB)\}=\half(\Tr(A+B)-\| A-B\|_1). \label{somsim}
\eeq
\end{lemma}
For example, taking $A=\rh_0$ and  $B=\rh_1$ (hence $u=0$) we obtain the explicit expression
\begin{equation}
\gm\equiv\gm(0)=\Tr(\rh_0[\rh_1\geq \rh_0])+\Tr(\rh_1[\rh_0>\rh_1])=
1-\half\| \rh_0-\rh_1\|_1,
\end{equation}
where the last equality can also be directly verified using \er{11.11} and \er{Sabs} for $S=\rh_1-\rh_0$.
\bex
Prove \er{somsim} and then  prove \er{HHL} from \er{preQNPL} and \er{somsim}.
\eex
For pure states $\rh_i=|\psi_i\ra\la\psi_i|$ this yields, after a $2\x 2$ matrix computation,\footnote{The eigenvalues of $\rh_0-\rh_1$ are $\lm_{\pm}=\sqrt{1-|\la\ps_0,\psi_1\ra|^2}$.}
\begin{equation}
\gm=1-\sqrt{1-|\la\ps_0,\psi_1\ra|^2}. \label{gammapurestates}
\end{equation}
In particular, if $\psi_0$ and $\psi_1$ are orthogonal, then $\gm=0$, and so the test 
\beq
T^*=[|\psi_0\ra\la\psi_0| - |\psi_1\ra\la\psi_1|>0]
\eeq
distinguishes $\rh_0$ and $\rh_1$ without any probability of error. Indeed, in the linear span of $\psi_0$ and $\psi_1$ (ignoring the rest of the \Hs) we take $\psi_0$ and $\psi_1$ as basis vectors, so that
\begin{align}
\rh_0=\left(
\begin{array}{cc}
 1 &   0  \\
 0 &   0
\end{array}
\right); && \rh_1=\left(
\begin{array}{cc}
 0 &   0  \\
 0 &   1
\end{array}
\right); && 
T^*=\left(\begin{array}{cc}
 1 &   0  \\
 0 &   0
\end{array}
\right),
\end{align} 
and hence $\al(T^*)=\Tr(\rh_0(1_2-T^*))=0$ and $\beta(T^*)=\Tr(\rh_1 T^*)=0$.

As in the classical case, we may then try to improve the precision of a test by independently repeating it $N$ times. This  corresponds to replacing $(H,\rh)$ by $(H^N,\rh^N)$,\footnote{Here $H^N$ is short for the $N$-fold tensor product $H^{\otimes N}$ and likewise $\rh^N$ denotes $\rh^{\otimes N}$.} and $T$ by $T_N$, some projection in $L(H^N)$ that may  be seen either as a test $N$ times repeated or as a single test. Hence 
     \begin{align}
   \al_N(T_N)=\Tr(\rh^N_0 (1_{H^N}-T_N)); && \beta_N(T)=\Tr(\rh^N_1 T_N),\label{11.4N}
   \end{align}
   where $T_N\in [0,1]_{H^N}\subset L(H^N)$, cf.\ \er{11.4}.  This does indeed decrease the error, because
   \begin{equation}
\gm_{N+1}(u)\leq \gm_N(u).
\end{equation}
Namely, for any $T_N\in [0,1]^{H^N}$ at level $N$ one may include $T_{N+1}=T_N\ot 1_H\in [0,1]^{H^{N+1}}$ at  $N+1$, so that the infimum at level $N$ can only be improved by including more tests at level $N+1$.

All this leads to satisfactory quantum analogues of Theorems \ref{Chernoff}, \ref{Stein}, and \ref{Hoeffding}, in which the classical relative entropies $S_t(p_0,p_1)$ and $S(p_0, p_1)$ which occur in 
\er{CIF}, \er{KL1}, and \er{HB}  are replaced by their naive quantum versions, viz.
\begin{align}
S_t(p_0,p_1)=  \log\left(\Sigma_{a\in A} p_0(a)^{1-t}p_1(a)^{t}\right) &\leadsto S_t(\rh_0,\rh_1):= \log \Tr(\rh_0^{1-t}\rh_1^t);
\label{11.16}\\
S(p_0, p_1) =\Sigma_{a\in A} p_0(a)\log\left(\frac{p_0(a)}{p_1(a)}\right)& \leadsto S(\rh_0,\rh_1)=\Tr(\rh_0(\log\rh_0 -\log\rh_1)),\label{11.15}
\end{align}
where in \er{11.15} we 
keep \er{KL2} and \er{QKLb} in mind, and  the   powers $\rh_i^s$ are defined by the functional calculus, that is, if
$\rh=\Sigma_i p_i|\ups_i\ra\la \ups_i|$, as in \er{sdrho}, then $\rh^s=\Sigma_i p^s_i|\ups_i\ra\la \ups_i|$.
We use the same notation as for the corresponding classical expression, since the arguments ($\rh_i$ instead of $p_i$) make clear whether it is  classical  or quantum. Likewise, the quantum relative R\'{e}nyi entropy is
\begin{align}
R_t(\rh_0,\rh_1)&:=\frac{1}{t-1}\log \Tr(\rh_0^{t}\rh_1^{1-t}) && \mbox{ if } \ker(\rh_1)\subseteq \ker(\rh_0)) && (t>0, t\neq 1);\label{defRtQ}\\
R_t(\rh_0,\rh_1)&:=\infty && \mbox{ otherwise} && (t>0, t\neq 1);\\
R_1(\rh_0,\rh_1)&:=S(\rh_0,\rh_1).  \label{R0S}
\end{align}
Note that \er{R0S} ``follows'' from \er{defRtQ} by continuity,\footnote{Unlike the classical case this is not entirely trivial, see Khatri \& Wilde (2024), Proposition 7.22.}
and that 
 \beq
 R_t(\rh_0,\rh_1):= \frac{S_{1-t}(\rh_0,\rh_1)}{t-1}.
 \eeq
 With some effort one may show that 
 \begin{equation}
 R_t(\rh_0,\rh_1)\geq 0,
\end{equation}
with equality iff $\rh_0=\rh_1$,\footnote{See Khatri \& Wilde (2024), Propositions 7.26 and 7.36.} and in fact that $R_t(\rh_0,\rh_1)$ enjoys all the good properties of the usual relative entropy $S(\rh_0,\rh_1)$.\footnote{See Khatri \& Wilde (2024), Corollary 7.25 (data-processing inequality) and Proposition 7.27 (joint convexity).}
For $t>1$ another interesting operator ordering choice turns out to be
\begin{align}
\til{R}_t(\rh_0,\rh_1)&:=\frac{1}{t-1}\log \Tr\left(\left(\rh_1^{\frac{1-t}{2t}}
\rh_0 \rh_1^{\frac{1-t}{2t}}\right)^t\right); \hspace{50pt} (t>0, t\neq 1);\label{defRildetQ}\\
R_1(\rh_0,\rh_1)&:=S(\rh_0,\rh_1), \label{R0tildeS}
\end{align}
 again provided $\ker(\rh_1)\subseteq \ker(\rh_0)$, and equal to $+\infty$ otherwise.\footnote{The proof that \er{R0tildeS} is the limit of 
\er{defRildetQ} as $t\downarrow 1$ is  laborious, see Khatri \& Wilde (2024), Proposition 7.30.}
This is called the \emph{sandwiched relative R\'{e}nyi entropy}.\footnote{For its good properties see  Khatri \& Wilde (2024), Theorem 7.33 and Propositions 7.35, 7.36, and 7.37, notably $\til{R}_t(\rh_0,\rh_1)\geq 0$ with equality iff $\rh_0=\rh_1$, 
data-processing inequality, joint convexity, etc.}
Finally, the \emph{quantum Hoeffding entropies} are  defined by
 \begin{align}
H(r,\rh_0,\rh_1)&:=\sup_{t\in(0,1)}\left\{\left(\frac{1-t}{t}\right)\cdot (R_t(\rh_1,\rh_0)-r)
\right\}; 
 \label{defH01Q}\\
 H^*(r,\rh_0,\rh_1)&:= \sup_{t>1}\left\{\left(\frac{t-1}{t}\right)\cdot (r-R_t(\rh_1,\rh_0))
\right\}, \label{defH01Q2}
\end{align}
cf.\   \er{defH01}.
Here, then, are  quantum analogues of Theorems \ref{Chernoff}, \ref{Stein}, and \ref{Hoeffding}:
\begin{theorem}[Quantum Chernoff theorem]
For any  $\rh_0, \rh_1\in D(H)$, the asymptotics of 
\beq
\gm_N(\rh_0, \rh_1)=\inf_{T_N\in [0,1]_{H^N}}\{\al_N(T_N)+\beta_N(T_N)\}, \label{10.5}
\eeq
see \er{11.4N}, 
 equals
\begin{equation}
\lim_{N\raw\infty}\frac{1}{N}\log \gm_N(\rh_0, \rh_1)=-C(\rh_0,\rh_1),
\label{10.6}\end{equation}
where
\begin{equation}
C(\rh_0,\rh_1)=-\inf_{t\in (0,1)} \{S_t(\rh_0,\rh_1)\}=\sup_{t\in (0,1)} \{(1-t)R_t(\rh_0,\rh_1)\}.
\end{equation}
\end{theorem}
\begin{theorem}[Quantum Stein lemma]\label{SteinQ}
For any  $\rh_0, \rh_1\in D(H)$, and 
\begin{align}
\beta_N^{\varep}(\rh_0, \rh_1)&:=\inf_{T_N\in [0,1]_{H^N}}\{\beta_N(T_N)\mid \al_N(T_N)\leq \varep\} \hspace{75pt} (0<\varep<1); \label{9.21Q}\\
\beta^*(\rh_0,\rh_1)&:= \inf_{(T_N)}\left\{ \lim_{N\raw\infty} \frac{1}{N}\log \beta_N(T_N)\mid  T_N\in [0,1]_{H^N},
\lim_{N\raw\infty}\al_N(T_N)<1
\right\};\label{QSL2} \\
\beta_*(\rh_0,\rh_1)&:= \inf_{(T_N)}\left\{ \lim_{N\raw\infty} \frac{1}{N}\log \beta_N(T_N)\mid  T_N\in [0,1]_{H^N},
\lim_{N\raw\infty}\al_N(T_N)=0
\right\}\label{QSL3},
\end{align}
taken over all \emph{families} $(T_N)$ of tests for which the limits inside the curly brackets exist,  
we have
\begin{align}
\lim_{N\raw\infty}\frac{1}{N}\log \beta^{\varep}_N(\rh_0, \rh_1)&=-S(\rh_0, \rh_1); \label{9.23Q}\\
\beta^*(\rh_0,\rh_1)=\beta_*(\rh_0,\rh_1)&=-S(\rh_0, \rh_1). \label{betabeta}
\end{align}
\end{theorem}
\begin{theorem}[Quantum Hoeffding theorem]\label{QHoeffding}
If for given $r>0$ we define
\begin{align}
\beta_N(r,\rh_0,\rh_1):=\inf_{T_N\in [0,1]_{H^N}}\{\beta_N(T_N)\mid\al_N(T_N)\leq e^{-Nr}\}, \label{9.21QH}
\end{align}
then  
\begin{align}
\lim_{N\raw\infty}\frac{1}{N}\log \beta_N(r,\rh_0,\rh_1)=-H(r,\rh_0,\rh_1).
\label{HBQ}
\end{align}
Moreover, the following  dichotomy arises: 
\begin{itemize}
\item 
for $0<r<S(\rh_1,\rh_0)$ we have $H(r,\rh_0,\rh_1)>0$, so that
\begin{equation}
\lim_{N\raw\infty} \beta_N(r,\rh_0,\rh_1)=0, \label{firstcase}
\end{equation}
\item for $r\geq S(\rh_1,\rh_0)$ we have $H(r,\rh_0,\rh_1)=0$ and
\begin{equation}
\lim_{N\raw\infty} \beta_N(r,\rh_0,\rh_1)=1. \label{secondcase}
\end{equation}
\end{itemize}
In \er{secondcase} the
approach to 1 is measured by the ``dual'' Hoeffding entropy \er{defH01Q2} according to
\begin{equation}
\lim_{N\raw\infty}\frac{1}{N}\log(1- \beta_N(r,\rh_0,\rh_1))=-H^*(r,\rh_0,\rh_1).
\end{equation}
where $H^*(r,\rh_0,\rh_1)>0$ if $r>S(\rh_1,\rh_0)$ whilst $H^*(r,\rh_0,\rh_1)=0$ for $0<r\leq S(\rh_1,\rh_0)$.
\end{theorem}

The interpretation of the second part of Hoeffding's theorem is the same as in the classical case. Also now.
in  a more general formulation of this theorem one defines, still for $r>0$, 
\begin{align}
h(r,\rh_0,\rh_1)&:= \inf_{(T_N)}\left\{ \lim_{N\raw\infty} \frac{1}{N}\log \beta_N(T_N)\mid T_N\in [0,1]_{H^N}, 
\lim_{N\raw\infty}\frac{1}{N}\log \al_N(T_N)\leq - r
\right\}\label{Hoef1Q},
\end{align} 
and claims that the infimum is  reached by some sequence $(T_N)$ of tests,\footnote{This version combines elements of  Audenaert et al.\ (2008), Theorem 5, and Hayashi (2017), Theorem 3.4.}
 and states \er{HBQ} as
\begin{equation}
h(r,\rh_0,\rh_1)=-H(r,\rh_0,\rh_1). \label{hHoef}
\end{equation}
The other claims in Theorem \ref{QHoeffding}  also hold, with $\lim_{N\raw\infty} \beta_N(r,\rh_0,\rh_1)$ replaced by $h(r,\rh_0,\rh_1)$.
\subsection*{Proof of quantum Chernoff theorem}
We first check the special case where $\rh_i=|\psi_i\ra\la\psi_i|$ are  pure states. As $N\raw\infty$, eq.\  \er{gammapurestates}
gives
\begin{equation}
\gm_N=1-\sqrt{1-|\la\ps^N_0,\psi^N_1\ra|^2}=1-\sqrt{1-|\la\ps_0,\psi_1\ra|^{2N}}\approx \half |\la\ps_0,\psi_1\ra|^{2N}=\half e^{-NC(\psi_0,\psi_1)},
 \label{gammaNpurestates}
\end{equation}
where $C(\psi_0,\psi_1):=-\log |\la\ps_0,\psi_1\ra|^2$. Then $\rh_0^{1-t}=\rh_0$ and $\rh_1^t=\rh_1$, since the eigenvalues of projections are 0 and 1. Hence $\Tr(\rh_0^{1-t}\rh_1^t)=\Tr(\rh_0\rh_1)=|\la\ps^N_0,\psi^N_1\ra|^2$, so that 
$C(\rh_0,\rh_1)=C(\psi_0,\psi_1)$ as just defined, consistent with \er{10.6}.

The general
proof consists of separate arguments for upper and lower bounds on the left-hand side of \er{10.6}. As in the classical proofs we omit the argument $(\rh_0,\rh_1)$ of the exponents $\al$, $\beta$, and $\gm$.
The upper bound follows from \er{HHL}, which gives an optimal test $T_N^*$, so that
\begin{align}
\gm_N=\Tr\left(\rh_0^N( 1_{H^N}-T_N^*)\right)+\Tr\left(\rh_1^N T_N^*\right);
&& T_N^*:=[\rh_0^N>\rh_1^N],
\label{10.7}
\end{align}
 and a quantum counterpart of the following fact about classical probability distributions:
 \beq
\Sigma_{a\in A}( f(a) 1_{f\leq g}(a)+g(a)1_{f>g}(a))\leq \Sigma_{a\in A} f(a)^{1-t}g(a)^t. \label{10.8C}
\eeq
 \begin{lemma}
 For all positive operators $A,B\geq 0$ and $t\in [0,1]$ we have
 \begin{equation}
\Tr (A [A\leq B]) +\Tr(B [A> B])\leq \Tr(A^{1-t}B^{t}),\label{10.8Q}
\end{equation}
 \end{lemma}
\emph{Proof.}\footnote{Given by   Jak\v{s}i\'{c} et al.\ (2012), Proposition 1.1, as well as by Hayashi (2017), Lemma 3.3, attributed to N. Ozawa.}
 From \er{11.11} and $[A>B]=1_H-[A\leq B]=1_H-[B\geq A]$, eq.\ \er{10.8Q} is the same as 
 \begin{equation}
\Tr(A) -\Tr(A^{1-t}B^{t})\leq \Tr(A-B)_-. \label{10.8pre}
\end{equation}
\bex Prove \er{10.8pre} by using (without proof)  the following operator inequality: 
\begin{align}
A\leq C\,\Raw\ A^t\leq C^t && (A,C\geq 0, \:\: t\in[0,1]),
\end{align} 
 with $C=B+(A-B)_+$. \QED
 \eex
From this lemma,  it is a simple exercise to show that
\begin{equation}
\lim\sup_{N\raw\infty}\frac{1}{N}\log \gm_N\leq -C(\rh_0,\rh_1).\label{10.9}
\end{equation}
\bex
Prove \er{10.9} from \er{10.8Q}.\eex
The crucial inequality providing a  lower bound on $\gm_N$ is
\begin{equation}
\inf_{T}\{ \Tr(\rh_0(1_H-T))+\Tr(\rh_1T)\}\geq\half\Sigma_{a\in A, b\in B} \min\{p_0(a),p_1(b)\} |\la \ups_a,\ups_b^{(1)}\ra|^2, \label{10.10}
\end{equation}
where $A=B=\{1, \ldots, \dim(H)\}$, and the spectral resolutions of $\rh_0$ and $\rh_1$ are given by \er{joint01}.
This inequality does not rely on \er{HHL}.
\bex
Try to prove \er{10.10}.\footnote{For help, look at the proof of Lemma 3.4 at the end of \S 3.2 in Hayashi (2017).}
\eex
Now introduce  probability distributions $P_0(\rh_0,\rh_1)$ and $P_1(\rh_0,\rh_1)$ on $A\x B$
 by
\begin{align}
P_0(a,b| \rh_0,\rh_1):= p_0(a) |\la \ups_a^{(0)},\ups_b^{(1)}\ra|^2; &&P_1(a,b|\rh_0,\rh_1):= p_1(b) |\la \ups_a^{(0)},\ups_b^{(1)}\ra|^2, \label{P0P1}
\end{align}
 where $\rh_0$ and $\rh_1$ are fixed labels and $(a,b)$ are arguments, suggestively follows by these labels.
Then the marginal of $P_0$ on $A$ equals $p_0$, whilst  the marginal of $P_1$ on $B$ equals $p_1$. Moreover,
\begin{align}
S(P_0, P_1)&=S(\rh_0, \rh_1); \label{10.12}\\
S_{t}(P_0,P_1)&=S_{t}(\rh_0,\rh_1).\label{10.13}
\end{align}
abusing notation as before, 
cf.\  \er{11.16} -  \er{11.15}. The point is that we may now rewrite \er{10.10} as
\begin{equation}
\gm\geq \half( P_0(P_0\leq P_1)+P_1(P_0>P_1)),
\end{equation}
where $P_0\leq P_1$ is the set 
\beq P_0\leq P_1=
\{(a,b)\in A\x B\mid P_0(a,b)\leq P_1(a,b)\}, \eeq
etc. Similarly, the spectral resolutions
\begin{align}
\rh_0^N=\Sigma_{\sg\in A^N} p^N_0(\sg) |\ups_{\sg}\ra\la \ups_{\sg}|; && \rh^N_1=\Sigma_{\ta\in A^N}^dp^N_1(\ta)|\ups'_{\tau}\ra\la \ups'_{\tau}|,
\end{align}
where $\ups_{\sg}=\ups_{\sg_0}\ot\cdots \otimes\ups_{\sg_{N-1}}$ are  eigenvectors of $\rh_0^N$ with eigenvalues $p_0^N(\sg)$, and likewise for $\rh_1$, give probability distributions $P_0(\rh_0^N,\rh_1^N)$ and $P_0(\rh_0^N,\rh_1^N)$ on $A^N\x A^N=(A\x A)^N$ as in \er{P0P1}:
\begin{align}
P_0(\sg,\ta|\rh_0^N,\rh_1^N):= p_0^N(\sg) |\la \ups_{\sg},\ups'_{\ta}\ra|^2; &&
P_1(\sg,\ta|\rh_0^N,\rh_1^N):= p_1^N(\ta) |\la \ups_{\sg},\ups'_{\ta}\ra|^2.
\end{align}
\bex
Show that
\begin{align}
P_0(\rh_0^N,\rh_1^N)=P_0(\rh_0,\rh_1)^N; && P_1(\rh_0^N,\rh_1^N)=P_1(\rh_0,\rh_1)^N.\label{10.18}
\end{align}\eex
The arguments from \er{10.10} onwards may now be repeated for $H\leadsto H^N$ and $\rh_i\leadsto \rh_i^N$, giving
\begin{equation}
\gm_N\geq \half( P^N_0(P^N_0\leq P^N_1)+P^N_1(P^N_0>P^N_1)),
\end{equation}
where now $P^N_0\leq P^N_1$ is the set 
\beq P^N_0\leq P^N_1=
\{(\sg,\ta)\in A^N\x B^N\mid P^N_0(\sg,\ta)\leq P^N_1(\sg,\ta)\},
\eeq 
and likewise $P^N_0>P^N_1$.
The classical case (based on Sanov's theorem), with $p_i$ replaced by $P_i$, combined with \er{10.13}, then gives 
\begin{equation}
\lim\inf_{N\raw\infty}\frac{1}{N}\log \gm_N\geq -C(P_0,P_1)=-C(\rh_0,\rh_1).\label{10.20}
\end{equation}
Combining \er{10.20} and \er{10.9} gives \er{10.6}. \QED
\subsection*{Proof of quantum Stein lemma}
Though a clean analogue of the classical Stein lemma, this quantum version is more difficult to prove than any previous result, and so we just provide sketches of the main ideas.  
In all approaches there are separate proofs of the upper and lower bounds. The original proof of the former, i.e.,
\beq
\lim\sup_{N\raw\infty}\frac{1}{N}\log \beta^{\varep}_N\leq-S(\rh_0, \rh_1),\label{9.23Qb}
\eeq
due to Hiai and Petz (1991), relied on the following operator version of Lemma \ref{CHiaiPetz}:
 \begin{lemma}\label{QHiaiPetz}
  For  any $\dl>0$ there is a sequence of tests $(T_N)$, $T_N\in [0,1]_{H^N}$, such that
 \begin{align}
   \lim_{N\raw\infty} \Tr(\rh_0^NT_N)&=1;\label{QHP2}\\
 \lim\sup_{N\raw\infty} \frac{1}{N} \log \Tr(\rh_1^N T_N) &\leq -S(\rh_0, \rh_1)+\dl. \label{QHP1}
 \end{align}
 \end{lemma}
This yields \er{9.23Qb} by the same reasoning as in the classical case, see the text after Lemma \ref{CON}. 
\smallskip

\noindent
\emph{Proof of Lemma \ref{QHiaiPetz}.} The required test is contrived, but it works.
 For $t\in [0,1]$ and $r>0$, take 
 \begin{align}
T_N^*(R):=[\rh_0^N>e^{NR}\rh_1^N]; && R=\frac{r+S_t(\rh_0,\rh_1)}{1-t}.\label{SHtest1} 
\end{align}
 Using \er{defH01Q}, the key inequality \er{10.8Q} then gives
 \begin{align}
 \Tr(\rh_0^N (1_{H^N}-T_N^*(R)))& \leq  e^{-N\left(\frac{-tr-S_t(\rh_0,\rh_1)}{1-t}\right)}\leq e^{-NH(r,\rh_0,\rh_1)}
 ;\label{11.55} \\
   \Tr(\rh_1^N T_N^*(R))& \leq e^{-Nr}.\label{11.56}
 \end{align}
 \bex
Using the definition \er{11.16} and the functional calculus, prove that
\begin{equation}
\frac{d}{dt}S_t(\rh_0,\rh_1)_{|t=0}=-S(\rh_0,\rh_1). \label{propertyS}
\end{equation}
\eex
Now $r<S(\rh_0, \rh_1)$ implies  $H(r,\rh_0,\rh_1)>0$. Namely, using \er{propertyS}, for  $0\leq t\leq 1$, take 
\beq
f(t):=\frac{tr+S_t(p_0,p_1)}{t-1}.
\eeq
Using \er{propertyS}, we find  $f(0)=0$ and $f'(0)=S(\rh_0, \rh_1)-r$, which is $>0$ provided $r< S(\rh_0, \rh_1)$. Hence the supremum in \er{defH01Q} is strictly positive, so that by \er{11.55} the test \er{SHtest1} satisfies \er{QHP2}. By \er{11.56}, 
if we write $r=S(\rh_0, \rh_1)-\dl$, for $\dl>0$, our test
 also satisfies \er{QHP1}.
\QED
\smallskip

 As in the classical case, others tests can do the job. For example, one can mimic the  test
\begin{equation}
\hat{T}_{N,\dl}(p_0,p_1):=\{\sg\in A^N\mid
e^{-N(S(p_0, p_1)+ S(p_0)+\dl)}\leq p_1^N(\sg)\leq e^{-N(S(p_0, p_1)+ S(p_0)-\dl)}\}. \label{odl2}
\end{equation}
To ``quantize'' this test, the idea is to use the spectral resolutions \er{joint01} of $\rh_0$ and $\rh_1$, given by
\begin{align}
\rh_i=\Sigma_{a\in A} p_i(a) |\ups_a^{(i)}\ra\la \ups_a^{(i)}|; && A:=\{1,\ldots, \dim(H)\}, &&
 i=0,1.
\end{align}
 Each basis $(\ups_a^{(i)})_{a\in A}$ of $H$ induces a basis $(\ups_{\sg}^{(i)})_{\sg\in A^N}$ of $H^N$, in terms of which
\begin{equation}
\rh_i^N=\Sigma_{\sg\in A^N}  p_i^N(\sg)|\ups_{\sg}^{(i)}\ra\la \ups_{\sg}^{(i)}|.
\label{older8.19}
\end{equation}
The quantum analogue of the classical test  $\hat{T}_{N,\dl}(p_0,p_1)\subset A^N$ is then the projection
\begin{align}
 \hat{T}_{N,\dl}(\rh_0,\rh_1)&:=\Sigma_{\sg\in \til{T}_{N,\dl}(\rh_0,\rh_1)}|\ups_{\sg}^{(1)}\ra\la \ups_{\sg}^{(1)}|; \label{8.25bis}\\
 \til{T}_{N,\dl}(\rh_0,\rh_1)&:= \left\{\sg\in A^N\mid
e^{-N(S(\rh_0, \rh_1)+ S(\rh_0)+\dl)}\leq p_1^N(\sg)\leq e^{-N(S(\rh_0, \rh_1)+ S(\rh_0)-\dl)}\right\}. \label{odl2b}
\end{align}
We also need a quantum version of the AEP set \er{AEPset}.but now in terms of $\rh_0$, namely 
\begin{align}
\hat{T}_{N,\dl}(\rh_0)&:=\Sigma_{\sg\in \til{T}_{N,\dl}(\rh_0)}|\ups_{\sg}^{(0)}\ra\la \ups_{\sg}^{(0)}|,
\end{align}
where $\til{T}_{N,\dl}(\rh_0)=T_{N,\dl}(p_0)$ as defined in \er{AEPset}, since $S(\rh_0)=S(p_0)$. The test of choice, then, is
\begin{equation}
T_{N,\dl}(\rh_0,\rh_1):= \hat{T}_{N,\dl}(\rh_0,\rh_1)  \hat{T}_{N,\dl}(\rh_0)  \hat{T}_{N,\dl}(\rh_0,\rh_1);
\end{equation}
indeed, nontrivial operator estimates show that $T_N=T_{N,\dl}(\rh_0,\rh_1)$ satisfies \er{QHP2} and \er{QHP1}, too.

The advantage of this  approach is that, much as the upper bound in \er{odl2b} forms the basis of the upper bound \er{QHP1}, the lower bound in  \er{odl2b} leads to a proof of the  lower bound
\beq
\lim\inf_{N\raw\infty}\frac{1}{N}\log \beta^{\varep}_N\geq-S(\rh_0, \rh_1).\label{9.23Qbc}
\eeq
This lower bound was originally proved  by contradiction,\footnote{See Ogawa \& Nagaoka (2002). A slightly more transparent proof may be found in Hayashi (2017), Lemma 3.7.} cf.\ the contrapositive of Lemma \ref{CON}:
\begin{lemma}\label{Japanese}
If  some sequence of tests $(T_N)$, $T_N\in [0,1]_{H^N}$,  satisfies 
\begin{equation}
\lim\inf_{N\raw\infty}\frac{1}{N}\log \beta_N(T_N)<-S(\rh_0, \rh_1),
\end{equation}
then $\lim\sup_{N\raw\infty}\al_N(T_N)=1$. Equivalently, if $\lim\sup_{N\raw\infty}\al_N(T_N)<1$, then \er{9.23Qbc} holds.
\end{lemma}
Since $0<\varep<1$ in \er{9.21Q}, this gives \er{9.23Qbc}. Moreover, Lemmas  \ref{QHiaiPetz}  and \ref{Japanese} jointly  prove all equalities in \er{betabeta}, just as in the classical case. It therefore remains to prove Lemma \ref{Japanese}. This proof is very length and we therefore omit it,\footnote{See Hayashi (2017), pp.\ 125--127.}  but it is based on the same trick as in the proof of the lower bound in the quantum Chernoff theorem, namely \er{10.10} and \er{P0P1}.
\QED
\begin{corollary}
The quantum Stein lemma implies  Theorem \ref{Ruskai}.3, that is,  the  property 
\begin{align}
S(\Phi(\rh), \Phi(\sg))\leq S(\rh,\sg) && (\rh,\sg\in D(H_A))
 \label{MSL}
\end{align}
of the relative entropy
under \emph{quantum channels}, i.e., completely positive  trace-preserving maps 
\beq
\Phi: L(H_A)\raw L(H_B).\label{PhiAB}
\eeq\end{corollary}
\emph{Proof}.\footnote{This corollary is due to Bjelakovi\'{c} \&  Siegmund-Schultze (2012), \S3.} 
The (\textsc{cpu}) dual map $\Phi^*: L(H_B)\raw L(H_A)$ is  defined by the property
\begin{align}
\Tr_A (T_A\Phi^*(T_B))=\Tr_B(\Phi(T_A)T_B),
\end{align}
where $T_A\in L(H_A)$ and $T_B\in L(H_B)$. Both $\Phi$ and $\Phi^*$ preserve the pertinent unit intervals, i.e.,
\begin{align}
\Phi([0,1]_{H_A})\subseteq [0,1]_{H_B}; &&  \Phi^*([0,1]_{H_B})\subseteq [0,1]_{H_A}.
\end{align}
All this may be expanded to $\Phi^N: L(H_A^N)\raw L(H_B^N)$ and $(\Phi^*)^N: L(H_B^N)\raw L(H_A^N)$. Then compute:
\begin{align}
\beta_N^{\varep}(\Phi(\rh), \Phi(\sg))&=\inf_{T_N\in [0,1]_{H_B^N}}\{\Tr_{H_B^N}(\Phi^N(\sg^N)T_N)\mid 
\Tr_{H_B^N}(\Phi^N(\rh^N)T_N)>1-\varep\} \nn \\
&=\inf_{T_N\in [0,1]_{H_B^N}}\{\Tr_{H_A^N}(\sg^N(\Phi^*)^N(T_N))\mid 
\Tr_{H_A^N}(\rh^N\Phi_N^*(T_N))>1-\varep\} \nn \\
& \geq
\inf_{T_N'\in [0,1]_{H_A^N}}\{\Tr_{H_A^N}(\sg^N T_N')\mid 
\Tr_{H_A^N}(\rh^N T_N')>1-\varep\} 
=\beta_N^{\varep}(\rh,\sg),
\end{align}
where the inequality in the third line comes from the fact that the infimum following it is taken over a potentially larger set of tests.  The quantum Stein lemma then at once gives \er{MSL}. \QED
\subsection*{Proof of quantum Hoeffding theorem}
This is also quite lengthy and hence we merely provide a sketch.\footnote{A complete unified proof seems lacking. See Hayashi (2007), Nagaoka (2006),  Audenaert et al.\ (2008), \S 5.4, Mosonyi \& Ogawa (2015), Hayashi (2017), \S\S 3.7--3.8, 
and Khatri \& Wilde (2024), \S7.10,  Propositions 7.79 and 7.80.}
The test \er{SHtest1} also works 
here:  a  computation similar to the one leading to \er{11.55} - \er{11.56} gives, for any $t\in (0,1)$,
  \begin{align}
 \Tr(\rh_1^N (T_N^*(e^{NR})))& \leq  e^{-N\left(\left(\frac{1-t}{t}\right)\cdot (R_t(\rh_1,\rh_0)-R)\right)}\leq e^{-NH(R,\rh_0,\rh_1)}
 ;\label{1155} \\
   \Tr(\rh_0^N (1_{H^N}-T_N^*(e^{NR})))& \leq e^{-NR},\label{1156}
 \end{align}
 which implies the upper (achievability) bound, valid for any $R>0$,
 \beq
 \lim_{N\raw\infty}\frac{1}{N}\log \beta_N(R,\rh_0,\rh_1)\leq -H(R,\rh_0,\rh_1).
 \eeq
 This enough to prove \er{firstcase}. On $[0,1]$ the function $t\mapsto R_t(\rh_1,\rh_0)$ is monotonically increasing from
 $R_0(\rh_1,\rh_0)=0$ to $R_1(\rh_1,\rh_0)=S(\rh_1,\rh_0)$,\footnote{See Khatri \& Wilde (2024), Proposition 7.23.}  so if $0<R<S(\rh_1,\rh_0)$, then there is some $t$ for which
 $0<R< R_t(\rh_1,\rh_0)<S(\rh_1,\rh_0)$, so that $R_t(\rh_1,\rh_0)-R>0$ and hence $H(R,\rh_0,\rh_1)>0$.
 This gives \er{firstcase}. For $R\geq S(\rh_1,\rh_0)$  the function $t\mapsto t\inv(1-t)$ is positive definite on $(0,1)$ and zero at $t=1$, whereas $t\mapsto R_t(\rh_1,\rh_0)-R$ is negative or zero, so that $H(R,\rh_0,\rh_1)=0$.
 
  The corresponding upper bound  is proved in a similar way as for the Chernoff theorem, since
  \begin{equation}
H(r, P_0,P_1)=H(r,\rh_0,\rh_1),
\end{equation}
cf.\ \er{defH01},  \er{defH01Q}. and \er{10.12} - \er{10.13}. One also shows  that for any test $T_N$ for which 
\begin{equation}
\Tr(\rh_0^N (1_{H^N}-T_N)) \leq e^{-NR},
\end{equation}
we have, this time for any $t>1$ after the first upper bound sign,
\begin{equation}
\Tr(\rh_1^N (T_N)) \geq 1- e^{-N\left(  \left(\frac{t-1}{t}\right)\cdot (R-R_t(\rh_1,\rh_0)) \right)}
\geq 1- e^{-NH^*(R,\rh_0,\rh_1)}.
\end{equation}
Similar reasoning then implies \er{secondcase} and surrounding claims in Theorem \ref{QHoeffding}. 

Finally, if both limits in \er{Hoef1Q} exist for some sequence $T=(T_N)$, call them $\beta(T)$ and $\al(T)$, respectively, then    the observation that  $R\mapsto H(R,\rh_0,\rh_1)$ is decreasing implies
that
   \begin{align}
   \al(T)\leq -R && \Raw&& \beta(T)\geq -H(R,\rh_0,\rh_1). \label{11.73}
   \end{align}
Now the tests $T=(T_N)$ in  \er{SHtest1} satisfy 
\begin{align}
\al(T)\leq -R; && \beta(T)\leq -H(R,\rh_0,\rh_1),
\end{align} so that
\beq
\beta(T)= -H(R,\rh_0,\rh_1).\label{11.74}
\eeq In other words, the test sequence  \er{SHtest1}  saturates the infimum in in \er{Hoef1Q}. \QED\smallskip
 
As in the classical case, the ``atomic'' binary quantum hypothesis testing considered in this section so far implies certain results on the asymptotic errors for compound hypotheses. If we replace $(A,p)$ by $(H,\rh)$, $B_i\subset \Pr(A)$ by $B_i\subset D(H)$, and $p_i$ by $\rh_i$, we can keep the notation of the supplement to \S\ref{Cht}, and ask if the equalities \er{5109C} - \er{5109H} are also valid in quantum theory. Of course, if $[\rh_0,\rh_1]=0$ for all $\rh_0\in B_0$ and $\rh_1\in B_1$ we recover all classical results. But otherwise the equalities only hold under additional hypotheses.\footnote{See  Mosonyi,   Szil\'{a}gyi, \& Weiner (2021) for a comprehensive survey.  Our eq.\ \er{QSTHI} is Theorem 1 in  N\"{o}tzel (2014). See also eq.\ (4) in  Hayashi \& Ito (2025).  Our Theorem \ref{CMP2005Q} is Theorem 1 in  Bjelakovi\'{c} et al.\ (2005). } The most interesting case here seems to be the quantum Stein lemma, 
where it has been shown that \er{5109} holds in case that $B_1=\{\rh_1\}$ is a singleton. In that case the probability \er{Steincompound} becomes
\begin{equation}
 \beta_N^{\varep}(B_0,\rh_1):=\inf_{T_N}\left\{\Tr(\rh_1^NT_N)\mid \sup_{\rh_0\in B_0} \{\Tr(\rh_0^N(1_{H^N}-T_N))\right\}\leq \varep\},\label{Steincompound2}
\end{equation}
and the quantum Stein lemma with compound null hypothesis reads
\begin{equation}
 \lim_{N\raw\infty}\frac{1}{N}\log  \beta_N^{\varep}(B_0,\rh_1)=-S(B_0,\rh_1):=-\inf_{\rh_0\in B_0}\{S(\rh_0,\rh_1)\}, \label{QSTHI}
\end{equation}
where we used the notation $S(B_0,\rh_1)$ familiar from large deviation theory. In the same spirit, Lemma  \ref{QHiaiPetz}  may be improved to what also has been called a
``quantum Sanov theorem':\footnote{The literature on ``quantum Sanov theorems'' and other generalizations of the results on quantum hypothesis testing in finite-dimensional \Hs\ in the main text is substantial, and growing rapidly. That said, recent literature includes
Fang, Fawzi, \&  Fawzi (2024);  Fang \& Hayashi (2025); Gao \& Rahaman (2024);  Hayashi (2025); Hayashi \& Ito (2025);  Hayashi \& Yamasaki (2025); 
Ji \& Regula (2025);  Lami (2025ab); Lizarribar-Carrillo, Vazquez-Vilar, \& Koch  (2026), and Janse van Rensburg (2026). See also footnote \ref{vnaSanov} for infinite-dimensional \Hs s.}
\begin{theorem}\label{CMP2005Q}
For any $\rh_1\in D(H)$ and $B_0\subset D(H)$ there is a sequence of tests $(T_N)$ such that
\begin{align}
\lim_{N\raw\infty}\Tr(\rh_0^N T_N)=1; &&
\lim_{N\raw\infty}\frac{1}{N}\log  \Tr(\rh_1^N T_N) =-S(B_0,\rh_1), \label{1599J}
\end{align}
for all $\rh_0\in B_0$. In particular,
for any $\rh_0,\rh_1\in D(H)$ there is a sequence of tests $(T_N)$ such that
\begin{align}
\lim_{N\raw\infty}\Tr(\rh_0^N T_N)=1; && 
\lim_{N\raw\infty}\frac{1}{N}\log  \Tr(\rh_1^N T_N) =-S(\rh_0,\rh_1), \label{9.78Q}
\end{align}
and, as another special case of \er{1599J}, if $B_0$ is closed and $\rh_1\notin B_0$, then $S(B_0,\rh_1)>0$ and hence
\begin{align}
\lim_{N\raw\infty}\Tr(\rh_0^N T_N)=1; &&
\lim_{N\raw\infty}\Tr(\rh_1^N T_N) =0.\label{1599Jb}
\end{align}
\end{theorem}
Compare with its classical version Theorem \ref{CMP2005C}, which isn't  a Sanov theorem either. Moreover, as acknowledged by its authors, Theorem \ref{CMP2005Q} lacks the \emph{universal} feature of the classical Sanov theorem, where the classical tests $T_N=(L_N\in B)$ can be chosen independently of the prior $\sg$.

One promising approach to a universal quantum Sanov theorem,\footnote{For details see Janse van Rensburg (2026), part of whose work we briefly summarize here. He also extends the result to infinite-dimensional \Hs s and type I \vna s.}   is to start from a \textsc{povm} 
\beq
F:A\raw [0,1]_H, \label{FPOVM}
\eeq cf.\ \er{POVM} and the entire supplement below,\footnote{We here write $A$ instead the usual $X$ in the supplement for better comparison with the classical Sanov theorem.} whose associated quantum-to-classical channel 
\beq
Q_F:D(H)\raw\Pr(A),
\eeq
defined by \er{QFchannel} is injective; such \emph{information-complete} \textsc{povm}s always exist (in finite dimension).
Apart from $Q_F$, the  \textsc{povm} $F$ also naturally induces the following  maps by expansion:
\begin{align}
F^N&:A^N\raw B(H^N); && F^N(\sg):=F(\sg_0)\ot\cdots\ot F(\sg_{N-1});\\
\hat{F}^N&:\CP(A^N)\raw B(H^N); && \hat{F}^N(S):=\sum_{\sg\in S} F^N(\sg). 
\end{align}
In the classical Sanov theorem \er{eqLDP1}, we now replace:
\begin{enumerate}
\item the classical event $B\in\Pr(A)$ by a quantum event $\mathcal{B}\subset D(H)$;
\item  the value set of the empirical distribution $(L_N\in B)=L_N\inv(B)\subset A^N$  by 
the operator 
\beq (\mathcal{L}_N\in\mathcal{B}):=
\hat{F}^N(L_N\in Q_F (\mathcal{B}))=\hat{F}^N(L_N\inv (Q_F (\mathcal{B}))) \in B(H^N);\label{15105}
\eeq
\item   the classical probability $q^N(\cdot)$ in  \er{eqLDP1},  by the quantum probability $\Tr(\sg^{N}\cdot)$;
\item the classical relative entropy $S_q(p)=S(p,q)$ by a hybrid classical--quantum relative entropy
\begin{equation}
S^F_{\sg}(\rh):=S(Q_F(\rh),Q_F(\sg)).\label{JJentropy}
\end{equation}
\end{enumerate}
We then  have the following quantum Sanov theorem (in which $H$ is finite-dimensional):
\begin{theorem}[Janse van Rensburg]\label{sanovJJ}
For any (fixed) information-complete \textsc{povm} \er{FPOVM}, any density operator
 $\sg\in D(H)$, and any measurable subset $\mathcal{B}\subset D(H)$, one has
\begin{align}
- S^F_\sg(\mathring{\mathcal{B}}) \leq \lim\inf_{N\raw\infty} \frac{1}{N} \log 
\Tr(\sg^N (\mathcal{L}_N\in\mathcal{B}))
\leq  \lim\sup_{N\raw\infty} \frac{1}{N} \log \Tr(\sg^N (\mathcal{L}_N\in\mathcal{B}))\leq - S^F_{\sg}(\mathcal{B}^-).  \label{eqLDPJJ} 
\end{align}
\end{theorem}
From the point of view of (quantum) hypothesis testing one may reconceptualize \er{15105} as a \emph{test}
\begin{equation}
T_N:=(\mathcal{L}_N\in\mathcal{B}), \label{15110}
\end{equation}
which turns out to satisfy all of Theorem \ref{CMP2005Q} except for two crucial differences:
\begin{enumerate}
\item The tests \er{15110} are now universal, that is, they  depend on $\mathcal{B}$ but not on the prior $\sg$;
\item One now has the new relative entropy \er{JJentropy} rather than the old (Umegaki) entropy.
\end{enumerate}
The appearance of a new, $F$-dependent entropy in 2 is the price for having the advantage of 1.  
 \subsection*{Supplement:  Positive-Operator Valued Measures}\addcontentsline{toc}{subsection}{Supplement: Positive-Operator Valued Measures}
  Since the 1970s the picture of quantum-mechanical measurement sketched in \S\ref{FromCQ} and this section has  been replaced by one in which projections (i.e., operators $E\in B(H)$ such that $E^*=E^2=E$, so that $\| E\|=1$ unless $E=0$) are generalized by so-called \emph{effects},\footnote{Standard textbooks about quantum mechanics based on effects are De Muynck (2002) and Busch et al. (2016). }
 that is, operators $0\leq T\leq 1_H$. This notation means both 
 $T\geq 0$ and $1_H-T\geq 0$, where we recall that for any $S\in B(H)$ we say that $S\geq 0$ iff $\la\psi, S\psi\ra\geq 0$ for each $\psi\in H$, or just for all unit vectors $\psi$. We therefore have
 \begin{align}
0\leq T\leq 1_H && \LRaw &&  0\leq \la \ps, T\psi\ra\leq 1 && (\psi\in H) && \LRaw && T\in [0, 1_H],
 \end{align}
 where the last item is just notation,
 This, in turn, can be shown to be  equivalent to $T^*=T$ and $\sg(T)\subset [0,1]$. Projections are obviously a special case. 
A non-projective example of an effect is the position operator $\hat{x}$ on $L^2([0,1], dx)$.
We restrict ourselves, however, to finite dimension.\footnote{In general, a \textsc{povm} is defined on a measure space $(X,\Sg)$ as a $\sg$-additive map $F:\Sg\raw [0,1]$ such that $F(X)=1_H$. Given the condition $F(X)=1_H$ it is enough to require $F(B)\geq 0$ for each $B\in\Sg$, or $F(x)\geq 0$ in the finite case.}

The spectral resolution \er{SRTfd} of a self-adjoint operator $T$, with eigenprojections $E_{\lm}$ defined in \er{defElm}, comes from the map $E:\lm\mapsto E_{\lm}$ from $\sg(T)$ to $\mathrm{Proj}(H)$, the set of projections in $L(H)$, with $E(\lm)=E_{\lm}$, in terms of which the probability of obtaining an outcome $\lm$ upon 
  measuring $T$ is given by the Born rule \er{BR1}.
   In the simplest case, then, where not only $\dim(H)<\infty$ but also the underlying set $X$ is finite, a  \emph{Positive-Operator Valued Measure} or \textsc{povm} on $X$ is a map
\begin{align}
F: X\raw [0, 1_H]; && \Sigma_{x\in X} F(x)=1_H. \label{POVM}
\end{align}
A \textsc{povm} $F: X\raw [0, 1_H]$ and a state $\rh\in D(H)$  induce a probability distribution $p_{\rh}$ on $X$  via 
\beq
p_{\rh}(x)=\Tr(\rh F(x)), \label{11.7a}
\eeq
which in physics is taken to be the probability of obtaining outcome $x\in X$ upon measuring $F$;  the idea that a measurement of a self-adjoint operator $T$ gives an element $\lm\in\sg(T)$ of its spectrum is generalized to the stipulation that a measurement of a \textsc{povm} on $X$ always yields an outcome $x\in X$ (see below).  Eq.\ \er{11.7a} turns a  \textsc{povm}  \er{POVM} into a \emph{quantum-to-classical channel}
\begin{align}
Q_F: D(H)\raw\Pr(X); && Q_H(\rh)=p_{\rh}, \label{QFchannel}
\end{align}
which may subsequently be expanded into a \textsc{cptp} (completely positive and trace-preserving) map
$L(H)\raw C(X)$, which is dual to a \textsc{cpu} (completely positive and unital) map in the opposite direction, see \S\ref{sec:QE}. Conversely,\footnote{See e.g.\ Waltrous (2018), Theorem 2.37. The root of such results is Busch's Theorem. 
A \emph{probability distribution on} $[0, 1_H]$ is defined as a function 
$P:[0, 1_H]\raw [0,1]$ such that  $P(1_H)=1$ and $P\left(\Sigma_n T_n\right)=\Sigma_n P(T_n)$ 
for any sequence $(T_n)$ in $[0, 1_H]$ such that $\Sigma_n T_n\leq 1_H$. This theorem then
states that any such $P$ takes the form $P(F)=\Tr(\rh F)$ for some $\rh\in D(H)$. See e.g.\ Landsman (2017), Theorem 2.44, which also relates Busch's theorem to Gleason's. 
Given a \textsc{povm} \er{POVM},  a probability distribution $P$ on $[0, 1_H]$
induces a probability distribution $p\in \Pr(X)$ by $p(x)=P(F(x))$, 
and so Busch's theorem shows that $p_{\rh}\in\Pr(X)$ as in \er{11.7a} necessarily derives from such a $P$.}  
 any such map is given by some \textsc{povm} \er{POVM}.

As we already suggested, a self-adjoint operator $T^*=T$ with spectral resolution  \er{SRTfd} 
is a special case of a \textsc{povm}, with $X=\sg(T)$ and $F(\lm)=E_{\lm}$ having the special feature that 
\begin{align}
F(x)F(y)=\dl_{xy}F(x) && (x,y \in X).
\end{align}
Such a special \textsc{povm} is called a \emph{Projection Valued Measure} or \textsc{pvm}. This,  in turn,  comes from a self-adjoint operator in the said way; just take $T=\Sigma_{x\in X} x F(x)$.  Via a famous theorem of Naimark and Stinespring,  any \textsc{povm}  comes from a \textsc{pvm} by a kind of compression:\footnote{This is a simple consequence of the Naimark--Stinespring  theorem, see e.g.\ Landsman (1998), Corollary 1.4.9.}  
\begin{theorem}
For a \textsc{povm} \er{POVM} there exist: a Hilbert space $K$, a projection $e:K\raw K$,  a unitary $u:H\raw eK$,  and a \textsc{pvm} $E:X\raw [0,1]_K$, such that $F$ is unitarily equivalent to $eFe$ via $u$:
\begin{align}
uF(x)u^*=e E(x)e && (x\in X).\label{NaiSti}
\end{align}\end{theorem}
Measurement of a self-adjoint operator (called a \emph{projective measurement}) is now seen as a special case of a measurement of some \textsc{povm} \er{POVM}. Our earlier idea that a (projective) measurement of $T$ comes down to randomly drawing a unit vector $\psi\in H$, restricted to $\ps\in E_{\lm}$ for some eigenvalue $\lm\in\sg(T)$, producing the latter as a numerical outcome,  is now dropped. Instead, a classical outcome $x\in X$   arises from an operator $F(x)\in [0, 1_H]$. In the spirit of the Copenhagen Interpretation, in the \textsc{povm} approach one refrains from any description of the measurement process and  focuses on the outcome probabilities,  seen as probabilities on some classical space (namely $X$).

Here is a simple example.\footnote{Taken from Busch, Grabowski, and Lahti (1996), \S I.1.2. De Muynck (2002), \S 8.3, gives a more detailed treatment.} A Stern--Gerlach measurement on a spin-$\half$ particle is naively described by the self-adjoint matrix $T=\half\sg_3$, i.e., by the \textsc{pvm} $F_n:\{-\half,\half\}\raw M_2(\C)$ given by
\begin{align}
F_n(-\half)= \left(
\begin{array}{cc}
 0 &   0  \\
 0 &   1
\end{array}
\right)\equiv |-\half\ra\la-\half|; &&
F_n(\half)= \left(
\begin{array}{cc}
 1 &   0  \\
 0 &     0
\end{array}
\right) \equiv |\half\ra\la\half|. \end{align}
So if the initial state of the particle is $\phv(0)=c_-\phv_-+c_+\phv_+$, then the outcome is $\pm\half$ with probability $|c_{\pm}|^2$.
However, an actual measurement (still vastly simplified!) goes as follows. The particle also has a spatial wavefunction $\ps$, which is initially $\psi(0)$, and which the magnet would send to some wavepacket $\ps_{\pm}(t)$ if the initial spin state were $\phv_{\pm}$, where $\psi_{\pm}$ is approximately localized on the top/bottom half of a screen. Thus the total initial state is $\Psi(0)=\phv(0)\ot \psi(0)$, which during the measurement process evolves into $
\Psi(t)=c_-\phv_-\ot\psi_- + c_+\phv_+\ot\psi_+$.
What is then really measured  is not $\phv$ but $\psi$, and this is done via an interaction with molecules on the screen, which we omit from the description except that it leads to an almost pointwise collapse of $\psi$ in either the upper half plane or the lower half plane, where  these outcomes are well separated, so that the measurement of $\ps$ is sharp. Thus we 
have projections  $P_{\pm}$ on  the \Hs\  $H=L^2(\R^3)$ that contains $\ps$, where $P_-+P_+=1_H$ and $P_-P_+=0$ (of course,  $\ps(t)$ is extremely localized in the plane containing the screen). But this does not mean that $\psi_{\pm}$ is strictly supported in either the upper or the lower half of the screen; it is a wavepacket that at best is concentrated in one of these areas. 

On this basis we look for a \textsc{povm} $F_r: \{-\half,\half\}\raw [0,1]_{\C^2}$ describing this measurement. 
Since
\begin{align}
\la \Psi(t), (1_{\C^2}\ot P_{\pm})\Psi(t)\ra =|c_-|^2 \la \ps_-, P_{\pm}\ps_-\ra+ |c_+|^2 \la \ps_+, P_{\pm}\ps_-\ra,
\end{align}
which by the Born rule equals the probability of the measurement outcome $\pm\half$, we obtain
\begin{align}
F_r(\pm\half)= \la \psi_-, P_{\pm}\ps_-\ra 
 |-\half\ra\la-\half| + \la \psi_+, P_{\pm}\ps_+\ra 
 |\half\ra\la\half|,
\end{align}
as we then have $F_r(\pm\half)\in  [0,1]_{\C^2}$ as well as $F_r(-\half)+ F_r(\half)=1_{\C^2}$, cf.\ \er{POVM}, and
finally
\begin{equation}
\la \Psi(t), (1_{\C^2}\ot P_{\pm})\Psi(t)\ra=\la \phv(0), F_r(\pm\half)\phv(0)\ra,
\end{equation}
for the probabilities of the two possible measurement outcomes.  This  matches the Naimark--Stinespring theorem \er{NaiSti}: take 
\begin{align}
H&=\C^2; && K=\C^2\ot L^2(\R^3); &&  u(c_-\phv_-+c_+\phv_+)=c_-\phv_-\ot\psi_-+c_+\phv_\ot\psi_+;\\
 E(\pm)&=1_{\C^2}\ot P_{\pm}; && e =[uH].
 \end{align}
It is crucial that the measurement setup makes clear how the outcome is $E(x)$ and hence $x$; this information is lost in the \textsc{povm} itself but it should be clear from the experimental setup that suggested it, as in the above example (where $x=\pm$ labels outcomes in the upper or lower half planes), and more generally in the Naimark--Stinespring setting, where it comes from the \textsc{pvm} $E$.
\section{Introduction to \vna s}\label{vNsection}
The most interesting entropies in quantum theory are \emph{entanglement entropies} \er{EABI}, and relative versions thereof. So far, these were defined (and computed) using reduced density operators and traces. But this approach only works for 
subsystems $M$ defined as factors in a tensor product, i.e., 
\begin{align}
M\subset B(H); && H=H_A\ot H_B; && M=B(H_A)\ot 1_{H_B},\label{111}
\end{align}
 cf.\ \er{naivefactor}.
Within the mathematical language we use this is all that can happen if we require that: 
\smallskip

\noindent (i) $H_{AB}$ is finite-dimensional; (ii) $M$ is itself a \ca\ containing the unit $1_H$; (iii) $M$ is a \emph{factor}.\vspace{1mm}

\noindent   The latter condition means that $M\cap M'=\C\cdot 1_H$, see \er{factor}. But 
 in general eq.\ \er{111} leaves out 
many interesting cases relevant to e.g.\  quantum field theory, quantum statistical mechanics, and black hole thermodynamics.  
More general subsystems are defined as more general subalgebras  $M\subset B(H)$, where $H$ is a separable infinite-dimensional \Hs. The above definition of a factor still makes sense, and this remains
the basic case,\footnote{In finite dimension every unital \ca\ $M\subset B(H)$ is a direct sum of factors; this  is  the basic structure theorem stating that every finite-dimensional \ca\ is isomorphic to a direct sum of matrix algebras. See e.g.\ Landsman (2017), Theorem C.163. In infinite dimension (under additional conditions making $M$ a \vna, see below) something similar happens, perhaps with ``direct integrals'' instead of sums. See e.g.\ Blackadar (2006), \S III.1.6. } 
\emph{but \er{111}  is no longer the only possibility: on an infinite-dimensional \Hs, factors need not be factors in a tensor product}.
 
  In connection with entropy we then need to deal with the problem  that a state $\om$ on $M\subset B(H)$ defined by a density operator $\rh\in D(H)$ via $\om(T)=\Tr(\rh T)$, $T\in M$, should have an \emph{intrinsic} associated entropy $S(\om)$, although initially we only have the von Neumann entropy \er{vNe} defined via $\rh$, which does not see the restriction to $M$; cf.\ the Course Overview around eq.\ \er{241I}.
Likewise for the relative entropy $S(\om_1,\om_2)$ for pairs of states on $M$, for which so far we only have the formulae \er{QKL} - \er{QKLb} for the relative entropy of the corresponding density operators on $H$.  If $M$  is a factor in a tensor product this may be resolved via the partial trace, since if \er{111} holds, then the entropy of $\om$ as a state on $M$ may be taken to be the entanglement entropy $S(\rh_A)$; Theorem \ref{bigdifference} then shows the huge difference between $S(\rh)$, which is irrelevant,  and $S(\rh_A)$. But the partial trace construction  does not work for general factors, and  it is major goal of the remainder of this course to solve this problem, that is, to associate  intrinsic density operators to states on $M$, with an associated (and computable) relative entropy for a pair of states. This leads us into deep waters.

Given the success of this assumption so far in 
quantum field theory and black hole thermodynamics, especially in connection with entropy,
 we assume that $M$ is a \emph{\vna}.\footnote{\label{OAbooks} Leading modern textbooks
are Kadison \& Ringrose (1983, 1986) and
Takesaki (2002, 2003ab). Blackadar (2006) and Sunder (2012) are shorter.  Connes (1994), Chapter V,  is a superb overview of the field (without proofs) by a master.}
 This  class of operator algebras  generalizes measure theory and hence probability theory in a noncommutative or quantum direction (much as \ca s generalize topology in that direction), and as such it promises to be natural setting for quantum theories of entropy--as it indeed will be.
\begin{definition}[definition and theorem]\label{defvna}
Let $M\subseteq B(H)$ be a unital $\mbox{}^*$-algebra within $B(H)$, i.e., $1_H\in M$ and $M$ is closed under linear operations, multiplication, and adjoint $T\mapsto T^*$. Then 
$M$ is a \vna\ iff one and hence all of the following equivalent conditions hold: 
\begin{enumerate}
\item $M''=M$;
\item $M=N'$, where $N\subseteq B(H)$ is some subset closed under the adjoint;
\item $M$ is closed under strong, or, equivalently,  weak operator limits;
\item $M$ is norm-closed and is the dual of some Banach space (called its \emph{predual} $M_*$).
\end{enumerate}
\end{definition}
\bex
Show  the equivalence between the first two points.
\eex
Although the equivalence $1\lraw 2$ is almost trivial, it is quite informative conceptually, since point 2 opens the perspective of seeing a \vna\ as an algebra of
\emph{symmetries}: to this end,  take $N=U(G)$ where $G$ is some group and $U:G\raw B(H)$ is a unitary \rep\ on $H$. 

The equivalence between 1 and 3,  called von Neumann's \emph{double commutant theorem}, was the very first and already  deep result in the theory. 
A proof must be included for that reason alone.

 Recall that $T_{\lm}\raw T$ \emph{weakly} if $\la\phv, T_{\lm}\ps\ra\raw \la\phv, T\ps\ra$ for all $\phv,\psi\in H$, and $T_{\lm}\raw T$  \emph{strongly} if $T_{\lm}\ps\raw T\ps$ for all $\psi\in H$, that is, $\|T_{\lm}\ps-T\ps\|\raw 0$. That said, despite all this technical topology
the essence of the proof is already contained in the finite-dimensional case $H=\C^n$.
\begin{lemma}
If $M$ is  a unital  $\mbox{}^*$-algebra in $M_n(\C)$, then $M''=M$.
\end{lemma}
\emph{Proof}
All we need to prove is $M''\subseteq M$, since the converse inclusion is obvious. Take $n$ arbitrary (and hence possibly linearly dependent) vectors
$\ups_1,\ldots,\ups_n$ in $H$, and, given $a\in M''$, find some $b\in M$ such that $a\ups_i= b\ups_i$ for all
$i=1,\ldots,n$. Hence $a=b$, so $a\in M$. To this end, we start with a single vector 
$\ups\in H$.
Form the linear subspace $M\ups=\{m\ups\mid m\in M\}$ of
$H$, with associated projection $e$ (i.e.\  $ew=w$ if $w\in M\ups$
and $ew=0$ if $w\in (M\ups)^{\perp}$).
Then $e\in M'$, and hence $a\in M''$ commutes with
$e$. Since $1_H\in M$, we have $\ups \in M\ups$, so
$\ups=e\ups$, and we compute $a\ups=ae\ups= ea\ups\in M\ups$. Hence $a\ups=b\ups$, for
some $b\in M$.
Now run the same argument with the following substitutions:
\begin{itemize}
\item  $H\rightsquigarrow
 H_n=H\oplus\cdots\oplus H$ (with $n$ terms).
  \item $M\rightsquigarrow M_n=\{\mathrm{diag}(m,\ldots,m)\mid m\in M\}$.
  \item  $\ups\rightsquigarrow\mathsf{v}=\oplus_i \ups_i\equiv (\ups_1,\ldots, \ups_n)$.
\end{itemize}
 We then have $(M_n)''=(M'')_n$, so for any matrix $\mathsf{a}=\mathrm{diag}(a,\ldots,a)$ in $(M'')_n$, the previous argument 
 yields a matrix $\mathsf{b}=\mathrm{diag}(b,\ldots,b)\in M_n$ such that $\mathsf{a}\ups=\mathsf{b}\ups$. But this is  $a\ups_i=b\ups_i$ for all $i=1,\ldots,n$, so that $a=b$ and hence  $M''\subseteq M$.\QED
 \bex
  If $H$ is  infinite-dimensional, adapt the above proof, assuming $M$ is strongly closed. Also prove the converse implications ($M''=M$) $\Raw$ $M$ is weakly closed $\Raw$ $M$ is strongly closed.
\eex
\begin{corollary}\label{vnaca}
The closures of a unital $\mbox{}^*$-algebra $M\subset  B(H)$ in the strong and weak operator topologies coincide with $M''$, which is therefore also norm-closed and hence is a \ca. 
\end{corollary}
For example, we now realize that $W^*(T)$ defined in \er{WTCT} is a \vna; one may always keep in mind $H=L^2([0,1], dx)$ and the bounded position operator 
\begin{align}
T=\hat{x}; && \hat{x}\psi(x):=x\psi(x),
\end{align} 
 so that $W^*(\hat{x})=L^{\infty}([0,1], dx)$. We say that $\hat{x}$ is a \emph{multiplication operator}, and more generally $f\in L^{\infty}([0,1], dx)$ is always meant to act on $L^2([0,1], dx)$ 
as a multiplication operator
\beq
\hat{f}\psi(x)=f(x)\psi(x),\label{defmult}
\eeq with  operator norm $\| \hat{f}\|$ equal to $\| f\|^{\mathrm{ess}}_{\infty}$, see \er{inftynorm}.
 Likewise for the more general measure spaces considered below.
For any (measurable) subset $\Dl\subset[0,1]$ the indicator function $1_{\Dl}$ is a projection  in $L^{\infty}([0,1], dx)$, and  these are in fact all its projections. Compare this with the \ca\  $C^*(T)=C([0,1])$, see \er{deffT} and preceding text, which also acts on $L^2([0,1], dx)$ via multiplication operators, and has no nontrivial projections (that is, other than $\hat{0}_{[0,1]}=0_H$ and $\hat{1}_{[0,1]}=1_H$).

More generally, for any reasonable measure space $(X,\mu)$,\footnote{The right technical condition is that $(X,\mu)$ must be \emph{locally finite}, which means that $X$ is a Hausdorff space with ensuing Borel structure and $\mu\in\Pr(X)$ is such that any $x\in X$ has a nbhd $U$ such that $\mu(U)<\infty$. For example, probability measures and more generally $\sg$-finite measures on locally compact Hausdorff spaces satisfy this condition. A difficult classification theorem guarantees that any commutative \vna\ is isomorphic to $L^{\infty}(X,\mu)$ for some locally compact Hausdorff space $X$ with Radon measure $\mu$. See Takesaki (2002), Theorem 1.18. If $H$ is separable the situation is much simpler: one may assume $X=[0,1]$ and $\mu$ to be one of the possibilities listed above \er{xomega}. See also Takesaki (2002), Theorem 1.22. If an abelian \vna\ $M$ is maximal, the isomorphism with one of these possibilities is even \emph{spatial}, i.e, given by a unitary transformation, see \S\ref{sectypes}.
The Harvard  lectures by Lurie (2011) give a  detailed survey of abelian \vna s, summarized in Landsman (2017), \S\S B.16--B.17. } the function space $M=L^{\infty}(X,\mu)$ whose elements act on $H=L^2(X,\mu)$ by multiplication operators, is a commutative \vna, even a maximal one, i.e., $M'=M$, see \er{AA} and \er{WTWT}. And also here, each (Borel measurable) subset $\Dl\subseteq X$ provides a projection $1_{\Dl}$ in $M$, and  each projection in $M$ takes this form. 

More generally, an important consequence of point 3 in Definition \ref{defvna}  is that not just commutative but all \vna s have lots of projections (i.e., operators $E$ satisfying $E^2=E^*=E$).
If we denote the set of all projections in a \vna\ $M$ by $\mathrm{Proj}(M)$, we have:
\begin{proposition}\label{richP}
A \vna\  $M$ is generated by $\mathrm{Proj}(M)$ in the following senses:
\begin{itemize}
\item $M$ equals $\mathrm{Proj}(M)''$, as well as the norm and strong closures of the linear span of $\mathrm{Proj}(M)$.
\end{itemize}
\end{proposition}
This follows from the previous examples, since $M$ consists of finite sums of elements from its commutative subalgebras $W^*(T)$, where $T=T^*\in M$, to which the statement applies.\footnote{See e.g.\ Landsman (2017), Corollaries B.104 and B.105 for details, especially about norm convergence. }
 
 We now come to the fourth characterization of a \vna\ in Definition \ref{defvna}, due to Sakai. This point 4 explains why apart from the norm topology and the weak and strong operator topologies, a \vna\ carries another topology, called the \emph{$\sg$-weak topology}, which is  even more important than the ones just mentioned in giving the dominant criterion for continuity.
 Short of proving the equivalence of this characterization to the others, which is quite difficult (and hence omitted),\footnote{
 For this, and also  for our Theorem \ref{vnad} below, see Blackadar (2006), \S III.2.4.2, Theorems III.2.1.4 and III.2.4.2.}
 let us review the two (classes of) examples known to us so far:
 \begin{itemize}
 \item In the abelian case  $M=L^{\infty}(X,\mu)\subset B(L^2(X,\mu))$ we have $M_*\cong L^1(X,\mu)$, see \er{Lpqd} for $p=1$ and $q=\infty$: in other words, every continuous linear map $\hat{T}:L^1(X,\mu)\raw\C$ is given by 
a unique function $T\in L^{\infty}(X,\mu)$ via $\hat{T}(\rh)=\int_X d\mu\, \rh T$, where $\rh\in  L^1(X,\mu)$.
 \item In the noncommutative case $M=B(H)$ we have $M_*\cong B_1(H)$, see \er{840} or \er{BpHBqH} with $p=1$ and $q=\infty$.
As we already saw, this means that every continuous map $\hat{T}: B_1(H)\raw \C$ is induced by some unique $T\in B(H)$ via $\hat{T}(\rh)=\Tr(\rh T)$, where this time $\rh\in B_1(H)$.
\end{itemize}
 Both are usually seen the other way round:\footnote{If $B$ is a Banach space with (Banach) dual $B^*$,
 then $B\hookrightarrow  B^{**}$ canonically via $\phv\mapsto \hat{\phv}$, $\hat{\phv}(T)=T(\phv)$.  Thus
 $B=M_*$ gives an inclusion $M_*\hookrightarrow  M^*$ via the detour $M_*^{**}\cong M^*$, whilst  $B=M$ at once gives $M\hookrightarrow M_*^*$, written however as $T\mapsto\hat{T}$. } 
 instead of seeing elements $T$ of $M$ as continuous linear functionals $\hat{T}$ on $M_*$, we rather regard elements $\rh$ of $M_*$ as continuous linear functionals $\hat{\rh}$ on $M$ via $\hat{\rh}(T)=\hat{T}(\rh)$.
This especially applies to states $\om$ on $M$,  i.e, $\om(T^*T)\geq 0$ and $\om(1_H)=1$. For $M=L^{\infty}(X,\mu)$
we thus obtain a bijection between normal states and densities (i.e., $\mu$-a.e.\ positive functions $\rh\in L^1(X,\mu)$ such that $\int_X d\mu\,\rh=1$), whereas for $M= B(H)$ we recover the  correspondence \er{normal} $\om=\hat{\rh}\lraw\rh$ 
 between \emph{normal} states and density operators, where normality was defined after \er{840}.  
 This definition of normality as $\sg$-weak continuity extends from $B(H)$ to linear maps defined on any \vna\ $M\subset B(H)$, which inherits the $\sg$-weak topology from $B(H)$ as defined after \er{840}, and such maps may be both functionals or homomorphisms onto some other \vna.  But the intrinsic meaning of the $\sg$-weak topology only emerges in the context of the property $M\cong M_*^*$ stated in Definition \ref{defvna}.4. Here are the main facts:
\begin{theorem}\label{vnad}
\begin{enumerate}
\item Let $M$ be a  unital $\mbox{}^*$-algebra within $B(H)$. Then $M$ is a \vna\  as defined by the property 
$M''=M$ iff $M$ is  closed in the $\sg$-weak  topology (as  defined, for the moment, via convergence, namely
 $T_{\lm}\raw T$  iff $\Tr(\rh T_{\lm})\raw \Tr(\rh T)$ for each $\rh\in D(H)$).
\item In that case, so for any \vna\ $M\subseteq B(H)$, define 
the predual $M_*$ as the space of all normal = $\sg$-weakly continuous linear functionals $\phv:M\raw \C$.
Then  $M_*$ is a norm-closed subspace of the Banach dual $M^*$ of $M$, and $M\cong M_*^*$ as Banach spaces via 
\begin{align}
M\raw M_*^*; && T\mapsto\hat{T}; && \hat{T}(\phv):=\phv(T) && (T\in M,\, \phv\in M_*),
\end{align}
which is a homeomorhism from the $\sg$-weak topology on $M$ to the weak$\mbox{}^*$-topology on $M_*^*$. 
\end{enumerate}
\end{theorem}
The point is that although initially the $\sg$-weak topology on a \vna\ $M\subseteq B(H)$ was defined via $B(H)$, Theorem \ref{vnad}.2 makes this topology, now (re)defined as the weak$\mbox{}^*$-topology on $M$ in its capacity as the Banach dual of $M_*$, and hence given by the convergence criterion  $T_{\lm}\raw T$  iff $\phv(T_{\lm})\raw \phv(T)$ for each $\phv\in M_*$, 
 a property of $M$ itself, independent of its embedding in $B(H)$. 
 
 Sakai's characterization 4 in Definition \ref{defvna} may now be restated as follows:
\begin{theorem}
A \vna\ is a \ca\ that is the dual of a Banach space.
\end{theorem}
This Banach space turns out to be unique (up to isomorphism, which by definition we take to be isometric), and hence it is the predual $M_*$ as just defined. This is a consequence of a deep result:\footnote{See Blackadar (2006), Corollary III.2.2.12.
An `algebraic' isomorphism $\phv:M\raw N$ is also meant to preserve the *, so it is an isomorphism of vector spaces such that 
$\phv(ST)=\phv(S)\phv(T)$ and $\phv(T^*)=\phv(T)^*$.}
\begin{theorem}\label{starnormal}
An algebraic  isomorphism  $\phv:M\raw N$ between \vna s  is automatically both an isometric  isomorphism of Banach spaces and  a $\sg$-weak  homeomorphism.
\end{theorem}
Since this implies that preduals of  algebraically isomorphic \vna s are isomorphic as Banach spaces,
\emph{the predual of a \vna\ is unique (up to isometric isomorphism).}

That said, in practice  \vna s $M$ usually do arise via  embeddings $M\subseteq B(H)$,  in terms of which 
each normal state $\om\in M_*$ on $M$ takes the familiar form:\footnote{See Blackadar (2006), Theorem III.2.1.4. part (vii).}
\begin{align} 
\om(T)=\Tr(\rh T); && \rh\in D(H). \label{omTDH}
\end{align}
 But, returning to the theme opening this section, it is only for $M=B(H)$ that $\rh$ is uniquely determined by $\om$. Otherwise, the association of a density operator $\rh$ to a state $\om$ is not intrinsic to the state, and much of our efforts will be devoted to solving this problem by producing some \emph{intrinsic} density-operator like object that implements $\om$ via some generalized trace in the spirit of \er{omTDH}.
  To see this lack of uniqueness more clearly, it may be helpful to see how the predual $M_*$ is constructed for some given \vna\ $M\subseteq B(H)$. Namely,\footnote{See Landsman (2017), proof of Theorem C.133 for details.} as expected one has
\begin{align}
M_* =  B(H)_*/M^{\perp}; && 
 M^{\perp}  =  \{\phv\in B(H)_*\mid \phv(T)=0\: \forall T\in M\},
\end{align}
which may show the obvious, i.e.,  that the lack of uniqueness of $\rh$ for given $\om$ comes from $M^{\perp}$.

 The abelian (= commutative) case is quite instructive in this context. According to our  analysis so far, this case may be reformulated and focused as follows. First, in writing
  \beq
 M=L^{\infty}(X,\mu)\subset B(L^2(X,\mu)),\label{167N}
 \eeq
 we identify $f\in L^{\infty}(X,\mu)$ with the corresponding  multiplication operator $\hat{f}$, see \er{defmult}, where
 \begin{equation}
\|\hat{f}\|=\|f\|^{\mathrm{ess}}_{\infty}.
\end{equation}
  Second, we identify $M_*$ with its image in $M^*$,\footnote{See Takesaki (2002), Theorem 1.27, for a description of $L^{\infty}(X,\mu)^*$ in terms of functions of bounded variation.} so that our earlier Banach space isomorphism 
    \beq
  L^{\infty}(X,\mu)_*\cong L^1(X,\mu)\label{119}
  \eeq
   means that
 each normal functional $\phv$ on $M$  takes the form $\phv=\phv_g$, where  $g\in L^1(X,\mu)$, and
  \begin{align}
  \phv_g(\hat{f})=\int_X d\mu\, gf.
\end{align}
As we saw,  a  \emph{normal state} $\phv_g\equiv\om_g$ is then given by a unique \emph{density} $g$ on $(X,\mu)$, i.e., 
 $g\in L^1(X,\mu)$, $g(x)\geq 0$ for $\mu$-a.e.\ $x\in X$,  and $\int_X d\mu\, g=1$.
 In contrast,
there are infinitely many density operators $\rh\in D(L^2(X,\mu))$ that  reproduce $\om_g$ via \er{omTDH}; yet in a typical case like $X=[0,1]$, $d\mu(x)=dx$, even within this multitude
one cannot make the natural choice $\rh=\hat{g}$ since generically $\hat{g}\notin D(H)$. In sum, producing a normal state on 
$M$ in \er{167N} via
\er{omTDH}, while possible in principle,  is a disaster.

 To explain more clearly why such densities are intrinsic to $M$ (whereas density operators are not), and also
to make the commutative and non-commutative cases look more like each other, and thus prepare for some concepts that will play an important role in the construction of quantum entropies, we  reformulated the commutative case via a trace, as follows. Define a linear map
\begin{align}
\ta: L^{\infty}(X,\mu)^+\raw[0,\infty]; && \ta(\hat{f}):=\int_X d\mu\, f, \label{traceL1}
\end{align}
where $M^+$ consist of all positive operators in $M$ (which here are simply the $\mu$-a.e.\  pointwise positive functions); this restriction is needed for the same reason as in $M=B(H)$, namely that the map \er{traceL1} may be both infinite and ill-defined on arbitrary elements of $L^{\infty}(X,\mu)$; on its positive part, $\ta$ is at least well defined, although it may be infinite.\footnote{For example, take $X=\R$ with Lebesgue measure and $f=1_{\R}$. Then $\hat{f}\in L^{\infty}(\R, dx)$ but
$\hat{f}\notin L^1(\R, dx)$ 
} We may also define $\ta$ on all of $L^1(X,\mu)$ by the same formula, and since $fg\in L^1(X,\mu)$ for $f\in L^{\infty}(X,\mu)$ and $g\in L^1(X,\mu)$,
\begin{equation}
\om_g(\hat{f})=\ta(\hat{g}\hat{f})\label{CCDO}
\end{equation}
is well defined and gives an intrinsic formula for $\om_g$ and hence for any normal state on $ L^{\infty}(X,\mu)$, on a par with \er{omTDH}. This may be taken one step further by extending the trace as defined in \er{traceL1} 
to the set of all (not necessarily essentially bounded) $\mu$-a.e.\ positive measurable functions $g:X\raw\C$, within which the ``trace-class'' $B_1(M)$ \emph{relative to the \vna\ $M=L^{\infty}(X,\mu)$} is given by all (possibly unbounded) multiplication operators $\hat{g}$ for which $\ta(|g|)<\infty$. Almost trivially,
\begin{equation}
B_1(L^{\infty}(X,\mu))\cong L^1(X,\mu),
\end{equation}
 acting on $L^2(X,\mu)$ by multiplication operators.\footnote{The domain of $g\in L^1(X,\mu)$ as an operator $\hat{g}$ on 
$L^2(X,\mu)$ consists of all $\psi\in L^2(X,\mu)$ for which $g\psi\in L^2(X,\mu)$.} But in general $L^1(X,\mu)$ may contain unbounded operators; see the example after \er{863}.  Thus we already see in a simple (abelian) case  that the realization of normal states via \emph{intrinsic} constructions like \er{CCDO} may involve unbounded operators; the case $M=B(H)$ is exceptional in $\rh\in D(H)$ in \er{omTDH} being bounded \emph{and} in $M$. 

These deliberations will be continued after we have gained more insight in the special nature of  $B(H)$ and 
$L^{\infty}(X,\mu)$ among all \vna s. The end result will be a beautiful theory, due to Haagerup, in which constructions like \er{omTDH} and \er{CCDO}, in which states are realized via density operators and traces intrinsically related to $M$, work for any \vna\ $M$.
\section{Types of \vna s and first classification of factors} \label{sectypes}
Returning to the opening paragraphs of the previous section, consider two \Hs s $H_A$ and $H_B$, now possibly infinite-dimensional, with tensor product $H_{AB}\equiv H_A\ot H_B$.\footnote{The simplest way to define this tensor product for separable \Hs\  is the same as for finite $A$ and $B$, namely
to take a basis $(\ups_a)_{a\in A}$ of $H_A$ and a basis $(u_b)_{b\in B}$ of $H_B$, and take $H_{AB}=\ell^2(A\x B)$.
If we define $\ups_a\ot u_b$ to be the function $\dl_{(a,b)}\in \ell^2(A\x B)$, this also makes $H_{AB}$ the \Hs\ with basis 
$(\ups_a\ot u_b)_{(a,b)\in A\x B}$, where we may even forget $\ell^2(A\x B)$ and take $\ups_a\ot u_b$ to be 
 just a notation for these basis vectors. Here  $\ell^2(X)$ for a countable set $X$ consists of all $\ps:X\raw\C$ for which $\Sigma_{x\in X}|\psi(x)|^2<\infty$, with (absolutely convergent)  inner product $\la\psi,\phv\ra=\Sigma_{x\in X}\ovl{\psi(x)}\phv(x)$. }
 \bex\label{factorex}
 Show that
 $M=B(H_A)\ot 1_{H_B}\subset B(H_{AB})$ has commutant $M'=1_{H_A}\ot B(H_B)$ and bicommutant $M''=M$ (so that $M$ is a \vna). Infer that  $M$ is a factor. \eex
In finite dimension all factors take this form up to unitary isomorphism (see Theorem \ref{Ifactors} below)
 but this is not the case in infinite dimension, where only  a so-called \emph{type I} factor $M\subseteq B(H)$ is essentially a factor in a tensor product $B(H_{AB})$ in the above way. 
 This is the main source of the immense richness of the theory of \vna s, but it also means trouble for entropy. In the type I case any normal state $\om_{AB}$ on $M$ (notably a pure state), which is always  given by a density operator $\rh_{AB}\in D(H_{AB})$,  induces not just a state $\om_A$ by restriction, but also a density operator $\rh_A\in D(H_A)$, namely via the partial trace construction. The latter, in turn, gives rise to a von Neumann entropy \er{vNe}. Likewise for pairs of states on $B(H)$, which give rise to a well-defined relative entropy \er{QKL} - \er{QKLb}. This is no longer the case for non-type I factors, which, seen as subsystems of $B(H)$,  require a much more elaborate construction for even the \emph{definition} (let alone the computation)  of the (relative) entropy induced by a state (or pair of states) on the ambient \vna\ $B(H)$, or on some other \vna\ containing $M$.
 
  Here and in what follows, it is important to distinguish: 
 \begin{itemize}
 \item An \emph{algebraic} isomorphism $\phv: M_A\raw M_B$ between \vna s $M_A$ and $M_B$  is an isomorphism of vector spaces such that  $\phv(ST)=\phv(S)\phv(T)$ and $\phv(T^*)=\phv(T)^*$.
 As we saw in Theorem \ref{starnormal}, such a map is automatically an isomorphism of $M_A$ and $M_B$ as Banach spaces as well as a homeomorphism of $M_A$ and $M_B$ carrying their $\sg$-weak topologies. 
\item Given embeddings $M_A\subseteq B(H_A)$ and $M_B\subseteq B(H_B)$, a \emph{spatial} (or  \emph{unitary}) isomorphism between $M_A$ and $M_B$ is a unitary operator 
$U:H_A\raw H_B$ such that $\phv(T)= UTU^*$ is an algebraic isomorphism from $M_A$ to $M_B$. This is  stronger than an algebraic isomorphism. For example, $L(H_A)$ acting on $H_A$ in the obvious way and $L(H_A)$ acting on $H_{AB}$ via $T_A\mapsto T_A\ot 1_{H_B}$ are algebraically isomorphic via $\phv(T)= T_A\ot 1_{H_B}$
but they are not spatially isomorphic.  
 \end{itemize}
 Both versions play a role in different circumstances, but they are related in the following way:\footnote{See Blackadar (2006), Theorem III.2.2.8. A subspace of a \Hs\ is by definition a \emph{closed linear} subspace.}
 \begin{proposition}\label{expred}
 If $M_A\subseteq B(H_A)$ and $M_B\subseteq B(H_B)$ are algebraically isomorphic, there is a \Hs\ $H_C$ such that $M_B$ is spatially isomorphic to the action of $M_A\ot 1_{H_C}$ on some subspace of $H_{AC}$.
 \end{proposition}
 
 What are the  possibilities beyond the ``tensor product'' factors  in Exercise \ref{factorex}? As a first introduction to the classification of factors we  state the preliminary
 classification of factors (up to algebraic isomorphism) into ``types'' due to von Neumann and his assistant Murray, developed in the 1930s. We restrict this to factors \emph{acting on a finite-dimensional or separable \Hs}.\footnote{\label{sepfootnote}
 Recall that a \Hs\ is separable if it has a countable basis. cf.\ footnote \ref{sepHs}. 
All \Hs s in the list following \er{WTWT}, or more generally, $L^2(X,\mu)$ for standard Borel spaces $X$, are separable.  A \vna\ $M\subseteq B(H)$ where $H$ is separable is  called \emph{$\sg$-finite}; equivalently, $M$ has a norm-separable predual $M_*$, see e.g.\ Jones \& Sunder (1997), Proposition A.2.1. There is an even stronger condition on $M$, namely  \emph{hyperfiniteness}, in $M$ being the $\sg$-weak closure of an increasing sequence of finite-dimensional subalgebras. These seem enough for practically all applications to mathematical physics studied so far. In any case,  we avoid non-separable Hilbert spaces.} 

\noindent  Namely, up to \emph{algebraic} isomorphism, any such factor
 occurs in the following list, and the different possibilities are (algebraically) inequivalent; to  be sure, this does \emph{not} imply that each entry in the list contains only one possibility, up to isomorphism! So this list is just a first step.
 \begin{itemize}
\item \emph{Type} $\mathrm{I}_n$,  $n\in\N$:  $M=L(\C^n)$, or more down to earth,  the $n\x n$ complex matrices $M_n(\C)$.
\item  \emph{Type} $\mathrm{I}_{\infty}$: $M=B(H)$ for some infinite-dimensional separable \Hs\ $H$.
\item Type $\mathrm{II}_1$:  we give two  characterizations of this intriguing case. The first is the conjunction:
\begin{enumerate}
\item[(i)] $M$ is
  infinite-dimensional but not isomorphic to a type $\mathrm{I}_{\infty}$ factor;
  \item[(ii)] Every isometry in $M$ is unitary.\footnote{That is, if $V\in M$ with  $V^*V=1_H$ and $VV^*$ a projection (making $V$ an isometry by definition), then $VV^*=1_H$.}
  \end{enumerate}
 This characterization turns out to be equivalent to a second, completely different one:
 \begin{enumerate}
\item  $M$ is not isomorphic to a type $\mathrm{I}_n$ factor;
\item  $M$ has a \emph{faithful normal finite trace}, that is, a normal linear map $\mathrm{tr}:M\raw \C$ such that 
\begin{align}
\tr(ST)=\tr(TS); && \tr(S^*S)\geq 0; && \tr(S^*S)= 0 \mbox{ iff } S=0.
\end{align}
\end{enumerate}
\item  Type $\mathrm{II}_{\infty}$: $M=N\otimes B(H)$, where $N$ is a type $\mathrm{II}_1$ factor and $H$ is infinite-dimensional.\footnote{Tensoring a type $\mathrm{II}_1$ factor with a type $\mathrm{I}_n$ factor gives a  type $\mathrm{II}_1$ factor.}
\item  Type $\mathrm{III}$: for any projection $E \in M$ there is $V\in M$ such that $V^*V=1_M$ and $VV^*=E$

 \end{itemize}
In type \tto\ factors the trace can be normalized such that $\tr(1_M)=1$, so that it is a (normal) state on $M$; clearly, tr
cannot be the usual trace Tr restricted to $M$ since $\Tr(1_M)=\Tr(1_H)=\infty$. As we shall see, all projections $E$ in
a type $\mathrm{II}_1$ factor are infinite-dimensional  with $\tr(E)\leq 1$. As already mentioned in the Historical Introduction, type $\mathrm{II}_1$ factors were von Neumann's favourite since he saw this trace as the maximum-entropy state which $B(H)$ lacks if $H$ is infinite-dimensional.\footnote{von Neumann (1937/1981) also built his fascinating ``continuous geometries'' on type $\mathrm{II}_1$ factors. Heuristically, tr may be seen as something like a limit as $n\raw\infty$ of $1/n$ times the usual trace $\Tr$ on $L(\C^n)$, the complex $n\x n$ matrices. \label{slfn}}
Type $\mathrm{II}_{\infty}$ factors also have a well-defined trace, but like the trace Tr on the type $I_{\infty}$ factor, from which it radically differs in many other ways, it may be infinite.  Type $\mathrm{III}$ factors, on the other hand, admit no trace at all (except one that on projections only takes the values $\{0,\infty\}$).

 The main tool in the classification of factors is the  lattice $\mathrm{Proj}(M)$ of all projections in $M$ (which, we recall, are the elements $E\in M$ for which $E^2=E^*=E$), partially ordered by 
 \begin{align}
 E\leq F && \mbox{ iff } && F-E \geq 0\:\:\:  (\mbox{in } M)  && \mbox{ iff }  && EF=FE=E&& \mbox{ iff } && EH\subseteq FH, \label{ProjM}
 \end{align}
 where the last condition assumes that $M\subseteq B(H)$; the first two conditions make $\leq$ intrinsic to $M$.
  \bex
  Show the equivalence between these three conditions defining the partial ordering.
  \eex
   This lattice is non-distributive unless $M$ is commutative.\footnote{This fact is the basis of \emph{quantum logic} as developed by von Neumann and Birkhoff (R\'{e}dei, 1998). In the author's view this logic has not achieved much and should be replaced by \emph{intuitionistic} logic, cf.\ Landsman (2017), Chapter 12.
   } It is
   \emph{complete} in the sense that suprema $\bigvee S$ of all subsets $S$ of $\mathrm{Proj}(M)$ exist;  if $M\subseteq B(H)$, then   $\bigvee S$ is the projection onto the closure of $\bigcup_{E\in S}EH$, which lies in $M$ because it is complete in the strong operator topology. However, it is not necessary to know lattice theory to define \emph{Murray--von Neumann} equivalence of projections:
\begin{definition} \label{MvNeq}
For any two projections $E$ and $F$ in a \vna\ $M$
we say that $E\sim F$ iff there exists $V\in M$ such that $V^*V=E$ and $VV^*=F$ (so that $V$ is a partial isometry).
\end{definition}
\bex
Show that $\sim$ is an equivalence relation.
\eex
 If  $M\subseteq B(H)$,
the operator $V$ has kernel $(EH)^{\perp}$, range $FH$, and is unitary 
from $EH$ to $FH$.  It is therefore  a partial isometry, with initial projection $E$ and  final projection $F$. Hence a \emph{necessary} condition for $E\sim F$ is  $\dim(EH)=\dim(FH)$, but unless $M=B(H)$,  this is not \emph{sufficient}, since  $V$ in Definition \ref{MvNeq}  \emph{is required to lie in $M$}. For example, if $H=\C\oplus\C$, then $e=\mathrm{diag}(1,0)$ is equivalent to
$f=\mathrm{diag}(0,1)$ for $M=M_2(\C)$, but not for the \emph{diagonal} $2\x 2$ matrices $M=D_2(\C)$. 

To see how natural this definition is, consider a unitary \rep\ $U$ of a group $G$ on $H$. If $H_i\subset H$ is stable under $U(G)$, $i=1,2$, then the restrictions  $U_i$ of $U$ to $H_i$ are unitarily equivalent precisely when $E_1\sim E_2$ (where $E_i$ is the projection onto $E_i$) \emph{with respect to $M=U(G)'$}.

 Let us apply this definition to our two familiar cases so far:
 \begin{itemize}
\item If $M$ is commutative, then $V^*V=VV^*$ for all $V\in  M$,  and therefore $E\sim F$ iff $E=F$.
For \er{167N}
 we  have $E=\hat{1}_{\Dl}$ and $F=\hat{1}_{\Dl'}$ for $\Dl,\Dl\subset X$, and $E\leq F$ iff $\Dl\subseteq\Dl'$ (up to measure zero). 
 \item If $M=B(H)$, then $E\sim F$ iff $\dim(EH)=\dim(FH)$ (which may be countably infinite).
\end{itemize} 
 The official first step in the  Murray---von Neumann classification of factors  is now as follows.
\begin{definition}\label{finfitefactors}
 A projection $E$ in a \vna\ $M\subseteq B(H)$ is called:
\begin{itemize}
\item  \emph{finite} if any projection $F\in \mathrm{Proj}(M)$ that satisfies
$F\sim E$ and $F\leq E$ equals $F=E$;
\item   \emph{infinite} if it is not finite;
\item  \emph{(in)finite-dimensional} if its range is (in)finite-dimensional. 
\item \emph{minimal} if $E\neq 0$ and  $F\leq E$ for some $F\in \mathrm{Proj}(M)$ implies either $F=E$ or $F=0$.
\end{itemize}
\end{definition}
This  is inspired by set theory, where a set $E$ is (Dedekind) finite iff for any subset $F\subseteq E$ (the counterpart of $F\leq E$ for projections) the existence of a bijection between $F$ and $E$ (the counterpart of $F\sim E$ for projections)  implies $F=E$. For $M=B(H)$ a projection $E$ is (in)finite iff $\dim(EH)$ is (in)finite, but we will see that  some \vna s (namely type II)  contain infinite-dimensional projections that are finite.
In a factor a minimal projection is the same as an \emph{abelian} projection $p$, meaning that $pMp$ is abelian.
Here is the key to the type classification of factors:
\begin{definition}\label{deftypes}
A factor $M$ is said to be:
\begin{itemize}
\item type $\mathrm{I}$ if $M$ contains a minimal (or, equivalently, an abelian) projection, refined to:
\begin{itemize}
\item 
 type $\mathrm{I}_{\aleph}$ if $1_M$ is the sum of $\aleph$ orthogonal (but equivalent) minimal projections, where $\aleph=n\in\N$ if the number of terms in this sum is $n$, and $\aleph=\infty$ if it is countable.\footnote{With higher infinities in the non-separable case. As already mentioned, we only discuss separable \Hs s.}
 \end{itemize}
\item  type $\mathrm{II}$ if $M$ has no minimal projections, but does have  nonzero finite projection, refined to:
\begin{itemize}
\item  type $\mathrm{II}_1$ if $M$ is type $\mathrm{II}$  and $1_M$ is  finite.
\item   type $\mathrm{II}_{\infty}$ if $M$ is type $\mathrm{II}$ and $1_M$ is  infinite.
\end{itemize}
\item  type $\mathrm{III}$ if all nonzero projections in $M$ are infinite. If $M\subset B(H)$ for \emph{separable} $H$, each nonzero projection is then equivalent to the unit and hence to any other projection in $M$.
\end{itemize}
\end{definition}
The (with hindsight confusing) name `factor' comes from the following important fact:
\begin{theorem}\label{Ifactors}
 A type $I_{\aleph}$ factor  is \emph{algebraically} isomorphic to $B(H_A)$ for some \Hs\ $H_A$, where $\aleph=\dim(H_A)$, and   \emph{spatially} isomorphic to $B(H_A)\ot 1_{H_B}\subset B(H_{AB})$, for a \Hs\  $H_B$.
 \end{theorem} 
 This theorem is not at all easy to \emph{prove},\footnote{See Blackadar (2006), \S\S III.1.5.12 to III.1.5.15, or Takesaki (2002), Corollary V.1.28.} but one way to \emph{understand} it is to use the fact that the commutant $M'$ of a type I factor $M$ is again type I (although it need not be a factor).\footnote{Here a \vna\ $M$ is said to be of type I if every nonzero projection in the center of $M$ majorizes an abelian projection in $M$,  type II if every central projection in $M$ majorizes a nonzero finite projection, and type III if $M$ has no nonzero finite projections (but unlike factors, arbitrary \vna s  need not be of one such type). Then  the  commutant of any given type of algebra has the same type. See Blackadar (2006), Theorem III.2.6.12. Ironically, what we here use as a ``fact'' is usually derived from Theorem \ref{Ifactors}, which we derived from the said fact!\label{typegen}
 } 
 Namely, take a maximal orthogonal family $(E_i)_{i\in I}$ of minimal projections in $M'$ such that 
 \begin{align}
 H=\bigoplus_{i\in I} H_i;  && M=\bigoplus_{i\in I} M_i; && H_i=E_iH, \label{Hidec}
 \end{align} where we denote $M_{|H_i}$ by $M_i\subseteq B(H_i)$; since $E_i\in M'$ the operators in $M$ map $H_i$ into itself, and so  \er{Hidec}  brings $M$ into block-diagonal form. 
Then $M_i'=\C\cdot 1_{H_i}$,  for otherwise the projections $E_i$ could not be minimal. Since $M_i''=M_i$ this implies $M_i=M_i''=(\C\cdot 1_{H_i})'=B(H_i)$.  Now take some partial isometry $V_{ij}$ with initial and final projections $V_{ij}^*V_{ij}=E_i$ and $V_{ij}V_{ij}^*=E_j$. Using the assumption that $M$ is a factor, 
 it can be shown that $V_{ij}$ is
  unitary from $H_i$ to $H_j$ and intertwines $M_i$ and $M_j$, i.e., $V_{ij}T_iV_{ij}^*=T_j$, where $T\in M$ is decomposed as $T=\oplus_i T_i$ according to \er{Hidec}.
   This implies that 
 all $H_i$ must be isomorphic, say $H_i\cong H_A$ for a single \Hs\ $H_A$, and hence that all $B(H_i)\cong B(H_A)$ are replicas of each other.
 Therefore, we have \emph{spatial} or unitary isomorphisms
  \begin{align}
  H\cong H_A\ot \ell^2(I); && M\cong B(H_A)\ot 1_{\ell^2(I)}.
  \end{align}
  
We also have, more or less straight from Definition \ref{deftypes}, a clarification about projections:
\begin{itemize}
\item    A type $I_{\aleph}$ factor $M=B(H)$ contains all projections on $H$;
\item All projections in type $\mathrm{II}$ factors are infinite-dimensional, yet in a $\mathrm{II}_1$ factor they are all finite, whereas in a $\mathrm{II}_{\infty}$ factor each infinite projection majorizes a nonzero finite projection.\footnote{This is true for both relevant partial orderings $\leq$ and $\lesssim$.}
\item All projections in type $\mathrm{III}$ factors are both infinite and infinite-dimensional (and \emph{vice versa}).
\end{itemize}
A  new perspective on these types comes from the idea of a \emph{dimension function}
\begin{align}
d&:\mathrm{Proj}(M)\raw [0,\infty] && \mbox{that satisfies:}\\ 
 d(E)&<\infty && \mbox{iff } E \mbox {is finite};\\
  d(E+F)&=d(E)+d(F) && \mbox{if } EF=0;\\
  d(E)& =d(F) && \mbox{if } E\sim F. \label{ifd}
  \end{align}
  Any \vna\ admits a nonzero dimension function $d$, which in a factor is unique up to rescaling.
   For the type I factors $M=B(H)$ we have $d(E)=\dim(EH)$, 
but since $1_M$ is finite in a $\mathrm{II}_1$ factor its dimension function must be something else. Likewise for the other  non-type I factors.
The significance of a dimension function emerges if we define a new partial ordering on $\mathrm{Proj}(M)$ by putting $E\lesssim F$ if there exists an $E'\in \mathrm{Proj}(M)$ with $E\sim E'$ and $E'\leq F$.\footnote{If we compare 
 projections in $M$ with sets and thence compare $\leq$ and $\sim$ with 
 $\subseteq$ (inclusion) and $\cong$ (isomorphism) of sets, respectively, then $E\lesssim F$ is analogous to the existence of an injective function between sets. 
  The Schr\"{o}der--Bernstein Theorem of set theory (which von Neumann knew  well) states that if $X\hookrightarrow Y$ and 
 $Y\hookrightarrow X$, then $X\cong Y$.} Then:\footnote{See Sunder (2012), Proposition 1.1.7 or Takesaki (2002), Proposition V.1.3, for a proof.}
 \begin{proposition}\label{SBN}
 In any \vna, if $E\lesssim F$ and $F\lesssim E$, then $E\sim F$.
 \end{proposition}
 The special role of factors with respect to the partial ordering $\lesssim$ now emerges:\footnote{See Blackadar (2006), Corollary III.1.1.11.}
 \begin{theorem}
 If $M$ is a factor, then  $\lesssim$ is a total ordering (i.e., $E\lesssim F$ or $F\lesssim E$), and the `if' in \er{ifd} is an `iff'. 
Consequently, $d(E)$ characterizes the equivalence class of $E$ under $\sim$ and hence induces an isomorphism between $\mathrm{Proj}(M)/\sim$ and some subset of $[0,\infty]$. One can then rescale the dimension function $d$ such that its range $\ran(d)$ completely classifies the type of $M$, as follows:\footnote{Continuing footnote \ref{slfn}, a heuristic way to understand the type \tto\ case is that the range of $d_n=(1/n)\cdot\Tr$ on the projections in $M_n(\C)$ is $\{k/n, k=0, 1, \dots, n\}$, which becomes ``increasingly dense'' in $[0,1]$ as $n\raw \infty$.
}
\begin{itemize}
\item $\ran(d)=\{0,1,2,\ldots, n\}$, for some $n\in\N$, which is the case iff $M$ is type $\mathrm{I}_n$; 
\item  $\ran(d)=\N\cup\infty$, which is the case iff $M$ is type $\mathrm{I}_{\infty}$;
\item $\ran(d)=[0,1]$, which is the case iff $M$ is type  $\mathrm{II}_1$;
\item  $\ran(d)=[0,\infty]$, which is the case iff $M$ is type  $\mathrm{II}_{\infty}$;
\item $\ran(d)=\{0,\infty\}$, which is the case iff $M$ is type  $\mathrm{III}$.
\end{itemize}
  \end{theorem}
In the last case only the zero projection has dimension zero; all other projections have dimension equal to infinity, which matches the fact that if $M\subseteq B(H)$, the range of these projections is infinite-dimensional in the usual sense. But beware of the terminology:  it cannot be overemphasized that all nonzero \emph{finite} projections in type II factors  have \emph{infinite}-dimensional range!

Dimension functions are closely related to generalized \emph{traces}:\footnote{As usual, lsc means  lower semicontinuous, which is the right sense of continuity if $\infty$ is an allowed value.}
\begin{definition}\label{deftrvna}
\begin{enumerate}
\item A \emph{(normal) trace} on a \vna\  $M$ is a ($\sg$-weakly lsc) map  
\begin{align}
\mathrm{tr}:M^+\raw[0,\infty]; && M^+:=\{T\in M\mid T=S^*S,\, S\in M\},
\end{align}
 that satisfies
 \begin{align}
\mathrm{tr}(\lm S+T)&=\lm\cdot \mathrm{tr}(S)+\mathrm{tr}(T) && (S,T\in M^+, \lm\geq 0);\label{tracial1}\\
\mathrm{tr}(SS^*)&=\mathrm{tr}(S^*S) && (S\in M), \label{tracial2}
\end{align}
or, equivalently, $\mathrm{tr}(USU^*)=\mathrm{tr}(S)$ for all $S\in M^+$ and unitary $U\in M$ (so that $USU^*\in M^+$). 
\item A trace is \emph{finite} if $\mathrm{tr}(T)<\infty$ for all $T\in M^+$,  \emph{semifinite} if  for any $T\in M^+$ there is a nonzero positive operator $S\leq T$ in $M^+$ for which $\mathrm{tr}(S)<\infty$, and  \emph{infinite} otherwise.\footnote{It can be shown that a finite trace on a factor is automatically  $\sg$-weakly continuous.}

\item A trace is \emph{faithful} if $\tr(S^*S)=0$ implies $S=0$ (the converse is trivial). 
\end{enumerate}
\end{definition}
For example, the usual trace $\Tr$  is a faithful normal trace on $B(H)$ in this new sense, which is finite iff $\dim(H)$ is finite, and semifinite otherwise.  So is
 the trace $\tr$ already discussed informally on a type  $\mathrm{II}$ factor: this trace is obviously finite on a
  $\mathrm{II}_1$ factor and semifinite on a type  $\mathrm{II}_{\infty}$ factor. 
  \begin{proposition} \label{MvNThm}
For any \vna\ $M$, the restriction of a faithful normal trace  to $\mathrm{Proj}(M)$  is a dimension
 function.
If $M\subseteq B(H)$ is a factor, with $H$ separable as always, then:
\begin{enumerate}
\item Any two  faithful normal  traces on $M$ are proportional.
\item  Any faithful normal trace on $M$ restricts to
a completely additive dimension function with the additional property that  $d(E)=d(E)$ if \emph{and only if} $E\sim F$.
\item Any dimension function with this additional property is the restriction of 
a faithful normal  trace, and hence is completely additive, and unique up to scaling (by some $\lm>0$)..
 \end{enumerate}
\end{proposition}
This gives yet another view on the type classification of factors $M$. 
\begin{proposition}\label{typefromtrace}
A factor $M$ has type:
\begin{itemize}
\item $\mathrm{I}_n$ or  $\mathrm{II}_1$ if $M$  has a finite faithful normal trace, and then $\mathrm{I}_n$ iff $M$ is finite-dimensional;
\item  $\mathrm{I}_{\infty}$ or   $\mathrm{II}_{\infty}$ if $M$ has a semifinite faithful normal trace but not a finite one,
 and then $\mathrm{I}_{\infty}$ iff $M$ has a minimal projection, or, equivalently, $M\cong B(H)$ for some infinite-dimensional $H$;
\item  $\mathrm{III}$ if $M$ has no  semifinite faithful normal trace. 
\end{itemize}
\end{proposition}
Thus type I or II factors (and, more generally, \vna s) are called \emph{semifinite}.

At last, after all this abstraction  it is high time for some concrete examples of non-type I factors!
The first example of a type  $\mathrm{II}_1$, due to Murray and von Neumann,  is as follows. Take a 
discrete group $G$.
For these, Haar measure is simple the counting measure.
Consider  operators on the \Hs\ $\ell^2(G)$ of the following kind, where $y\in G$ and $f:G\raw\C$ has finite support:
\begin{align}
u(y)\psi(x)=\psi(y\inv x); && \pi(f)=\Sigma_{y\in G} f(y)u(y); &&
\pi(f)\psi(x)=\Sigma_{y\in G} f(y)\ps(y\inv x), \label{uintus}
\end{align}
The \emph{group von Neumann algebra}
$W^*(G)$ is the bicommutant of the set of all such $u(y)$, $y\in G$, or, equivalently,  the bicommutant of the set of all $\pi(f)$, $f\in C_c(G)$.
\begin{proposition}
\label{iccthm}
If all nontrivial  conjugacy classes in $G$ (i.e., all  except $\{e\}$) are infinite, then 
the group \vna\ $W^*(G)$ of a countable group is a  type  $\mathrm{II}_1$ factor.
\end{proposition}
In that case, we say that $G$ has (or ``is'') \emph{icc}, i.e., $G$ has \emph{infinite conjugacy classes}.\footnote{Each free group on $n>1$ generators is i.c.c. So is the group $\GS_{\infty}$ of finite permutations of $\N$. A $j$-\emph{cycle} is a cyclic permutation of $j$ objects (called the \emph{carrier} of the cycle). Any element $p$ of  $\GS_{\infty}=\cup_n \GS_n$ is a finite product 
of $j$-cycles with disjoint carriers, and for each $j\in\N$, the number of $j$-cycles in such a decomposition of $p$ is uniquely determined by $p$.
Two permutations in $\GS_{\infty}$, then,  are conjugate iff they have the same number of $j$-cycles, for all $j\in\N$.}
\smallskip

\noindent \emph{Proof}.
To study $W^*(G)\cap W^*(G)'$ we check when  $\pi(f)\pi(g)=\pi(g)\pi(f)$ for all $g\in C_c(G)$. Define
\begin{align}
f*g(x)=\Sigma_{y\in G} f(xy\inv) g(y); && f^*(x)=\ovl{f(x\inv)},\label{CG}
\end{align}
so that $\pi(f)\pi(g)=\pi(f*g)$ and $\pi(f)^*=\pi(f^*)$. Now $\pi$ is obviously injective, so  
\begin{align}
\pi(f)\pi(g)=\pi(g)\pi(f) && \LRaw&& f*g=g*f&& \LRaw&& f(yxy\inv)=f(x)\:\forall x,y\in G.
\end{align} Hence  $f$ lies in the center of at least $\pi(C_c(G))\subset W^*(G)$ iff $f$  is constant on each conjugacy class of $G$. If $G$ is icc, this implies that  $f=\lm\cdot\dl_e$, $\lm\in\C$. 
Since $\pi(\dl_e)=1_{\ell^2(G)}$,  this proves that $W^*(G)$ is a factor, except for the extension of this argument from $C_c(G)$ to $W^*(G)$. We omit this  step. 

We now define a map $\mathrm{tr}:W^*(G)\raw\C$ by linear and (e.g.\ weak) continuous extension of
 \begin{align}
 \mathrm{tr}(f):=\la\dl_e, f*\dl_e\ra =f(e) && \LRaw && \tr(u(y)):=0\:\: (y\neq e); && \tr(u(e)):=\tr(1_{\ell^2(G)})=1, \label{trwg}
 \end{align}
 which by \er{CG}  satisfies \er{tracial1} - \er{tracial2} and 
defines a normal faithful finite trace on $W^*(G)$. 
 Since $G$ must  be infinite in order to be icc, 
 $W^*(G)$ is infinite-dimensional, and hence $W^*(G)$ cannot be $M_n(\C)$ for any $n\in\N$. The claim then follows from Proposition \ref{typefromtrace}.\QED\smallskip

Murray and von Neumann also gave some  examples of type III factors using ergodic theory, but it is easier to base examples of both type II and III on the CAR-algebra familiar from physics.\footnote{CAR means \emph{Canonical Anticommutation Relations.}
Bratteli \& Robinson (1981) is a good reference for this.} 
\begin{definition}
Let $H$ be a separable \Hs\ and let $\mathcal{F}_A(H)$ be the
associated fermionic Fock space.\footnote{This is the subspace of the 
total Fock space $\mathcal{F}(H)=\bigoplus_{n=0}^{\infty} H^n$ in which  each subspace $H^n\equiv H^{\ot n}$ is restricted to its totally antisymmetrized part. In physics the fermionic Fock space describes systems of identical fermions.}
 The C*-algebra  $\mathrm{CAR}(H)$ is the smallest \ca\ in $B(\mathcal{F}_A(H))$ that contains all operators $a^*(f_0)$, $f_0\in H$, that are defined by linear extension (and closure) of
 \begin{align}
a^*(f_0)f_1\wed\cdots \wed f_n:=\sqrt{n+1} f_0\wed f_1\cdots\wed f_n && (f_0, f_1, \ldots, f_n\in H).
\end{align}
\end{definition}
Such operators  satisfy the \emph{canonical anticommutation relations}, where $[S,T]_+:= ST+TS$,
 \begin{align}
 [a(f),a^*(g)]_+=\la f,g\ra_H\cdot 1_{\mathcal{F}(H)}; &&  [a(f),a(g)]_+= [a^*(f),a^*(g)]_+ =0.
 \end{align}
 Picking a basis $(e_n)$ of $H$, these assume the perhaps more familiar form, with $a_n=a(e_n)$, etc.:
  \begin{align}
 [a_n,a^*_m]_+=\dl_{nm};  &&  [a_n,a_m]_+= [a^*_n,a^*_m]_+ =0.\label{car}
 \end{align}
 If $H=\C$, then $\mathrm{CAR}(\C)\cong M_2(\C)$ on the identifications
  \begin{align}
  a=
\left(
\begin{array}{cc}
 0 &  1    \\
  0   &  0 
\end{array}
\right);  &&  a^*=
\left(
\begin{array}{cc}
 0 &  0    \\
  1   &  0 
\end{array}
\right);
&&  a^*a=
\left(
\begin{array}{cc}
 0 &  0    \\
  0   &  1 
\end{array}
\right);
&&  aa^*=
\left(
\begin{array}{cc}
 1 &  0    \\
  0   &  0 
\end{array}
\right),
 \label{1220}
 \end{align}
 and this may be expanded into a realization of  $\mathrm{CAR}(\ell^2(\N))$ as an infinite tensor product
 \begin{equation}
\mathrm{CAR}(\ell^2(\N))\cong \bigotimes_{n=1}^{\infty} M_2(\C), \label{1221}
\end{equation}
but this is more difficult to define. 
However, mathematicians unfamiliar with fermions might like the following alternative realization. Let $2^*=\cup_{N\in\N_*}2^N$, where $\N_*=\{1,2,3,\ldots \}$,  be the \emph{countable} set of all \emph{finite} binary strings, with associated \emph{separable} \Hs\ $\ell^2(2^*)$. We embed $2^*$ in the \emph{uncountable} set $2^\N$ of all \emph{infinite} binary sequences by adding an infinite number of zeros to each $\sg\in 2^N$, and hence embed $\ell^2(2^*)$ in 
the \emph{non-separable} \Hs\ $\ell^2(2^{\N})$; elements of $\ell^2(2^*)$ are then those functions $\ps:2^\N\raw\C$ that vanish on any $s\in 2^\N$ outside $2^*$ (and are square-summable over $2^*$).
Now take operators $T$ on $\ell^2(2^\N)$ for which there is some $N$ so that $T$ only depends on the first $N$ digits of the argument $s$ of any $\psi$ on which it acts.\footnote{This means that 
for any $\psi\in \ell^2(2^{\N})$ we have  $T\psi(s)=T\ps(s')$ whenever $s_{|N}=s'_{|N}$ (i.e., $s_0\cdots s_{N-1}=s'_0\cdots s'_{N-1}$).} 
 Such operators map $\ell^2(2^*)\subset \ell^2(2^\N)$ into itself and generate a \ca\ $C^*(2^*)$ within $B(\ell^2(2^*))$.
This \ca\
is isomorphic to $\mathrm{CAR}(\ell^2(\N))$ in such a way that $a_n$ corresponds to the operator 
\begin{align}
T_n\in C^*(2^*); && 
T_n\psi(s):=0\:\:\: \mbox { if } s_n=1; && T_n\ps(s_0\cdots 0_n\cdots):= \ps(s_0\cdots 1_n\cdots),
\end{align}
where  $0_n$ means that $s_n=0$, etc. Equivalently, $\ell^2(2^*)$ has a basis $(\dl_{\sg})_{\sg\in 2^*}$, which may be written as
$\dl_{\sg}=|\sg_0\cdots \sg_{N-1}0\cdots 0\cdots\ra$, and then $a_n$ acts on the $n$th entry in the obvious manner.

Either way, $\mathrm{CAR}(\ell^2(\N))\cong C^*(2^*)$  is clearly infinite-dimensional, yet it has a finite trace!  Let us give a more general description of this trace that will later also lead us to type III factors. 
Namely, for any $\lm\in(0,1]$, consider the map 
\begin{align}\phv_{\lm}: M_2(\C)\raw \C; &&
\phv_{\lm}(b)= \frac{\lm}{1+\lm}b_{11}+\frac{1}{1+\lm}b_{22}, \label{1223}
\end{align}
so that $\lm=1$ gives the normalized trace $\tr_2$. This can be expanded to a map $\om_{\lm}: \mathrm{CAR}(\ell^2(\N))\raw\C$ 
by linear and continuous extension of its action on monomials $\mu$ in the $a_n$ and $a^*_n$, as follows: 
\begin{align}
\om_{\lm}(1_{\mathcal{F}(H)})=1; && \om_{\lm}\left(\prod_{i=1}^N a^*_{n_i}a_{n_i}\right)=(1+\lm)^{-N}; &&\om_{\lm}(\mu_{\mathrm{\footnotesize{odd}}})=0, \label{Philm}
\end{align}
where in the middle expression all $n_i$ are different from each other; using the CAR \er{car}, any monomial with at most one $a^*_n$ and exactly one $a_n$ for each  $a_n^*$ that occurs can be brought into the stated form. 
The third possibility $\mu_{\mathrm{\footnotesize{odd}}}$, then,  is any monomial not of that form.
 In particular, the realization  \er{1220} for a single index makes \er{1223} compatible with \er{Philm}, which implies
\begin{align}
\om_{\lm}(a^*_na_n)=\frac{1}{1+\lm}; && \om_{\lm}(a_na^*_n)=\frac{\lm}{1+\lm}; &&
  \om_{\lm}(a_n)= \om_{\lm}(a_n^*)=0.\label{1227}
 \end{align}
For $\lm=1$ this functional is a trace, which in the second realization  can be defined directly via
\begin{align}
\mathrm{tr}: C^*(2^*)\raw\C; && \mathrm{tr}(T)=\lim_{N\raw\infty}2^{-N}\Sigma_{\sg\in 2^N}\la \dl_{\sg}, T\dl_{\sg}\ra,
\end{align}
where, as above, $\sg\in 2^N\subset 2^*\subset 2^\N$ by adding zeros. For the operators $T$ just considered this sequence stabilizes at some $N$, and for those in the norm-closure defining $C^*(2^*)$ the limit exists by continuity.\footnote{Since tr is positive and normalized it has unit norm and hence is continuous.}
In all cases, $\om_{\lm}$ is a state on $\mathcal{A}\equiv \mathrm{CAR}(\ell^2(\N))$, 
 and so we may perform the  {\sc gns}-construction,\footnote{The weak closure or bicommutant of
 $\mathrm{CAR}(H)$ itself is just the boring type I factor $B(\mathcal{F}_A(H))$.}
  cf.\  the proof of Theorem \ref{GNT}. Then the  \vna\ $M=\pi_{\tr}(\mathcal{A})''$:
  \begin{itemize}
\item is a type  $\mathrm{II}_1$ factor for $\lm=1$ (called $R$, which in fact is the unique hyperfinite \tto\ factor).
\item is  a type  III factor for $\lm\in (0,1)$--it is the unique hyperfinite type $\mathrm{III}_{\lm}$ factor of Connes.\footnote{See Connes (1994), \S V.9. Of these, $\lm=1$ is crucial for theoretical physics, see footnote \ref{typeIIIfn} below. \label{nogIII}}.
\end{itemize}
As to the first case, the cyclic vector $\Om_{\om_1}$ of the {\sc gns}-construction defines a normal faithful finite trace on $M$
   via the property  \er{877GNS}, that is, by a weak operator topology extension of 
   \beq
   \tr(T)=\la\Om_{\om_1}, \pi_{\om_1}(T)\Om_{\om_1}\ra.
   \eeq
The fact that $M$ is a factor is most easily seen by truncating $\ell^2(\N)$ to finite $N$ and performing the  {\sc gns}-construction component-wise; this makes each copy of $M_2(\C)$ act on itself from the left, so that its commutant is $M_2(\C)$  acting on the right, and this is clearly a factor. Unfortunately, the fact that for $\lm\in (0,1)$ we obtain a type III factor is beyond the scope of these notes.\footnote{See however Blackadar (2006), \S III.3.1.9 for a heroic attempt to explain this without modular theory.}

 Without explanation, here is  an incomplete short survey of factors appearing in mathematical physics, all of which are hyperfinite.\footnote{I.e., $M$ is the $\sg$-weak closure of an increasing sequence of finite-dimensional subalgebras, cf.\ footnote \ref{sepfootnote}.}
The general rule is that finite-dimensional quantum systems are described by type I factors and infinite-dimensional ones by type III, but there are some exceptions: type $\mathrm{II}_1$, von  Neumann's favourite, occurs (as we shall see) for infinite quantum systems in thermal equilibrium states at infinite temperature, whereas
 type $\mathrm{II}_{\infty}$ was  recently shown to play a role in black hole thermodynamics in the presence of observers or quantum reference frames.\footnote{See the references in footnote \ref{Kudler}. \label{Vuyst}}
 The oldest physically relevant examples of type III factors, known since the early 1960s, are equilibrium states of infinitely extended non-relativistic free Bose and Fermi gases. In relativistic (algebraic) quantum field theory  the  local \vna\ algebras $\mathcal{M}(\CO)$ are also
 type $\mathrm{III}$ factors.\footnote{See e.g.\ Haag (1992), Yngvason (2005), Fredenhagen (2015), and Fewster \& Verch (2015). In the Connes classification of hyperfinite factors, see footnote \ref{nogIII}, they are all isomorphic
 to the unique hyperfinite $\mathrm{III}_1$ factor. \label{typeIIIfn}} at least 
  in vacuum \rep s in Minkowski spacetimes, and more generally in curved spacetime with Hadamard states instead of vacuum states,  provided  suitable axioms expressing locality etc.\ as well as the existence of a regular scaling limit hold. 
  Type III factors also occur in the quantum-information theoretic task of embezzlement of entanglement.\footnote{See van Luijk et al.\ (2024), where all the hyperfinite types $\mathrm{III}_{\lm}$ appear in a beautiful manner.} 
\section{The standard form of a \vna}
We repeat that our goal is to find a useful expression for a normal state $\om$ on a \vna\ $M\subseteq B(H)$, perhaps (but not necessarily) given by some density operator  $\rh\in D(H)$ via $\om(T)=\Tr(\rh T)$,  where `useful' means that one can define and compute at least the more common entropies like \er{vNe} and \er{QKL}. Theorem \ref{Ifactors} solves this problem for type I factors, since given that $M$ is spatially isomorphic to $B(H_A)\ot 1_{H_B}$ we can restrict $\om$ to the first factor, i.e., take the partial trace of $\rh=\rh_{AB}$ over the second factor, to obtain a canonical density operator $\rh_A$ on $M$. 

There is  another way of solving the problem for type I factors, whose underlying technique applies to all \vna s, and in whose modern theory it plays a central role. This technique is the \emph{standard form},\footnote{This technique was developed in the 1970s by Araki, Connes, Takesaki, and Haagerup.} which will  solve our problem at once for type II factors, and, with considerable extra work, finally for type III factors as well. The theory of the standard form simplifies if we repeat our standing assumption that $M$ acts on a separable \Hs\ $H$.
Before even defining what it means exactly, let us first give the standard form for a type I factor $M=B(H')$; more precisely, we give two unitarily equivalent version of it, which is instructive. 
\begin{itemize}
\item \emph{Standard form of $B(H')$, version 1.} Take the \Hs\ $H=B_2(H')$ of Hilbert--Schmidt operators on $H'$,
 whose elements are  $T\in B(H')$ for which $\Tr(T^*T)<\infty$.
Operators in $B_2(H')$ now play a double role: they can be either vectors in $H=B_2(H')$ or operators thereon. 
To distinguish these roles  we will henceforth denote elements of $H$ by $\xi$ and other  Greek letters, and operators in $B(H')$ by $A, B$, etc. Using this, the inner product on $H$ is 
\beq
\la \xi,\eta\ra_2=\Tr(\xi^*\eta), \label{HSIPB2H}
\eeq
and $B(H')$ acts on $H$ by left multiplication via $L(A)\xi=A\xi$.
 Since $B_2(H')$ is a two-sided ideal in $B(H')$, this defines a bounded operator: that is, if $A\in B(H')$ and $\xi \in B_2(H')$, then $A\xi\in B_2(H')$. Likewise for right-multiplication defined by $R(A)\xi=\xi A\in B_2(H')$.

Now consider the \emph{antilinear} map $J:B_2(H')\raw B_2(H')$ defined by 
\beq
J\xi=\xi^*. \label{JtypeI}
\eeq
 This map satisfies 
\begin{align}J^*=J; && J^2=1_H && \LRaw && J\inv=J,
\end{align}
so that $J$ is \emph{anti-unitary}.\footnote{\label{fnau} The adjoint of an antilinear map $J$ is defined by the property $\la \ps, J\phv\ra=\ovl{\la J\ps,\phv\ra}$, and so an anti-unitary map satisfies $\la J\ps, J\phv\ra=\ovl{\la \ps,\phv\ra}$, and is invertible.
This still comes down to the usual equations $J^*J=JJ^*=1_H$ for unitarity.} For any $A\in B(H')$, acting from the left on $B_2(H')$, we have
\begin{align}
JAJ=R(A^*),
\end{align}
from which, writing $M=L(B(H'))\cong B(H')$, and computing that $M'=R(B(H'))$, we infer 
\beq
JMJ=M'. \label{JMJ1}
\eeq
The map $j(A):=JAJ$ defines an \emph{antilinear isomorphism} of algebras in the sense that  $j(AB)=j(A)j(B)$
  but $j(\lm A)=\ovl{\lm}j(A)$, whereas the map $j^*(A):=JA^*J$ gives a \emph{linear} \emph{anti-isomorphism} of algebras, in the sense that 
$j^*(\lm A)=\lm j^*(A)$ but $j^*(AB)=j^*(B)j^*(A)$. So the commutant $M'$ of $M=L(B(H'))$ is both antilinearly isomorphic and linearly anti-isomorphic to $M$.\footnote{In this case, $M$ is even linearly isomorphic (= isomorphic) to $M$, since $B(H')$ happens to be anti-isomorphic to itself: take a basis $(e_i)$ and then put $T\mapsto T^t$ with respect to that basis, where 
$T^t_{ij}= T_{ji}$, with  $T_{ij}=\la e_i, Te_j\ra$.  But a famous paper by Connes (1975) constructed a type III factor on a separable \Hs\ that is not anti-isomorphic to itself, so in general $JMJ$ will not be isomorphic to $M$ even in the standard form of $M$ (see below). }

 Moreover, $B_2(H')$ has a \emph{(positive) cone}  with spectacular properties.
\begin{definition} \label{Defsdc}
A  \emph{cone} in a \Hs\  $H$ is a convex subset $P\subset H$ such that:
\begin{enumerate}
\item  if $\psi\in H$, then $t\psi\in H$ for all $t\geq 0$, or briefly: $\R^+P=P$;
\item $P\cap -P=\{0\}$.
\end{enumerate}  Moreover, defining the \emph{dual cone}  $P^\dagger:=\{\psi\in H\mid\, \forall_{\eta\in P}\la\psi,\eta\ra\geq 0\}$, a cone is \emph{self-dual} if
\begin{enumerate}[resume]
\item $P=P^{\dagger}$.
\end{enumerate}\end{definition}
For $M= B(H')$ acting on  $H=B_2(H')$ from the left, as above, we take $P$ to be
\begin{equation}
P=B_2(H')^+=\{A^*A\mid A\in B_2(H')\}= \{AA^*\mid A\in B_2(H')\}=\{AJA\mid A\in B_2(H')\}. \label{PtypeI}
\end{equation}
\bex Show that
 this is  a self-dual cone, with the following additional properties:
\begin{align}
J\xi=\xi &&  (\xi \in P); && AJA\xi\in P && ( A\in B(H'), \xi\ \in P).\label{137}
\end{align}\eex
Here is the main point about $M=B(H')\subset B(H)$, $H=B_2(H')$. This  will soon be generalized  to all \vna s, so that we write $\la\cdot,\cdot\ra$ for the inner product $\la\cdot,\cdot\ra_2$ on $B_2(H')$. For any \vna\ $M$ with predual $M_*$, we denote the set of \emph{positive} normal functionals on $M$ by $M_*^+$; the normal states on $M$ are those $\om\in M_*^+$ with $\om(1_M)=1$.
\begin{proposition}\label{SFkey}
There is a homeomorphism $M_*^+\cong P$, and hence in particular a bijection between the normal states $\om$ on $M$ and the unit vectors $\Om$ in $P$, given by
\begin{equation}
\om(A)=\la\Om, A\Om\ra. \label{138}
\end{equation}
\end{proposition}
Here the topologies are the norm-topologies on both $M_*=B_1(H')$ and $B_2(H')$ with respect to their canonical norms $\|\cdot\|_1$ and $\|\cdot\|_2$, respectively.\footnote{These norms are canonical in the sense that $\|\cdot\|_1$ is the norm on $B_1(H')$ in its capacity as the predual of $B(H')$, whereas $\|\cdot\|_2$ is the norm on $B_2(H')$ as a \Hs\ with inner product \er{HSIPB2H}.}  The proof is easy because from Theorem \ref{theorem88} we know that the relation \er{normal} gives a bijective correspondence between normal states 
$\om$ on $H$ and density operators $\rh\in D(H')$.  The relation
\begin{equation}
\Om=\rh^{1/2} \label{189v}
\end{equation}
gives an isometric bijection
$\rh\mapsto\rh^{1/2}$  from $B_1(H')^+$ to $B_2(H')^+$ with inverse $\rh=\Om^2$.
\bex
Show that  if we assume \er{189v},  then \er{138} and  \er{normal}  reproduce \er{omTDH}.\eex

In particular, $D(H')\subset B_1(H')^+$ is  mapped to the set $P_1$ of unit vectors in $P$. 
Running this  in the opposite direction, Proposition \ref{SFkey} implies the bijective correspondence \er{normal} between normal states $\om$ on $B(H')$ and density operators $\rh$ on $H$, including the topology.

We are not yet done even with this first example of the standard form. There is a canonical way of getting it from a generalization of the \textsc{gns}-construction, starting from the trace. This is not a state and if $H$ is infinite-dimensional it cannot be rescaled so as to become a state (which for $H=\C^n$ could be done by replacing $\Tr$ by $(1/n)\cdot\Tr$). The appropriate replacement of a state on a \vna\ for a situation like this is a \emph{weight}.
\begin{definition}\label{defweight}
A \emph{weight} on a \vna\ $M$ is a map $\phv:M^+\raw[0,\infty]$ that is linear for positive coefficients. It is \emph{normal} if it is $\sg$-weakly lower semicontinuous,\footnote{This is equivalent to the property $\phv(\bigvee_i A_i)=\sup_i\phv(A_i)$ for any bounded increasing net $(A_i)$ in $M_+$. See for example Blackadar (2006), Theorem III.2.2.18. Lower semicontinuity is optimal for maps into $[0,\infty]$ including $\infty$!.}
  \emph{semifinite} if the set of all $A\in A^+$ for which $\phv(A)<\infty$ is $\sg$-weakly dense in $M^+$, and \emph{faithful} if $\phv(A)=0$ implies $A=0$ ($A\in M^+$). We often write `nfsf' for normal faithful semifinite'.
\end{definition}
Compare Definition \ref{deftrvna};  a (normal) trace is a (normal) weight that in addition satisfies \er{tracial2}. 
In particular, the usual trace Tr on $B(H')$ is a nfsf weight. Hence a weight is nothing to be scared of: 
in the commutative (and $\sg$-finite) case $M=L^{\infty}(X,\mu)$ a weight is just a possibly unbounded measure on $X$.
 Furthermore, like a state, a weight $\phv$ on $M$ induces a \rep\ $\pi_{\phv}$ of $M$ on a \Hs\ $H_{\phv}$, and $\pi_{\phv}(M)$ is a \vna\ if $\phv$ is normal. We just sketch this construction for an nfsf weight (like the trace Tr). Let 
 \beq
 N_{\phv}=\{ A\in M\mid \phv(A^*A)<\infty\};
 \eeq for the trace on $B(H')$ this is of course $B_2(H')$. 
Then put an inner product on $N_{\phv}$ by 
\beq
\la A,B\ra_{\phv}=\phv(A^*B);
\eeq for Tr this is \er{HSIPB2H}. Since $\phv$ is assumed to be faithful, this is positive definite, and $H_{\phv}$ is the completion of $N_{\phv}$ in the corresponding norm.  For the trace on $M=B(H')$ taking the completion is unnecessary since $B_2(H')$ is already complete in the norm $\|A\|_2=\sqrt{\Tr(A^*A)}$, but having no need for a completion, like here, is exceptional. Then $M$ first acts on $N_{\phv}$ via 
\beq
\pi_{\phv}(A)B=AB;
\eeq this operator is bounded, and hence may be extended to all of $H_{\phv}$. Unless the weight is actually a state, what is missing compared with the \textsc{gns}-construction is a unit vector $\Om_{\phv}$ that is cyclic for $\pi_{\phv}(M)$, and which returns $\phv$ via something like \er{877GNS}.  In any case, for the trace Tr on $B(H')$ this clearly reproduces version 1 of the standard form as just described. 
\item \emph{Standard form of $B(H')$, version 2.} In order to see the connection with partial traces, which technique is applicable to type I factors, we replace the \Hs\ $B_2(H')$ by $H'\ot \ovl{H}'$, where $\ovl{H}'$ is the same real vector space as $H'$ but equipped with the complex conjugate scalar multiplication and inner product. The map $U$ defined by (finite) linear extension of 
\begin{align}
U: B_2(H')\raw H'\ot \ovl{H}'; && U|\phv\ra\la\psi|=\phv\ot\ovl{\psi},
\end{align}
where  $\ovl{\psi}$ is the same as $\psi\in H'$ but now seen as an element of $\ovl{H}'$,
and  then further extended to $B_2(H')$ by (operator norm) continuity,  is unitary.
\bex
\begin{enumerate}
\item Show that
\begin{align}
UAU^*=A\ot 1_{\ovl{H}'} && (A\in B(H')).
\end{align}
\item 
Show that $J'=UJU^*$ and $P'=UP$ are given by
\begin{align}
J'(\phv\ot\psi)=\psi\ot\phv; &&P'=\left\{\Sigma_i \psi_i\ot \ovl{\psi}_i, \psi_i\in H'\right\}^-,
\end{align}
 where  $-$ denotes closure the closure of the underlying set in $H'\ot \ovl{H}'$.
 \item Show that 
 if a normal state $\om$ on $B(H')$ is given by \er{normal} with 
$\rh=\Sigma_i p_i|\ups_i\ra\la\ups_i|$,
then  
\begin{align}
\om(A)=\la\Om', (A\ot 1_{\ovl{H}'}) \Om'\ra_{H'\ot \ovl{H}'}; && \Om'=\Sigma_i \sqrt{p_i}\ups_i\ot\ovl{\ups}_i.
\end{align}
\item Show that if we then relabel $H'$ as $H_A$ and $\ovl{H}'$ as $H_B$, so that $H'\ot \ovl{H}'=H_{AB}$, and 
consider $\rh_{AB}=|\Om\ra\la\Om|$, then the partial trace $\rh_A$ of $\rh_{AB}$ over $H_B$ is $\rh_A=\rh$.
\end{enumerate}\eex
\item \emph{Standard form of $B(H')$, version 3.} We may also start with a faithful normal state $\om$ on $B(H')$ with associated density operator $\rh\in D(H')$ satisfying \er{normal}. The \textsc{gns}-construction then gives $H_{\om}$ as the completion of $B(H')$ in the norm 
\beq
\| A\|^2_{\om}=\Tr(\rh A^*A)
\eeq derived from the inner product
\beq
\la A,B\ra_{\om}=\om(A^*B)=\Tr(\rh A^*B).
\eeq
 One has 
 \beq
 \pi_{\om}(A)B=AB,
 \eeq and 
$H_{\om}$ contains  the usual
 cyclic and separating vector $\Om_{\om}=1_H$, so that for all $A\in B(H')$,
 \begin{equation}
\la \Om_{\om}, \pi_{\om}(A)\Om_{\om}\ra_{\om}=\la 1_H, A\ra_{\om}=\Tr(\rh A)=\om(A).
\end{equation}
The map 
\begin{align}
U: H_{\om} \raw B_2(H'); && UA=A\rh^{1/2}
\end{align}
 is isometric with respect to the inner products $\la\cdot,\cdot\ra_{\om}$ on $B(H')$ and $\la \cdot, \cdot\ra_2$ on $B_2(H')$, cf.\ \er{HSIPB2H}, as well as invertible with inverse $U\inv \xi=\xi\rh^{-1/2}$ (were $\rh^{-1/2}$ is typically an unbounded operator, see supplement to \S\ref{TTsec}). Hence the operator $J''=U^*JU$ is given by
\beq
J''A=\rh^{1/2}A^* \rh^{-1/2},
\eeq
 so that $P''=U^*P$ consists of all vectors $A\rh^{1/2}A^* \rh^{-1/2}$. 
 \end{itemize}
 This is  quite messy, so that we better stick with version 1, i.e., $H=B_2(H')$ with its $\rh$-independent inner product \er{HSIPB2H}, and inject $\Om_{\om}=1_H$ into it via $U$ by $U\Om_{\om}=\rh^{1/2}\in B_2(H')$.
 Given that $\rh$ is faithful, this vector is both cyclic and separating for $B(H')$.\footnote{Recall that $\Om$ is \emph{cyclic} for $M$ if $M\Om$ is dense in $H$, and \emph{separating} for $M$ if $A\Om=0$ for $A\in M$ implies $A=0$.}  In  the next section we will develop the standard form for all $M$ including $B(H')$ from that starting point. 

So in what sense has any of these constructions brought $M=B(H')$ into ``standard form''?
\begin{definition}\label{defstandardform}
A \vna\ $M\subset B(H)$ is said to be in \emph{standard form} if:
\begin{enumerate}
\item There is an antiunitary map $J:H\raw H$ such that $JMJ=M'$;
\item For any $A\in M\cap M'$ one has $JAJ=A^*$;
\item $H$ contains a self-dual cone $P$ such 
\begin{align} 
J\xi=\xi; && AJA\xi\in P; && (\xi \in P,\, A\in M).
\end{align}
\end{enumerate}
\end{definition}
Compare \er{137}.
For factors the second property is automatic since $J$ is antilinear. Our three constructions already show that there isn't one standard form for any given \vna, but all standard forms are unitarily (i.e., spatially) equivalent (as we already saw in our example).\footnote{See Blackadar (2006), Theorem III.2.6.6, or Hiai (2021a), Theorem 3.13.} 

As usual, the commutative case is instructive:  $M=L^{\infty}(X,\mu)$  acting on $H=L^2(X,\mu)$ by multiplication operators already happens to be in standard form, with
\begin{align}
J\psi(x)=\ovl{\psi(x)}; && P= L^2(X,\mu)^+=\{\psi\in L^2(X,\mu)\mid \psi(x)\geq 0\: \mu\mbox{-a.e.}\}.\label{1314N}
\end{align}
Since $M$ is maximal abelian, i.e., $M'=M$, cf.\ \er{AA}, and $J\hat{f}J=\hat{\ovl{f}}$, we  have 
\beq
JMJ=M=M',
\eeq
 and since $\hat{f}^*=\hat{\ovl{f}}$, also point 2 of Definition \ref{defstandardform} is met. 
Also here, the homeomorphism $M_*^+\cong P$ in  Proposition \ref{SFkey} holds: we know that $M_*=L^1(X,\mu)$, and so $M^+_*=L^1(X,\mu)^+$ consists of the $\mu$-a.e.\ positive $L^1$-functions, whose square roots are in $P$, and \emph{vice versa}.  
If $\om_{\rh}\in M^+_*$ is given by
\begin{align}
\om_{\rh}(f)=\int_X d\mu \rh f; &&  (\rh\in L^1(X,\mu)^+),
\end{align}
then $\Om_{\rh}=\sqrt{\rh}$, and hence
 \er{138} simply comes down to
\begin{equation}
\om_\rh(f)=\la \Om_{\rh}, \hat{f} \Om_{\rh}\ra=\int_X d\mu \rh f.\label{1315N}
\end{equation}

A  construction similar to $B(H')$ brings a type II factor $M$ into standard form. If $M$ is type $\mathrm{II}_1$  the trace tr is finite and if it is normalized by $\tr(1_M)=1$, as we always assume for type $\mathrm{II}_1$, then tr is  a normal \emph{state}, and we can perform the  \textsc{gns}-construction. This endows $M$ with an inner product
\begin{align}
\la \xi,\eta\ra_2=\tr(\xi^*\eta) && (\xi,\eta\in M), \label{L2IP}
\end{align}
where again we write $\xi$ etc.\ for an element of $M$ seen as an element of a \Hs\ on which $M$ acts, and $A$ if it thusly acts. The only complication is that $M$ is not complete in the ensuing norm 
\begin{equation}
\| A\|_2=\sqrt{\tr(A^*A)}. \label{2norm}
\end{equation}
Indeed,\footnote{See e.g.\ Landsman (2017), Theorem B.150 and eq.\ (B.493) for the case of $B(H')$.} the inequalities 
\beq
\|A\|\leq \|A\|_2\leq\|A\|_1
\eeq
 for the type I case $M=B(H')$ are replaced in the above type \tto\ situation by the reverse inequalities
\begin{equation}
\|A\|_1\leq \|A\|_2\leq\|A\|,
\end{equation}
where $\|A \|_1=\tr(|A|)$ with $|A|=\sqrt{A^*A}$ as usual. Indeed,  Cauchy--Schwarz for  \er{L2IP} gives
\begin{align}
\| A\|^2_1=\la |A|, 1_M\ra^2\leq \| A\|_2^2\cdot  \|1_M\|_2^2=\|A \|_2^2=\| A1_M\|_2^2\leq \|A\|^2 \|1_M\|_2^2=\|A\|^2.
\end{align}
Therefore, much as $B_2(H')$ is complete in the Hilbert--Schmidt norm $\|\cdot\|_2$ but incomplete in the operator norm $\|\cdot\|$, because the inequalities now run in the opposite direction our type $\mathrm{II}_1$ factor $M$ is complete in its norm topology 
but incomplete in its \Hs\ norm $\|\cdot\|_2$. Its completion in the latter is a \Hs\ denoted by $L^2(M)$ or $L^2(M,\tr)$, which carries the inner product \er{L2IP}.

Defining $J$ as in the type I case by $J\xi=\xi^*$, see \er{JtypeI}, there is a slight complication: 
this is initially valid for $\xi\in M\subset L^2(M)$ only and must be extended from $M$ to all of $L^2(M)$ by continuity. Similarly, 
$P$ is defined like \er{PtypeI}, but now as the \emph{closure} (in the  \Hs\ norm $\|\cdot\|_2$) of 
\beq
\tilde{P}=\{A^*A,\,  A\in M\}.\label{1320}
\eeq 
As a special case of Theorem \ref{SFkey2} below, we then have a homeomorphism $M^+_*\cong P$ given by \er{138}, almost exactly as in Proposition \ref{SFkey}, but with a subtle difference: if $\Om\in M$, then 
\begin{align}
\om(A)=\la\Om, A\Om\ra_2=\tr(\Om^2 A)\equiv\tr(\rh A), \label{1319}
\end{align}
with $\rh=\Om^2$ and hence $\rh\in M$. But if $\Om\notin M$, so that $\Om=\lim_m\Om_n$ with $\Om_n\in M$ as a limit in $L^2(M)$, we obtain $\rh=\lim_n \Om_n^2$ as a limit in $L^1(M)\equiv L^1(M,\tr)$, defined as the completion of $M$ in the norm $\|\cdot\|_1$ already discussed. Like $M\subset L^2(M)$, also $M\subset L^1(M)$ is a proper subspace and hence $\rh\in L^1(M)$ as in \er{1319} may not be in $M$, even when $\om$ is a normal state. In any case,\footnote{See Hiai (2021a), Proposition 4.45, and more generally his entire \S 4.3 for $L^p(M)$ spaces over type II factors $M$.} one has a similar property as in  type I, namely that $\rh\in L^1(M)$ and $A\in M$ imply $\rh A\in L^1(M)$.

The standard form of a type $\mathrm{II}_{\infty}$ factors may again be realized on $L^2(M,\tr)$, whose construction combines
 the disadvantages  of  the $\mathrm{I}_{\infty}$ and  $\mathrm{II}_1$ cases:\footnote{These problems are  obvious if $M$ is a hyperfinite $\mathrm{II}_{\infty}$ factor, in which case it is (algebraically) isomorphic to $R\ot B(H)$, where $R$ is the unique hyperfinite $\mathrm{II}_1$ factor, e.g., $R=W^*(G)$ for some icc group $G$, $H$ is a separable \Hs\ , and $\ot$ is the appropriate tensor product for \vna s; see e.g.\ Blackadar (2006), \S III.1.5. }
 first,  one needs a subspace $M_2\subset M$ consisting of all $A\in M$ for which $\tr(A^*A)<\infty$, as in the type  $\mathrm{I}_{\infty}$ case, where this was $B_2(H')$; second,  as in the $\mathrm{II}_1$ case (but unlike the $\mathrm{I}_{\infty}$ case) this space strictly needs to be completed to obtain the \Hs\  $L^2(M,\tr)$. This also means that $J$ and $P$ must be defined by continuity and completion, as in the $\mathrm{II}_1$ case, so that the complications below \er{1319} happen \emph{verbatim} in all  type II factors. 
 
 Thus our crucial  Proposition \ref{SFkey} generalizes to type II factors, and in fact it is always true:\footnote{See 
 Bratteli \&  Robinson (1987), Theorem 2.5.31, or
 Hiai (2021a), Theorem 3.12, 
 for  (lengthy) proofs. The main difficulty  is to prove surjectivity of the map $\Om\mapsto \om$; injectivity and hence uniqueness of $\Om\in P$ for given $\om\in M_*^+$. Injectivity  follows from the neat but nontrivial inequality $\|\Om_1-\Om_2\|^2\leq \|\om_1-\om_2\|$, and given surjectivity the homeomorphism property
follows by supplementing  the previous inequality with $\|\om_1-\om_2\|\leq \|\Om_1-\Om_2\|\cdot \|\Om_1+\Om_2\|$. }
\begin{theorem}\label{SFkey2}
If $M\subset B(H)$ is in standard form there is a homeomorphism $M_*^+\cong P$, and hence in particular a bijection between the normal states on $M$ and the unit vectors in $P$, given by \er{138}.
\end{theorem}
For us, this ``canonical purification'' of a normal state $\om$ by a unique unit vector $\Om\in P$ that returns $\om$ via \er{138} is the main point of the standard form, since as in the type I and II cases above, it is what leads to the association of a ``canonical'' density operator $\rh$ to a normal state $M$, where $\rh\in M$ for type I and $\rh$ ``almost''  lies in $M$ for type II (that is, $\rh$ is a limit of elements of $M$). 
\begin{theorem}\label{SFkey3}
For type I and II factors, and more generally for semifinite \vna s, the homeomorphism $M_*^+\cong P$ extends to an isomorphism of Banach spaces
\begin{align}
M_*\cong L^1(M); && \om(A)=\tr(\rh A)&& (\om\in M_*, \: \rh\in L^1(M)).
 \label{1322}
\end{align}
\end{theorem}
Here $L^1(M)$ has already been defined above for  $\mathrm{II}_1$ factors; for type I we defined $L^1(M)=B_1(H')$, and for  $\mathrm{II}_{\infty}$ it is the completion of $M_1$, the space all $A\in M$ for which $\tr(|A|)<\infty$, in the trace norm
\begin{equation}
\| A\|_1=\tr(|A|).\label{1323}
\end{equation}
Hence in all cases one obtains $L^1(M)$ and $L^2(M)$ as the completions of the spaces
\begin{align}
M_1=\{A\in M\mid \tr(|A|)<\infty\}; && M_2=\{A\in M\mid \tr(|A|^2)<\infty\},
\end{align}
 in the norms  $\|\cdot\|_1$ and  $\|\cdot\|_2$, respectively, see \er{2norm}, where $|A|^2=A^*A$, 
so that 
\beq
L^1(M)^+\cong P,
\eeq
 where $P$ is now defined also in general as the closure of $M_2^+$ in  $\|\cdot\|_2$, cf.\ \er{1320}.
Starting from 
\beq
\Om\in M^+_2\subset P,
\eeq and passing to $\rh=\Om^2\in M_1$ as before,
 Theorem \ref{SFkey2} and a density argument then give 
 \beq
 M_*^+\cong L^1(M)^+.
 \eeq
  The general case  \er{1322} follows from this by an (admittedly highly nontrivial) polar decomposition  of normal functionals on a \vna.\footnote{See Blackadar (2006), Proposition III.2.3.1, or Takesaki (2002), Theorem III.4.2. The density argument in question, as part of a complete  proof of Theorem \ref{SFkey3}, may be found in Hiai (2021a), Theorem 4.50 and Corollary 4.51.}
 Theorem \ref{SFkey3} is part of a neat theory of noncommutative $L^p$ spaces over semifinite \vna s.\footnote{See  Hiai (2021a), \S 4.3, for a detailed exposition of this theory, also called `noncommutative integration theory'.} To start, we stipulate:
 \begin{definition} 
 For $1\leq p<\infty$, $L^p(M)$ is the  completion of $M_p=\{A\in M\mid \tr(|A|^p)<\infty\}$  in the norm  $\|A\|_p=\left(\tr(|A|^p)\right)^{1/p}$.  For $p=\infty$ one defines $L^{\infty}(M):= M$ with normm$\|\cdot\|_{\infty}=\|\cdot\|$.
 \end{definition}
  These are all Banach spaces. 
  The  commutative $L^p(X,\mu)$ spaces \er{defLp}, where $\tr=\ta$ as defined in \er{traceL1}, are a special case, as are  the Schatten classes, since we have isometric isomorphisms
  \begin{align}
  L^p(L^{\infty}(X,\mu))=L^p(X,\mu); && L^p(B(H))\cong B_p(H).
  \end{align}
As in the commutative theory we have $L^p$ - $L^q$ duality: for  $1\leq p<\infty$ and $1<q\leq \infty$, $p\inv + q\inv=1$, we have   
  $L^p(M)^*\cong L^q(M)$
cf.\  \er{Lpqd}, 
where the isomorphism is now given by the trace, i.e., Tr for type I factors,  tr for type II factors, and $\int_X d\mu$ for the commutative case $L^{\infty}(X,\mu)$, all written as $\tau$ in what follows. More precisely, 
any normal linear functional $\phv: L^p(M)\raw\C$ takes the form $\phv=\phv_{\rh}$ for some $\rh\in L^q(M)$ such that $\phv_{\rh}(A)=\ta(\rh A)$, $A\in L^p(M)$.  This includes \er{1322} as the case $p=1$, $q=\infty$, which specializes to normal states by adding the condition 
\beq\rh\in D(M):=\{\rh\in L^1(M)\mid \rh\geq 0, \tr(\rh)=1\}.
\eeq
Also, the
 case $p=q=2$ returns the \Hs\ $L^2(M)$ which brings $M$  in standard form. 

For type III new techniques are needed, brought about by the revolution in \vna s during 1967--1975, to which the 
next section provides an introduction. The connection with the present section is given by the following corollary of Theorem \ref{TTSF} in the next section:
\begin{corollary}\label{gnssf}
 If $M\subset B(H)$ for some separable \Hs\ $H$, then $M$ is in standard form iff $H$ contains a unit vector $\Om$ that is both cyclic and separating for $M$. In particular, the \textsc{gns}-construction from any  faithful normal state brings a \vna\ into standard form.
\end{corollary}
Indeed, a normal state gives a \textsc{gns}-\rep\ with a cyclic vector, and if the state is faithful, then this vector is also separating. Conversely, a cyclic vector defines a normal state, and if the vector is also separating, then this state is faithful.
Since any \vna\ on a separable \Hs\ actually has a  faithful normal state,\footnote{See Blackadar (2006), Corollary III.2.6.9.
\S III.2.6. Conversely, Hiai (2021a), Lemma 3.9,  constructs such a vector if $M$  is already in standard form. 
 Blackadar (2006), \S III.2.6, shows that any \vna, even one initially defined on a non-separable \Hs, has an  an nfsf weight and hence
 can be brought into standard form, since the Tomita--Takesaki theory in the next section also works for  nfsf weights  instead of  normal faithful states.}
  this means that any such \vna\ can be brought into standard form. 
We just saw this for the semifinite case. In general, the fact that the mere existence of a cyclic and separating vector for $M$ yields the objects $J$ and $P$ that brings $M$ into standard form 
are provided by the  Tomita--Takesaki theory discussed in the next section. 
 \section{Tomita--Takesaki (modular) theory of \vna s}\label{TTsec}
 The new techniques alluded to at the end of the previous section are  the contents and consequences of the \emph{Tomita--Takesaki theory} (or \emph{modular theory}), widely seen as the most important development in \vna s since the work of von Neumann himself (with Murray) in the 1930s, with a huge impact on both the abstract theory and on mathematical physics.\footnote{Takesaki (2003a) is the leading textbook treatment of modular theory  by one of its creators.  Bratteli \& Robinson (1987) also contains a complete discussion. Stratila (1981) is even entirely devoted to this theory. The general texts by
 Blackadar (2006) and Sunder (2012) include efficient brief treatments. 
 Sorce (2023) is a nice introduction for physicists. For applications to mathematical physics Bratteli \& Robinson (1981) and Borchers (2000) remain very good starting points for quantum statistical mechanics and quantum field theory, respectively, as is Haag (1992) for both. }
 Some history is contained in our  Overview of the Course (\S\ref{section2}). 
In these notes the theory will be used in two ways:  one is to define a relative entropy for normal states on any \vna\ (following Araki and Uhlmann);  the other is to relate this definition to density operators and traces, especially in the type III case, which \emph{a priori} lacks a trace. This second step is based on Haagerup's noncommutative $L^p$ spaces.
We assume that the reader has reviewed the introduction to this theory in \S\ref{section2}, so that we can proceed relatively quickly, and add the necessary details. 
One complication is that both modular theory and Haagerup's theory are full of unbounded operators. See the supplement to this section for those unfamiliar with these (or in need of refreshment).\footnote{See Landsman (2017), \S\S B.13, B.21, for a brief introduction, and Reed \& Simon (1972) for a complete treatment.}

The simplest (though not the original) version of the  Tomita--Takesaki theory starts from a \vna\
$M\subset B(H)$ with  a \emph{cyclic} and \emph{separating} unit vector $\Om$ (with $H$ separable).\footnote{The theory also works from an nfsf weight $\phv$ on $M$. With $N_{\phv}\subset M$ and $H_{\phv}$ as specified after Definition \ref{defweight}, the Tomita operator $S$ is initially defined on the domain $N_{\phv}\subset H_{\phv}$ by $SA=A^*$ (as if $\Om=1_M$, which however is not in $H_{\phv}$).\label{fnTTw}}
 For example, for $M=B(H')$ acting on $H=B_2(H')$ by left-multiplication one may take $\Om=\rh^{1/2}$ for any $\rh\in D(H')$ such that $\om(A)=\Tr(\rh A)$ is a faithful state on $B(H')$ (which meant: $\om(A)=0$ for $A\geq 0$ implies $A=0$). This means that $\rh$ has a (typically unbounded) inverse. For a  $\mathrm{II}_1$ factor $M$ with normalized trace $\tr$ one may take $H=L^2(M)$ with $\Om=1_M$, as in the previous section. 
 
 Amazingly, the entire theory is built from the \emph{antilinear} (Tomita) operator $S$ on $H$, defined by
 \begin{align}
SA\Om:=A^*\Om && (A\in M); \label{NdefSTT}
\end{align}
\begin{itemize}
\item since $\Om$ is \emph{separating} for $M$ this operator is \emph{well} defined (as $A\Om=B\Om\Raw A=B$);
\item  since $\Om$ is \emph{cyclic} for $M$ it is \emph{densely} defined on an initial domain $D_0(S)=M\Om\subset H$. 
\end{itemize}
Despite the simplicity of \er{NdefSTT} the operator $S$ may be unbounded,\footnote{
In fact, $S$ is bounded iff $M$ is semifinite,  but it requires almost the entire Tomita--Takesaki theory to see this!} but it is closable and we denote its closure also by $S$. 
  The key point is that (the closure) $S$ has a unique \emph{polar decomposition}
 \begin{align}
S=J\Dl^{1/2}; && \Dl=S^*S, \label{NpolarS}
\end{align}
where $J$ is anti-unitary and $\Dl$ is self-adjoint (and possibly unbounded).\footnote{See e.g.\ Landsman (2017),
p.\ 510 for polar decompositions. Here, for a general closed operator $S$ we first \emph{define}  $\Dl^{1/2}=|S|$ (i.e., $\Dl=S^*S$), with the same domain as $S$, upon which $J$ is initially defined on $\ran(|S|)$ by $J|S|\psi:=S\psi$ and on $\ran(|S|)^{\perp}=\ker(|S|)$ by $J\psi=0$. The former is well defined since $\| |S|\psi\|=\|S\ps\|$, which makes $S$ isometric (and hence a partial isometry), and therefore bounded, so that it can be extended to the closure of
$\ran(|S|)$ by continuity.  In the case at hand,  $S$ has dense range (again because $\Om$ is cyclic for $M$ and hence for $M^*=M$), so that the closure of $J$ would be unitary if $S$ were linear, but since $S$ is antilinear $J$ is anti-unitary.
See also footnote \ref{fnau} about anti-unitarity. \label{polarfn}} 
\bex
Show that  $J^*=J\inv=J$ and hence $J^2=1_H$ (i.e., $J$ is an involution), as well as 
\begin{align}
J\Om=\Om; && \Dl\Om=\Om.
\end{align}\eex
In both of our semifinite examples, viz.\ $M=B(H'), H=B^2(H'), \Om=\rh^{1/2}$ for invertible $\rh$, and $M$ type \tto\ with $H=L^2(M,\tr)$ with $\tr(1_M)=1$ and $\Om=1_M$,  we obtain the ``generic'' form of $J$, viz.
\begin{equation}
 J\xi=\xi^*,\label{Jcan}
\end{equation}
using the notation $\xi\in H$. 
However, in the second case we find $\Dl=1_H$, whereas in the first we get 
 \begin{align}
 \Dl \xi=\rh \xi\rh\inv. \label{Dlcan}
  \end{align}
  This  comes out for \tto\ factors, too, if we replace $\Om=1_M$ by $\Om=\rh^{1/2}$ for some $\rh\in L^1(M,\tr)$, as defined after 
  \er{1319}, with $\rh\geq 0$ and $\tr(\rh)=1$. We repeat that unlike the type I case, for type \tto\ the density operator $\rh$ may not lie in $M$ (although it is always \emph{affiliated} to $M$ as originally defined on some \Hs\ $H'$, see the supplement to this section, and hence $\rh$ is ``almost in $M$''). 
  
  Since $\Dl$ is a positive self-adjoint operator we may form the unitary operators $\Dl^{it}$ for $t\in \R$ via the continuous functional calculus; if $\Dl=e^{\hat{h}}$ for some (typically unbounded) operator $\hat{h}$, we have
  \begin{equation}
\Dl^{it}=e^{it\hat{h}}.
\end{equation}
A key move in the Tomita--Takesaki theory is then the introduction of a ``time-evolution'' on $M$ via
\begin{align}
\sg_t(A):=\Dl^{it}A\Dl^{-it} && (t\in\R, \, A\in M). \label{NmgTT}
\end{align} 
This is, of course, defined in principle for all $A\in B(H)$, but  the restriction $A\in M$ is justified:\footnote{\label{AutMfn}
An \emph{automorphism} of a \vna\ $M$ is a normal linear map $\al:M\raw M$ such that $\al(B)=\al(A)\al(B)$ and $\al(A^*)=\al(A)^*$; if $\al$ is invertible, as will always be the case for the maps $\al=\sg_t$ considered in the main text and the maps $\al_g$ below, then it is automatically normal. 
For example, any unitary $U\in M$ provides such a map via $\al(A)=UAU^*$, and if $M=B(H')$ this exhausts the possibilities,  see e.g.\ Blackadar (2006), Example II.5.5.14,   but for other types this is no longer the case. Let $\mathrm{Aut}(M)$ be the group of all automorphisms of $M$, and let $G$ be a  locally compact group (just keep $G=\R$ in mind). A \emph{group action} of $G$ on $M$ is then a  homomorphism $\al:G\raw \mathrm{Aut}(M)$, $g\mapsto\al_g$ that is continuous in the following natural way: for each $g\in G$ we also have a dual map $\al_g^*$ on $M_*$ given by $\al^*_g(\om)(A)=\om(\al_{g\inv}(A))$, and we require the map $G\x M_*$ given by $(g,\om)\mapsto \al^*_g\om$ to be continuous (with respect to the norm topology on $M_*$). See Blackadar (2006), Theorem
 III.3.2.2, for equivalent definitions of this continuity condition. 
}
\begin{theorem}\label{TTT1}
Let $M\subset B(H)$ have a cyclic and separating (unit) vector $\Om\in M$, with associated 
time-evolution $\sg_t$ as in \er{NmgTT}.
For any $t\in\R$ and $A\in M$ we have $\sg_t(A)\in M$, so that $t\mapsto \sg_t$ defines a group action of $\R$  on $M$ (where $\R$ is seen as an abelian group under addition in the usual way).
\end{theorem}
This is one of the three main theorems of the Tomita--Takesaki theory. It may be unexpected, since $\Dl^{it}$ does not lie in $M$, as we see already
for  $M=B(H')$ from \er{Dlcan}. To analyze this further,  we put $\rh=e^h$ for some (typically  unbounded) self-adjoint operator $h$ on $H'$, with the right properties to guarantee that $\rh\in D(H)$ (for which $h$ should at least have discrete spectrum with finite multiplicities). Using our old notation $L(h)\xi=h\xi$ and $R(h)\xi=\xi h$, eq.\ \er{Dlcan} then yields
 \begin{equation}
\hat{h}=L(h)-R(h).
\end{equation}
Although this confirms the fact that $\Dl$ combines operators in $M$ and its commutant $M'$, it also shows why in this case
$\sg_t(A)\in M$: because $R(h)$ commutes with  $A\in M=L(B(H'))$, we have 
\beq
\sg_t(L(A))=L(e^{ith}Ae^{-ith})\in M. \label{19999}
\eeq
If we simplify  the notation such that $L(A)$ for $A\in B(H')$ is written as $A\in B(H)$, eq.\ \er{19999} means that
for type I the automorphism $\sg_t$ is \emph{inner}, in the sense that there is a unitary $U_t\in M$ such that
\begin{equation}
\sg_t(A)=U_t AU_t^*.
\end{equation}

This is true for all semifinite \vna s (see Proposition \ref{innersemifinite}), and it is in fact a criterion for $M$ being type III that $\sg_t$ is \emph{outer} (i.e., not inner). The key distinction is not between $\Dl$ and hence $\sg_t$ being \emph{trivial} or not (we already saw that this can go either way), but  between $\sg_t$ being \emph{inner} or not. Either way, the operators $\Dl^{it}$ (or  the associated maps $\sg_t$) form the so-called \emph{modular group} of $M$, defined, crucially, with respect to the initial cyclic and separating vector $\Om$, or more generally, the initial normal faithful state (or even more generally, the nfsf weight) $\om$ that starts the construction.\footnote{The name `modular' comes from some of Tomita's examples, which involved locally compact groups that 
are \emph{not} unimodular (i.e., whose left-invariant and right-invariant Haar measures are different;) these are related by a `modular function' that traditionally was also written as $\Dl$ and recurred in Tomita's $\Dl$. See Takesaki (2003a), \S VII.3.}
 To indicate this dependence, one often writes $\sg_t^{\om}$ instead of $\sg_t$.

The second main theorem of the Tomita--Takesaki theory states the following:
\begin{theorem} \label{TTT2}
Let $M\subset B(H)$ have a cyclic and separating (unit) vector $\Om\in M$, with associated `modular conjugation'$J$ defined via \er{NdefSTT} and \er{NpolarS}. Then
\begin{equation}
M'=JMJ. \label{JMJ2}
\end{equation}
\end{theorem}
As already noted after the special case \er{JMJ1}, one  has a choice between an antilinear isomorphism (of algebras) $M'\cong M$ via $A\mapsto j(A):=JAJ$ and a linear anti-isomorphism $j^*(A):=JA^*J$. We have also seen this theorem in action in both the type I and type II cases; these cases were known  to Tomita and Takesaki, and so the main thrust of their theorem is that it holds also for type III. 

The third theorem of the theory, which should be attributed to Takesaki and Winnink,\footnote{This third theorem first  appeared in the unpublished PhD thesis of Marinus Winnink (1936–2023).} concerns the \emph{KMS condition}, see \er{KMS1}, which turns out to characterize the `modular flow' $\sg_t$:
\begin{theorem}   \label{TTTW} Let $M\subset B(H)$ have a cyclic and separating unit vector $\Om\in M$, with associated 
time-evolution $\sg_t$ on $M$, cf.\ Theorem \ref{TTT1}. Then
for any $A,B\in M$ the two-point function
\beq
f(z):=\om(A\sg_z(B))\eeq
has an analytic extension from $z\in\R$ to the strip $-1<\mathrm{Im}(z)<0$ with continuous boundary values 
\begin{align}
f(t)=\om(A\sg_t(B)); && f(t-i)=\om(\sg_t(B)A) && (t\in\R), \label{KMSAB}
\end{align}
where $\om(A)=\la\Om,A\Om\ra$.
 Conversely, if some time-evolution $\sg_t$ on $M$ satisfies  the above (KMS) condition, 
then $\sg_t$ must coincide with the modular group \er{mgTT} defined by $\Om$ as in Theorem \ref{TTT1}.
\end{theorem}
Although, as for the previous two theorems, we refrain from even beginning to give a proof, we do wish to present the key computation that leads to \er{KMSAB}. Writing $B(t)$ for $\sg_t(B)$, we have:
\begin{align}
f(t)&= \om(AB(t))=\la\Om, A\Dl^{it}B \Dl^{-it}\Om\ra=\la\Om, \Dl^{-it/2} A \Dl^{it/2}\Dl^{it/2}B \Dl^{-it/2}\Om\ra\nn \\ &=\la\Om, A(-t/2)B(t/2)\Om\ra, \label{KMShalf}
\end{align}
so that, analytically continuing \er{NmgTT}  to imaginary values within the strip of analyticity of $f$,
\begin{align}
f(t-i)&= \la\Om, A(-t/2)\Dl^{1/2}\Dl^{1/2}B(t/2)\Om\ra=\la \Dl^{1/2}A(-t/2)^*\Om, \Dl^{1/2}B(t/2)\Om\ra\nn \\
&=\ovl{\la J\Dl^{1/2}A(-t/2)^*\Om, J\Dl^{1/2}B(t/2)\Om\ra}= \ovl{\la SA(-t/2)^*\Om, SB(t/2)\Om\ra}\nn \\
&=\ovl{\la A(-t/2)\Om, B(t/2)^*\Om\ra}=\la B(t/2)^*\Om, A(-t/2)\Om\ra= \la\Om, B(t/2)A(-t/2)\Om\ra\nn \\
&=\la \Om, B(t)A,\Om\ra=\om(B(t)A).
\end{align}

The third main theorem of the theory  is that the mere existence of a cyclic and separating vector $\Om$ for $M$ is enough to show that $M\subset B(H)$ is in standard form, cf.\ Definition \ref{defstandardform}:\footnote{For more details and a proof see Hiai (2021a), Theorem 3.2, and basically his entire Chapter 3..}
\begin{theorem}\label{TTSF}
 Let $M\subset B(H)$ have a cyclic and separating unit vector $\Om\in M$, with associated modular conjugation $J$, cf.\ \er{NdefSTT} and \er{NpolarS}. Then $M$is in standard form relative to $J$ and the cone
 \begin{equation}
P=\{AJA\Om \mid A\in M\}^-=(\Dl^{1/4}M^+\Om)^-\equiv \{\Dl^{1/4}A^*A \Om \mid A\in M\}^-. \label{1412}
\end{equation}
\end{theorem}
Here the closure is taken in $H$ (with norm topology). To check the second equality we trust that $M$ contains a dense set of ``analytic'' elements for which the \emph{operator} $\sg_t(A)$, and not merely the  two-point functions as in the KMS condition, can be analytically continued to the strip  $-1<\mathrm{Im}(z)<0$,
so that $\sg_{-i/4}(A)$ is defined and  lies in $M$ by Theorem \ref{TTT1}. Since this gives the same closure, we may then replace $A$ in the first expression for $P$ by 
\beq
\sg_{-i/4}(A)=\Dl^{1/4}A\Dl^{-1/4},
\eeq
 so as to obtain
\begin{align}
\sg_{-i/4}(A)J \sg_{-i/4}(A)\Om&=\Dl^{1/4}A\Dl^{-1/4}J \Dl^{1/4}A\Dl^{-1/4}\Om=\Dl^{1/4}A\Dl^{-1/4}J \Dl^{1/2}\Dl^{-1/4}A\Dl^{1/4}\Om\nn \\ &= \Dl^{1/4}A\Dl^{-1/4}S\Dl^{-1/4}A\Dl^{1/4}\Om= \Dl^{1/4}A\Dl^{-1/4}\Dl^{1/4}A^*\Dl^{-1/4}\Om\nn \\
&= \sg_{-i/4}(AA^*)\Om= \Dl^{1/4}AA^* \Dl^{-1/4}\Om= \Dl^{1/4}AA^*\Om.
\end{align}
 In our example $M=B(H')$ acting on $H=B_2(H')$ from the left, with $\Om=\rh^{1/2}$ for some invertible $\rh\in D(H)$ as usual, so that \er{Jcan} and \er{Dlcan} hold, the first formula for $P$ returns \er{PtypeI}, viz.
 \begin{align}
 P&=\{ AJA\rh^{1/2},\, A\in B(H)\}^-=\{ A\rh^{1/2}A^*,\, A\in B(H)\}^- =\{ A\rh^{1/4}(A\rh^{1/4})^*,\,  A\in B(H)\}^-
\nn \\ &=\{ AA^*, \, A\in B_4(H)\}^-=
 B_2(H)^+,
 \end{align}
since for $\rh\in B_1(H)$ we have $\rh^{1/4}\in B_4(H)$ and hence $A\rh^{1/4}\in B_4(H)$, likewise $(A\rh^{1/4})^*\in B_4(H)$,
and the product of two operators in $B_4(H)$ is in $B_2(H)$, and is obviously positive.

Our final theorem is an essential part of the Tomita--Takesaki theory, although it is due to Connes and was proved a few years after the establishment  of the theory in 1970.
To state the simplest version of his result, assume that $H$ contains two  unit vectors $\Om_1$ and $\Om_2$, each of which is cyclic and separating for ${M}$. We write $\sg_t^{(i)}$ for the modular group derived from $\Om_i$, $i=1,2$.
\begin{theorem}\label{CCRNT}
There is a family $U_t$ of unitary operators in ${M}$ ($t\in\mathbb{R}$), such that for all $A\in M$,
\begin{align}
\sg_t^{(1)}(A)&=U_t\sg_t^{(2)}(A)U_t^*; \label{cc1}\\
U_{t+s}&=U_s\sg_s^{(2)}(U_t).  \label{cc2}
\end{align}
\end{theorem}
These unitaries are often written as 
\begin{align}
U_t=(D\Om_1:D\Om_2)_t && \mbox{or} &&  U_t=(D\om_1:D\om_2)_t.\label{DDnotation}
\end{align} For example, for $B(H')$ acting on $H=B_2(H')$ from the left, with $\Om_i=\rh_i^{1/2}$, it is easy to show from \er{Dlcan} and \er{NmgTT} that
\begin{equation}
(D\om_1:D\om_2)_t=\rh_1^{it}\rh_2^{-it}.
\end{equation}
 The inner automorphisms of $M$ form a normal subgroup $\mathrm{Inn}(M)$ of the group $\mathrm{Aut}(M)$ of all automorphisms of $M$, with quotient $\mathrm{Out}(M)=\mathrm{Aut}(M)/\mathrm{Inn}(M)$.
Theorem \ref{CCRNT} shows that the image of the modular group $\sg_t(\R)$  in  $\mathrm{Out}(M)$ under the canonical projection $\mathrm{Aut}(M)\raw \mathrm{Out}(M)$ is independent of $\Om$, so that invariants of this image will be invariants of $M$. This insight eventually  led Connes to his classification of hyperfinite type III factors into subtypes $\mathrm{III}_{\lm}$, $\lm\in [0,1]$.
\smallskip

\emph{Proof.}
 The proof of Theorem \ref{CCRNT} is Connes's favourite (as he declared in an interview),\footnote{See Cornelissen, Landsman, \& van Suijlekom (2010).}  so we sketch it in some detail. It is based on the following idea. Extend ${M}$ to $\mathrm{Mat}_2({M})$, i.e., the \vna\ of $2\times 2$ matrices with entries in ${M}$, and let $\mathrm{Mat}_2({M})$ act on $H_2=H\oplus H$ 
  in the obvious way. 
  Subsequently, let  $\mathrm{Mat}_2({M})$ act on
 $H_{(4)}=H\oplus{H}\oplus{H}\oplus{H}=H_2\oplus{H}_2$ by 
 \begin{align}
 \mathbf{A}=\left(
\begin{array}{cc}
  A_{11} & A_{12}     \\
  A_{21} &    A_{22}
\end{array}
\right)\in \mathrm{Mat}_2({M}) &&\leadsto&&  \mathbf{A}_{(4)}:= \left(
\begin{array}{cc}
 \mathbf{A}&   \mathbf{0}    \\
   \mathbf{0} &   \mathbf{A}
\end{array}
\right)\in B(H_4),
\end{align}
 The vector $\Om_{(4)}:=(\Om_1,0,0,\Om_2)\in{H}_4$ is then cyclic and separating for $\mathrm{Mat}_2({M})$,
 since 
 \begin{equation}
\mathbf{A}_{(4)}\Om_{(4)}=\left(
\begin{array}{c}
 a_{11}\Om_1   \\
a_{21}\Om_1\\
a_{12}\Om_2\\
a_{22}\Om_2
\end{array}
\right).
\end{equation}
The Tomita operator $S_{(4)}$ on $H_{(4)}$ with respect to $\mathrm{Mat}_2({M})$ and $\Om_{(4)}$ is then given by
\begin{align}
S_{(4)}=\left(
\begin{array}{cccc}
S_{11} & 0&0&0     \\
 0 & 0&  S_{12}&  0\\
 0& S_{21} & 0 & 0\\
 0&0&0& S_{22},
\end{array}
\right)
\end{align}
in terms of the \emph{relative} Tomita operators $S_{ij}$ on $H$, $i=1,2,$ defined by (the closures of)
\begin{equation}
 S_{ij}A\Om_j=A^*\Om_i, \label{relTO}
\end{equation}
so that in fact $S_{11}=S_1$ and $S_{22}=S_2$. 
The corresponding modular operator is then given by 
\begin{align}
\Delta_{(4)}=S_{(4)}^*S_{(4)}=\mathrm{diag}(\Delta_{11},\Delta_{21},\Delta_{12},\Delta_{22}); && \Dl_{ij}=S^*_{ij}S_{ij},
\end{align}
 so that in particular  $\Dl_{11}=\Dl_1$ and $\Dl_{22}=\Dl_2$. This yields
\begin{align}\label{Conneseq}
{\Delta_{(4)}^{it}\left(
\begin{array}{cc}
 \mathbf{A}&   \mathbf{0}    \\
   \mathbf{0} &   \mathbf{A}
\end{array}
\right)\Delta_{(4)}^{-it}}=\left(
\begin{array}{cccc}
 \Delta_1^{it} A_{11} \Delta_1^{-it} &  \Delta_1^{it}A_{12} \Delta_{21}^{-it}  &0&0   \\
   \Delta_{21}^{it} A_{21}    \Delta_1^{-it}&   \Delta_{21}^{it}\ A_{22}\Delta_{21}^{-it}& 0& 0\\
0&0&  \Delta_{12}^{it} A_{11} \Delta_{12}^{-it} &  \Delta_{12}^{it}A_{12} \Delta_2^{-it}  \\
0&0  &  \Delta_2^{it} A_{21}    \Delta_{12}^{-it}&   \Delta_2^{it}\ A_{22}\Delta_2^{-it}
\end{array}
\right)
\end{align}
But by Theorem \ref{TTT1}, the right-hand side of \er{Conneseq} must be in $\mathrm{Mat}_2({M})$ and hence must take the form $\mathrm{diag}(\mathbf{B},\mathbf{B})$ for some $\mathbf{B}\in \mathrm{Mat}_2({M})$. So both blocks must coincide, from which we conclude:
\begin{enumerate}
\item Taking $A_{12}=1_M$ gives $\Delta_1^{it} \Delta_{21}^{-it}=\Delta_{12}^{it} \Delta_2^{-it}$, either of which is our $U_t$. To be specific, take
\begin{equation}
U_t\equiv (D\Om_1:D\Om_2)_t:=\Delta_{12}^{it} \Delta_2^{-it}, \label{1430U}
\end{equation}
which lies in $M$ because all nonzero elements of  \er{Conneseq} lie in $M$. This also
gives \er{cc2}.
\item Taking $A_{11}=A$ gives $\Delta_1^{it} A \Delta_1^{-it} = \Delta_{12}^{it} A\Delta_{12}^{-it}$. Using the previous point, we recover \er{cc1}:
\begin{align*}
U_t\sg_t^{(2)}(A)U_t^*=\Delta_{12}^{it} \Delta_2^{-it}\Dl_2^{it}A\Dl_2^{-it}\Delta_2^{it}\Delta_{12}^{-it} =\Delta_{12}^{it} A\Delta_{12}^{-it} =\Delta_1^{it} A \Delta_1^{-it} =\sg_t^{(1)}(A).  \tag*{$\Box$}
\end{align*}
\end{enumerate}
The relative  Tomita operators $\Dl_{ij}$ will be crucial for the definition of relative quantum entropy.
\subsection*{Supplement: Unbounded operators on \Hs}\addcontentsline{toc}{subsection}{Supplement: Unbounded operators on \Hs}
Tomita's operator $S$ in \er{NdefSTT}, around which the entire  theory revolves,  is potentially unbounded. So are  the position operator $\hat{x}$ and the momentum operator $\hat{p}=-id/dx$ on $L^2(\R)$ of elementary \qm.  The key features of an unbounded operator $A$ on a \Hs\ $H$ are:
\begin{enumerate}
\item $A$ is initially defined merely on a (proper) \emph{subspace} $D_0(A)$ of $H$ in order for its ``formula'' to make sense in the first place; we always assume this initial \emph{domain} of $A$ to be dense in $H$.
\item $A$ cannot be extended to all of $H$ by continuity because it is not continuous.
Equivalently,\footnote{Recall that a linear operator on a Hilbert space (or even a Banach space) is continuous iff it is bounded.}
\beq
\sup\{\|A\ps\|, \ps\in D_0(A), \|\psi\|\leq 1\}=\infty.
\eeq
\end{enumerate}
\bex
Check this for both $\hat{x}$ and $\hat{p}$ by taking $D_0(\hat{x})=D_0(\hat{p})=\cci(\R)$, for example. 
\eex
Yet a good (spectral) theory of unbounded operators  still emerges if $D_0A)$ can be enlarged to some domain $D(A)$  on which $A$ is still defined (which may coincide with $D_0(A)$ if one was prescient), coinciding with its previous values on $D_0(A)\subset D(A)$, and is
\emph{closed}, which means that its graph 
\beq
G(A):=\{(\ps, A\ps)\in H\oplus H, \ps\in D(A)\}
\eeq
 is closed in the direct sum \Hs\ $H\oplus H$, whose elements $(\psi,\phv)$, $\psi,\phv\in H$, have norm 
 \beq
 \|(\psi,\phv)\|=\|\psi\|+\|\phv\|.\eeq
\bex Show that $A$ is closed iff the properties $\ps_n\raw\ps$ \emph{and} $A\ps_n\raw \phv\in H$ for some sequence $(\ps_n)$ in $D(A)$ imply $\ps\in D(A)$ and $A\psi=\phv$. Check that a bounded operator is closed.
\eex
For example, $\hat{x}$ is not closed on $D_0(\hat{x})=\cci(\R)$, but it is closed on
 \beq
 D(\hat{x})=\{\psi\in L^2(\R)\mid \hat{x}\ps\in L^2(\R)\}.
 \eeq
  More generally,\footnote{Taking the Fourier transform, it follows that if we define the derivative $\psi'$ in a distributional sense, then  $\hat{p}$ is not closed on $D_0(\hat{p})=\cci(\R)$ either, but it is closed on the larger domain 
$D(\hat{p})=\{\psi\in L^2(\R)\mid \ps'\in L^2(\R)\}$.}
a multiplication operator $\hat{f}$ on $H=L^2(X,\mu)$ is closed on its (maximal) domain
\beq
 D(\hat{f})=\{\psi\in L^2(X,\mu)\mid \hat{f}\psi\in L^2(X,\mu). \label{maxhatf} 
 \eeq
The minimal requirement for finding such a closed extension of an initial attempt 
\beq
A_0: D_0(A)\raw H, \label{A0}
\eeq
where for clarity we have changed the name of the initial operator from $A$ to $A_0$, 
 is that the closure of $G(A_0)$ in $H\oplus H$  with respect to its initial domain $D_0(A_0)$ is  the graph of an operator 
 \beq
 A:D(A)\raw H,
 \eeq
   which is then trivially closed.
 In that case $A_0$ is called \emph{closable}, and $A\equiv A_0^-$ is called its \emph{closure}. 
\bex
Show that closability of $A_0$ 
 is equivalent to the following property:  if $\ps_n\raw 0$ for some  sequence $(\ps_n)$ in $D_0(A_0)$, \emph{and} $(A_0\ps_n)$ also converges, then $A_0\ps_n\raw 0$. 
 \eex
 There may not be a \emph{unique} closed extension of $A_0$, if it is closable: but this will not concern us.

 Our position operator  $\hat{x}$ defined on  $D(\hat{x})$, and more generally real-valued multiplication operators $\hat{f}$ defined on the domain \er{maxhatf}, are not only closed; they are \emph{self-adjoint}. For a bounded operator $A$ this would simply mean that if we define the adjoint $A^*$  by the property  
 \beq
 \la A^*\ps,\phv\ra=\la\ps, A\phv\ra, \label{familiarformula}
 \eeq
  for all $\ps,\phv\in H$, then $A$ is self-adjoint if $A^*=A$. For an unbounded operator $A: D(A)\raw H$ we should first define $D(A^*)$ so that \er{familiarformula} can even be \emph{stated}; thus $D(A^*)$ consists of all $\ps\in H$ for which the functional $\phv\mapsto\la \psi, A\phv\ra$ is bounded on $D(A)$, so that (by the Riesz--Fischer theorem) there exists a unique $\eta\in H$ such that $\la \psi, A\phv\ra=\la\eta,\phv\ra$. Then $A^*$ is defined on $D(A^*)$ by 
   \begin{align}
 \la \psi, A\phv\ra=\la\eta,\phv\ra && \Raw &&  A^*\psi:=\eta,
   \end{align}
    so that one retrieves the formula  \er{familiarformula}, but now for all $\phv\in D(A)$ and $\psi\in D(A^*)$.  If one has \emph{both}
\begin{align}
D(A^*)=D(A); && A^*\psi=A\psi && (\psi\in D(A^*)=D(A)), \label{citeHHW}
\end{align}
then $A^*$ is called \emph{self-adjoint}, written $A^*=A$.
A self-adjoint operator is necessarily closed.

\emph{Positivity} of $A$, written $A\geq 0$, requires self-adjointness, and is defined in practically the same way as in the bounded case, namely by $\la\ps, A\ps\ra\geq 0$, now for all $\psi\in D(A)$. This is equivalent to:
\begin{align}
 A^*=A; &&  \sg(A)\subseteq[0,\infty),
 \end{align}
  where the spectrum $\sg(A)$ is  defined, as usual, as the complement in $\C$ of the set of all $\lm\in\C$ for which $A-\lm\cdot 1_H$ has a \emph{bounded} inverse. Here is an interesting and important example:
\bex \label{inverseunbounded}  
\begin{enumerate}
\item 
Let $\rh$ be a  positive bounded operator with $\ker(\rh)=\{0\}$. Show that the (naive) inverse $\rh\inv$ of 
$\rh$, defined on $D(\rh\inv)=\ran(\rh)$ by $\rh\inv (\rh\psi)=\ps$, is self-adjoint.
\item Now drop the condition $\ker(\rh)=\{0\}$. The \emph{generalized inverse} of 
 \beq
 \rh=\int_0^{\infty}dE_{\lm}\cdot\lm
 \eeq
  is  
  \beq \rh\inv=\mathrm{s}-\lim_{\varep\raw 0}\int_{\varep}^{\infty}dE_{\lm}\cdot\lm\inv,
  \eeq
   defined on  all $\ps\in H$ where the strong operator limit exists.\footnote{This definition implies that $\rh\inv\xi=0$ for all $\xi\in\ker(\rh)$, and that the previous definition applies on $\ker(\rh)^{\perp}$.
   } Show that $\rh\inv$ is self-adjoint.
\end{enumerate}
\eex

In the remainder of these notes we will also need to concept of a closed  operator $A$ (and especially a positive one) \emph{affiliated} to a \vna\ $N\subseteq B(H)$, written $A\eta N$ (this notation $A\eta N$ is also meant to imply that $A$ is closed).    This is defined in two steps.
\begin{definition}
Let $A:D(A)\raw H$ be a closed (possibly unbounded) operator (where $D(A)=A$ iff $A$ is bounded) and $N\subseteq B(H)$ a \vna.
\begin{enumerate}
\item  If $A$ is self-adjoint we say that it is affiliated to $N$  if
all its spectral projections 
\beq
E(\Dl)=1_{\Dl}(A)
\eeq
 lie in $N$, where $\Dl\in\sg(A)\subseteq\R$ is measurable.
\item For general $A$ one uses the (unique) polar decomposition $A=V|A|$, where $V$ is a (necessarily bounded) partial isometry and $|A|\geq 0$ ``carries the possible unboundedness of $A$''; we then define $A\eta N$ iff $V\in N$ and $|A|\eta N$ in the  sense of the previous point (since $|A|$ is self-adjoint).
 \end{enumerate}\end{definition}
\bex \begin{enumerate}
\item Show (from this definition) that if $A$ is bounded, then  $A\eta N$ iff $A\in N$.
\item
Show that this definition is equivalent to the following one: 
 $A\eta N$ iff 
 \begin{align}
 B'A\subseteq AB'&& (B\in N'), \label{affcomm}
 \end{align} where the notation means that if $\ps\in D(A)$, then $B'\psi\in D(A)$ and $AB'\psi=BA\psi$. 
 \end{enumerate}
 \eex
  Roughly speaking, eq.\ \er{affcomm} means that $A$ commutes with all $B'\in N'$, so that if $A$ were bounded, then $A\in N''=N$; if $A$ is unbounded, this condition suggests that $A$ is at least ``close'' to $N$. 
 
 We can freely add and multiply \emph{bounded} operators; this is the basis of the theory of operator algebras. In the unbounded case, if $A:D(A)\raw H$ and $B:B(H)\raw H$ are closed (always assumed with dense domains), then the best we can do is the following:
 \begin{itemize}
\item define $A+B$ as the closure of the operator defined on $D_0(A+B):=D(A)\cap D(B)$ by 
\beq
(A+B)\psi=A\ps+ B\ps.\label{AplusB} \eeq
\item  likewise define $AB$  on $D_0(AB):=\{\psi\in D(B)\mid B\psi\in D(A)\}$ as the closure of the operator 
\beq
(AB)\ps=A(B\psi) .\label{AkeerB} \eeq
\end{itemize}
 But this only works well if these domains are densely defined and the operators in question are closable! Already the first clause may go wildly wrong:  in principle, even if $D(A)$ and $D(B)$ are dense in $H$, the above initial domains $D_0(A+B)$ and $D_0(AB)$ may only consist of the zero vector.
But von Neumann saw that if $N$ is a $\mathrm{II}_1$ factor, sums and products and adjoints of operators affiliated to $N$, defined in the above way, are densely defined and closable, so that \emph{the set $\ovl{N}$ of all closed operators affiliated to $N$  is an algebra with unit and involution} (which  extends $N$). 
 For \emph{any} semifinite \vna\ with an nfsf trace $\ta$, this  holds for  the possibly smaller space $\til{N}$ of all operators  $A\eta N$ that are
\emph{$\ta$-measurable}.\footnote{For $\ta$-measurable operators see Takesaki (2003a), \S IX.2, and Terp (1981), based on work of Haagerup, which is not easy to find (although the author has a copy).  Hiai (2021a), Chapter 4, quite closely follows Terp's presentation.} With $|A|=\int_0^{\infty} dE_{\lm}\cdot\lm$, this means that
\beq
\lim_{\lm\raw\infty} \ta(E([0,\lm])^{\perp})=0, \label{144}
\eeq
which condition
 can be shown to be equivalent to the existence of some $\lm>0$ for which 
\beq
\ta(E([0,\lm])^{\perp})<\infty. \label{tameasurable}
\eeq  
Thus 
\emph{the space $\til{N}$ of $\ta$-measurable operators affiliated with a semifinite \vna\ with a trace $\ta$ forms an algebra (with involution and unit) under natural operations \er{AplusB} - \er{AkeerB}}. 
\bex
\begin{enumerate}
\item Show that every operator affiliated with a $\mathrm{II}_1$ factor is $\tr$-measurable.
\item Show that an operator affiliated with $B(H)$ is \emph{Tr}-measurable iff it is bounded.
\item Show that an operator $A$ is affiliated with $M=L^{\infty} (X,\mu)\subset B(L^2(X,\mu))$ iff 
$A=\hat{f}$ for some measurable function $f:X\raw\C$  (in the usual sense), and that $\hat{f}$ is $\mu$-measurable 
iff $f$ is bounded on the complement $X\backslash A$ for some subset $A\subset X$ with $\mu(A)<\infty$. 
\end{enumerate}
\eex
 \section{Entropy of normal states on \vna s}
Let $M\subset B(H)$ be in standard form, which for the moment just means that $H$ has a cyclic and separating unit vector $\Om$;  as we saw in Theorem \ref{TTSF}, this gives us all. In particular, Theorem \ref{SFkey2} applies, so that for any two normal states $\om_1,\om_2$ on $M$ we can find unique unit vectors $\Om_1, \Om_2$ in $P$ such that 
$\om_i(A)=\la\Om_i, A\Om_i\ra$ for $i=1,2$. We first assume that both $\Om_1$ and $\Om_2$ are  cyclic and separating for $M$; equivalently, that $\om_1$ and $\om_2$ are faithful normal states on $M$.
To define a relative entropy, recall  the \emph{relative modular operators} $S_{12}$ and $S_{21}$ on $H$ from the proof of Theorem \ref{CCRNT}:
 \begin{align}
S_{12}A\Om_2:=A^*\Om_1; && S_{21}A\Om_1:=A^*\Om_2 && (A\in M), \label{relativeSTT}
\end{align}
with associated polar decompositions as in \er{polarS}, that is,
\begin{align}
S_{12}&=J_{12}\Dl_{12}^{1/2}=J\Dl_{12}^{1/2}; &&S_{21}=J_{21}\Dl_{21}^{1/2}=J\Dl_{21}^{1/2};\label{152N}\\
\Dl_{12}&=S_{12}^*S_{12}; && \Dl_{21}=S_{21}^*S_{21},\label{153N}
\end{align}
where $J$ is the modular conjugation for $\Om$, and we have anticipated the fact that 
\beq
J_{12}=J_{21}=J,
\eeq
 whenever $\Om_1\in P$ and $\Om_2\in P$. This is a consequence of either one of \er{137} or \er{1412}, which implies that all cyclic and and separating vectors in $P$ define the same cone $P$. More formally:\footnote{See Hiai (2021a), Lemma 3.9 and Proposition 3.10. In addition, $\Om'\in P$ is cyclic for $M$ iff it is separating for $M$.} 
\begin{proposition}
If $M\subset B(H)$ has a cyclic and separating vector $\Om$ with associated self-dual cone $P$ (see Definitions \ref{Defsdc} and \ref{defstandardform}) and modular conjugation $J$, 
and $\Om'\in P$ is also  cyclic and separating  for $M$, 
with associated self-dual cone $P'$ and modular conjugation $J'$,
then 
\begin{align}
J'=J; && P'=P.
\end{align}  
\end{proposition}
\bex
Show that \er{relativeSTT} implies that $S_{12}=S_{21}\inv$, and that consequently
\begin{equation}
\Dl_{12}=J\Dl_{21}\inv J. \label{153A}
\end{equation}\eex
If $M=B(H')$ and $H=B_2(H')$, with $\Om_i=\rh_i^{1/2}$ ($i=1,2$) for invertible $\rh_i\in D(H')$, we find
\begin{align}
J_{12}=J_{21}=J; &&
\Dl_{12}\xi=\rh_1 \xi\rh_2\inv;  && \Dl_{21}\xi=\rh_2\xi\rh_1\inv, \label{relativeDeltaI}
\end{align}
where $J$ is given by \er{JtypeI}. The last two equalities then obviously confirm \er{153A}.
\begin{definition}
For  normal faithful states (for the moment), \emph{Araki's relative entropy} is given by
 \begin{equation}
 S(\om_1,\om_2):=\la \Om_2,\Dl_{12}\log \Dl_{12}\Om_2\ra.
\label{Araki1}
\end{equation}\end{definition}
One also has a  somewhat simpler expression that turns out to be equal to \er{Araki1}, viz.\footnote{Araki's entropy $S(\om_1,\om_2)$ is a special case of a \emph{standard quantum $f$-divergence}, defined as follows (where we restrict ourselves to faithful states). In  the same setting as in the main text, i.e., a \vna\  $M\subset B(H)$ in standard form, 
let $f:(0,\infty)\raw \R$ be a convex function, and let $\om_1$ and $\om_2$ be  faithful normal states on $M$ implemented by unit vectors 
$\Om_1\in P$ and $\Om_2\in P$, respectively. Then  $S_f(\om_1,\om_2):=\la \Om_2,f(\Dl_{12})\Om_2\ra$, so that \er{Araki1} arises as the special case $f(x)=x\log x$.
The equivalence between \er{Araki1} and \er{Araki2} is then a special case of the equality
 $S_f(\om_1,\om_2)= S_{\til{f}}(\om_2,\om_1)$, that is, $\la \Om_2,f(\Dl_{12})\Om_2\ra= \la \Om_1,\til{f}(\Dl_{21})\Om_1\ra$,
 where $\til{f}(x):=xf(x\inv)$. The easy proof is again based on \er{153A}. 
See Hiai (2018) as well as Hiai( 2021b), Chapter 2, also for the definition of $S_f$ for general normal states.}
\beq
S(\om_1,\om_2)=-\la\Om_1, \log\Dl_{21}\Om_1\ra. \label{Araki2}
\eeq
Indeed, using the first part of \er{relativeSTT} both with $A=1_H$ and $A=\log\Dl_{12}$ as well as \er{153A} we compute
\begin{align}
\la \Om_2,\Dl_{12}\log \Dl_{12}\Om_2\ra&= \la \Dl_{12}^{1/2}\Om_2,\Dl_{12}^{1/2}\log \Dl_{12}\Om_2\ra=
\ovl{\la J\Dl_{12}^{1/2}\Om_2,J\Dl_{12}^{1/2}\log \Dl_{12}\Om_2\ra}\nn \\
&=\ovl{\la S_{12}\Om_2,S_{12}\log \Dl_{12}\Om_2\ra}=\ovl{\la \Om_1,\log \Dl_{12}\Om_1\ra}\nn \\ &=\ovl{\la J\Om_1,\log \Dl_{12}J\Om_1\ra}=\la \Om_1,J^*\log \Dl_{12}J\Om_1\ra=\la \Om_1,J\log \Dl_{12}J\Om_1\ra\nn \\ &=
\la \Om_1,\log \Dl_{12}\inv\Om_1\ra=-\la \Om_1,\log \Dl_{12}\Om_1\ra,
\end{align}
where we also used the property $J\xi=\xi$ for all $\xi\in P$ and hence $J\Om_1=\Om_1$, cf.\ Definiton
\ref{defstandardform}.
A third expression for the relative quantum entropy, due to Uhlmann, is, cf.\ \er{DDnotation} and \er{1430U},
\begin{equation}
S(\om_1,\om_2)=i\frac{d}{dt} \la\Om_1,(D\Om_2:D\Om_1)_t\Om_2\ra_{|t=0}=i\frac{d}{dt} \la\Om_2, \Dl_{21}^{it}\Dl_1^{-it}\Om_1\ra_{|t=0}. \label{Uhl}
\end{equation}
\bex Show that $S(\om_1,\om_2)$ in \er{Uhl} equals $S(\om_1,\om_2)$ in  \er{Araki2}.
\eex

Finally, an Araki-style relative R\'{e}nyi entropy is defined, initially for $0<t<1$ as usual, by
\begin{equation}
 R_t(\om_1,\om_2):=\frac{1}{t-1}\log \la \Om_2,\Dl^t_{12}\Om_2\ra, \label{ArakiR}
\end{equation}
cf.\ \er{RRE}.  As in \er{LRE}, for $t=0$ and $t=1$ we then have, either by definition or by continuity,\footnote{The second case it not so easy. Let $\Dl_{12}=\int_0^{\infty} dE_{\lm}\cdot\lm$ with $\int_0^{\infty} dE_{\lm}=1_H$
be the spectral resolution of $\Dl_{21}$, so that
$F(t):=\la\Om_2,\Dl_{12}^t\Om_2\ra=\int_0^{\infty} dE_2(\lm) \cdot\lm^t$ with $dE_2(\lm)=d(\|E_{\lm}\Om_2\|^2)$, i.e., for $A\subset \R^+$ we have $E_2(A)=\| 1_A(\Dl_{12})\Om_2\|$. Then $\frac{F(t)-F(1)}{t-1}=\int_0^{\infty} dE_2(\lm) \cdot\ \frac{\lm^t-\lm}{t-1}$, whose limit as $t\uparrow 1$ equals $\int_0^{\infty} dE_2(\lm) \cdot \lm\log\lm=\la\Om_2,\Dl_{12}\log \Dl_{12}\Om_2\ra=S(\om_1,\om_2)$.
Since $F(1)=1$, we also have $\lim_{t\uparrow 1} \frac{\log F(t)}{t-1}= \lim_{t\uparrow 1} \frac{\log F(t)-\log F(1)}{t-1}=\frac{1}{F(1)} \lim_{t\uparrow 1} \frac{F(t)- F(1)}{t-1}= S(\om_1,\om_2)$.}
\begin{align}
R_0(\om_1,\om_2)=0; &&
R_1(\om_1,\om_2)=S(\om_1,\om_2).
\end{align} 
 To motivate these expressions we compute\er{Araki2} and \er{ArakiR} for our familiar semifinite factors:
\begin{theorem}
\begin{enumerate}
\item For $M=B(H')$ and  $H=B_2(H')$ with $\om_i(A)=\Tr(\rh_i A)$ and $\Om_i=\rh_i^{1/2}$, 
\begin{align}
S(\om_1,\om_2)&=\Tr(\rh_1(\log\rh_1-\log\rh_2))=:S(\rh_1,\rh_2);\\
 R_t(\om_1,\om_2)&= \frac{1}{t-1} \log \Tr (\rh_1^t\rh_2^{1-t})=:R_t(\rh_1,\rh_2), \label{ERex1}
\end{align}
which recovers our earlier definitions \er{QKL} - \er{QKLb} and \er{defRtQ}, respectively. 
\item For a type II factor $M$ with trace $\tr$ acting on $H=L^2(M,\tr)$ with $\om_i(A)=\tr(\rh_i A)$ and $\Om_i=\rh_i^{1/2}$, cf.\ Theorem \ref{SFkey3}, we obtain the new results (with $\rh_i$ possibly unbounded!):
\begin{align}
S(\om_1,\om_2)&=\tr(\rh_1(\log\rh_1-\log\rh_2)); \label{SRex}, \\
 R_t(\om_1,\om_2)&= \frac{1}{t-1} \log \tr (\rh_1^t\rh_2^{1-t}).\label{ERex2}
\end{align}
\item For $M=L^{\infty}(X,\mu)$ acting on $L^2(X,\mu)$, with $\om_i(f)=\int_X d\mu\, \rh_i f$ and $\Om_i=\rh^{1/2}_i$ as in \er{1315N},
\begin{align}
S(\om_1,\om_2)&=\int_X d\mu_2\, \frac{d\mu_1}{d\mu_2}\log\left( \frac{d\mu_1}{d\mu_2}\right)=:S(\mu_1,\mu_2);\\
 R_t(\om_1,\om_2)&= \frac{1}{t-1}\log \int_X d\mu\, \left(\frac{d\mu_1}{d\mu}\right)^t  \left(\frac{d\mu_2}{d\mu}\right)^{1-t}=:R_t(\mu_1,\mu_2) \label{ERex3}
\end{align}
where $S(\mu_1,\mu_2)$ is the Kullback--Leibler divergence  \er{DefSq1}, and the measures $\mu_i$ ($i=1,2$) in the
Radon--Nikodym derivatives $d\mu_1/d\mu_2=\rh_1/\rh_2$ and $d\mu_i/d\mu=\rh_i$. are defined with respect to $\mu$ and the measures $\mu_i:= \mu\rh_i$ on $X$ given by
$\mu_i(\Dl)=\int_{\Dl} d\mu\, \rh_i$.
\end{enumerate}
\end{theorem}
Note that $\rh_1/\rh_2$ is well defined (though it may be unbounded) since both $\om_i$ are assumed faitful.
\smallskip

\noindent
 \emph{Proof.} The first case is
\begin{align}
S(\om_1,\om_2)&=-\la\Om_1, \log\Dl_{21}\Om_1\ra_2=-\la \rh_1^{1/2}, \log (L(\rh_2)R(\rh_1\inv)\rh_1^{1/2}\ra_2\nn \\
&= -\la \rh_1^{1/2}, (L(\log \rh_2)-R(\log \rh_1))\rh_1^{1/2}\ra_2\nn \\ &=\la \rh_1^{1/2},\, \rh_1^{1/2}\log \rh_1\ra_2-
\la \rh_1^{1/2}, (\log \rh_2)\rh_1^{1/2}\ra_2\nn \\ &=
\Tr(\rh_1(\log\rh_1-\log\rh_2))=S(\rh_1,\rh_2).
\end{align}
For the last case, it easily follows from \er{relativeSTT} - \er{153N}, see also  \er{1314N} - \er{1315N}, that
\begin{align} J\psi=\ovl{\psi}; && \Dl_{21}=\frac{\rh_2}{\rh_1}= \frac{d\mu\cdot \rh_2}{d\mu\cdot \rh_1}\equiv\frac{d\mu_2}{d\mu_1}, \label{1515N}
\end{align}
where we omit the hats making these functions multiplication operators. With \er{1315N}, this gives
\begin{align}
S(\om_1,\om_2)&=-\la\Om_1, \log\Dl_{21}\Om_1\ra_{L^2}=-\la \rh_1^{1/2}, \log (\rh_2/\rh_1)\rh_1^{1/2}\ra_{L^2}\nn \\
&=-\int_X d\mu\, \rh_1 \log (\rh_2/\rh_1)=\int_X d\mu_2\, \frac{d\mu_1}{d\mu_2}\log\left( \frac{d\mu_1}{d\mu_2}\right).
\end{align}
\bex
Prove \er{ERex1}, \er{SRex},  \er{ERex2}, and \er{ERex3}.  \QED
\eex

For finite sets $X=A$, $\mu(\{a\})=1$ for all $a\in A$, and $\rh_i(a)=p_i(a)$, we also recover the special case \er{KL1} central to large deviation theory.

For the other types eq.\ \er{Araki1} brings us into new territory, where we are comforted however by the fact that $S(\om_1,\om_2)$ shares the ``good'' properties of the special case $M=B(H')$ just studied:\footnote{See Hiai (2018), Theorem 4.1. Similarly for the relative R\'{e}nyi entropy \er{ArakiR}, cf.\ Hiai (2018), Proposition 5.3.}
\begin{enumerate}
\item  $S(\om_1,\om_2)\geq 0$ with equality iff $\om_1=\om_2$; 
\item  $S(\om_1,\om_2)$ is jointly lower semicontinuous in the weak topology on $M_*$;\footnote{This topology is defined with respect to its pairing with $M$, i.e., $\om_{\lm}\raw\om$ iff $\om_{\lm}(A)\raw\om(A)$ for all $A\in M$.}
\item $S(\om_1,\om_2)$ is jointly convex, that is,
$S\left(\Sigma_{i=1}^n p_i\om_i , \Sigma_{i=1}^n p_i\om'_i\right)\leq \Sigma_{i=1}^n p_iS(\om_i,\om'_i)$,
cf.\ \er{102.1};
\item $S(\om_1,\om_2)$ is monotone under quantum channels $\Phi_*: M_*\raw N_*$, in that 
\begin{equation}
S(\Phi_*\om_1,\Phi_*\om_2)\leq S(\om_1,\om_2), \label{MSQC}
\end{equation}
for all normal states $\om_1,\om_2$ on $M$. Here a quantum channel is the dual of a  completely positive unital normal maps $\Phi:N\raw M$, cf.\ Definition \ref{defQC} in finite dimension.\footnote{Eq.\ \er{MSQC} even holds for \emph{Schwarz maps}, i.e.,
  positive unital normal maps $\Phi:N\raw M$ that satisfy 
$\Phi(A^*A)\geq \Phi(A)^*\Phi(A)$ for all $A\in N$. This inequality follows from complete positivity by \emph{Kadison's Schwarz inequality} (Kadison, 1952, Theorem 1), and implies positivity. See Reible (2025), Theorem IV.3.7, for a clear and detailed proof of \er{MSQC}.} 
\end{enumerate}
Further confidence in \er{Araki2} comes from the \tto\ case.  Since the trace is a normal state, we may take  $\om_1$ defined by $\om_1(A)=\tr(\rh_1 A)$ for $\rh_1\in L^1(M)$,  see \er{1322}, and $\om_2=\tr$. Then, cf.\ \er{likeD1}, 
\begin{align}
S(\om,\tr)=-S(\rh); && S(\rh):=-\tr(\rh\log\rh). \label{nieuw}
\end{align}
This is an easy computation: for $\rh_1\in M\subset L^1(M)$ we have $\Om_1=\rh_1^{1/2}$ (taking $\Om=1_M$ as usual), so that 
$J\xi=\xi^*$ and $\Dl_{21}\xi=R(\rh_1\inv)\xi\equiv \xi\rh_1\inv$, as can be checked from \er{relativeSTT} - \er{153N}. Hence
\begin{equation}
S(\om_1,\tr)=-\la\rh_1^{1/2}, \log (R(\rh_1\inv))\rh_1^{1/2}\ra_2=\la\rh_1^{1/2}, \log (R(\rh_1))\rh_1^{1/2}\ra_2=\tr(\rh_1\log\rh_1).
\end{equation}
\bex Derive the general case $\rh\in L^1(M)$  from this by taking limits $\rh_n\raw\rh$ within $L^1(M)$.\eex

Finally, all of this can be generalized to arbitrary normal states $\om_1, \om_2$ on $M$. Let
\begin{equation}
s'(\om_i):=\inf\{E\in\mathrm{Proj}(M)\mid \om_i(E)=1\}, \label{supproj}
\end{equation}
be  the \emph{support projection} of any normal state $\om_i$ on $M$, where the infimum is taken in the complete lattice $\mathrm{Proj}(M)$, see \er{ProjM}. For example, if $\om_i$ is faithful, then $s'(\om)=1_M$, and on the other hand, if $M=B(H)$ and $\om_i(A)=\la\ps, A\ps\ra$ for some unit vector $\ps\in H$, then $s'(\om_i)=|\psi\ra\la\psi|$. In general,
\begin{equation}
s'(\om_i)=[M'\Om_i],
\end{equation}
i.e., the projection on the closure of the linear subspace $M'\Om_i=\{A'\Om_i, A'\in M'\}$. We also define
\begin{equation}
s(\om_i):=[M\Om_i] =Js'(\om_i)J.
\end{equation}
\bex Prove the second equality.\eex
Compared with  \er{relativeSTT},  the relative Tomita operators $S_{ij}$ are now defined by
decomposing $\psi\in H$ as $\ps=A\Om_j+\ps'$ with $A\in M$ and $\ps'\in  (M\Om_2)^{\perp}$, and then putting
\begin{equation}
S_{ij}(A\Om_j+\ps'):=s'(\om_j)A^*\Om_i.\label{Sij}
\end{equation}
Although $\Om_j$ need neither be cyclic nor separating for $M$, this operator is well defined: 
if 
\beq
\ps=A_1\Om_j+\ps_1'=A_2\Om_j+\ps_2',
\eeq
 then $A_1\Om_j-A_2\Om_j=\ps_2'-\ps_1'$, which must be zero,
since the left-hand side lies in $M\Om_j$ whilst the right-hand side lies in its orthogonal complement.
Hence $A_1\Om_j=A_2\Om_j$, so that 
\begin{align}
A_1s'(\om_j)=A_2s'(\om_j) && \Raw && s'(\om_j)A_1^*=s'(\om_j)A_2^* && \Raw &&
s'(\om_j)A_1^*\Om_j=s'(\om_j)A_2^*\Om_j. 
\end{align}Thus 
the projection $s'(\om_j)$ makes $S_{ij}$ well defined on $M\Om_j$ (although $\Om_j$ no longer needs to be separating for $M$),  upon which the projection $s(\om_j)$, which is implicit in \er{Sij} as it kills $\ps'$, extends $S_{ij}$ from
$M\Om_j$ to $M\Om_j\oplus (M\Om_j)^{\perp}$, which is dense in $H$. Then $S_{ij}$ remains closable, with closure still  written $S_{ij}$, and we can define the positive operator
 $\Dl_{ij}$ via \er{153N}. Finally, we define
\begin{align}
S(\om_1,\om_2)&:=\la \Om_2,\Dl_{12}\log \Dl_{12}\Om_2\ra=-\la\Om_1, \log\Dl_{21}\Om_1\ra && (s'(\om_1)\leq s'(\om_2));
\label{Araki3} \\
S(\om_1,\om_2)&:=\infty && \mbox{(otherwise).}
 \label{Araki4}
\end{align}
In our two familiar cases the condition $s'(\om_1)\leq s'(\om_2)$ 
  recovers the clauses $\ker(\rh_2)\subseteq \ker(\rh_1)$ 
  in \er{QKL} and $p\ll q$, or rather $\mu_1\ll \mu_2$, in   \er{DefSq1}, so that \er{Araki4} comes down to \er{QKLb} or \er{DefSq2}:
 \begin{itemize}
\item For $M=B(H')$ and $H=B_2(H')$ etc.\ we have $s'(\om_i)=[\rh_i^{1/2}B(H)]=[\ran(\rh_i)]$, so that
\begin{align}
s'(\om_1)\leq s'(\om_2) && \LRaw && [\ran(\rh_1)]\leq [\ran(\rh_2)] && \LRaw && \ker(\rh_2)\subseteq \ker(\rh_1),
\end{align}
since $[\ran(A)]=[\ker(A)]^{\perp}$ for self-adjoint operators $A$, 
and $E\leq F$ iff $F^{\perp}\leq E^{\perp}$. Since $s'(\om)$ is the projection onto $\ker(\rh_1)^{\perp}$,
formulae like \er{relativeDeltaI} remain valid if $\rh_1\inv$ is defined as the generalized inverse of $\rh_1$ (see Exercise \ref{inverseunbounded}). 
\item For $M=L^{\infty}(X,\mu)$ and $H=L^2(X,\mu)$, etc., we find $s'(\om_i)=[L^{\infty}(X,\mu)\rh_i^{1/2}]=1_{\rh_i>0}$, 
so that 
\begin{align}
s'(\om_1)\leq s'(\om_2) && \LRaw && 1_{\rh_1>0}\leq 1_{\rh_2>0}&& \LRaw && \mu_1\ll \mu_2,
\end{align}
with $d\mu_i=d\mu\cdot \rh_i$ as before. Furthermore, eq.\ \er{1515N} becomes $\Dl_{21}=(d\mu_2/d\mu_1)1_{\rh_1>0}$.
\end{itemize}
Finally, for $0<t<1$ the  relative R\'{e}nyi entropy \er{ArakiR} is defined without any condition, since for  for $0<t<1$ the vector
$\Om_2$ is in the domain of $\Dl_{12}^t$ (given the correct definition \er{Sij} of $S_{ij}$ and hence of $\Dl_{12}$). The extra condition in \er{Araki3} is only necessary for the existence of the limit $t\uparrow 1$.
\section{Magic without magic: traces for any \vna}\label{MWM}
As we have seen, only semifinite \vna s $M$ (and hence especially type I and type II factors) have a trace, which allows the realization of normal states $\om$ on $M$ by density operators $\rh\in L^1(M)$ via the pairing provided by the trace. See Theorem \ref{SFkey3}, and especially the formula 
\begin{align}
\om(A)=\tr(\rh A) && (A\in M), \label{16.1}
\end{align}
where, as in Definition \ref{deftrvna}, ``tr'' includes both the usual trace Tr on $B(H')$ and the unusual one for type II.
This was achieved through the construction of a ``canonical'' standard form $M\subset B(H)$:
\begin{itemize}
\item For type I factors $M=B(H')$ we had $H=B_2(H')\equiv L^2(M,\Tr)$, constructed through Tr on $B(H')$, now seen as a tracial \emph{weight}, see Definition \ref{defweight}, with ensuing \textsc{gns}-construction.
\item For a type \tto\ factor $M$ with  trace  such that $\tr(1_M)=1$, the \Hs\ $H=L^2(M,\tr)$ was simply obtained from the \textsc{gns}-construction by regarding $\tr$ as a faithful normal state on $M$. 
\item   A type \tti\ factor $M$ shares features of both cases: like the \tto\ case
it uses the special (``renormalized'') trace tr, but like the type I case this is now a (tracial) weight rather than a state.
\end{itemize}
What about type III? Following Haagerup,\footnote{See footnote \ref{Haageruprefs} for references to Haagerup's work; Hiai (2018, 2021a) contains the applications to entropy.} we will achieve the seemingly impossible: 
for \emph{any} \vna\ $M$ we are going to construct a \Hs\ $H=L^2(M)$ such that:
\begin{itemize}
\item The inner product is given by a \emph{bona fide} trace $\tr_H$ via $\la \xi,\eta\ra=\tr_H(\xi^*\eta)$.
\item $M$ acts on $L^2(M)$ ``from the left''  and is in standard form with $J\xi=\xi^*$ and $P=L^2(M)^+$.
\item For any normal state $\om$ on $M$ one has $\om(A)=\la\Om, A\Om\ra=\tr_H(\rh A)$ for a unique unit vector $\Om\in P$ and a unique density operator $\rh\in L^1(M)^+$, where $L^1(M)\cong M_*$ is also well defined.
\item The relative modular operator $\Dl_{ij}$, the relative entropy \er{Araki1}, and  the relative R\'{e}nyi entropy \er{ArakiR}. are given by the same formulae as for type I, so that, at least for faithful states,
\begin{align}
S(\om_1,\om_2)&=\tr_H(\rh_1(\log\rh_1-\log\rh_2)); \\
R_t(\om_1,\om_2)&= \frac{1}{t-1} \log \tr_H (\rh_1^t\rh_2^{1-t}).
\end{align}
\end{itemize}

How can this be? The key difference between Haagerup's construction and the earlier constructions for type I and II is that his \Hs\ $L^2(M)$ is constructed from $M$ in a completely different way, where $M$, or some dense subspace of it, is no longer contained in $L^2(M)$; in fact,
\begin{equation}
M\cap L^2(M)=\{0\}.
\end{equation}
Similarly, $L^1(M)$, the home of density operators,  intersects trivially with $M$ and even with $L^2(M)$:
\begin{equation}
M\cap L^1(M)=L^2(M)\cap L^1(M)=\{0\}.
\end{equation}
As we shall see, $L^1(M)$, $L^2(M)$, and all the other $L^p(M)$ spaces, $1\leq p\leq\infty$, with $L^{\infty}(M)\cong M$,  
are Banach spaces contained in a single space $\ovl{N}$ of possibly unbounded operators on a \Hs\ 
\beq
L^2(\R,H)\cong L^2(\R)\ot H,
\eeq
 where $M\subset B(H)$ is in standard form.\footnote{Here $L^2(\R,H)$ consists of all suitably measurable functions $\ps:\R\raw H$ such that $\int_\R dt \la\ps(t),\ps(t)\ra_H<\infty$, with inner product $\la\ps,\phv\ra=\int_\R dt \la\ps(t),\phv(t)\ra_H$. An isomorphism with $L^2(\R)\ot H$ is given by $U:L^2(\R)\ot H\raw L^2(\R,H)$, defined by  linear and continuous extension of $U(f\ot \Psi)=(t\mapsto f(t)\Psi)$, where $f\in L^2(\R)$ and $\Psi\in H$. 
 } 
 This \Hs, of which we will  use the first form $L^2(\R,H$), carries a normal \rep\ $\pi$ of $M$ as well as a continuous unitary \rep\ $\lm$ of $\R$ (as an additive) group, defined as follows.\footnote{One may  look at this construction in a  more general way. Let $G$ be a locally compact group and $\al:G\raw\mathrm{Aut(M)}$  a $G$-action on $M$, 
 see footnote \ref{AutMfn}.  A \emph{covariant \rep} of the pair $(G,M)$ with respect to  this group action consists of a continuous unitary \rep\ $U$ of $G$ and a normal \rep\ $\pi$ of $M$, both on the same \Hs, such that 
 $U(g)\pi(A)U(g)^*=\pi(\al_g(A))$ for all $g\in G$ and $A\in M$;
 in other words, the group action is implemented by $U$. In the main text we have $G=\R$, the modular group, $\al=\sg$, and $U=\lm$, and \er{16.8} is a special case pf the following: take a left-invariant Haas measure $dg$ on $G$, and define a \Hs\ $L^2(G,H)$ as consisting of all $\ps:G\raw H$ for which $\int_G dg\la\ps(g),\ps(g)\ra<\infty$ with inner product
 $\la\ps,\phv\ra=\int_G dg\la\ps(g),\phv(g)\ra$. Then take $\lm(h)\ps(g):=\psi(h\inv g)$ and $\pi(A)\psi(g)=\al_{g\inv}(A)\psi(g)$.
 This satisfies the covariance condition, and one has an ensuing crossed product $N=M\rtimes_{\al} G$ generated by the $\lm(g)$, $g\in G$, and $\pi(A)$, $A\in M$.
 Using C*-algebras instead of \vna s, most of both canonical and deformation quantization theory can be developed from this point of view (Landsman, 2017).  \label{Gcp}
 }
 Take some cyclic and separating unit vector $\Om\in H$ for $M$, with modular automorphism group $(\sg_t)_{t\in\R}$, see \er{NmgTT}. For $A\in M$ and $s\in\R$,  define
 \begin{align}
 \pi(A)\psi(t)&:=\sg_{-t}(A)\psi(t); && \lm(s)\psi(t):=\ps(t-s).\label{16.8}
 \end{align}
 \bex
Show that $\pi$ is an injective normal \rep\ $\pi$ of $M$, that $\lm$ is a continuous unitary \rep\ of $\R$, and that $\pi(M)$ and $\lm(\R)$ are connected by the \emph{covariance condition}
\begin{align}
\lm(s)\pi(A)\lm(s)^*=\pi(\sg_s(A)) && (A\in M, s\in\R).\label{16.9}
\end{align}
\eex
The \vna\ generated by the unitaries $\lm(s)$ and $\pi(A)$ is the \emph{crossed product}
\begin{align}
N:=M\rtimes_{\sg}\R=\{\lm(s), \pi(A), s\in\R, A\in M\}''; && N\subset B(L^2(\R,H)).\label{16.10}
\end{align}
By construction $N$ contains $\pi(M)\cong M$, and $\pi(M)$ will be our \vna\ $L^{\infty}(M)$. To locate $\pi(M)$ within $N$, we introduce the \emph{dual action} $\hat{\sg}$ of $\R$ on $N$; denoting elements of this new $\R$ by $E$, we first define a unitary \rep\ $\hat{\lm}$ of $\R$ on $L^2(\R,H)$, and then $\hat{\sg}$ in terms of $V$, by
\begin{align}
\hat{\lm}(E)\psi(t):=e^{-iEt}\psi(t); && \hat{\sg}_E(B):=\hat{\lm}(E)B\hat{\lm}(E)^* && (E\in\R,\,  B\in N). \label{16.11}
\end{align}
\bex Show that for all $E\in\R$, $A\in M$, and $t\in\R$, one has
\begin{align}
\hat{\lm}(E)\pi(A)\hat{\lm}(E)^*=\pi(A); &&\hat{\lm}(E)\lm(t)\hat{\lm}(E)^*=e^{-iEt}\lm(t).\label{16.12}
\end{align}\eex
Hence $\pi(M)$ is invariant under all automorphisms $\hat{\sg}_E$, 
and a bit more work in fact yields
\begin{equation}
N^{\hat{\sg}}:=\{B\in N\mid  \hat{\sg}_E(B)=B\:\:\: \forall\: E\in\R\}=\pi(M). \label{16.13}
\end{equation}
Since $N$ will play a crucial role in what follows, it may be reassuring that up to spatial equivalence its construction is independent of the choice of $\Om$ and the ensuing time evolution $(\sg_t)$. Namely, let $\Om'\in H$ be another cyclic and separating unit vector for $M\subset B(H)$, with associated unitaries 
\begin{align}
U_t=(D\Om':D\Om)_t; && U\psi(t):=U_{-t}\psi(t),
\end{align}
where $U_t:H\raw H$,  cf.\ \er{DDnotation}, and $U: L^2(\R,H)\raw L^2(\R,H)$. Then $U$ does the job:
\bex Prove that $U(M\rtimes_{\sg}\R)U^*=M\rtimes_{\sg'}\R$, where $(\sg_t')$ is the modular group of $M$ with respect to $\Om'$, 
 and that $ \mathrm{Ad}(U)\circ \hat{\sg}_E= \hat{\sg}'_E\circ\mathrm{Ad}(U)$, where $\sg_E'$ in \er{16.11} is now seen as an automorphism of $N'$,  and the equality means that for any $B\in N$ 
we have $U \hat{\sg}_E(B)U^*=\hat{\sg}'_E(UBU^*)$.\eex
 A much more difficult result, whose proof will provide the  tools for the  $L^p(M)$ spaces, is this:
\begin{theorem}
For any \vna\ $M$ the crossed product $N$ in  \er{16.10} is semifinite.
\end{theorem}
\emph{Proof (sketch).} The proof unfortunately involves weights on $N$ in an unavoidable way, and then uses the fact that the entire Tomita--Takesaki works for nfsf (normal faithful semifinite) weights (instead of faithful normal states). Moreover, Theorem \ref{CCRNT} is also valid for nfsf weights, so that, if we can construct an nfsf weight $\til{\om}$ on $N$ whose modular automorphisms $\sg_t^{\til{\om}}$ are inner (that is, implemented by unitaries $U_t$ in $N$ via $\sg_t^{\til{\om}}(B)=U_tBU_t^*$, $B\in N$), then this is also the case for the modular automorphisms defined  by
any other nfsf weight or faithful normal state on $N$. We will, indeed, construct such an nfsf weight weight $\til{\om}$ on $N$ whose modular automorphisms are  given by
\begin{align}
\sg_t^{\til{\om}}(B)=\lm(t)B\lm(t)^* && (t\in\R,\,  B\in N), \label{16.16}
\end{align}
where the unitaries $\lm(t)$ are given by \er{16.8}. The next step is the construction of an nfsf trace on $N$, which proves that $N$ is semifinite, cf.\ Proposition \ref{typefromtrace}. This is  easy: by Stone's theorem, 
\beq
\lm(t)=e^{iht},
\eeq
 where, although $\lm(t)\in N$ is bounded, the operator $h$ is typically unbounded and is merely affiliated to $N$ (see the supplement to \S\ref{TTsec}).
Writing $\rh=e^h$, define a new weight
\begin{align}
 \ta:N^+\raw[0,\infty]; && 
\ta(C):=\til{\om}(\rh^{-1/2} C\rh^{-1/2}) && (C\in N^+).\label{deftauN}
\end{align}
The KMS condition \er{KMSAB} for the weight $\til{\om}$ and its modular group $\sg_t^{\til{\om}}$
 then yields
\begin{equation}
\ta(B^*B)=\ta(BB^*), \label{tautrace}
\end{equation}
so that $\ta$ is a trace, cf.\ Definitions \ref{deftrvna} and \ref{defweight}.
To see this, we rewrite \er{KMSAB}  for  $\til{\om}$ and $\sg_t^{\til{\om}}$ as
\begin{align}
f(t)=\til{\om}(B^*(-t/2) B(t/2)); && f(t-i)=\til{\om}(B(t)B^*) && (t\in\R), \label{KMSBB}
\end{align}
where we wrote $B(t)$ for $\sg_t^{\til{\om}}(B)$, put $A=B^*$, and used \er{KMShalf},
The quickest way to derive \er{tautrace} from the KMS condition \er{KMSBB}  is to use the fact that for  ``analytic'' elements $B\in N^+$ we have 
\beq
f(-i)=\til{\om}((\Dl^{1/2}B\Dl^{-1/2})^* \Dl^{1/2}B\Dl^{-1/2})=\til{\om}((\rh^{1/2}B\rh^{-1/2})^* \rh^{1/2}B\rh^{-1/2}),
\eeq 
where we used  \er{16.16}. Using \er{deftauN} in the form $\til{\om}(C)=\ta(\rh^{1/2}C\rh^{1/2})$,
the second equation in  \er{KMSBB} for $t=0$, that is, 
\beq
f(-i)=\til{\om}(BB^*)=\tau(\rh^{1/2} BB^*\rh^{1/2}),
\eeq
 then becomes
 \begin{equation}
\ta(B^* \rh B)=\tau(\rh^{1/2} BB^*\rh^{1/2}).
\end{equation}
Writing the left-hand side as $\ta((\rh^{1/2} B)^*\rh^{1/2} B)$ and replacing $\rh^{1/2}B$ by $B$ gives \er{tautrace} for a  $\sg$-weakly dense set of elements in $N^+$, which  makes \er{tautrace} valid on all of $N^+$ by normality of $\til{\om}$ and hence of $\ta$. Within the domain of this trace we have $\ta(AB)=\ta(BA)$, so that eq.\ \er{deftauN} gives 
\begin{equation}
\til{\om}(B)=\ta(\rh B).  \label{16.22}
\end{equation}

So what is this special nfsf weight $\til{\om}$ on $N$? Let us first define an ``operator-valued weight'' 
\begin{align}
T:N^+\raw N^+\cup\{\infty\}; && T(B):=\int_\R dE\, \hat{\sg}_E(B), \label{16.23}
\end{align}
see \er{16.11}. One really needs $\infty$ as a possible value, since for any $A\in M$ we have 
\beq
\hat{\sg}_E(\pi(A))=A.
\eeq
On the other hand, operators on which $T$ gives a finite result  include the following: take some $F\in C_c(\R,M)$, where continuity is meant with respect to the $\sg$-weak topology on $M$, and define
\begin{equation}
B(F):=\int_\R dt\, \pi(F_t)\lm(t). \label{16.24}
\end{equation}
Using \er{16.12} and  elementary distribution theory, this gives, with apologies for the different $\pi$'s, 
\begin{equation}
T(B(F)^*B(F))=2\pi \int_{\R}dt\, \lm(t)^* \pi(F_t^*F_t)\lm(t)=2\pi \int_{\R}dt\, \pi(\sg_{-t}(F_t^*F_t)),
\end{equation}
where we used the covariance condition \er{16.9}.
So this lies in $\pi(M)$, which is true in general: by the invariance of the Haar measure $dE$ on $\R$ we have $\hat{\sg}_E\circ T=T$, so that by \er{16.13} we have
 $T(B)\in M^+$ or $T(B)=\infty$ for all $B\in N^+$. Consequently, for any weight $\psi$ on $M$ we may define
 \begin{align}
\til{\ps}:N^+\raw[0,\infty]; && \til{\ps}:=\ps\circ\pi\inv\circ T, \label{Haag1}
\end{align}
where $\til{\ps}(B)=\infty$ if $T(B)=\infty$; then $\til{\psi}$ is a normal and semifinite weight if $\psi$ has these properties, and, perhaps surprisingly, $\til{\psi}$ is faithful if $\psi$ is faithful. For example, eq.\ \er{Haag1} gives
\begin{equation}
\til{\psi}(B(F)^*B(F))=2\pi \int_{\R}dt\,\psi(\sg_{-t}(F_t^*F_t)). \label{16.27}
\end{equation}
Then the left-hand side vanishes iff $F_t=0$ for each $t\in\R$, which implies $B(F)=0$.

In particular, take the cyclic and separating unit vector $\Om\in H$ with associated faithful normal state $\om$ that produced the modular theory for $M\subset B(H)$. Then $\til{\om}=\om\circ\pi\inv\circ T$, as in \er{Haag1}, is the nfsf weight on $N$ doing the job, in the sense that \er{16.16} is satisfied, from which the rest followed. This can be proved by showing that 
 $\sg_t^{\til{\om}}$ \emph{as defined in \er{16.16}} satisfies the KMS condition for $\til{\om}$, and invoking Theorem \ref{TTTW}. It is sufficient 
 to check this on operators of the kind \er{16.24}. Writing $\sg^{\om}_t$ for $\sg_t$ (as we should), and once again using the covariance condition \er{16.9}, we first have
 \begin{align}
 \sg_{-t/2}^{\til{\om}}(B(F))^*=\int_\R ds\, \pi(\sg^{\om}_{-t/2}(F_s)^*)\lm(s); &&  \sg_{t/2}^{\til{\om}}(B(F))=
 \int_\R ds\, \pi(\sg^{\om}_{t/2}(F_s))\lm(s).
 \end{align}
Eq.\ \er{16.27} and the invariance property $\om(\sg^{\om}_s(A))=\om(A)$ then give
 \begin{align}
f(t):= \til{\om} (\sg_{-t/2}^{\til{\om}}(B(F))^*\sg_{t/2}^{\til{\om}}(B(F)))&=2\pi \int_{\R}ds\,  \om(\sg^{\om}_{-t/2}(F_s)^*\sg^{\om}_{t/2}(F_s)).
\end{align}
Hence the KMS condition for $\sg_t^{\om}$ in the form \er{KMSBB} yields the  KMS condition for $\sg_t^{\til{\om}}$, since
 \begin{align}
f(t-i)&=2\pi \int_{\R}ds\,  \om(\sg^{\om}_{t}(F_s)F_s^*)= \til{\om} (\sg_{t}^{\til{\om}}(B(F))B(F)^*),
\end{align}
where the second equality is obtained by reading \er{16.27} with $\psi\leadsto\om$ from right to left. \QED
\smallskip

The argument leading to \er{tautrace}, which modified the original state $\om$ into a trace provided the modular automorphisms defined by $\om$ are inner, works more generally to prove a key result:\footnote{See Takesaki (2003a), Theorem VIII.3.1.14.}
\begin{proposition}\label{innersemifinite}
A \vna\ $M$ is semifinite iff the modular automorphism group defined by some normal faithful state (or, equivalently, some cyclic and separating vector) is inner.
\end{proposition}
Moreover,  the crossed product of $N$ of $M$ with $\R$ with respect to $\hat{\sg}$  satisfies \emph{Takesaki--Takai duality}:
\begin{equation}
N\rtimes_{\hat{\sg}}\R=(M\rtimes_{\sg}\R)\rtimes_{\hat{\sg}}\R\cong M\otimes B(L^2(\R)),
\end{equation}
\emph{spatially}. Moreover, if $M$ is a type III factor, then $M\otimes B(L^2(\R))\cong M$, and we can say  more:\footnote{See Takesaki (2003a), Theorem XII.1.1.}
 \begin{proposition}
If $M$ is a type III factor,  then $N$ is a type \tti\ \vna\ (though not necessarily a factor),\footnote{
See footnote \ref{typegen} for types of general \vna s. $N$ is a factor iff $M$ has type $\mathrm{III}_1$ as per Connes.\label{CIII1}} and
$N\rtimes_{\hat{\sg}}\R\cong M$ \emph{algebraically} (moving back from type \tti\ to type III).
 \end{proposition}
Switching between type III factors $M$ and type \tti\ \vna s $N$ in this way was a key tool in the classification of type III factors and  has also become  popular in theoretical physics.\footnote{See the references in footnotes \ref{Kudler} and \ref{Vuyst}. Continuing footnote \ref{Gcp}, the second crossed product construction can  be generalized to any \emph{abelian} locally compact group. Let $\hat{G}$ be the dual of $G$, i.e., the group of characters (= one-dimensional continuous unitary \rep s) $\chi:G\raw\mathbb{T}$.
For example, $\hat{\R}\cong \R$ and each $\chi$ is $\chi_E$ for some $E\in\R$, defined by $\chi_E(t)=e^{iEt}$. Another example is $\hat{\Z}\cong \T$, where $\chi=\chi_z$ for some $z\in\T$, given by $\chi_z(n)=z^n$; conversely, $\hat{\T}\cong\Z$ with $\chi_n(z)=z^n$ (this is a special case of Pontrjagin duality $\hat{\hat{G}}\cong G$). For $E\in\hat{G}$ we then have a unitary \rep\ $\hat{\lm}(E)\psi(g)=\ovl{\chi_E(g)}\psi(g)$ of $\hat{G}$ on $L^2(G,H)$, which in turn defines an action $\hat{\al}$ of $\hat{G}$ on $N=M\rtimes_{\al}G$. This time we have $N\rtimes_{\hat{\al}}\hat{G}\cong M\ot B(L^2(G))$. See  Takesaki (2003a), Theorem X.2.8.}

We now continue the construction of the Haagerup $L^p$-spaces. First,  the trace \er{deftauN} satisfies
\begin{align}
\ta(\hat{\sg}_E(B))=e^{-E}\ta(B) && (E\in\R,\, B\in L^1(N,\ta)). \label{tauE}
\end{align}
To prove  \er{tauE}, first note that \er{16.11}, \er{16.12}, and the definition $\rh=\lm(-i)$ at once give
\begin{align}
\hat{\sg}_E(\rh)=e^{-E}\rh && \Raw&& \hat{\sg}_{-E}(\rh^{-1/2})=e^{-E/2}\rh^{-1/2}. 
\end{align}
Furthermore, since $T\circ \hat{\sg}_E=T$ by \er{16.23}, we also have $\til{\om}\circ \hat{\sg}_{-E}=\til{\om}$. Therefore,
\begin{align}
\ta(\hat{\sg}_E(B))&=\til{\om}(\rh^{-1/2}\hat{\sg}_E(B)\rh^{-1/2})=\til{\om}(\hat{\sg}_{-E}(\rh^{-1/2}\hat{\sg}_E(B)\rh^{-1/2}))\nn \\
&= \til{\om}(\hat{\sg}_{-E}(\rh^{-1/2})B\hat{\sg}_{-E}(\rh^{-1/2}))=e^{-E} \til{\om}(\rh^{-1/2}B\rh^{-1/2})\nn \\ &=e^{-E}\ta(B). 
\end{align}
Second, we we recall that for any semifinite \vna\ with a nfsf trace $\ta$:
\begin{itemize}
\item  the Banach space $L^1(N,\ta)$ is 
the completion of 
\beq
N^1_{\ta}:=\{B\in N\mid \ta(|B|)<\infty\}, \eeq
where $|B|=\sqrt{B^*B}$, 
 in the norm 
 \beq
 \| B\|_1=\ta\left(\sqrt{B^*B}\right), \label{B1again}
 \eeq 
 and $\ta$ extends from $N^1_{\ta}$ to $L^1(N,\ta)$ by continuity, with $|\ta(B)|\leq \| B\|_1$;
\item the \Hs\  $L^2(N,\ta)$ is the completion of 
\beq
N^2_{\ta}:=\{B\in N\mid \ta(B^*B)<\infty\}
\eeq
 in the norm 
 \beq
 \| B\|_2=\sqrt{\ta(B^*B)},
 \eeq
compare \er{B1again}, and for any $A,B\in L^2(N,\ta)$ we have $AB$ and $BA$ in $L^1(N,\ta)$, with
\beq
\ta(AB)=\ta(BA). \label{16.32}
\eeq 
\end{itemize}
Moreover,  $\ta$ can also be shown to be the unique trace on $N$ (up to rescaling) with this property.\footnote{This is relevant when $N$ is not a factor, see footnote \ref{CIII1}, in which case the trace as such is generally not unique. }

We return to \er{Haag1}, in which we now assume that $\psi$ is a 
normal \emph{state} on $M$; even in that case $\til{\psi}$ is a normal semifinite \emph{weight} on $N$. Compared to Theorem \ref{SFkey}, we therefore need an extension of the bijective correspondence $N_*\cong L^1(N,\ta)$: for all $B\in N^+$ in the domain of $\til{\ps}$ we still have 
\begin{equation}
\til{\ps}(B)=\ta(\rh_{\ps} B), \label{16.36}
\end{equation}
for some ``density operator'' $\rh_{\ps}$ bijectively corresponding to $\psi$, but we may longer assume that $\rh_{\ps}$ is in $L^1(N,\ta)$, noting that already that case generally involves unbounded operators. Instead:
\begin{itemize}
\item  if $\ps$ is an nsf \emph{weight} on $M$, then $\rh_{\ps}$ is \emph{affiliated with} $N$ (hence is not necessarily \emph{in} $N$);
\item if $\psi$ is a normal \emph{state}, then  $\rh_{\ps}$ is in addition \emph{$\ta$-measurable} (see the supplement to \S\ref{TTsec}).
\end{itemize}
\emph{Proof of the  second point.} We show that \er{tameasurable} holds for $\lm=1$, where $E(\Dl)$ are the spectral projections of  $\rh_{\psi}=\int_0^{\infty}dE_{\lm}\cdot\lm$, so that $E[0,1]^{\perp}=\int_1^{\infty}dE_{\lm}$.
  Starting from  \er{16.36} in the form 
  \begin{align}
  \ta(B)=\til{\psi}(\rh_{\psi}\inv B) &&  (B=E[0,1]^{\perp}),
  \end{align}
   and 
  $\rh_{\psi}\inv=\int_{0^+}^{\infty}dE_{\lm}\cdot\lm\inv$ (the generalized inverse as explained in part 2 of Exercise \ref{inverseunbounded}), 
 and using  \er{Haag1} with \er{16.23} as well as \er{16.37} below in the form $\hat{\sg}_s(\rh_{\psi}\inv)=e^s\rh_{\psi}\inv$, we compute: 
  \begin{align}\ta\left(E[0,1]^{\perp}\right)& = \til{\psi}\left(\rh_{\psi}\inv E[0,1]^{\perp}\right)=\til{\psi}\left(\rh_{\psi}\inv \int_1^{\infty}dE_{\lm}\right)=\til{\psi}\left(\int_1^{\infty}dE_{\lm}\cdot\lm\inv\right)
\nn \\ &=\psi\left(\pi\inv\left(\int_{\R}ds\, \hat{\sg}_s \left(\int_1^{\infty}dE_{\lm}\cdot\lm\inv\right)\right)\right)
\nn \\ &=\psi\left(\pi\inv\left(\int_\R ds\, \int_1^{\infty}dE_{e^s\lm}\cdot\lm\inv\right)\right)
\nn \\ &=\ps\left(\pi\inv\left(\int_{0^+}^{\infty} dE_{\lm}\right)\right)=\psi(\pi\inv(1_{(0,\infty)}(\rh_{\psi}))),
\end{align} where  $1_{(0,\infty)}(\rh_{\psi}))$ is in $\pi(M)$ since $(0,\infty)$ is invariant under scaling by any $e^{-s}$, $s\in\R$, 
and hence 
\beq
\hat{\sg}_s(1_{(0,\infty)}(\rh_{\psi}))=1_{(0,\infty)}(\rh_{\psi}).
\eeq
 Now $1_{(0,\infty)}(\rh_{\psi})$ is the projection onto the range of $\rh_{\psi}$, which equals the support projection
 $s'_N(\til{\psi})$ of the weight $\til{\psi}$ on $N$ defined by $\rh_{\psi}$ via \er{16.36}, cf.\ \er{supproj}.\footnote{For a weight $\til{\psi}$ the support projection is defined by $s_N'(\til{\psi})=1-\sup\{E\in\mathrm{Proj}(N)\mid\til{\psi}(E)=0\}$.}
 \bex
 Check that  if $s'_M(\psi)$ is the support projection of the state $\ps$ on $M$, then 
 \beq
 s'_N(\til{\psi})=\pi(s'_M(\psi)).
 \eeq
\eex
Hence 
\beq
\psi(\pi\inv(1_{(0,\infty)}(\rh_{\psi})))=\ps(s'_M(\ps))<\infty,
\eeq so that $\rh_{\psi}$ is $\ta$-measurable.\QED\smallskip

Furthermore,\footnote{The following argument uses the fact that the automorphisms
$\hat{\sg}_E$ can be extended from $N$ to $\ovl{N}$ by continuity.} since $\til{\ps}\circ \hat{\sg}_{E}=\til{\ps}$, as in the special case $\psi=\om$, eqs.\ \er{tauE} and \er{16.36} give
\begin{equation}
\hat{\sg}_E(\rh_{\ps})=e^{-E}\rh_{\ps}. \label{16.37}
\end{equation}
Let $\til{N}_E^+$ be the space of $\ta$-measurable positive operators on $L^2(\R,H)$ that are affiliated with $N$ and satisfy \er{16.37}.
Sending $\psi\in M_*^+$ to $\rh_{\psi}\in \til{N}_E^+$ is an injective map, since $\til{\phv}=\til{\psi}$ iff $\phv=\psi$.
This map is also surjective, though that is  more difficult to prove.\footnote{See Hiai (2021a), Theorem 9.5, which relies 
on deep results in modular theory by Pedersen \& Takesaki (1973).}
 Hence 
 \beq
 M_*^+\cong \til{N}_E^+,
 \eeq and by linearity,
\begin{equation}
M_*\cong L^1(M):=\{\rh\in \til{N}\mid \hat{\sg}_E(\rh)=e^{-E}\rh\:\forall E\in\R\}. \label{16.38}
\end{equation}
More generally, for any $1\leq p\leq\infty$ the \emph{Haagerup $L^p(M)$ spaces} are defined by
\begin{align}
 L^p(M)&:=\{B\in \til{N}\mid \hat{\sg}_E(B)=e^{-E/p}B\:\:\:\forall E\in\R\} \:\:\: (1\leq p<\infty)
 ; \label{Haap}\\
 L^{\infty}(M)&:= \pi(M).  \label{Haainfty}
\end{align}
Note that heuristically the $p=\infty$ case follows from $p<\infty$ by some kind of continuity, since for $p\raw\infty$ the condition $\hat{\sg}_E(B)=e^{-E/p}B$ becomes $\hat{\sg}_E(B)=B$, so that \er{16.13} returns \er{Haainfty}.
This notation suggests that $L^p(M)$ is a Banach space in some norm $\|\cdot\|_p$, and this is indeed the case:
\begin{itemize}
\item For $p=1$ we first define  the \emph{Haagerup trace} $\tr_H: L^1(M)\raw\C$ by linear extension of 
\begin{align}
\tr_H: L^1(M)^+\raw [0,\infty); && \tr_H(\rh_{\psi}):=\psi(1_M), \label{16.41}
\end{align}
where we recall that $\psi\in M_*^+$ induces $\rh_{\psi}\in L^1(M)^+$, and that any $\rh\in L^1(M)^+$ equals $\rh_{\psi}$ for some unique $\psi\in M_*^+$. For  all $\rh\in L^1(M)$ we then put 
\beq
\|\rh\|_1:=\tr_H(|\rh|).
\eeq  Since for positive elements $\psi$ of $M_*$ we have $\|\psi\|=\psi(1_M)$, this makes the isomorphism  \er{16.38} isometric.
\item For $1\leq p<\infty$,  just as for the usual $L^p$-spaces we define (generalizing the case $p=1$ above)
\begin{equation}
\| B\|_p:= \tr_H(|B|^p)^{1/p}. \label{16.42}
\end{equation}
\item In particular, $L^2(M)$ is a \Hs\ in the Haagerup--Hilbert--Schmidt  inner product 
\beq
\la \xi,\eta\ra:=\tr_H(\xi^*\eta).\eeq
\item For $p=\infty$ we just take the  norm to be the operator norm:
\beq
\|A\|_{\infty}:=\|\pi(A)\|_{B(L^2(\R,H))}=\| A\|_{B(H)} .
\eeq
\end{itemize}
We omit the  proof that $(L^p(M), \|\cdot\|_p)$ is  a Banach space,\footnote{See Hiai (2021a), \S 9.2, for all proofs we omit.} and  prove only one crucial case of:
\begin{theorem}
If $1\leq p<\infty$ and $1< q\leq \infty$ 
with $p\inv + q\inv=1$, then
\begin{align}
\tr_H(PQ)=\tr_H(QP) && (P\in L^p(M),\, Q\in L^q(M)). \label{16.44}
\end{align}
Moreover, we have $L^p(M)^* \cong L^q(M)$ as Banach spaces under the Haagerup trace: any (norm) continuous linear functional $\phv: L^p(M)\raw\C$ takes the form $\phv=\phv_Q$ for a unique $Q\in L^q(M)$, with
\begin{align}
\phv_Q(P)=\tr_H(QP) && (P\in L^p(M)). \label{16.45}
\end{align}
\end{theorem}
Note that \er{Haap} implies that $PQ$ and $QP$ are in $L^1(M)$, so their trace is finite. 
\smallskip

\noindent \emph{Proof for $p=1$ and $q=\infty$}.  Because of \er{16.38}, \er{Haainfty}, and the property $\pi(M)\cong M$, see \er{16.8}, 
the isomorphism $L^1(M)^* \cong L^{\infty}(M)$ takes the form $M_*^*\cong  M$, see Theorem 11.5.2. 
The embedding $M_*\subset M^*$ then turns  \er{16.45}, with $\phv_Q(P)$ reinterpreted as $\phv_{\rh}(A)$, into
the  expression
 \begin{align}
\ps(A)=\tr_H(\rh_{\ps} \pi(A)) && (\ps\in M_*,\,  A\in M,\, \rh_{\ps}\in L^1(M)), \label{16.45b}
\end{align}
which comes close to the usual ``$\om(A)=\Tr(\rh A)$''.
Both \er{16.44} and\er{16.45b} follow from \begin{align}
\rh_{A\psi C^*}=\pi(A)\rh_{\psi}\pi(C^*) && (\psi\in M_*,\, A,\, C\in M), \label{16.46} 
\end{align}
where, for given $\psi\in M_*$ the functional $A\psi C^*\in M_*$ is defined by $A\psi C^*(B):=\psi(C^*BA)$. Note that by \er{16.13} both sides of \er{16.46} are in $L^1(M)$. 
Eq.\  \er{16.46} follows from the special case
\begin{align} \rh_{A\psi A^*}=\pi(A)\rh_{\psi}\pi(A^*) && (\psi\in M_*^+,\, A\in M), \label{16.47} 
\end{align}
by decomposing $\psi=\psi_1-\psi_2+i\psi_3-i\psi_4$ with all $\ps_i\in M_*^+$, and likewise polarizing
\begin{equation}
A\psi C^*=\quar\Sigma_{k=0}^3i^k(A+i^kC) \psi (A+i^kC)^*,
\end{equation}
where we omitted $\pi(\cdot)$.
To prove \er{16.47}, using \er{Haag1}, \er{16.32}, \er{16.36}, and the equality 
\beq
\pi(A^*)T(B)\pi(A)=T(\pi(A^*)B\pi(A)),
\eeq 
cf.\ \er{16.13}, we find $\ta(\rh_{A\psi A^*}B)= \ta(\pi(A)\rh_{\psi}\pi(A^*) B)$ for any $B\in N$, which implies \er{16.47}. Then:
\begin{itemize}
\item  Eq.\ \er{16.44}, as said  for $p=1$ and $q=\infty$,  follows from  \er{16.41} and \er{16.46} by noting that 
\beq
\psi(A)=\tr_H(\rh_{\ps A})=\tr_H(\rh_{A\ps}),
\eeq
 and then using \er{16.46}. This gives $\tr_H(\rh_{\ps}\pi(A))=\tr_H(\pi(A)\rh_{\ps})$, which proves \er{16.44} for $B\geq 0$ (since every $B\in L^1(M)^+$ equals $\rh_{\psi})$ for some $\psi\in M_*^+$), upon which a decomposition of $B$ into (at most) four positive elements finishes the proof.
\item Eq.\ \er{16.45b}  follows from the definition \er{16.41} of the Haagerup trace by taking 
\begin{equation}
\psi(A^*A)=\tr_H(\rh_{A\psi A})=\tr_H(\pi(A)\rh_{\psi}\pi(A^*))=\tr_H(\rh_{\psi}\pi(A^*A)),
\end{equation}
where we  used \er{16.44}.This gives  \er{16.45b} on $M^+$, from which the general case once again follows by a linear decomposition of general elements of $M$ into positive ones.\QED
\end{itemize}
The case $p=q=2$ is equivalent to $L^2(M)$ being a \Hs.
It  follows from \er{Haap} and from the fact that the $\ta$-measurable operators affiliated to $N$ form an algebra that
$\pi(M)$ acts from the left on $L^2(M)$, since if $A\in M$, so that $\hat{\sg}_E(\pi(A))=\pi(A)$, cf.\ \er{16.13}, 
 and $\xi\in L^2(M)$, so that if $\hat{\sg}_E(\xi)=e^{-E/2}\xi$ (for all $E\in\R$), then $\pi(A)\xi\in L^2(M)$; indeed, since $\hat{\sg}_E$ is an automorphism, 
 \beq
 \hat{\sg}_E(\pi(A)\xi)= \hat{\sg}_E(\pi(A)) \hat{\sg}_E(\xi)= e^{-E/2}\pi(A)\xi.
 \eeq
 \begin{theorem}\label{HaagerupSF}
 The left action of $\pi(M)$ on $L^2(M)$ just stated brings $M$ in standard form, with 
 \begin{align}
 J\xi=\xi^*; && P=L^2(M)^+. 
 \end{align}
 \end{theorem}
 Thus  the construction of density matrices for semifinite \vna s can be generalized to arbitrary \vna s, 
 from which we can also rederive \er{16.45b} in the usual way, cf.\ the proof of Theorem \ref{SFkey3}: by Theorem \ref{SFkey2}
 any normal state $\psi$ on $M$ bijectively corresponds to a unit vector $\Om_{\psi}\in L^2(M)^+$ such that, writing $\rh_{\psi} :=\Om_{\psi}^2$ (so that $\rh_{\psi}\in L^1(M)$ since $\Om_{\psi}\in L^2(M)$),
 \beq
 \ps(A)=\la \Om_{\psi}, A\Om_{\psi}\ra=\tr_H(\Om_{\psi}^* A \Om_{\psi})=\tr_H(\Om_{\psi}^2 A)=\tr_H(\rh_{\psi} A).\label{old1654}
 \eeq
 Here products like $\Om_{\psi}^2 A$ are taken within $\ovl{N}$, the algebra of operators on $L^2(\R,H)$ affiliated to $N$.
  \begin{corollary}\label{HaagerupCor}
The pairing \er{16.45b} defined by the Haagerup trace gives a bijective correspondence between normal states $\psi$ on $M$  and ``density operators'' $\rh_{\psi} \in D(M)\subset L^1(M)$, where
 \begin{align}
D(M):=\{\rh\in L^1(M)\mid \rh\geq 0,\,  \tr_H(\rh)=1\}. \label{16.52}
\end{align} \end{corollary}
That said, despite these perks the situation is also quite different from the semifinite case:
\begin{itemize}
\item Except for $p=\infty$ cf.\ \er{Haainfty}, all nonzero operators in $L^p(M)$ are unbounded.\footnote{Here strictly meant as: not bounded (the term `unbounded'  often includes bounded operators). This claim follows from the scaling property in \er{Haap}.
For $P\geq 0$ in $L^p(M)$ with spectral resolution $P=\int_0^{\infty} dE_{\lm}\cdot\lm$, where $E(\Dl)=1_{\Dl}(P)$ for $\Dl\subset\sg(P)$, 
eq.\ \er{Haap} implies $\hat{\sg}_F(E_{\lm})=E_{e^{F/p}\lm}$ and hence $\hat{\sg}_F(E(\Dl))=E(e^{F/p}\Dl)$. If $P\neq 0$ then there is some nonempty $\Dl\subset \sg(P)$ with $E(\Dl)\neq 0$ and hence $E(e^{F/p}\Dl)=
 \hat{\sg}_{F}(E(\Dl))\neq 0$ for all $F\in\R$, so that  $E(t\Dl)\neq 0$ for all $t>0$.}
 \item If $p\neq q$, then  $L^p(M)\cap L^q(M)=\{0\}$, as is immediate from \er{Haap}. In particular, for any type of $M$, the spaces $L^1(M)$ and $L^2(M)$ are no longer ($\sg$-weakly) dense subspaces of $M$ (as in $L^p(B(H),\Tr)=B_p(H)$ for type I), or  completions of $M$ (as in $L^p(M,\tr)$ for type \tto), or combinations thereof (as for type \tti). Thus the density operators $\rh_{\psi}$ in $L^1(M)$ or the unit vectors $\Om_{\psi}\in L^2(M)$ associated to
 the normal states $\psi$ on $M$ are no longer even close to $M$.
\end{itemize}
This begs the question how our earlier  spaces $L^p(M,\tr)$ for a semifinite \vna\ with a trace are related to the Haagerup spaces $L^p(M)$. 
To answer this question we take the normal faithful state $\om$ on $M$ that initiated the entire construction to be the (normalized) trace $\tr:M\raw\C$. This is a state if $M$ is finite (i.e., type \tto\ or type $\mathrm{I}_n$), as the reader may want to keep in mind in what follows, and  a tracial weight otherwise (i.e., in type \tti\ or type $\mathrm{I}_{\infty}$), in which case the following argument needs some very slight modifications (but with the same result). 
Either way, for $\om=\tr$ the modular group is trivial, so that the action \er{16.8} of $A$ on $\psi\in L^2(\R,H)$ is simply given by 
\beq
\pi(A)\psi(t)=A\psi(t). \label{16.56}
\eeq
We now diagonalize  the  (unitary) action $\lm(s)$ of $s\in\R$, cf.\ \er{16.8}, via a (unitary)  Fourier transform
\begin{align}
\check{\psi}(k):=\int_\R dt\, e^{-ikt}\psi(t); &&\check{\lm}(t)\check{\psi}(k)=e^{-ikt} \check{\psi}(k).\label{16.57}
\end{align}
For $f\in C_c(\R)$ we define $\lm(f):=\int_\R dt\, f(t)\lm(t)$, which Fourier-transforms into the action
\begin{align}
\check{\lm}(f)\check{\psi}(k)=\check{f}(k)\check{\psi}(k); && \check{f}(k):=\int_\R dt\, e^{-ikt}f(t);
 \label{16.58}
\end{align}
whilst \er{16.56} turns into
\begin{equation}
\check{\pi}(A){\psi}(k)=A\check{\psi}(k).
\end{equation}
These operators are defined on a new $L^2(\R,H)$, whose elements are now functions $\check{\psi}:\R\raw H$ with inner product
$\la \check{\psi}, \check{\phv}\ra=\int_\R dk\, \la \check{\psi}(k), \check{\phv}(k)\ra_H$. For $A\in M$ and $f\in C_c(\R)$, define the operator
\begin{align}
A\ot f:  L^2(\R,H)\raw  L^2(\R,H); && (A\ot f)\check{\psi}(k):=\check{\pi}(A)\check{f}(k) \check{\psi}(k)=\check{f}(k)A \check{\psi}(k),
\label{Aotf}  \end{align}
so that $N$ is the \vna\ generated by the operators $A\ot f$. Symbolically, \footnote{If $L^2(\R),H)$ is seen as $L^2(\R)\ot H$, then the meaning of  \er{16.60} is immediate.}
 we have
\beq
N\cong M\otimes L^{\infty}(\R)\label{16.60}
\eeq
spatially, since the operators $\check{\pi}(A)$ and $\check{\lm}(f)$ have now decoupled entirely,  and  $L^{\infty}(\R)$ is the \vna\ generated by the multiplication operators $\check{f}$  on $L^2(\R)$. To compute the trace $\ta$ on $N$ in this Fourier-transformed description of $N$, we start from \er{deftauN}, which simply becomes 
\begin{equation}
\ta(B)=\til{\tr}(\rh\inv B),\label{16.62}
\end{equation}
cf.\ \er{16.22}. It follows from \er{16.57} and the description of $\rh$ between \er{16.16} and \er{deftauN} that
\begin{align}
\check{\rh}\check{f}(k) =e^{-k}\check{f}(k); && \check{\rh}\inv\check{f}(k) =e^{k}\check{f}(k).
 \label{16.63}
\end{align}
We also need the Fourier transform of the dual modular group $(\hat{\sg}_E)_{E\in\R}$ defined in \er{16.11}. Since
\begin{align}
\hat{\sg}_E(\lm(f))&=\hat{\sg}_E\left( \int_\R dt,\  f(t)\lm(t)\right)= \int_\R dt\,  f(t)\hat{\sg}_E(\lm(t))= \int_\R dt\,  f(t)e^{-iEt}\lm(t),
\nn\\
&=\lm(e^{-iEt}f),
\end{align}
where $e^{-iEt}f$ denotes the function $t\mapsto e^{-iEt}f(t)$, for its Fourier transform we obtain from \er{16.58}
\begin{equation}
\hat{\sg}_E(\check{\lm}(f))(k)=\int_\R dt\, e^{-ikt}e^{-iEt}f(t)=\check{f}(k+E), \label{16.65}
\end{equation}
cf.\ \er{16.58}. Moreover, from \er{16.12} we infer that
\begin{equation}
\hat{\sg}_E(A\ot f)=\check{\pi}(A)\ot \hat{\sg}_E(\check{\lm}(f)). \label{16.66}
\end{equation}
Using  \er{Haag1}, \er{16.23}, \er{16.62}, and \er{16.63}, we find, writing $e^k\check{f}$ for the function $k\mapsto e^k\check{f}(k)$,
\begin{align}
\ta(A\ot f)&=\til{\tr}(\rh\inv (A\ot f))=\tr\circ \hat{\pi}\inv\left( \int_\R dE\, \hat{\sg}_E(\rh\inv(A\ot f))
 \right)\nn \\ &=\tr(A) \int_\R dE\, \hat{\sg}_E(e^{k}\check{f})= \tr(A)\int_\R dE\,e^{k+E}\check{f}(k+E)\nn \\ &=\tr(A)\int_\R dk\, e^k\check{f}(k).\label{16.67}
\end{align}
Note that this computation confirms the scaling property \er{tauE} of $\ta$, since \er{16.65} gives
\begin{equation}
\ta(\hat{\sg}_E(A\ot f))=\tr(A)\int_\R dk\, e^k\check{f}(k+E)=e^{-E} \tr(A)\int_\R dk\, e^k\check{f}(k)=e^{-E} \ta(A\ot f).
\end{equation}
We return to \er{Haap}. Using \er{16.65} and \er{16.66},  the constraint $\hat{\sg}_E(A\ot f)=e^{-E/p}A\ot f$ becomes
\begin{equation}
\check{f}(k+E)=e^{-E/p}\check{f}(k),  \label{16.69}
\end{equation}
with no condition on $A\in M$. For measurable $\check{f}$, eq.\ \er{16.69} is  solved by $\check{f}(k)=C\cdot e^{-k/p}$, so that 
\begin{equation}
L^p(M)=\{A\ot f\mid A\in M, \check{f}(k)=e^{-k/p}\}.\label{16.70}
\end{equation}
The Haagerup trace $\tr_H$ on $L^1(M)$ can be computed from its definition \er{16.41}, where $\rh_{\ps}\in L^1(M)$ is defined by \er{16.36} with the weight $\til{\psi}$ on $N$ defined by \er{Haag1}, \er{16.23}, and \er{16.11}.
First, a computation similar to \er{16.67}  yields
\begin{equation}
\til{\psi}(A\ot f)=\psi(A)\int_\R dk\, \hat{f}(k). \label{16.71}
\end{equation}
Second, since $M$ is (semi)finite in our present argument, we may define $\ps\in M_*^+$ by 
\begin{align}
\ps(A)=\tr\left(\rh_{\ps}^{(M)}A\right); && \rh_{\ps}^{(M)}\in L^1(M,\tr).
\end{align}
Third, the definition  \er{16.36} of $\rh_{\psi}$ here comes down to
\begin{align}
\til{\psi}(A\ot f)=\ta(\rh_{\psi}(A\ot f)) && (A\in M,\,  \hat{f}\in L^{\infty}(\R)). \label{16.73}
\end{align}
Using \er{16.71}, another computation similar to \er{16.67} then shows that \er{16.73} is solved by
\begin{equation}
\rh_{\psi}=\rh_{\ps}^{(M)}\ot e^{-k},
\end{equation}
so that
 the Haagerup trace $\tr_H:L^1(M)^+\raw\R^+$ on positive operators is given by
\begin{equation}
\tr_H\left(\rh_{\ps}^{(M)}\ot e^{-k}\right)=\psi(1_M)=\tr\left(\rh_{\ps}^{(M)}\right),\label{16.75}
\end{equation}
which formula can immediately be extended to all of $L^1(M)$ by linearity. Consequently, the map
\begin{align}
L^1(M,\tr)\raw L^1(M); && \rh\mapsto \rh\ot e^{-k},
\end{align} 
which is well defined and a linear bijection because of \er{16.70}, 
 is also trace-preserving by \er{16.75}, and hence is
an isometry because of the definition of the norms \er{16.42} and \er{1323}.
Similarly, 
\begin{align}
L^p(M,\tr)&\raw L^p(M); && \rh\mapsto \rh\ot e^{-k/p}, \\
M&\raw L^{\infty}(M)=\pi(M); && A\mapsto \hat{\pi}(A)
\end{align} 
is an isometric isomorphism for any $1\leq p\leq\infty$. This example clearly displays the unpleasant phenomena listed after \er{16.52}, but also puts them into perspective: in the semifinite case all the unboundedness of operators in $L^p(M)$, realized as operators on $L^2(\R,H)$, is in the function $e^{-k/p}$, and likewise all the disjointness of the $L^p(M)$ for different values of $p$ is in these functions, rather than in anything related to $M$ acting on $L^2(\R,H)$. For type III things are not so simple, of course.

In any case, a huge advantage of the Haagerup spaces is that by Theorem \ref{HaagerupSF} and Corollary \ref{HaagerupCor} all formulae for the Tomita--Takesaki theory, and hence also those for Araki's relative entropy \er{Araki1} and relative R\'{e}nyi entropy \er{ArakiR}, are \emph{the same} as for the type I case, provided the density operators are now assigned to the normal states $\om_1$ and $\om_2$ via \er{16.45b} or \er{old1654}, and the Haagerup trace $\tr_H$ is used in all entropic formulae, instead of the usual trace.\footnote{See Hiai (2021a), Chapter 10, and Hiai (2021b), Chapters 2 and 3, for details.}
 With proofs adapted to the specific modular form of Araki's entropy, this implies that the basic theorems on quantum hypothesis testing as presented earlier in these lectures, namely the theorems of Chernoff, Stein, and Hoeffding, remain valid for general \vna s of any type.\footnote{\label{vnaSanov} This was already  shown without the Haagerup theory by
Jak\v{s}i\'{c},   Ogata,  Pillet, \&  Seiringer (2012). See  Hiai \&  Mosonyi  (2023), Pautrat \& Wang (2023), Junge \& Laracuente (2025), 
 and Fawzi, Gao, \& Rahaman (2026) for further developments in hypothesis testing that do use the Haagerup theory.
 }
\bex \label{superex}
Derive the Bekenstein--Hawking formula $S=A/4G$ for the entropy $S$ of a black hole in terms of the area $A$ of its event horizon (and Newton's constant $G$) on this basis.
\eex
\newpage
\section{Entropy of quantum dynamical systems}\label{CNTE}
In this final section we outline a quantum analogue of the Kolmogorov--Sinai entropy for classical dynamical systems, as developed by Connes in collaborations with  St\o rmer (for finite \vna s)  as well as with  Narnhofer and Thirring (for arbitrary \ca s).\footnote{The basic references are Connes \& St\o rmer (1975) and  Connes, Narnhofer, \& Thirring (1987). 
Connes (1994), \S V.6.$\beta$, reviews the basic ideas and history. Neshveyev \&  St\o rmer (2006) is a monograph on the subject, whose early chapters we roughly follow. See also the brief survey by  St\o rmer (2000). Benatti (2023)  gives a detailed treatment of the theory, with many applications, and also explains a different entropy for quantum dynamical systems due to Alicki \& Fannes (2001), which also deserves attention but unfortunately is beyond the scope of these lectures.} We restrict ourselves to  \vna s here in so far as the definition is concerned; practical computability of the entropy also requires hyperfiniteness, which fortunately is enough for most applications to mathematical physics. With this restriction we shall see that the  construction of the quantum dynamical entropy combines various ideas that played a role in these lectures so far. 

The first step is a reformulation of the classical setting of \S\ref{EDS} that paves the way for the subsequent noncommutative generalization. So let $(X,P,T)$ be a dynamical system (cf.\ Definition \ref{defDS}), where we now require $T:X\raw X$ to be invertible and $(X,\Sg)$,  where we  include the $\sg$-algebra $\Sg$ on which the probability measure $P$ is defined in our notation, to be a standard Borel space. 
 The probability space $(X,\Sg,P)$ then gives rise to the commutative \vna\ 
 \beq
 M=L^{\infty}(X,\Sg,P), \label{221CC}
 \eeq
  on which the map $T$ induces an automorphism $\al_T$ defined by 
 \beq
 \al_T(f)=f\circ T\inv.\label{alT}
 \eeq
  Since $T$ preserves $P$, this automorphism preserves the ``tracial'' normal state $\phv$ on $M$, given by 
  \begin{align}
  \phv(f)=\int_X dP\, f; &&  \phv\circ\al_T=\phv.
   \label{222C}
  \end{align}
Conversely,\footnote{See e.g.\ Takesaki (2003b), Prop.\ XIII.1.2, which, for commutative $M$, even reconstructs $(X,P,T)$ from $(M,\phv,\al)$.
} any automorphism $\al$ of $M$ that preserves $\phv$ is given by $\al=\al_T$ as in \er{alT} for some invertible Borel isomorphism $T:X\raw X$,  so that, roughly speaking:\footnote{However, the assumption that $T$ be invertible is not normally made and unnecessary in classical ergodic theory!}
\begin{quote}\begin{small}
classical ergodic theory is, after translation, the same thing as the study, up to conjugacy, of the automorphisms of $M$ that fix $\phv$. (Connes, 1994, p.\ 475)
\end{small}
\end{quote}

The next ingredient to be described via $M$ is the partition $\pi=(U_a)_{a\in A}$ of $X$, see \er{Xdu}, which formed the starting point of the construction of the Kolmogorov--Sinai entropy of $(X,P,T)$. If
\beq
M_0:=L^{\infty}(X,\Sg/\pi,P)\label{223C}
\eeq
 is the ($|A|$-dimensional) subalgebra generated by the characteristic functions $1_a\equiv 1_{U_a}$ (so that $\Sg/\pi$ is the smallest $\sg$-algebra on $X$ making each $1_a$ measurable),
 then 
$1_M=\Sigma_{a\in A} 1_a$ is a resolution of the unit by projections in $M_0$, which is the same as
a \textsc{pvm} (projection-valued measure) $E:A\raw M$ taking values in $N$, viz.\ $E(a)=1_a$. 
 Conversely, any \textsc{pvm} on some finite set $A$ taking values in a finite-dimensional von Neumann subalgebra $M_0\subset M$ 
comes from a finite (measurable) partition of $(X,\Sg,P)$ as defined by \er{Xdu}. From \er{ppKS2} with $\gm\leadsto\pi$, or \er{HPpiN} with $N=1$, we then have:\footnote{Here  $\eta(x):=-x\log x$ for $x>0$ and $\eta(0):=0$.}
\begin{equation}
H_P(\pi)=\Sigma_{a\in A}\eta(\phv(1_a)). \label{224C}
\end{equation}

A key step towards the Kolmogorov--Sinai entropy was the introduction of the \emph{finite} partition $\pi^N$ in \er{defpiN} - \er{Xsg0}, which is generated by the partitions $\pi_n=T^{-n}\pi$, $n=0,\ldots, N-1$, with
\begin{align}
H_P(\pi^N)=\Sigma_{\sg\in A^N}\eta(\phv(1_{\sg})); &&(1_{\sg}\equiv 1_{U_{\sg}}).\label{225C}
\end{align}
Each $\pi_n$  corresponds to the finite-dimensional subalgebra $M_n\subset M$ generated by the indicator functions $1_{T^{-n}U_a}$ with $n$ fixed and $a\in A$ varying, 
and $\pi^N$ corresponds to the subalgebra $\vee_{n=0}^{N-1} M_n$ of $M$ generated by the subalgebras $M_n$ and hence by the functions $1_{\sg}$, where $\sg\in A^N$.\footnote{Tthroughout this section a \emph{subalgebra} $M_n\subset M$ means a von Neumann subalgebra, that is, $M_n$ is a \vna\ in its own right under the operations inherited from $M$.} 

However, this does not work for noncommutative algebras $M$, since even two finite-dimensional subalgebras may generate an infinite-dimensional subalgebra of $M$. In addition, 
because 
\beq
1_{\sg}=1_{U_{\sg}}=1_{\bigcap_{n=0}^{N-1} T^{-n}U_{\sg(n)}}=\prod_{n=0}^{N-1}1_{T^{-n}U_{\sg(n)}},\label{226C}
\eeq 
what we have done in \er{225C} is combine projections from different subalgebras by multiplying them so as to obtain new projections. This does not work for noncommuting subalgebras (and hence noncommuting projections) either. But we know how to avoid projection-valued measures! We can try \textsc{povm}s instead of  \textsc{pvm}s. To practice this manoeuvre, we start with $N=1$. In that case,
\begin{equation}
H_P(\pi)=\sup_F\left\{\Sigma_{a\in A}(\eta(\phv(F_a))-\phv(\eta(F_a)))\right\}, \label{227C}
\end{equation}
where the supremum is over all \textsc{povm}s $F:A\raw [0,1]_{L^{\infty}(X,\Sg/\pi,P)}$, that is, over all finite sets $A$ and functions $F:A\raw L^{\infty}(X,\Sg,P)$ that take values in $L^{\infty}(X,\Sg/\pi,P)$ and satisfy $0\leq F(a)\leq 1_X$.
Here the main point of interest is the somewhat mysterious second term, versions of which will appear throughout this construction. 
To prove \er{227C}, we first note that taking $F_a=1_a$ already returns $H_P(\pi)$ via \er{224C}, and indeed 
$\eta(1_a)=0$, so that the second term in \er{227C} vanishes. 
\bex
Finish the proof by showing that the supremum in \er{227C} is bounded by $H_P(\pi)$.
\eex
One can also rewrite \er{227C} by allowing arbitrary \textsc{povm}s $F:A\raw  L^{\infty}(X,\Sg,P)$ and forcing these to take values in $L^{\infty}(X,\Sg/\pi,P)$ by using the (unique) conditional expectation 
\begin{align}
E_{\pi} \equiv E(\cdot\mid \Sg/\pi): L^{\infty}(X,\Sg,P)\raw L^{\infty}(X,\Sg/\pi,P); && \phv\circ E_{\pi}=E_{\pi}.\label{228C}
\end{align}
This gives
\begin{equation}
H_P(\pi)=\sup_F\left\{\Sigma_{a\in A}(\eta(\phv(F_a))-\phv(\eta(E_{\pi}(F_a)))\right\}, \label{229C}
\end{equation}
where this time the supremum is over all \textsc{povm}s 
\beq
F:A\raw [0,1]_{L^{\infty}(X,\Sg,P)};
\eeq
 in the first term we could have written $E_{\pi}(F_a)$ instead of $F_a$ but this gives the same result because of the last part of \er{228C}.

Without undue extra effort the generalization of \er{227C} to $N>1$ can be stated for arbitrary (finite) partitions $(\pi_0,\ldots, \pi_{N-1})$, i.e., without assuming that $\pi_n=T^{-n}\pi$. We label these as
\begin{align}
\pi_n=(U^{(n)}_{\sg(n)})_{\sg(n)\in A_n}; && \sg=(\sg(0), \ldots, \sg(N-1)),
\end{align}
 for certain finite sets $(A_n)$, so previously we had $U^{(n)}_{\sg(n)}=T^{-n}U_{\sg(n)}$ where $\sg(n)\in A_n=A$, the same index set for each $n$.  For brevity we still write $A_0\x\cdots\x A_{N-1}$ as $A^N$, and correspondingly,
 \begin{align}
 \Sigma_{\sg\in A^N} F_{\sg}&:=\Sigma_{\sg(0)\in A_0, \ldots, \sg(N-1)\in A_{N-1}} F(\sg(0), \ldots, \sg(N-1));\\
  F_{\sg(n)}&:=\Sigma_{\sg(0)\in A_0, \ldots, \sg(n-1)\in A_{n-1}, \sg(n+1)\in A_{n+1}, \ldots\sg(N-1)\in A_{N-1}} F(\sg(0), \ldots, \sg(N-1)),\label{2212C}
 \end{align}
 so that $\Sigma_{\sg\in A^N} F(\sg)=\Sigma_{\sg(n)\in A_n}  F_{\sg(n)}$ for each fixed $n\in N$. As in \er{225C} - \er{226C}, 
 \begin{align}
 H_P\left(\vee_{n=0}^{N-1}\pi_n\right)=\Sigma_{\sg\in A^N}\eta(\phv(1_{\sg})); &&\left(1_{\sg}=1_{\bigcap_{n=0}^{N-1}U^{(n)}_{\sg(n)}}\right). \label{2213C}
 \end{align}
 The key result, then, is
 \begin{equation}
 H_P\left(\vee_{n=0}^{N-1}\pi_n\right)=\sup_F\left\{\Sigma_{\sg\in A^N}\eta(\phv(F_{\sg}))-\Sigma_{n=0}^{N-1}\Sigma_{\sg(n)\in A_n}
 \phv(\eta(E_{\pi_n}(F_{\sg(n)})))\right\}, \label{2214C}
\end{equation}
where $E_{\pi_n}$ is defined as in \er{228C} with $\pi\leadsto\pi_n$, and the supremum is now over all  \textsc{povm}s \beq
F:A^N\raw [0,1]_{L^{\infty}(X,\Sg,P)}.\eeq
Note that we could have replaced $F$ in \er{2214C} by $E_{\vee_{n=0}^{N-1}\pi_n}(F)$, since 
\begin{align}
\phv\circ E_{\vee_{n=0}^{N-1}\pi_n}=\phv; && E_{\pi_n}\circ E_{\vee_{n=0}^{N-1}\pi_n}=E_{\pi_n};
\end{align}
the first is similar to \er{228C}, and the second is because we have $\pi_n\leq \vee_{n=0}^{N-1}\pi_n$.
\bex
Prove \er{2214C} by applying \er{227C} with $\pi\leadsto\vee_{n=0}^{N-1}\pi_n$ (difficult).\footnote{ Answer:
 Neshveyev \&  St\o rmer (2006), Proposition 1.3.2.}
\eex

In a possible generalization of \er{2214C} to arbitrary \vna s $M$ each partition $\pi_n$ still corresponds to a finite-dimensional (von Neumann) subalgebra $M_n\subset M$, but instead of replacing the partition $\vee_{n=0}^{N-1}\pi_n$ by the subalgebra  $\vee_{n=0}^{N-1}M _n$ generated by the $M_n$, which as already mentioned may be infinite-dimensional, we first try to define some generalization of the entropy
\beq
H_{\phv} (\pi_0, \ldots, \pi_{N-1}):=  H_P\left(\vee_{n=0}^{N-1}\pi_n\right).
\eeq
This quantity depends on some  normal state $\phv$ on $M$, which in the commutative case was given by \er{222C}, and on some finite-dimensional subalgebras $(M_0, \ldots, M_{N-1})$, which replace the partitions $(\pi_0, \ldots, \pi_{N-1})$, to which these subalgebras reduce in the commutative case. The simplest case (and historically also the first that was understood in this light) is when $M$ is a finite factor with normalized trace $\ta$, so $\ta=(1/n)\cdot\Tr$ for $M=M_n(\C)$ and $\tau=\tr$ for a \tto\ factor. In that case, try:
 \begin{equation}
 H_{\ta}(M_0,\ldots, M_{N-1}):=\sup_F\left\{\Sigma_{\sg\in A^N}\eta(\ta(F_{\sg}))-\Sigma_{n=0}^{N-1}\Sigma_{\sg(n)\in A_n}
 \ta(\eta(E_{M_n}(F_{\sg(n)})))\right\}, \label{2218C}
\end{equation}
which is the obvious generalization of \er{2214C}, in which the supremum is over all  \textsc{povm}s
\beq
F:A^N\raw[0,1]_M,\label{2219C}
\eeq
and $E_{M_n}: M\raw M_n$ is the unique normal conditional expectation that preserves $\ta$.\footnote{A \emph{conditional expectation} on a  \vna\ $M$ relative to a (von Neumann) subalgebra $N\subset M$ is a normal linear map $E:M\raw N$ such that $E(B)=B$ for all $B\in N$ and $\| E(A)\|\leq \|A\|$ for all $A\in M$; this implies (and is equivalent to the property) that $E:M\raw N$ is an idempotent  of norm 1 (so $E^2=E$ and $\|E\|=1$), and either way $E$ satisfies $E(BAC)=BE(A)C$ for all $A\in A$ and $B,C\in N$; $E$ is completely positive; and $E(A)^*E(A)\leq E(A^*A)$ for all $A\in A$. If $\psi$ is a state (or more generally a weight) on $M$, then $E$ is said to preserve $\psi$ if $\psi\circ E=\psi$. See Takesaki (2003a), \S IX.4. Provided the restriction of $\psi$ to $N$ is semifinite, 
 such a $\psi$-preserving normal conditional expectation exists and is unique iff $\sg_t^{\psi}(N)=N$ for all $t\in \R$, where $\sg_t^{\psi}$ is the modular group of $\psi$, cf.\ \S\ref{TTsec}. For a finite (or even a semifinite) factor $M$ with $\psi=\ta$, as in the main text,  the modular group is trivial and our $N=M_n$ is finite-dimensional, 
 and so one always has a unique $\ta$-preserving conditional expectation $E:M\raw M_n$. This can also be proved directly, see
 Takesaki (2002), Proposition V.2.36 for existence and  Blackadar (2006), Corollary II.6.10.8, for uniqueness. }
 
This still needs some doctoring in order to allow further generalization, if only because for arbitrary (normal) states $\phv$ on $M$ there may not be any $\phv$-preserving conditional expectation onto each $N_n$.  First, the \textsc{povm}s $F$ are replaced by state decompositions:\footnote{From the point of view of quantum information theory a quantum-to-classical channel defined by our \textsc{povm} \er{2219C}, which maps the normal state space $\mathcal{S}_n(M)$ of $M$ to $\Pr(A^N)$ via $\om\mapsto P_{\om}$ with
$P_{\om}(\sg)=\om(F_{\sg})$, 
 is replaced by a classical-to-quantum channel, which in the opposite direction maps $P\in \Pr(A^N)$ to $\mathcal{S}_n(M)$ via
 $P\mapsto \Sigma_{\sg\in A^N} P(\sg)\ta_{\sg}$.}
 defining $\ta_{\sg}\in M_*^+$  by 
\begin{align}
\ta_{\sg}(A):=\ta(F_{\sg}A),
\label{2220C}
\end{align}
so that, for some index set in which $\sg$ takes values (here $A^N$) one has elements $\ta_{\sg}\in M_*$ such that
\begin{align}
\ta_{\sg}\in M_*^+; && \Sigma_{\sg\in A^N} \ta_{\sg}=\ta. \label{statedec}
\end{align}
 Furthermore, 
using Theorem \ref{SFkey3} and  \er{likeD1}, which applies because each $M_n$ is finite-dimensional by assumption,\footnote{Because of \er{nieuw} this construction would actually work for arbitrary subalgebras of $M$.}
for $\rh\in M^+_n$ we have 
\beq
\ta(\eta(\rh))=-\ta(\rh\log\rh)=-S(\ta_{| M_n}(\rh\cdot),\ta_{| M_n}),
\eeq
where the relative entropy $S$ is defined by \er{SRex}, extended to arbitrary positive arguments, 
and the positive functional $\ta_{| M_n}(\rh\cdot)$ on $M_n$ is defined by $B\mapsto \ta_{| M_n}(\rh B)$,  $B\in M_n$.
Taking \beq
\rh=E_{M_n}(F_{\sg(n)}),
\eeq we then find, using the properties of our $\ta$-invariant conditional expectation,
\begin{align}
\ta_{| M_n}(E_{M_n}(F_{\sg(n)}) B)&=\ta (E_{M_n}(F_{\sg(n)})B)=\ta (E_{M_n}(F_{\sg(n)}B))=\ta(F_{\sg(n)}B)=\ta_{\sg(n)}(B)\nn \\ &=
\ta_{\sg(n)|M_n}(B), 
\end{align}
so that, also using the obvious property $\ta(F_{\sg})=\ta_{\sg}(1_M)$, eq.\  \er{2218C} may be rewritten as
 \begin{equation}
 H_{\ta}(M_0,\ldots, M_{N-1}):=\sup_{\ta=\Sigma_{\sg}\ta_{\sg}}\left\{\Sigma_{\sg\in A^N}\eta(\ta_{\sg}(1_M))+\Sigma_{n=0}^{N-1}\Sigma_{\sg(n)\in A_n}
 S(\ta_{\sg(n)|M_n}, \ta_{|M_n})
 \right\}, \label{2223C0}
\end{equation}
where $\ta_{\sg(n)}\in M_*^+$ is defined similarly to \er{2212C}, and the supremum is taken over all decompositions 
of $\ta$ according to \er{statedec}, from which indeed the \textsc{povm} has disappeared.

 At last, we have brought the entropy defined by a partition into a form that can be generalized to arbitrary \vna s $M$ with a faithful normal state $\phv$ and a finite collection of finite-dimensional subalgebras $M_n\subset M$, $n=0, \ldots, N-1$, viz.
 \begin{equation}
 H_{\phv}(M_0,\ldots, M_{N-1}):=\sup_{\phv=\Sigma_{\sg}\phv_{\sg}}\left\{\Sigma_{\sg\in A^N}\eta(\phv_{\sg}(1_M))+\Sigma_{n=0}^{N-1}\Sigma_{\sg(n)\in A_n}
 S(\phv_{\sg(n)}\circ \iota_n, \phv\circ \iota_n)
 \right\}, \label{2223C}
\end{equation}
where the supremum is over all state decompositions of $\phv$, that is, with the same notation as before,
\begin{align}
\phv=\Sigma_{\sg\in A^B} \phv_{\sg}; && \phv_{\sg}\in M_*^+,\label{phvdec}
\end{align}
 and 
$\iota_n:M_n\hookrightarrow M$ is the inclusion map;  our notation $\phv_{\sg(n)}\circ \iota_n$ and
$\phv\circ \iota_n$ instead of the equivalent $\phv_{\sg(n)|M_n}$  and $\phv_{|M_n}$, as in \er{2223C0}, is useful for understanding generalizations we do not discuss. 
The finishing touch to $ H_{\phv}(M_0,\ldots, M_{N-1})$  is a reformulation of \er{2223C} in terms of \emph{abelian models}:\footnote{Both Connes, Narnhofer, \& Thirring (1987), Definition III.2, and  Neshveyev \&  St\o rmer (2006), p.\ 36, call this an abelian model for $(M, \phv, M_0, \ldots, M_{N-1})$, but in fact the model does not depend on the subalgebras $M_n$.}
\begin{definition}\label{defABM}
An \emph{abelian model} $(X,\, P,\,  (C_n),\, \Phi)$ of $(M,\phv)$ consists of:
\begin{itemize}
\item A finite probability space $(X,P)$, or the corresponding \ca\ $C(X)$ with state $\om$;\footnote{As usual this means that $\om(f)=\Sigma_{x\in X} P(x)f(x)$.}
\item A collection $(C_n)_{n=0}^{N-1}$ of  subalgebras of $C(X)$, or, equivalently, of partitions $(\pi_n)_{n=0}^{N-1}$ of $X$;
\item A completely positive unital map $\Phi: M\raw C(X)$ such that 
\begin{align}
\Phi^*\om=\phv; && \Phi^*\om:=\om\circ\Phi. \label{Abmodel3}
\end{align}
\end{itemize}
 \end{definition}
 To clarify the  equivalence, which we have seen before in the more general context of a standard Borel space $(X,\Sg)$,
 we note that each subalgebra $C\subset C(X)$ corresponds to a  partition $\pi=(U_a)_{a\in A}$ of $X$, such that $C$ consist of all functions that are constant on each cell $U_a$ of $\pi$.\footnote{Equivalently,  each subalgebra $C$ of $C(X)$ 
 is given by a surjection $q:X\raw Y$ for some finite set $Y$ via $C=q^*C(Y)$. Hence $C$ consists of all functions 
$x\mapsto f(q(x))$, where $f\in C(Y)$, so that $C$ consists of all function on $X$ that are constant along the fibers $q\inv(\{y\})$ of $q$, for $y\in Y$. The corresponding partition then has cells equal to these fibers. } In particular, the atomic projections in $C$ are the indicator functions $p_a=1_{U_a}$. 
Thus an abelian model yields a state decomposition of $\phv$, where $A_n=(1,\ldots, \dim(C_n))$, and $\sg(n)\in A_n$ labels the atomic projections 
\beq
p^{(n)}_{\sg(n)}=1_{U^{(n)}_{\sg(n)}}\eeq
 in $C_n$, where the $U^{(n)}_{\sg(n)}$ are the cells of the partition $\pi_n$ of $X$ corresponding to $C_n$. 
Namely, we put
\begin{equation}
\phv_{\sg}(A)=\om\left(\Phi(A) p^{(0)}_{\sg(0)}\dots p^{(N-1)}_{\sg(N-1)}\right).\label{2228}
\end{equation}
\bex\label{224E}
\begin{enumerate}
\item Show that this rule indeed satisfies \er{phvdec}.
\item Show that, conversely, each state decomposition comes from an abelian model.
\end{enumerate}
\eex
Since $\Phi(1_M)=1_X$ and 
\begin{equation}
\om\left(p^{(0)}_{\sg(0)}\dots p^{(N-1)}_{\sg(N-1)}\right)=P\left(U^{(0)}_{\sg(0)}\bigcap\dots\bigcap U^{(N-1)}_{\sg(N-1)}\right),
\end{equation}
in \er{2223C} we have, as in \er{2213C},
\begin{equation}
\Sigma_{\sg\in A^N} \eta(\phv_{\sg}(1_M))=H_P\left(\vee_{n=0}^{N-1}\pi_n\right)=S\left(\om_{|\bigvee_{n=0}^{N-1}C_n}\right):=-\Sigma_{y\in Y} p(y)\log p(y),\label{2229C} 
\end{equation}
which is the Boltzmann--Shannon entropy of the pair $(Y,p)$ given by the Gelfand isomorphism 
\beq
\vee_{n=0}^{N-1}C_n\cong C(Y),
\eeq
 where $Y$ is any set that labels the cells of the partition $\vee_{n=0}^{N-1}\pi_n$,
 and the probability distribution $p\in\Pr(Y)$ corresponding to the restriction of the state $\om$ to $C(Y)$.
  Hence the first term in \er{2223C}, viz.\ \er{2229C},  is  independent of the subalgebras $M_n$; it  depends on the pair $(M,\phv)$ through the condition \er{Abmodel3}.
Towards the second term: for each $m\in N$ we have 
\beq
\Sigma_{m=0}^{N-1}p^{(m)}_{\sg(m)}=1_M,
\eeq
 so that, for any $A\in M$,
\begin{equation}
\phv_{\sg(n)}(A)=\Sigma_{\sg(0)\in A_0, \ldots, \sg(n-1)\in A_{n-1}, \sg(n+1)\in A_{n+1}, \ldots\sg(N-1)\in A_{N-1}} \phv_{\sg}(A)=\om\left(\Phi(A)p^{(n)}_{\sg(n)}\right),
\end{equation}
Hence the summand in the second term of \er{2223C} equals
\begin{equation}
S(\phv_{\sg(n)}\circ \iota_n, \phv\circ \iota_n)=S\left(\om\left(\Phi(\cdot) p^{(n)}_{\sg(n)}\right)_{|M_n},  \phv_{|M_n}\right),
\end{equation}
where $\om\left(\Phi(\cdot) p^{(n)}_{\sg(n)}\right)_{|M_n}$  is the state on $M_n$ given, for $B\in M_n$,  by 
\beq
B\mapsto \om\left(\Phi(B)p^{(n)}_{\sg(n)}\right).\eeq
 Finally, this can be brought into a neater form, which leads to the expression favoured by Connes:\footnote{See Connes (1994), p.\ 480. `It took 10 years before Connes saw what had to be done.' (St\o rmer, 2000, p.\ 12).}
\bex\label{225E}
\begin{enumerate}
\item For any state $\psi=\Sigma_a\psi_a$ on an abelian \vna\ $N$,  show that 
\begin{equation}
\Sigma_a S(\psi_a,\psi)=S(\psi)-\Sigma_a S(\ps_a),
\end{equation}
where $\psi$ is seen as a probability distribution on the spectrum $Z$ of $N$ (i.e., $N\cong C(Z)$).
\item Taking $N=M_n$, $\psi=\phv_{|M_n}$, $a=\sg(n)$, and $\psi_{\sg(n)}(B)=\om(\Phi(B)p_a)$ for $B\in M_n$, 
cf.\ \er{2228}, and also using \er{2229C}, show that the interior part of \er{2223C} may be (re)written as
\begin{equation}
\Sigma_{\sg\in A^N}\eta(\phv_{\sg}(1_M))+\Sigma_{n=0}^{N-1}\Sigma_{\sg(n)\in A_n}
 S(\phv_{\sg(n)}\circ \iota_n, \phv\circ \iota_n)=S\left(\om_{|\bigvee_{n=0}^{N-1}C_n}\right)-\Sigma_{n=0}^{N-1}S_{\om}(\Phi_n),
\end{equation}
where the last term is Connes's `entropy defect'
\begin{equation}
S_{\om}(\Phi_n):=S(\om)-\Sigma_{x\in X} P(x)S(\Phi^*_n\om, \Phi^*_n\dl_x),
\end{equation}
in which $\dl_x$ is the state on $C(X)$ given by $\dl_x(f)=f(x)$, and $\Phi_n: M_n\raw C_n$ is defined, in terms of the unique $\om$-preserving conditional expectation $E_n:C(X)\raw C_n$, by 
\beq
\Phi_n:=(E_n\circ\Phi)_{|M_n}.
\eeq  
\item Show that this entropy defect vanishes for $M=C(X)$, $C_n=M_n$, and $\Phi=\mathrm{id}_M$.
\end{enumerate}
\eex
In view of Exercises \ref{224E} and \ref{225E}, we are thereby  led to the following reformulation of \er{222C}:
\begin{definition}\label{defHphv}
Let $M$ be a \vna\ with  finite-dimensional  subalgebras  $M_n\subset M$ ($n=0, \ldots, N-1$), and 
$\phv$ a  normal state on $M$. The associated entropy  $H_{\phv}(M_0,\ldots, M_{N-1})$ is
\begin{equation}
H_{\phv}(M_0,\ldots, M_{N-1}):=\sup_{(X,\, P,\, (C_n),\, \Phi)}\left\{S\left(\om_{|\bigvee_{n=0}^{N-1}C_n}\right)-\Sigma_{n=0}^{N-1}S_{\om}(\Phi_n)\right\}, \label{2237C}
\end{equation}
where 
 the supremum is taken over all abelian models $(X,\, P,\,  (C_n),\, \Phi)$ of $(M, \phv)$.
\end{definition}
\begin{quote}
\begin{small}
In other words, one optimizes a \emph{commutative translation} of the situation $(M, (M_n), \phv)$ that one compares with the commutative situation $(X,\, P,\, (C_n),\, \Phi)$. In the commutative case the natural quantity is the quantity of information or entropy $S\left(\om_{|\bigvee_{n=0}^{N-1}C_n}\right)$ of the partition generated by the $C_n$. However, because of the loss of information in the translation, one must subtract the quantity $\Sigma_n S_{\om}(\Phi_n)$, which yields the formula \er{2237C}.
(Connes, 1994, pp.\ 480--481)
\end{small}
\end{quote}
\bex Show that in the commutative case \er{221CC},  eq.\ \er{2237C} recovers \er{2214C}, where the partitions $(\pi_n)$ correspond to the subalgebras $(M_n)$ as explained around \er{223C}.
\eex
Even in simple cases like the quantum Bernoulli shift below, the following tool is indispensable:
\begin{lemma}\label{QKSlemma}
In the situation of Definition \ref{defHphv}, suppose $M$ has the following subalgebras:
\begin{enumerate}
\item $B$, containing each $M_n$ and admitting a $\phv$-preserving condition expectation $E:M\raw B$;
\item $(C_n)_{n=0,\ldots, N-1}$, each abelian and the family mutually commuting, such that:\footnote{The centralizer $B_{\phv}\subset B$ of a state $\phv$ on $B$ consists of all $b\in B$ for which $\phv(ab)=\phv(ba)$ for all $a\in B$. In the context of Tomita--Takesaki theory this is equivalent to  $B_{\phv}$ being the fixed-point algebra under the modular group defined by $\phv$. This can be used to prove that there exists a $\phv$-preserving conditional expectation $F:B\raw B_{\phv}$, and also onto any subalgebra $C\subset B_{\phv}$, cf.\ Blackadar (2006), Theorem III.4.7.7. Since in our case $C$, being generated by \emph{commuting} finite-dimensional abelian \vna s, is finite-dimensional, this can also be shown by  elementary means!}
\begin{enumerate}
\item $C:=\bigvee_{n=0}^{N-1}C_n$ is a maximal abelian subalgebra of the centralizer $B_{\phv}\subset B$ of $\phv_{|B}$;
\item  $C_n\subset M_n\subset B$.
\end{enumerate}
\end{enumerate}
Then 
\begin{equation}
H_{\phv}(M_0,\ldots, M_{N-1})=S\left(\phv_{|C}\right). \label{2238C}
\end{equation}
\end{lemma}
\bex Prove this by first finding a $\phv$-preserving conditional expectation $F:B\raw C$; then showing that the finite-dimensional commutative \vna\ $C$ with state $\om=\phv_{|C}$, subalgebras $(C_n)$, and CPU map $\Phi=F\circ E$, is an abelian model for $(M,\phv)$;  and finally proving that  the entropy defect form this model vanishes so that the supremum in \er{2237C} is attained.\footnote{Answer:  Neshveyev \&  St\o rmer (2006), Proposition 3.1.6. Perhaps the exercise is to explain their proof!}
\eex

Our goal of defining a \emph{computable} noncommutative analogue of \er{preKS} and subsequently \er{defKSE} is now almost achieved, since the subadditivity property \er{KSsa} generalizes to:\footnote{See  Neshveyev \&  St\o rmer (2006), Proposition 3.1.3 (iii) for a proof.}
\begin{equation}
H_{\phv}(M_0, \ldots, M_{N_1-1},  M_{N_1},\ldots, M_{N_1+N_2-1})\leq H_{\phv}(M_0, \ldots, M_{N_1-1})+ H_{\phv}(
M_{N_1}, \ldots, M_{N_1+N_2-1}).
\end{equation}
Therefore, by the same arguments as in the classical case, if for some $\al\in\mathrm{Aut}(M)$ we fix a finite-dimensional subalgebra $M_0\subset M$ as before, and now take $M_n=\al^n(M_0)$, $n=0, \ldots, N-1$, then 
\begin{equation}
h_{\phv}(M_0,\al):=\lim_{N\raw\infty}\frac{1}{N} H_{\phv}(M_0, \al(M_0), \ldots, \al^{N-1}(M_0))
\end{equation}
exists. Its (possibly infinite) supremum over all finite-dimensional subalgebras $M_0\subset M$,
\begin{equation}
h(M,\phv,\al):=\sup_{M_0}\{h_{\phv}(M_0,\al)\},
\end{equation}
is the  \emph{Connes--Narnhofer--Thirring entropy} of $(M,\phv,\al)$.  One also  has a counterpart of Theorem \ref{KSGen}.\footnote{See Connes, Narnhofer \& Thirring (1987), Theorem VII.4; also cf.\ Connes (1994), Theorem V.6.8. } Recall that $M$ is \emph{hyperfinite} if there is an increasing sequence $(M_m)_m$ of finite-dimensional subalgebras $\CM_m\subset M$ such that $\bigcup_m \CM_m$ is $\sg$-weakly dense in $M$; symbolically, 
\beq
M=\lim_{n\raw\infty}\CM_m. \label{symhf}
\eeq
\begin{theorem}\label{QSinai}
If  $M$ is hyperfinite as given by \er{symhf}, then
\begin{equation}
h\left(M,\phv,\al\right)=\lim_{m\raw\infty} h_{\phv}(\CM_m,\al).
\end{equation}
\end{theorem}
Again as in the classical case, this theorem allows one to compute the entropy of the \emph{non-commutative Bernoulli shift}, which is defined as follows. Let $\ps$ be a state on the $l\x l$ matrices $M_l(\C)$, so $\psi(X)=\Tr (\rh X)$, $X\in M_l(\C)$, for some $\rh\in D(\C^l)$, with von Neumann entropy 
\beq
S(\psi)\equiv S(\rh)=-\Tr(\rh\log\rh).
\eeq
We now define $M$ as an infinite tensor product.\footnote{This construction goes back to von Neumann.
See e.g.\ Landsman (2017), \S 8.4, for details and explanations.}  Take the \textsc{gns}-triple $(H_{\psi},\pi_{\psi},\Om_{\psi})$ for $M_l(\C)$, so that $\la \Om_{\psi},\pi_{\psi}(X)\Om_{\psi}\ra=\ps(X)$. Then construct a $\Z$-fold  tensor product $H_{\psi}^{\ot\Z}$ of $\Z$ of $H_{\psi}$ with itself relative to $\Om_{\psi}$: the starting elements of $H_{\psi}^{\ot\Z}$ are sequences $\eta=(\eta_k)_{k\in\Z}$ in $H_{\psi}$  that satisfy 
\begin{align}
\Sigma_{k\in\Z}|\la\eta_k,\eta_k\ra_{H_{\psi}}-1|<\infty; && \Sigma_{k\in\Z}|\la\eta_k,\Om_{\psi}\ra_{H_{\psi}}-1|<\infty,
\end{align}
with inner product $\la \eta, \eta'\ra:=\prod_{k\in\Z}\la \eta_k,\eta'_k\ra_{H_{\psi}}$.\footnote{This product is convergent on the convention that if $\prod_k z_k$ diverges but $\prod_k |z_k|$ converges, we set $\prod_k z_k:=0$.} 
 The Hilbert space $H_{\psi}^{\ot\Z}$ is then defined as the completion of the linear span of these sequences with respect to this inner product; we write $\eta$ as $\ot_{k\in\Z}\eta_k$.
  For each $k\in\Z$, any $X\in M_l(\C)$ defines an operator $X_k$ on $H_{\psi}^{\ot\Z}$ by acting on $\eta_k$ alone;
  likewise, $X_k\otimes Y_m$ acts with $X$ in $\eta_k$ and with $Y$ on $\eta_m$, etc.
 The \vna\ generated by all these $X_k$, $k\in\Z$ is called $M_l(\C)^{\ot\Z}_{\psi}$, since it  depends on $\psi$, which also defines a normal state 
 \begin{align}
 \psi^{\ot \Z}\equiv \ot_{k\in\Z}\ps_k; &&  \psi^{\ot\Z}(X^{(1)}_{k_1}\ot\cdots\ot X^{(N)}_{k_N}):=\psi(X^{(1)})\cdots \psi(X^{(N)}),
\end{align}
 extended to all of $M_l(\C)^{\ot\Z}_{\psi}$ by linear and ($\sg$-weakly) continuous extension.
All that remains is an interesting automorphism on $M_l(\C)^{\ot\Z}_{\psi}$, which as in the commutative case we take to be the \emph{shift} 
\begin{align}
\sg:M\raw M; && \sg(X_k):=X_{k+1},
\end{align}
extended by linearity, the automorphism property (note that $X_k\ot Y_m=X_kY_m$) , and continuity. 

The quantum counterpart of Sinai's result \er{SinaiSp} for the usual Bernoulli shift, then, is
\begin{equation}
h\left(M_l(\C)^{\ot\Z}_{\psi}, {\psi^{\ot\Z}},\sg\right)=S(\ps).\label{2247C}
\end{equation}
\bex Prove this from Lemma \ref{QKSlemma} and Theorem \ref{QSinai}, using the fact that
\begin{align}
 M_l(\C)^{\ot\Z}_{\psi}=\lim_{m\raw\infty}\CM_m; && \CM_m:=\bigotimes_{k\in [-m,m]} M_l(\C)_k, 
\end{align} 
where  $[-m,m]$ is meant as an interval in $\Z$, and $M_l(\C)_k$ consist of all $X_k$, for $X\in M_l(\C)$, so that
\begin{equation}
h\left(M_l(\C)^{\ot\Z}_{\psi}, {\psi^{\Z}},\sg\right)=\lim_{n\raw\infty}\lim_{N\raw\infty}\frac{1}{N} H_{\phv}(\CM_m, \sg(\CM_m),\ldots, \sg^{N-1}(\CM_m)).\label{2249C}
\end{equation}
Show that $\sg^n(\CM_m)=\bigotimes_{k\in [-m+n,m+n]} M_l(\C)_k$, and that in Lemma \ref{QKSlemma} one can take
\begin{align}
B=\bigotimes_{k\in [-m,m+N-1]} M_l(\C)_k; && C_n=\bigotimes_{k\in [-m+n,m+n]} D_k,
\end{align}
where $D\subset M_l(\C)$ is a maximal abelian subalgebra in the centralizer of $\psi$. Finally, show that
\begin{equation}
S\left(\psi^{\ot \Z}|C\right)=(2m+N)S(\psi),
\end{equation}
and use  \er{2238C} and \er{2249C} to finish the proof of \er{2247C}. \QED
\eex
 \newpage
 \appendix
 \section{Convexity}\label{AppC}
  The concept of convexity pervades thermodynamics and associated theories of entropy, as well as the theory of operator algebras (mostly through the structure of state spaces). Far from a structured course on convexity or even some small part of it (although there will be some exercises!),     this appendix is just a list of definitions and results that are relevant for the main text.\footnote{Courses in convexity tend towards either analysis or geometry.
  In the former direction high-end classics are Rockafellar (1970), Simon (2011), and Borwein \& Vanderwerff (2010). Borwein \& Zhu (2005) and Benot (2013, 2016) are easier. 
 In the latter direction, which is less relevant here (though not irrelevant), see e.g.\ Barvinok (2002).}

 \begin{definition}\label{conbasicdef}
 Let $X$ be a vector space.
 \begin{enumerate}
\item A subset  $C\subseteq X$ is \emph{convex} if for any $x,y\in C$ 
the straight line segment between $x$ and $y$ entirely lies in $C$ (i.e., $\lm x+(1-\lm)y\in C$ for all $\lm\in(0,1)$).\footnote{
 Equivalently,
for any finite set of probabilities $(p_i)_{i\in I}$, i.e.\ $p_i\geq 0$ and $\Sigma_i p_i=1$, any any set $(x_i)_{i\in I}$  of points in $\mathcal{D}_f$ (i.e.\ as many points as there are probabilities), we have $\Sigma_{i\in I} p_ix_i\in C$. }
\item Let $C\subseteq X$ be convex and let $f:C\raw\R$ be some function, with  \emph{epigraph}  
\beq
\mathrm{epi}(f):= \{(x,t)\in C\x\R\mid f(x)\leq t\}. 
\eeq
Then $f$ is convex (as a function) if its epigraph $\mathrm{epi}(f)$ is convex (as a subset of $X\x \R$).

More generally,  suppose that $C\subseteq X$ is still convex but $f:C\raw (-\infty,\infty]=\R\cup\{\infty\}$ takes values in the \emph{extended} reals.\footnote{As is often the case in both large deviation theory and operator algebras!}
 The \emph{domain} $\mathcal{D}_f$ of $f$ is  the subset of $C$ where $f$ is finite, i.e.,
 \beq
 \mathcal{D}_f:=\{x\in C\mid f(x)<\infty\},
 \eeq
 and, consistent with the previous definition, we now define its epigraph as 
 \beq
\mathrm{epi}(f):= \{(x,t)\in \mathcal{D}_f\x\R\mid f(x)\leq t\}.
\eeq
Then $f$ is (again)  \emph{convex} if its epigraph is convex.
 \item We say that  $f:C\raw (-\infty,\infty]$ is \emph{lower semicontinuous} (lsc) iff $\mathrm{epi}(f)$ is closed in $X\x \R$.   
 \item  A function $g:C\raw [-\infty,\infty)$ is  \emph{concave} if $-g$ is convex, and \emph{upper semicontinuous} (usc)  if 
$-g$ is lower semicontinuous.  
\end{enumerate}
 \end{definition}
 Hence
 $f$ is convex and lsc iff $\mathrm{epi}(f)$ is convex and closed; it should also be clear from the main text that 
 lower semicontinuity and convexity is the ``right'' combination. A simple example of an lsc function is 
 $f:\R\raw\R$ defined by
 $f(x)=0$ if $x\leq 0$ and $f(x)=1$ if $x>0$ (but  $f(x)=1$ for $x\geq 0$ is not lsc). Also, $f:\R\raw [0,\infty]$ with $f(x)=0$ if $x\leq 0$ and $f(x)=\infty$ if $x>0$ is equally well lsc. 
 \bex \label{epigraphconvex}
 \begin{enumerate}
\item Show that if $f:C\raw (-\infty,\infty]=\R\cup\{\infty\}$ is convex, then $\mathcal{D}_f$ is convex.
\item Show that $f:X\raw(-\infty, \infty]$ is convex iff   for all $x,y\in \mathcal{D}_f$ and $\lm\in(0,1)$ we have 
 \begin{equation}
f(\lm x+(1-\lm)y))\leq \lm f(x)+(1-\lm)f(y),\label{deffconvex} 
\end{equation}
and that this is the case iff for all $(p_i)$ with $p_i\geq 0$ with $\Sigma_i p_i=1$ and all $(x_i)$ in $\mathcal{D}_f$, we have
\begin{equation}
f\left(\Sigma_i p_i x_i\right)\leq \Sigma_i p_i f(x_i).\label{Jen0}
\end{equation}
\end{enumerate}
\eex Here the number of $p_i$ and $x_i$ is supposed to be finite, 
but \er{Jen0} has an mportant generalization:
\begin{proposition}[Jensen's inequality]
Let $(X,\Sg,P)$ be some probability space, $F:X\raw\R$ in 
$L^1(X,\Sg,P)$, and $f:I\raw\R$ measurable and convex (more generally, it is enough that $f:I\raw\R$ is defined on some convex subset $I\subseteq\R$ such that $F(X)\subseteq I$). Then 
\begin{align}
f(\la F\ra_P)\leq \la f\circ F\ra_P && \LRaw &&
f\left(\int_X dP\,  F\right)\leq \int_X dP\, f\circ F. \label{Jensen} 
\end{align}
If we  define $f:C\raw\R$ to be \emph{strictly convex} if, for all $x\neq y$ and $\lm\in (0,1)$, eq.\ \er {deffconvex} 
 holds with strict inequality $<$ instead of $\leq$. then in \er{Jensen} we have equality iff $F$ is constant $P$-a.e. 
 \end{proposition}
  The definition  \er{Jen0} is now the special ``finite'' case of Jensen's inequality given by  
 \begin{align}
 X=\{x_1, \ldots, x_{|I|}\}; && P(x_i)=p_i; && F(x)=x.
\end{align}
Strict convexity can also be defined via the epigraph, but for general vector spaces $X$ 
this is rather complicated.\footnote{See Simon \& Verovic (2018), also for a complicated result beyond $X=\R^n$.  The restriction to $\R^n$ often suffices.} For $X=\R^n$ with $\mathrm{epi}(f)\subset\R^{n+1}$, however, the result is just as expected:
\begin{enumerate}
\item We say that  $C\subset\R^d$ is \emph{strictly} convex if all open line segments 
\begin{align}
\{\lm x +(1-\lm)y, \lm\in (0,1)\} && (x,y\in C, \, x\neq y)
\end{align} lie in the (topological) interior of $C$ (and hence in $C$, so that $C$ is strictly convex if it is convex in such a way that the topological boundary $\partial C$ contains no proper line segments). If $C$ is closed and convex,\footnote{Which, as we saw, is the case  if  $C= \mathrm{epi}(f)$ for some convex lsc function $f:C\raw (-\infty,\infty]=\R\cup\{\infty\}$ with $C\subset\R^n$.} then $C$ is strictly convex iff
 $\partial C\subseteq \partial_e C$ (where $\partial C$ is the topological boundary of $C$ and $\partial_e C$ its extreme boundary, cf.\ Definition \ref {defPK}).\footnote{For example, open or closed  unit balls in any dimension are strictly convex (and hence convex); the open or closed upper half plane in $\R^2$ is convex but not strictly convex. The latter example is equivalent to $f(x)=0$ from $\R$ to $\R$ being convex but not strictly convex; this is more generally the case for affine functions.} 
\item If  $C\subseteq\R^n$ is convex,  $f:C\raw (-\infty,\infty]$ is strictly convex iff $\mathrm{epi}(f)\subset\R^{n+1}$ is strictly convex.
\end{enumerate}
Jensen's inequality implies \emph{Chebyshev's (or Markov's) inequality}, which we sometimes use in the main text and therefore mention here (although as such it is hardly a result in convexity theory):
\begin{lemma}[Chebyshev--Markov]\label{CMlemma} 
For any real-valued random variable $X$ distributed by $P$, any real-valued  non-decreasing function $f$ on (at least) the range of $X$, and any $s\in\R$  with $f(s)>0$, 
\begin{equation}
P(X\geq s)\leq\frac{\la f(X)\ra_P}{f(s)}. \label{CMineq}
\end{equation}
\end{lemma}
\bex Prove this, and derive the  \emph{exponential Chebyshev inequality}: for $s\in\R$, $t\geq 0$, 
 \begin{equation}
P(X\geq s)\leq e^{-st}\la e^{t X}\ra_P. \label{CMcor}
\end{equation}
\eex
The best-known application is the weak law of large numbers, which follows by taking $|X|$ instead of $X$ and $f(x)=x^2$ just defined on $\R^+$, where it is indeed non-decreasing, so that if $\la X\ra_P<\infty$,
\begin{align}
P(|X-\la X\ra_P|\geq\varep)\leq \frac{\mathrm{Var}(X)}{\varep^2}; && \mathrm{Var}(X):=\la |X-\la X\ra_P|^2\ra_P=\la X^2\ra-\la X\ra_P^2.
\end{align}
 For i.i.d.\ $(X_n)$  we then take $S_N$ for $X$, and since 
$\mathrm{Var}(S_N)=\mathrm{Var}(X_1)/N$.
 for all $\varep>0$ we obtain
 \begin{align}
 \lim_{N\raw\infty} P(|S_N-\mu|\geq\varep)= 0 && \LRaw &&
 \lim_{N\raw\infty} P(|S_N-\mu|<\varep)= 1.
 \end{align}
 
Back to convexity! For differentiable functions convexity is easy to establish, as follows:
 \begin{proposition}
 \begin{enumerate}
\item If $I\subset\R$ is an interval, and  then $f:I\raw\R$ is $C^1$, then $f$ is convex iff $f'$ is nondecreasing on $I$, and strictly convex iff $f'$ is strictly increasing on $I$.
\item In particular, if $f$ is $C^2$, then $f$ is convex iff $f''(x)\geq 0$ on $I$ and strictly convex iff $f(x)>0$.
\item If $I\subset\R$ is an open interval and $f:I\raw\R$ is convex, then $f$ is differentiable almost everywhere on $I$ (more precisely: the set of $I$ where $f'(x)$ does not exist is at most countable).
\item If $U\subset\R^d$ is open and convex and $f:U\raw\R$ is $C^1$, then $f$ is convex iff for all $x,y\in U$,
\begin{equation}
\la \nabla f(x)-\nabla f(g), x-y\ra\geq 0.
\end{equation}
\item In particular, if $f$ is $C^2$ then $f$ is convex iff $\nabla^2f(x)\geq 0$ for all $x\in U$ and $h\in\R^d$.
\end{enumerate}
 \end{proposition}
 Here $\nabla^2f(x)$ is the \emph{Hessian} at $x$, which in the $C^2$ case is simply the  matrix with entries 
 $\partial^2 f/\partial x^i \partial x^j$.
 
  Going down in regularity, here is an efficient way to check if an lsc function on $\R$ is convex:
 \begin{lemma}\label{halfconvex}
 For some interval $C\subset \R$, let $f:C\raw (-\infty,\infty]$ be lsc.
Then $f$ is convex iff
 \begin{equation}
f(\half x+\half y)\leq \half f(x)+\half f(y)\:\:\:\: \mathrm{for\: all\:} x,y\in C. \label{halfconvex3}
\end{equation}
 \end{lemma}
 Thus  \er{deffconvex} only needs to be checked for $\lm=\half$.  
The proof  iterates \er{halfconvex3}, which gives
 \er{deffconvex}for all dyadic $\lm\in(0,1)$, upon which lsc gives it for all $\lm\in(0,1)$.  
 \begin{proposition}\label{BV-2-1-12}
 If $X=\R^d$ and  $f:C \raw (-\infty,\infty]$ is convex, then $f$ is continuous (and even locally Lipschitz) on 
 $\mathring{\mathcal{D}}_f$ (i.e.\ the interior of $\mathcal{D}_f$). This remains true if $X$ is a possibly infinite-dimensional Banach space, under the additional assumption that $f$ is continuous at \emph{some} point in $\mathcal{D}_f$.
 \end{proposition}
 
There  exists a purely topological characterization (or definition) of lower semicontinuity:
\begin{lemma}\label{lsclemma}
\begin{enumerate}
\item A function $f:X\raw(-\infty,\infty]$ is lower semicontinuous iff for each $x\in X$ and each $t<f(x)$ there is an open neighbourhood $U$ of $x$ such that $t<f(y)$ for all $y\in U$. 
\item
For metric spaces,\footnote{This is even true for general topological spaces if sequences are replaced by nets.}
this is the case iff for any convergent sequence $x_n\raw x$ one has
 \begin{equation}
\lim\inf_{n\raw\infty} f(x_n)\geq f(x). \label{lscliminf}
\end{equation}\end{enumerate}
\end{lemma}
 
 We now give a general characterization of convex lsc functions, starting from the observation that constant functions are convex and lsc. As a next step, if $X$ is  a normed space, define 
 \begin{align}
 \ell_{(\phi,r)}:X\raw\R; && \ell_{(\phi,r)}(x):=\phi(x)+r; && (\phi\in X^*, r\in X).\label{defaffine}
 \end{align}
 This is an \emph{affine function} on $X$, which is continuous because $\phi\in X^*$, the dual space of $X$ consisting of all \emph{continuous} linear maps $\phi: X\raw\R$. If $X=\R^d$, all such maps are given, abusing notation, by
 \begin{align}
 \phi(x)=\la\phi,x\ra && (\phv\in\R^d,\, x\in\R^d),
 \end{align}the Euclidean inner product.
 More generally, \er{defaffine} makes sense if $X$ is a topological vector space and $\phi\in X'$ is a continuous linear functional on $X$ (where $X'$ is the topological dual of $X$, i.e., the space of all continuous linear functionals on $X$); also in that case one often writes $\la\phi,x\ra$ for $\phi(x)$.
Affine functions are  convex and continuous functions are  lsc. The point is that although one does not obtain all continuous lsc functions in that way, taking suprema thereof one does:
 \begin{proposition}\label{superaffine}
 Let $X$ be a topological vector space and let $f:X\raw(-\infty,\infty]$ be a function. 
  \begin{enumerate}
\item If $f$ is the supremum of a nonempty family of continuous affine functions on $X$, then either $f$ is  identically equal to $\infty$ or $\mathcal{D}_f\neq\emptyset$ and $f$ is convex and lsc.
\item Conversely, if $\mathcal{D}_f\neq\emptyset$ and $f$ is convex and lsc,  then $f$ is the supremum of a family
$(f_{\gm})$ of continuous affine functions, namely the ones that are majorized by $f$. That is:
\begin{align}
f&=\sup\{ \ell_{(\phi,r)}\mid \phi\in X', r\in X,  \ell_{(\phi,r)}\leq f\}.
 \label{A7}
\end{align}
\item The supremum of any family $(f_{\gm})$  of convex lsc functions $f_{\gm}:X\raw(-\infty,\infty]$ is convex lsc. 
\end{enumerate}
 \end{proposition} One can also try \er{A7} for arbitrary functions $f:X\raw (-\infty,\infty]$. This gives the \emph{convex envelope}
 \begin{align}
f_*:&=\sup\{ \ell_{(\phi,r)}\mid \phi\in X', r\in X,  \ell_{(\phi,r)}\leq f\}\nn \\ &=\sup\{ g\mid g\leq f, 
g\: \mathrm{convex\: and\: lsc},\, \mathcal{D}_g\neq\emptyset\}\leq f,
 \label{A8}
\end{align}
which in general differs from $f$; but if $f$ is convex and lsc with $\mathcal{D}_f\neq\emptyset$, we clearly have $f_*=f$, and $\mathrm{epi}(f_*)$, which in general equals the closed convex hull  $\ovl{\mathrm{co}}(\mathrm{epi}(f))$, coincides with  $\mathrm{epi}(f)$.\footnote{The \emph{convex hull} $\mathrm{co}(S)$ of any subset $S$ of a vector space $Y$ is the smallest convex set in $Y$ that contains $S$; this set exists and equals the intersection of all convex subsets of $Y$ that contain $S$. If $Y$ is a topological vector space, then the  \emph{closed convex hull} $\ovl{\mathrm{co}}(S)$
is the closure of $\mathrm{co}(S)$. Here analytic and geometric approaches to convexity meet.}

The following construction gives a beautiful and universal way to write a convex lsc function as a supremum of affine functions. Here,
 $X$ may be taken to be a normed space with $Y=X^*$ its (Banach) dual, but any two topological vector spaces $(X,Y)$  in separating duality may be used.\footnote{This means that $Y\subseteq X'$ is a linear space of continuous linear functionals on $X$ such  that $\phv(x)=\phv(y)$ for all $\phv\in Y$ implies $x=y$. If $X$ is normed, this is even true if $Y$ is just a dense subspace of the Banach space dual $X^*$ of $X$.} It is often enough to keep $X=Y=\R^n$ in mind, with $\phv(x)=\la \phv,x\ra$ given by the inner product.
\begin{definition}\label{DefF}
Let $f:X\raw (-\infty,\infty]$ be any function.
  The \emph{Fenchel transform} of $f$ is
 \begin{align}
 f^*: Y\raw (-\infty,\infty]; && f^*(\phi):=\sup_{x\in X}\{\phi(x)-f(x)\}. \label{FeT}
 \end{align}\end{definition}
 Since $f^*$ is a supremum of continuous affine functions, by Proposition \ref{superaffine} it is either identically equal to $\infty$, or convex and lsc with $\mathcal{D}_{f^*}\neq\emptyset$. The second case therefore applies iff $f$ majorizes \emph{some} affine function, as  Definition \ref{DefF}  implies that for $\ph\in Y$ and hence $(\phi,r)\in Y\x\R$ we have
\begin{align} 
(\phi,r)\in\mathrm{epi}(f^*) && \Leftrightarrow&& \ell_{(\phi,r)}\leq f,
\end{align}
where the $\leq$ is pointwise. Hence there is a pair $(\phi,r)\in Y\x\R$ such that $\ell_{(\phi,r)}\leq f$ iff there exists some $\phi\in Y$ for which $f^*(\phi)<\infty$. 
This leads to a nice geometric interpretation of the Fenchel transform, most easily visualized for $X=Y=\R$, so that $\phi(x)=\phi x$.
If $f^*(\phi)<\infty$, the definition \er{FeT} implies that $\phi(x)-f(x)\leq f^*(\phi)$ for all $x\in X$, and hence
\beq
\phi(x)-f^*(\phi)\leq f(x), \label{FY1}
\eeq
 for all $x$.
By definition of a supremum, $f^*(\phi)$ is therefore the smallest number $r\in\R$ such that $\phi(x)-r\leq f(x)$ for all $x$. In case that $\phi(x)\leq f(x)$ for all $x$, $f^*(\phi)$ is therefore the smallest number such that the graph of $x\mapsto \phi(x)$ in $X\x\R$ moved upward by $-f^*(\phi)$  just touches the graph (or epigraph) of $f$. If $\phi(x)> f(x)$ for some $x$, then $f^*(\phi)$ is  the smallest number such that the graph of $x\mapsto \phi(x)$  moved downward by $f^*(\phi)$  just touches the graph of $f$.
\bex
Prove (i.e.\ compute) the following examples in $X=\R$ (we write $y$ instead of $\phv$):
\begin{enumerate}
\item $f(x)=e^x$ $\leadsto$  $f^*(y)=y\log y-y$ for $y>0$, $f^*(y)=0$ for $y=0$, and $f^*(y)=\infty$ for $y<0$.
\item $f(x)=-\log x$ for $x\in (0,\infty)$ and $f(x)=\infty$ otherwise $\leadsto$ $f^*(y)=-1-\log(-y)$ for $y\in (-\infty, 0)$ and $f^*(y)=\infty$ otherwise. 
\item $f(x)=p\inv |x|^p$ with $p>1$ on $\R$ $\leadsto$ $f^*(y)=q\inv |y|^q$ on $\R$  for $q$ such that $p\inv + q\inv=1$. 
\item $f(x)=-p\inv x^p$ with $p<1$ on $[0,\infty)$ and $f(x)=\infty$ otherwise $\leadsto$ $f^*(y)=-q\inv (-y)^q$
for $y\in (-\infty, 0)$ and $f^*(y)=\infty$ otherwise
(again for the above ``conjugate'' value of $q$).
\item What is $f^*(y)$ for $f(x)=|x|$?
\end{enumerate}\eex
Note that seemingly strange signs are chosen so as to make $f$  (and hence $f^*$) convex and lsc.

In principle, the double Fenchel transform $f^{**}$ would be a function on some dual $Y'$ (such as the Banach dual $X^{**}$ in case that $Y=X^*$ for a normed space $X$), but we regard it
as a function on $X$ via the canonical embedding  $X\hookrightarrow Y'$ given by 
$x\mapsto\hat{x}$, where $\hat{x}\in Y'$ is defined by $\hat{x}(\phi)=\phi(x)$, where $x\in X$ and $\phi\in Y$. Thus replacing $\hat{x}$ by $x$  gives, as a formula or simply as a definition,
\begin{equation}
f^{**}(x)=\sup_{\phi\in Y}\{\phi(x)-f^*(\phi)\}.
\end{equation}
\begin{theorem}\label{FDT}Let $f:X\raw (-\infty,\infty]$ with  $\mathcal{D}_f\neq\emptyset$ and also suppose that $\mathcal{D}_{f^*}\neq\emptyset$.\footnote{So this is the case iff $f$ majorizes \emph{some} affine function.}
Then 
\beq
f^{**}=f_*.
\eeq
Consequently, $f^{**}(x)\leq f(x)$ for all $x\in X$, and if $f$ is convex and lsc (so that $f_*=f$), then  
\beq
f^{**}=f.\label{A19}
\eeq
Therefore, if $f$ is convex and lsc (which is automatic for $f^*$ for any $f$), we have a \emph{Fenchel dual pair}
\begin{align} 
f(x)=\sup_{\phi\in Y}\{\phi(x)-f^*(\phi)\}; &&  f^*(\phi)=\sup_{x\in X}\{\phi(x)-f(x)\}. \label{FeDP}
\end{align}
\end{theorem}
 Thus $f$, if it is convex and lsc, is again a supremum of a family affine functions maojorized by $f$, as in \er{A7}, but we now have an explicit description of this family.

\smallskip
 
\emph{Proof.} Since $\phi(x)+r\leq f(x)$ for all $x$ is the same as $-r\geq f^*(\phi)$, eq.\ \er{A8} comes down to
\begin{align}
f_*(x)&=\sup\{\phi(x)+r\mid \phi\in Y, r\in X,  -r\geq f^*(\phi)\}=\sup_{\phi\in Y}\{\phi(x)-r\mid r\in X, r\geq f^*(\phi)\}\nn \\ &=\sup_{\phi\in Y}\{\phi(x)-r\mid  r= f^*(\phi)\}=
\sup_{\phi\in Y}\{\phi(x)-f^*(\phi)\}=f^{**}(x), 
\end{align}
since any $r>f^*(\phi)$ can only lower $\phi(x)-r$ compared to $\phi(x)-f^*(\phi)$ and will therefore not contribute to the supremum, whereas $r<f^*(\phi)$ is excluded by  the constraint $r\geq f^*(\phi)$.\QED
\smallskip

The following result (more specifically its corollary) forms the ultimate explanation of the equality between the two different expressions \er{IMEP} and \er{IqC1} for the rate function in Cram\'{e}r's theorem, and more generally is a very abstract form of maximum entropy principles.\footnote{Good sources are Borwein \& Vanderwerff, \S4.4.2,  Borwein \& Zhu, \S4.4, and Penot (2016),  \S6.3.  Borwein c.s.\  assume that $X$ and $Z$ are Banach spaces, but completeness is not used in the proof and indeed Penot, Theorem 6.18, just assumes normed spaces like we do.
Dohmatob (undated) also provides a nice (but unfortunately unfinished) introduction. }
\begin{theorem}[ Fenchel--Rockafellar duality]\label{FRDT}
Let $X$ and $Z$ be normed spaces, and let 
\begin{align}
f:X\raw(-\infty,\infty]; && g:Z\raw(-\infty,\infty]; && T:X\raw Z
\end{align}
 be two convex lsc functions, and  a bounded linear operator, respectively.  Define
\begin{align}
p&:=\inf_{x\in X}\{f(x)+g(Tx)\}; \label{defp}\\
d&:=\sup_{\phv\in Z^*}\{-g^*(-\phv)-f^*(T^*\phv)\} \label{defd}
\end{align}
to be the so-called \emph{primal} and \emph{dual} optimization problems. Then
\begin{equation}
p\geq d, \label{pgeqd}
\end{equation}
and if either $T(\mathcal{D}_f)$ contains a point of continuity of $g$, or $\mathrm{int}(\mathcal{D}_g-T(\mathcal{D}_f))$ contains 0, then
\begin{equation}
p=d. \label{pequalsd}
\end{equation}
 \end{theorem}
 Before proving this, let us derive a crucial consequence for the theory of entropy:
 \begin{corollary}\label{forMEP}
 For $f$ and $T$ as in Theorem \ref{FRDT}, and any $z\in Z$, we  have
 \begin{equation}
\inf_{x\in X}\{f(x)\mid Tx=z\}\geq \sup_{\phv\in Z^*}\{\phv(z)-f^*(T^*\phv)\}, \label{dME}
\end{equation}
with equality under the additional assumption $z\in \mathrm{int}(T(\mathcal{D}_f))$.
 \end{corollary}
 \emph{Proof of Corollary \ref{forMEP}.} 
Take $g=\iota_{\{z\}}$ in Theorem \ref{FRDT}, where, for any subset $S\subset Z$, we define
 \begin{align}
\iota_S:Z\raw (-\infty,\infty]; && \iota_S(z)=0\:\:\: (z\in S); && \iota_S(z)=\infty\:\:\: (z\notin S).
\end{align} 
  Its Fenchel transform is easy to compute, viz.\ $1^*_S(\phv)=\sup_{z\in S} \{\phv(z)\}$, so for $S=\{z\}$ we obtain
 \begin{align}
 1_{\{z\}}^*: Z^*\raw(-\infty,\infty]; &&
1_{\{z\}}^*(\phv)=\phv(z).
\end{align}
The term $g(Tx)$ in \er{defp} then obviously becomes the constraint $Tx=z$ on the left-hand side of \er{dME}, whereas the term $-g^*(-\phv)$ in \er{defd} comes down to $\phv(z)$ on the right-hand side of \er{dME}.

Finally, since obviously $\mathcal{D}_{\iota_{\{z\}}}=\{z\}$, the condition $0\in \mathrm{int}(\mathcal{D}_g-T(\mathcal{D}_f))$ in Theorem \ref{FRDT} becomes the assumption $z\in \mathrm{int}(T(\mathcal{D}_f))$ in the second claim of  Corollary \ref{forMEP}.\QED
\smallskip

  \emph{Proof of Theorem \ref{FRDT}.}  For \er{pgeqd}, we note that $g^{**}(x)\leq g(x)$ (see Theorem \ref{FDT}) and hence:
   \begin{align}
p\geq\inf_{x\in X}\{f(x)+g^{**}(Tx)\}&= \inf_{x\in X}\sup_{\phi\in Z^*} \{f(x)+\phi(Tx)-g^*(\phi)\} \nn \\ &\geq \sup_{\phi\in Z^*}  \inf_{x\in X}\{f(x)+\phi(Tx)-g^*(\phi)\} 
\nn \\ &= \sup_{\phi\in Z^*}  \inf_{x\in X}\{f(x)-T^*\phi(x)-g^*(-\phi)\} \nn \\
&= \sup_{\phi\in Z^*}  \inf_{x\in X}\{ -(T^*\phi(x)-f(x)+g^*(-\phi))\} \nn \\
&= \sup_{\phi\in Z^*} \{-g^*(-\phi) -\sup_{x\in X}\{ (T^*\phi(x)-f(x) \}\} \nn \\
&=\sup_{\phv\in Z^*}\{-g^*(-\phv)-f^*(T^*\phv)\} =d.
\end{align}   
 For the equality \er{pequalsd}, we first note that if $p=-\infty$, then $p\geq d$ forces $p=d$ and we are ready. 
We may therefore  assume that $p\neq -\infty$, in which case we
  introduce an auxiliary function
 \begin{align}
 h:Z\raw [-\infty,\infty]; && h(z):=\inf_{x\in X} \{f(x)+g(Tx+z)\}.
\end{align}
The additional assumptions stated above \er{pequalsd} are used to show that $\partial h(0)\neq\emptyset$; we omit this highly technical part of the proof.\footnote{See Lemma 4.3.1 in Borwein \& Zhu. It would be helpful of this step could be simplified.} Granting this, there exists $-\phi\in\partial h(0)$ and hence, 
by definition of the subdifferential, $h(0)\leq h(z)+\phi(z)$ for all $z\in Z$. Hence for all $z\in Z$ we have:
   \begin{align}
   p&\leq \inf_{x\in X}\{f(x)+g(Tx+z)+\phi(z)\} \nn \\
  \Raw \forall_{x\in X, z\in Z} \: p&\leq f(x)+g(Tx+z)+\phi(z)\nn \\
 \stackrel{z\leadsto z-Tx}{\Longrightarrow} \forall_{x\in X, z\in Z} \: p&\leq f(x)-T^*\phi(x) -(-\phi(z)-g(z))\nn \\
  \stackrel{\inf_z}{\Longrightarrow}   \forall_{x\in X} p&\leq -(T^*\phi(x)-f(x)) -g^*(-\phv)
  \nn \\
  \stackrel{\inf_x}{\Longrightarrow} p&\leq -f^*(T^*\phi)-g^*(-\phv)  \nn \\
  \stackrel{\sup_{\phi}}{\Longrightarrow} p&\leq d,
\end{align} 
where we repeatedly used ``$-\inf = \sup -$'' and \er{FeT}. \QED
\smallskip

We now explore the relationship between convexity and the search for minima. Since our convex functions are often just lsc rather than differentiable, we need a replacement for the derivative. 
\begin{definition}
Let $X$ be a normed space and let $f:X\raw(-\infty,\infty]$ be some function. The \emph{subdifferential} $\partial f(x)\subset X^*$ is defined by
\begin{align}
\partial f(x)& :=\{\phi\in X^*\mid \forall_{x'\in X}( \phi(x')-\phv(x)\leq f(x')-f(x))\} \:\:\: (f(x)<\infty);\\
\partial f(x)& :=\emptyset \:\:\: (f(x)=\infty).
\end{align}
\end{definition}
\bex\label{A15ex} 
 Show that if $f:\R\raw\R$ is in $C^1(\R)$ then  $\partial f(x)=\{f'(x)\}$.  \eex
\bex
Let $f(x)=|x|$ on $\R$. Show that 
\begin{align}\partial f(x)=\{-1\}\:\:\: (x<0); && \partial f(0)=[-1,1]; && \partial f(x)=\{1\}\:\:\: (x>0).
\end{align}\eex
This example generalizes as follows to normed spaces: if $f(x)=\|x\|$, then
\begin{align}
\partial f(0)=\{\phi\in X^*\mid \|\phi\|\leq 1\}; && \partial f(x)=\{\phi\in X^*\mid \|\phi\|= 1, \phi(x)=\|x\|\}\:\:\: (x\neq 0).
\end{align}
The subdifferential may also be (nontrivially) empty: a textbook example is the convex lsc function 
\begin{align}
f:\R\raw(-\infty,\infty]; && f(x)=-\sqrt{1-x^2}\:\:\: (-1\leq x\leq1); && f(x)=\infty\:\:\: (x\notin[-1,1]).
\end{align}
\bex Show that $\partial f(x)=f'(x)=\{x/\sqrt{1-x^2}\}$ for $x\in (-1,1)$ but $\partial f(x)=\emptyset$; not only for $x\notin[-1,1])$, which is true by definition, but even, within its domain, at $x=\pm 1$.
\eex
The subdifferential is defined also  if $f$ is not convex and/or lse, but it has better properties if $f$ is convex and lsc. ~For example, it
interacts nicely with the Fenchel transform. 
\begin{proposition}[Fenchel--Young inequality]
Let $X$ be a normed space and let $f:X\raw(-\infty,\infty]$ be an arbitrary function. Then for  any $x\in\mathcal{D}_f$ and $\phi\in X^*$ we have
\begin{equation}
f(x)+f^*(\phv)\geq \phi(x),
\end{equation}
with equality iff $\phi\in\partial f(x)$. The latter implies $x\in\partial f^*(\phv)$. If $f$ is convex and lsc, we even have
\begin{align}
\phi\in\partial f(x)&& \LRaw && x\in\partial f^*(\phv).\label{subsstar}
\end{align}
\end{proposition}
\bex Prove this proposition.\eex
\begin{proposition}\label{PropA13}
A function $f:X\raw(-\infty,\infty]$ on a normed space $X$ attains a local minimum at $x$  iff $0\in\partial f(x)$, and if $f$ is convex this is  a global minimum. 
 If in addition $f$ is lsc and $C$ is compact, then $f$ has a local and hence a global minimum.
   \end{proposition} 
This is clear from the definitions. Returning to Exercise \ref{A15ex}: in higher dimension the connection between the subdifferential and  the derivative 
requires the \emph{G\^{a}teaux derivative}:
\begin{definition}
A function $f:X\raw(-\infty,\infty]$ on a normed space $X$ is \emph{G\^{a}teaux differentiable} at $x\in\mathcal{D}_f$ if the following limit exists for all $x'\in X$ and defines (the \emph{G\^{a}teaux derivative}) $Df(x)\in X^*$:
\begin{align}
Df(x): X\raw\R; && y\mapsto Df(x,y):=
\lim_{h\raw 0} \frac{f(x+hy)-f(x)}{h}.
\end{align}
Similarly,  (one-sided) \emph{directional derivatives} $D^{\pm}f(x,y)$ are defined, if the limits exists,  by
\begin{equation}
D^{\pm}f(x,y):=\pm \lim_{h\downarrow 0} \frac{f(x\pm hy)-f(x)}{h}. \label{Dfxy}
\end{equation}
\end{definition}
\begin{proposition}
If  $f:X\raw(-\infty,\infty]$ is convex, as well as G\^{a}teaux differentiable at $x\in \mathcal{D}_f$, then
\beq
Df(x)\in \partial f(x).
\eeq
 If under the same assumptions $x$ is in the \emph{interior} of $\mathcal{D}_f$, then $\partial f(x)$ contains a single element:
\beq
\partial f(x)=\{Df(x)\}. \label{pfs}
\eeq
Conversely, if $f$ is convex, as well as \emph{continuous} at $x\in\mathcal{D}_f$, then \er{pfs} holds iff $f$ is G\^{a}teaux differentiable at $x$. 
\end{proposition}
Invoking Proposition \ref{PropA13}, we obtain:
\begin{corollary}\label{CorA17}
If $f$ is convex and lsc, and $f(0)$ is finite, then the set of minimizers of $f^*$ is $\partial f(0)$. Likewise, if $f^*(0)$ is finite, then   the set of minimizers of $f$ is $\partial f^*(0)$. Moreover, if $f(x)<\infty$, then
\begin{equation}
\partial f(x)=\{\phv\in X^*\mid \forall_{y\in X} \phv(y)\leq Df(x,y)\},
\end{equation}
\end{corollary}
\begin{proposition}
If $I\subset \R$ is an open interval (possibly $I=\R)$ and $f:I\raw (-\infty,\infty]$ is convex, then both the right (+) and left (-) derivatives 
\beq D^{\pm}f(x,1)\equiv D^{\pm}f(x)=\pm \lim_{h\downarrow 0} \frac{f(x\pm h)-f(x)}{h} \label{DfDf1}
\eeq
exist and are finite at each $x\in I$, and satisfy
 \begin{align}
D^+f(x)\leq D^-f(y)\leq Df^+(y)&& (x<y). \label{DfDf2}
 \end{align}
\end{proposition}
In the main text we need the case $X=X^*=\R$ and $f$  convex and lsc, and continuous on some compact interval $[a,b]\subset\R$ (if $f$ is convex and continuous on all of $\R$, then what follows remains valid
on $(-\infty,\infty)$ instead of $[a,b]$).\footnote{The following material is taken from Jak\v{s}i\'{c}, Ogata,  Pillet, \&  Seiringer (2012), \S 2.1.}
For consistency with the main text we switch notation 
\begin{align}
f\leadsto \Pi; && x\leadsto t; && \phv\leadsto x; &&f^*\leadsto I,
\end{align}
 and also find it convenient to define $\Pi(t):=\infty$ whenever $t\notin[a,b]$, which makes $\Pi:\R\raw(-\infty,\infty]$  lsc and convex.
 We then have the Fenchel dual pair
\begin{align}
I(x)&=\sup_{t\in\R}\{xt-\Pi(t)\} =\sup_{t\in[a,b]}\{xt-\Pi(t)\}; \\
\Pi(t)&= \sup_{x\in\R}\{ xt-I(x)\}. 
\end{align}
Recall that $D^-\Pi(t)\leq D^+\Pi(t)$, see \er{DfDf2}.
The last part of Corollary \ref{CorA17} implies that:
\begin{align}
\partial\Pi(a)&=(-\infty, D^+\Pi(a)]; && \partial\Pi(b)=[D^-\Pi(b),\infty);\label{A42}\\
\partial\Pi(t)&= [D^-\Pi(t), D^+\Pi(t)]; && (t\in (a,b)); \label{A43}\\
\partial\Pi(t)&= \emptyset;&& (t\notin [a,b]);\label{A44}\\
\partial I(x)&=[D^-I(x), D^+I(x)]; && (x\in\R).
\end{align}
If $[a,b]$ is replaced by $\R$ and $\Pi$ is finite, convex, and continuous on $\R$, we likewise simply have
\begin{align}
\partial \Pi(t)=[D^-\Pi(t), D^+\Pi(t)] && (t\in\R), 
\end{align}
and if $\Pi$ is $C^1$, this further simplifies to
\begin{equation}
\partial \Pi(t)=\{\Pi'(t)\}. \label{Piprime}
\end{equation}
It then easily follows from Proposition \ref{PropA13} and \er{subsstar} that:\footnote{For 1, note that for any $y$ and $t\in [a,b]$ we have $I(y)\geq ty-\Pi(t)$, so if $t\geq a \geq 0$ and $y>x$,  then 
$I(y) \geq ty-\Pi(t)=t(y-x) +tx-\Pi(t)\geq tx-\Pi(t)$, so taking $\sup_{t\in[a,b]}$ of the right-hand side we obtain $I(y)\geq I(x)$.
For 3, , coming from $-\infty$, $x$ reaches the  minimum of $I(x)$ iff $0\in \partial I(x)$ and hence $x\in\partial \Pi(0)$, which, if $0\in (a,b)$, means that $x=D^-\Pi(0)$, since
  \er{A43} gives $\partial \Pi(0)=[D^-\Pi(0), D^+\Pi(0)]$.
  Then $I(x)$ is constant $I(x)=-\Pi(0)$ 
  through $x\in \partial\Pi(0)$, and leaves this constant minimum value 
at $x\geq D^+\Pi(0)$, where it then necessarily starts increasing, initially strictly so. By convexity $I(x)$ cannot strictly decrease and since $x$ no longer lies in $\partial\Pi(0)$ it cannot stay constant either (strict increase may stop as soon as $I(x)$ takes the value $+\infty$).  If $\Pi$ is differentiable at $t=0$, this just means that $I(x)$ has a unique minimum at
$x=\Pi'(0)$. For 4, using \er{A42} and \er{subsstar}, we see that
$x\in\partial\Pi(a)$ and hence $a\in\partial I(x)$. This yields the claim by a somewhat deeper result in convexity: Let
$F(x)=\sup_t f_t(x)$. Then $\partial F(x)$ is the convex hull of all sets $\partial f_t$ for which $f_t(x)=F(x)$, i.e., for those $t$ where the supremum is attained. See e.g. Penot (2013), Proposition 3.38. Now clearly, if $f_t(x)=tx-\Pi(t)$,  then $\partial f_t(x)=\{t\}$, cf.\ \er{pfs}. 
 Hence if $a\in\partial I(x)$, then
$\sup_{t\in [a,b]} f_t$ must be attained at $t=a$.
}
\begin{enumerate}
\item \emph{If $a\geq0$ then $I(x)$ is   increasing for all $x$}.
\item \emph{If $b\leq 0$ then $I(x)$ is decreasing for all $x$}.
\item  \emph{If $0\in (a,b)$, then $I(x)$ is decreasing for  $x\leq D^-\Pi(0)]$ and increasing for  $x\geq D^+\Pi(0)]$. 
This is always true if $\Pi(t)$ is finite, convex, and continuous on $\R$.}
\item \emph{If $x\leq D^+\Pi(a)$, then $I(x)=ax-\Pi(a)$.}
\item \emph{If $x\geq D^-\Pi(b)$, then $I(x)=bx-\Pi(b)$}. 
\end{enumerate}
\section{Stochastic processes and Markov chains}\label{MarC}
A \emph{stochastic process} is a function $X$ of two variables $t\in T$ and $\om\in\Om$, where $T$ is (just) a set, $(\Om,\Sg, P)$ a probability space, and $X_t(\om)\in A_t$, for a measure space $(A_t,\Sg'_t)$, is such that each map
\beq
X_t:\Om\raw A_t
\eeq
 is measurable. In what follows we assume $(A_t,\Sg'_t)=(A,\Sg')$ to be independent of $t$.
 The simplest cases have discrete time, that is, $T=\N$ (the one-sided case) or $n\in\Z$  (two-sided), and 
$X_n:\Om\raw A$, where $A$ is finite. This vast generality is somewhat reduced by 
the \emph{Kolmogorov representation theorem} discussed below, which  states that without loss of generality one may assume that:
\begin{align}
\Om=A^{\N}\: \mbox{ or }\: \Om=A^{\Z};&& X_n(s)=s_n; && s:\N\raw A \: \mbox{ or }\: s:\Z\raw A.
\end{align}
The diagram in the Historical Introduction then displays a (truncated) sample path of a discrete-time stochastic process, and indeed all other interpretations given are special cases of this. 
This  theorem relies on \emph{Kolmogorov's extension lemma} (and is often conflated with it).  It takes no extra effort to discuss these theorems for arbitrary index sets $T$  and quite general state spaces $(A,\Sg')$, which for us are ``at worst'' Polish spaces.\footnote{
See e.g.\ Dudley (1989), \S 8.2 and 12.1, or Klenke (2020), \S 14.1, which textbooks we highly recommend. In the most general setting (Dudley, 1989, Theorem 12.1.2; Klenke, 2020, Theorem 14.39) each measure space $(A_t,\Sg'_t)$  is merely supposed to be isomorphic to some Borel set in $\R$. This includes all standard Borel spaces.} The reader may just think of finite $A$ with 
\beq
\Sg'=\CP(A).
\eeq
As usual, $A^T$ consists of all functions $s:T\raw A$. We define first the ``right'' $\sg$-algebra:\footnote{ Likewise, the canonical (= product) topology on $A^T$, where $(A,\CO)$ is a topological space and $T$ is just a set, is the coarsest (= smallest) topology that makes all coordinate functions continuous. This is the topology in which, famously, by Tychonoff's theorem $A^T$ is compact whenever $A$ is compact (for any $T$!).}
\begin{definition}\label{defF}
The \emph{cylindrical $\sg$-algebra} $\F$ on $A^T$ is the smallest $\sg$-algebra that makes each coordinate function (= evaluation map) $s\mapsto s(t)\equiv s_t$ from $A^T$ to $A$ measurable ($t\in T$).
\end{definition}
Equivalently,\footnote{If $(X_1,\Sg_1)$ and $(X_1, \Sg_2)$ are measure spaces,  $f:X_1\raw X_2$ is measurable if for each $B\in \Sg_2$ we have $f\inv(\Sg_2)\in\Sg_1$.}  as follows from the definition of measurability of functions, for each $B\in\Sg'$ and $t\in T$ the so-called \emph{cylinder set}
\beq
[B]_t:=\{s\in A^T\mid s(t)\in B\}
\eeq
 must be measurable, so that  $\F$ is the $\sg$-algebra generated by these cylinder sets.\footnote{Any collection of subsets $S\subset \CP(X)$ of $X$ generates a $\sg$-algebra $\Sg(S)$ on $X$, namely the smallest $\sg$-algebra on $X$ (i.e.,  in $\CP(X)$) that contains $S$; thus $\Sg(S)$ is just the intersection of all $\sg$-algebras on $X$ that contain $S$, and this is non-empty because $\CP(X)$ obviously contains $S$ and is a $\sg$-algebra. We say that $\Sg(S)$ is the $\sg$-algebra \emph{generated by $S$}. 
The best-known example is the  \emph{Borel} $\sg$-algebra, which is generated by the open sets of some topology on $X$.  In fact, the cylindrical $\sg$-algebra $\F$ on $A^T$ is a special case of this, using the product topology (which has the same basis).
} If
 for any \emph{finite} subset $F\subset T$ we define $\F_F\subset\CP(A^T)$ as the $\sg$-algebra generated by the \emph{rectangular sets} 
 \begin{align}
\left[\prod_{t\in F} B_t\right]_F:= \{s\in A^T\mid \forall_{t\in F} s_t\in B_t\}=\left\{s\in A^T\mid s_{|F}\in \prod_{t\in F} B_t\right\} && (B_t\in\Sg',\, t\in F),
 \label{33}
 \end{align}  
 then $\F$  is the $\sg$-algebra generated by all  $\CF_F$ ($F\subset T$ finite). This is useful, if only because of:
 \begin{theorem}\label{Cara}
 Any probability measure $P$ on $\CF$ is uniquely determined by its values on $\bigcup_F\CF_F$, and even by its values on all rectangular sets \er{33}, where $F\subset T$ is finite and $B_t\in\Sg'$ ($t\in F$). 
 \end{theorem}
 Since $\bigcup_F\CF_F$ is an algebra, the first part follows from  the Carath\'{e}odory extension theorem,\footnote{See e.g.\ Klenke (2020), Theorem 1.41. Carath\'{e}odory's theorem states  that if a $\sg$-algebra $\Sg$ is generated by an algebra $R\subset\Sg$ (see footnote \ref{DK1}), then any 
 countably additive set function on $R$ taking values in $[0,\infty]$ (i.e.\ any premeasure) uniquely extends to a 
 measure on $\Sg$; this trivially implies that $P$ is uniquely determined by its values on $R$.} but the very useful second part does not.\footnote{See Klenke (2020), Theorem 14.12.} 
If we (tacitly) define the appropriate $\sg$-algebra $\Sg'_F$ on $A^F$ as  the one from Definition \ref{defF}, with $T\leadsto F$, we may also define more general cylinder sets in $A^T$ by replacing the rectangular sets $\prod_{t\in F} B_t$ in \er{33}, by arbitrary sets $\CB\in\Sg'_F$. Then $\CF_F$ simply consists of all $[\mathcal{B}]_F$, where $\CB\in\Sg'_F$; in particular, this is already a $\sg$-algebra. 
 
 This simplifies if $A$ is finite with $\Sg'=\CP(A)$: for any $\sg\in A^F$, define
 \beq
[\sg]_F:=\{s\in A^T\mid \forall_{t\in F} (s(t)=\sg(t))\}= \{s\in A^T\mid s_{|F}=\sg\}. \label{33b}
\eeq
Then it is easy to see that $\CF_F$ is the $\sg$-algebra generated by all such sets, and, invoking Theorem \ref{Cara}, any probability measure $P$ on $\CF$ is uniquely determined by its values on the sets $[\sg]_F$.
 If also $T=\N$, one may even restrict attention to the sets $[\sg]_F$ where $F\subset\N$ is restricted to
\beq
F=N=\{0, 1, \ldots, N-1\}\subset\N,
\eeq 
so that $[\sg]_N$ is our familiar cylinder set \er{cylinder}. Indeed, the $\sg$-algebras $\F_N$ generated by these  $[\sg]_N$
generate $\CF$,
since both $\bigcup_N \F_N$ and
$\bigcup_F \F_F$ (the union now again being taken over all finite $F\subset\N$) contain the following elementary cylinder sets which by Definition \ref{defF} generate $\CF$:
\begin{equation}
[a]_n:=\{s\in A^\N\mid s_n=a\}, \:\:\: n\in\N,\, a\in A\}.
\end{equation}
In fact, $\bigcup_N \F_N$ consists of all finite unions of the sets $[\sg]_F$ (and is an {algebra).}\footnote{See footnotete \ref{DK1} for the definition of al algebra of subsets.}
The simplest example of this construction is $A=2=\{0,1\}$ and $P=p^\N$ for any $p\in\Pr(2)$, cf.\ \er{pompN}.
\bex \label{diffex} For $f(0)=f(1)=1/2$, show that the binary expansion \er{xomega} induces an isomorphism \emph{of probability spaces} (where $\mu_L$ is Lebesgue measure defined on the Borel sets $\mathcal{B}$):\footnote{This is false  in topology, where $2^\N$ is a Cantor space! The proof requires a generalization of the theorem  in footnote \ref{DK1}.
 Let $(X_1,\Sg_1, P_1)$ and $(X_2,\Sg_2, P_2)$ be probability spaces (or more generally just measure spaces), let $R_2\subset \Sg_2$ be a \emph{generating semi-algebra} for $\Sg_2$, that is,
$\emptyset\in R_2$; if $A,B\in R_2$ then $A\cap B\in R_2$; and if $B\in R_2$ then $X\backslash B$ is a \emph{countable} disjoint union of elements of $R_2$; and $R_2$ generates $\Sg_2$. Also assume that $R_2$ contains an increasing sequence $(B_n)$ such that $X_2=\bigcup_n B_n$.
Then: if $f:X_1\raw X_2$ is a map that satisfies $f\inv(B)\in \Sg_1$ and $P_1(f\inv(B))=P_2(B)$ for each $B\in R_2$,  then $f$ is measurable and $P_1(f\inv(B))=P_2(B)$ for each $B\in \Sg_2$.
 See e.g.\ Dajani, \& Kalle (2021), Theorem 12.3.1.\label{DK2}
}
\begin{align}
(2^\N, \F, f^\N)\cong ([0,1], \mathcal{B}, \mu_L). \label{nietinlit}
\end{align}
\eex
Let us return to Theorem \ref{Cara}. Note that $A^F$ is not a subset of $A^T$, but that there 
are natural maps 
\begin{align}
\pi_F: A^T\raw A^F; && \pi_F(s)=s_{|F};  \label{defpiF}\\
 \pi_{GF}: A^G\raw A^F; &&   \pi_{GF}(\sg)=\sg_{|F}, \label{defpiGF}
\end{align}
where $F\subseteq G\subset T$, \emph{in which $F$  and $G$ are always finite} in what follows, although the maps $\pi_{GF}$ are defined more generally. It is clear from \er{defpiF} that any $P\in\Pr(A^T)$ induces $p_F\in\Pr(A^F)$ via
\begin{align}
p_{F}:=\pi_{F}\inv P,  \label{Kcc}
\end{align}
 for each  $F\subset T$. For $\CB\in\Sg'_F$, e.g.\ the rectangular set $\prod_{t\in F}B_t$ as in \er{33},  eq.\ \er{Kcc} gives
 \begin{equation}
p_F(\CB)=P([\CB]_F)=P(s_{|F}\in \CB). \label{310}
\end{equation}
For example, for $A$ finite, $T=\N$,  $F=N$, and $\CB=\{\sg\}$ with $\sg\in A^N$, eq.\  \er{310} simply reads
\begin{equation}
p_N(\sg)\equiv p_N(\{\sg\})=P([\sg]_N). \label{311N}
\end{equation}
In  general, since $\pi_F=\pi_{GF}\circ \pi_G$ if $F\subseteq G\subset T$, the  $p_F$ 
 satisfy the \emph{consistency condition}
  \begin{equation}
p_{F}=\pi_{GF}\inv p_G\label{38} .
\end{equation} 
The  \emph{Kolmogorov extension lemma} turns the  consistency condition \er{38} on its head:
\begin{lemma}[Kolmogorov]\label{KolET}
If a  family of probabilities  $(p_F\in\Pr(A^F,\Sg'_F))$  satisfies  \er{38},  there is a unique probability measure $P$ on $(A^T,\F)$  that induces the given family $(p_F)$ via \er{Kcc}.
\end{lemma}
The idea of the proof is  simple: the cylinder sets $[\CB]_F$, where $\CB\in\Sg'_F$, form an algebra $\bigcup_F \CF_F$ in $\CP(A^T)$, and one defines $P$ on this algebra by reading \er{310} from right to left; this move is of course only valid if the consistency condition \er{38} holds. Since $\CF$ is by definition equal to the $\sg$-algebra generated by $\bigcup_F \CF_F$,
 the Carath\'{e}odory extension theorem does the rest \emph{provided $P$  is countably additive on $\bigcup_F \CF_F$}. This is the difficult  part we omit;  some extra assumptions on the measure space $(A,\Sg')$, e.g., being Polish,  are needed for this step.\footnote{
 Since each $p_F$ is given as a probability measure, we know that $P$ is countably additive on each $\CF_F$, which as we already mentioned is a $\sg$-algebra, but this is not enough for countable additivity on the union $\bigcup_F \CF_F$. The proof is even difficult for $A$ finite and $T=\N$, which case is spelled out by Calude (2002), Theorem 1.7.}
 
 Recall that, for a finite set $A$, the set $A^*=\bigcup_{N\in\N} A^N$ consists of  all finite $A$-valued strings\footnote{We here include $A^0$ consisting of the empty string $\ep$.}
 \begin{exercise}\begin{enumerate}
\item
Show that for  finite $A$ and $T=\N$ the consistency condition \er{38} implies
\beq
p_N(\sg)= \Sigma_{a\in A} p_{N+1} (\sg a), \label{cc}
\eeq
 for all $N\in\N$, $\sg\in A^N$ and $a\in A$,
where $\sg a \in A^{N+1}$ is the string defined by 
\begin{align}
(\sg a)_n=\sg(n)\equiv\sg_n\:\:\: (n=0, \ldots, N-1); && (\sg a)_{N}=a. 
\end{align}
\item Show that a  family of functions $(p_N)$ that satisfies \er{cc} is equivalent to a single function 
\begin{align}
p:A^*\raw [0,1]
\end{align}
 that satisfies 
 \begin{align}
 p(\sg)=\Sigma_{a\in A} p(\sg a); && p(\ep)=1.
 \end{align} 
\item Using Lemma \ref{KolET}, show that such a function $p$, or the equivalent $(p_N)$ in part 1,  already uniquely determines a probability measure $P\in\Pr(A^\N)$ that returns the $p_N$ via \er{311N}. \end{enumerate}
\end{exercise}
Here, then, is the fundamental \emph{Kolmogorov representation theorem}: 
\begin{theorem}[Kolmogorov]\label{KolRT}
Let $(X_t)_{t\in T}$ be a stochastic process taking values in (for example) a Polish state space $(A,\Sg')$,
so that $X_t:\Om\raw A$ for some  probability space  $(\Om,\Sg,P)$.
Then
 without loss of generality one may realize the process as evaluation maps $X_t'$ on $(A^T,\CF)$, that is,
\begin{align}
X'_t: A^T\raw A;&& X'_t(s)=s_t; && s: T\raw A, \label{Kct}
\end{align}
with a probability measure $P'\in\Pr(A^T,\CF)$ that returns the joint probabilities of the $(X_t)$ via 
\begin{equation}
P(X_t\in B_t, t\in F)=P\left(\left[\prod_{t\in F} B_t\right]\right)=P'(s_t\in B_t, t\in F), \label{pFdef} 
\end{equation}
for each finite  $F\subset T$ and  $(B_t)_{t\in F}$, ($B_t\in\Sg'$), with ensuing product $\prod_{t\in F} B_t\in\Sg'_F$, cf.\ \er{33}.
\end{theorem}
Here the symbolic notation  $P(X_t\in B_t, t\in F)$ stands for 
\beq
P(X_t\in B_t, t\in F)\equiv
P(\{\om\in\Om\mid \forall_{t\in F}(X_t(\om)\in B_t)\}),
\eeq
 and likewise $P'(s_t\in B_t, t\in F)$ abbreviates 
 \beq P'(s_t\in B_t, t\in F)\equiv P'(\{s\in A^T\mid  \forall_{t\in F}(s_t \in B_t)\}).\eeq
\emph{Proof}. 
For each finite $F\subset T$, define  $p_F\in\Pr(A^F,\Sg'_F)$ via a little variation on \er{310}, namely
\beq
p_F\left(\prod_{t\in F} B_t\right):= P(X_t\in B_t, t\in F). \label{311}
\eeq
The difference between \er{311} and \er{310} lies in the fact that the latter defines $p_F$ via $P\in\Pr(A^T)$, whereas in the former we are given $P\in \Pr(\Om)$ and attempt to construct $P'\in\Pr(A^T)$ from this $P$, namely via \er{311}.
This $p_F$ then has a unique extension to $p_F\in\Pr(A^T,\Sg'_F)$.\footnote{See Klenke (2020), Theorem 14.12.(iii).}
\bex
Show that the ensuing family $(p_F)$ satisfies the consistency condition \er{38}.
\eex
Lemma \ref{KolET}  gives us the  distribution $P'$, upon which \er{pFdef} follows from \er{310}. \QED\smallskip

 \noindent For finite $A$ and $T=\N$, for each $N\in\N$ and $\sg\in A^N$
eqs.\ \er{pFdef} - \er{311} simply come down to \begin{equation}
P(X_0=\sg_0, \ldots, X_{N-1}=\sg_{N-1}) = P(s_{|N}=\sg)=P([\sg]_N)=p_N(\sg). \label{1.52}
\end{equation}

This remains quite general. We now define a number of important special cases (with $A$ always finite), where as usual  conditional probabilities are defined as follows:   provided $P(C)>0$,
\beq
P(B|C):=\frac{P(B\cap C)}{P(C)}.\eeq  
\begin{itemize}
\item In the original description of our stochastic process, the random variables  $(X_n)$ are called \emph{independent} (with respect to $P$)
iff any finite subset of it is independent (\emph{idem dito}). In particular,
 for each $n\in\N$, $a\in A$ and $\sg=(\sg_0, \ldots, \sg_{N-1})\in A^N$, we then have
\begin{equation}
P(X_N=a|X_0=\sg_0, \ldots, X_{N-1}=\sg_{N-1})=P(X_N=a). \label{XNi}
\end{equation}
In the Kolmogorov model the condition \er{XNi} means  that
\begin{equation}
P(s_N=a|s_{|N}=\sg)\equiv P(s_N=a|s_0=\sg_0, \ldots, s_{N-1}=\sg_{n-1})=P(s_N=a). \label{iidK}
\end{equation}
\item The random variables  $(X_n)$ are \emph{identically distributed} (i.d.)  if $P(X_N=a)$ is independent of $N$ and hence equal, for all $N\in\N$,  to $p(a)$ for some $p\in\Pr(A)$. The $(X_n)$ are called 
\emph{i.i.d.} if they are both independent and identically distributed. In that case,\footnote{This may be seen as a repeated sampling of $(A,p)$, where $X_n$ is the $n$'th try.
If $(X,P,T)$ is a dynamical system and $f:X\raw\R$ or $f:X\raw A$ is a random variable, then the process $X_n=f\circ T^n$ is i.d.\ (since $T$ preserves $P$) but not i.i.d.}
 \beq
 P=p^\N.\eeq
\item In the next step of generality, the $(X_n)$ form a \emph{Markov chain} if, in the original model,
\begin{equation}
P(X_N=a|X_0=a_0, \ldots, X_{N-1}=a_{N-1})=P(X_N=a\mid X_{N-1}=a_{N-1}), \label{MC}
\end{equation}
or, similarly to \er{iidK}, in the Kolmogorov model, for all $N\in\N$, $a\in A$, and $\sg\in A^N$,
\begin{equation}
P(s_N=a|s_{|N}=\sg)=P(s_N=a| s_{N-1}=\sg_{N-1}). \label{MCK}
\end{equation}
\item 
A Markov chain is  \emph{stationary} iff  $P(X_N=a\mid X_{N-1}=b)$
does not depend on $N$, so that
\begin{equation}
P(X_0=a_0, X_1=a_1,\ldots,  X_{N}=a_{N})=P(X_M=a_0, X_{M+1}=a_1,\ldots,  X_{M+N}=a_{N}),
\end{equation}
for all $M, N>0$.
 We may then conveniently introduce \emph{transition probabilities} by
\begin{equation}
P_{ab}:=P(X_{N+1}=b\mid X_N=a)=P(X_1=b\mid X_0=a). \label{defTP}
\end{equation}
This is a \emph{stochastic matrix}, in that all entries are non-negative and 
\beq
\Sigma_b P_{ab}=1 \label{sumb}
\eeq
for each $a$.
In a stationary Markov chain, $P(X_n=a)$ is independent of $n$, and we write
\begin{equation}
p(a)\equiv p_a:= P(X_n=a)=P(X_0=a). \label{defpaMC}
\end{equation}
For example, for $P=p^\N$ we have $P_{ab}=p_b$ for all $a$.
\bex 
\begin{enumerate}
\item 
Show that in a stationary Markov chain (with finite state space $A$):\footnote{Physicists: be warned that repeated indices are \emph{not} summed unless there is an explicit sumation sign!}
\begin{align}
P(X_{N}=b\mid X_0=a)&=(P^N)_{ab}; \label{pTP}\\
P(X_0=a_0, X_1=a_1,\ldots, X_{N-1}=a_{N-1}, X_{N}=a_{N})&=p_{a_0} P_{a_0a_1}\cdots P_{a_{N-1}a_{N}}; \label{SMC}
\\
P(X_1=b, X_2=c)&=p_bP_{bc}; \\
\Sigma_a p_a P_{ab}&=p_b.\label{paP}
\end{align}
\item Conversely, show that any stochastic matrix $(P_{ab})$ plus some probability distribution $p\in\Pr(A)$ that satisfies \er{paP} defines a stationary Markov chain via \er{SMC}.
\end{enumerate}
\eex
 Stationarity also implies that if $p(a)=0$ for some $a\in A$, then any occurrence $X_n=a$ has zero probability, so that we may remove $a$ from $A$. In what follows we therefore assume that  $p(a)>0$ for each $a\in A$. As a check on \er{paP}, we already noted that for the i.i.d.\ case $P=p^\N$ we have 
 \beq
 P_{ab}=p_b,
 \eeq so that \er{paP} reduces to $\Sigma_ap_ap_b=p_b$.
\item
 The existence of a left-eigenvector $p$ of a given stochastic matrix $(P_{ab})$ is guaranteed by the \emph{Perron--Frobenius theorem}, but  this is not enough to conclude  $p\in\Pr(A)$. In the cleanest case, which is central to both ergodic theory and large deviation theory,   $(P_{ab})$ is \emph{irreducible}: \end{itemize}
 \begin{definition}\label{IMC}\begin{enumerate}
\item A stationary Markov chain is called \emph{irreducible} if for all $a,b\in A$ there is $N\in\N$ such that
\beq
P(X_N=b|X_0=a)>0.
\eeq
\item  Equivalently, its stochastic matrix $(P_{ab})$ (and  generally a matrix with nonnegative real entries)  is  \emph{irreducible} if for all $a,b\in A$ there is $N\in\N$ such that $(P^N)_{ab}>0$.
\item And this is equivalent to the existence, for all $a,b\in A$,  of indices $(a_0, \ldots, a_N)$ with $a_0=a$ and $a_N=b$ such that
\begin{align} 
P_{a_ia_{i+1}}>0 && (i=0, \ldots, N-1).
\end{align}
\end{enumerate}
\end{definition}
\bex
\begin{enumerate}
\item   Prove that these criteria are indeed equivalent.
\item Is  a Markov chain defined by  the i.i.d.\ case $P=p^\N$ irreducible?
\end{enumerate}
 \eex
 An interesting stationary irreducible Markov chain that is not i.i.d.\ is the  \emph{Ehrenfest urn model}.\footnote{This model was introduced by Ehrenfest \& Ehrenfest (1907ab) as a toy model for the Boltzmann equation. The literature on the Ehrenfest model (not to speak of its wider context) is huge;  Bricmont (2022), \S8.7.1, is a good starting point. See also Dekkers \& Landsman (2025) for a link with algorithmic randomness as discussed in Appendix \ref{EKR} below.} This model has $A=\{0,1,\ldots, M\}$, so that $a\in A$ is a number $0\leq m\leq M$, seen
as the number of balls (or fleas) on one of two urns (or dogs), labeled $U_1$ (the other is called $U_0$). The random variable $X_n$ describes the number of balls in $U_1$ a time $n$ (so that if $X_n=m$, then $M-m$ is the number of balls in $U_0$). The transition probabilities are given by:
\begin{align}
P_{m,m-1}:=\frac{m}{M} \:\:\: (m\geq 1); && P_{m,m+1}:=\frac{M-m}{M} \:\:\: (m<M); && P_{mn}=0\:\:\: (n\neq m\pm 1).
\end{align}
The idea is that at each time step $n$ some ball  is randomly and fairly chosen from the  set of $M$ balls, and is then moved to the other urn. If $U_1$ contains $m$ balls the probability that the chosen ball is in $U_1$ is  $m/M$, in which case the total number of balls in $U_1$ after the jump equals $m-1$. Conversely, the probability that the chosen ball is in $U_0$ equals $1-(m/M)$, and after its jump to $U_1$ the total number of balls in $U_1$ is $m+1$. For example, for $M=4$ the transition probability matrix  equals
\begin{equation}
(P_{mn})=\left(
\begin{array}{ccccc}
0  & 1  &0 &0 &0   \\
 1/4 & 0  & 3/4 & 0& 0 \\
 0 & 1/2  & 0  & 1/2 & 0\\
  0  & 0  &  3/4 & 0&1/4 \\
  0    &   0& 0  &1 & 0\\
\end{array}
\right).
\end{equation}
Check that this matrix is indeed stochastic: each row adds up to unity, cf.\ \er{sumb}.
Irreducibility is easy to see without any computation from part 1 of Definition \ref{IMC}: one simply imagines  possible transfers needed to get from any number of balls on $U_1$ to any other number. 
The key result is that the stationary probability $p\in\Pr(M)$ is given by the binomial distribution
\begin{equation}
p_m=2^{-M}
\left(
\begin{array}{c}
M\\
m
\end{array}
\right), \label{binEF} 
\end{equation}
which  is the probability distribution if the balls were randomly distributed over the two urns in the first place. If $M$ is even, the value $m=M/2$ has the highest probability and for large $M$ all  values far away are increasingly unlikely.  
\begin{exercise}
Prove \er{binEF}.
\end{exercise} There is also a finer version of the Ehrenfest model, in which $A=2^{M}$ and each $a\in A$ gives the detailed distribution of the balls (i.e., its microstate): a state $a$ now specifies 
$a_k\in\{0,1\}$  for each $k=0, \ldots M-1$, which means that ball number $k$ is in urn $a_k$. We now have
$P_{ab}> 0$ iff $b_k=a_k$ for all $k$ except for one value $k'$ (so that the jump flips the bit $a_{k'}$), in which case $P_{ab}=1/M$. Since for each $b$ there are $M$ microstates $a$ that differ from $b$ by exactly one bit, this matrix duly satisfies \er{sumb}. For the same reason the stationary probability $p\in\Pr(2^M)$ is given by the flat one, i.e., 
 \beq
 p_a=2^{-M} \:\:\: \mbox{for each}\:\: a\in 2^M.
 \label{pa2M} 
 \eeq\vspace{-5mm}
  \begin{exercise}
  Show that \er{pa2M}  satisfies \er{paP}.
  \end{exercise}

In studying the relationship between irreducibility, ergodicity and large deviations  the following well-known linear algebra theorem is often used; but it also clarifies Definition \ref{IMC}. 
\begin{theorem}[Perron--Frobenius] \label{limlemma}
 Let  $(P_{ab})$ be an irreducible  real  matrix  with each 
 \beq
 P_{ab}\geq 0.
 \eeq 
\begin{enumerate}
\item The matrix $(P_{ab})$ has a unique nondegenerate real eigenvalue $\rh$ with a corresponding left eigenvector  $p$ having strictly positive entries  
\beq
p_a>0.
\eeq
\item If the matrix  $(P_{ab})$ is stochastic, then $\rh=1$, and we may normalize its eigenvectors $p$ to achieve $p\in\Pr(A)$; the stationarity condition  \er{paP} then holds by construction. 
\item For every strictly positive vector $v\in\R^{|A|}$ (i.e.\  $v_a>0$ for each $a$), and arbitrary $b\in A$,
\begin{equation}
\lim_{N\raw\infty}\frac{1}{N} \log\left(  \Sigma_{a\in A} v_a(P^N)_{ab}\right)=\log\rho. \label{PFTlog}
\end{equation}
\item The corresponding (one-dimensional) spectral projection $E^{(\rh)}$ for $\rh$ is given by
\begin{equation}
E^{(\rh)}_{ab}=\lim_{N\raw\infty}\frac{1}{N} \Sigma_{n=1}^N\rh^{-n} P^n_{ab}.
\end{equation} 
 In particular,
in the irreducible stochastic case  $p\in\Pr(A)$ is given
 for arbitrary $a\in A$ by
\begin{equation}
p_b=\lim_{N\raw\infty}\frac{1}{N} \Sigma_{n=1}^NP^n_{ab}.\label{limP}
\end{equation}
\item If there exists $M>0$ such that 
$P^M_{ab}>0$ for all $a,b\in A$ (in which case $(P_{ab})$ is irreducible),\footnote{We have $P^M_{ab}>0$ for all $a,b$ for some $M>0$ iff $(P_{ab})$ is irreducible and \emph{aperiodic}, i.e., for each $a\in A$ we have
$d(a):=\mathrm{gcd}\{k\in\N_*\mid P^k_{aa}>0\}=1$.  In the presence of irreducibility, $d(a)=1$ needs to be checked just for a single $a$.}
 then the C\'{e}saro limit in \er{limP} may be replaced by the ordinary limit
\begin{equation}
p_b=\lim_{N\raw\infty}(P^N)_{ab}.\label{limP2}
\end{equation}
\item Conversely, if $(P_{ab})$ is  stochastic  then
the limit \er{limP} exists;  if the resulting vector $p$: \\  (i) lies in  $\Pr(A)$
and (ii) solves \er{paP}, then $(P_{ab})$ is irreducible.
\end{enumerate}
\end{theorem} 
For example, in the i.i.d.\ case $P=p^\N$ we have 
\begin{align}
P_{ab}=p_b && (a\in A),
\end{align} so that $P^2=P$ and hence the limit in \er{limP} equals $p$. The ensuing Markov chain  is irreducible provided $p_a>0$ for each $a\in A$.
 \section{Ergodic theory}\label{ErgApp}
Ergodic theory goes back to Boltzmann in his work on statistical mechanics, but the corresponding mathematical theory was primarily developed outside this specific context (and inside measure theory and dynamical systems)
by Birkhoff (senior) and von Neumann.\footnote{See Halmos (1958), Mackey (1974), von Plato (1994) and Moore (2015) for some history of ergodic theory. For the theory itself we recommend Tao (2007) and in more detail and polish Dajani \& Kalle (2021) for student-friendly  first introductions, and Viana \& Oliveira (2016) for a full meal. Walters (1982) is intermediate and still useful. } From a probabilistic point of view, we  saw that 
 if $(X,P,T)$ is a dynamical system and $f:X\raw\R$ is a random variable, then the process $X_n=f\circ T^n$ is i.d.\ but not i.i.d. The pointwise ergodic theorem will generalize the strong law of large numbers from i.i.d.\  to i.d.\ variables, using the following key concept.
 \begin{definition}\label{defE1}
Let $(X,P,T)$ be a dynamical system, i.e.\ a probability space $(X,\Sg,P)$ with measure-preserving map $T:X\raw X$. This map is \emph{ergodic} if
 $T\inv(B)= B$ implies
$P(B)=0$ or $P(B)=1$.
\end{definition}
Note that the logical structure is: $\forall_{B\in\Sg} ((T\inv B=B)\raw (P(B)=0\vee P(B)=1))$, and hence ergodicity is equivalent to the non-existence of some $B\in\Sg$ for which $T\inv (B)=B$ and $0<P(B)<1$.  Up to measure zero sets, $P(B)=0$ means that $B=\emptyset$, whereas $P(B)=1$ means that $B=X$, so that ergodicity essentially means that the only nonempty $T$-invariant subset $B\subseteq X$ is $X$. 

By iteration of the antecedent, Definition \ref{defE1} implies that $(X,P,T)$ is ergodic iff $T^{-n}B=B$ for all $n\in\N$ implies $P(B)=0$ or $P(B)=1$. If $T$ is invertible, it gives rise to a $\Z$-action on $X$ via $n\cdot x:= T^n x$ (where $\Z$ is seen as an additive group). From that point of view, ergodicity (for invertible $T$) is equivalent to:
any  $\Z$-invariant set $B$ satisfies either $P(B)=0$ or $P(B)=1$. 
\bex
Show that ergodicity is equivalent to any of the following properties:\footnote{Recall that  the \emph{symmetric difference} of  $B,C\subset X$ is
$B\Delta\, C:=(B\backslash C)\cup (C\backslash B)$.}
\begin{enumerate}
\item  $T\inv(B)\subseteq B$ implies
$P(B)=0$ or $P(B)=1$.
\item $T\inv(B)\supseteq B$ implies
$P(B)=0$ or $P(B)=1$.
\item $P(T\inv(B)\Delta\, B) =0$ implies
$P(B)=0$ or $P(B)=1$.
\end{enumerate}
\eex
The simplest nontrivial examples of  ergodic systems are given by certain operations on  the torus 
\beq
\T:=\{z\in\C\mid |z|=1\},
\eeq parametrized by $z=\exp(ix)$, and equipped with  normalized Lebesgue measure (i.e.\ Haar measure) 
\begin{align}
dP(z)=\frac{dz}{2\pi i z}; && \LRaw &&
dP(x)=\frac{dx}{2\pi},
\end{align}
so that $P(\T)=1$.
 The following result is similar to the doubling map and the tent map, see \S\ref{EDS}. 
 \bex
 Show that  Lebesgue measure on $\T$ (which is the group-theoretic Haar measure) is invariant under rescaling by an integer $p\neq 0$.
 \eex
 The cases of interest are \emph{irrational rotations} and \emph{scalings},  where $\al\in[0,2\pi)$ and $p\in\Z_*$:
\begin{align}
R_{\al}: \T\raw\T; && z\mapsto e^{i\al} z; && x\mapsto x +\al\:\:\: \mathrm{mod}\, 2\pi\:\:\: (x\in[0,2\pi); \label{irr}\\
M_p: \T\raw\T; &&z\mapsto z^p; &&  x\mapsto px \:\:\: \mathrm{mod}\, 2\pi\:\:
\end{align}\vspace{-5mm}
 \begin{proposition}  $(\T,P,R_{\al})$ is ergodic iff $\al/(2\pi)$ is irrational, and $(\T,P,M_p)$ is ergodic iff $|p|\geq 2$. \end{proposition}
  Although this  can be proved directly from Definition \ref{defE1},\footnote{See e.g.\ 
 Shields, Prop.\ I.2.14. His  proof relies on \emph{Kronecker's theorem}:
 the orbit $(R_{\al}^N)_{N\in\N}$ is dense in $\T$ iff $\al$ is irrational. This was also Boltzmann's original intuition about ergodicity.}  it is instructive to give an analytic proof based on a reformulation of ergodicity that will often be used.
 \begin{proposition}\label{EDE}
 Let $(X,P,T)$ be a dynamical system. The following conditions are equivalent, and hence $(X,P,T)$ is ergodic iff any of these conditions hold:
 \begin{enumerate}
\item For all measurable $B\subseteq X$: If $T\inv(B)= B$, then
$P(B)=0$ or $P(B)=1$.
\item For all $f\in L^2(X,P)$: If  $f(Tx)=f(x)$ $P$-a.e., then $f$ is constant $P$-a.e.
\item For all $f\in L^1(X,P)$: If  $f(Tx)=f(x)$ $P$-a.e., then $f$ is constant $P$-a.e.
\item For all measurable $f:X\raw\R$:  If  $f(Tx)=f(x)$ $P$-a.e., then $f$ is constant $P$-a.e.
\item There exists no nontrivial convex decomposition (the trivial case being $P_1=P_2=P$)
\beq
P=tP_1+ (1-t)P_2, \label{convexsum}
\eeq
 where $t\in (0,1)$ and $P_1$ and $P_2$ are also $T$-invariant probability measures.\footnote{
Here the space $\Pr(X)$ of all probability measures on $X$ is seen as a convex set, see Definition \ref{defPK}.
If $X$ is compact, the so is $\Pr(X)$ (in the weak or vague topology), so that its closed subspace $\Pr^T(X)$ of all $T$-invariant probability measures is also a compact convex set. Within the triple $(X,P,T)$, we here vary $P$ rather than $T$.} That is, 
 \begin{equation}
P\in\partial_e \Pr^T(X). \label{44}
\end{equation}
\end{enumerate}
 \end{proposition}
 Here $\Pr^T(X)$  is the set of $T$-invariant probability measures on $X$, and
we take the $L^p$-spaces to consist of real-valued, but could equally well use $\C$ instead of $\R$. 
 \medskip

\noindent \emph{Proof.} 
 The  implication $4\raw 3$ is trivial, and $3\raw 2$ follows from easy functional analysis:
 \bex \label{E15} Prove $3\raw 2$ by showing that $L^2(X,P)\subseteq L^1(X,P)$ 
  for probability measures $P$.
  \eex
The real idea is in the implications $2\raw 1$ and $1\raw 4$, which rely on three simple facts:
\begin{enumerate}
\item $1_B\circ T=1_{T\inv B}$.
\item  $1_{T\inv B}=1_B$ $P$-a.e.\ iff  $T\inv B =B$ (up to $P$-measure zero). 
\item $1_B$ is constant $P$-a.e.\ iff $P(B)=0$ or $P(B)=1$.
\end{enumerate}
This  gives the implication $2\raw 1$ (see exercise).
The implication $1\raw 4$ contains a technical part we omit, and an easy part based on the above facts.
The easy part concerns the case  $f= \Sigma_i c_i 1_{B_i}$. The technical part 
(see e.g.\ Walters, 1982, Theorem 1.6) is that since $f$ is measurable, it may be approximated by such (finite) sums, and the easy part survives this approximation procedure.
\bex
Prove the above facts, and from these, prove   $2\raw 1$ as well as the easy part of $1\raw 4$.
\eex
We prove $5\raw 1$  contrapositively (so we prove $\neg 1\raw \neg 5$). If $P$ is not ergodic ($\neg 1$),
 by Definition \ref{defE1} there is 
is $B\subset X$ such that $0<P(C)<1$. But this implies $\neg 5$, since we may put:
\begin{align}
P=tP_1+(1-t)P_2; && t:=P(B); && P_1(C):=P(C|B); && P_2(C):=P(C|B^c).  
\end{align} 
The converse $1\raw 5$ relies on a lemma of independent interest (with a luxury bonus part).
\begin{lemma}\label{VOLemma}
If $P,Q\in\Pr^T(X)$ with $Q$ ergodic, then $P\ll Q$ iff $P=Q$. If $P$ is  ergodic, too, then either $P=Q$ or $P\perp Q$ in  that there is a measurable $B\subset X$ such that $P(B)=1$ and $Q(B)=0$). 
\end{lemma}
\bex Prove  this lemma. 
\eex
Since  \er{convexsum}  implies that
$P_1\ll P$ and $P_2\ll P$, the first part of the lemma gives $1\raw 5$.\QED
\smallskip

\noindent The equivalence $1\lraw 5$  is interesting even in the trivial case $T=\mathrm{id}_X$, i.e., $T(x)=x$.
Since every $B\in\Sg$ then satisfies $T\inv B=B$, a probability measure  $P$ is ergodic iff $P(B)=0$ or 1 for any $B\in\Sg$. According to criterion 5, this should be equivalent to $P\in\partial_e \Pr(X)$, as it is by Proposition \ref{P1310}. 

We now prove condition 2 of Proposition \ref{EDE} for the irrational rotation \er{irr}.\footnote{We follow Shields, end of \S I.2.b, and Viana \& Oliveira, \S 4.2.1.} Any $f\in L^2(\T)$ has a Fourier series, so that, realizing $L^2(\T)$ as $L^2([0,2\pi], dx)$,  we have
\begin{align} f(x)=\Sigma_{n\in\Z}c_ne^{inx}; && f(R_{\al}x)=
\Sigma_{n\in\Z}c_ne^{in(x+\al)}. \label{FSf}
\end{align}
 Then $f(R_{\al}x)=f(x)$ a.e.\ implies that $\exp(in\al)=1$ for all $n\in\Z$ for which $c_n\neq 0$.
If $\al\notin 2\pi\Q$, then $\exp(in\al)=1$ implies $n=0$
 Hence the Fourier series only has a constant term $f(x)=c_0$  a.e.\ (note that the equality signs in \er{FSf} are  a.e.), so that $(\T, P, R_{\al})$ is ergodic. 
  On the other hand, if $\al\in 2\pi\Q$, say $\al= 2\pi (p/q)$ for $p,q\in\Z_*$, then clearly there are many $n\neq 0$ such that $\exp(in\al)=1$, namely all integral multiples of $q/p$. In that case
 $f$ is \emph{not} constant, and hence $(\T, P, R_{\al})$ is \emph{not} ergodic.   \QED
 \bex
Prove (similarly) that condition 2 holds for $M_p$ iff $|p|\geq 2$. 
\eex
\smallskip

For $P=p^\N$ or $P=p^\Z$ (i.e.\ the i.i.d.\ case) the shift map is ergodic. 
This follows from  a more general result for stationary Markov chains (Theorem \ref{IMCE}), but it is  instructive to also prove it from a different stronger result. Both to this end and for other purposes, we define some  stronger properties than ergodicity  which, when they hold, are nonetheless often easier to prove:
\begin{definition}
A dynamical system $(X,P,T)$ is called:
\begin{itemize}
\item  \emph{mixing} if for each measurable $C,D\subseteq X$ we have
\begin{equation}
\lim_{n\raw\infty}P((T^{-n}C)\cap D)=P(C)P(D); \label{mix}
\end{equation}
\item \emph{weakly mixing} if for each measurable $C,D\subseteq X$ we have
\begin{equation}
\lim_{N\raw\infty}\frac{1}{N}\Sigma_{n=0}^{N-1}| P((T^{-n}C)\cap D)-P(C)P(D)|=0. \label{wmix}
\end{equation}
\item \emph{`ergodic'} if for each measurable $C,D\subseteq X$ we have
\begin{equation}
\lim_{N\raw\infty}\frac{1}{N}\Sigma_{n=0}^{N-1} P((T^{-n}C)\cap D)=P(C)P(D). \label{emix}
\end{equation}
\end{itemize}
\end{definition}
Eq.\ \er{mix} means that for large $n$ the events $T^{-n}C$ and $D$ become independent (relative to $P$). These definitions (in connection with chaos) are part of the \emph{ergodic hierarchy}.\footnote{See   Frigg,  Berkovitz, \& Kronz (2025) for an introduction with plenty of references.}
The implication:  mixing $\Raw$ weakly mixing follows from Analysis, since convergence implies  C\'{e}saro convergence. 
\begin{exercise}Prove the  implications: weakly mixing $\Raw$ `ergodic' $\Raw$ ergodic.
\end{exercise}
In fact, `ergodicty' 
is \emph{equivalent} to ergodicity\footnote{Eq.\ \er{wmix} implies \er{emix}, since $|N\inv \Sigma_{n=1}^{N}f(n)-a|=|N\inv \Sigma_{n=1}^{N}(f(n)-a)|\leq N\inv \Sigma_{n=1}^{N}|f(n)-a|$.}, but the converse implication requires the weak ergodic theorem (see below). 
The irrational rotation \er{irr} is ergodic but not even weakly mixing: taking small intervals for $C,D$ one sees that $\lim_{n\raw\infty}P((T^{-n}C)\cap D)$ may not even exist, since it continues to jump between zero and nonzero values that do not converge to zero. Positively: 
\begin{proposition}\label{BSME}
For  any $p\in\Pr(A)$, the dynamical system $(A^\N, p^\N, S)$, cf.\ \er{ULS}, is mixing, and hence ergodic. This also holds for the two-sided case (Bernoulli shift), with $\N$  replaced by $\Z$.
\end{proposition}
\emph{Proof.} It can be shown that it is enough to verify \er{mix} for all generators of the $\sg$-algebra $\Sg$ on which $P$ is defined.\footnote{See Viano \& Oliveira, 2016, Lemma 7.1.2.} 
So let $C=[\sg]_L$ and $D=[\ta]_M$, for some $\sg\in A^L$ and $\ta\in A^M$. Since
\begin{align}
S^{-n}[\sg]_L&=\{s\in A^\N\mid s_n=\sg_0, s_{n+1}=\sg_1, \ldots, s_{n+L-1}=\sg_{L-1}\};\\
[\ta]_M&= \{s\in A^\N\mid s_0=\ta_0, s_1=\ta_1, \ldots, s_{M-1}=\ta_{M-1}\}, \label{taM}
\end{align}
for $n>M$ we simply have
\begin{equation}
(S^{-n}C)\cap D=\{s\in A^\N\mid s_0=\ta_0, \ldots, s_{M-1}=\ta_{M-1}, s_n=\sg_0, \ldots, s_{n+L-1}=\sg_{L-1}\}.\label{TCD}
\end{equation}
Thus $P=p^\N$  gives $P((S^{-n}C)\cap D)=P(C)P(D)$ already for $n>M$.
Likewise for $\N\leadsto\Z$.\QED
\smallskip

\noindent This is a special case of the following  result, which also holds for the two-sided shift in $(A^\Z, P, S)$.
\begin{theorem}\label{IMCE}
Let $P\in \Pr(A^\N)$ describe a stationary Markov chain and let $S:A^\N\raw A^\N$ be the  unilateral shift \er{ULS}, so that
 $P\circ S\inv =P$.  Then $(A^\N, P, S)$  is ergodic iff $P$ is irreducible. 
\end{theorem}
We just sketch the proof of ergodicity from irreducibility.\footnote{We follow Shields (1996), Example I.2.8; see Viana \& Oliveira (2016), Theorem 7.2.8 for the converse.} Using \er{TCD}, for $k>0$ we find
\begin{align}
P((S^{-M-k}C)\cap D)&=P(S^{-M-k}C|D)P(D)\nn \\
&=P(s_{M+k}=\sg_0, \ldots, s_{M+k+L-1}=\sg_{L-1}| s_0=\ta_0, \ldots, s_{M-1}=\ta_{M-1})P(D)\nn \\
&=P(s_{M+k}=\sg_0, \ldots, s_{M+k+L-1}=\sg_{L-1}|s_{M-1}=\ta_{M-1})P(D)
\nn \\
&=P(s_{M+k}=\sg_0, \ldots, s_{M+k+L-1}=\sg_{L-1}) P(s_{M+k}=\sg_0| s_{M-1}=\ta_{M-1})P(D)
\nn \\
&=P(s_{M+k}=\sg_0, \ldots, s_{M+k+L-1}=\sg_{L-1}) P^{k+1}_{\ta_{M-1}\sg_0}P(D)
=\nn \\
&=
 P^{k+1}_{\ta_{M-1}\sg_0} P_{\sg_0\sg_1} \cdots P_{\sg_{L-2}\sg_{L-1}} P(D).
\end{align}
 Since $M$ is fixed for given $D$, cf.\ \er{taM},  from \er{limP} we then obtain
 \begin{align}
&\lim_{N\raw\infty}\frac{1}{N}\Sigma_{n=0}^{N-1} P((S^{-n}C)\cap D)=\lim_{N\raw\infty}\frac{1}{N}\Sigma_{n=M+1}^{N+M} P((S^{-n}C)\cap D)=\lim_{N\raw\infty}\frac{1}{N}\Sigma_{k=1}^{N} P((S^{-M-k}C)\cap D)\nn \\
&=\lim_{N\raw\infty}\frac{1}{N}\Sigma_{k=1}^{N}P(D) P^{k+1}_{\ta_{M-1}\sg_0} P_{\sg_0\sg_1} \cdots P_{\sg_{L-2}\sg_{L-1}}
=P(D) p_{\sg_0} P_{\sg_0\sg_1} \cdots P_{\sg_{L-2}\sg_{L-1}}\nn \\ &=P(D)P(C),
\end{align}
where, since $P$ is assumed irreducible, we used \er{limP}. This proves \er{emix} and hence 
ergodicity.\footnote{
There is also a criterion for mixing we will not use and hence state without proof (see Viana \& Oliveira, 2016, Theorem 7.2.11). It  uses a strengthening of irreducibility to the effect that there is $N\in\N$ such that $(P^N)_{ab}>0$ for all $a,b\in A$; whereas 
irreducibility meant that for all $a,b\in A$ there is $N\in\N$ such that $(P^N)_{ab}>0$, cf.\ Definition \er{IMC}.
A  stationary Markov chain, then,  is mixing iff its stochastic matrix has this stronger property.}
\smallskip

 Using some additional assumptions, one can prove the existence of ergodic measures. Given some measure space $(X,\Sg)$ and some measurable map $T:X\raw X$, we write $\Pr^T(X)$ for the set of all $T$-invariant probability measures $P$ on $X$ (so that $T$ is a measure-preserving map for $(X,P)$).
\begin{theorem}[Krylov--Bogoliubov]\label{KB}  If $X$ is a compact metric space (with Borel structure),\footnote{Equivalently, $X$ is compact and separable. Among the compact spaces this is the non-pathological case.
}  then 
$\Pr^T(X)$ is not empty, and neither is its extreme boundary $\partial_e\Pr^T(X)$ of ergodic measures.
\end{theorem}
\emph{Proof.} The second claim follows from the first via the Krein--Milman Theorem,\footnote{See footnote \ref{Kreinfn}.}since if $X$ is compact and metrizable, then $\Pr(X)$ is a compact convex space (in the weak$\mbox{}^*$ = weak = vague topology).\footnote{See e.g.\ Denker, Grillenberger, \& Sigmund (1976), Proposition (2.8), or Walters (1982), \S6.2.}
 The first claim follows from the fact that for any $P\in\Pr(X)$ the sequence 
 \begin{align}
 (P_N)_N; && 
P_N=\frac{1}{N}\Sigma_{n=0}^{N-1}T^{-n}P,
\end{align}
has an accumulation point (since $\Pr(X)$ is compact), which is necessarily $T$-invariant.\footnote{See e.g.\ Denker, Grillenberger, \& Sigmund (1976), Theorem (3.6).}\QED\smallskip

 It is an interesting question if there is a \emph{unique} ergodic measure:
\bex
\begin{enumerate}
\item  Let $X=[0,1]$ with $T(x)=x^2$. What is 
$\partial_e\Pr^T(X)$? 
\item Same question for $T(x)=1-x$.
\item For any $X$, what is or what are the ergodic measure(s) for $T(x)=x$?
\end{enumerate}
\eex
 Under the same assumption (i.e.\ $X$ compact metric) both $\Pr(X)$ and $\Pr^T(X)$ are  \emph{(Choquet) simplices}. These generalize the familiar $n$-simplices $\Dl_N=\Pr(N)$.
\begin{definition} \label{Choquet} 
A (Choquet) simplex is a convex compact metrizable  space $K$ (in a locally convex vector space) in which any element is, in a suitable unique sense, a limit of a sequence consisting of finite convex sums of extremal boundary points (i.e.,  points in $\partial_e K$). More precisely: each $x\in K$ is the barycenter of a \emph{unique} probability measure $\mu$ supported on $\partial_e K$, and this in turn means that \beq
x=\int_{\partial_e K} d\mu(y)\, y,
\eeq  or, equivalently, $\mu(f)=f(x)$ for all affine continuous functions on  $K$.
\end{definition}
 To see that this captures the intuitive idea of a barycenter, take $K=[0,1]$, so that $\partial_eK=\{0,1\}$,
and $x\in[0,1]$. Then 
\begin{align}\mu=(1-x)\dl_0+x\dl_1; && \mu(f)=(1-x)f(0)+xf(1).
\end{align} 
 For any affine function $f(y)=ay+b$ this indeed gives $\mu(f)=f(x)$. Here $\mu$ is unique (for given $x$), as befits a simplex.
But the following compact convex set  is not a simplex:
\beq
B^3:=\{(x_1,x_2,x_3)\in\R^3\mid x_1^2+x_2^2+x_3^2\leq 1\}\eeq
\bex
Show that 
\beq
\partial_eB^3=\partial B^3=S^2:= \{(x_1,x_2,x_3)\in\R^3\mid x_1^2+x_2^2+x_3^2=1\},
\eeq
 and give an example of some $x\in \mathring{B}^3$ with more than one $\mu\in \Pr(S^2)$ whose barycenter is $x$.
\eex
 
 A finite-dimensional simplex is affinely isomorphic to some $\Dl_N$, see \er{A48b}, where $\partial_e \Dl_N$ consists of all unit vectors in $\R^{N+1}$, but if $X$ is infinite surprising examples may occur. The strangest of these is the \emph{Poulsen simplex}, in which $\partial_e K$ is dense in $K$. This is the case already in the simple example $K=\Pr^T(\mathbb{T})$, where $\mathbb{T}$ is the unit circle in $\C$ and $T$ is the doubling map $T(z)=z^2$ (though this is not at all simple to prove). Another example was described after Theorem \ref{5theorem}. Up to affine isomorphism, there is  just one Poulsen simplex (i.e., a Choquet simplex with the said property).\footnote{Simon (2011), Chapters 10 and 11, is an introduction to Choquet theory. Example 9.7;
 For the Poulsen simplex see Lindenstrauss, Olsen, \& Sternfeld (1978); Gelfert \& Kwietniak (2018); and 
 Simon (2011), Example 9.7.} The theory of simplices (i.e.\ Choquet theory)  comes down to the theorem that (for $X$ compact and metrizable) the set $\Pr^T(X)$  of $T$-invariant probability measures on $X$ is a Choquet simplex, so that according to the theory just summarized each $P\in \Pr^T(X)$ is the barycenter of a probability measure supported on the set $\partial_e \Pr^T(X)$ of ergodic probability measures on $X$.

We now discuss the \emph{ergodic theorem}, which comes in two versions (ironically, ergodicity is not assumed in either of these; but if it is assumed the ergodic theorem is considerably sharpened):
\begin{theorem}\label{ET}
Let $(X,P,T)$ a dynamical system, let $f\in L^1(X,P)$, and consider the sum
\begin{equation}
f_N(x):=\frac{1}{N}\Sigma_{n=0}^{N-1} f(T^nx). \label{ETsum}
\end{equation}
Then $\lim_{N\raw\infty} f_N=f^*$ for some $T$-invariant function $f^*\in L^1(X,P)$, in both of the following senses:
\begin{enumerate}
\item $f_N(x)\raw f^*(x)$ pointwise for $P$-a.e.\ $x\in X$ (\emph{Birkhoff's pointwise ergodic theorem});
\item $f_N\raw f^*$ in $L^1$; and if $f\in L^2$, $f_N\raw f^*$  in $L^2$ (\emph{von Neumann's mean ergodic theorem}).\footnote{ For finite measure spaces $L^2$-convergence  implies $L^1$ convergence, cf.\ Exercise \ref{E15}.
von Neumann's ergodic theorem even holds in $L^p$ for all $1\leq p<\infty$, cf.\ Walters (1982), Corollary 1.14.1.}
\end{enumerate}
\end{theorem}
\bex
Show that $L^1$-convergence (part 2) implies
\begin{equation}
\la f^*\ra_P\equiv \int_X dP\, f^*=\int_X dP\, f \equiv \la f\ra_P. \label{preerg}
\end{equation}
\eex
Before proving Theorem \ref{ET}, we note that it combines with ergodicity to give:
 \begin{corollary}\label{C1ET}
 If $(X,P,T)$ is ergodic and $f\in L^1(X,P)$, then $f^*(x)=\la f\ra_P$ for $P$-a.e.\ $x\in X$, i.e.,
 \begin{equation}
\lim_{N\raw\infty} \frac{1}{N}\Sigma_{n=0}^{N-1} f(T^nx)=\int_X dP\, f. \label{BET}
\end{equation}
 \end{corollary}
 \bex
 Prove this. Hint: show that $f^*(Tx)=f^*(x)$ $P$-a.e.\ and use Proposition \ref{EDE}.3.
 \eex
 This corollary was Boltzmann's  idea: \emph{time average = space average}, in  that the average of $f$ over a ``generic'' trajectory in time equals its average over $X$. Of course, asking this for ``every'' function $f$ requires such trajectories to exhaust $X$, at least in some (measure-theoretic) sense.\footnote{Eq.\ \er{BET} was essentially
 Boltzmann's  ``ergodic hypothesis'', stated for
   average values of physical observables  $f$ in equilibrium states--notably in the microcanonical ensemble, which involves a hypersurface $X$ of constant energy in phase space, with  flat prior $P$. See Darrigol (2018), pp.\ 100--104, 553--560, and Uffink (2024), \S2, \S5.1.}
 
 For each $x\in X$ the point measure $\dl_x$ on $X$ is defined by $\dl_x(B)=1$ if $x\in B$ and $\dl_x(B)=0$ if $x\notin B$;  as functionals of $C_b(X)$ we have $\dl_x(f)=f(x)$. Since weak convergence $P_n\raw P$ in $\Pr(X)$  is defined by $P_n(f)\raw P(f)$ for each $f\in C_b(X)$, eq.\ \er{BET} may be restated as follows:
   \begin{corollary}\label{C1ETbis}
 If $(X,P,T)$ is ergodic, then, weakly in $\mathrm{Prob}(X)$ for $P$-a.e.\ $x\in X$,
  \begin{align}
  \lim_{N\raw\infty}\frac{1}{N} \Sigma_{n=0}^{N-1}\dl_{T^nx}=P. \label{Erg1}
\end{align}
 \end{corollary}
 Indeed, each term in the sum on the left is a random variable 
 \begin{align}X \raw\mathrm{Prob}(X);
&&  x\mapsto \dl_{T^nx}, 
 \end{align} and since $\dl_{T^nx}\in\Pr(X)$, for each $N$ also 
 the convex combination 
 \beq
 P_N(x):=N\inv \Sigma_{n=0}^{N-1}\dl_{T^nx}
 \eeq is in $\Pr(X)$. Eq.\ \er{Erg1} then states that $P_N(x)\raw P$, weakly in $\mathrm{Prob}(X)$ for $P$-a.e.\ $x\in X$. A nice way of looking at \er{Erg1}, read the other way round,  is to write it as 
 \begin{equation}
P(B)=\ta(x,B):=\lim_{N\raw\infty} \frac{1}{N}|\{n\in \{0,1,\ldots, N-1\}\mid T^nx \in B\}|,  \label{sojourn1}
\end{equation}
 so that the probability of $B\in\Sg$ is equal to the \emph{sojourn time} of a generic path in $B$ (i.e.\ 
the fraction of time a particle on such a path spends in $B$).

  In our favourite case
 $X=A^\N$,  $T=S$,  with $P\circ S\inv =P$, the evaluation 
 map  at $n=0$, i.e.,
   \begin{align}
X_0: A^\N\raw A, && X_0(s)= s_0,
     \end{align}
 canonically
 induces a  map $X_0\inv$ between probabilities, viz.
  \begin{align}
X_0\inv:  \Pr(A^\N)\raw \Pr(A);&& X_0\inv P(B)=P(\{s\in A^\N\mid s_0\in B\})=P(X_0\in B). \label{P0map}
  \end{align}
  Since $(\dl_{S^ns})_0=\dl_{s_n}$, the image of $N\inv\Sigma_{n=0}^{N-1}\dl_{S^ns}$ under \er{P0map} is our familar
empirical measure
\begin{equation}
L_N(s)=\frac{1}{N}\Sigma_{n=0}^{N-1} \dl_{s_n}, \label{LN}
\end{equation}
where  $\dl_{s_n}\in\Pr(A)$ is a point measure on $A$ (since $s_n\in A$). For fixed $N$, this is a random variable
 \begin{align}L_N: A^\N \raw\mathrm{Prob}(A);
&&   s\mapsto L_N(s),
 \end{align}
whose value $L_N(s)$ depends only on $s_{|N}\in A^N$. It is easy to show (from the Portmanteau theorem) that $X_0\inv$ is weakly continuous,\footnote{See footnote \ref{PMT2}. This theorem  gives $P_n\raw P$ weakly in $\Pr(A^\N)$ iff $P_n(B)\raw P(B)$ for all cylinder sets $[\sg]_N$, $\sg\in A^N$. Taking $N=1$ gives weak continuity of $X_0\inv$.} so that, for any $S$-invariant $P\in\Pr(A^\N$), 
 \er{Erg1} implies that
\begin{align}
  \lim_{N\raw\infty}L_N(s)=p; && p(a):=P(X_0=a),\label{EMlimit0}
\end{align}
for $P$-almost every $s\in A^{\N}$, provided that $(A^\N,P,S)$ is ergodic. The simplest situation where we know this to be the case is the  i.i.d.\ probability $P=p^\N$ for some prior $p\in\Pr(A)$, cf.\ Proposition \ref{BSME}.  Note that the $p$ in \er{EMlimit0} is now simply the given $p$ in $P=p^\N$.
As we shall see shortly, eq.\  \er{EMlimit0}  implies  the strong law of large numbers, which therefore follows from the (pointwise) ergodic theorem. But it holds more generally than for i.i.d.  Eq.\   \er{EMlimit0} and
Theorem \ref{IMCE} give:
\begin{proposition}\label{wascor} 
Let $p\in\Pr(A)$ be the  unique stationary probability in an irreducible stationary Markov chain with finite state space $A$, i.e.\ $p(a)=P(X_0=a)$,  cf.\ Theorem \ref{limlemma}. Then
\begin{equation}
  \lim_{N\raw\infty}L_N(s)=p,\label{EMlimit}
\end{equation}
  for $P$-almost every sequence $s\in A^\N$ (realizing the chain in the Kolmogorov model).  
\end{proposition}
 \begin{corollary}\label{C2ET}
Given an injective function $E:A\raw\R$, define random variables 
\begin{align}
S_N:A^\N\raw\R; && S_N(s):=\frac{1}{N}\Sigma_{n=1}^{N-1} E(s_n). \label{defEfroms} 
\end{align}
If $P\in\Pr(A^\N)$ satisfying $P\circ S\inv =P$ defines
 an irreducible stationary Markov chain, then
   \begin{align}
  P\left(\lim_{N\raw\infty} S_N=\la E\ra_p\right)=1 && \Leftrightarrow&&
S_N\raw \la E\ra_p\:\:\: (P-\mathrm{almost\:\: surely}),
   \label{LLNsum}   \end{align}
 where $p\in\Pr(A)$ is the   unique stationary probability of the Markov chain, cf.\ Theorem \ref{limlemma}.
 
    In particular, this is true if if the $X_n$ are i.i.d.\ with respect to $P$ (and hence $P=p^\N$). 
   \end{corollary}
   This  follows from Proposition \ref{wascor} by taking the average of $E$ on both sides of \er{EMlimit}. First,
   \beq
   \la E\ra_{L_N(s)}=\Sigma_{a\in A} \frac{1}{N}\Sigma_{n=0}^{N-1} \dl_{s_n}(a)E(a)=
   \frac{1}{N}\Sigma_{n=0}^{N-1}\Sigma_{a\in A}\dl_{s_na}E(a)=
   \frac{1}{N}\Sigma_{n=1}^{N-1} E(s_n)=S_N(s),
   \eeq
  whereas on the right-hand side this averaging trivially gives $\la E\ra_p$. This averaging is weakly continuous  by definition of the weak topologies on $\Pr(A^\N)$ and $\Pr(A)$, and hence, with more ado, 
the restriction to finite sets $A$ so far may be lifted; any ``reasonable'' measure space works.

 The case of repeated sampling from a probability space $(A,p)$ is worth spelling out: 
 for any measurable function $E:A\raw\R$ with finite mean $\la E\ra_p$, we have
\begin{equation}
\lim_{N\raw\infty}\frac{1}{N}(E(s_1)+\cdots + E(s_N))=\la E\ra_p \equiv \int_{A}dp\, E,\label{LRF}
\end{equation}
for $p^\N$-almost all  $s\in A^\N$. For $E=1_B$ (with $B\subseteq A$ measurable), eq.\ \er{LRF}  gives, for $p^\N$-a.e.\ $s\in A^\N$,
\begin{equation}
\ta(s,B):=\lim_{N\raw\infty} \frac{1}{N}|\{n\in \{0,1,\ldots, N-1\}, s_n\in B\}|
= p(B).  \label{nuas5}
\end{equation}
Thus one may verify (but not: define!) the probabilities  $p$  from long-term frequencies in a long series of observations. These need not even be independent! It is enough if they form an irreducible stationary Markov chain.
 Looking at the sampling as a path in time, \er{nuas5} is once again a formula for the \emph{sojourn time} $\ta(s,B)$; i.e.\  the average time spent in $B$ during a journey $s$, cf.\  \er{sojourn1}.\footnote{Eq.\ \er{nuas5} also gives convergence of the \emph{Monte Carlo method} for computing volumes $\mu(B)$ of  subsets $B\subseteq A=[0,1]^d$ (and hence also integrals). In its simplest form: use a random generator to provide $N\cdot d$ elements of $[0,1]$ and hence $N$ elements ($d$-tuples) of $[0,1]^d$; count the number of points in $B$, divide this number by $N$, and let $N\raw\infty$. The result is (almost surely) $\mu(B)$.}
\smallskip

We will not prove part 1 of Theorem \ref{ET}, i.e., the pointwise ergodic theorem, since all its proofs are too long and  technical for these notes.\footnote{Our favourite proof is the one  by Kamae \& Keane (1997) as presented in  Dajani \& Kalle (2021), \S 3.1.}
Fortunately, part 2 (the mean ergodic theorem) has an easy proof,  due to von Neumann himself, which also
historically preceded Birkhoff's.

\noindent \emph{Proof of Theorem \ref{ET}.2.}
 The main technique of the proof is the use of the  \emph{Koopman operator} 
\begin{align}
U_T: L^2(X,P)\raw L^2(X,P); && U_Tf:= f\circ T, 
\end{align}
which by $T$-invariance of $P$ is an isometry,\footnote{this means that
$U_T^*U_T=1_H$. But $U_T$ is unitary only if $T$ is invertible.} in that it satisfies
\begin{equation}
\la U_T f, U_Tg\ra=\la f,g\ra,
\end{equation}
where $\la f, g\ra:=\int_X dP (x)\ovl{f(x)} g(x)$ is the inner product in $L^2(X,P)$. A key role will be played by 
\beq
H_T:=\ker(U_T-1_H)=\{f\in L^2\mid U_Tf=f\}\subset L^2(X,P)\equiv H,
\eeq
where $1_H$ is the unit operator on $H$. Being the kernel of a continuous (i.e.\ bounded) linear operator, $H_T$ is a closed linear subspace of $H$.  We now compute its orthogonal complement  $H_T^{\perp}$, consisting of all $g\in L^2$ such that
$\la g,f\ra=0$ for all $f\in H_T$.  The following observation is crucial to this end:
\bex
Show that $U_Tf=f$ iff $U_T^*f=f$.
\eex 
For any bounded operator $A:H\raw H$  we have $\ker(A^*)^{\perp}=\ran(A)^-$,
where the bar denotes closure. For $A=U_T-1_H$, by the exercise we  have $\ker(A^*)=\ker(A)$ and hence $\ker(A)^{\perp}=\ran(A)^-$, i.e., 
\begin{equation}
H_T^{\perp}=\ran(U_T-1_H)^-=\{U_Tg-g, g\in L^2\}^-. \label{vNcr} 
\end{equation}
Trivially, 
\beq
f= p_Tf+(1_H-p_T),
\eeq where  $p_T:H\raw H_T$
is the  orthogonal projection
defined by 
\begin{align}
p_Tf=f\:\:\: (f\in H_T); &&  p_Tf=0\:\:\: (f\in H_T^{\perp}),
\end{align}
Since $(1_H-p_T)f\in H_T^{\perp}$, if we ignore the closure in \er{vNcr} for the moment 
  there is $g\in H$ such that
 \begin{equation}
f=p_Tf+ U_Tg-g. \label{simpler}
\end{equation}
By definition of $f_N$, see \er{ETsum}, we  have 
\begin{align}
(p_Tf)_N=p_Tf; &&
(U_Tg-g)_N=\frac{1}{N}(U_T^N-1_H)g.
\end{align} Hence
\begin{align}
\|f_N-p_Tf\|_2=\frac{1}{N}\|U_T^Ng-g\|_2\leq \frac{1}{N}\| U_T^N-1_H\| \|g\|_2\leq\frac{1}{N}(\| U^N_T\| +\|1_H\|)\|g\|_2\leq \frac{2}{N}\|g\|_2,
\end{align}
by  the triangle inequality for the operator norm $\|\cdot\|$, and the fact that $\|U\|=1$ for any (nonzero)  isometry $U$.
It follows that $ \|f_N-p_Tf\|_2\raw 0$, so that $f_N\raw p_Tf$ in $L^2$.  \QED
\smallskip

\noindent \emph{The proof  identifies $f^*=\lim_N f_N$ as $p_Tf$},  at least in $L^2$ (where orthogonal projections are defined).
\bex
Finish the proof by taking the closure  in \er{vNcr} into account; this affects \er{simpler}.
\eex
 
 To close this section we illustrate the coherence of our material by returning to \er{SMB}, which turns out to be a special case of what in ergodic theory is still called the  Shannon--McMillan--Breiman (SMB) theorem.  Recall the setting of (probabilistic) dynamical systems, where we start with a partition \er{parpi} of some probability space $X$, which is subsequently refined to \er{defpiN}. The latter partition $\pi^N$ of $X$ then gives rise to the entropy \er{HPpiN}. As we see from \er{ppKS2}, this entropy is the \emph{average} of the information function  $I_P(\pi^N)$, see \er{Imany}, w.r.t.\ $P$.
The SMB theorem shows that as $N\raw\infty$ one may replace this average by judiciously chosen \emph{pointwise} values of $I_P(\pi^N)$:
\begin{theorem}\label{SMBs}
If $(X,P,T)$ is ergodic with finite partition \er{parpi}, then for $P$-almost every $x\in X$,
\beq
h_P(\pi)=\lim_{N\raw\infty} \frac{1}{N}  I_P(\pi^N)(x),\label{SMB2}
\eeq 
where  $\pi^N(x)$ is the cell  of the partition $\pi^N$ that contains $x$, cf.\ \er{pinot}.
\end{theorem} 
\bex
Show that  \er{SMB} is a special case of \er{SMB2}.
\eex
 \emph{Proof of Theorem \ref{SMBs}.} 
Consider the function $I_P(\pi^N)$. Recalling \er{pi1},  for $N>1$ we abbreviate
 \beq
 f_N:= I_P(\pi|\wed_{n=1}^{N-1}T^{-n}\pi)=-\Sigma_{A\in \pi}\, \Sigma_{B\in\pi_1^N} 1_{A\cap B}  \log \left(\frac{P(A\cap B)}{P(B)}\right),
 \eeq
 cf.\ \er{IPc}. A nontrivial measure-theoretic argument shows that the sequence $(f_N)$ has a  limit  
 \beq
 f(x)=\lim_{N\raw\infty} f_N(x) , \label{deffNf} 
 \eeq
 pointwise  $P$-a.e.\footnote{See e.g.\ Dajani \& Kalle, \S 9.4. The point is to rewrite $I_P(\al|\beta)$ as conditional expectation with respect to the $\sg$-algebra $\sg(\beta)$ generated by the partition $\beta$, namely
  $I_P(\al|\beta)(x)= -\Sigma_{A\in\pi} 1_A(x)\log E_P(1_A|\sg(\beta))$. This gives the pointwise limit function as $f(x)=I_P(\pi|\sg(\cup_{n=1}^{\infty}T^{-n}\pi))$. } This is the function we  now apply the pointwise (Birkhoff) ergodic theorem to (in the last line below), as well as  the dominated convergence theorem. Using  \er{cE} and \er{HPalt},
 \begin{align}
h_P(\pi)&=\lim_{N\raw\infty} H_P(\pi|\pi_1^N)=\lim_{N\raw\infty} \int_X dP\, I_P(\pi|\pi_1^N)=\lim_{N\raw\infty} \int_X dP\, f_N=\int_X dP\, \lim_{N\raw\infty} f_N\nn \\&=
\int_X dP f=\lim_{N\raw\infty} \frac{1}{N} \Sigma_{n=0}^{N-1}f(T^nx), \label{576}
\end{align}
 for $P$-almost every $x\in X$. 
To relate this to  Theorem \ref{SMBs}, we iterate \er{ex941} and use the property 
\beq
I_P(T\inv\pi)(x)=I_P(\pi)(Tx). \label{C52e}
\eeq
\bex Prove \er{C52e} from \er{PTB} and use it to compute
 \begin{align}
 I_P(\pi^N)= \cdots= I_P(\vee_{n=0}^{N-3}T^{-n}\pi)\circ T^2+f_{N-1}\circ T+f_N=\cdots =\Sigma_{n=0}^{N-1} f_{N-n}\circ T^n, 
 \end{align}
 where $f_1:= I_P(\pi)$ by convention (here the dots replace a routine proof by induction).
 \eex This is not quite the same as the right-hand side of \er{576}; the proof ends by showing that $P$-a.e.,
 \begin{equation}
\lim_{N\raw\infty} \frac{1}{N} \Sigma_{n=0}^{N-1}(f(T^nx)-f_{N-n}(T^n(x))=0.
\end{equation}
This may be unsurprising in view of \er{deffNf}, but the argument is  technical and we omit it.\footnote{See  Dajani \& Kalle (2021), \S 9.4.} \QED
\smallskip

The final part of this appendix supports the theory of Gibbs measures in \S\ref{GM}. 
 So far,  we just defined ergodicity in the setting of dynamical systems $(X,P,T)$, where $(X,\Sg,P)$ is a probability space and $T:X\raw X$ is measurable. But if $T$ is invertible this amounts to a $\Z$-action on $X$, given by $n:x\mapsto T^nx$. A more abstract setting of ergodic theory is then given by measurable $G$-actions on a measure space, which we always take to be a probability space $(X,\Sg,P)$, in which $P$ is $G$-invariant.\footnote{This more \emph{abstract} setting is not more \emph{general} than the previous one, since $T$ need not be invertible and hence may not define a group action. In the abstract setting one requires $G$ to be \emph{amenable} (and \emph{a priori} locally compact).  This means that every continuous $G$-action on a compact metrizable space $\Om$ has an invariant probability measure, and may also be defined by the property that
for any compact subset $K\subset G$ and $\dl > 0$ there is a Borel set $B \subset G$ with compact closure such that 
$\mu(B\Delta (K\cdot B)) < \dl \mu(B)$, where $\mu$ is the (left) Haar measure on $G$. Every compact group and every locally compact abelian group is amenable (but there are others). 
In the ergodic theorem sequences of the type $\Lm_N\subset \Z^d$ are replaced by
so-called \emph{F\o lner sequences}, defined as compact subsets $B_N\subset G$ that (at least for sufficiently large $N$) have the above property $\mu(B_N\Delta (K\cdot B_N)) < \dl \mu(B_N)$ for every compact $K\subset G$ and $\dl>0$.
}
 We take $G=\Z^d$ and its action on $\Om=A^{\Z^d}$ the shift action
induced by the left-action \er{925} of $\Z^d$ onto itself, i.e., 
\begin{equation}
(y\cdot \om)_x:=\om_{x+y}.\label{9106}
\end{equation}
Cf.\ the shift \er{ULS}, which (defined on $A^{\Z}$) is a special case; but we also consider general compact metrizable spaces $\Om$, with the ensuing Borel structure $\Sg$. This class of spaces overlaps with $A^{\Z^d}$ provided $A$ is compact and $A^{\Z^d}$ carries the product topology (in which it is compact by Tychonoff's theorem).\footnote{For any  set $T$ and any space $A$ the \emph{product topology} on 
$\Om=A^T$ is the coarsest topology that makes all evaluation maps $A^T\raw A$, $\om\mapsto\om_x$, continuous; it is generated by the cylinder sets (like the $\sg$-algebra $\mathcal{F}$). For $T=\Z^d$ the product topology is metrizable, for example by the metric 
$d(\om,\om'):= \lm^{-\inf \{\|x\|,\, x\in\Z^d,\, \om_x\neq\om'_x\}}$, where $\|x\|:=
\max_{i=1, \ldots d}\{|x^i|\}$ or any other norm, and any $\lm>1$, with the convention or theorem that $\inf\emptyset=\infty$ as usual, so that $d(\om,\om)=0$.  For $d=1$ and $A=\{0, 1, \ldots, |A|-1\}$, an inequivalent metric that nonetheless gives the same topology would be $d'(\om,\om')=\Sigma_{n=-\infty}^{\infty} \lm^{-|n|} |\om_n-\om'_n|$, again for any $\lm>1$. See Katok \& Hasselblatt (1995), \S1.9.
} We say that a measurable set $B\in\Sg$ is \emph{$G$-invariant} iff $x\cdot B=B$ for all $x\in G$, where 
\beq
x\cdot B=\{x\cdot\om, \om\in B\}.
\eeq These $G$-invariant measurable sets form the \emph{invariant $\sg$-algebra} $\mathcal{I}$, and clearly
measurable functions are $G$-invariant iff they are $\mathcal{I}$-measurable.
Instead of Definition \ref{defE1} we say that the 
 $G$-action is \emph{ergodic} iff $B\in \mathcal{I}$ implies $P(B)=0$ or $P(B)=1$. 
Proposition \ref{EDE} then remains valid if we replace $f(Tx)=f(x)$ by $f(x\cdot\om)=f(\om)$ for all $x\in G$. For compact $\Om$ it is enough to require this for all \emph{continuous} functions $f$, so that the $G$-action is ergodic iff any $G$-invariant function $f\in C(\Om)$ is constant, where $G$-invariance now means that
 $f(x\cdot\om)=f(\om)$ for all $x\in G$. We only state the pointwise ergodic theorem for $G=\Z^d$ acting on $\Om=A^{\Z^d}$:\footnote{See Georgii (2011), Theorem 14.A8. For more general (amenable) groups and spaces see Lindenstrauss (2001). } 
 \begin{theorem}\label{ETZd}
 Let $\Om$ be compact metrizable space equipped with both a $\Z^d$-action and a $\Z^d$-invariant probability measure $P$. For any $f\in L^1(\Om,P)$,  the limit
 \begin{equation}
f^*(\om):=\lim_{N\raw\infty} \frac{1}{|\Lm_N|}\Sigma_{x\in\Lm_N} f(x\cdot\om)
\end{equation}
exists pointwise $P$-almost surely and equals a $G$-invariant function $f^*\in L^1(\Om,P)$, given by
\begin{equation}
f^*=E_P(f\mid \mathcal{I}).\label{fstarEf2}
\end{equation}
In particular, if the $G$-action is ergodic, then  for $P$-almost every $\om\in\Om$ we have
\begin{equation}
\lim_{N\raw\infty} \frac{1}{|\Lm_N|}\Sigma_{x\in\Lm_N} f(x\cdot\om)=\int_{\Om} dP\, f,
\end{equation}
or, equivalently,  for $P$-almost every $\om\in\Om$, as weak convergence of probability measures,
 \begin{equation}
 \lim_{N\raw\infty} \frac{1}{|\Lm_N|}\Sigma_{x\in\Lm_N}\dl_{x\cdot\om}=P.
 \label{ETMZd}
\end{equation}
 \end{theorem}

  \section{Entropy and Kolmogorov randomness}\label{EKR}
  Since entropy is closely related to randomness, it is no surprise that the most sophisticated concept of randomness known to date, namely \emph{Kolmogorov randomness}, aka \emph{algorithmic randomness}, is related to entropy.\footnote{This section is mainly based on Cover \& Thomas (2006), 
  chapter 14. The standard reference for algorithmic randomness of strings is 
  Li \&  Vit\'{a}nyi, (2008), of which \S 8.6 discusses some connections with entropy. See also 
   Gr\"{u}nwald \& Vit\'{a}nyi (2003).
  Algorithmic randomness of sequences is the main topic of Calude (2002) and Downey \& Hirschfeldt (2010),
  but this is a matter of emphasis: all three books discuss both  strings and sequences (we repeat our convention that a  \emph{string} $\sg$ is finite row of bits or letters from some finite alphabet, whereas a \emph{sequence} $s$ is an infinite one). 
 For first introductions Gr\"{u}nwald \& Vit\'{a}nyi (2008) and  Dasgupta (2011) are recommended. The pointwise results in the main text are more advanced; see Towsner (2020) for a survey and original references. Franklin \& Porter (2020) is a recent survey of various aspects of algorithmic randomness. The place of  algorithmic randomness in a broader spectrum of theories of randomness is discussed for example in Porter (2012), Eagle (2019),  Landsman (2020), and Dekkers \& Landsman (2025).  For the history of the subject see 
van Lambalgen (1987), Porter (2012), as well as the `History and References' sections at the end of each chapter in   Li \&  Vit\'{a}nyi, (2008). 
 } 
  Kolmogorov's problem (of which his new notion of randomness was supposed to be a solution), which was noticed already by Laplace and perhaps even earlier probabilists, was that, specializing to a 50-50 Bernoulli process for simplicity,
  any  binary string $\sg$ of length $N$ has probability $P(\sg) = 2^{-N}$ and any (infinite) binary sequence $x$ has probability $P(x) = 0$, although say $\sg=0011010101110100$ looks much more random than $\sg=111111111111111$. In other words, their \emph{probabilities} say little or nothing about the \emph{randomness} of individual outcomes.
  
   Imposing statistical properties helps but is not enough to guarantee randomness. We first explain this in base 10. We call a real number $x\in [0,1]$ 
    \emph{Borel normal} if in its decimal expansion 
    \beq
    x=\lim_{N\raw\infty} \Sigma_{n=0}^N s_n 10^{-n-1},
    \eeq
     where $s_n\in 10=\{0,1,\ldots, 9\}$, and hence $s\in 10^\N$, 
    each decimal string $\sg\in 10^*$ has (asymptotic) frequency $10^{-\ell(\sg)}$ in $s$, so that each digit $0, \ldots, 9$ occurs 10\% of the time ($\ell(\sg)=1$),  each block $00$ to $99$ occurs 1\% of the time
    ($\ell(\sg)=2$),
    etc. \footnote{The literature on Borel normality is large. Khoshnevisan (2006) is a nice introduction.} More precisely and more generally,  in any base $b=\{0,\ldots, b-1\}$,
  a  $b$-valued sequence $s\in b^\N$  is Borel normal if 
 for any $b$-valued string $\sg\in b^*$,
 \begin{equation}
\lim_{N\raw\infty} \frac{1}{N} |\{ n\in N\mid s_{|n}\in b^*\sg\}|=b^{-\ell(\sg)},
\end{equation}
and hence $x\in[0,1]$ is Borel normal in base $b$ if the $b$-valued sequence $s\in b^\N$ in its $b$-ary expansion  
\beq
x=\lim_{N\raw\infty} \Sigma_{n=0}^N s_n b^{-n-1}
\eeq
 is Borel normal in the above sense (this may depend on $b$!).\footnote{The $b$-ary expansion of $x\in [0,1]$ is unique unless $x=p/b^k$ for some $k>0$ and $p\in\N$ satisfies $0< p< b^k$. But rationals are not Borel normal anyway, so this lack of uniqueness does not jeopardize the definition.}
  For example, for $b=10$ consider
    \begin{align}
    0.123456789101112131415161718192021222324252629 \ldots \nn\\
    0.235711317192329313741434753596167717379838997\ldots
\end{align} The first is  
 \emph{Champernowne's number}, which just lists all positive integers, whereas the second is the \emph{Copeland--Erd\"{o}s number}, which lists all primes.\footnote{These numbers are supposed to have an infinite number of digits, so that we precede them by $0.$ to get a finite real.} Both are Borel normal!
  \bex
 Prove that  Champernowne's number is Borel normal.
 \eex
 Once you know the pattern behind these  numbers, it is obvious how they continue. 
 What about
 \begin{equation}
0.30927562832084531584652001027797235612923012605863\ldots?
\end{equation}
This looks really random, but these are the first few digits of $\pi$ after the first  million ones!  Without this information, no one would recognize their origin. Unlike the previous two numbers, the decimal expansion of $\pi$ is merely \emph{conjectured} to be Borel normal,  but this has been empirically verified in billions of decimals,
so let us assume it is. Thus Borel randomness cannot be a good notion of randomness, since all three numbers can be generated by short formulae or algorithms, and on any reasonable intuition about randomness, this means that they can't be random. We may also argue that these numbers are  \emph{computable},  which seems the very opposite to being \emph{random}. 

 Kolmogorov's  paradoxical idea was to define randomness \emph{through} computability! 
 Roughly speaking, his idea was that a string $\sg$ is random iff the shortest computer program $p$ that computes $\sg$ has about the length of $\sg$ itself (in which case $p$  simply stores $\sg$ in its memory and prints it):
 \begin{center}
\emph{An object $O$ is random iff the shortest computable description of $O$ is $O$ itself.}
\end{center}
This idea apparently exceeds the world of binary strings, although it is limited to objects for which one has a notion  of computability. Indeed,  adding \emph{computable}, or at least some other definition of what is meant by a  ``description'' is essential in view of \emph{Berry's paradox}:
\begin{center}
\emph{The Berry number is the smallest positive integer that cannot be described in $<18$ words.}
\end{center}
The paradox, then, is that on the one hand this number must exist, since only finitely many integers can be described in less than eighteen words and hence the set of such numbers must have a lower bound, while on the other hand Berry's number cannot exists by its own definition. This is, of course, one of innumerable paradoxes of natural language (like the liar's paradox).  
 
 To make this idea precise 
 we assume basic familiarity with the theory of computation, or at least with the concept of a Turing machine, or at the very least with computers and computer programs. A Turing machine $T$ is nothing but a particular physical model of a computer, which in turn may be seen as a 
 computable (partial) function $f:D_f\raw \N$, whose domain $D_f\subseteq\N$  consists of all $n\in\N$ for which $f(n)\in \N$ is defined. 
 Using some (computable) bijection $\N\cong 2^*$ we
 may identify $n\in\N$ as the binary code of a computer program $p$ run by $T$, in which case 
 $n\in D_f$ iff $T$ terminates (or ``halts'') on the corresponding program $p$ (written $T(p)\daw$) and produces $f(n)$ or $T(p)$ as its output; using the same (or some other) bijection also  on this end, we may and will consider $T(p)$ to be a binary string, now written as $\sg\in 2^*$. Thus $T(p)=\sg$ and we consider the Kolmogorov complexity of binary strings.\footnote{One may equally well look at integers or strings over some finite alphabet $A$. }
For technical reasons,\footnote{Namely: (1) a good relationship with coding theory, and (2): a smooth extension of the theory from strings to sequences. Briefly, if $T(p)=\sg$ we may regard $p$ as a code-word for $\sg$, i.e.\ $C(\sg)=p$, which is decoded by $T$, and so a prefix Turing machine produces a uniquely decodable code.}
 we assume that the codes $p$ of all programs for which $T$ halts (i.e.\ produces an output $T(p)=\sg$) form a prefix set; as before, this means that if $T(p)\daw$ and $T(q)\daw$ for $p,q\in 2^*$, then $p$ cannot be a prefix of $q$ (or \emph{vice versa}).
 
  As in the usual case, one can prove that there are universal prefix Turing machines $U$, in the sense that for any prefix Turing machine $T_n$ (based on some computable enumeration $(T_n)$ of all prefix Turing machines) one has 
  \beq
  T_n(p)=U(\la n,p\ra),
  \eeq
   where 
  $\la \cdot,\cdot\ra$ is some computable map from $\N\x 2^*$ into a prefix subset of $2^*$. \emph{We often omit ``prefix'' in what follows, but it is always meant.}
\begin{definition}\label{def111}
Let $U$ be a universal prefix Turing machine. 
\begin{enumerate}
\item The \emph{Kolmogorov complexity} of $\sg\in 2^*$ (relative to $U$) is given by
\begin{equation}
K_U(\sg):=\min\{\ell(p), U(p)=\sg\} .\label{defKC1}
\end{equation}
\item
The \emph{conditional Kolmogorov complexity} of $\sg\in 2^*$ (relative to $U$) is given by
\begin{equation}
K_U(\sg\mid \ell(\sg)):=\min\{\ell(p), U(\la \ell(\sg), p\ra)=\sg\}.\label{defKC2}
\end{equation}
\end{enumerate}
\end{definition}
How do these definitions depend on the choice of the universal prefix Turing machine $U$?
 \begin{proposition}\label{Kk1}
 For any two universal prefix Turing machines $U,V$ there is a constant $C(U,V)\in\N$ such that
 for all $\sg\in 2^*$ we have
 \begin{equation}
|K_U(\sg)-K_V(\sg)|< C(U,V), \label{Kolineq}
\end{equation}
and similarly (for a different constant) for the conditional Kolmogorov complexity.
 \end{proposition}
 \emph{Proof (sketch).} First assume $V$ is merely a (prefix) Turing machine (so not necessarily universal, like $U$). Then any program $p_V$ for $V$ that produces $\sg$, i.e., $V(p_V)=\sg$, can be transferred to $U$ via a program $q_{U,V}$ on $U$ that simulates $V$, so that $p_U=q(U,V)p_V$ with length
 $\ell(p_U)=\ell(q_{U,V})+\ell(p_V)$ produces $\sg$ via $U$. Hence 
 \beq
 K_U(\sg)\leq K_V(\sg)+\ell(q_{U,V}),\eeq uniformly in $\sg$. If $V$ is also universal this works the other way round, too, and one obtains \er{Kolineq}
 with $C(U,V)=q_{U,V}+q_{V,U}$. \QED\smallskip
 
 In view of this it is common practice to omit the label $U$ and just talk about \emph{the} Kolmogorov complexity $K(\sg)$ of strings $\sg$, which, then, is defined up to a $\sg$-independent constant. 
 As such, we may define a string to be  $c$-\emph{random} (relative to $U$)  for some $c\in\N$ if
\begin{equation}
K_U(\sg)\geq \ell(\sg)-c.\label{KUc} 
\end{equation}
This is a useful criterion  for \emph{strings} whose length far exceed the $U$-dependent constant $c$, and hence it enables us to define randomness of \emph{sequences} independently of $U$:
\begin{definition}\label{LC}
 A sequence $s\in 2^\N$ is \emph{random} if 
 there is a constant $c\in\N$ such that for all $N>1$, 
\begin{equation}
K(s_{|N})\geq N-c. \label{CL}
\end{equation}
\end{definition}
More precisely, one should say that for any universal prefix Turing machine $U$ there is a constant $c(U)$ such that $K_U(s_{|N})\geq N-c(U)$, but the point is that although the (minimum) \emph{value} of constant $c(U)$ does depend on $U$,  by  Proposition \ref{Kk1} the \emph{existence} of such a constant does not. 
Definition \ref{LC} (and its reformulation by Martin-L\"{o}f that we shall not discuss here) forms the basis of the theory of algorithmic randomness. To get an idea about this definition, as well as about Kolmogorov complexity in general,  we analyze the case of strings in some more detail.
\begin{proposition} \label{114}
There are constants $c, c'$ such that, for all $\sg\in 2^*$,
\begin{align}
K(\sg)&\leq \ell(\sg)+2\log_2\ell(\sg)+c; \label{le2}\\
K(\sg\mid\ell(\sg))&\leq \ell(\sg)+c'. \label{le1}
\end{align}
 \end{proposition}
  \emph{Proof (sketch).} In the worst case, storing $\sg$ costs $\ell(\sg)$ bits. This is preceded by some
 $\sg$-independent  general printing instruction of length $c$. But the set of all $\sg\in 2^*$ should be turned into a prefix set, for otherwise the programs $p$ thus constructed do not form a prefix. A simple (but expensive) way to accomplish this is 
 to double each digit in $\sg$ and close with 01 (for example, $\sg=1010$ becomes $1100110001$), which is then read as a stop sign. This doubling act costs $2\log\ell(x)$ extra bits, which, absorbing the extra two bits 01 in $c$, gives \er{le2}.

 If $\ell(\sg)$ is given to $U$, we could use the same map $\la\cdot,\cdot,\ra: \N\x 2^*\raw 2^*$ (whose image is a prefix set) that was mentioned before Definition \ref{def111}; this would effectively 
incorporate $\ell(\sg)$  into the printing instruction. Hence the need for the above move disappears.\footnote{Using a more efficient (recursive) coding of $\ell(\sg)$, one may improve the term $2\log_2\ell(\sg)$ to
 $\log_2^*\ell(\sg)$, where, for $n\in\N$, one defines $\log_2^*n=\log_2 n+ \log_2\log_2n + \log_2\log_2\log_2n+\cdots$, where the last term is the last possible \emph{positive} term (e.g.\ $\log_2^*7$ has the above three terms); but this makes no difference to the relevant asymptotics.}
\QED\smallskip
 
 Combining \er{KUc}, \er{CL},  and \er{le2}, we see that $s$ is random if $K(s_{|N})\approx N$ as $N\raw\infty$, and more precisely,\footnote{See   Li \&  Vit\'{a}nyi, (2008), page 220.}
  one can show that $s$ is random iff
 \begin{equation}
N+O(1) \leq K(s_{|N})\leq N +K(N)+O(1), \label{lowerupperK}
\end{equation}
where $K(N)$ is defined as in \er{defKC1} but now with $N\in\N$ instead of $\sg\in 2^*$ as the output. Like $K(\sg)$, this makes sense up to a $U$-dependent but $N$-independent constant, which disappears into the $O(1)$ terms. Consistency with \er{le2} comes from the estimate $K(N)\leq 2\log_2 N +c$, which is obvious: if the program $p$ is given the length $\ell(N)\approx\log_2 N$ of the binary expansion $N$ then at worst it needs to store these bits (giving it a length $\log_2N$), and if not, providing  $\ell(N)$ takes another $\log_2N$ bits. In fact, the upper bound in \er{lowerupperK} ``wins'',\footnote{See Calude (2002), Theorem 6.38 (attributed to Chaitin).} since it can also be shown that  $s$ is random iff:
\begin{equation}
\lim_{N\raw\infty} (K(s_{|N})-N)=+\infty.\label{Calude}
\end{equation}

  Similarly, a string $\sg$ is random if $K(\sg)\approx \ell(\sg)$, at least for long strings.
 On the other hand, long strings that are  computable from short programs have $K(\sg)\approx  \log_2(\ell(\sg))$, since the length of $\sg$ plus some short instructions are sufficient for some efficient program $p$ to compute it. Also,
 \begin{equation}
|\{\sg\in 2^*\mid K(\sg)<k\}| <2^k,
\end{equation}
since we may simply count the number of programs $p$ of length $\ell(p)<k$ to be
\beq
\Sigma_{n=0}^{k-1}2^n=2^k-1.
\eeq
 On the other hand, the total number of strings $\sg$ of length $\ell(\sg)\leq k$ is $2^{k+1}-1$ (this is just the previous number plus $2^k$), and hence only a fraction $\frac{2^k-1}{2^{k+1}-1}<\half$ (which for large $k$ of course approaches $\half$) of these strings can be described in a way that saves just one bit! 
More generally, the fraction of strings of length $\ell(\sg)\leq k$ that can be described saving at least $n$ bits is 
 $\frac{2^{k-n+1}}{2^{k+1}-1}<2^{-n}$.
 
  Hence the large majority of strings is random, and the same is true for sequences:\footnote{See Calude (2002), Theorem 6.31. This is proved using Martin-L\"{o}f's own reformulation of randomness.}
 \begin{theorem}[Martin-L\"{o}f]\label{PML}
$f^\N$-almost every sequence $s\in 2^\N$ is random.
  \end{theorem} Here $f$ is the flat prior on $A=2$ and $f^\N$ is the associated Bernoulli probability measure on $2^\N$. 
  If we define $x\in [0,1]$ to be random iff the sequence $s$ in its binary expansion \er{xomega} is random, it follows from Exercise \ref{diffex} that almost every $x\in[0,1]$ is random with respect to Lebesgue measure. Borel already proved this for Borel normality in 1909, and Theorem \ref{PML} implies his result:\footnote{For details and proofs see Calude (2002), Corollary 6.32 in \S 6.3 and almost all of
 \S 6.4. } 
\begin{proposition}\label{116}
\begin{enumerate}
\item A random sequence  is Borel normal in any base.
\item  A random sequence contains any  string infinitely often.
\item  A random sequence $s$ satisfies the strong law of large numbers, in the sense that 
\begin{equation}
\lim_{N\raw\infty}\frac{1}{N} \Sigma_{n=0}^{N-1} s_n=\half.\label{SLLN4}
\end{equation}
\end{enumerate}
 \end{proposition}
 The second part follows from the first, as is easily proved by contradiction, and so does the third. We see a considerable improvement over the usual strong law of large numbers (SLLN): whereas the latter merely states that \er{SLLN4} is true for $f^\N$-almost every sequence $2\in 2^\N$,
Proposition \ref{116}.3 \emph{explicitly identifies} sequences for which \er{SLLN4} holds, namely the random ones; and  Theorem \ref{PML} adds that, consistent with the usual SLLN, these form a subset $R$ of $2^\N$ of probability $f^\N(R)=1$.

On the other hand one may wonder how ``explicit'' this identification is:\footnote{More precisely, if $s\in2^{\N}$ is random, only finitely many true statements of the form: `the $n$'th bit $s_n$ of $s$ equals its actual value' (i.e.\ 0 or 1) are provable in $T$. See Calude (2002), Theorem 8.7, which is stated for Chaitin's $\Omega$ but whose proof holds for any random sequence.  As in G\"{o}del's  theorems,
  one also assumes that $T$ is formalized  as an axiomatic-deductive system in which proofs could in principle be carried out mechanically by a computer.
   See Chaitin (1987) for his own presentation and analysis of his two incompleteness theorems, i.e.\
   our Theorems \ref{Chaitin2} and \ref{Chaitin1}. 
    Raatikainen (1998) also gives a detailed presentation of the second theorem, including a devastating critique of  Chaitin's ideology.}
\begin{theorem}[Chaitin]\label{Chaitin2}
If $s\in 2^{\N}$ is random, then any   consistent and sufficiently comprehensive mathematical theory $T$ (like ZFC) can compute only finite many digits of $s$.\end{theorem}
This  excludes defining a random number by somehow listing its digits, but some can be described by a formula. The most famous example is Chaitin's number $\Omega$, or more precisely $\Om_U$,\footnote{There  exists a $U$ for which not a single digit of $\Om_U$ can be known, see  Calude (2002), Theorem 8.11.}
 which is the halting probability of some fixed universal prefix  Turing machine $U$, given by
\begin{equation}
\Om_U:=\Sigma_{\sg\in 2^*\mid U(\sg)\downarrow}2^{-\ell(\sg)}.
\end{equation}
There is also an analogue of this theorem for random strings:\footnote{The proof is based on the existence of a computably enumerable (c.e.)  list  $\mathsf{T}=(\ta_1,\ta_2,\ldots)$ of the theorems of $T$, and on the fact that after G\"{o}delian encoding by numbers,
    theorems of any given grammatical form can be computably searched for in this list and will eventually be found.  In particular, there exists a program $p$  such that $p(n)$ halts iff there exists a string $\sg$
for which $K(\sg)>n$ is a theorem of $T$. If there is such a theorem the output is $p(n)=\sg$, where $\sg$ appears in the first such theorem of the kind (according to the list  $\mathsf{T}$). By definition of  $K(\cdot)$, this means that $K(\sg)\leq |P|+|n|$. 
 Now suppose that no $C$ as in the above  statement of the theorem exists. Then there is $n\in\N$ large enough that $n>\ell(p)+|n|$ and there is a string $\sg\in 2^*$ such that $T$ proves $K(\sg)>n$. Since $T$ is consistent and hence sound (i.e., it only proves true theorems) this is actually true, which  gives a contradiction between $K(\sg)>n>\ell(p)+|n|$ and $K(\sg)\leq \ell(p)+|n|$; this contradiction can be made more dramatic by taking $n$ such that $n>>\ell(p)+|n|$.
Note that this proof shows that a \emph{proof} in $T$ of $K(\sg)>n$ (if true) would also \emph{identify} $\sg$.}
\begin{theorem}\label{Chaitin1}
For any  consistent and sufficiently comprehensive mathematical theory $T$ (like ZFC)  there 
 is a constant $C\in\N$ such that $T$ cannot prove any sentence of the form $K(\sg)>C$ (although infinitely many such sentences are true), and as such $T$ can only prove (Kolmogorov) randomness of finitely many strings (although infinitely many strings \emph{are} in fact random).
\end{theorem}

At last, we now turn to the connection between Kolmogorov complexity and entropy! So far, we only have the resources to understand the following result for $A=2$, but we state the general case. To this end we do note that for finite sets $A$ one may extend Kolmogorov complexity $K(\sg)$ to $\sg\in A^*$, for example by using a computable bijection $A^*\cong 2^*$ (since both are $\cong \N$). 
 \begin{theorem}\label{Kentropy}
 For any finite set $A$ and prior $p\in\Pr(A)$, and any integer $N>0$, we have 
 \begin{equation}
S_2(p)\leq \frac{1}{N}\Sigma_{\sg\in A^N} p^N(\sg)K(\sg\mid N)\leq S_2(p)+\frac{(|A|-1)\log_2 N}{N}+O(1/N),
\label{boundsK}
\end{equation}
so that
\begin{equation}
\lim_{N\raw\infty} \frac{1}{N}\Sigma_{\sg\in A^N} p^N(\sg)K(\sg)\equiv
\lim_{N\raw\infty} \frac{1}{N}\la K\ra_{p^N}
=S_2(p). \label{lbK}
\end{equation}
 \end{theorem}
 In other words, for large $N$ the ``average'' complexity $\la K\ra_{p^N}$ of strings in $A^N$ goes like $NS_2(p)$.
  
 Eq.\  \er{lbK} follows from \er{boundsK}  in view of the fact that 
 \beq
 K(\sg\mid N)-K(\sg)= O(\log_2 N),
 \eeq as explained in the proof of Proposition \ref{114}; hence  \er{lbK} also holds for $K(\sg\mid N)$ instead of $K(\sg)$. 
 \smallskip
 
 \emph{Proof.} Theorem \ref{NCT} suggests looking at a program $p=p_{\sg}$ reaching the minimum in \er{defKC2} as an encoder of $\sg$, so that
 $\ell(p_{\sg})=K(\sg\mid N)$ is the length $C(\sg)$ of the codeword. That this suggestion is correct follows by noting that,\footnote{ Note that we use Lemma \ref{KraftIn} with $A\leadsto A^N$, $a\leadsto \sg$, and $p\leadsto p^N$.} being a subset of the prefix set of all $p$, the shortest $\sg$-producing programs $p_{\sg}$ thus defined form a prefix set in $2^*$  and hence their lengths satisfy 
 \beq
 \Sigma_{\sg\in A^N} e^{-\ell(p_{\sg})}= \Sigma_{\sg\in A^N} e^{-K(\sg\mid N)}
 \leq 1,
 \eeq 
 by the Kraft inequality \er{Kraft}.
 The lower bound in \er{boundsK} then follows from the lower bound in \er{boundsS}. 
 We  prove the upper bound first for $A=2$. In that case it follows from the estimate
 \begin{equation}
K(\sg\mid N)\leq NS_2\left(\frac{1}{N}\Sigma_{n=0}^{N-1}\sg_n\right) +\log_2 N+c.\label{kestimate}
\end{equation}
Granting this for the moment, we take the expectation value of this inequality under $p^N$ and use Jensen's inequality for the concave function $S_2$, followed by  $\la \sg_n\ra_{p^N}=\la\sg_0\ra_p=p$. This gives
\begin{equation}
\Sigma_{\sg\in A^N} p^N(\sg)S_2\left(\frac{1}{N}\Sigma_{n=0}^{N-1}\sg_n\right)\leq  S_2\left(\frac{1}{N}\Sigma_{n=0}^{N-1}\Sigma_{\sg\in A^N} p^N(\sg)\sg_n\right) =
S_2\left(\frac{1}{N}\Sigma_{n=0}^{N-1}p\right) =
S_2(p).
\end{equation}
To derive \er{kestimate}, take $\sg\in 2^N$ and let  $k=\Sigma_{n=0}^{N-1}\sg_n$, so that $\sg$ contains  $k$ copies of 1. Storing $k$ in a program $p$ (that knows $N$) takes $\log_2 k\leq \log_2 N$ bits. There are  $ 
\left(\begin{array}{c} N   \\  k   \end{array}\right)$ binary strings of length $N$ with $k$ copies of 1, which may be listed in some computable way.
 Identifying  $\sg$ from these possibilities by its list number
  requires $p$ to also store this list number, which is at most  $ 
\left(\begin{array}{c} N   \\  k   \end{array}\right)$ and hence this takes at most $\log_2  
\left(\begin{array}{c} N   \\  k   \end{array}\right)$ bits. Hence the length of $p$ is at most
\begin{equation}
\ell(p)\leq \log_2 N+ \log_2  
\left(\begin{array}{c} N   \\  k   \end{array}\right)+c'.
\end{equation}
Eq.\  \er{kestimate} then follows from the upper bound in \er{est1}, see exercise below. \QED
\bex
Explain that, in the notation of \er{est1},
 for $A=2$ we have \beq
 |T_N(k/N)|=\left(\begin{array}{c} N   \\  k   \end{array}\right).\eeq
On this basis, extend the proof to  finite sets $A$: 
 instead of the  number $\frac{1}{N}\Sigma_{n=0}^{N-1}\sg_n=\frac{k}{N}$, which defines an element 
 $p\in\Pr(2)$ via $p(1)=k/N$, use the empirical measure \er{LNb}, and instead of 
 $\left(\begin{array}{c} N   \\  k   \end{array}\right)$, use $|T_N(k/N)|$, cf.\ \er{TNp}.
\eex
Theorem \ref{Kentropy} describes the asymptotics of the \emph{average} Kolmogorov complexity of \emph{strings}. The pointwise ergodic theorem (in a computable context) eventually leads to a \emph{pointwise}
 version of \er{lbK} for \emph{sequences}.\footnote{See Towsner (2020)  and references therein.} 
 We first extend the notion of randomness for sequences $s\in A^\N$ to randomness with respect to a prior $p\in\Pr(A)$ and associated Bernoulli measure $p^\N$ (as we will see, the original definition then corresponds to the flat prior $p=f$). We refine Definition \ref{LC} to:\footnote{This notion of $p^\N$-randomness is a special case of a generalized Martin-L\"{o}f style randomness concept introduced for arbitrary computable probability spaces by Hertling \& Weihrauch (2003) and Hoyrup \& Rojas (2009). } 
 \begin{definition}\label{LCp}
 A sequence $s\in A^\N$ is $p^\N$-\emph{random} if 
 there is $c\in\N$ such that for all $N>1$, 
\begin{equation}
K(s_{|N})\geq -\log_2 p^N(s_{|N})
-c. \label{CLp}
\end{equation}
\end{definition}
For $A=2$ and $p=f$ this recovers \er{CL}, since $f^N(\sg)=|A|^{-N}$ for all $\sg\in A^N$.
For general (finite) $A$ and $p=f$, one has $K(s_{|N})\geq N\log_2(|A|) -c$, which is also correct since $N\log_2(|A|)$ is the length of $\sg\in A^N$ \emph{measured in bits} (as opposed to: $A$-valued digits).\footnote{It is also interesting to notice that even in the original case  $A=2$ and $p=f$, we \emph{could} have stated Definition \ref{LC} in the form \er{CLp}, which shows that even though Kolmogorov randomness at first sight appears to be an entirely non-probabilistic concept, the flat prior $p=f$ is somehow lurking in the background. }
\begin{definition}
A real $x\in\R$ is \emph{computable} if there is a computable function $f:\N\raw\Q$ such that
for each $n\in\N$,
\beq
x\in \left[\frac{f(n)-1}{n}, \frac{f(n)+1}{n}\right].
\eeq
A probability distribution  $p\in\Pr(A)$ is \emph{computable} if each $p(a)\in [0,1]$ is a computable real.\end{definition}
\begin{theorem}\label{Kentropy2}
For all computable  $p\in\Pr(A)$ and all $p^\N$-random sequences $s\in A^{\N}$ we have
\begin{equation}
 \lim_{N\raw\infty}\frac{1}{N} K(s_{|N})= S_2(p). \label{folklore} 
\end{equation}
\end{theorem}
This is  similar in spirit to the Shannon--McMillan--Breiman theorem \er{SMB}. The result can be clarified via the Asymptotic Equipartition Property, see \er{pone2} and subsequent text, which suggests that if $s$ is $p$-random, then $s_{|N}$ lies in the generic set whose elements all have probability  
\beq
p^N(s_{|N})\approx e^{-NS(p)}=2^{-NS_2(p)},
\eeq
so that \er{CLp} gives $K(s_{|N})\geq NS_2(p)-c$. This gives the  lower bound in \er{boundsK}, but now pointwise. The upper bound  was proved (for general $A$, as in the exercise) by replacing the empirical measure $L_N(\sg)\in\Pr(A)$ by its average $p$, which also works for $\sg=s_{|N}$ and $p$-random $s$. 

Theorems \ref{Kentropy} and \ref{Kentropy2}, then,  give a new perspective on the Shannon entropy $S_2(p)$ by relating it to the Kolmogorov complexity of strings produced by a random source distributed by $p$.
\newpage
\addcontentsline{toc}{section}{References}
\begin{small}

\end{small}
\end{document}